\documentclass[12pt,preprint]{aastex}
\usepackage{graphicx}
\usepackage{color}
\usepackage{float}
\usepackage{amsmath}

\RequirePackage{ifthen}
\newboolean{make_heavy}
\newboolean{make_very_heavy}

\setboolean{make_heavy}{true}
\setboolean{make_very_heavy}{false}

\tightenlines

\newcommand \RoneCone {}
\newcommand \RoneCtwo {}
\newcommand \RoneCthree {}
\newcommand \RtwoCone {}
\newcommand \RtwoCtwo {}
\newcommand \RtwoCthree {}
\newcommand \RthreeCone {}
\newcommand \RthreeCtwo {}
\newcommand \RthreeCthree{}
\newcommand \RfourCone {}
\newcommand \RfourCtwo {}
\newcommand \RfourCthree {}

\newcommand \FirstNormal {
\includegraphics[width=5.6cm,height=4.8cm,clip=true,trim=0.0cm 2.4cm 0.0cm 0.3cm]{\RoneCone}
\hspace*{-4.9cm}
\includegraphics[width=4.9cm,height=4.8cm,clip=true,trim=3.6cm 0.0cm 0.0cm 0.0cm]{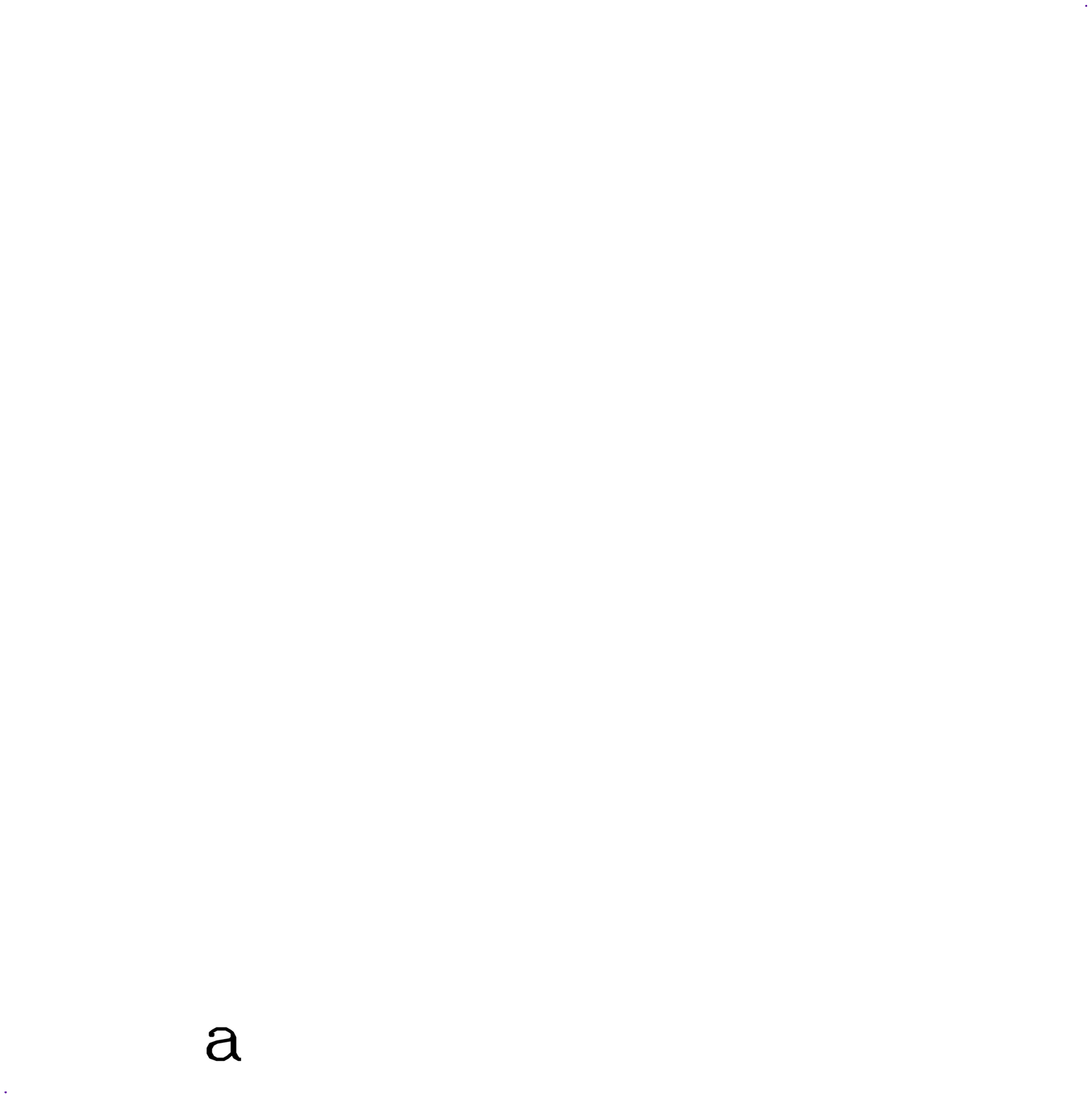}
&
\includegraphics[width=4.9cm,height=4.8cm,clip=true,trim=2.8cm 2.4cm 0.0cm 0.3cm]{\RoneCtwo}
\hspace*{-4.9cm}
\includegraphics[width=4.9cm,height=4.8cm,clip=true,trim=3.6cm 0.0cm 0.0cm 0.0cm]{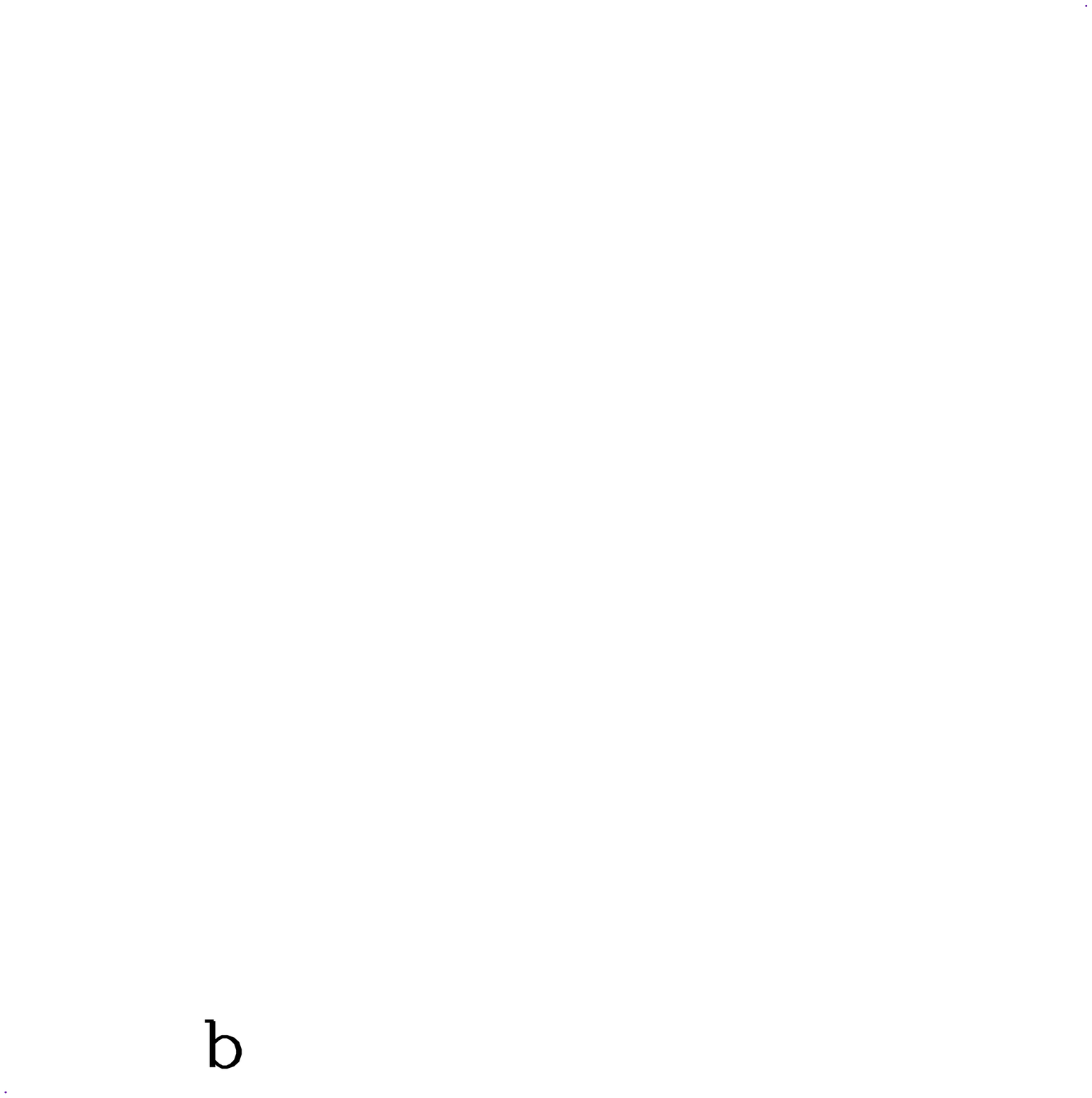}
& 
\includegraphics[width=4.9cm,height=4.8cm,clip=true,trim=2.8cm 2.4cm 0.0cm 0.3cm]{\RoneCthree}
\hspace*{-4.9cm}
\includegraphics[width=4.9cm,height=4.8cm,clip=true,trim=3.6cm 0.0cm 0.0cm 0.0cm]{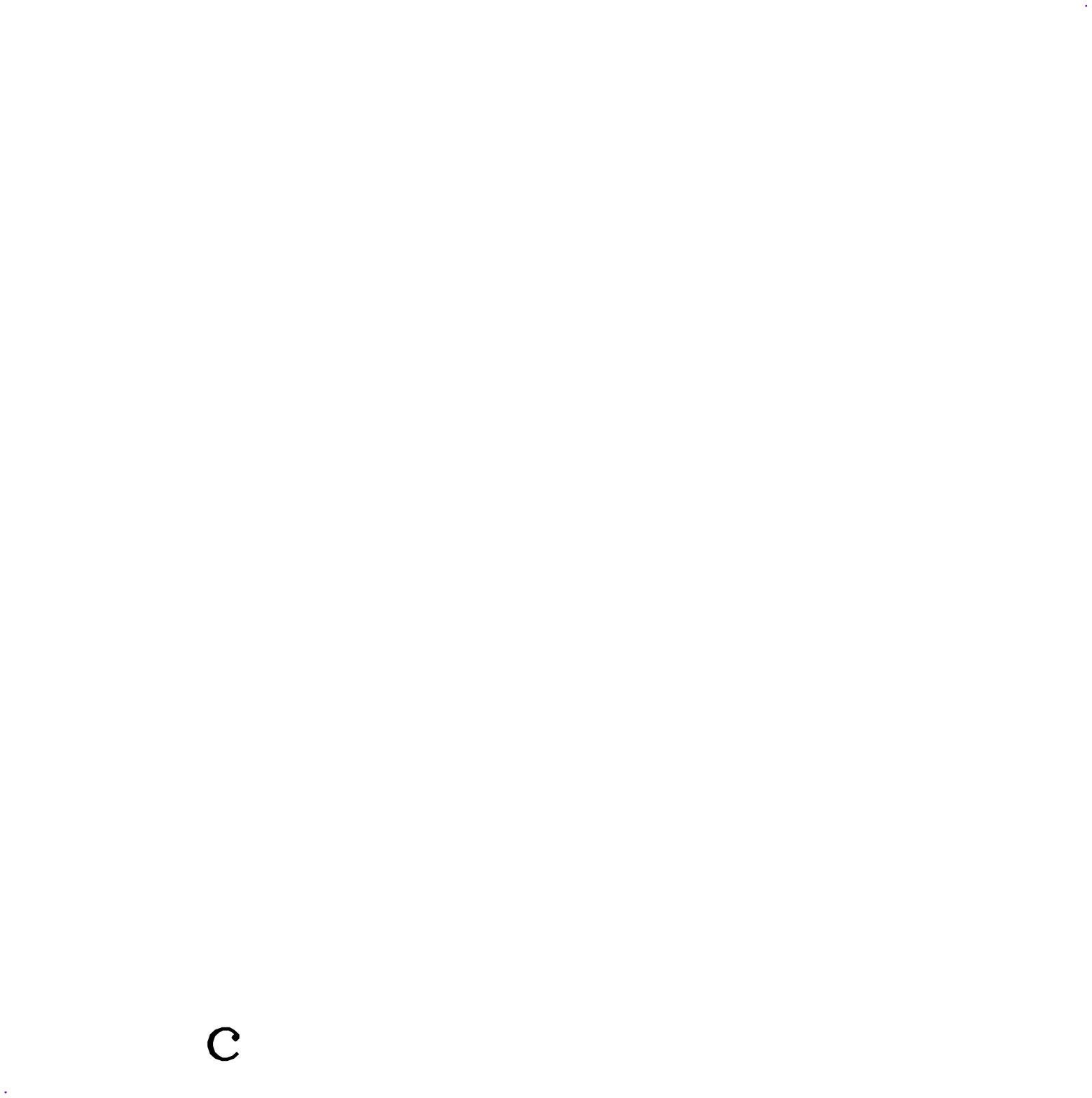}
\\ }

\newcommand \FirstLast {
\includegraphics[width=5.6cm,height=5.2cm,clip=true,trim=0.0cm 0.5cm 0.0cm 0.3cm]{\RoneCone}
\hspace*{-4.9cm}
\includegraphics[width=4.9cm,height=5.2cm,trim=3.6cm  -2.5cm 0.0cm 0.0cm]{labela.eps}
&
\includegraphics[width=4.9cm,height=5.2cm,clip=true,trim=2.8cm 0.5cm 0.0cm 0.3cm]{\RoneCtwo}
\hspace*{-4.9cm}
\includegraphics[width=4.9cm,height=5.2cm,trim=3.6cm  -2.5cm 0.0cm 0.0cm]{labelb.eps}
& 
\includegraphics[width=4.9cm,height=5.2cm,clip=true,trim=2.8cm 0.5cm 0.0cm 0.3cm]{\RoneCthree}
\hspace*{-4.9cm}
\includegraphics[width=4.9cm,height=5.2cm,trim=3.6cm  -2.5cm 0.0cm 0.0cm]{labelc.eps}
\\ }

\newcommand \SecondNormal {
\includegraphics[width=5.6cm,height=4.8cm,clip=true,trim=0.0cm 2.4cm 0.0cm 0.3cm]{\RtwoCone}
\hspace*{-4.9cm}
\includegraphics[width=4.9cm,height=4.8cm,trim=3.6cm 0.0cm 0.0cm 0.0cm]{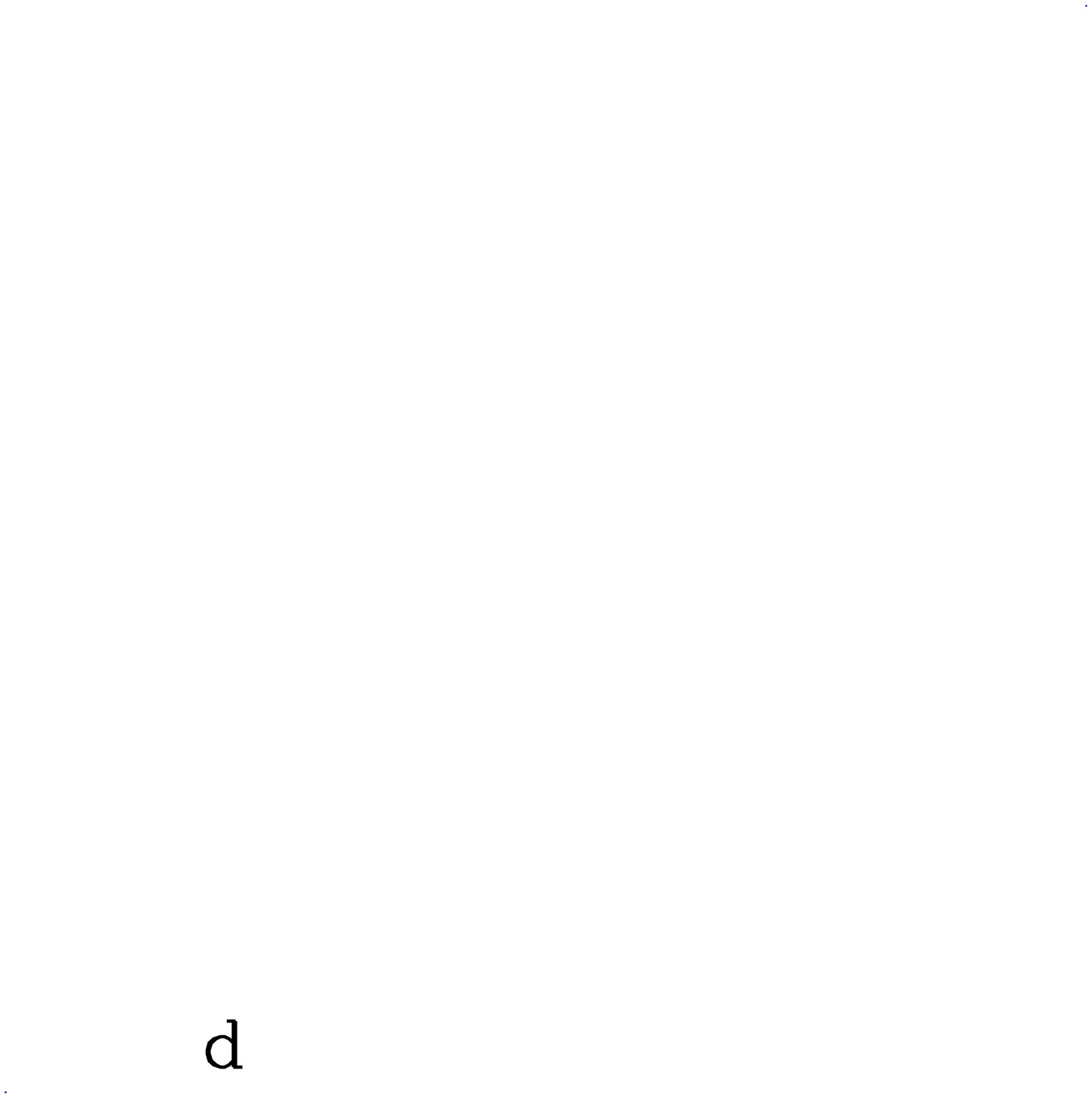}
&
\includegraphics[width=4.9cm,height=4.8cm,clip=true,trim=2.8cm 2.4cm 0.0cm 0.3cm]{\RtwoCtwo}
\hspace*{-4.9cm}
\includegraphics[width=4.9cm,height=4.8cm,trim=3.6cm 0.0cm 0.0cm 0.0cm]{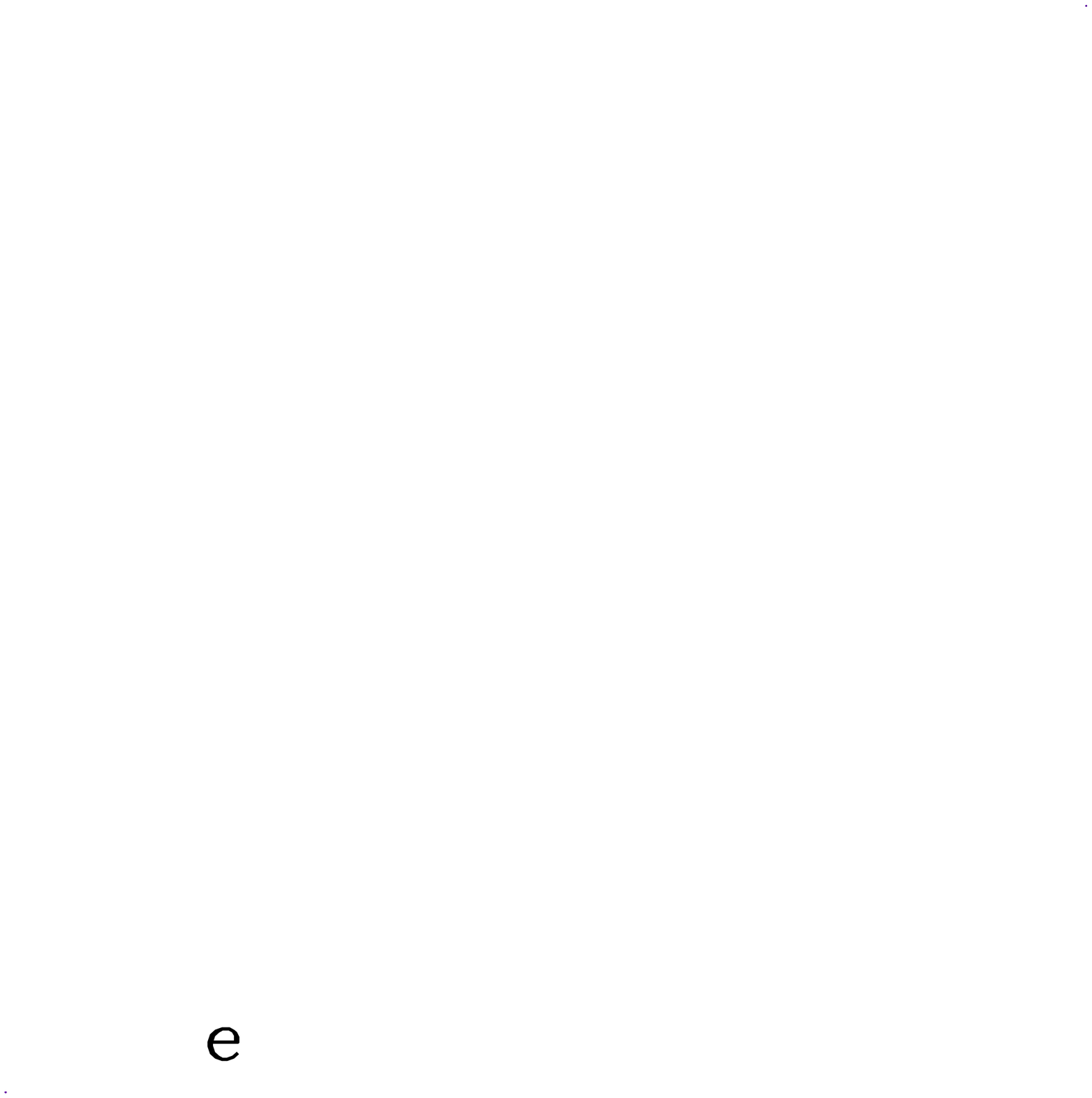}
& 
\includegraphics[width=4.9cm,height=4.8cm,clip=true,trim=2.8cm 2.4cm 0.0cm 0.3cm]{\RtwoCthree}
\hspace*{-4.9cm}
\includegraphics[width=4.9cm,height=4.8cm,trim=3.6cm 0.0cm 0.0cm 0.0cm]{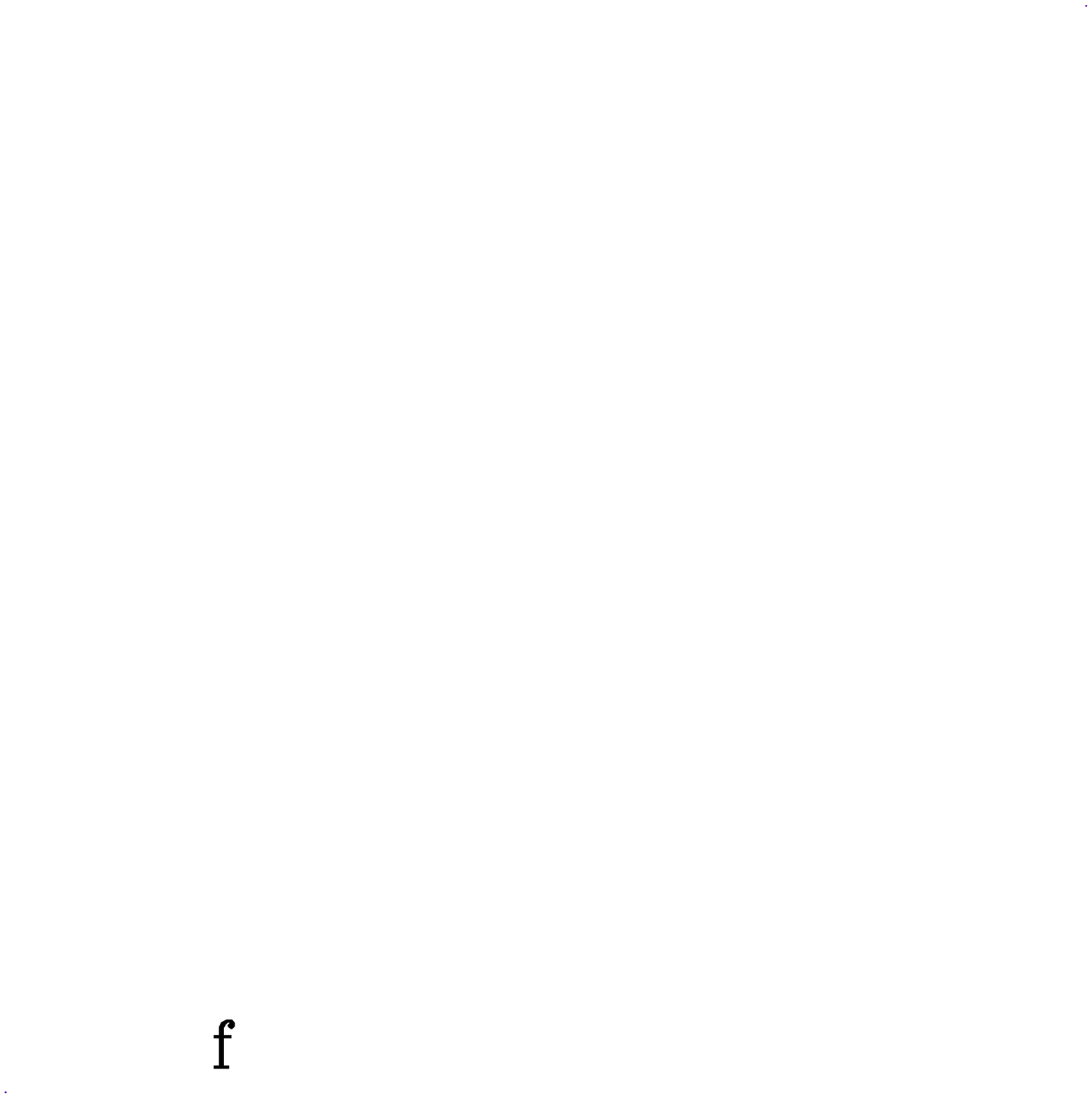}
\\ }

\newcommand \SecondLast {
\includegraphics[width=5.6cm,height=5.2cm,clip=true,trim=0.0cm 0.5cm 0.0cm 0.3cm]{\RtwoCone}
\hspace*{-4.9cm}
\includegraphics[width=4.9cm,height=5.2cm,trim=3.6cm -2.5cm 0.0cm 0.0cm]{labeld.eps}
&
\includegraphics[width=4.9cm,height=5.2cm,clip=true,trim=2.8cm 0.5cm 0.0cm 0.3cm]{\RtwoCtwo}
\hspace*{-4.9cm}
\includegraphics[width=4.9cm,height=5.2cm,trim=3.6cm -2.5cm 0.0cm 0.0cm]{labele.eps}
& 
\includegraphics[width=4.9cm,height=5.2cm,clip=true,trim=2.8cm 0.5cm 0.0cm 0.3cm]{\RtwoCthree}
\hspace*{-4.9cm}
\includegraphics[width=4.9cm,height=5.2cm,trim=3.6cm -2.5cm 0.0cm 0.0cm]{labelf.eps}
\\ }

\newcommand \ThirdNormal {
\includegraphics[width=5.6cm,height=4.8cm,clip=true,trim=0.0cm 2.4cm 0.0cm 0.3cm]{\RthreeCone}
\hspace*{-4.9cm}
\includegraphics[width=4.9cm,height=4.8cm,trim=3.6cm 0.0cm 0.0cm 0.0cm]{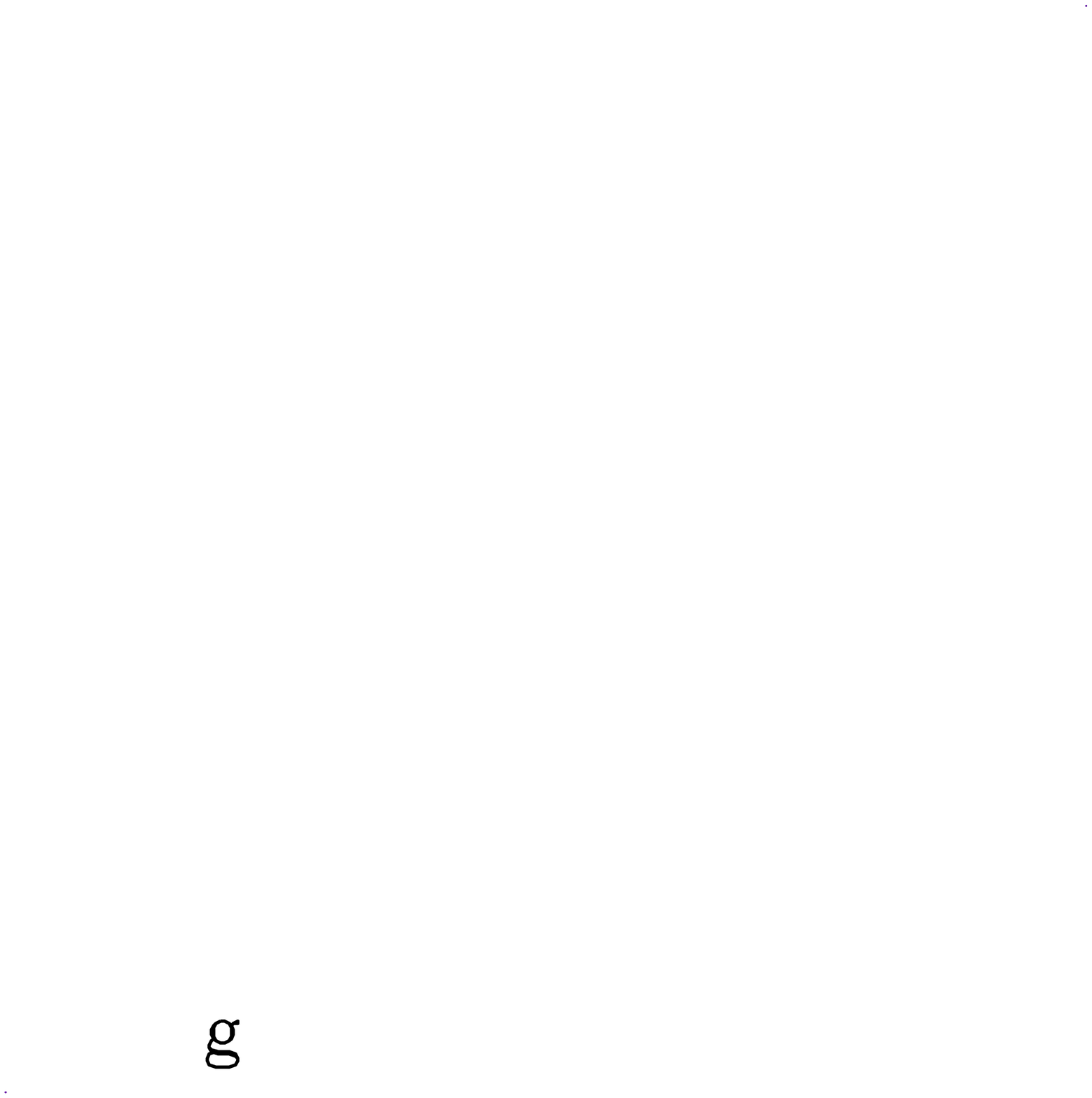}
&
\includegraphics[width=4.9cm,height=4.8cm,clip=true,trim=2.8cm 2.4cm 0.0cm 0.3cm]{\RthreeCtwo}
\hspace*{-4.9cm}
\includegraphics[width=4.9cm,height=4.8cm,trim=3.6cm 0.0cm 0.0cm 0.0cm]{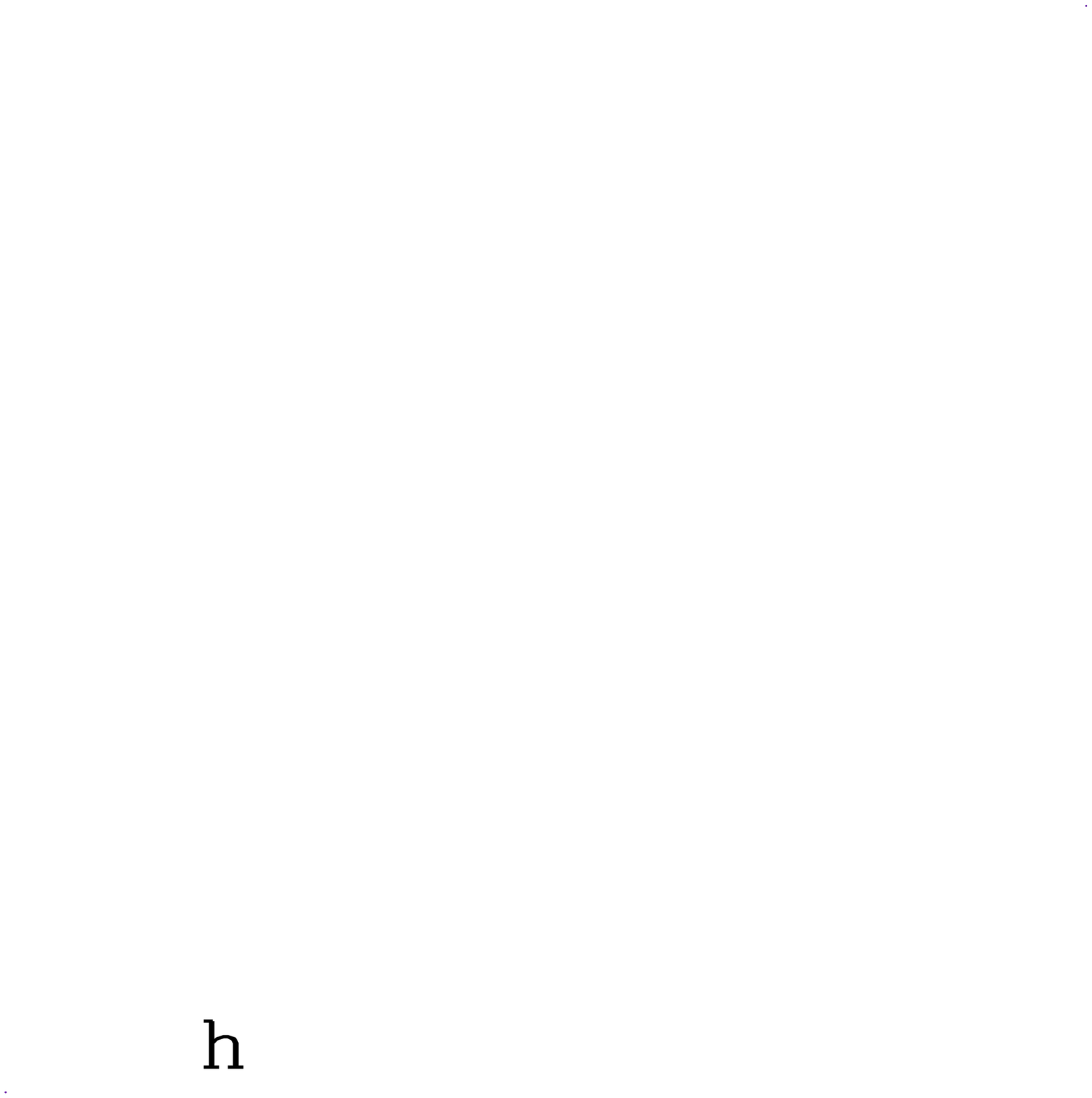}
& 
\includegraphics[width=4.9cm,height=4.8cm,clip=true,trim=2.8cm 2.4cm 0.0cm 0.3cm]{\RthreeCthree}
\hspace*{-4.9cm}
\includegraphics[width=4.9cm,height=4.8cm,trim=3.6cm 0.0cm 0.0cm 0.0cm]{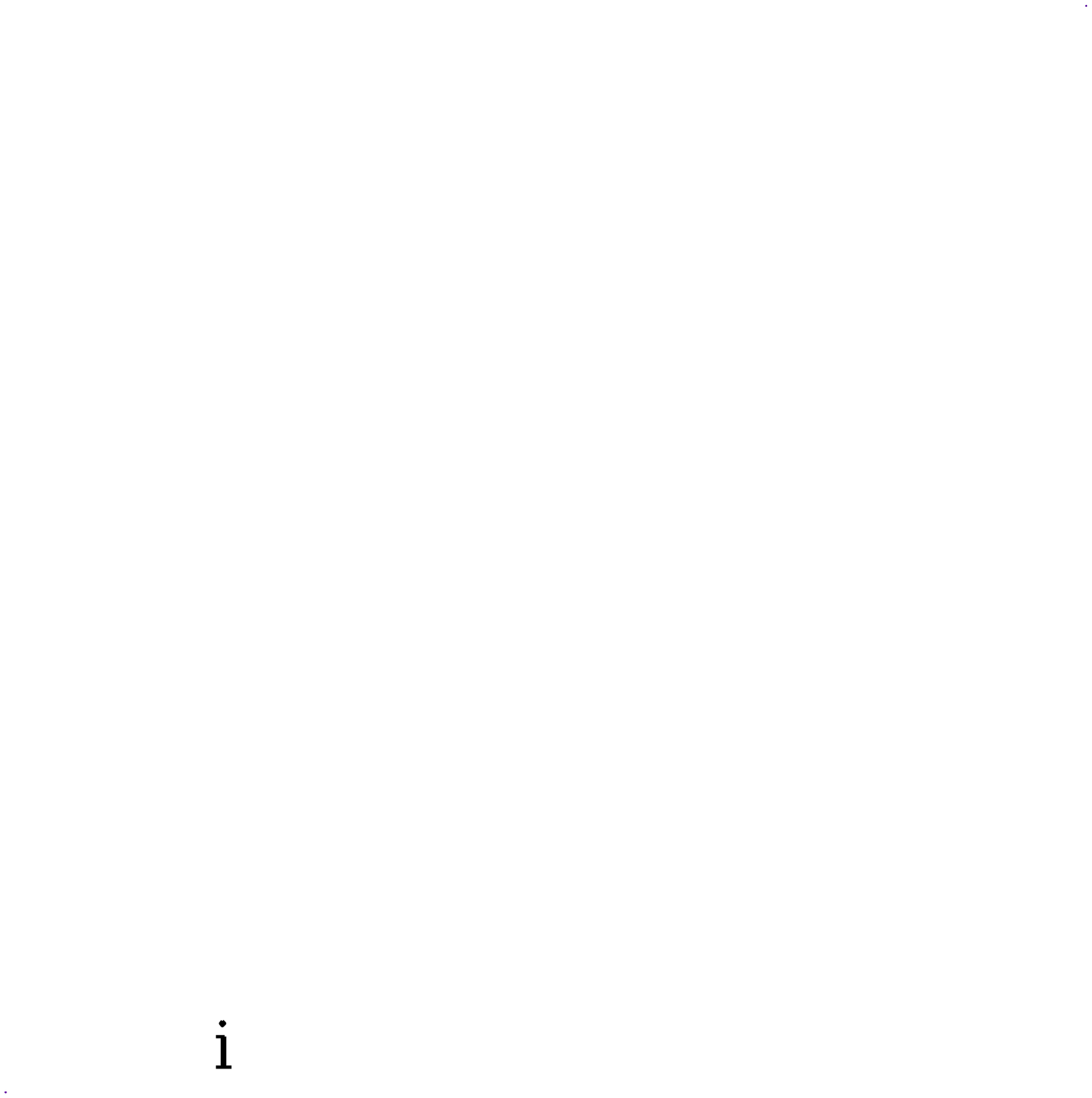}
\\ }

\newcommand \ThirdLast {
\includegraphics[width=5.6cm,height=5.2cm,clip=true,trim=0.0cm 0.5cm 0.0cm 0.3cm]{\RthreeCone}
\hspace*{-4.9cm}
\includegraphics[width=4.9cm,height=5.2cm,trim=3.6cm -2.5cm 0.0cm 0.0cm]{labelg.eps}
&
\includegraphics[width=4.9cm,height=5.2cm,clip=true,trim=2.8cm 0.5cm 0.0cm 0.3cm]{\RthreeCtwo}
\hspace*{-4.9cm}
\includegraphics[width=4.9cm,height=5.2cm,trim=3.6cm -2.5cm 0.0cm 0.0cm]{labelh.eps}
& 
\includegraphics[width=4.9cm,height=5.2cm,clip=true,trim=2.8cm 0.5cm 0.0cm 0.3cm]{\RthreeCthree}
\hspace*{-4.9cm}
\includegraphics[width=4.9cm,height=5.2cm,trim=3.6cm -2.5cm 0.0cm 0.0cm]{labeli.eps}
\\ }

\newcommand \FourthLast {
\includegraphics[width=5.6cm,height=5.2cm,clip=true,trim=0.0cm 0.5cm 0.0cm 0.3cm]{\RfourCone}
\hspace*{-4.9cm} 
\includegraphics[width=4.9cm,height=5.2cm,trim=3.6cm 0.0cm 0.0cm 0.0cm]{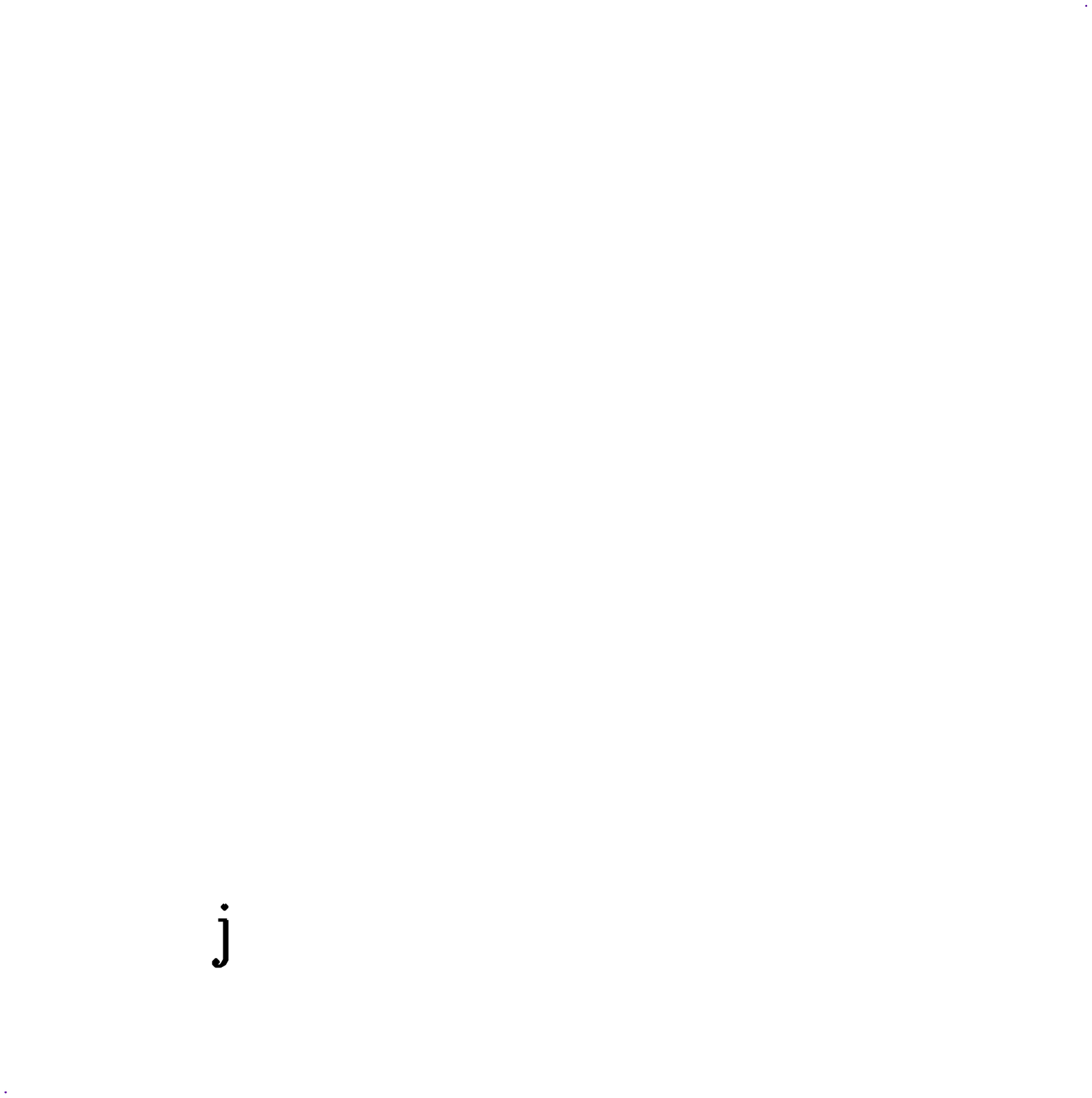}
&
\includegraphics[width=4.9cm,height=5.2cm,clip=true,trim=2.8cm 0.5cm 0.0cm 0.3cm]{\RfourCtwo}
\hspace*{-4.9cm}
\includegraphics[width=4.9cm,height=5.2cm,trim=3.6cm 0.0cm 0.0cm 0.0cm]{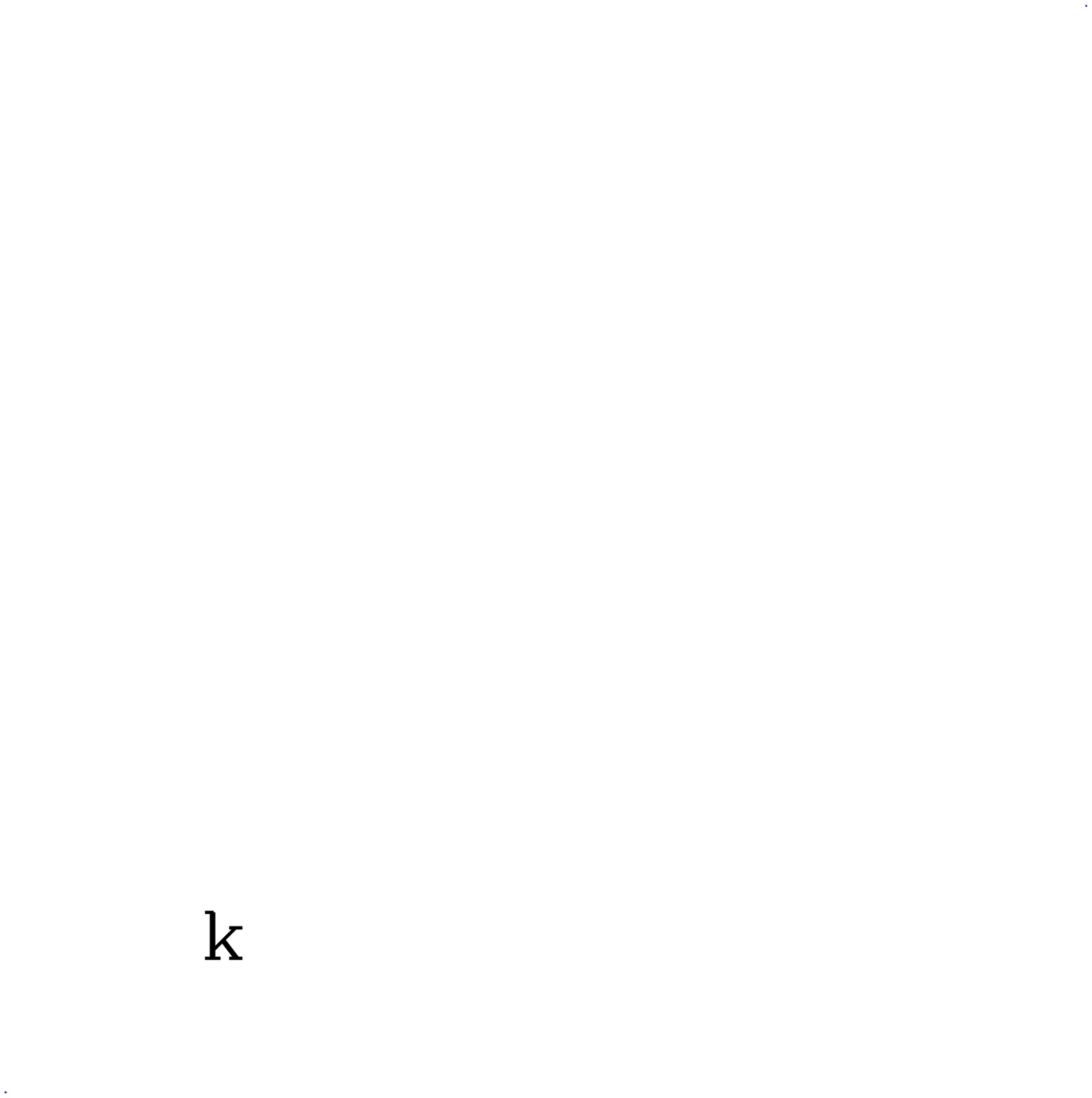}
& 
\includegraphics[width=4.9cm,height=5.2cm,clip=true,trim=2.8cm 0.5cm 0.0cm 0.3cm]{\RfourCthree}
\hspace*{-4.9cm}
\includegraphics[width=4.9cm,height=5.2cm,trim=3.6cm 0.0cm 0.0cm 0.0cm]{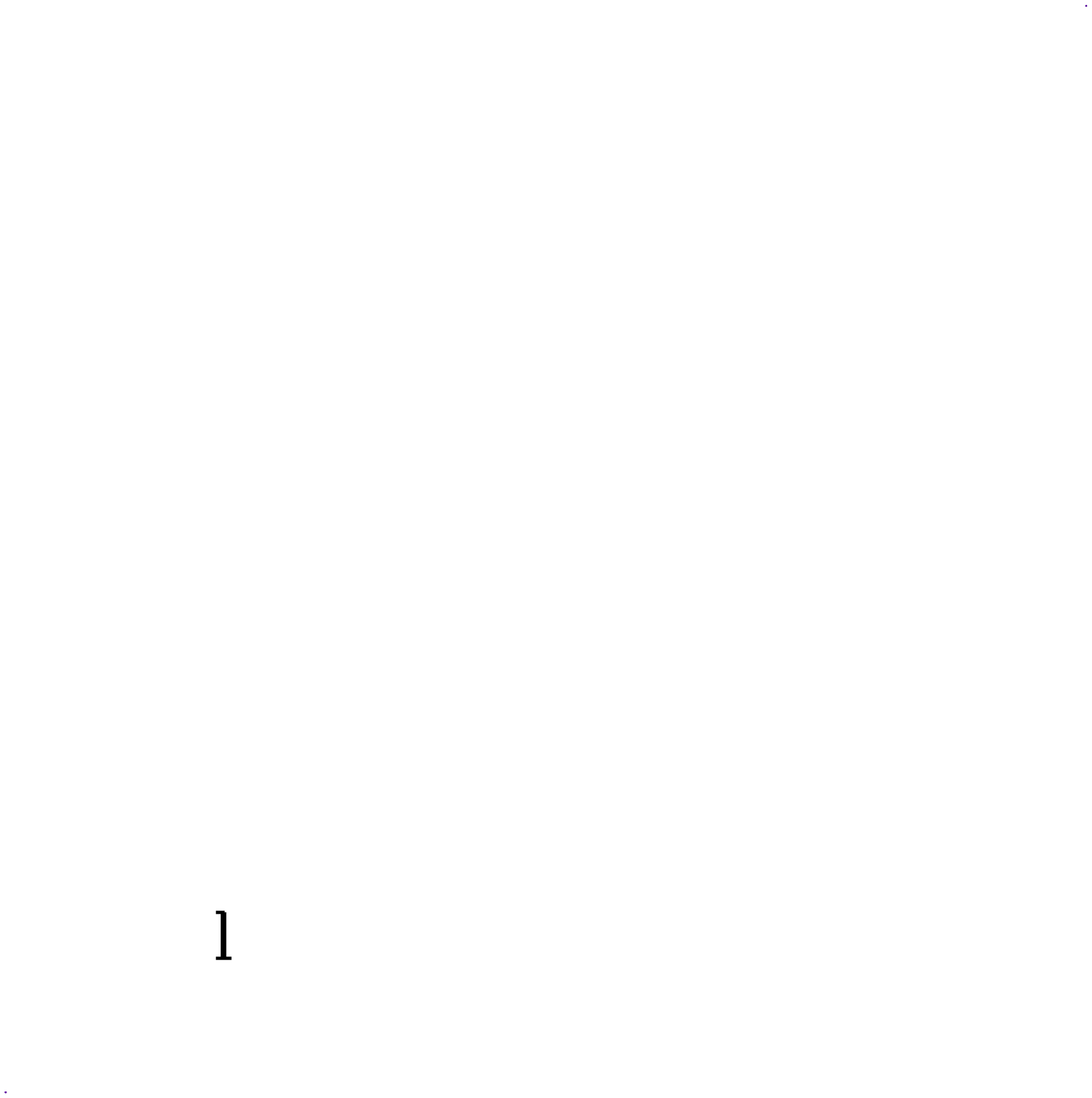}
\\ }

\newcommand \FirstNoTrim {
\includegraphics[width=5.6cm,height=4.8cm,clip=true,trim=0.0cm 2.4cm 0.0cm 0.3cm]{\RoneCone} &
\includegraphics[width=4.9cm,height=4.8cm,clip=true,trim=2.8cm 2.4cm 0.0cm 0.3cm]{\RoneCtwo}& 
\includegraphics[width=5.0cm,height=5.0cm]{\RoneCthree}\\ }

\newcommand \SecondNoTrim {
\includegraphics[width=5.6cm,height=4.8cm,clip=true,trim=0.0cm 2.4cm 0.0cm 0.3cm]{\RtwoCone} &
\includegraphics[width=4.9cm,height=4.8cm,clip=true,trim=2.8cm 2.4cm 0.0cm 0.3cm]{\RtwoCtwo}& 
\includegraphics[width=5.0cm,height=5.0cm]{\RtwoCthree}\\ }

\newcommand \ThirdNoTrim {
\includegraphics[width=5.6cm,height=4.8cm,clip=true,trim=0.0cm 2.4cm 0.0cm 0.3cm]{\RthreeCone} &
\includegraphics[width=4.9cm,height=4.8cm,clip=true,trim=2.8cm 2.4cm 0.0cm 0.3cm]{\RthreeCtwo}& 
\includegraphics[width=5.0cm,height=5.0cm]{\RthreeCthree}\\ }

\newcommand \FourthNoTrim {
\includegraphics[width=5.6cm,height=5.2cm,clip=true,trim=0.0cm 0.7cm 0.0cm 0.3cm]{\RfourCone} &
\includegraphics[width=4.9cm,height=5.2cm,clip=true,trim=2.8cm 0.7cm 0.0cm 0.3cm]{\RfourCtwo}& 
\includegraphics[width=5.0cm,height=5.0cm]{\RfourCthree}\\ }

\newcommand \FirstColor {
\includegraphics[width=5.6cm,height=4.8cm,clip=true,trim=0.0cm 2.4cm 0.0cm 0.3cm]{\RoneCone} &
\includegraphics[width=4.9cm,height=4.7cm,clip=true,trim=0cm 0cm 0.0cm 0.9cm]{\RoneCtwo}& 
\includegraphics[width=4.9cm,height=4.8cm,clip=true,trim=0cm 0cm 0.0cm 0.9cm]{\RoneCthree}\\ }

\newcommand \SecondColor {
\includegraphics[width=5.6cm,height=4.8cm,clip=true,trim=0.0cm 2.4cm 0.0cm 0.3cm]{\RtwoCone} &
\includegraphics[width=4.9cm,height=4.7cm,clip=true,trim=0cm 0cm 0.0cm 0.9cm]{\RtwoCtwo}& 
\includegraphics[width=4.9cm,height=4.7cm,clip=true,trim=0cm 0cm 0.0cm 0.9cm]{\RtwoCthree}\\ }

\newcommand \ThirdColor {
\includegraphics[width=5.6cm,height=4.8cm,clip=true,trim=0.0cm 2.4cm 0.0cm 0.3cm]{\RthreeCone} &
\includegraphics[width=4.9cm,height=4.7cm,clip=true,trim=0cm 0cm 0.0cm 0.9cm]{\RthreeCtwo}& 
\includegraphics[width=4.9cm,height=4.7cm,clip=true,trim=0cm 0cm 0.0cm 0.9cm]{\RthreeCthree}\\ }

\newcommand \FourthColor {
\includegraphics[width=5.6cm,height=5.2cm,clip=true,trim=0.0cm 0.5cm 0.0cm 0.3cm]{\RfourCone} &
\includegraphics[width=4.9cm,height=5.1cm,clip=true,trim=0cm 0cm 0.0cm 0.9cm]{\RfourCtwo}& 
\includegraphics[width=4.9cm,height=5.1cm,clip=true,trim=0cm 0cm 0.0cm 0.9cm]{\RfourCthree}\\ }

\newcommand     \um     {$\mu$m\,}        
\newcommand     \mum    {\,\mu{\rm m}\,}  
\newcommand     \as      {$^{\prime \prime}$}        
\newcommand     \mas    {^{\prime \prime}\,}  

\newcommand     \beq    {\begin{equation}}
\newcommand     \beqa   {\begin{eqnarray}}
\newcommand     \cm     {\,{\rm cm}}

\newcommand     \dust   {{\rm d}}
\newcommand     \eeq    {\end{equation}}
\newcommand     \eeqa   {\end{eqnarray}}

\newcommand     \fpdr   {f_{\rm PDR}}

\newcommand     \gtsim  {\gtrsim}                
\newcommand     \Ha     {{\rm H}}
\newcommand     \HI     {{\rm H~I}}

\newcommand     \HH     {{\rm H}_2}

\newcommand     \K              {\,{\rm K}}
\newcommand     \kms    {\,{\rm km~s}^{-1}}
\newcommand     \kpc    {\,{\rm kpc}}

\newcommand     \Lsol   {L_{\odot}}
\newcommand     \ltsim  {\lesssim}               
\renewcommand   \mho    {\,{\rm mho}}

\newcommand     \Mpc    {\,{\rm Mpc}}
\newcommand     \Msol   {M_{\odot}}

\newcommand     \Mdust  {M_{\rm dust}}

\newcommand     \model  {{\rm model}}

\newcommand     \NH     {N_{\rm H}}

\newcommand     \PAH    {{\rm PAH}}
\newcommand     \PDR    {{\rm PDR}}
\newcommand     \pc     {\,{\rm pc}}
\newcommand     \qpah   {q_{\rm PAH}}
\newcommand     \qpahj   {q_{{\rm PAH},j}}
\newcommand     \rand   {r}           

\newcommand     \s      {\,{\rm s}}

\newcommand     \taub   {\tau_{\rm bulk}}

\newcommand     \Ubar   {\overline{U}}
\newcommand     \UPDR   {U_{\rm PDR}}
\newcommand     \Umax   {U_{\rm max}}
\newcommand     \Umaxj   {U_{{\rm max},j}}
\newcommand     \Umin   {U_{\rm min}}
\newcommand     \Uminj  {U_{{\rm min},j}}
\newcommand     \XCO    {X_{\rm CO}}
\newcommand     \XCOxx  {X_{{\rm CO},20}}
\newcommand     \yr     {\,{\rm yr}}

\newlength{\figwidth}
\newlength{\figwidthw}
\newlength{\figwidthww}
\newlength{\figwidthd}
\begin{document}
\title{
       Modeling Dust and Starlight in Galaxies
       Observed by Spitzer and Herschel:
       NGC~628 and NGC~6946.
       }

\slugcomment{ This ``light'' version of the manuscript lack most of the figures, please download the full version at www.astro.princeton.edu/$\sim$ganiano/Papers . Manuscript accepted by ApJ, to be publish in the September 2012 issue.}

\author{
G. Aniano\altaffilmark{1}, 
B.~T.~Draine\altaffilmark{1}, 
D. Calzetti\altaffilmark{2}, 
D. A. Dale\altaffilmark{3}, 
C. W. Engelbracht\altaffilmark{4}, 
K. D. Gordon\altaffilmark{5}, 
L. K. Hunt\altaffilmark{6}, 
R. C. Kennicutt\altaffilmark{7}, 
O. Krause\altaffilmark{8}, 
A. K. Leroy\altaffilmark{9}, 
H-W. Rix\altaffilmark{8}, 
H. Roussel\altaffilmark{10}, 
K. Sandstrom\altaffilmark{8}, 
M. Sauvage\altaffilmark{11}, 
F. Walter\altaffilmark{8}, 
L. Armus\altaffilmark{12}, 
A. D. Bolatto\altaffilmark{13}, 
A. Crocker\altaffilmark{2}, 
J. Donovan Meyer\altaffilmark{14}, 
M. Galametz\altaffilmark{7}, 
G. Helou\altaffilmark{15}, 
J. Hinz\altaffilmark{4}, 
B. D. Johnson\altaffilmark{10}, 
J. Koda\altaffilmark{14}, 
E. Montiel\altaffilmark{4}, 
E. J. Murphy\altaffilmark{16}, 
R. Skibba\altaffilmark{4}, 
J.-D.T. Smith\altaffilmark{17},
M. G. Wolfire\altaffilmark{13}}

  
\altaffiltext{1}{%
Princeton University Observatory, Peyton Hall, Princeton, NJ 08544-1001,USA; ganiano@astro.princeton.edu, draine@astro.princeton.edu}
\altaffiltext{2}{%
Department of Astronomy, University of Massachusetts, Amherst, MA 01003, USA} 
\altaffiltext{3}{%
Department of Physics and Astronomy, University of Wyoming, Laramie, WY 82071, USA} 
\altaffiltext{4}{%
Steward Observatory, University of Arizona, Tucson, AZ 85721, USA} 
\altaffiltext{5}{%
Space Telescope Science Institute, 3700 San Martin Drive, Baltimore, MD 21218, USA} 
\altaffiltext{6}{%
INAF - Osservatorio Astrofisico di Arcetri, Largo E. Fermi 5, 50125 Firenze, Italy} 
\altaffiltext{7}{%
Institute of Astronomy, University of Cambridge, Madingley Road, Cambridge CB3 0HA, UK} 
\altaffiltext{8}{%
Max-Planck-Institut fur Astronomie, Konigstuhl 17, D-69117 Heidelberg, Germany} 
\altaffiltext{9}{%
National Radio Astronomy Observatory, 520 Edgemont Road, Charlottesville, VA 22903, USA} 
\altaffiltext{10}{%
Institut dÕAstrophysique de Paris, UMR7095 CNRS, Universit«e Pierre and Marie Curie, 98 bis Boulevard 
Arago, 75014 Paris, France} 
\altaffiltext{11}{%
CEA/DSM/DAPNIA/Service dÕAstrophysique, UMR AIM, CE Saclay, 91191 Gif sur Yvette Cedex} 
\altaffiltext{12}{%
Spitzer Science Center, California Institute of Technology, MC 314-6, Pasadena, CA 91125, USA} 
\altaffiltext{13}{%
Department of Astronomy, University of Maryland, College Park, MD 20742, USA} 
\altaffiltext{14}{%
Department of Physics and Astronomy, SUNY Stony Brook, Stony Brook, NY 11794-3800, USA} 
\altaffiltext{15}{%
NASA Herschel Science Center, IPAC, California Institute of Technology, Pasadena, CA 91125, USA} 
\altaffiltext{16}{%
Observatories of the Carnegie Institution for Science, 813 Santa Barbara Street, Pasadena, CA 91101, USA} 
\altaffiltext{17}{%
Department of Physics and Astronomy, University of Toledo, Toledo, OH 43606, USA} 



\begin{abstract}

We characterize the dust
in NGC\,628 and NGC\,6946, two nearby spiral galaxies in the
KINGFISH sample.
With data from 3.6\um\ to 500\um,
dust models are strongly constrained.
Using the \citet{Draine+Li_2007} dust model,
(amorphous silicate and carbonaceous grains),
for each pixel in each galaxy we estimate
(1) dust mass surface density,
(2) dust mass fraction 
contributed by polycyclic aromatic hydrocarbons (PAH)s, 
(3) distribution of starlight intensities heating the dust,
(4) total infrared (IR) luminosity emitted by the dust,
and 
(5) IR luminosity originating in regions with high starlight intensity.
We obtain maps for the dust properties, which trace the
spiral structure of the galaxies.  The dust models successfully
reproduce the observed global and resolved spectral energy
distributions (SEDs).
The overall dust/H mass ratio is
estimated to be $0.0082\pm0.0017$ for NGC~628, and $0.0063\pm0.0009$
for NGC~6946, consistent with what is expected for galaxies of
near-solar metallicity.  Our derived dust masses are larger (by up to
a factor 3) than estimates based on single-temperature modified
blackbody fits.  We show that the SED fits
are significantly improved if the starlight intensity distribution includes a
(single intensity) ``delta function'' component.  We find no
evidence for significant masses of cold dust $(T\lesssim12\rm{K})$.
Discrepancies between 
PACS and MIPS photometry in both low and high surface brightness areas
result in large uncertainties when the modeling is done at PACS
resolutions, in which case SPIRE, MIPS70 and MIPS160 data cannot
be used. We recommend against attempting to model dust 
at the angular resolution of PACS.

\end{abstract}

\keywords{ISM: dust, extinction ---
          ISM: general ---
	  galaxies: abundances ---
	  galaxies: general ---
	  galaxies: ISM ---
          infrared: galaxies
	  }


\section{\label{sec:intro}Introduction}

Interstellar dust affects the appearance of galaxies,
by attenuating short wavelength radiation from stars and ionized gas, and
contributing IR, 
submm, mm, and microwave emission
[for a review, see \citet{Draine_2003a}].
Dust is also an important agent in the
fluid dynamics, chemistry, heating, cooling, and even ionization balance
in some interstellar regions,
with a major role in the process of star formation.
Despite the importance of dust, determination of the physical properties of
interstellar dust grains has been a challenging task -- even the overall amount
of dust in other galaxies has often been very uncertain.

Many previous studies have used far-infrared photometry to estimate
the dust properties of galaxies.
For example, 
\citet{Draine+Dale+Bendo+etal_2007}
used global photometry of 65 galaxies in the ``Spitzer Infrared
Nearby Galaxies Survey'' (SINGS) galaxies 
to estimate the total dust mass and PAH abundance in each galaxy, and
to characterize the intensity of the starlight heating the dust.
For most of these galaxies the photometry
extended only to 160$\micron$, although ground-based
global photometry at 850$\micron$ was also available
for 17 of the 65 galaxies.\footnote{Photometry at 850$\micron$ was available for 26 galaxies in the SINGS sample, but the data were only reliable (and used) for 17 galaxies.}
\citet{Munoz-Mateos+Gil_de_Paz+Boissier+etal_2009} used images of the
SINGS galaxies to examine the radial distribution of the dust surface
density. 
\citet{Sandstrom+Bolatto+Draine+etal_2010} studied the PAHs in the Small Magellanic Cloud SMC on a pixel-by-pixel basis with a very similar dust model as the present work.
Very recently, \citet{Totani+Takeuchi+Nagashima+etal_2011} 
used global photometry in 6 bands -- 9, 18, 65, 90,
140, and 160$\micron$ -- to estimate dust masses for a sample of more than
1600 galaxies in the Akari All Sky Survey
\citep{Ishihara+Onaka+Kataza+etal_2010,
       Yamamura+Makiuti+Ikeda+etal_2010}. 
In the present work, we develop state-of-the-art image processing and dust modeling techniques aiming to reliably determine the dust properties in resolved studies.

The present study makes use of combined imaging by the Spitzer Space Telescope
and Herschel Space Observatory, covering wavelengths from 3.6$\micron$
to 500$\micron$, to produce well-resolved maps of the dust in two nearby
galaxies.  The present study is focused on two well-resolved spiral galaxies, NGC~628 and
NGC~6946, as examples to illustrate the methodology.   
Subsequent work
(Aniano et al.\ 2012, in preparation)
will extend this analysis to all 61 galaxies in the KINGFISH sample
\citep{Kennicutt+Calzetti+Aniano+etal_2011}.

The paper is organized as follows.
In \S \ref{sec:datasources} we
discuss the data sources.
Background subtraction and data processing
are discussed in \S \ref{sec:dataproc}.
The dust model is summarized in \S \ref{sec:dustmodel}, and 
the fitting procedure 
is described in \S \ref{sec:SEDFIT}.
Results for NGC\,628 and NGC\,6946 are given in \S \ref{sec:results}.
The sensitivity of the derived parameters to the set of cameras
used to constrain the dust models is explored in
\S \ref{sec:importance of SPIRE}, where we compare dust mass estimates
obtained at high spatial resolution (without using MIPS160 or the 
longest-wavelength SPIRE bands) with estimates made at lower spatial
resolution (using all the cameras available).
We also investigate the reliability of the photometry 
by comparing MIPS70 and MIPS160 images with PACS70 and PACS160 images.
In \S \ref{sec:maps vs global} we compare dust mass estimates
based on spatially-resolved images with dust mass estimates based on
global photometry, as would apply to distant, unresolved galaxies.
The ``goodness of fit'' of different dust models is discussed in \S 9.
The principal results are discussed in \S 10 and
summarized in \S \ref{sec:summary}.

Appendices A-D describe the method for image segmentation (i.e., galaxy and background recognition), background subtraction, estimation of photometric uncertainties in the images and estimation of dust modeling uncertainties. Appendix E is a comparison of PACS and MIPS photometry.


\section{\label{sec:datasources}Observations and Data Reduction}

NGC~628 and NGC~6946 are part of the SINGS galaxy sample 
\citep{Kennicutt+Armus+Bendo+etal_2003}, and were imaged by the Spitzer Space
Telescope \citep{Werner+Roellig+Low+etal_2004}.
A large subset of the SINGS galaxies are also included in the Herschel Space
Observatory Open Time Key Project KINGFISH
\citep{Kennicutt+Calzetti+Aniano+etal_2011} and
were observed with the Herschel Space Observatory \citep{Pilbratt+Riedinger+Passvogel+etal_2010}.

We will use ``camera'' to identify each optical configuration of the observing instruments, i.e., each different channel or filter arrangement of the instruments will be referred to as different ``camera''. With this nomenclature, each ``camera'' has a characteristic optical resolution, spectral response, and point spread function (PSF).\footnote{For example, the 70\um and 100\um channels of the PACS instrument use the same optical path in the telescope, but differ in the filter used, and thus have different PSFs, so we will consider them as different ``cameras''.}
We will refer to the IRAC, MIPS, PACS and SPIRE
cameras using their nominal wavelengths
in microns: IRAC3.6, IRAC4.5, IRAC5.8, IRAC8.0, 
MIPS24, MIPS70, MIPS160, PACS70, PACS100, PACS160,
SPIRE250, SPIRE350, and SPIRE500.
For our standard modeling, the
PACS and SPIRE data were reduced by HIPE followed by 
Scanamorphos (see Subsection \ref{PACS} for details).
In Appendix E, and Table 6, where we compare PACS data with and without
Scanamorphos processing,
we use PACS(H)70, PACS(H)100, PACS(H)160 to denote PACS data that was
processed by the HIPE pipeline only. 

Table 1 summarizes the optical resolution of the cameras.


\subsection{Spitzer}

The Infrared Array Camera (IRAC) and the Multiband Imaging Photometer for {\it Spitzer} (MIPS) cameras on the Spitzer Space Telescope were used to observe all 75 galaxies in the SINGS sample, including NGC~628 and NGC~6946, following the observing strategy described by \citet{Kennicutt+Armus+Bendo+etal_2003}. 
Spectroscopic observations of selected regions were also obtained, although not used in the current study. 
Aniano, Reyes, Draine et al (2012, in preparation; hereafter ARD12)
use the spectroscopic observations of selected regions to further constrain the dust modeling.

\subsubsection{IRAC}

IRAC \citep{Fazio+Hora+Allen+etal_2004}
imaged the galaxies in 4 bands, centered at 3.6\um,
4.5\um, 5.8\um and 8.0\um.  The images were processed by the SINGS
Fifth Data Delivery pipeline.\footnote{Details can be found in the data release documentation, 
  http://data.spitzer.caltech.edu/popular/sings/20070410-enhanced-v1/Documents/sings-fifth-delivery-v2.pdf}
The IRAC images are calibrated for point sources.
Photometry of extended sources requires so-called ``aperture corrections''.
We multiply the intensities in each pixel by the asymptotic (infinite radii) value of the aperture correction (i.e., the aperture correction corresponding to an infinite radius aperture). We use the factors 0.91, 0.94, 0.66 and 0.74 for the 3.6\um, 4.5\um, 5.8\um and
8.0\um bands, respectively, as described in the IRAC Instrument
Handbook (V2.0.1)\footnote{
http://irsa.ipac.caltech.edu/data/SPITZER/docs/irac/iracinstrumenthandbook/IRAC\_Instrument\_Handbook.pdf}.

\subsubsection{MIPS}

Imaging with MIPS
\citep{Rieke+Young+Engelbracht+etal_2004} was carried out
following the observing strategy described in
\citet{Kennicutt+Armus+Bendo+etal_2003}.  The data were reduced using
the LVL (Local Volume Legacy) project pipeline.\footnote{Details can
  be found in the data release documentation,
  http://data.spitzer.caltech.edu/popular/lvl/20090227-enhanced/docs/LVL-DR3-v1.pdf}
A correction for nonlinearities in the MIPS70 camera was applied by
the team, as described by \citet{Dale+Cohen+Johnson+etal_2009} and \citet{Gordon+Meixner+Meade+etal_2011}.


\subsection{Herschel}

The Photodetector Array Camera and Spectrometer for {\it Herschel} (PACS) and the Spectral and Photometric Imaging Receiver (SPIRE) cameras on the Herschel Space Observatory are being used to observe the 61 galaxies in the KINGFISH sample, in particular NGC~628 and NGC~6946, following the observing strategy described by
\citet{Kennicutt+Calzetti+Aniano+etal_2011}. 
The maps were designed to cover a region out to $\gtsim 1.5$ times the optical radius $R_{opt}$, with good signal-to-noise (S/N) and redundancy. The depth of the PACS images at 70\um and 160\um is less than that of MIPS, but the higher resolution of PACS is able to better single out compact star-forming regions.

\subsubsection{PACS\label{PACS}}

NGC~628 and NGC~6946 were observed with the PACS instrument on Herschel
\citep{Poglitsch+Waelkens+Geis+etal_2010} on 2010 Jan.\ 28 (NGC~628)
and Mar.\ 10 (NGC~6946), using the ``Scan Map'' observation mode.  
The PACS images were first reduced to ``level 1'' (flux-calibrated brightness time series, with attached sky coordinates) using HIPE \citep{Ott_2010} version  5.0.0, and maps (``level 2'') were created using  the Scanamorphos data reduction
pipeline \citep{Roussel_2012}, version 16.9. 
This reduction strategy includes the latest PACS calibration available \citep{Muller+Nielbock+Balog+etal_2011}, and aims to preserve the low surface brightness diffuse emission.

Additionally, PACS data was reduced completely (i.e., to ``level 2'') using HIPE.
We used HIPE version 5.0.0, and further divide the fluxes by 1.119, 1.151, and 1.174 for the cameras PACS70, PACS100, and PACS160 respectively to account for the latest calibration of the cameras.
Both PACS data reduction strategies present strong discrepancies with the corresponding MIPS photometry, as is discussed in \S \ref{sec:importance of MIPS}. The HIPE pipeline removes a large fraction of the flux in the low surface brightness areas, and in our work it was only used for comparison, i.e., we found (Appendix F) that the Scanamorphos pipeline is more reliable, and we only employ the Scanamorphos reductions when PACS data are used in the dust modeling.

\subsubsection{SPIRE}

The 2 galaxies were observed with the SPIRE instrument
\citep{Griffin+Abergel+Abreu+etal_2010}
on 2009 Dec. 31 (NGC~6946) and 2010 Jan.\ 18 (NGC~628).
The data were first reduced to ``level 1'' using HIPE 
version spire-8.0.3287, followed by Scanamorphos version 17.0\footnote{See also ftp://hsa.esac.esa.int/URD$\_$rep/KINGFISH$\_$DR1$\_$SPIRE/}. 
The assumed beam sizes are 435.7,  773.5, and 1634.6$\,{\rm arcsec}^2$ for SPIRE250, SPRE350, and SPIRE500, respectively.  
Additionally, we excluded discrepant bolometers from the map, and adjusted the pointing to match the
MIPS24 map. Data reduction details can be found in \citet{Kennicutt+Calzetti+Aniano+etal_2011}. 


\subsection{Atomic and molecular gas\label{sec:gas}}

NGC~628 and NGC~6946 are in The \ion{H}{1} Nearby Galaxy Survey 
\citep[THINGS;][]{Walter+Brinks+deBlok+etal_2008}, 
with $\sim$$6\arcsec$ resolution
\ion{H}{1} 21-cm imaging by the NRAO Very Large Array.
The \ion{H}{1} 21-cm emission is assumed to be optically-thin, in which
case the \ion{H}{1} surface density is directly proportional to the
21-cm line intensity.

The CO$\,J = 2 \rightarrow1$ transition was observed with
$\sim$$13\arcsec$ angular resolution by the HERA CO
Line Extragalactic Survey 
\citep[HERACLES;][]{Leroy+Walter+Bigiel+etal_2009} using the HERA multipixel
receiver on the IRAM 30-m telescope, with estimated uncertainties of $\pm20\%$.

As is usual, the $\HH$ mass surface density is taken to be proportional to the CO$\,J = 2 \rightarrow1$ line intensity.
$\XCO$ is the ratio of $\HH$ column density to 
CO$\,J = 1 \rightarrow0$ intensity integrated over the line profile.
In what follows, we define
\beq
\XCOxx\equiv \frac{\XCO}{10^{20}\,{\rm H}_2\cm^{-2}\,(\K\kms)^{-1}},
\eeq
and we assume a $J\!=\!2\!\rightarrow1/J\!=\!1\!\rightarrow0$ antenna temperature ratio $R_{21}=0.8$ \citep{Leroy+Walter+Bigiel+etal_2009}.

The conversion factor $\XCO$ is uncertain.  $\XCOxx\approx2$ is the
value normally adopted for molecular gas in the Milky Way
\citep{Dame+Hartmann+Thaddeus_2001,Okumura+Kamae+FLAT_2009}.
\citet{Planck_dust_2011} found $\XCOxx=2.54\pm0.13$ for the Milky Way.
\citet{Draine+Dale+Bendo+etal_2007} found that $\XCOxx\approx4$
appeared to give the most reasonable dust/H mass ratios for
the SINGS galaxy sample.  \citet{Blitz+Fukui+Kawamura+etal_2007} found
$\XCOxx\approx4$ to be the best overall value for galaxies in the
Local Group.  \citet{Leroy+Bolatto+Gordon+etal_2011} found $\XCOxx =
1.2\,-\,4.2$ for M~31, M~33, and the Large Magellanic Cloud (LMC), and
the very high values $\XCOxx = 14$ and $\XCOxx = 32$ for NGC~6822 and
the Small Magellanic Cloud (SMC).  In the present work, we take
$\XCOxx=4\pm1$ to be the best overall value for NGC~628 and NGC~6946.
In \S \ref{sec:results} (Fig.\ \ref{fig:Xco}) we show the H gas maps
and dust/H mass ratio for $\XCOxx=2,\,3,\,4$.

The \ion{H}{1} and $\HH$ masses are added to generate maps of the total H surface density $\Sigma_{\rm{H}}$. 
Including Helium would give   $\Sigma_{\rm{gas}} \approx 1.38 \times \Sigma_{\rm{H}}$, but in our discussion will use  $\Sigma_{\rm{H}}$ as it is the ``observable'', avoiding uncertainties in the Helium abundance.


\section{\label{sec:dataproc}Image Analysis}

Before modeling the spatially resolved SEDs, it is necessary to adjust the images in a number of ways to ensure meaningful results.  
In the following sections we describe the image analysis steps in detail.  
These steps include background estimation and subtraction, convolution to a common PSF, and resampling of the convolved images to a common pixel grid.  
We also correct the final images for missing or bad data in the original images and estimate the uncertainties on the flux in each pixel.
The pixel sizes used in the final maps are chosen to Nyquist-sample the final-map PSFs.

All camera images are first rotated so that North is up, 
and then trimmed to a common sky region.
We then estimate and remove a background from each image, as described
in Appendix \ref{app:background_estimation}.  Following this, we
convolve the images to a common point spread function (PSF), and
resample all the images on a common final-map grid.  We correct the
final images for the bad pixels (or missing data) in the original
images, as described in Appendix \ref{app:missing_data}.  Finally we
use the background pixel dispersion to estimate the pixel flux
uncertainties, as described in Appendix \ref{app:unc_estimation}.

\subsection{Background recognition and Subtraction}

We use a multistep algorithm to generate a {\em Background Mask},
consisting of the area not covered by either the target galaxy or
other discrete sources (e.g., recognizable background galaxies or
foreground stars).  This algorithm is described in detail in Appendix
\ref{app:masks} and avoids overestimating the background in cameras
with low S/N.

After the background mask is generated, for each image we estimate the
best-fit background ``tilted plane'', as described in Appendix
\ref{app:masks}.  After the final best-fit background ``tilted plane''
has been found for each camera, it is subtracted from the original
images.  The background subtraction is performed on each original map
independently, using its native pixel grid.

\begin{figure}
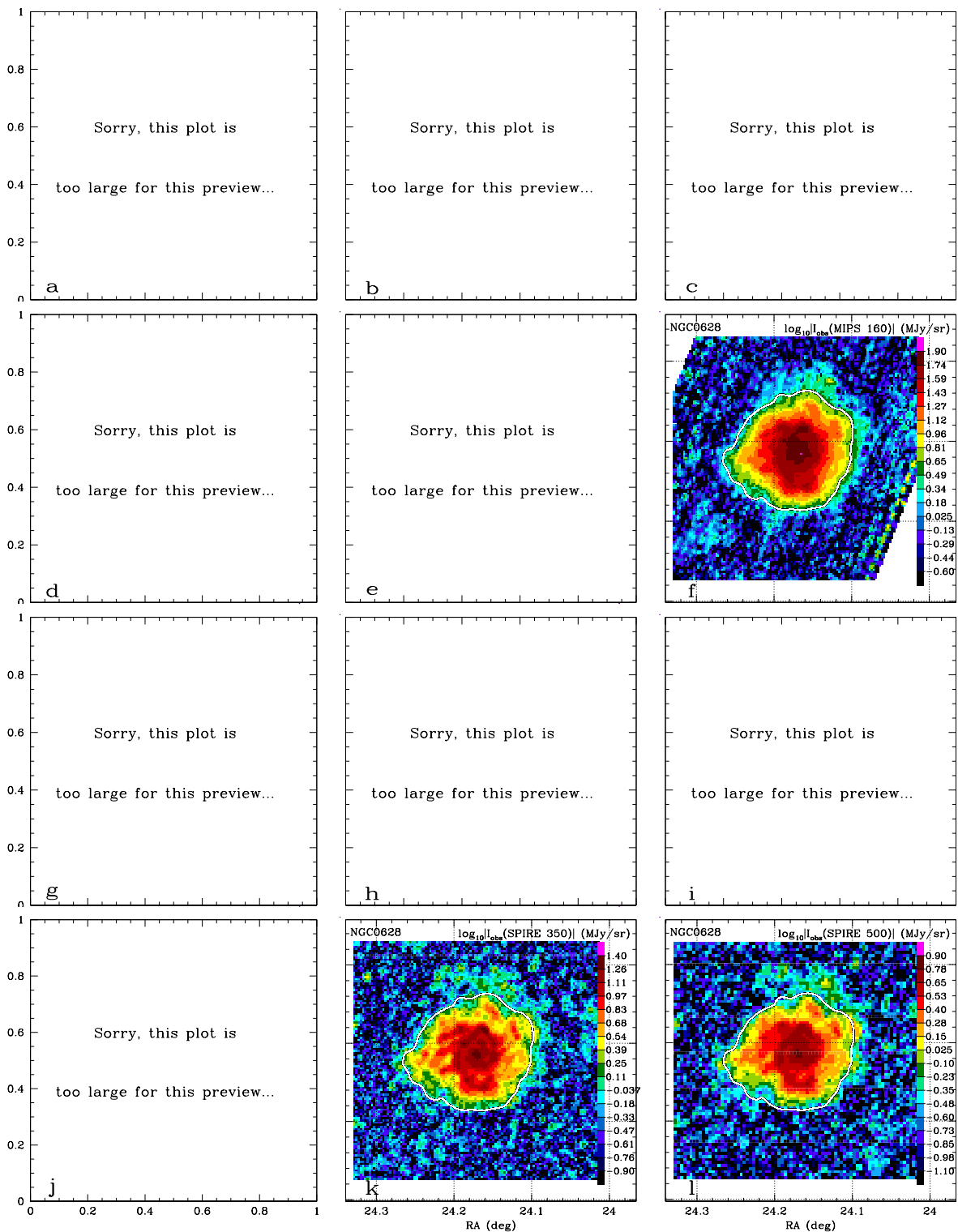
 
\centering
\begin{tabular}{c@{$\,$}c@{$\,$}c} 
\FirstNormal
\SecondNormal
\ThirdNormal
\FourthLast
\end{tabular}
\caption{\footnotesize
Background-removed images of NGC~628. 
Top row: IRAC3.6, 5.8, 8.0 (Left, middle, and right row respectively).
Second row: MIPS24, 70, 160.
Third row: PACS70, 100, 160.
Bottom row: SPIRE250, 350, 500.
The white contour is the boundary of the galaxy mask, within which the data allow reliable estimation of the dust properties.\label{Fig_0628_ori}}
\end{figure}

\begin{figure}
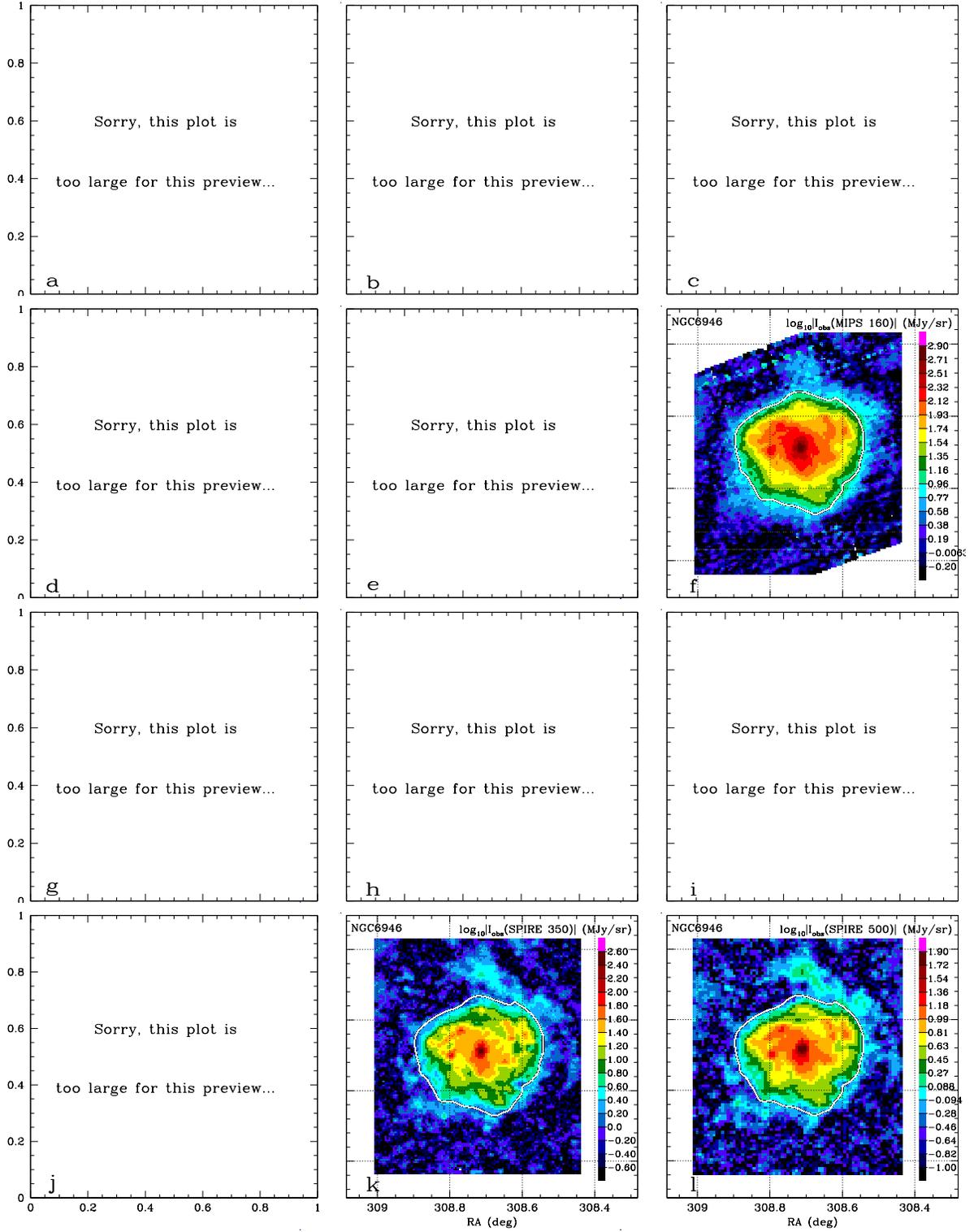
 
\centering
\begin{tabular}{c@{$\,$}c@{$\,$}c} 
\FirstNormal
\SecondNormal
\ThirdNormal
\FourthLast
\end{tabular}
\caption{\footnotesize \label{Fig_6946_ori}
Same as Figure \ref{Fig_0628_ori}, but for NGC~6946.}
\end{figure}

The result is, for each camera, a background-subtracted image with its
native pixel grid and PSF.  Figures \ref{Fig_0628_ori} and
\ref{Fig_6946_ori} show the resulting background-subtracted images.

\renewcommand \RoneCone    {NGC0628_Bck_Removed_IRAC_3oo6_std_res_Original.eps}
\renewcommand \RoneCtwo    {NGC0628_Bck_Removed_IRAC_5oo8_std_res_Original.eps}
\renewcommand \RoneCthree {NGC0628_Bck_Removed_IRAC_8oo0_std_res_Original.eps}
\renewcommand \RtwoCone    {NGC0628_Bck_Removed_MIPS_24_std_res_Original.eps}
\renewcommand \RtwoCtwo    {NGC0628_Bck_Removed_MIPS_70_Original.eps}
\renewcommand \RtwoCthree  {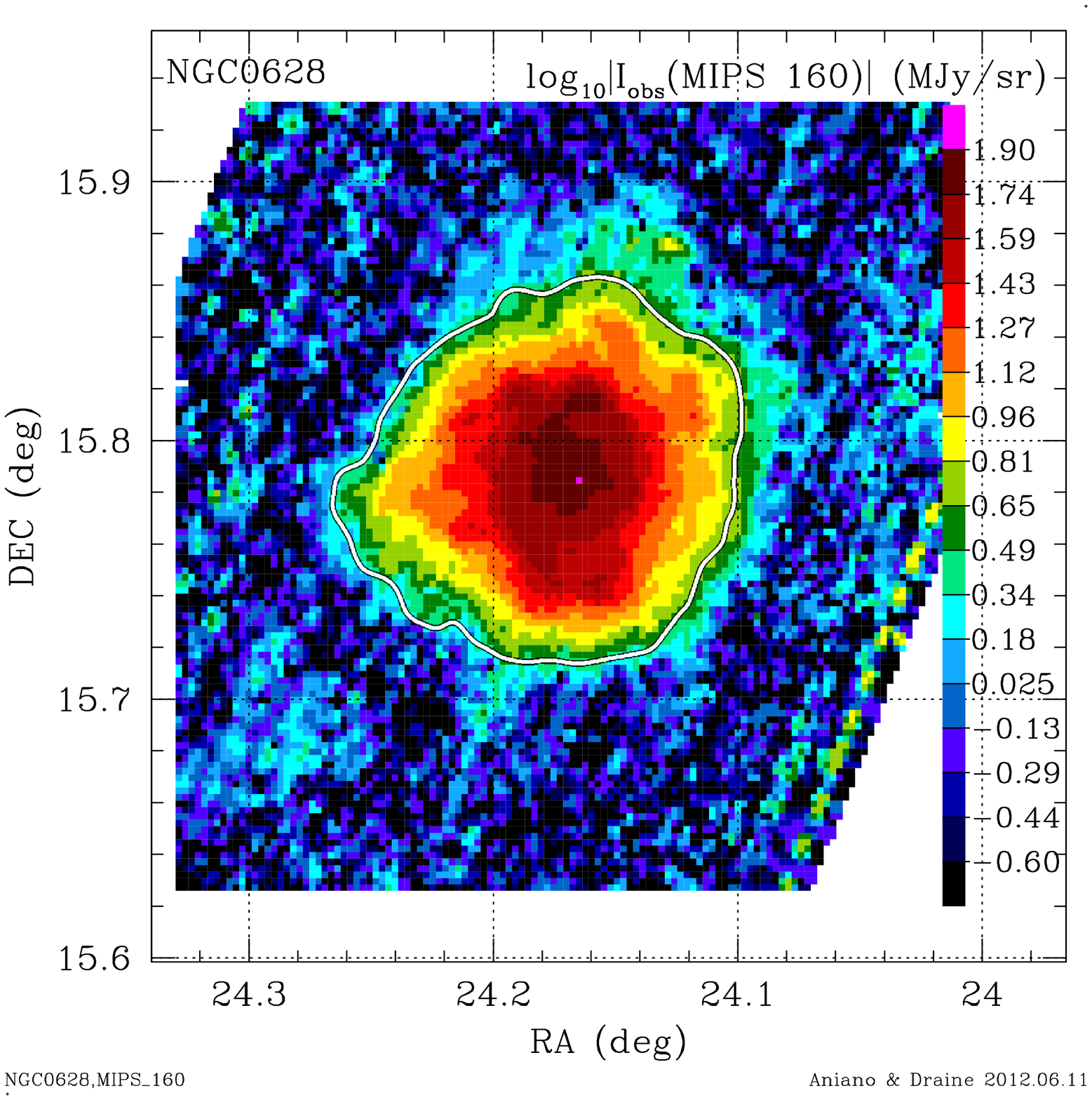}
\renewcommand \RthreeCone {NGC0628_Bck_Removed_PACSS_70_std_res_Original.eps}
\renewcommand \RthreeCtwo  {NGC0628_Bck_Removed_PACSS_100_std_res_Original.eps}
\renewcommand \RthreeCthree{NGC0628_Bck_Removed_PACSS_160_Original.eps}
\renewcommand \RfourCone    {NGC0628_Bck_Removed_SPIRE_250_Original.eps}
\renewcommand \RfourCtwo    {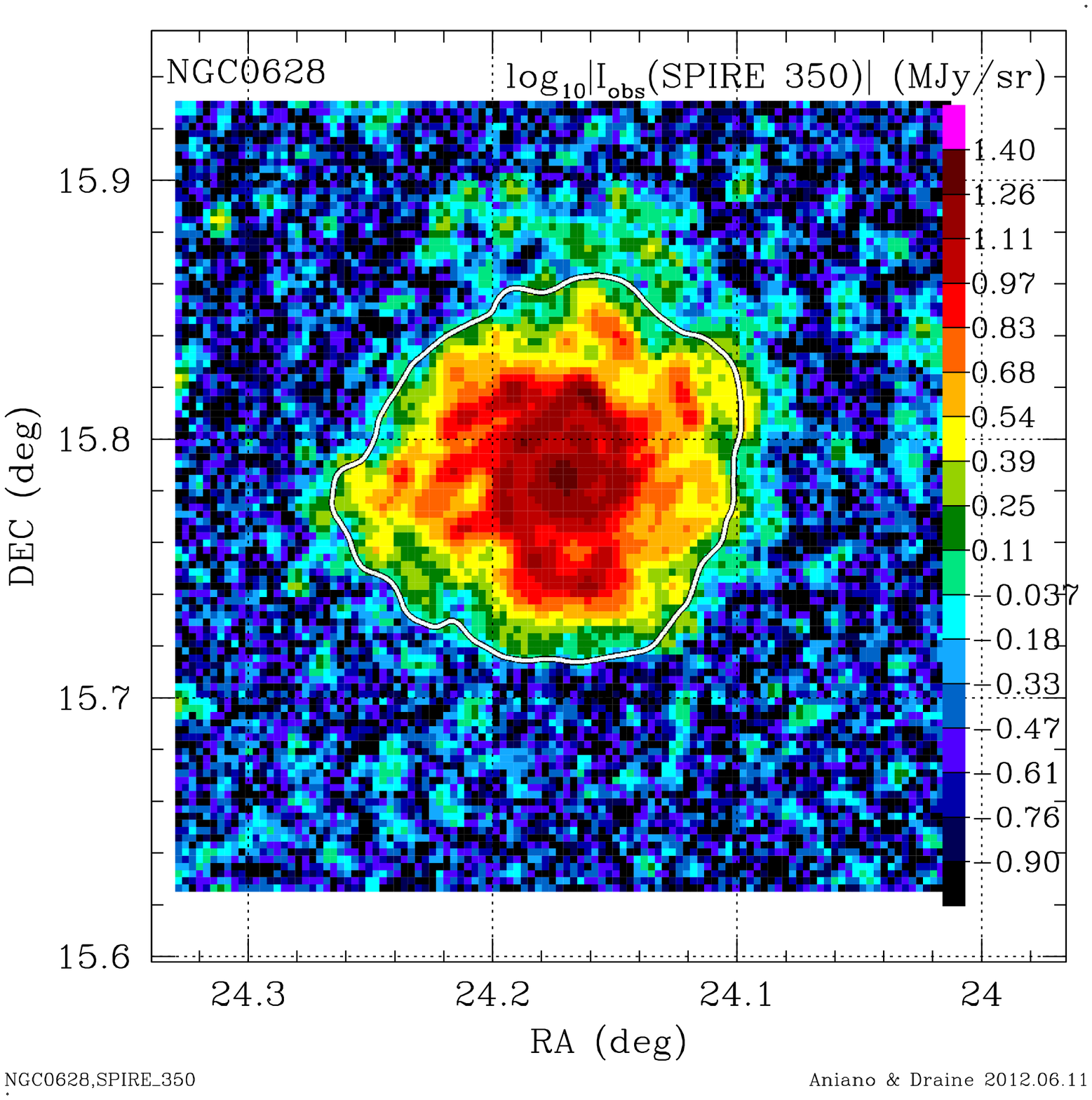}
\renewcommand \RfourCthree  {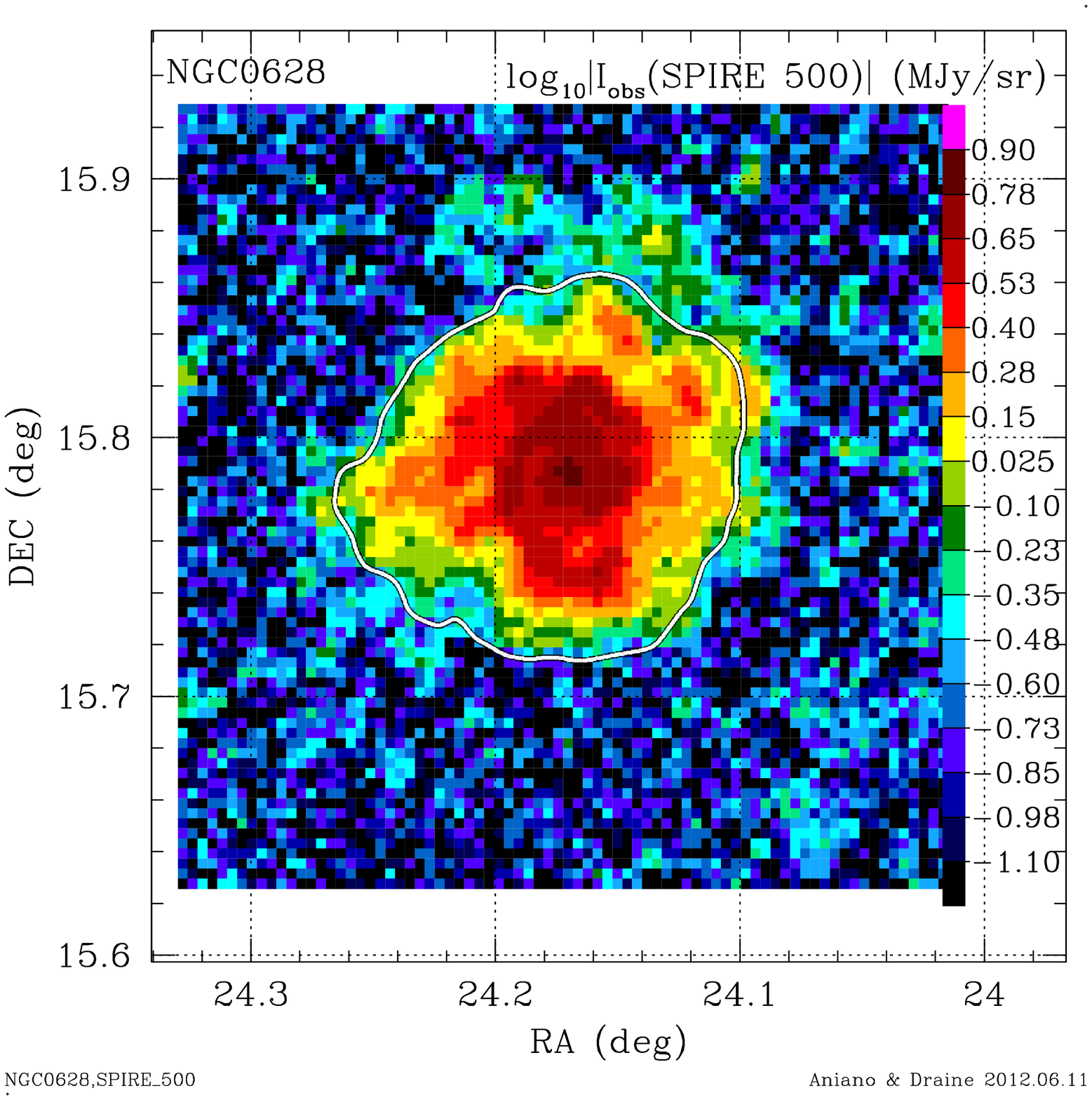}
\ifthenelse{\boolean{make_very_heavy}}{ }
{ \renewcommand \RoneCone    {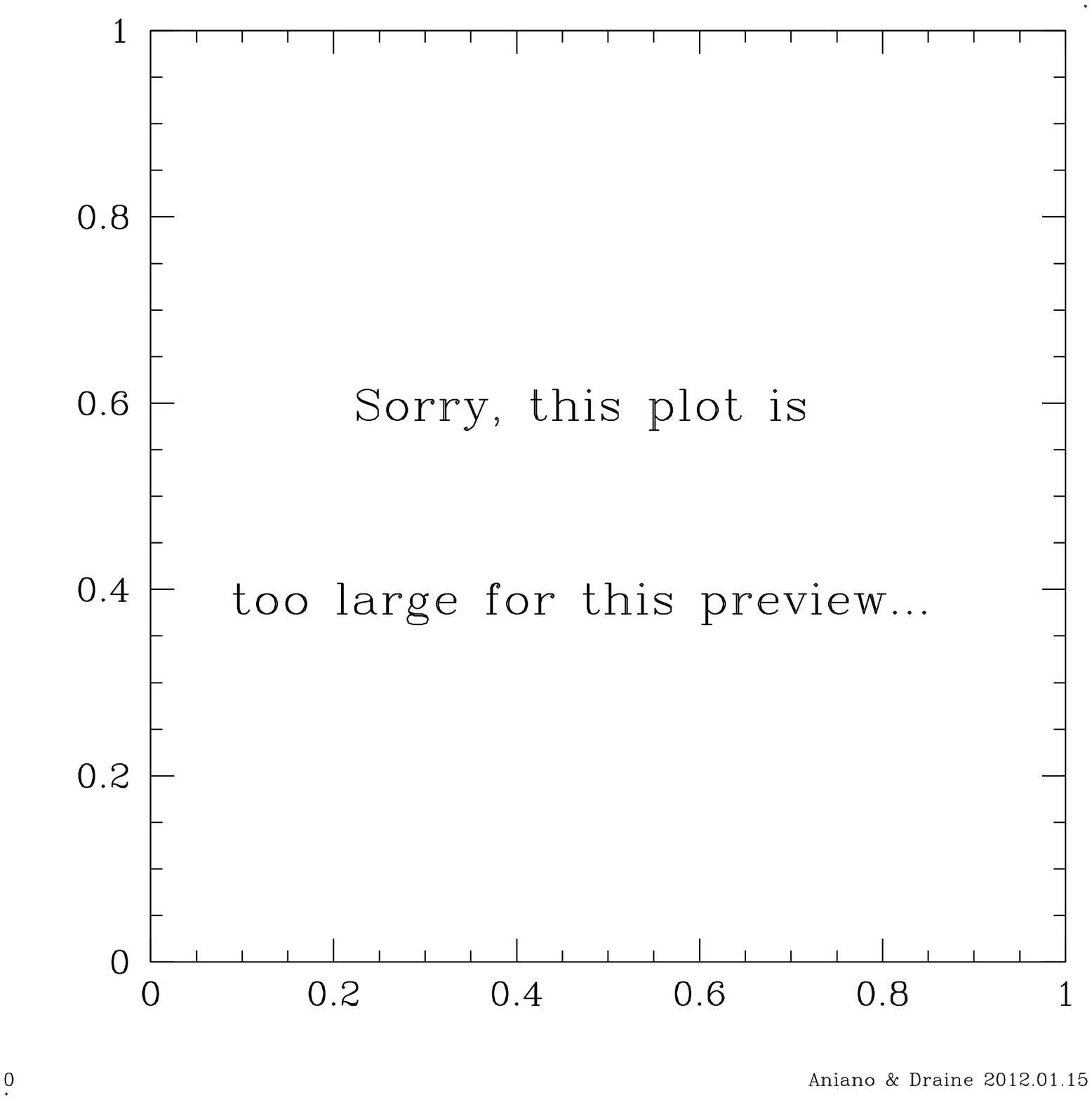}
\renewcommand \RoneCtwo    {No_image.eps}
\renewcommand \RoneCthree  {No_image.eps}
\renewcommand \RthreeCone {No_image.eps}
\renewcommand \RthreeCtwo  {No_image.eps}
\renewcommand \RthreeCthree{No_image.eps}
\renewcommand \RtwoCone    {No_image.eps}
\renewcommand \RtwoCtwo    {No_image.eps}
\renewcommand \RfourCone    {No_image.eps}}

\renewcommand \RoneCone    {NGC6946_Bck_Removed_IRAC_3oo6_std_res_Original.eps}
\renewcommand \RoneCtwo    {NGC6946_Bck_Removed_IRAC_5oo8_std_res_Original.eps}
\renewcommand \RoneCthree {NGC6946_Bck_Removed_IRAC_8oo0_std_res_Original.eps}
\renewcommand \RtwoCone    {NGC6946_Bck_Removed_MIPS_24_std_res_Original.eps}
\renewcommand \RtwoCtwo    {NGC6946_Bck_Removed_MIPS_70_Original.eps}
\renewcommand \RtwoCthree  {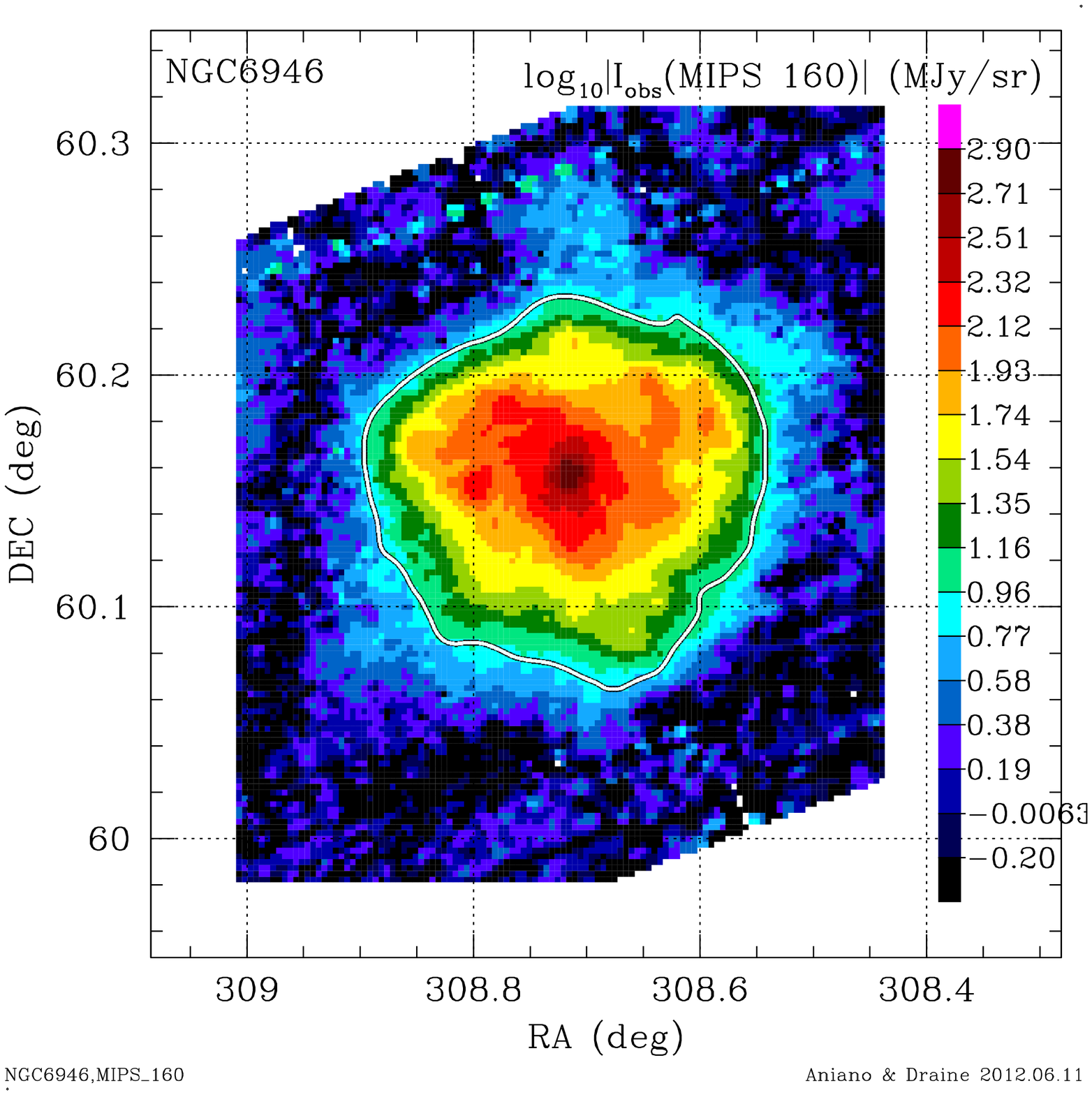}
\renewcommand \RthreeCone {NGC6946_Bck_Removed_PACSS_70_std_res_Original.eps}
\renewcommand \RthreeCtwo  {NGC6946_Bck_Removed_PACSS_100_std_res_Original.eps}
\renewcommand \RthreeCthree{NGC6946_Bck_Removed_PACSS_160_Original.eps}
\renewcommand \RfourCone    {NGC6946_Bck_Removed_SPIRE_250_Original.eps}
\renewcommand \RfourCtwo    {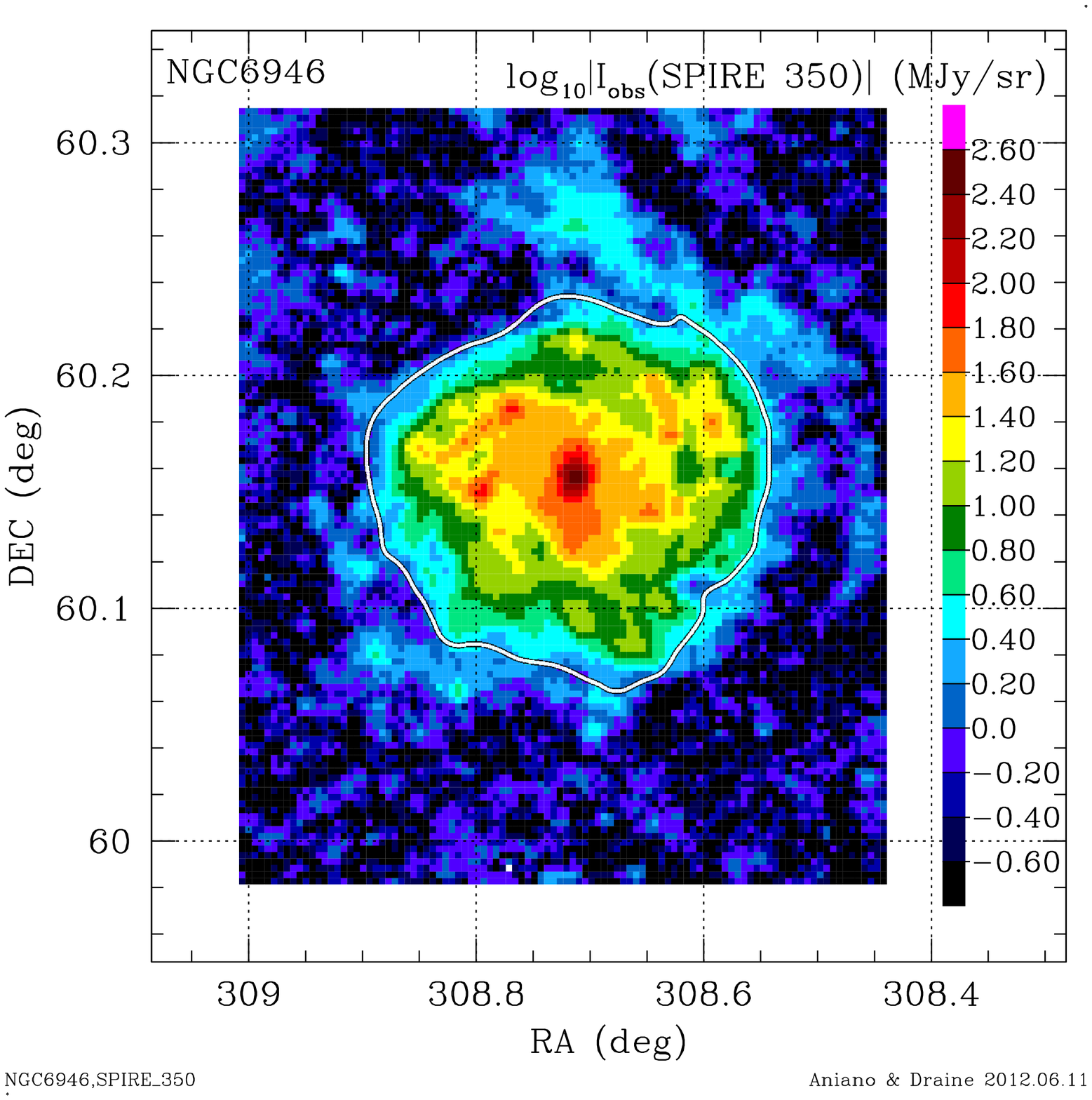}
\renewcommand \RfourCthree  {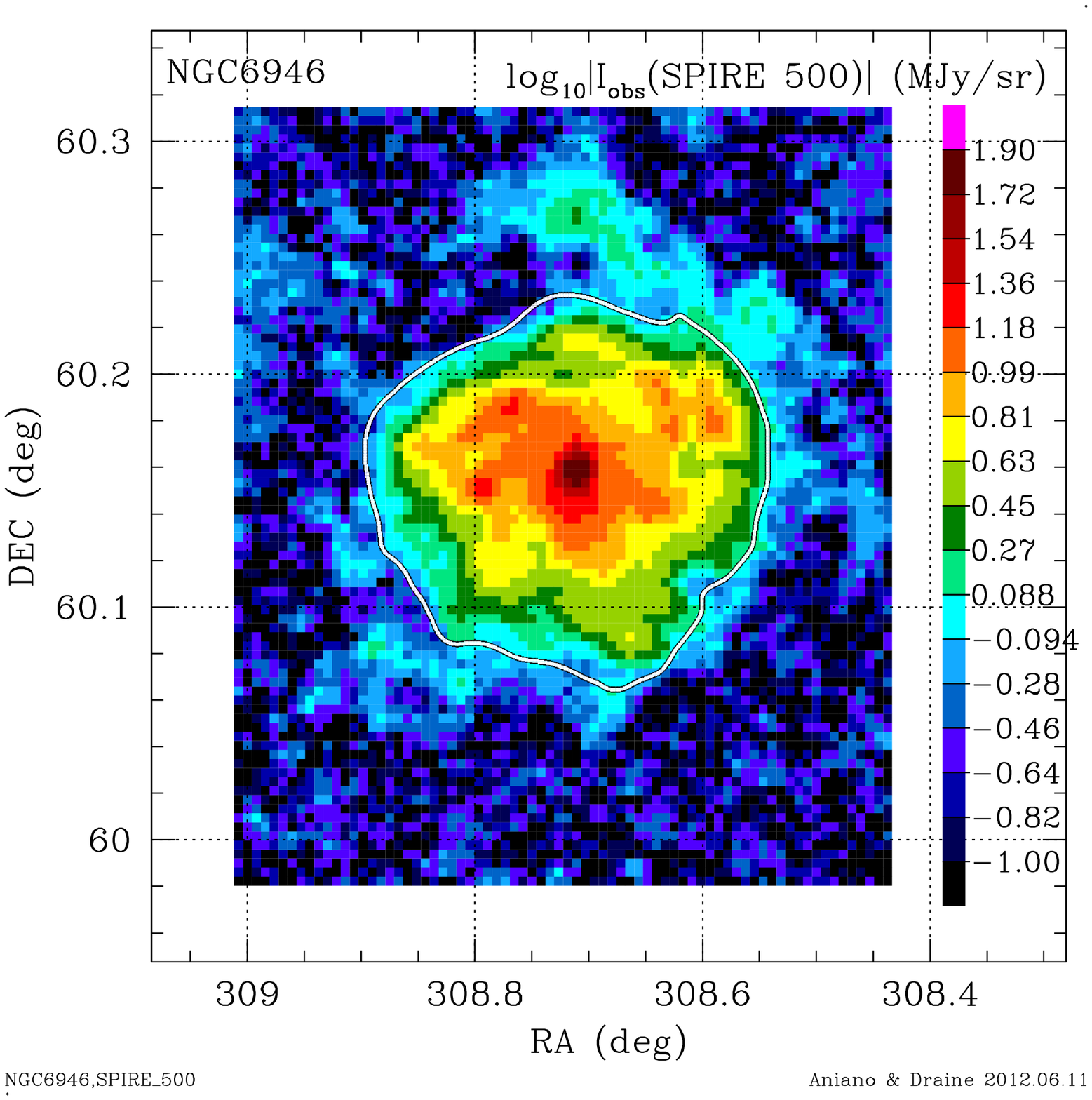}
\ifthenelse{\boolean{make_very_heavy}}{ }
{ \renewcommand \RoneCone    {No_image.eps}
\renewcommand \RoneCtwo    {No_image.eps}
\renewcommand \RoneCthree  {No_image.eps}
\renewcommand \RthreeCone {No_image.eps}
\renewcommand \RthreeCtwo  {No_image.eps}
\renewcommand \RthreeCthree{No_image.eps}
\renewcommand \RtwoCone    {No_image.eps}
\renewcommand \RtwoCtwo    {No_image.eps}
\renewcommand \RfourCone    {No_image.eps}}

\subsection{Convolution to a common PSF}

In order to perform any resolved dust study, it is necessary to
convolve all the images to a common PSF.  To generate maps with
appropriate wavelength coverage to perform the dust modeling, the
natural final-map PSFs to use are those of the PACS160, SPIRE250,
MIPS70, SPIRE350, SPIRE500, and MIPS160 cameras.  For a given
final-map PSF, only a subset of cameras may be transformed into it
reliably, and we proceed to investigate the most reasonable compatible
camera combinations, considering the tradeoff between (1) angular
resolution and (2) availability of long-wavelength data to constrain
the dust models.  After choosing the appropriate PSF for a given set
of cameras, we transform all the background-subtracted images to this
common PSF using convolution kernels described by
\citet{Aniano+Draine+Gordon+Sandstrom_2011}.  In \S 6 we will focus on
maps generated at three final-map PSFs: PACS160, SPIRE250, and MIPS
160.  PACS160 is the PSF with smallest FWHM (full width at half
maximum) that allows use of enough cameras to constrain the dust SED
(IRAC, MIPS24, and PACS70, 100, 160).  SPIRE250 allows use of the same
cameras as PACS160 plus the SPIRE250 camera. Adding the 250\um
constraint produces more reliable maps.  The MIPS160 PSF allows
inclusion of all the cameras (IRAC, MIPS, PACS, SPIRE), therefore
producing the most reliable dust maps; this will be our ``gold standard''.

The convolution kernels assume that the PSFs can be approximated by
rotationally symmetric functions.  In general, a convolution kernel
will relocate flux in the images to transform them to a desired PSF.
\citet{Aniano+Draine+Gordon+Sandstrom_2011} developed a criterion for
camera compatibility, to determine which cameras can be reliably
transformed into a given PSF.  Essentially, a PSF can be safely
transformed into another PSF with similar extended wings provided that
the final FWHM is larger than the original.  When the extended wings
of the two PSFs are dissimilar, the criterion involves quantifying the
amount of energy that a kernel should remove from the extended wings
of the first PSF, as this power removal is correlated with the risk of
introducing artifacts in the convolved image.  The performance of the
convolution kernels is excellent: for ``safe'' pairs of PSFs, the
discrepancies between the convolved narrower PSF and the broader PSF
is smaller than the uncertainties in determining the PSFs
themselves\footnote{Table 4 of
  \citet{Aniano+Draine+Gordon+Sandstrom_2011} quantifies the
  corresponding PSF mismatch: all the kernels employed have $D<0.064$.}.
\citet{Arab+Abergel+Habart+etal_2012} implemented a method to
construct convolution kernels for non rotationally symmetric PSFs,
using the theoretical PSFs for the Herschel cameras.
This method relies on the theoretical PSFs, whereas in the present method
we are able to use rotational averaging to empirically characterize the
extended wings of the actual PSFs measured using observations of saturated
point sources.
Table \ref{tab:resolutions} lists the resolutions of the cameras, the pixel size in the final-map grids used, and the other cameras that can be used at this resolution.
 
The CO$\,J = 2 \rightarrow1$ maps (used to generate the  $\HH$ maps) are provided in rotationally symmetric gaussian PSFs with 13.4\as FWHM, which can be safely transformed into all the final-map PSFs used. 
The original \ion{H}{1} maps have non-circular gaussian PSFs. (FWHM=$9.30\mas \!\times\! 11.88\mas$ for NGC~628, and FWHM=$5.61\mas \!\times\! 6.04\mas$ for NGC~6946). When convolving the \ion{H}{1} maps into the final map resolutions,  we will use kernels generated for rotationally-symmetric gaussian PSFs
with FWHM=$\sqrt{9.30\!\times\!11.88}\,\mas$ and 
$\sqrt{5.61\!\times\!6.04}\,\mas$
for NGC~628 and NGC~6946, respectively.

\begin{table}
\begin{center}
\caption{\label{tab:resolutions}Image resolutions}
\begin{tabular}{c c c c c}
\hline 
                    & FWHM$^a$         & 50\% power$^a$                    &  Final grid                & Compatible\\
Camera       & ($\arcsec$) & diameter ($\arcsec$)   & pixel$^b$ ($\arcsec$) & cameras$^c$\\
\hline 
IRAC3.6  &  1.90 &  2.38 &  --- & not used as a final-map PSF\\
IRAC4.5  &  1.81 &  2.48 &  --- & not used as a final-map PSF\\
IRAC5.8  &  2.11 &  3.94 &  --- & not used as a final-map PSF\\
IRAC8.0  &  2.82 &  4.42 &  --- & not used as a final-map PSF\\
PACS70  &  5.67 &  8.46 &  --- & not used as a final-map PSF\\
MIPS24   &  6.43 &  9.86 & --- & not used as a final-map PSF\\
PACS100  &  7.04 &  9.74 & --- & not used as a final-map PSF\\
PACS160  & 11.2  & 15.3  & 5.0 & IRAC; MIPS24; PACS\\
SPIRE250 & 18.2  & 20.4  & 8.0 & IRAC; MIPS24; PACS;SPIRE250 \\
MIPS70   & 18.7  & 28.8  & 10.0 & IRAC; MIPS24,70; PACS; SPIRE250 \\
SPIRE350 & 24.9  & 26.8  & 10.0 & IRAC; MIPS24,70; PACS; SPIRE250,350\\
SPIRE500 & 36.1  & 39.0  & 15.0 & IRAC; MIPS24,70; PACS; SPIRE\\
MIPS160  & 38.8  & 58.0  & 16.0 & IRAC; MIPS; PACS; SPIRE\\
\hline
\multicolumn{5}{l}{$^a$ Values from \citet{Aniano+Draine+Gordon+Sandstrom_2011} for the circularized PSFs.}\\
\multicolumn{5}{l}{$^b$ The pixel size in the final-map grids is chosen to Nyquist-sample the PSFs.}\\
\multicolumn{5}{l}{$^c$ Other cameras that can be convolved into the camera PSF (see text for details).}
\end{tabular}
\end{center}
\end{table}


\subsection{Uncertainty estimation}

For each camera, after the image processing (rotation to RA-Dec,
background subtraction, convolution to a common PSF, and resampling to
the final grid) the flux in each final pixel is a (known) linear
combination of the flux of (in principle) all the original pixels of
the camera.  If the statistical properties of the uncertainties in the
original pixel fluxes were known, it would be possible to propagate
these uncertainties (and their statistical properties) to each final
pixel.  The original maps oversample the beam, and have artifacts that
extend over several pixels, so realistic statistical properties of the
uncertainties are difficult to determine.  We therefore estimate the
uncertainties directly in the final (post-processed) image.

Using the pixels of the background mask (adapted to the final-map
grid), we measure the dispersion of the background pixels (which
includes noise coming from unresolved undetected background sources,
image artifacts and detector noise) as described in Appendix
\ref{app:unc_estimation}.  By comparing the MIPS and PACS images, we
can also estimate a calibration uncertainty, as described in Appendix
\ref{app:unc_estimation}.


\section{\label{sec:dustmodel}Dust Model}

The composition of interstellar dust remains uncertain, but models
based on a mixture of amorphous silicate grains and carbonaceous
grains have proven successful in reproducing the main observed properties
of interstellar dust.
We employ the dust model of \citet[][hereafter DL07]{Draine+Li_2007}, 
using ``Milky Way'' grain size distributions \citep{Weingartner+Draine_2001a}.
The DL07 dust model has a mixture of amorphous silicate grains and carbonaceous
grains, with a distribution of grain sizes, chosen to reproduce the wavelength dependence of interstellar extinction within a few kpc of the Sun
\citep{Weingartner+Draine_2001a}. 
The silicate and carbonaceous content of the dust grains was constrained by observations of the gas phase depletions in the interstellar medium.

The bulk of the dust in the diffuse ISM is heated by a general diffuse radiation field contributed by many stars. 
However, some dust grains will happen to be located in regions close to luminous stars, such as photodissociation regions (PDRs) near OB stars, where 
the starlight heating the dust will be much more intense than the 
diffuse starlight illuminating the bulk of the grains.
Since our pixels have a large physical size ($\approx$500 pc side for MIPS160 PSF), we will assume that, in each pixel, there is dust exposed to a distribution of starlight intensities.


\subsection{Carbonaceous Grains}

The carbonaceous grains are assumed to have the properties of
polycyclic aromatic hydrocarbon (PAH) molecules or clusters
when the number of carbon atoms per grain $N_{\rm C}\ltsim10^5$, 
but to have the properties of graphite when $N_{\rm C}\gg 10^5$, with an ad-hoc
smooth transition between the two regimes.

A carbonaceous particle of equivalent radius $a$ is taken to have
absorption cross section 
\beq C_{\rm abs}(a,\lambda)
= 
(1-f_{\rm g})N_{\rm C}\sigma_{\rm PAH}({\rm H:C},\lambda) + 
f_{\rm g} C_{\rm abs}({\rm graphite},a,\lambda) ~,
\eeq 
where $\sigma_{\rm PAH}({\rm H\!:\!C},\lambda)$ 
is the absorption cross section per C for PAH
material with given H:C ratio, 
$C_{\rm abs}({\rm graphite},a,\lambda)$
is the absorption cross section calculated for randomly-oriented
graphite spheres of radius $a$, and the graphite ``weight'' 
is taken to be
\beqa
f_{\rm g} &=& 0.01 ~{\rm for}~ a \leq a_c \\ &=& 0.01 +
0.99\left[1-(a_c/a)^3\right] ~{\rm for}~ a > a_c ~,
\eeqa 
where we take the transition radius $a_c=0.0050\micron$.  Carbonaceous grains with
$a > a_c$ are, therefore, treated as having a graphitic component,
with the graphite weight $f_{\rm g}\rightarrow 1$ for $a\gg a_c$.
However, with $f_g=0.01$ for $a<a_c$, 
even very small PAHs are assumed to have a small continuum
opacity underlying the PAH features.

The PAH absorption cross section can be represented as the sum
of a number of vibrational features with specified central wavelength,
FWHM, and band strength.  \citet[][hereafter LD01]{Li+Draine_2001b} 
presented a set of resonance parameters that appeared to be consistent with
pre-Spitzer observations.   
\citet{Smith+Draine+Dale+etal_2007} used the PAHFIT fitting software with
the SINGS spectra to improve observational determinations of
central wavelengths, shapes, and overall strengths of the
PAH emission profiles; DL07 used these results to adjust the LD01
profile parameters.  Subsequent modeling of the SINGS nuclear spectra
(ARD12) led to some
additional small changes in some of the PAH band strengths.
In the present study we employ the PAH cross sections
from DL07 and ARD12.

\citet{Draine+Li_2007} adopted the model put forward by 
\citet[][hereafter DL84]{Draine+Lee_1984} for the
far-infrared properties of graphite.  
Graphite is a highly anisotropic material, with
very different responses for $\vec{E}\parallel \vec{c}$ and 
$\vec{E}\perp \vec{c}$, where the $\vec{c}$ axis is 
normal to the ``basal plane''.
DL84 included ``free-electron''
contributions $\delta\epsilon_\perp^f$, $\delta\epsilon_\parallel^f$ 
to the dielectric tensor, using a 
simple Drude model for the free electron response,
\beq \label{eq:Drude free electron form}
\delta\epsilon^f(\omega) = \frac{-(\omega_p\tau)^2}{(\omega\tau)^2+i\omega\tau}~,
\eeq
where $\omega=2\pi c/\lambda$, and to allow for size effects the
``mean free time'' $\tau$ is taken to be
\beq \label{eq:mft}
\tau^{-1} = \taub^{-1} + v_{\rm F}/a~,
\eeq
where $\taub$ is the mean free time in bulk material,
$v_{\rm F}$ is the Fermi speed, and $a$ is the grain radius.

For $\vec{E}\perp \vec{c}$ we continue to use the graphite dielectric function
from DL84.
However, for $\vec{E}\parallel \vec{c}$,
the DL84 free electron model for $\delta\epsilon_\parallel^f$
resulted in
an opacity at $\lambda\gtsim100\micron$ that
gave somewhat more emission than observed in the Milky Way cirrus by
\citet{Finkbeiner+Davis+Schlegel_1999}.
In addition, the free-electron model used by DL84 for $\vec{E}\parallel \vec{c}$
produced an opacity peak near 33$\micron$ that does not
give a good match to InfraRed Spectrograph (IRS) observations of 
emission from regions where the grains
are hot enough to radiate near 33$\micron$.
To broaden the opacity peak, we now take the free-electron contribution for  $\vec{E}\parallel \vec{c}$ to be
\beq \label{eq:epsilon_parallel}
\delta\epsilon_\parallel^f(\omega) = 
- \sum_{j=1}^2 \frac{(\omega_{p,j}\tau_j)^2}{(\omega\tau_j)^2+i\omega\tau_j}~,
\eeq
with 
$\omega_{p,1}=1\times10^{14}\s^{-1}$, and
$\omega_{p,2}=2\times10^{14}\s^{-1}$.
The $\tau_j$ are obtained from eq.\ (\ref{eq:mft}) with
$\tau_{{\rm bulk},1}=3.51\times10^{-14}\s$,
$\tau_{{\rm bulk},2}=0.88\times10^{-14}\s$.
This gives a d.c. electrical conductivity 
$\sum_j (\omega_{p,j}^2\tau_j)/4\pi
= 5.62\times10^{13}\s^{-1} = 62.5\mho\cm^{-1}$, within the range reported for high-quality graphite crystals at $300\K$
[$\sim1\mho\cm^{-1}$ \citep{Klein_1962} to 
$\sim200\mho\cm^{-1}$ \citep{Primak_1956}].
This d.c. conductivity is larger than the value $30\mho\cm^{-1}$ adopted by DL84; 
the increased conductivity lowers the FIR emission and brings the overall emission spectrum into better agreement with  the observed spectrum from \citet{Finkbeiner+Davis+Schlegel_1999}.
We take $v_{\rm F}=3.7\times10^{6}\left(1+T/255\K\right)^{1/2}\cm\s^{-1}$.
The two-component free-electron form of
eq.\ (\ref{eq:epsilon_parallel}) is not intended
to have physical significance.  It is adopted because it is analytic,
satisfies the Kramers-Kronig relations, gives a reasonable
value for the d.c. electrical conductivity, 
and results in an opacity that is less
peaked than the original single-component form 
(\ref{eq:Drude free electron form}).
The resulting graphite opacity varies as $\lambda^{-2}$ for
$\lambda\gtsim 200\micron$.


\subsection{\label{sec:qpah}PAH Abundance $\qpah$}

As discussed above, the PAHs are part of the carbonaceous grain population.
The PAH abundance is measured by the parameter $\qpah$, defined to
be the fraction of the total grain mass contributed
by PAHs containing $N_{\rm C}< 10^3$ C atoms.
The PAH size distribution used in the DL07 models extends up to PAH particles containing 
$N_{\rm C} > 10^5$ C atoms ($a>6.0\times 10^{-7}$cm). 
However, IRAC photometry at 5.8\um and 8.0\um is sensitive primarily to PAHs with 
$N_{\rm C}\ltsim10^3$ C atoms, small enough so that single-photon heating can result in significant 8\um emission \citep[see, e.g., Fig.\ 7 of][]{Draine+Li_2007}.
For the size distribution in the DL07 models, the mass fraction contributed by PAH particles with $N_{\rm C} < 10^6$ is 
$1.478\, \qpah$.

WD01 constructed grain size distributions with different values of
$\qpah$ that were compatible with the average extinction curve
in local diffuse clouds.  Such models are possible for
$\qpah\ltsim 0.046$, with part of the 2175\AA\ extinction feature
contributed by the PAHs, and part contributed by small graphitic
grains.  For $\qpah\gtsim 0.046$ the predicted 2175\AA\ feature
from the PAHs alone would be stronger than the {\it average} 
observed 2175\AA\ feature in local diffuse clouds\footnote{\citet{Draine+Dale+Bendo+etal_2007}
in a global study of 61 galaxies in the SINGS sample found a median value of $\qpah = 0.034$.}. 

Nevertheless, because the PAH abundance in other regions 
could conceivably exceed the value in the local Milky Way, 
the WD01 dust models have been extended by simply adding PAHs to the
$\qpah=0.046$ model, with no adjustment to the populations of
silicate or larger carbonaceous grains.   These models 
produce stronger emission in the PAH emission features, particularly
in the IRAC8.0 band.
The models were extended to $\qpah=0.10$.
The models with $\qpah > 0.046$ have a 2175\AA\ feature strength larger than the average value in the local ISM.\footnote{In our modeling, we find best-fit $\qpah < 0.05$ for virtually all of the galaxy pixels, compatible with the {\it average} 
observed 2175\AA\ feature in local diffuse clouds, i.e., models with $\qpah >0.05$ are not needed, except for error estimation.}
The model set was also extended down to $\qpah=0$ (the smallest value
of $\qpah$ considered by WD01 was 0.0047). 
Models were computed in a grid of $\qpah=0,\,0.01,\,0.02,\,...\,0.10$, and linearly interpolated to a grid with spacing 
$\Delta\qpah=0.001$.


\subsection{\label{sec:amorphous silicate}Amorphous Silicate Grains}

In the DL84 dust model, the amorphous silicate absorption in the infrared
was modeled by a set of damped Lorentz oscillators, resulting in
an opacity varying as $\lambda^{-2}$ for $\lambda \gg 25\micron$.
However, the COBE-FIRAS measurements of the $\lambda > 110\micron$ 
emission spectrum of
dust at high galactic latitudes
\citep{Wright+Mather+Bennett+etal_1991,
       Reach+Dwek+Fixsen+etal_1995,
       Finkbeiner+Davis+Schlegel_1999}
were not accurately reproduced by the $\lambda^{-2}$ opacity
of the DL84 graphite-silicate model.
\citet{Li+Draine_2001b} therefore made an ad-hoc modification to
${\rm Im}(\epsilon)$ for amorphous silicate
at $\lambda >250\micron$, so that it is no
longer a simple power-law
[with the corresponding changes to ${\rm Re}(\epsilon)$
required by the Kramers-Kronig relations].
The adjustments to ${\rm Im}(\epsilon)$
were not large -- the modified amorphous silicate 
${\rm Im}(\epsilon)$ \citep{Li+Draine_2001b} is within
$\pm12\%$ of that for DL84 amorphous silicate 
for $\lambda < 1100\micron$ -- but these modest adjustments brought
the emission spectrum for the dust model into fairly good agreement
with observations of the emission spectrum of high-latitude dust
\citep[see Fig.\ 9 of][]{Li+Draine_2001b}.
The adopted opacity has no dependence on the grain temperature $T$.

It is possible that the amorphous
silicate opacity may in actuality be $T$-dependent
\citep{Meny+Gromov+Boudet+etal_2007}, and some authors have argued that
this is indicated by observations 
\citep{Paradis+Veneziani+Noriega-Crespo+etal_2010,
       Paradis+Bernard+Meny+Gromov_2011}.
The ``two-level-system'' model of 
\cite{Meny+Gromov+Boudet+etal_2007},
with the standard parameters recommended by
\citet{Paradis+Bernard+Meny+Gromov_2011},
has the far-infrared spectral index $\beta\equiv d\ln\kappa/d\ln\nu$
near $\lambda=500\micron$ varying from $\sim2$ to $\sim$1.3 as $T$ increases
from 10K to 50K.
However, in the present study
we find that the DL07 dust model is able to
satisfactorily reproduce the observed 
spatially-resolved SEDs, as well as the global emission.
At least for near-solar metallicity galaxies such as NGC~628 and NGC~6946,
dust models with $T$-dependent opacities do not appear to be required.


\subsection{Dust Heating\label{DustHeating}}

Each dust grain is assumed to be heated by radiation with energy density per unit frequency
\beq
u_\nu=U\times u_\nu^{\rm{MMP83}}
\eeq
where $U$ is a dimensionless scaling factor and $u_\nu^{\rm{MMP83}}$ is the 
interstellar radiation field (ISRF) estimated by \citet{Mathis+Mezger+Panagia_1983} for the solar neighborhood.
We ignore variations in the spectral shape.
 
Each pixel in our modelling will be larger than $2\times10^4\,\pc^2$, so it will contain ISM in a variety of physical enviroments.
A fraction $(1-\gamma)$ of the dust mass is assumed to be heated
by starlight with a single intensity $U=\Umin$ (i.e., heated by a diffuse
ISM radiation field), while the remaining
fraction $\gamma$ of the dust mass is
exposed to a power-law distribution of starlight intensities between
$\Umin$ and $\Umax$ with $dM/dU\propto U^{-\alpha}$.

The starlight heating intensities are thus characterized by four
parameters: $\gamma$, $\Umin$, $\Umax$, and $\alpha$,
where the fractional dust mass $dM_\dust(U)$ heated by starlight intensities in
$(U,U+dU)$ is
\beq \label{eq:dMd/dU}
\frac{1}{M_{d,tot}} \left(\frac{dM_{\dust}}{dU}\right)
=
(1-\gamma) \delta(U-\Umin)
+
\gamma 
\frac{(\alpha-1)}{\Umin^{1-\alpha}-\Umax^{1-\alpha}}U^{-\alpha}
~{\rm for~} \Umin\leq U \leq \Umax
\eeq
for $\alpha \ne 1$,and
\beq \label{eq:dMd/dU1}
\frac{1}{M_{d,tot}} \left(\frac{dM_{\dust}}{dU}\right)
=
(1-\gamma) \delta(U-\Umin)
+
\gamma \frac{1}{
\ln\left(\Umax/\Umin\right)} U^{-1}
~{\rm for~} \Umin\leq U \leq \Umax
\eeq
for $\alpha = 1$, where $M_{d,tot}\equiv{\int^{\Umax}_{\Umin}(dM_{\dust}/dU')dU'}$.
More complicated starlight heating distributions could be contemplated,
but we find that the simple 4-parameter $(\gamma, \Umin, \Umax,
\alpha)$ model
of eq. (\ref{eq:dMd/dU} - \ref{eq:dMd/dU1})
appears able to usually provide an acceptable fit to observed SEDs in
star-forming galaxies with near-solar metallicities.

Galliano et al (2011) recently claimed that the emission from the Large Magellanic Cloud (LMC) can be reproduced using a starlight distribution function that lacks the ``delta function'' component of eq (\ref{eq:dMd/dU} - \ref{eq:dMd/dU1}), i.e., fixed $\gamma = 1$. However, we show in \S \ref{sec:chi2} that adding the ``delta function'' component significantly improves the quality of the fit for the pixels in NGC~628 and NGC~6946.
 
Many authors choose to fit the $\lambda \gtsim 70 \mum$ emission using a blackbody $B_\nu(T_\dust)$ multiplied
by a power law opacity $\propto \lambda^{-\beta}$. 
The best-fit value of $T_\dust$ is closely related to our heating parameters $\Umin, \,\gamma,\,\alpha$. 
In a subsequent work (Aniano \& Draine 2012, in prep.) we show that $T_\dust \approx 20 \,\Umin^{0.15}\,\rm{K}$, when the DL07 SED is approximated by a blackbody multiplied by a power law opacity.
 
Given that there may be significant regional variations in the starlight spectrum, $U$ should be interpreted not as a measure of the starlight energy
density, but rather as the ratio of the actual dust heating rate to the heating rate for the MMP83 radiation field.

The fraction of dust luminosity emerging in the PAH features does depend on the spectrum of the starlight heating the dust.
\citet{Draine_2011b} showed that the fraction of the dust emission appearing at 8\um increases by a factor of 1.57 as the starlight spectrum is changed from MMP83 to a 20 kK blackbody (cut off at 13.6eV).
If the actual $h\nu < 13.6\rm{eV}$ starlight spectrum is harder (softer) than the MMP83 spectrum assumed in the models, 
we will overestimate (underestimate) $\qpah$.


\subsection{Contribution of Direct Starlight\label{direct}}

Starlight enters in the dust modeling in two ways: via dust heating (as discussed in \S \ref{DustHeating}), and as a direct starlight component (i.e., direct starlight escaping the observed region). 
Our main goal in the present work is to study the properties of the dust and the starlight heating the dust, so we adopt a simple model for the direct starlight component.
Following
\citet{Bendo+Dale+Draine+etal_2006} and
\citet{Draine+Dale+Bendo+etal_2007},
we approximate the $\lambda > 3\micron$ stellar emission from the galaxy as simply 
\beq \label{eq:cs_spec} 
F_{\star}(\lambda) = \Omega_\star
B_\nu(\lambda,T_\star) 
~, 
\eeq
where $\Omega_\star$ is the
solid angle subtended by the stars, $B_\nu$ is the blackbody function,
and $T_\star=5000\K$ is a representative photospheric temperature to
approximate the integrated stellar emission at $\lambda > 3\micron$.
The direct starlight contribution is thus adjusted by only the parameter
$\Omega_{\star}$. 
Direct starlight will only contribute significantly to the IRAC bands, and very marginally to MIPS24.
In ARD12 corrections arising from photospheric absorption as well as emission from hot circumstellar dust
around AGB stars are studied.
These corrections are only important in elliptical galaxies with
little interstellar medium, and the results in the present paper
would be virtually unchanged if included. 
We also neglect possible reddening at the wavelengths ($\lambda\geq 3.6\micron$) in the
present study.

Although in principle the direct starlight and heating starlight parameters should be connected, the uncertain and complex distribution of dust and stars within the galaxies make such connection very complex\footnote{The mean stellar brightness
 emerging from each pixel is not the same as the starlight intensity seen by the dust grains, and therefore we do not expect a tight
correlation between the dust heating and $\Omega_\star$}.
 Our simplified treatment of the direct starlight should only be regarded as a way of ``removing'' the direct starlight component from the near infrared photometry so we can have an estimate of the dust emission.


\subsection{Dust Model Emission}

For each given set of dust parameters $(\qpah,\gamma, \Umin, \Umax,\alpha)$, and the given chemical composition, grain size distribution and grain properties, the dust emission spectrum is computed from first principles.

First, for each given starlight heating parameter $U$,
the temperature distribution of the dust grains (including the PAH component)
is computed as described elsewhere
\citep{Draine+Li_2001,Draine+Li_2007}. 
This is performed for a logarithmically-spaced grid of 41 $U$ values from 0.01 to $10^8$. 
From the temperature distribution functions, model spectra are computed and stored.
To obtain spectra for intermediate $U$ values, we interpolate.

Secondly, for each starlight heating distribution, the specific power spectrum per unit dust mass $p_\nu(\model)$ is computed.
We essentially have two independent heating starlight intensity distributions, the ``delta function component'' (i.e., the dust exposed to $U=\Umin$)
 and the ``power law component'' (i.e., the dust heated by starlight with $\Umin < U <\Umax$).
Lastly each $p_\nu(\model)$ is convolved with the various spectral response functions\footnote{
We employ the appropriate spectral response function for each camera.
For SPIRE, we use the spectral response functions appropriate
for extended sources, as described in the SPIRE observers manual.} 
to obtain the predicted photometry
$\langle p(\model)\rangle _k$ for each camera $k$, with nominal wavelength $\lambda_k$.
We construct a library of model emission for a finely-sampled grid of parameters $(\qpah, \Umin, \Umax,\alpha)$ and $(\qpah, \Umin)$.


\section{\label{sec:SEDFIT}Determining the Dust and Starlight Heating Parameters.}

NGC~628 and NGC~6946 are well resolved, with each galaxy providing many independent pixels, even at MIPS160 resolution.
For each pixel $j$, we find the model of dust and starlight that best
reproduces the observed SED, within the modeling scheme described by DL07.  

As discussed in \S \ref{direct}, starlight enters the fitting in two ways: via direct starlight in the pixel, and by heating the dust.
The direct starlight contribution to pixel $j$ is adjusted by varying only the parameter
$\Omega_{\star,j}$. 
The heating starlight intensity is characterized by four
parameters: $\gamma_j$, $\Uminj$, $\Umaxj$, and $\alpha_j$,
where the dust mass $dM_\dust$ heated by starlight intensities in
$(U,U+dU)$ is given by equation (\ref{eq:dMd/dU}).
For each pixel $j$ we adjust the total dust mass $M_{\dust,j}$,
the PAH abundance parameter $q_{\PAH,j}$ (PAH mass fraction),
and the characteristics of the starlight heating the dust in that pixel. 
If mid-IR photometry is unavailable, one loses the ability to constrain 
$\qpah$, 
but (adopting some arbitrary value of 
$\qpah$ for the modeling), 
the dust mass estimation itself would be largely unaffected. 
Fortunately, we do have mid-IR coverage of our galaxies, so we can obtain full $\qpah$ maps for them.
The present modeling assumes the grains to be heated 
by a standard starlight spectrum (corresponding to the starlight
in the local ISM),
and to have a fixed balance between PAH neutrals and ions.
When fitting to IRAC, MIPS, PACS, and SPIRE photometry, the best-fit
value of $\qpah$
is then essentially proportional to the strength of the (nonstellar)
IRAC8.0 band power
relative to the total IR power.

We will find that $\gamma_j\ll 1$ in nearly all regions where the dust luminosity surface density $\Sigma_{L_\dust,j}>10^7 \Lsol\kpc^{-2}$: here, $\Uminj$ is presumed
to represent the diffuse interstellar medium, or the counterpart to the ``infrared cirrus'' component of the galaxy \citep{Low+Young+Beintema+etal_1984}, accounting
for the bulk of the dust mass in pixel $j$.
The small fraction $\gamma_j$ of the dust mass exposed to starlight
intensities $U>\Uminj$ is presumed to correspond primarily to dust
in star-forming regions.

The model flux density in camera $k$ is
\beq
F(\model,\lambda_k)=\langle F_{\star}(\lambda)\rangle _k 
+\frac{M_\dust}{4\pi D^2}\langle p(\model)\rangle _k 
~~~,
\eeq
where $\langle F_{\star}(\lambda)\rangle _k $ is the direct contribution of starlight given by Eq. (\ref{eq:cs_spec}) convolved with the instrumental response function.
The dust model is characterized by 
$\{\Omega_{\star},M_\dust,\qpah,\gamma,\Umin\,,\Umax\,,\alpha\}$.

In principle, $\Umaxj$ could be treated as an adjustable parameter. 
Previous work \citep{Draine+Dale+Bendo+etal_2007} has shown that the
quality of the global fit to the SED is relatively insensitive to the
choice of $\Umax$. 
We experimented by allowing $\Umaxj$ to be fitted\footnote{We try $\log_{10}\Umax\in\{3,4,5,6,7,8\}$.}
 in the resolved maps, and found that the best value for most of the pixels is $\Umaxj=10^7$.
In the pixels where the resulting best-fit value is not $10^7$, fixing $\Umaxj=10^7$ did not decrease the quality of the fit significantly, i.e. the total $\chi^2$ is essentially the same, as shown in \S 9. Allowing $\Umax$ to be fitted in the range $10^3 \le \Umax \le 10^7$ or fixing it to  $\Umax = 10^7$ does not produce appreciable changes in the inferred dust masses.
We therefore fix $\Umaxj=10^7$ in our modeling.

The limits on adjustable parameters are given in Table \ref{tab:param limits}.
The allowed range for $\Umin$ is determined by the wavelength coverage
of the data used in the fit.  For the SINGS galaxy sample, it was
found that if the photometry extends to $\lambda_{\rm max}=160\micron$, 
models with
$\Umin\geq0.6$ are well-constrained.  However, if longer wavelength
data are available, we allow the possibility of cooler dust, heated
by starlight intensities $U<0.6$, down to $U_{\rm min}=0.06$ if
$\lambda_{\rm max}=250\micron$, and down to $U_{\rm min}=0.01$ if
$\lambda_{\rm max}\geq350\micron$.

\begin{table}
\begin{center}
\caption{\label{tab:param limits}Allowed Ranges for Adjustable Parameters}
\begin{tabular}{c c c c c}
\hline
Parameter & min & max & & Parameter grid used\\
\hline
$\Omega_\star$ & 0 & $\Omega_j$ & &continuous fit \\
$M_\dust$ & 0 & $\infty$ & &continuous fit \\
$\qpah$ & 0.00 & 0.10 & &in steps $\Delta\qpah=0.001$ \\
$\gamma$ & 0.0 & 1.00 & &continuous fit \\
$\Umin$ & 0.7 & 30 & when $\lambda_{\rm max}=160\micron$ &unevenly spaced grid$^a$ \\
        & 0.07 & 30 & when $\lambda_{\rm max}=250\micron$ &unevenly spaced grid$^a$ \\
        & 0.01 & 30 & when $\lambda_{\rm max}=350\micron$&unevenly spaced grid$^a$ \\
        & 0.01 & 30 & when $\lambda_{\rm max}\geq 500\micron$&unevenly spaced grid$^a$\\
$\alpha$ &    1.0      & 3.0        & &in steps $\Delta\alpha=0.1$\\
$\Umax$ & $10^7$ & $10^7$ & &not adjusted \\
\hline
\multicolumn{5}{l}
{$^a$ The fitting procedure uses pre-calculated spectra for $\Umin \in \{$0.01, 0.015,}\\
\multicolumn{5}{l}
{  0.01, 0.02, 0.03, 0.05, 0.07, 0.1, 0.15, 0.2, 0.3, 0.4, 0.5, 0.6, 0.7, 0.8, 1.0, 1.2, 1.5,}\\
\multicolumn{5}{l}
{ 2.0, 2.5, 3.0, 4.0, 5.0, 6.0, 7.0, 8.0, 10, 12, 15, 20, 25, 30$\}$.}
\end{tabular}
\end{center}
\end{table}

We observe that for a given set of parameters $\{\qpah,\Umin,\Umax,\alpha\}$ the model emission is multi-linear in 
$\{\Omega_{\star},\Mdust,\gamma\}$. 
This allows us to easily calculate the best values of $\{\Omega_{\star},\Mdust,\gamma\}$ for a given parameter set 
$\{\qpah,\Umin,\Umax,\alpha\}$. 
Therefore, when looking for the best-fit model in the 7-dimensional model parameter space $\{\Omega_{\star},M_\dust,\qpah,\gamma,\Umin\,\Umax,\alpha\}$, we only need to do a search over the 4-dimensional subspace spanned by $\{\qpah,\Umin,\Umax,\alpha\}$. 
In the case of a fixed $\Umax$, the search is performed over a 3-dimensional space spanned by $\{\qpah,\Umin,\alpha\}$.
In any case, for the computed grid of $\{\qpah,\Umin,\Umax,\alpha\}$, the multidimmensional search for optimal parameters can be performed by brute force,
rather than needing to rely on a nonlinear minimization algorithm.

With $\Umax$ fixed, for each pixel $j$, 
the model library is used to search for the model parameter vector
$\xi_j=\{\Omega_{\star},M_\dust,\qpah,\gamma,\Umin,\alpha\}$ 
that minimizes
\beq \label{eq:chi2}
\chi_j^2 \equiv \sum_{k} 
\frac{[F_{{\rm obs},j}(\lambda_k)_j-F(\model,\lambda_k)]^2}
     {\sigma_{\lambda_k,j}^2}
~~~,
\eeq
where $F_{{\rm obs},j}(\lambda_k)$ is the observed flux density and 
$\sigma_{\lambda_k,j}$ is the 1-$\sigma$ uncertainty in the measured flux
density for pixel $j$ at wavelength $\lambda_k$ (see Appendix D for a detailed discussion on how 
$\sigma_{\lambda_k,j}$ is obtained).\footnote{
In some regions where we do not have complete data coverage or the S/N is low, we may further fix some of the parameter values. This situation typically arises in background areas without IRAC coverage, in which case we fix $\Omega_{\star}=0$ }

The above procedure yields ``best-fit'' estimates for the
model parameter vector 
$\xi_j=\{\Omega_\star,M_\dust,\qpah,\gamma,\Umin,\alpha\}$ 
for each pixel $j$. 
Each pixel is fitted independently of the remaining pixels in the galaxy. 
Since the final-map pixel size are chosen to Nyquist sample the final-map PSF, the camera images are smooth on a pixel scale.
The fact that we obtain smooth parameter maps is an indication of the stability of the fitting procedure, i.e., even though every pixel is modeled independently of its neighbors, the continuity in the images lead to continuity in the results. 
The quoted ``best-fit'' parameter maps arise from fitting the observed flux in each pixel\footnote{We do not use the median maps that could be constructed from modeling adding random noise to the observations to estimate the parameter uncertainties}.

\subsection{Properties Derived from the Models}

The infrared luminosity $L_{\dust,j}$ for the dust model in the pixel $j$ is:
\beq
L_{\dust,j} = P_0(q_{{\rm PAH},j}) \,M_{\dust,j}\, \overline{U}_j\,,
\eeq
where $P_0(\qpah)$ (which depends only weakly on $\qpah$) is the total power radiated per unit dust mass by the model
when heated by starlight with intensity $U=1$, 
and $\overline{U}_j$  is the mass-weighted mean starlight heating intensity, given by:
\beq
\overline{U}_j = (1-\gamma_j) \Uminj +  \gamma_j
\left\{
\begin{array}{l l}
      \left(\frac{\alpha_j-1}{\alpha_j-2}\right)
      \left[\frac{U_{\rm max}^{2-\alpha_j}-\Uminj^{2-\alpha_j}}
                 {U_{\rm max}^{1-\alpha_j}-\Uminj^{1-\alpha_j}}
      \right]
      &
      {\rm for~}\alpha_j\neq 1,\alpha_j\neq 2\\
      ~\\
     \Umax\, \frac{1-\Uminj/\Umax}{\ln\left(\Umax/\Uminj\right)} 
      &
      {\rm for~}\alpha_j = 1\\
      ~\\
      \Uminj\, \frac{\ln\left(\Umax/\Uminj\right)}{1-\Uminj/\Umax}
      &
      {\rm for~}\alpha_j=2~.
\end{array}
\right.
\eeq
Star-forming regions have significant starlight power absorbed
by dust grains in regions of high starlight intensity, which generally
correspond to photodissociation regions (PDRs).
We will refer to the luminosity radiated by dust in regions with
$U>U_\PDR$ as $L_\PDR$, given by:
\beq
L_{{\rm PDR},j} =  P_0(q_{{\rm PAH},j}) \,M_{\dust,j}  \,\gamma_j  \times
\left\{
\begin{array}{l l}
      \left(\frac{\alpha_j-1}{\alpha_j-2}\right)
      \left[\frac{U_{\rm max}^{2-\alpha_j}-\UPDR^{2-\alpha_j}}
                 {U_{\rm max}^{1-\alpha_j}-\Uminj^{1-\alpha_j}}
      \right]
      &
      {\rm for~}\alpha_j\neq 1,\alpha_j\neq 2\\
      ~\\
      \frac{\Umax -\UPDR}{\ln\left(\Umax/\Uminj\right)} 
      &
      {\rm for~}\alpha_j = 1\\
      ~\\
     \frac{\ln\left(\Umax/\UPDR\right)}{\Uminj^{-1}-\Umax^{-1}}
      &
      {\rm for~}\alpha_j=2\\
\end{array}
\right
.
\eeq
We take $U_\PDR=10^2$ as a plausible cutoff to select dust in high intensity regions (choosing another cutoff value would change only the inferred $L_{{\rm PDR},j}$, leaving all the remaining dust parameters unaltered). We further define $f_{{\rm PDR},j}$ as:
\beq
f_{{\rm PDR},j} \equiv {L_{{\rm PDR},j} \over L_{\dust,j}}\,.
\eeq

The region observed is at a distance $D$ from the observer and $\Omega_j$
is the solid angle of pixel $j$.
For each pixel $j$, the best-fit model vector
$\{\Omega_\star,M_\dust,\qpah,\gamma,\Umin,\alpha\}_j$
corresponds to a dust mass surface
density: 
\beq
\Sigma_{M_\dust,j} \equiv \frac{1}{D^2\Omega_j} M_{\dust,j}
~.
\eeq
Similarly, we can compute the infrared luminosity surface
density $\Sigma_{L_\dust,j}$ and the surface density of 
dust luminosity from regions with $U>U_\PDR$, $\Sigma_{L_\PDR,j}$, as:
\label{eq:total}
\beq
\Sigma_{L_\dust,j} \equiv \frac{1}{D^2\Omega_j} L_{\dust,j}, \quad\quad \Sigma_{L_\PDR,j} \equiv \frac{1}{D^2\Omega_j} L_{\PDR,j}
~.
\eeq

The DL07 dust models used here are consistent with the Milky Way ratio
of visual extinction to H column, 
$A_V/\NH=5.34\times10^{-22}{\rm mag}\cm^2/\Ha$, for a dust/H ratio
$\Sigma_{M_\dust}/\NH m_\Ha = 0.010$ 
\citep[see Table 3 of][]{Draine+Dale+Bendo+etal_2007}.
The dust surface density corresponds to a visual extinction (through the disk)
\beq
A_V = 0.67 \left(\frac{\Sigma_{M_\dust}}{10^5\Msol\kpc^{-2}}\right) {\rm mag}
~.
\eeq

\subsection{Global Quantities}

After the resolved (pixel-by-pixel) modeling of the galaxy is performed, we compute a set of global quantities by adding or taking weighted means (denoted as $\langle...\rangle$) of the quantities in each individual pixel of the map. 
The total dust mass $M_{\dust,tot}$, total dust luminosity $L_{\dust,tot}$, and total dust luminosity radiated by dust in regions with $U>U_\PDR$, $L_{\dust,tot}$, are given by:
\beq
\label{eq:mean_mdust}
M_{\dust,tot} \equiv \sum_{j=1}^N M_{\dust,j}
\,,\quad
L_{\dust,tot} \equiv \sum_{j=1}^N L_{\dust,j}
\,,\quad
L_{\PDR,tot}\equiv \sum_{j=1}^N L_{\PDR,j}
\,,
\eeq
where the sums extend over all the pixels $j$ that correspond to the target galaxy (see Appendix A for the galaxy segmentation procedure).
The dust-mass weighted PAH mass fraction $\langle \qpah \rangle$, and mean starlight intensity $\langle \overline{U} \rangle$, are given by:
\beq
\label{eq:mean_1}
\langle \qpah\rangle  \equiv 
\frac{\sum_{j=1}^N \qpahj \, M_{\dust,j}}{\sum_{j=1}^N M_{\dust,j}}
\,,\quad
\langle \overline{U}\rangle  \equiv 
\frac{\sum_{j=1}^N \overline{U}_j \,  M_{\dust,j}}{\sum_{j=1}^N M_{\dust,j}}
\,.
\eeq
The dust mass-weighted minimum starlight intensity $\langle \Umin \rangle$ is given by:
\beq
\label{eq:mean_2}
\langle \Umin\rangle \equiv 
\frac{\sum_{j=1}^N (1-\gamma_j)\, \Uminj \,M_{\dust,j}}{\sum_{j=1}^N (1-\gamma_j)M_{\dust,j}}
\,.
\eeq
The  dust-luminosity weighted value of $\fpdr$, $\langle \fpdr\rangle$ is:
\beq
\label{eq:mean_fpdr}
\langle \fpdr\rangle  \equiv 
\frac{L_{\PDR,tot}}{L_{\dust,tot}}
\,.
\eeq

Alternatively, we can fit the dust model to the global photometry for each
galaxy\footnote{
We take global photometry to be the photometry within the galaxy mask, i.e., the galaxy regions where we can reliably determine the dust luminosity. There is undoubtedly additional emission from material outside the galaxy mask, but it cannot be measured reliably.}
(i.e., a single-pixel dust model). 
The dust parameters obtained from the global-photometry (single-pixel) model fit will be compared to the corresponding resolved modeling global quantities defined in equations (\ref{eq:mean_mdust} - \ref{eq:mean_fpdr}) in \S \ref{sec:maps vs global}. 

Clearly, the emission summed over the map will not correspond to the emission of dust exposed to starlight of the form given by eq. (\ref{eq:dMd/dU} - \ref{eq:dMd/dU1}) since different pixels will have different values of $\Uminj, \gamma_j$ and $\alpha_j$.
For completeness, we could define the dust mass-weighted mass fraction heated by a power-law ($U>\Umin$) component $\langle \gamma\rangle $ as:
\beq
\label{eq:mean_3}
\langle \gamma\rangle  \equiv 
\frac{\sum_{j=1}^N \gamma_j \, M_{\dust,j}}{\sum_{j=1}^N M_{\dust,j}}\,,
\eeq
and an ``effective power-law exponent'' $\langle\alpha\rangle$, defined as the value that satisfies:
\beq
\label{eq:mean_alpha}
L_{\PDR,tot} =
\frac{ L_{\dust,tot}} {\langle \overline{U}\rangle }  \,
\langle\gamma\rangle  \,
      \left(\frac{\langle\alpha\rangle-1}{\langle\alpha\rangle-2}\right)
      \left(\frac{U_{\rm max}^{2-\langle\alpha\rangle}-\UPDR^{2-\langle\alpha\rangle}}
                 {U_{\rm max}^{1-\langle\alpha\rangle}-\langle\Umin\rangle^{1-\langle\alpha\rangle}}
\right).
\eeq
Unfortunately, $\langle\gamma\rangle$ and $\langle\alpha\rangle$ do not turn out to be useful global quantities, and will, in general, differ significantly from the values obtained from the global-photometry (single-pixel) model fit.

\subsection{Parameter Uncertainty Estimates}

To estimate uncertainties in the derived dust parameters for each pixel $j$,
we simulate data by adding zero-mean random noise 
$\delta F(\lambda_k)_{j,\rand}$, for $\rand=1,2,3,...,{N_r}$, to the observed
flux $F(\lambda_k)_j$ in each band, and fit the simulated noisy data. 
In Appendix \ref{app:unc_estimation} we describe the statistical construction of the sample $F(\lambda_k)_{j,\rand}$ of ${N_r}$ random realizations.
For the fit parameters $\{a,b,...\}\in\{\Omega_{\star},M_{\dust},\qpah,\gamma,\Umin,\alpha\}$ we have a set of ${N_r}+1$ values and we calculate the covariance matrix:
\beq
V_{ab} = V_{ba} \equiv N_r^{-1}\sum_{\rand=1}^{N_r}(a_\rand-a_0)(b_\rand-b_0)
~~~,
\label{eq.var}
\eeq
where $a_0,b_0$ are the best-fit parameter values for the observed
fluxes, and $a_\rand,b_\rand$ are the best-fit values for the $\rand$-th
random noise realization. 
For each model parameter $a$, the 1--$\sigma$ uncertainty is taken to
be
\beq
\sigma(a) = V_{aa}^{1/2}\,\,\,.
\label{eq.var1}
\eeq

Each random noise realization $\rand=1,2,...N_r$ produces a global quantity via eq. (\ref{eq:mean_mdust}), (\ref{eq:mean_1}), and (\ref{eq:mean_2}), (\ref{eq:mean_fpdr}), and (\ref{eq:mean_alpha}). 
We proceed to calculate uncertainties for global quantities using equations \ref{eq.var} and \ref{eq.var1}.
In Appendix \ref{app:unc_parameters} we describe the details of the parameter uncertainty estimation.


\clearpage
\section{\label{sec:results}Results}

We construct dust maps with several different angular resolutions.
For a given resolution, one can only use cameras with the PSF FWHM smaller than the reference PSF.
In a normal star-forming galaxy, most of the
dust mass is at temperatures $T_\dust \approx 15-25\K$, with $\nu
L_\nu$ peaking at $\lambda \approx hc/6kT_\dust \approx 100 - 160\micron$.
Reliable estimates of the dust mass therefore require long wavelength
data, at least out to 160$\micron$.  In the present study we will
compare dust models using maps at the resolution of the PACS160
(FWHM=$11.2\arcsec$), SPIRE250 (FWHM=$18.2\arcsec$), and MIPS160
(FWHM=$38.8\arcsec$) cameras.  As discussed by
\citet{Aniano+Draine+Gordon+Sandstrom_2011} the MIPS160 PSF cannot be
convolved safely into the SPIRE500 PSF (with FWHM=$36.1\arcsec$).  This
implies that MIPS160 is the narrowest PSF that can be used if we want
to include MIPS160 data into the modeling.  A drawback of using the
MIPS160 PSF is its extended wings,
which can cause power radiated by the bright central regions to
contribute significantly to the observed surface brightness of the faint
outer regions.
However, we will see in \S 7 that
this effect does not significantly affect dust mass estimates, and discrepancies between PACS and MIPS photometry make it important to include the 
MIPS160 camera in the dust modeling.

In order to generate the $\Ha$ surface density maps (used in the dust
mass \!/\! H mass ratio maps), a value of $\XCOxx$ needs to be
chosen. Figure \ref{fig:Xco} shows the total H maps (first and third
rows) for both galaxies, for 
$\XCOxx=2$, 3, and 4, convolved to the MIPS160 PSF.
Note
that the CO maps cover almost all of both galaxy masks, but do not cover
the full field of view, leading to the box-like step in the H mass
maps.

Using dust maps based on all of the available photometry (with the
MIPS160 PSF -- see Figures \ref{fig:ngc0628-1}c and
\ref{fig:ngc6946-1}c below), Figure \ref{fig:Xco} shows the dust/H
mass ratios for the different values of $\XCOxx$, for NGC~628 (second
row) and NGC~6946 (fourth row). The choice $\XCOxx=4$ gives the
smoothest dust/H mass ratios over the galaxies (outside of
the central region of NGC~6946, which is further discussed in
\S\ref{sec:n6946_dustmaps}).  
As discussed in Sec. \ref{sec:gas}, we take $\XCOxx=4$ for both NGC~628
and NGC~6946.


\renewcommand \RoneCone {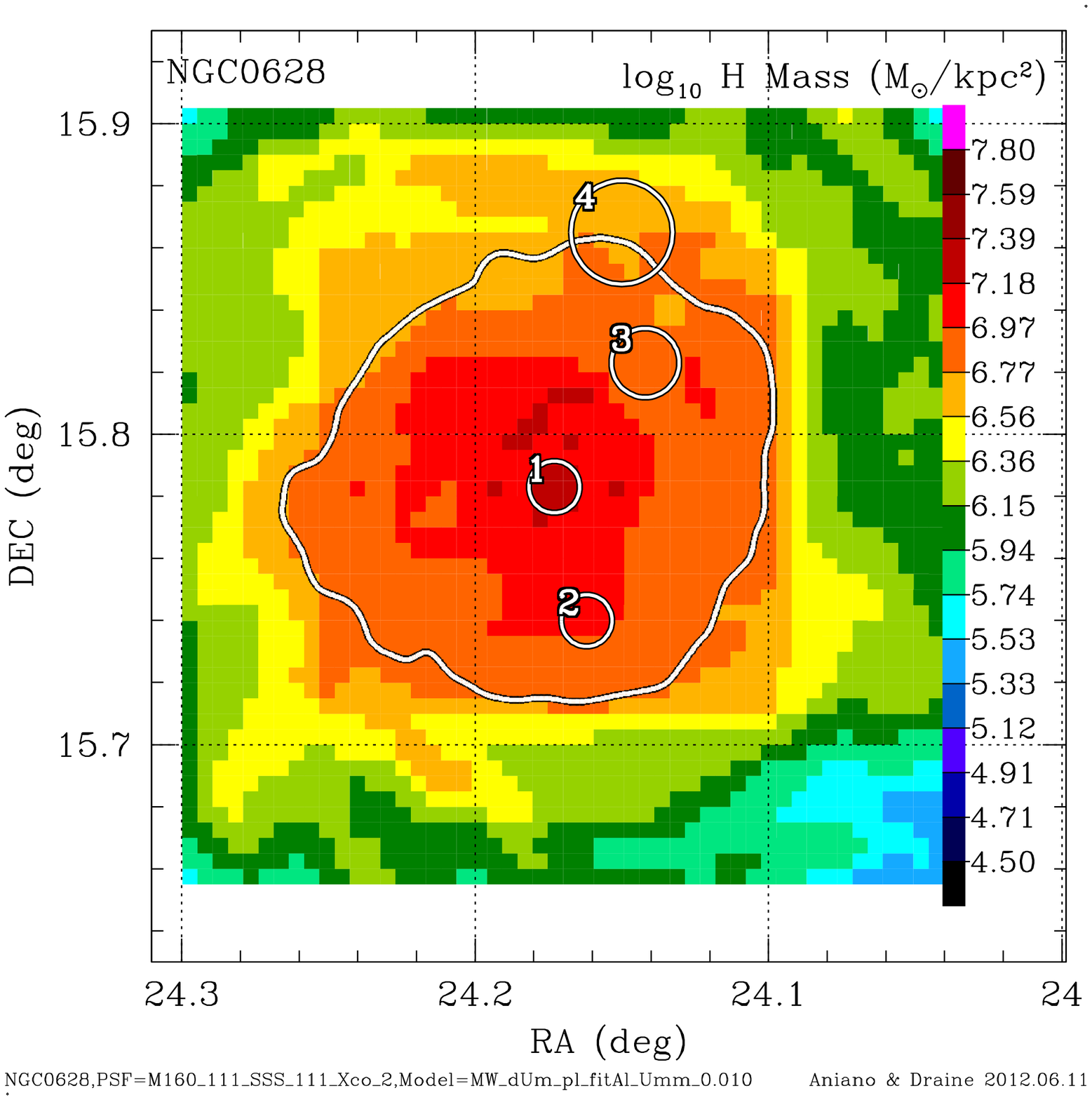}
\renewcommand \RoneCtwo {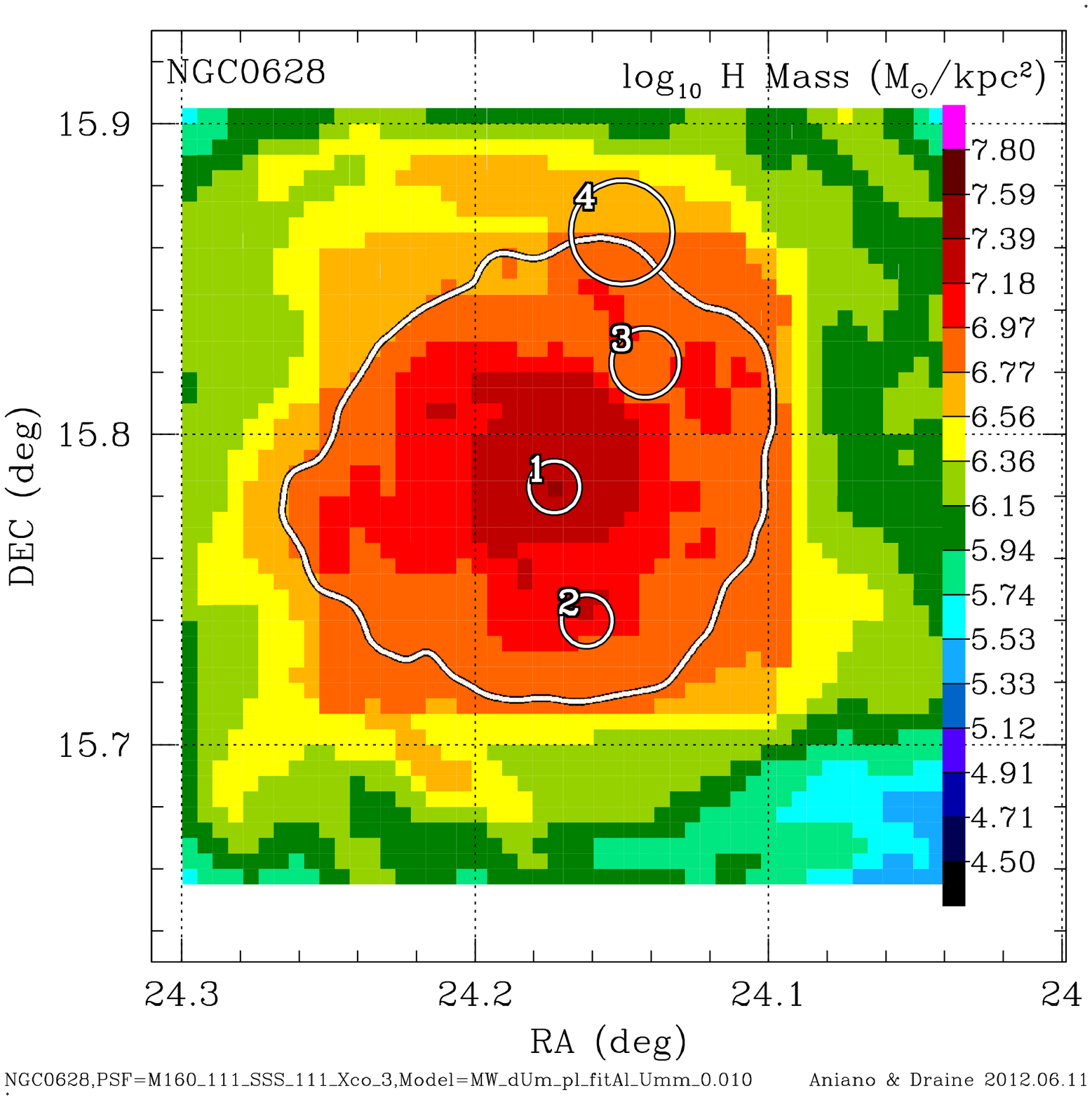}
\renewcommand \RoneCthree {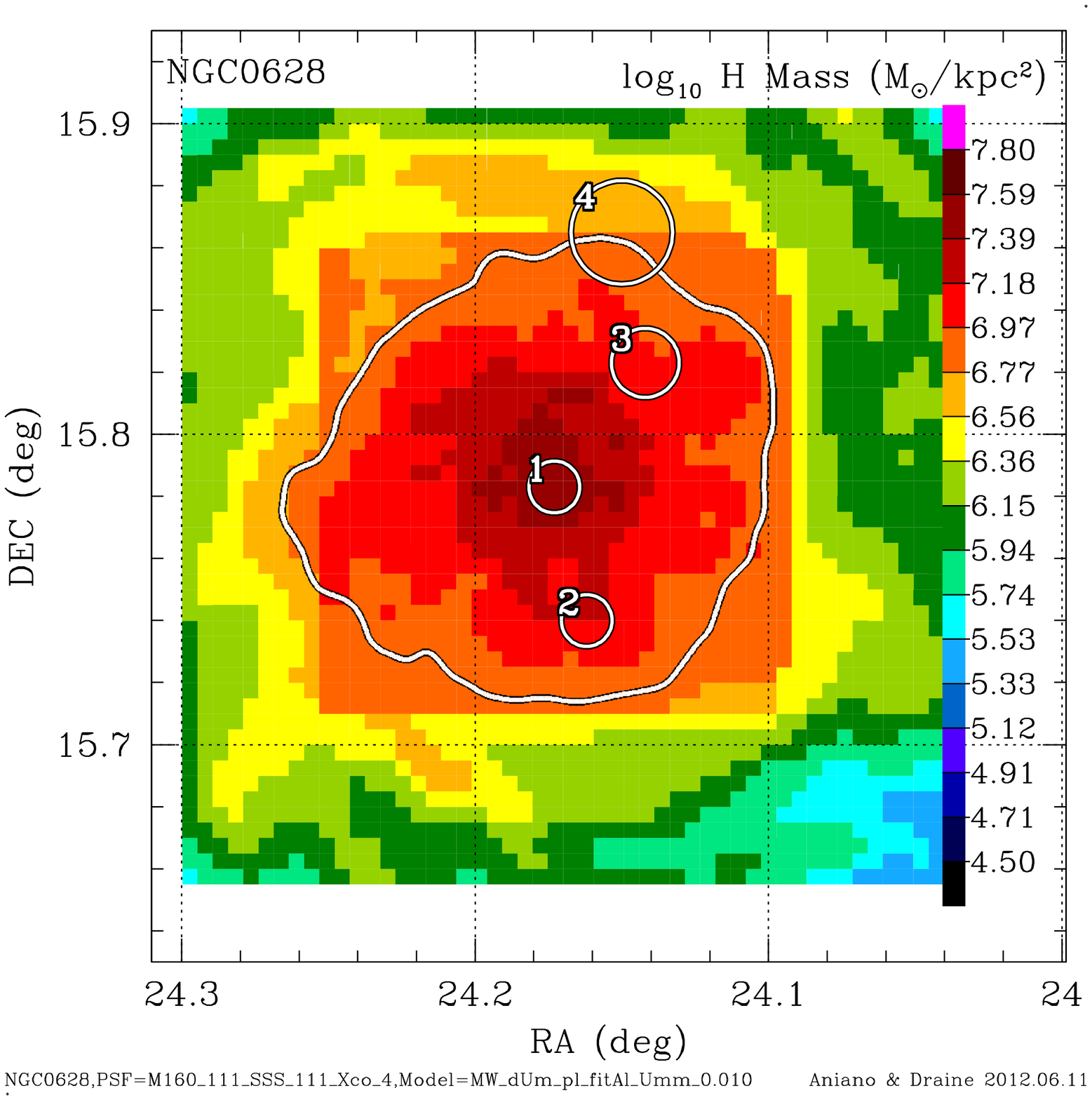}
\renewcommand \RtwoCone {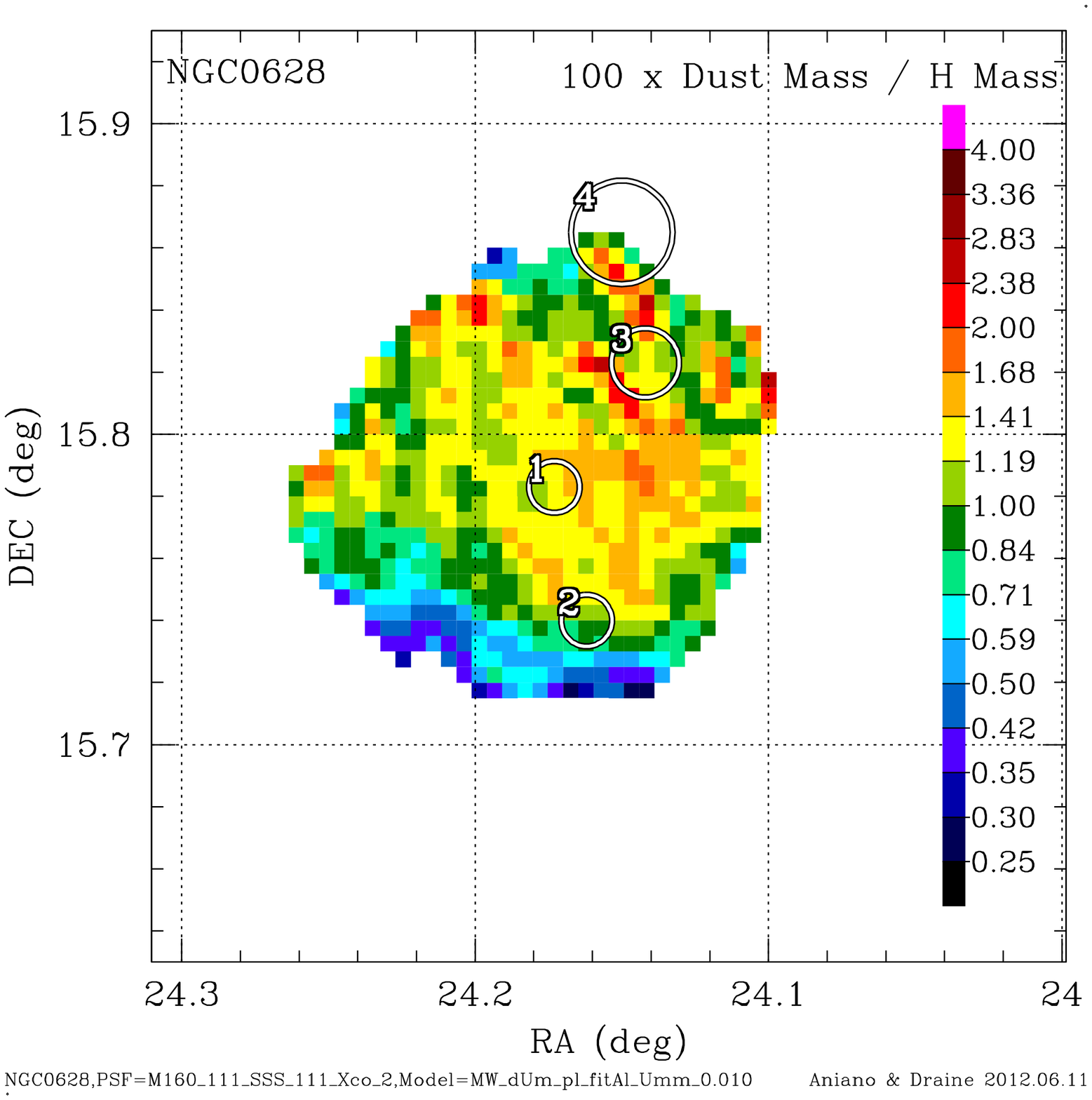}
\renewcommand \RtwoCtwo {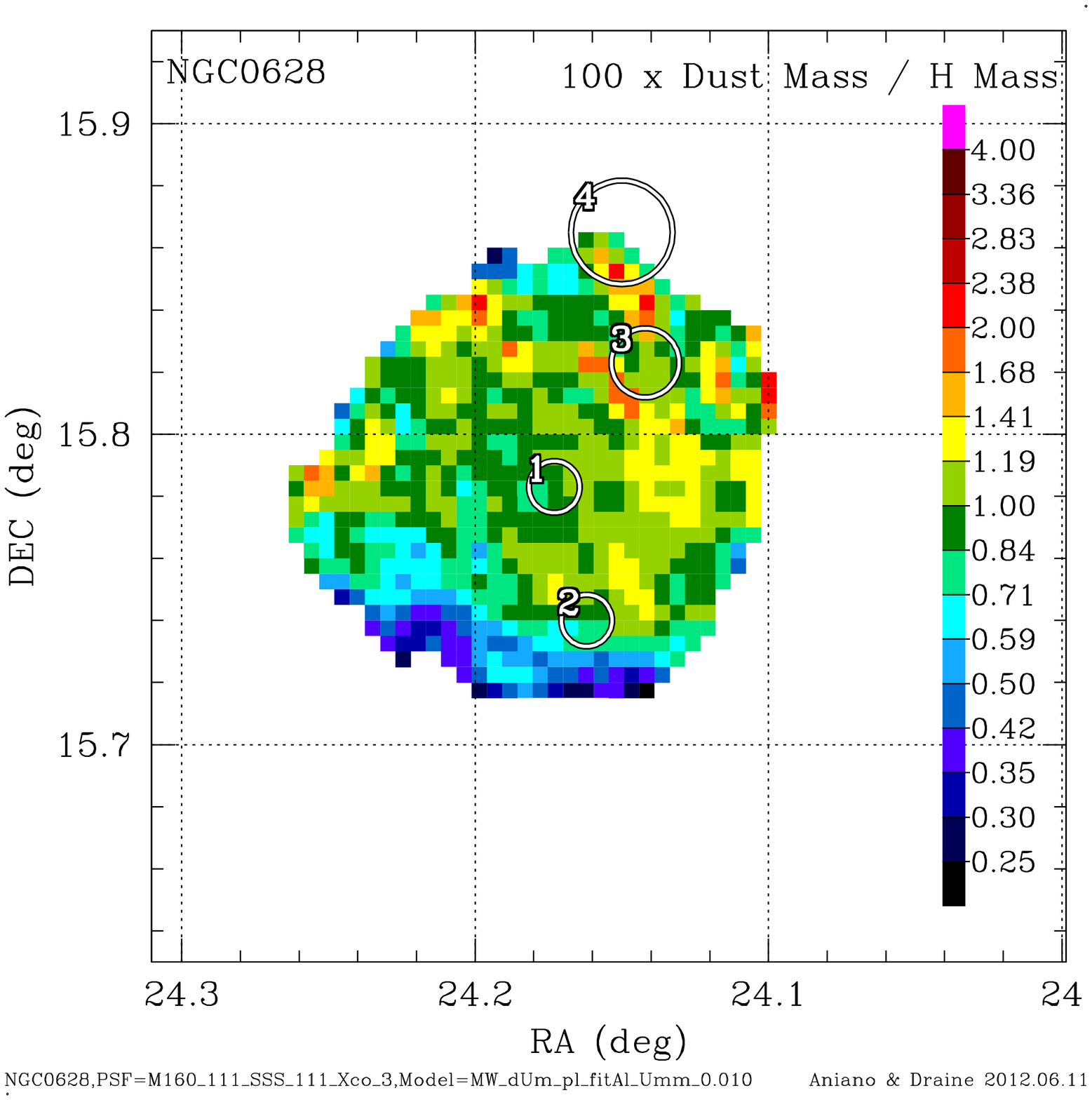}
\renewcommand \RtwoCthree {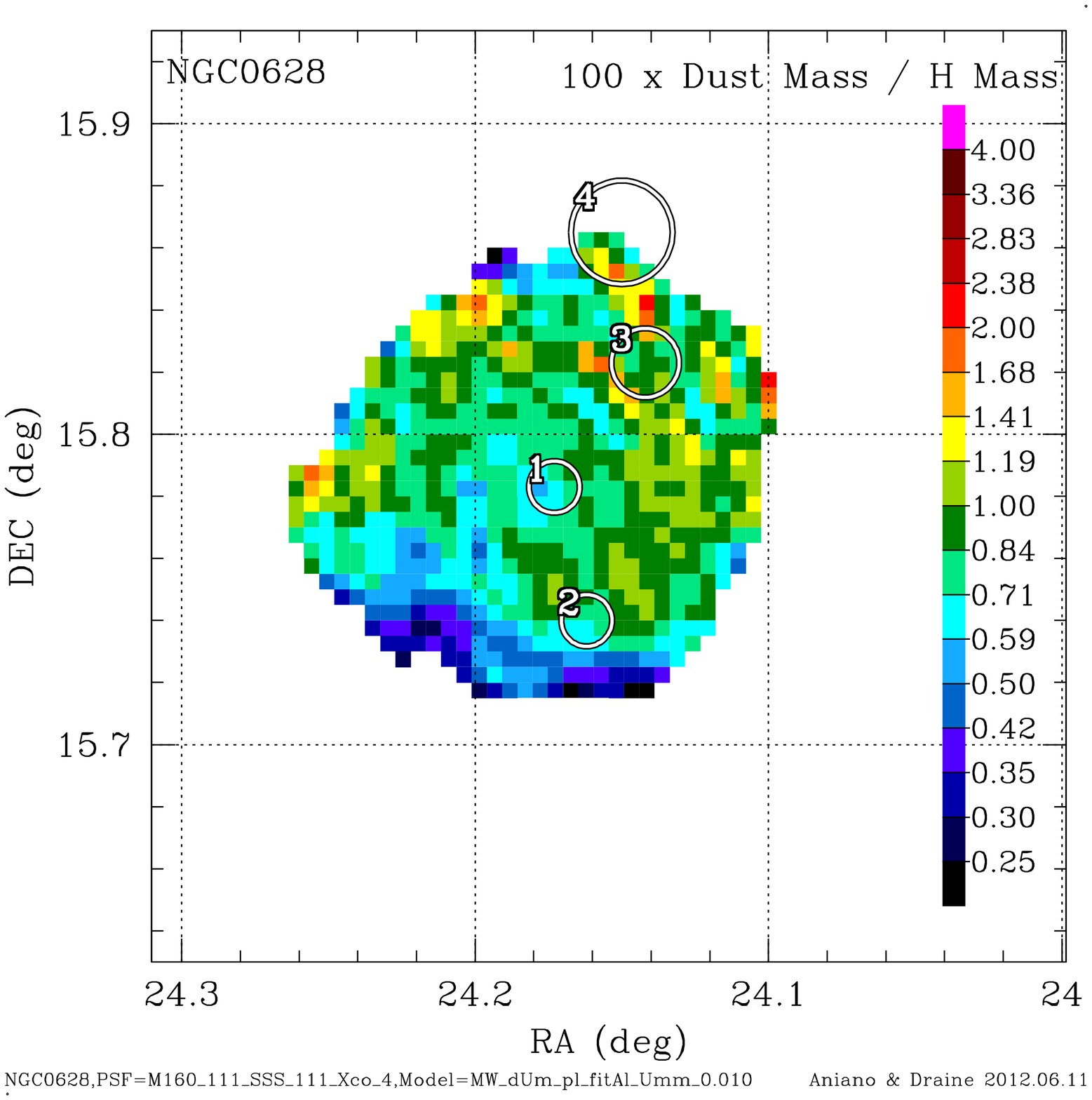} 
\renewcommand \RthreeCone {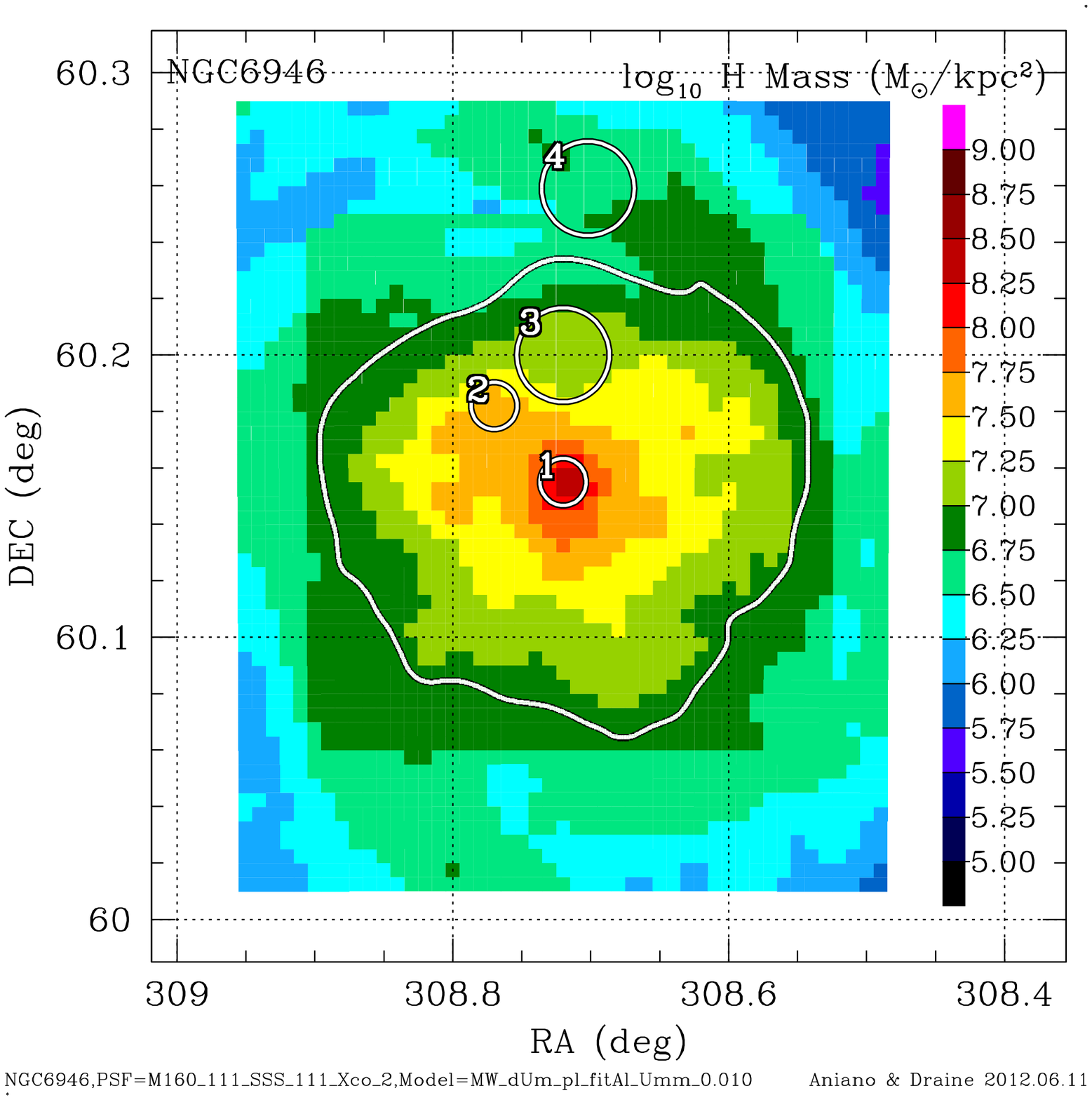}
\renewcommand \RthreeCtwo {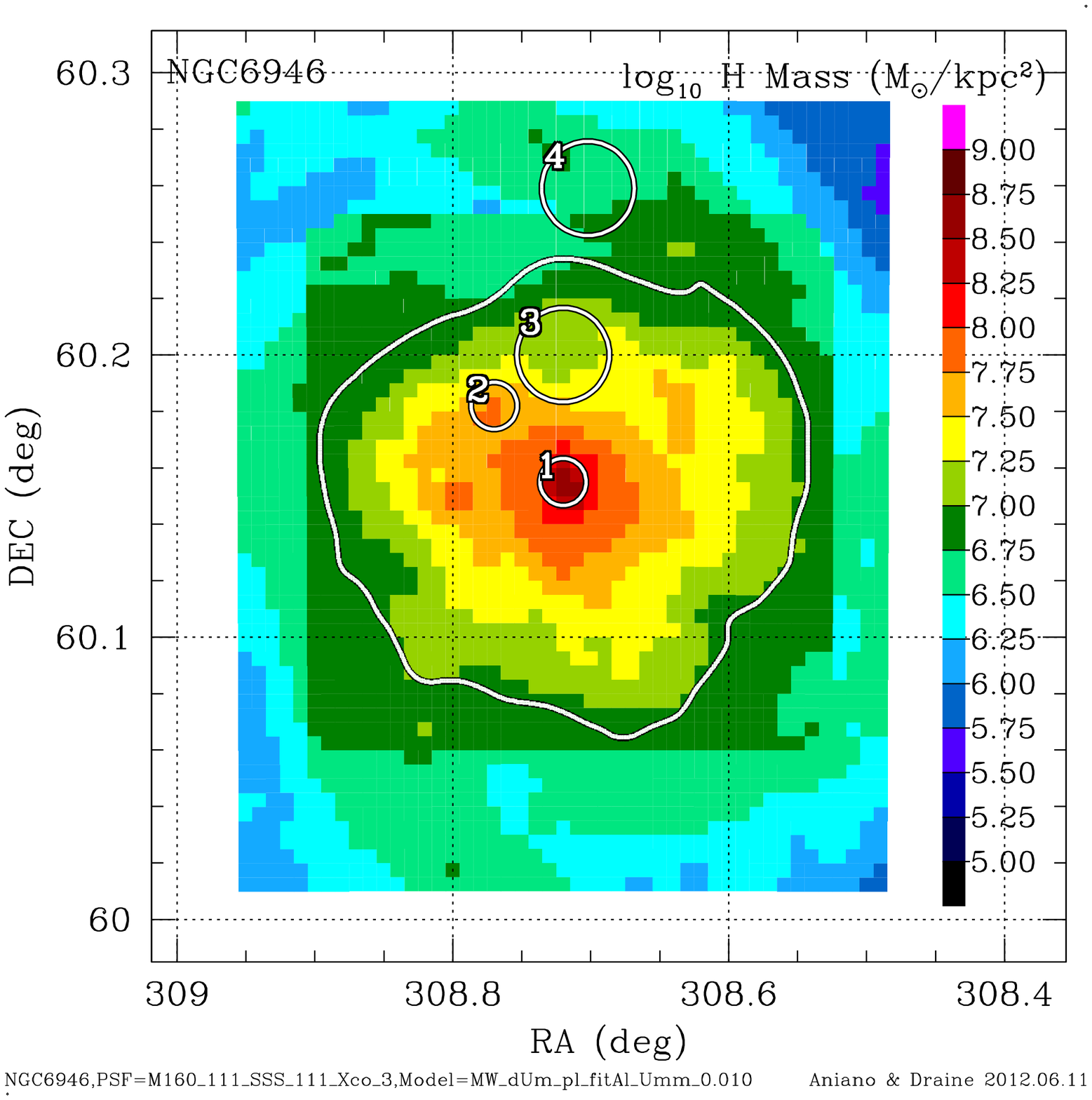}
\renewcommand \RthreeCthree {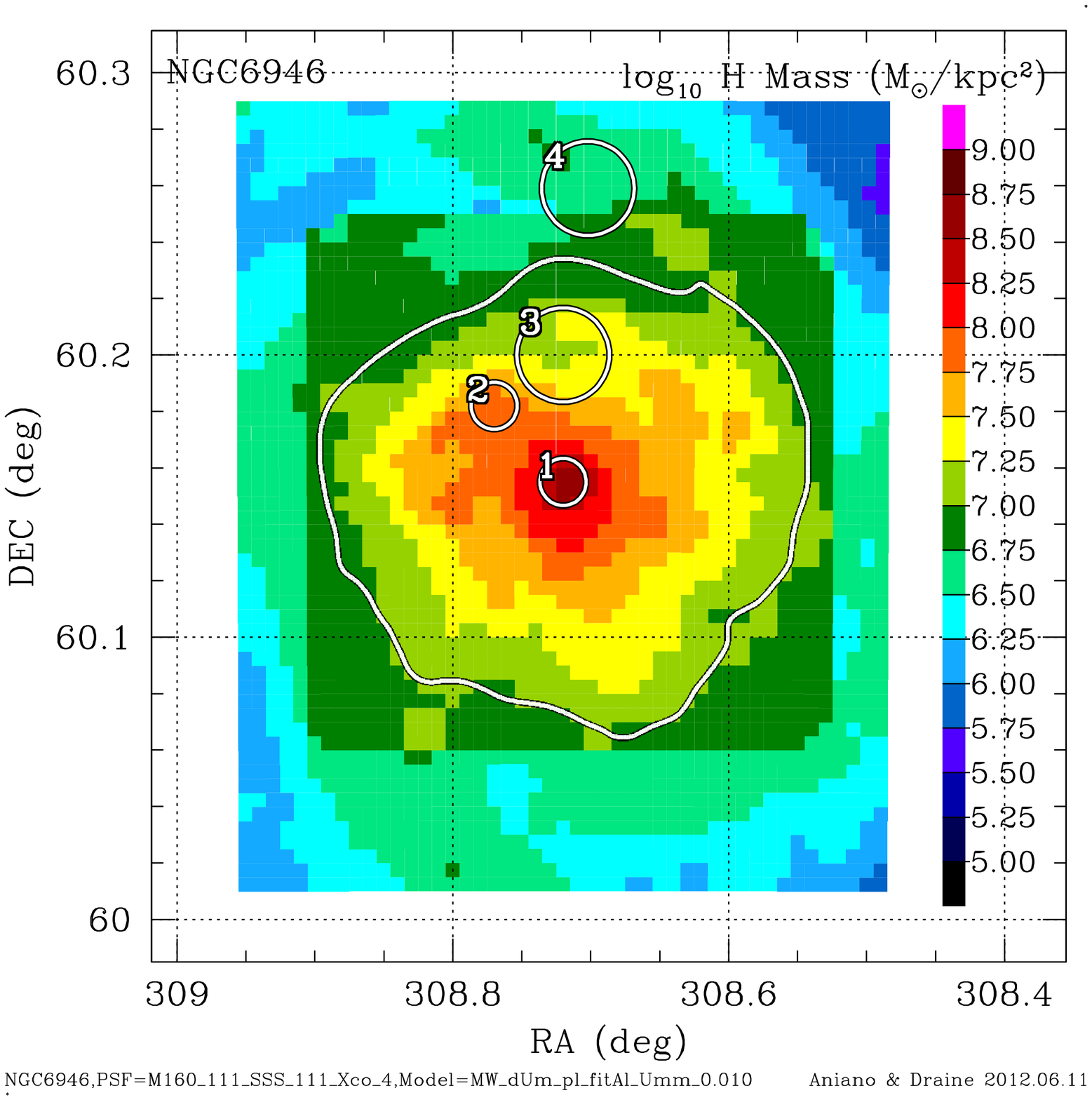}
\renewcommand \RfourCone {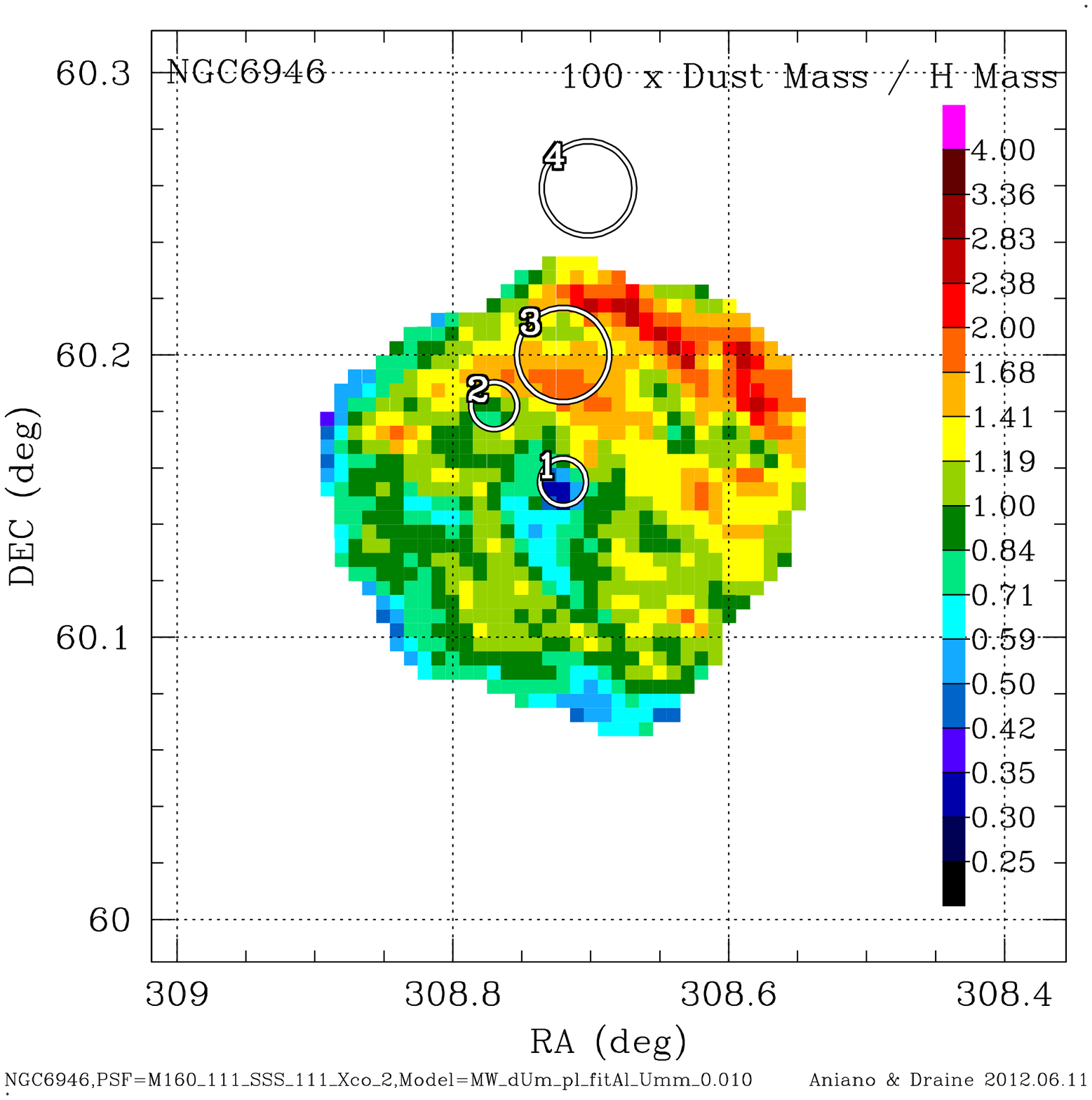}
\renewcommand \RfourCtwo {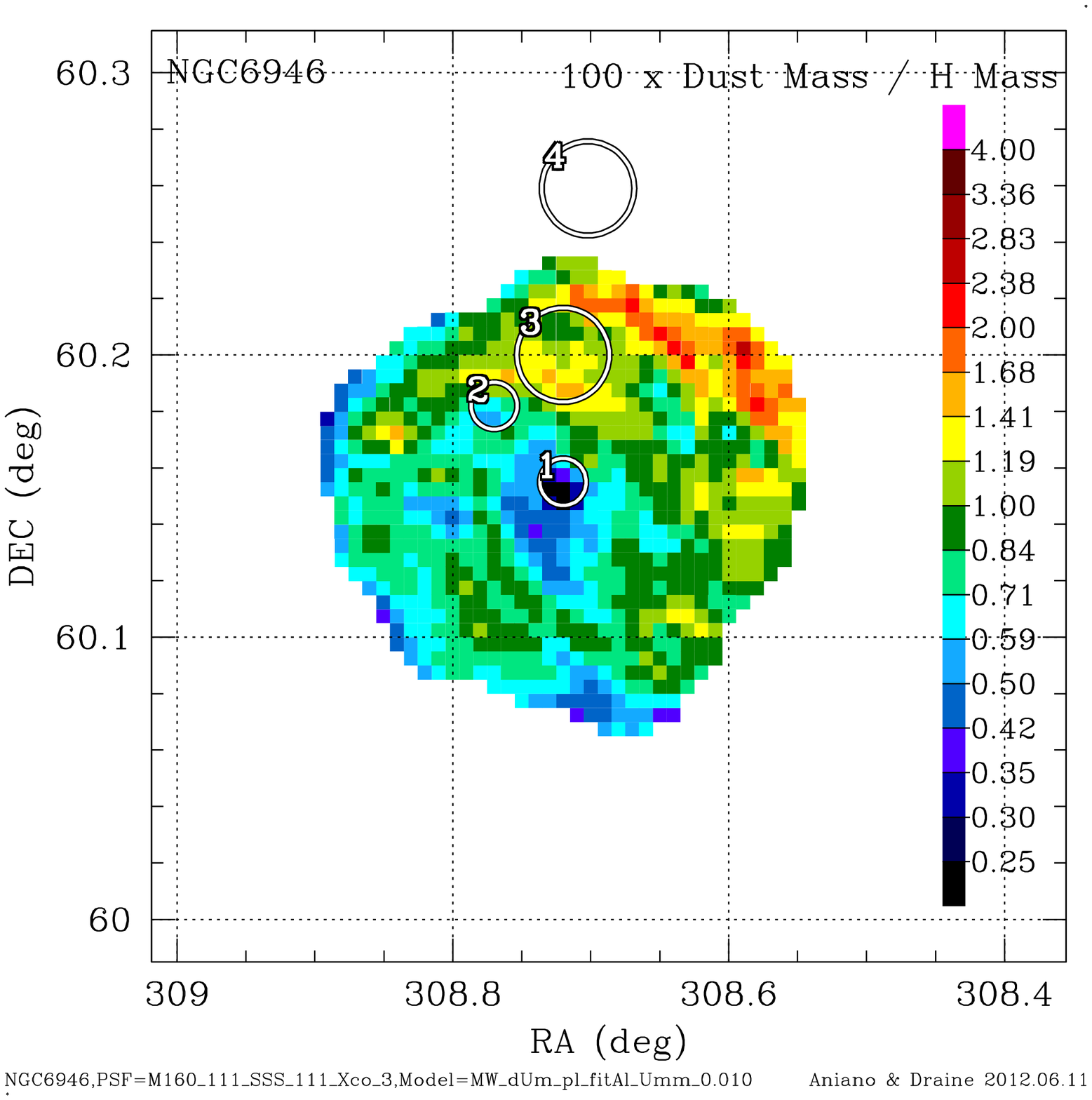}
\renewcommand \RfourCthree {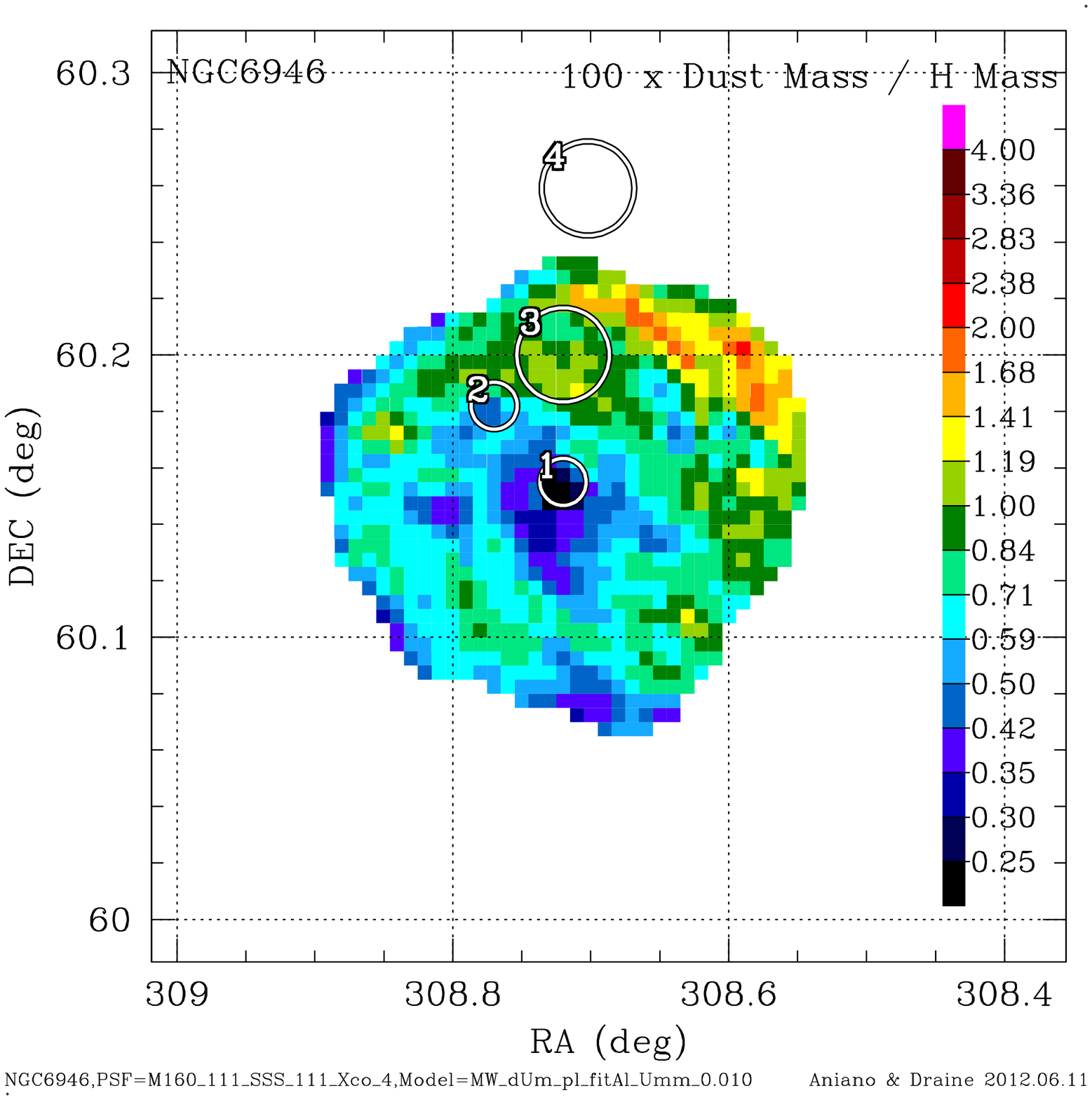}

\begin{figure} 
\centering 
\begin{tabular}{c@{$\,$}c@{$\,$}c} 
\footnotesize $\XCOxx = 2 $ & \footnotesize  $\XCOxx = 3 $ & \footnotesize  $\XCOxx = 4 $ \\
\FirstNormal
\SecondLast
& & \\
\ThirdNormal
\FourthLast
\end{tabular}
\vspace*{-0.5cm}
\caption{\footnotesize\label{fig:Xco}
Row 1: Total H surface density for NGC~628 at MIPS160 resolution
for $\XCOxx=2$, 3, and 4.
Row 2: Dust/gas mass ratio derived from the dust map in 
Figure \ref{fig:ngc0628-1}c, and the gas maps in row 1.
Row 3: Same as row 1, but for NGC~6946.
Row 4: Dust/gas mass ratio derived from the dust map in 
Figure \ref{fig:ngc6946-1}c, and the gas maps in row 3.
For NGC~628 the dust/gas maps seem best-behaved for $\XCOxx=4$ (Figure 3f).
For NGC~6946 the dust/gas map outside the central region also seem
best-behaved for $\XCOxx=4$ (Figure 3l).
}
\end{figure} 

\begin{table}
\newcommand \KKf {a}
\newcommand \size {b}
\newcommand \things  {c}
\newcommand \heracles{d}
\newcommand \mkt {e}
\newcommand \klf {f}
\newcommand \CWH {g}
\newcommand \sch {h}
\newcommand \mur {k}
\footnotesize
\begin{center}
\caption{\label{tab:galaxy properties}
         Galaxy Properties}
\begin{tabular}{l c c}
\hline
                                             & NGC~628                     & NGC~6946                   \\
\hline
Type$^\KKf$                                  & SAc                         & SABcd                      \\
$D$$^\KKf$  (Mpc)                            & 7.2                         & 6.8                        \\
Resolution at $D$                            & 1\as=34.9pc                 & 1\as=33.0pc                \\
galaxy optical size$^\size$ (kpc)            & 11.0 $\times$ 9.9           & 11.4 $\times$ 9.7          \\
galaxy mask area $A$ (kpc$^2$)               & 274                         & 307                        \\
galaxy mask radius $\sqrt{A/\pi}$ (kpc)      & 9.4                         & 9.9                        \\
$M(\HI)$$^\things$ ($\Msol$), total in map   & $(3.7\pm0.2)  \times10^9$   & $(5.5\pm0.3)  \times10^9$  \\
$M(\HH)$$^\heracles$ ($\Msol$), total in map & $(2.5\pm0.6)  \times10^9$   & $(9.7\pm2.7)  \times10^9$  \\
$M(\HI)$ ($\Msol$), within mask              & $(1.5\pm0.1)  \times10^9$   & $(2.1\pm0.1)  \times10^9$  \\
$M(\HH)$$^\heracles$ ($\Msol$), within mask  & $(2.0\pm0.2)  \times10^9$   & $(8.6\pm0.5)  \times10^9$  \\
$M_\Ha$  ($\Msol$), within mask              & $(3.5\pm0.3)  \times10^9$   & $(10.7\pm0.6) \times10^9$  \\
$\log_{10}({\rm O/H})+12$$^\mkt$             & $8.27\pm0.12$               & $8.37\pm0.06$              \\
$L(\Ha\alpha)$$^\klf$  ($\Lsol$), total      & $1.88\times10^7$            & $1.00\times10^8$           \\
SFR$^\CWH$ ($\Msol\yr^{-1}$), total          & $0.7\pm0.2$                 & $4.5\pm1.2$             \\
$F_\nu({\rm IRAC3.6})$ (Jy), within mask     & $0.83  \pm  0.10$           & $3.13 \pm  0.52$ \\
$F_\nu({\rm IRAC4.5})$ (Jy), within mask     & $0.57  \pm  0.06$           & $2.22 \pm  0.32$ \\
$F_\nu({\rm IRAC5.8})$ (Jy), within mask     & $0.98  \pm  0.23$           & $4.9  \pm  1.2$  \\
$F_\nu({\rm IRAC8.0})$ (Jy), within mask     & $2.61  \pm  0.44$           & $13.2 \pm  2.4$  \\
$F_\nu({\rm MIPS24})$ (Jy), within mask      & $3.04  \pm  0.33$           & $18.9 \pm  2.0$  \\
$F_\nu({\rm MIPS70})$ (Jy), within mask      & $31.5  \pm  9.8$            & $196  \pm  55$   \\
$F_\nu({\rm MIPS160})$ (Jy), within mask     & $104   \pm  20$             & $420  \pm  127$  \\
$F_\nu({\rm PACSS70})$ (Jy), within mask     & $39    \pm  12$             & $245  \pm  58$   \\
$F_\nu({\rm PACSS100})$ (Jy), within mask    & $73    \pm  21$             & $433  \pm  122$  \\
$F_\nu({\rm PACSS160})$ (Jy), within mask    & $110   \pm  20$             & $526  \pm  126$  \\
$F_\nu({\rm SPIRE250})$ (Jy), within mask    & $59.5  \pm  7.0$            & $247  \pm  29$   \\
$F_\nu({\rm SPIRE350})$ (Jy), within mask    & $27.4  \pm  3.3$            & $101  \pm  12$   \\
$F_\nu({\rm SPIRE500})$ (Jy), within mask    & $10.5  \pm  1.4$            & $35.4 \pm  4.2$  \\
$M_\dust$ ($\Msol$), within mask             & $(2.9\pm0.4)\times10^7$     & $(6.7\pm0.6)\times10^7$    \\
$100\times M_\dust/M_{\rm H}$, within mask   & $0.82\pm0.17$               & $0.63\pm0.09$              \\
$L_\dust$ ($\Lsol$), within mask             & $(6.8\pm0.4)\times10^{9}$   & $(3.3\pm0.3)\times10^{10}$ \\
$L_\PDR$ ($\Lsol$), within mask              & $(7.9\pm1.4)\times10^{8}$   & $(4.7\pm0.8)\times10^{9}$  \\
$\langle \Umin\rangle $, within mask         & $1.6\pm0.3$                 & $3.1\pm0.6$                \\
$\overline{U}$, within mask                  & $1.7\pm0.3$                 & $3.6\pm0.6$                \\
$\langle f_\PDR\rangle $, within mask        & $(11.6\pm1.3)\%$            & $(14.2\pm2.8)\%$           \\
$\langle \qpah\rangle $, within mask         & $(3.7\pm0.3)\%$             & $(3.6\pm0.4)\%$            \\
\hline
\multicolumn{3}{l}{$^\KKf$ From \cite{Kennicutt+Calzetti+Aniano+etal_2011}.}\\
\multicolumn{3}{l}{$^\size$ Major and minor radii at $\mu_B$=25 mag arcsec$^{-2}$ isophote, from \cite{Kennicutt+Calzetti+Aniano+etal_2011}.}\\
\multicolumn{3}{l}{$^\things$ We use the ``Natural'' (NA) weighting maps (see  \citet{Walter+Brinks+deBlok+etal_2008}  for details).}\\
\multicolumn{3}{l}{$^\heracles$ From \citet{Leroy+Walter+Bigiel+etal_2009},
                   for $\XCOxx=4$ (see text).}\\
\multicolumn{3}{l}{$^\mkt$ PT05 H\,II region abundances, unweighted average, from
                   \citet{Moustakas+Kennicutt+Tremonti+etal_2010}.}\\
\multicolumn{3}{l}{$^\klf$ From \citet{Kennicutt+Lee+Funes+etal_2008}, uncorrected for internal extinction.}\\
\multicolumn{3}{l}{$^\CWH$ From \citet{Calzetti+Wu+Hong+etal_2010}, based on H$\alpha$+24$\micron$.}\\
\end{tabular}

\end{center}
\end{table}

\clearpage
\subsection{\label{sec:0628}NGC~628}

\subsubsection{Maps of Gas and Dust}

NGC\,628 (= M\,74), at a distance $D=7.2\Mpc$, is classified as SAc.
With major and minor optical diameters of 10.5 and 9.5 arcmin, 
it is well resolved even by the MIPS160 camera.

The star formation rate is estimated to be $0.7\pm0.2\Msol\yr^{-1}$
\citep{Calzetti+Wu+Hong+etal_2010}.
Two supernovae have been observed in NGC\,628: SN~2002ap (Type Ic)
and SN~2003gd (Type II-P).

The total H\,I mass from 21-cm observations
is $M({\rm H\,I})=(3.7\pm0.2)\times10^9\Msol$ 
\citep{Walter+Brinks+deBlok+etal_2008}
and the total H$_2$ mass estimated from observations of CO 2-1
is $M(\HH)=(2.5\pm0.6)\times10^9(\XCOxx/4)\Msol$
\citep{Leroy+Walter+Bigiel+etal_2009}.
The fact that the adopted value of 
$\XCO$ is larger than the value $\XCOxx\approx2$
found for resolved CO clouds in the Milky Way
will be discussed in Section \ref{sec:discussion} below.

\renewcommand \RoneCone {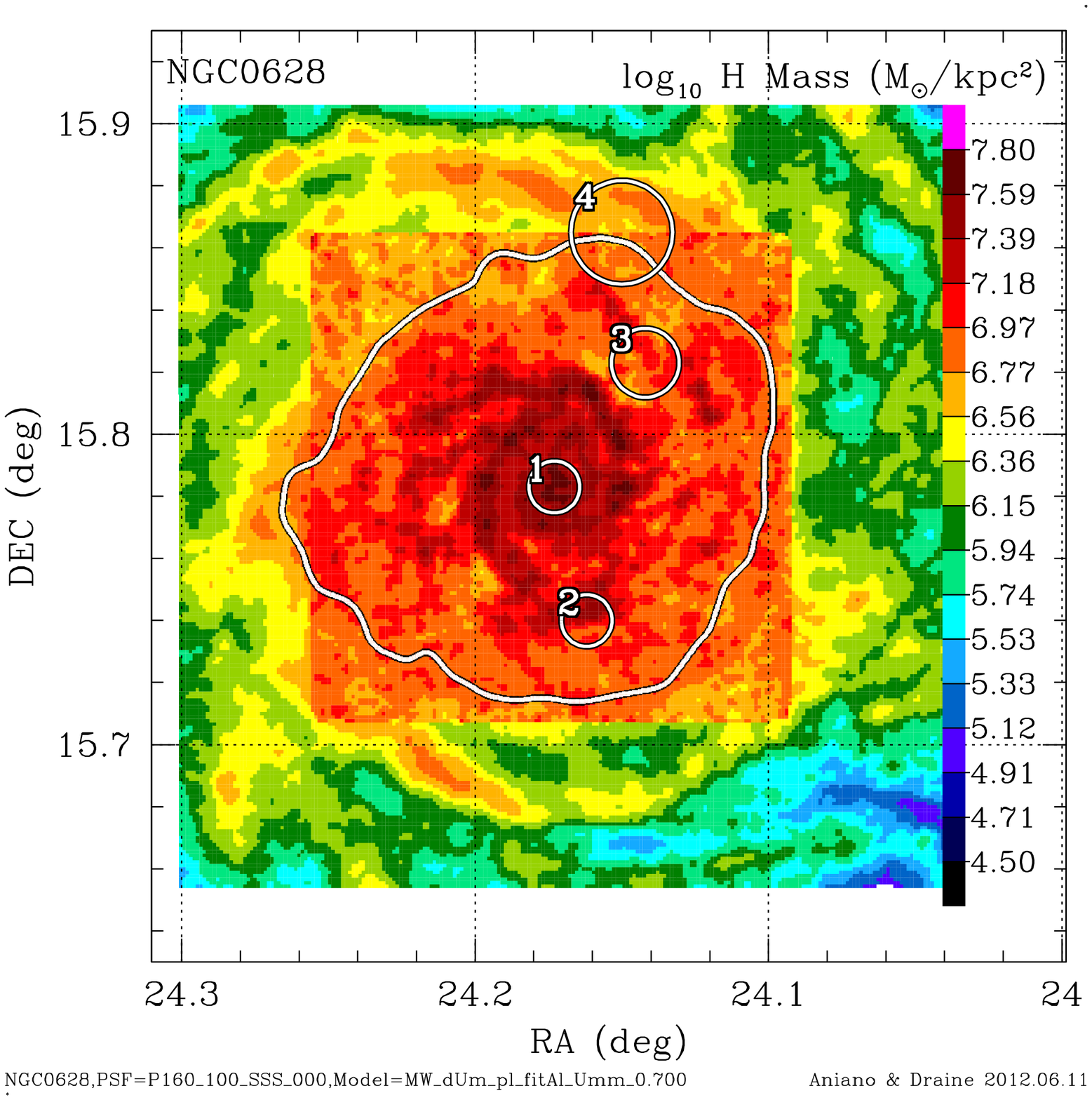}
\renewcommand \RoneCtwo {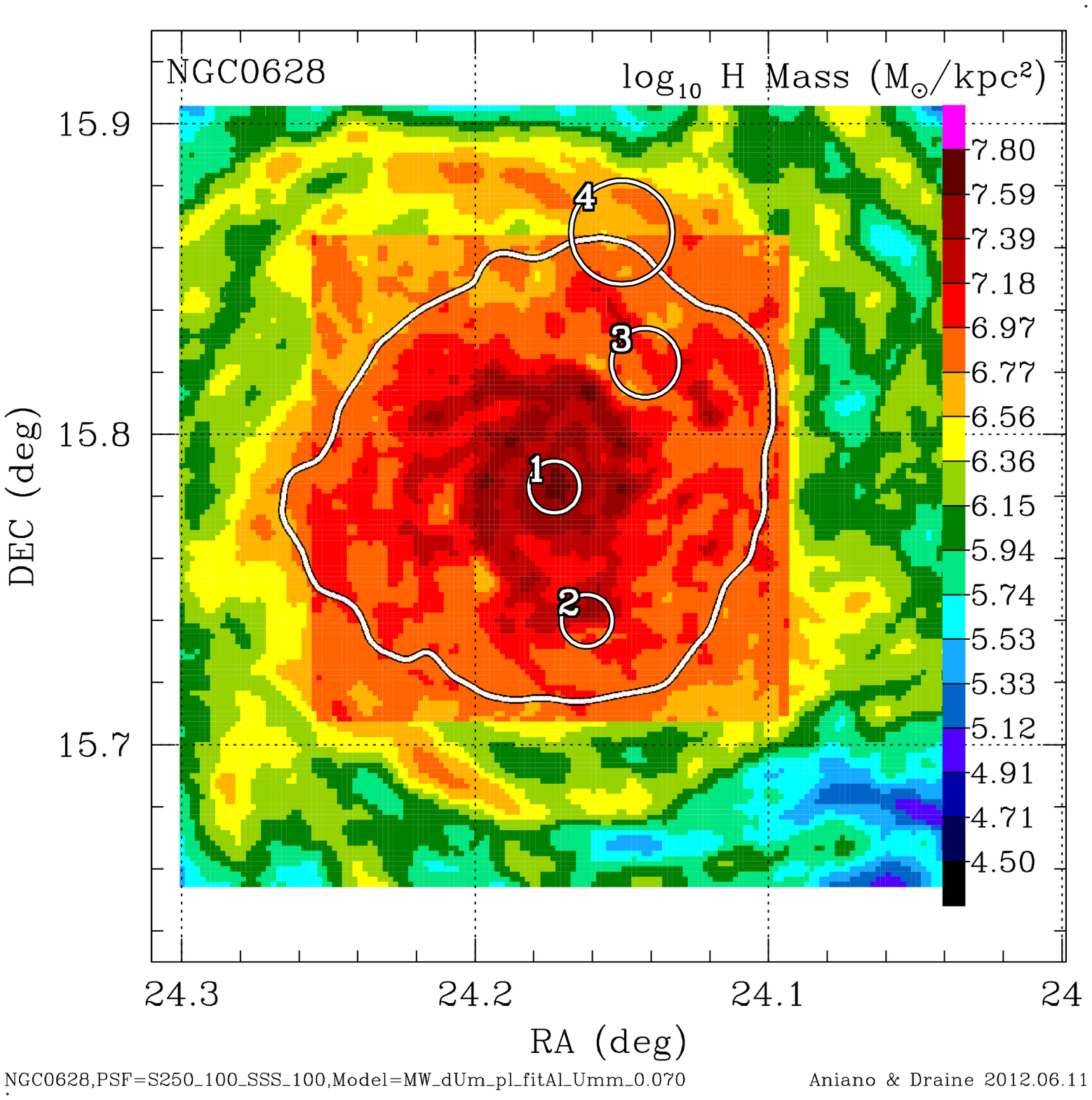}
\renewcommand \RoneCthree {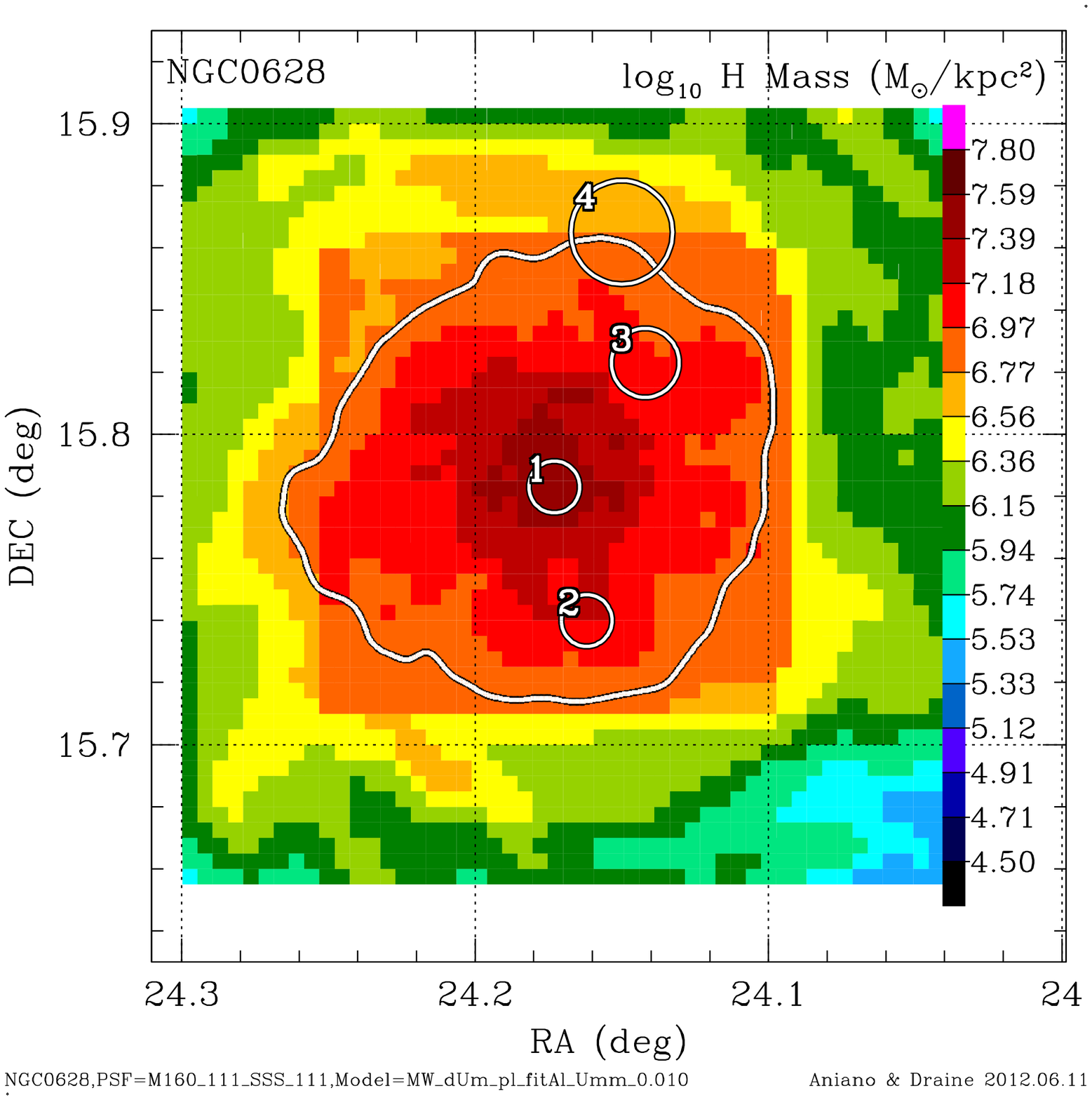}
\renewcommand \RtwoCone {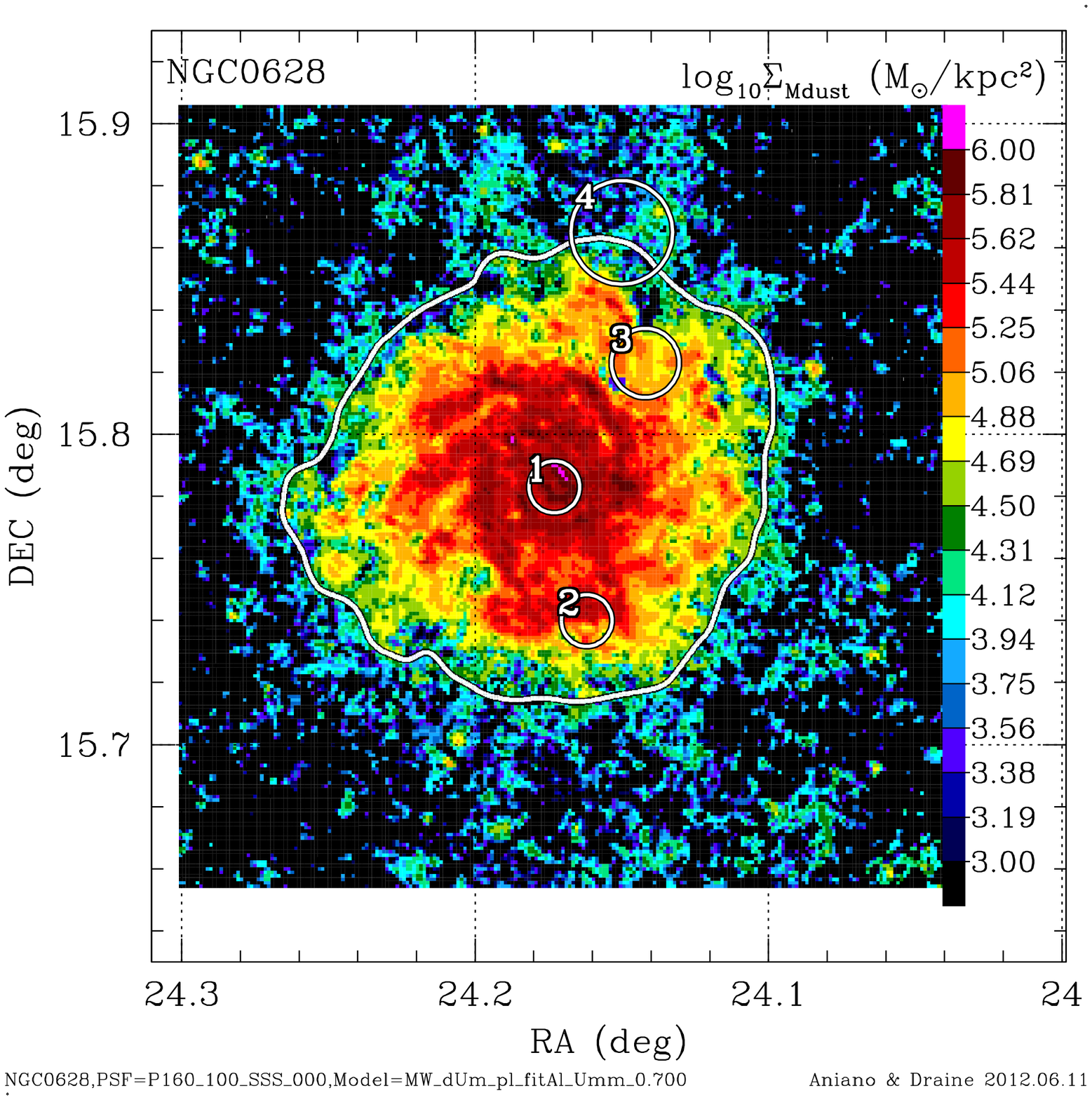}
\renewcommand \RtwoCtwo {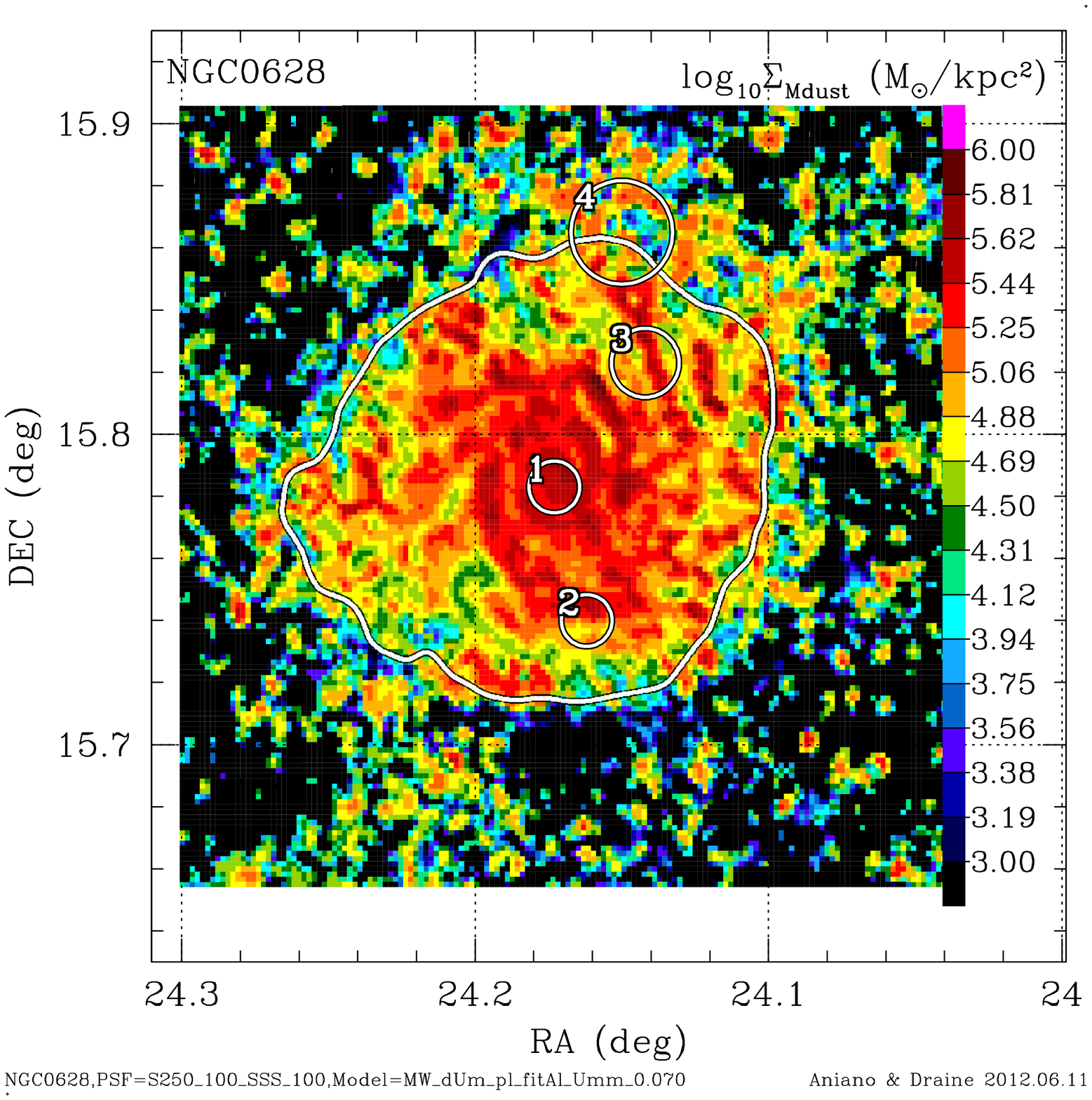}
\renewcommand \RtwoCthree {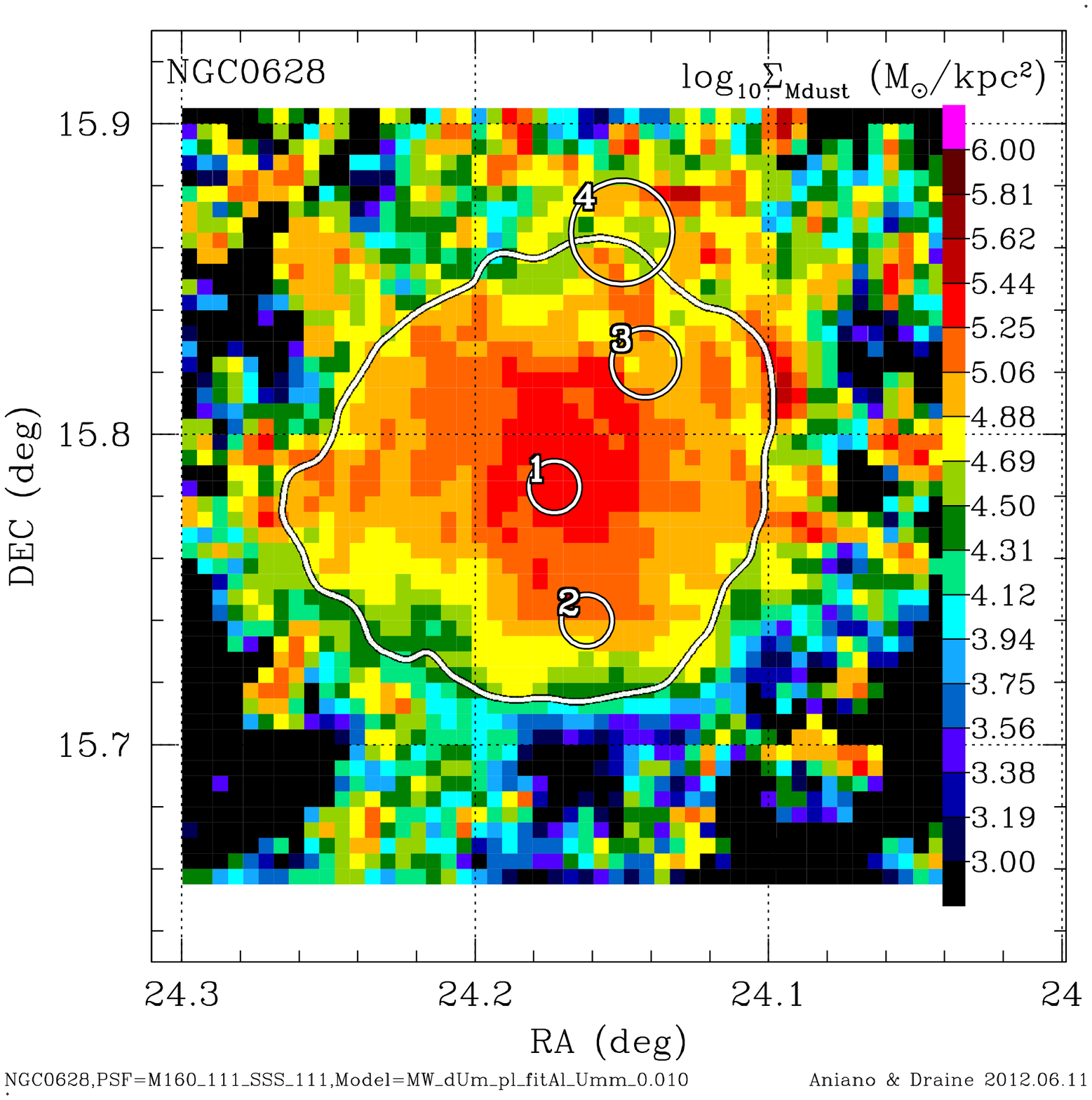} 
\renewcommand \RthreeCone {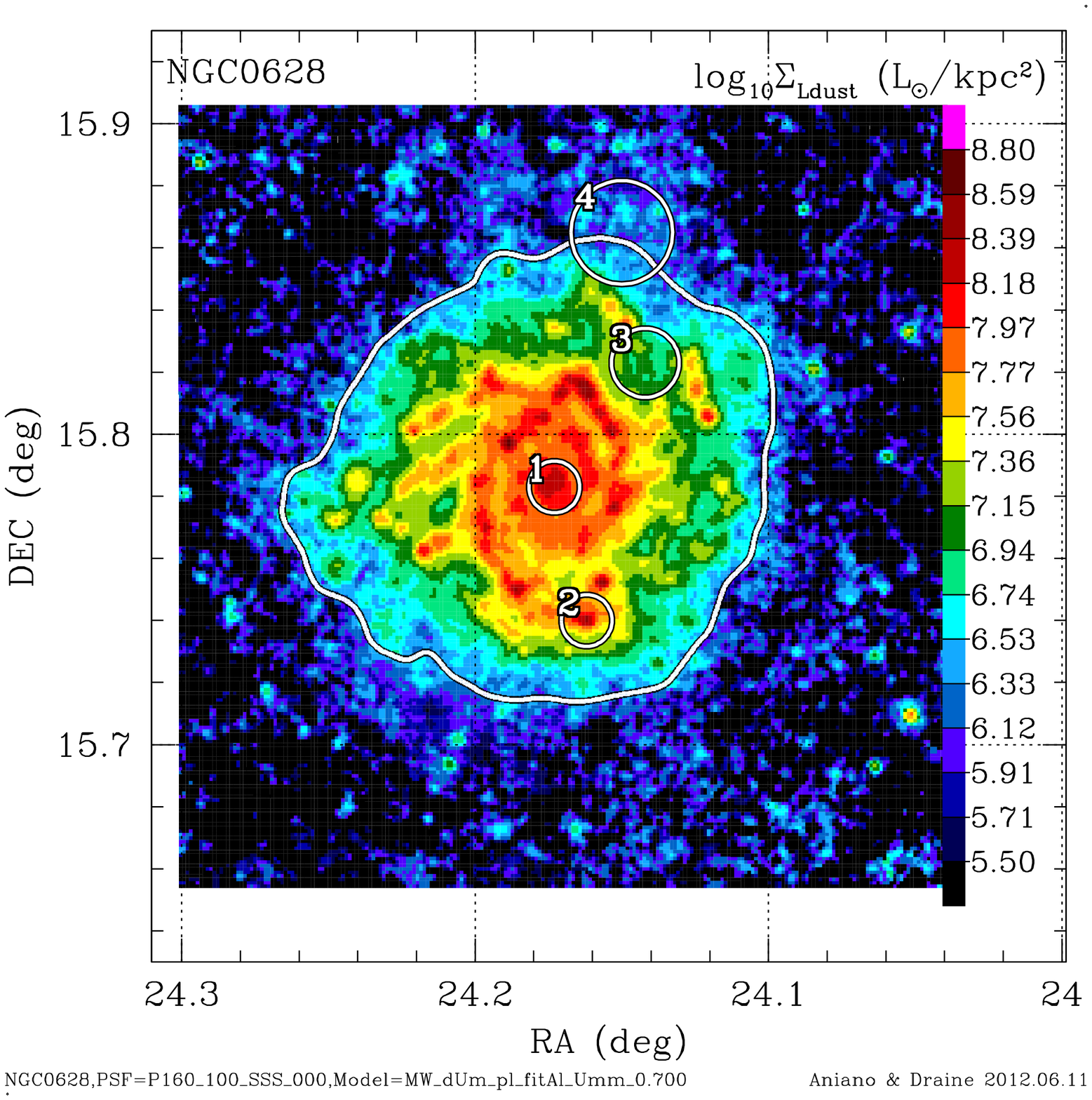}
\renewcommand \RthreeCtwo {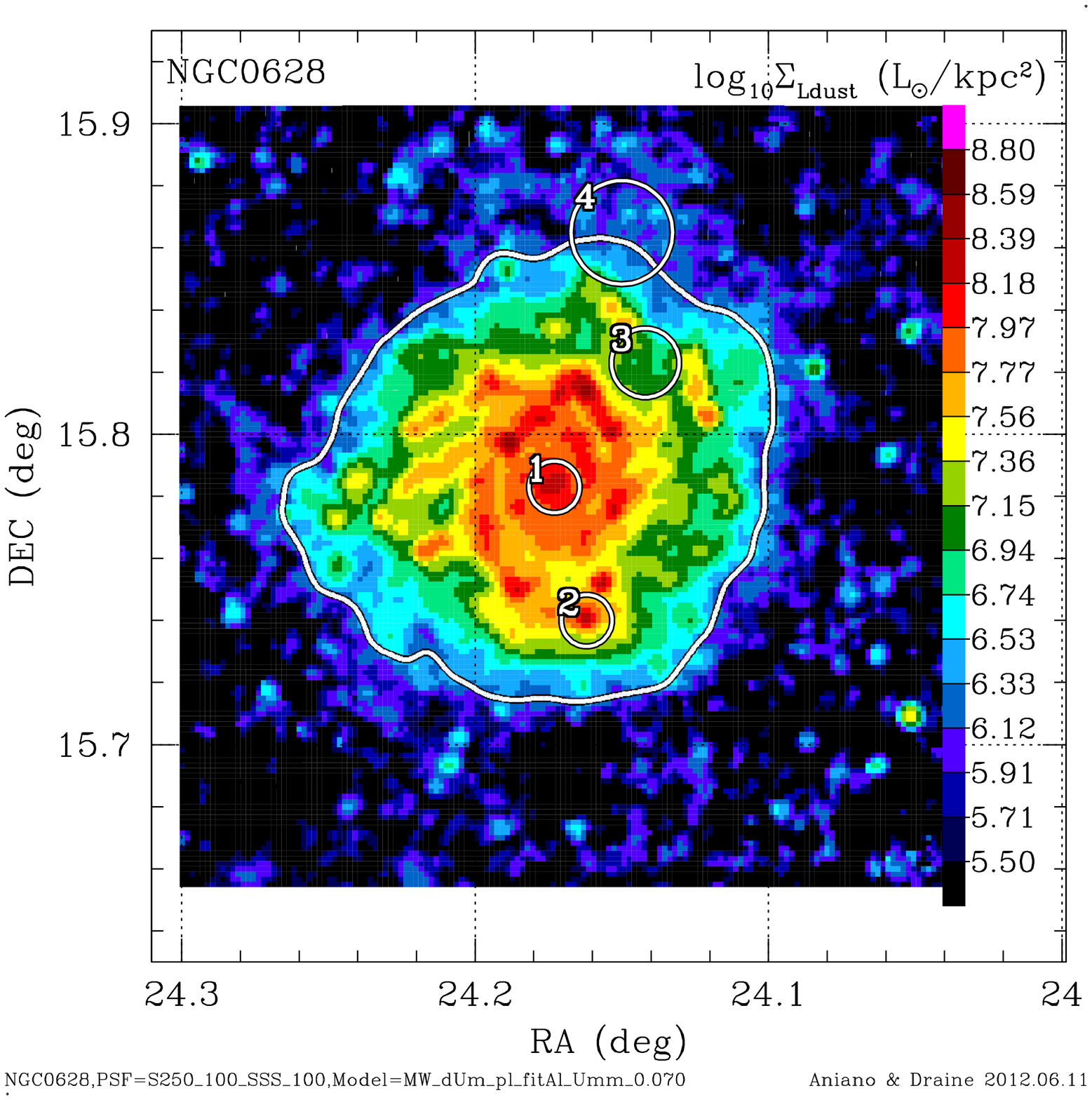}
\renewcommand \RthreeCthree {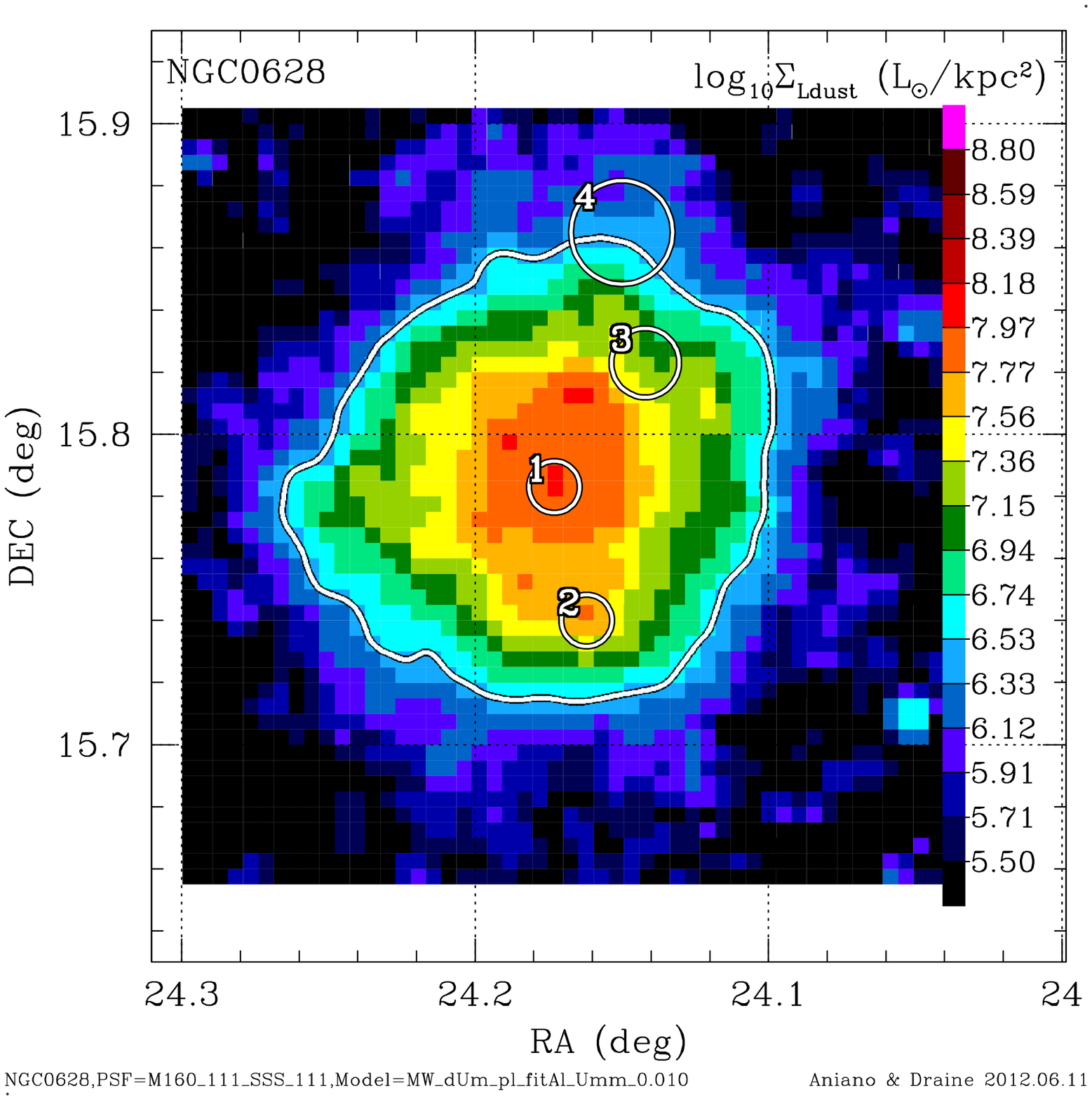}
\renewcommand \RfourCone {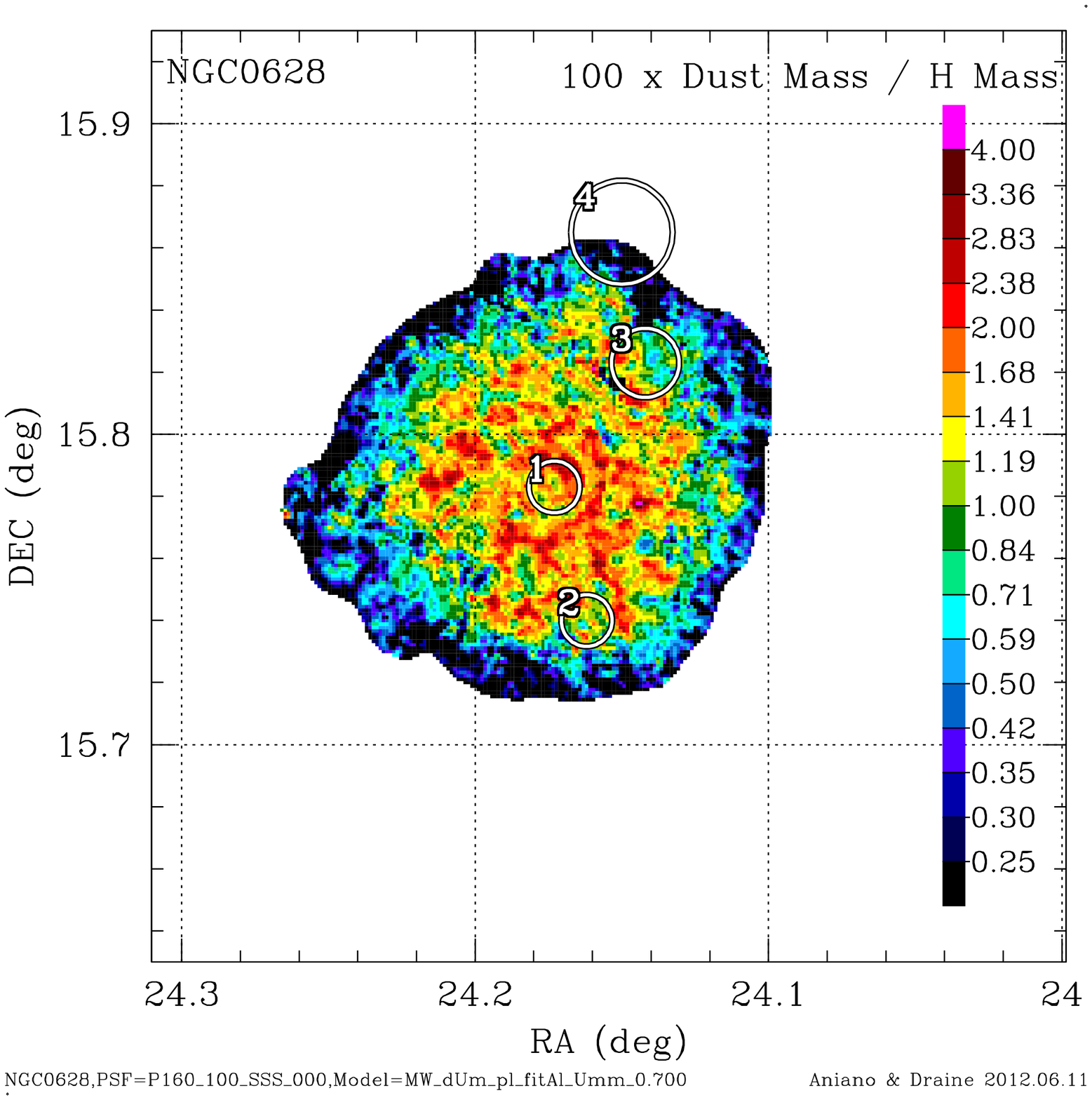}
\renewcommand \RfourCtwo {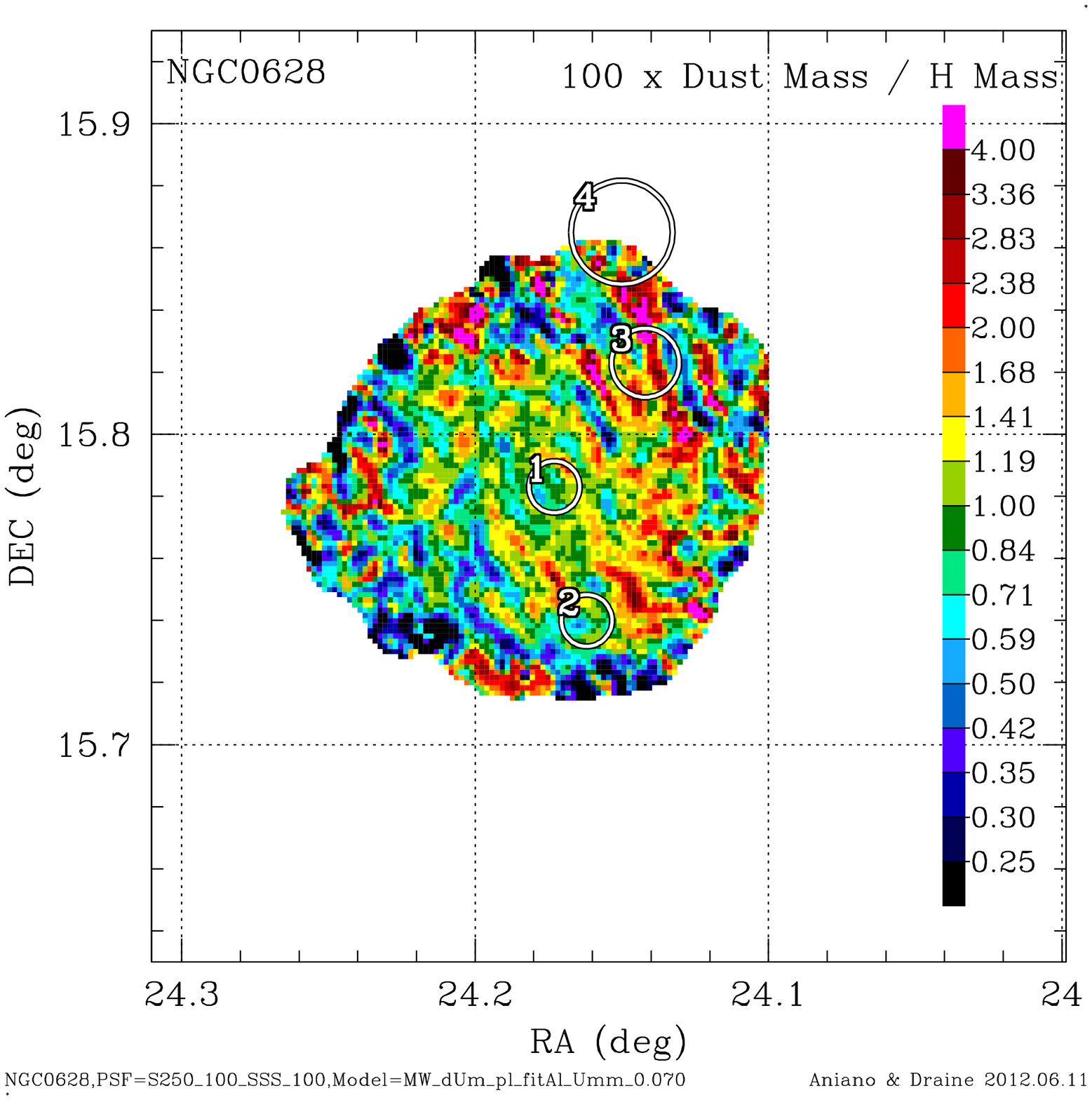}
\renewcommand \RfourCthree {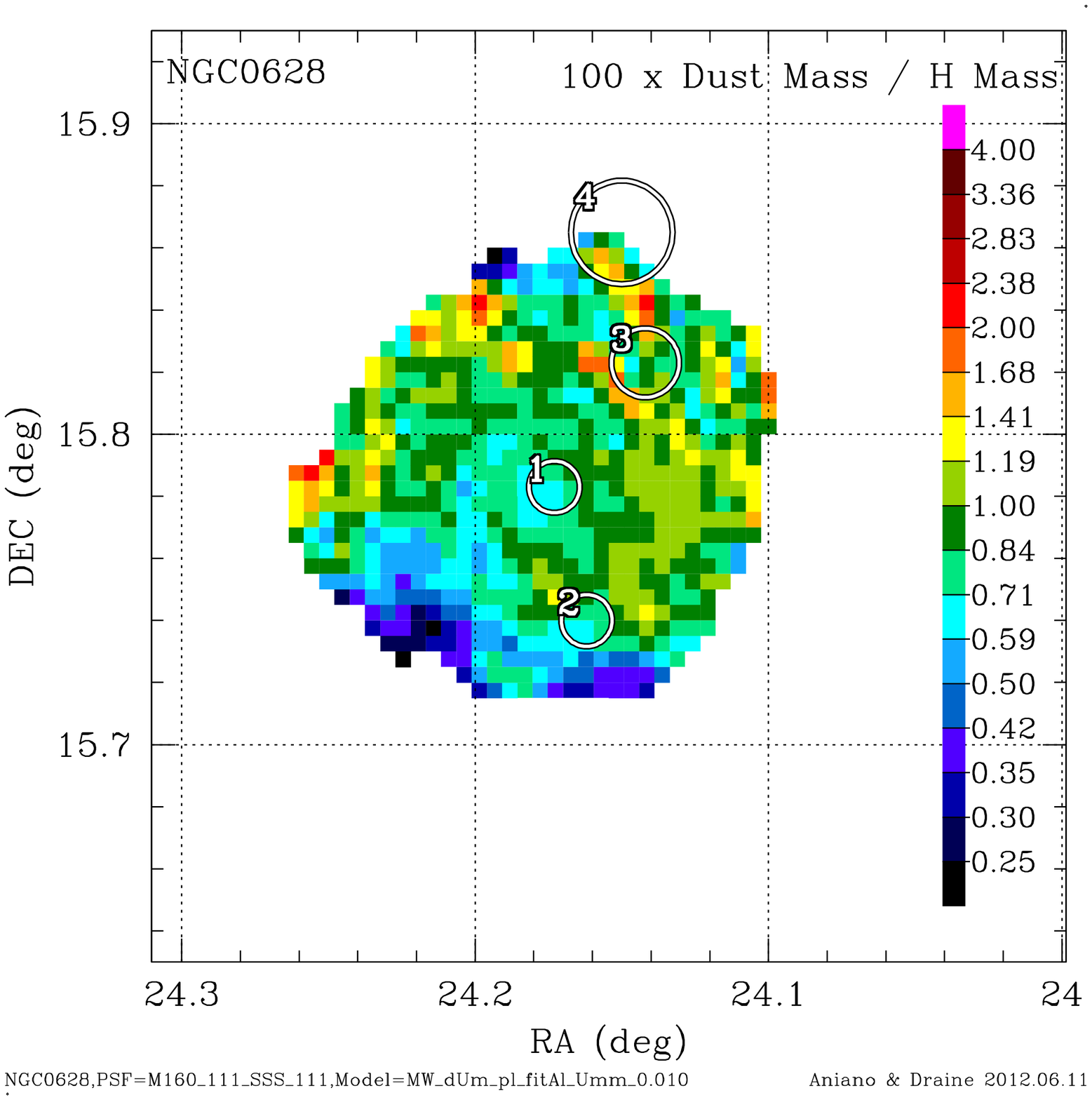}
\ifthenelse{\boolean{make_heavy}}{ }
{ \renewcommand \RoneCone    {No_image.eps}
\renewcommand \RtwoCone    {No_image.eps}
\renewcommand \RthreeCone {No_image.eps}
\renewcommand \RfourCone {No_image.eps}
\renewcommand \RoneCtwo    {No_image.eps}
\renewcommand \RtwoCtwo    {No_image.eps}
\renewcommand \RthreeCtwo {No_image.eps}
\renewcommand \RfourCtwo {No_image.eps}}

\begin{figure}
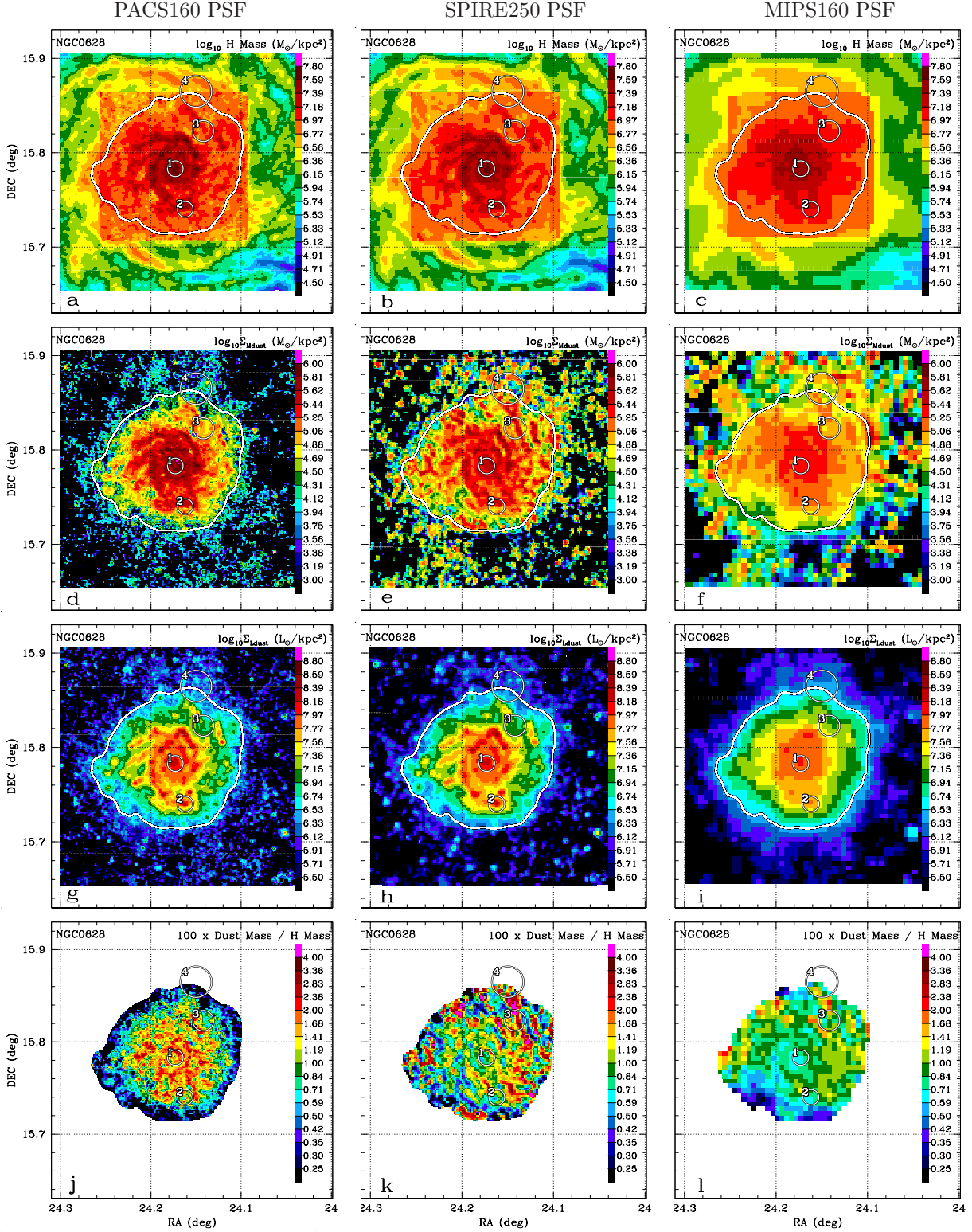
 
\centering 
\begin{tabular}{c@{$\,$}c@{$\,$}c} 
\footnotesize PACS160 PSF & \footnotesize SPIRE250 PSF & \footnotesize MIPS160 PSF \\
\FirstNormal
\SecondNormal
\ThirdNormal
\FourthLast
\end{tabular}
\vspace*{-0.5cm}
\caption{\footnotesize\label{fig:ngc0628-1}
NGC~628 at the resolution of PACS160 (left), SPIRE250 (center), and MIPS160
  (right).  PACS160 resolution models are based only on IRAC, MIPS24,
and PACS data; SPIRE250 resolution models are based on IRAC, MIPS24, PACS
and SPIRE250 data; MIPS160 resolution models are based on all (IRAC, MIPS,
PACS, SPIRE) data.
Row 1: surface density of H (both H\,I and $\HH$) for $\XCOxx=4$
(see text).  
Row 2: estimated dust surface density $\Sigma_{M_\dust}$ (see text).
Row 3: dust luminosity surface density $\Sigma_{L_\dust}$.
Row 4: dust/H mass ratio over the main galaxy.
The irregular white contour is the boundary of the ``galaxy mask'' (see text).
White circles are selected apertures (see Fig.\ \ref{fig:ngc0628-5}). }
\end{figure} 

Figure \ref{fig:ngc0628-1} shows the resulting maps of $\Ha$ surface
density (panels a-c), dust surface density $\Sigma_{M_\dust}$ (panels
d-f), dust luminosity surface density $\Sigma_{L_\dust}$ (panels g-i),
and dust/H mass ratio (panels j-l), obtained by fitting photometry
with PACS160, SPIRE250, and MIPS160 resolution, using all the
compatible cameras in each case.  The dust models with PACS160
resolution show clear spiral structure, but the noise in the smaller
pixels is such that the dust is only reliably detected at surface
densities $\Sigma_{M_\dust}\gtsim 10^{5.0}\Msol\kpc^{-2}$,
corresponding to $A_V\gtsim 0.7\,$mag.  If the mapping is done at the
resolution of the SPIRE250 camera, the dust is reliably detected for
$\Sigma_{M_\dust} \gtsim 10^{4.5}\Msol\kpc^{-2}$, corresponding to
$A_V\gtsim 0.2\,$mag; at this resolution, the spiral structure is
still visible.  Maps made at MIPS160 resolution are the most
sensitive, because of the larger pixel size, and the fact that they
use data from all of the cameras, including the MIPS160 camera.
Unfortunately, the lower resolution of these maps ($36\arcsec$ FWHM)
largely washes out the spiral structure which is visible in higher
resolution maps of NGC~628.  Nevertheless, imaging at MIPS160
resolution allows reliable detection of dust at surface densities as
low as $10^{4.3}\Msol\kpc^{-2}$, or $A_V\gtsim 0.14\,$mag.

We note that the dust/H mass ratio maps change
significantly when the modeling is done at the different
resolution/camera combination (see last row of
Fig.\ \ref{fig:ngc0628-1}).  These discrepancies are mainly due to the
low sensitivity of the PACS cameras in the low-surface brightness
areas (compare the MIPS and PACS flux in the outer parts of the
galaxies in Figures \ref{Fig_0628_ori} and \ref{Fig_0628_ori}), and
discrepancies in the high surface brightness areas. In \S 7 we discuss
the PACS-MIPS photometry discrepancies further.

\renewcommand \RoneCone {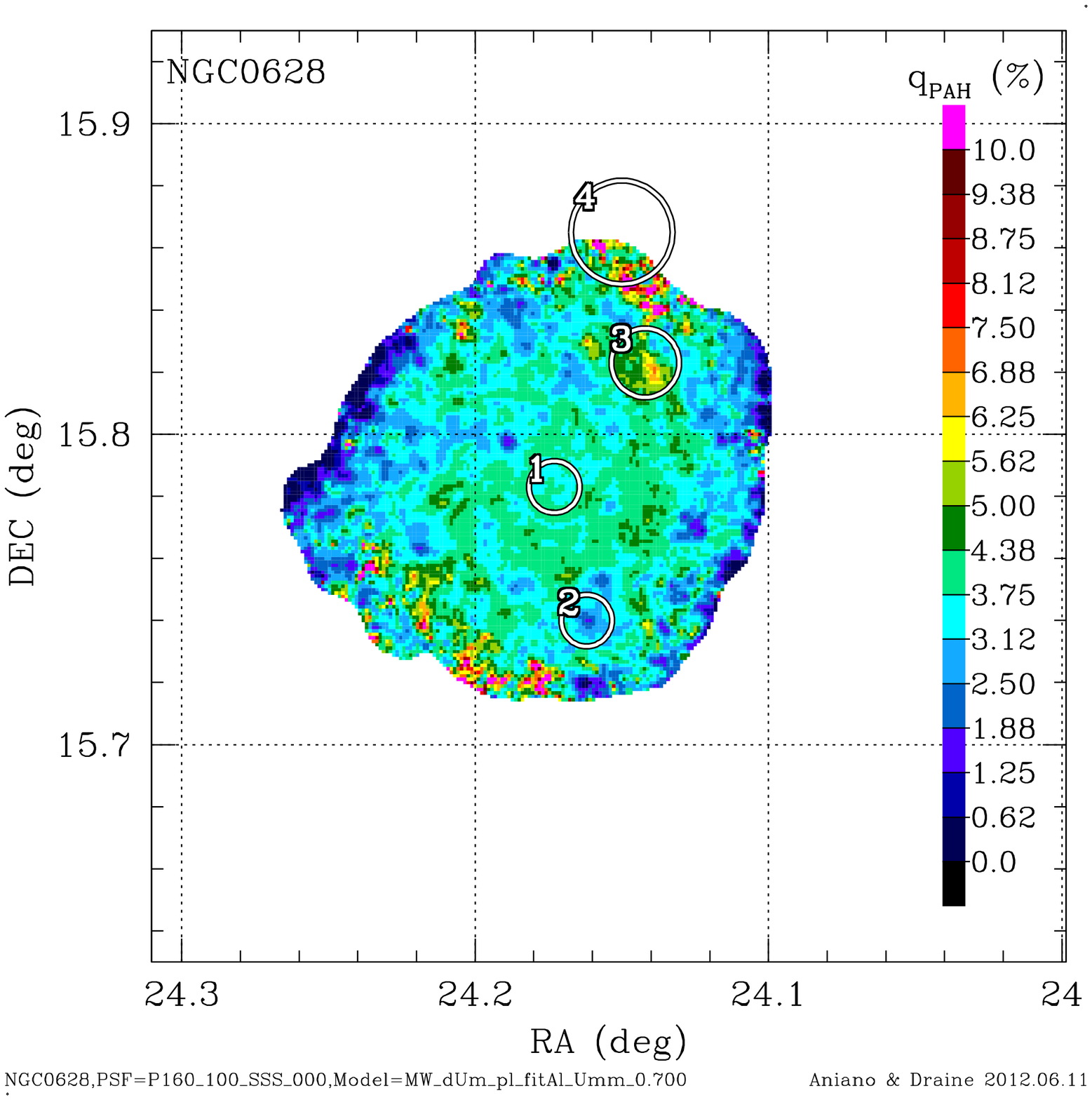}
\renewcommand \RoneCtwo {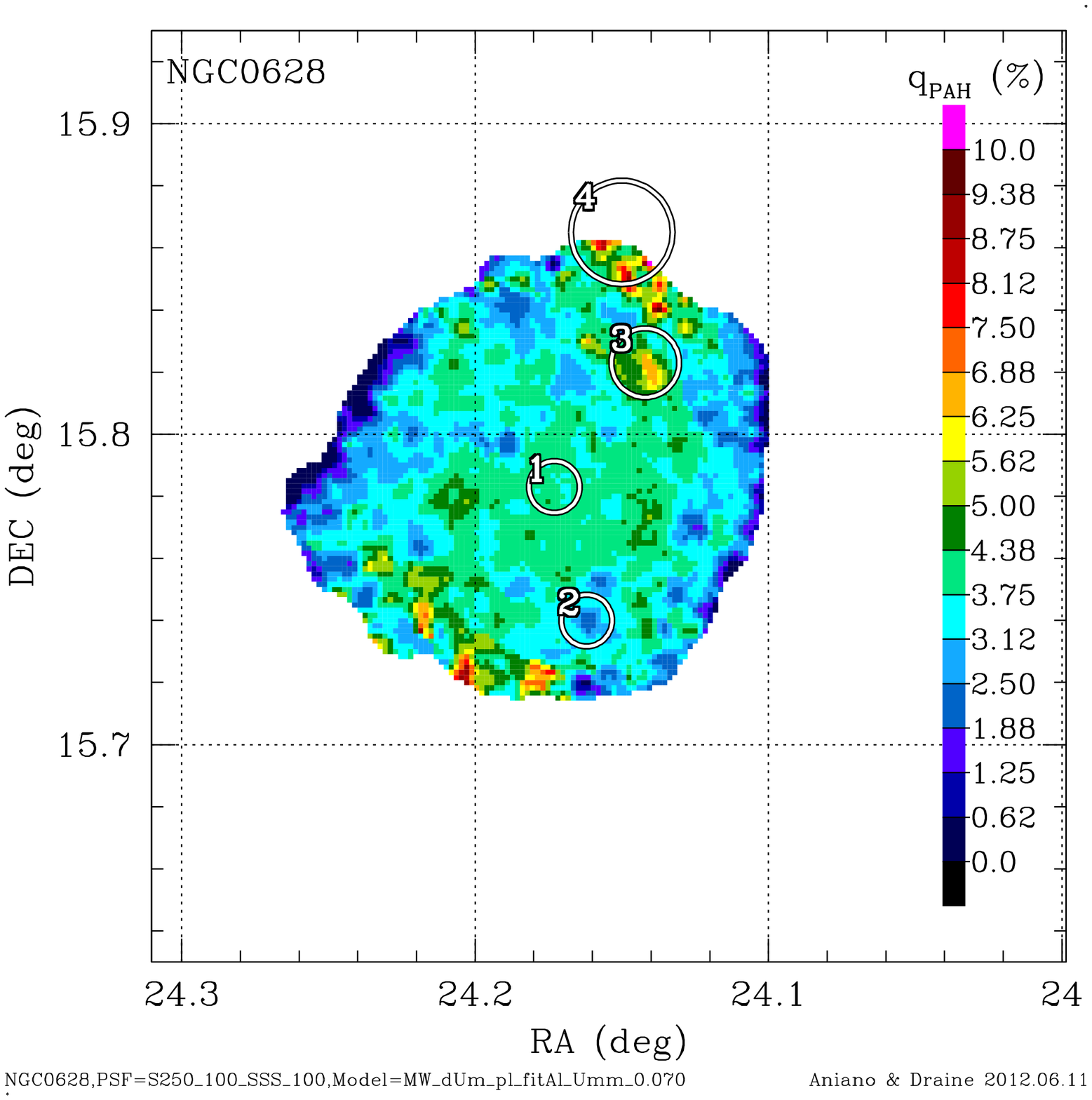}
\renewcommand \RoneCthree {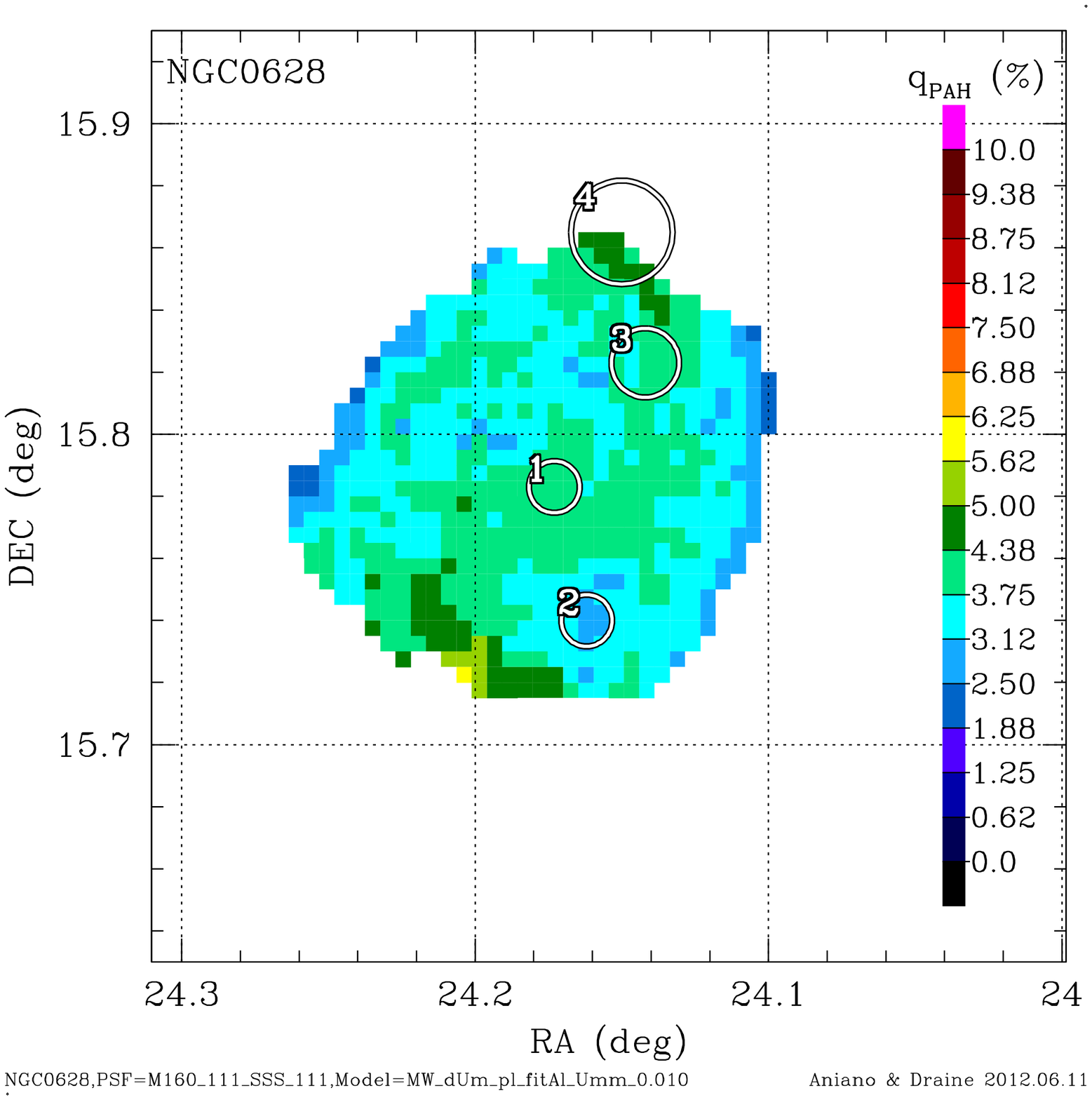}
\renewcommand \RtwoCone {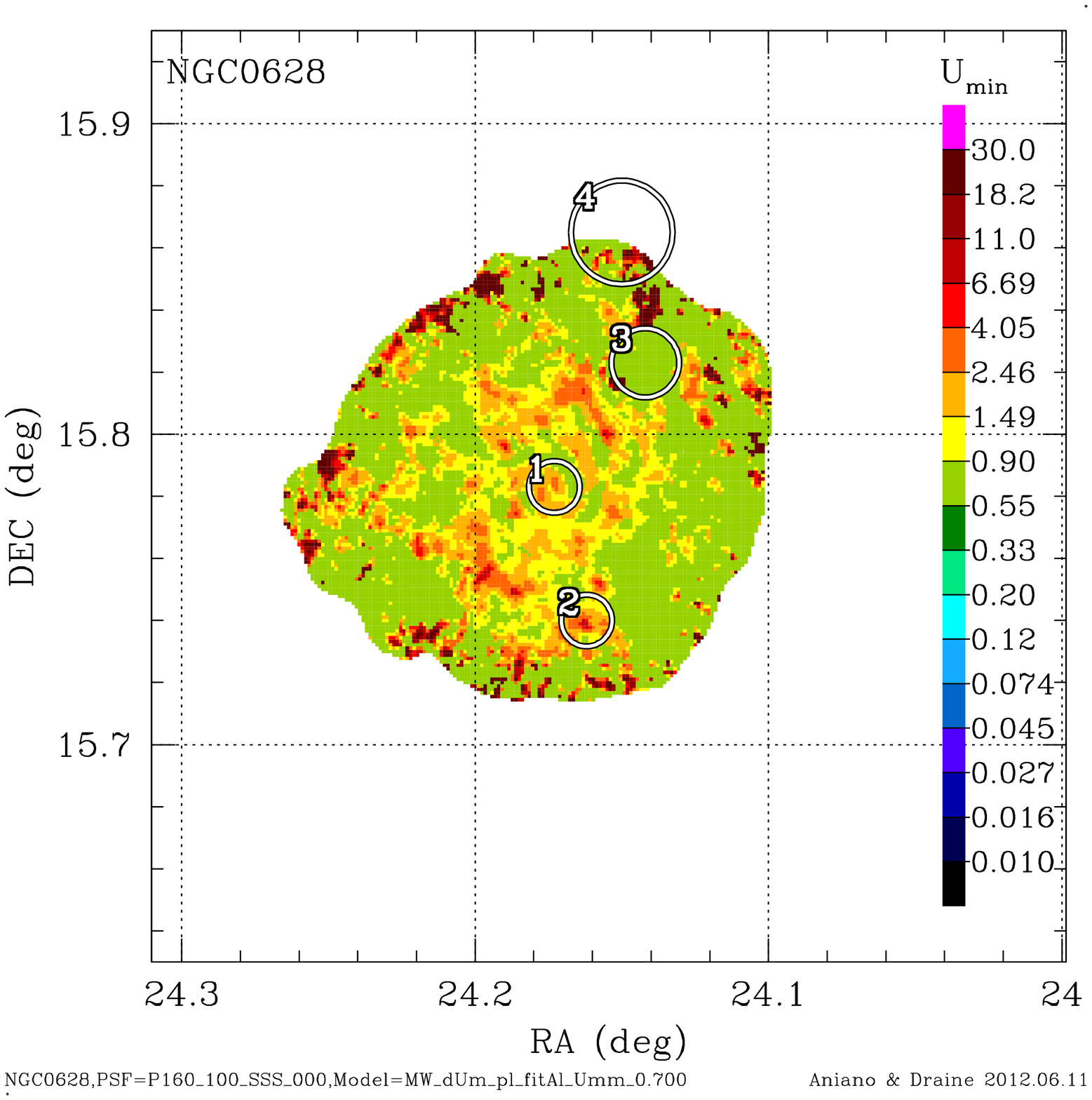}
\renewcommand \RtwoCtwo {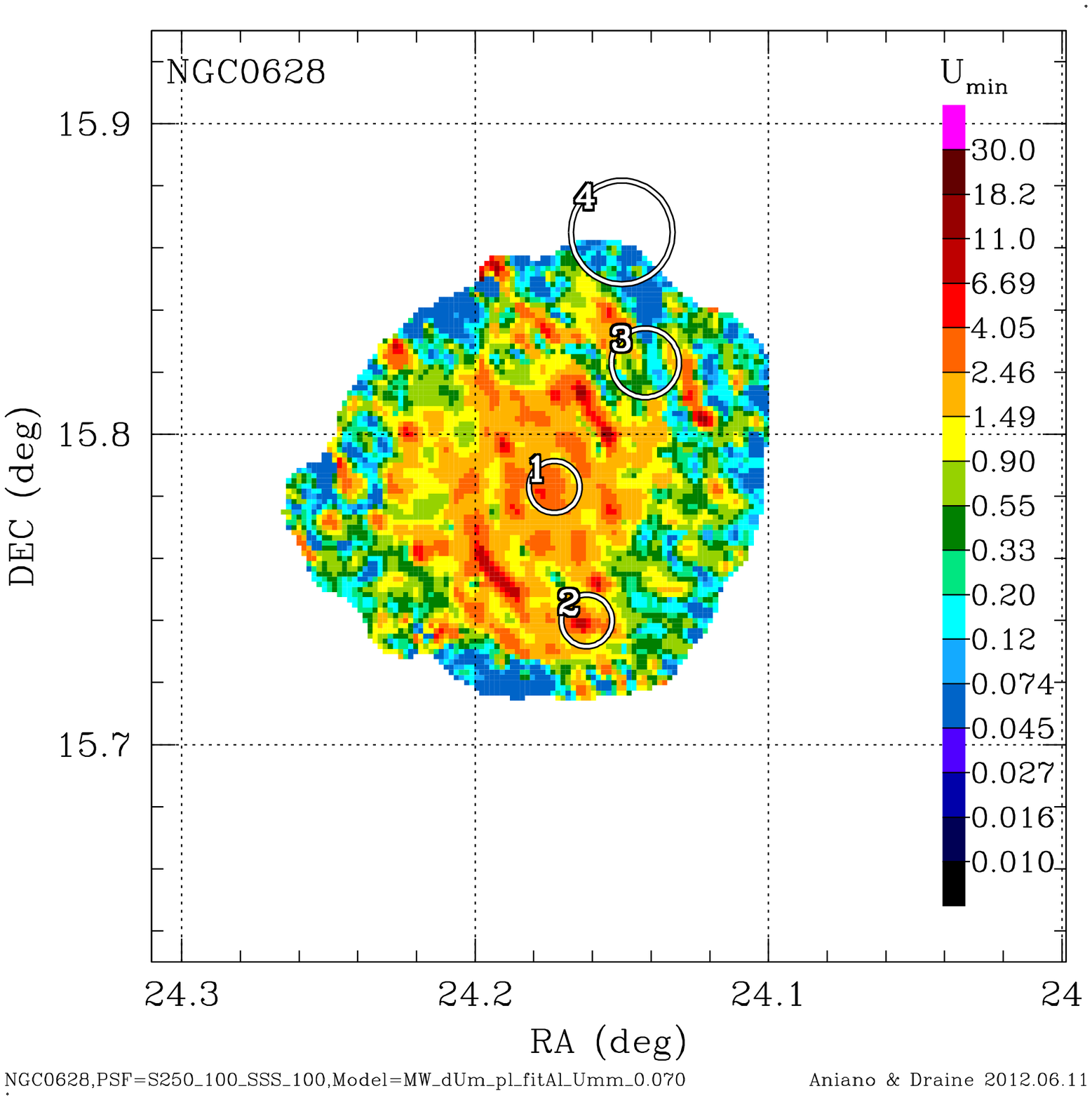}
\renewcommand \RtwoCthree {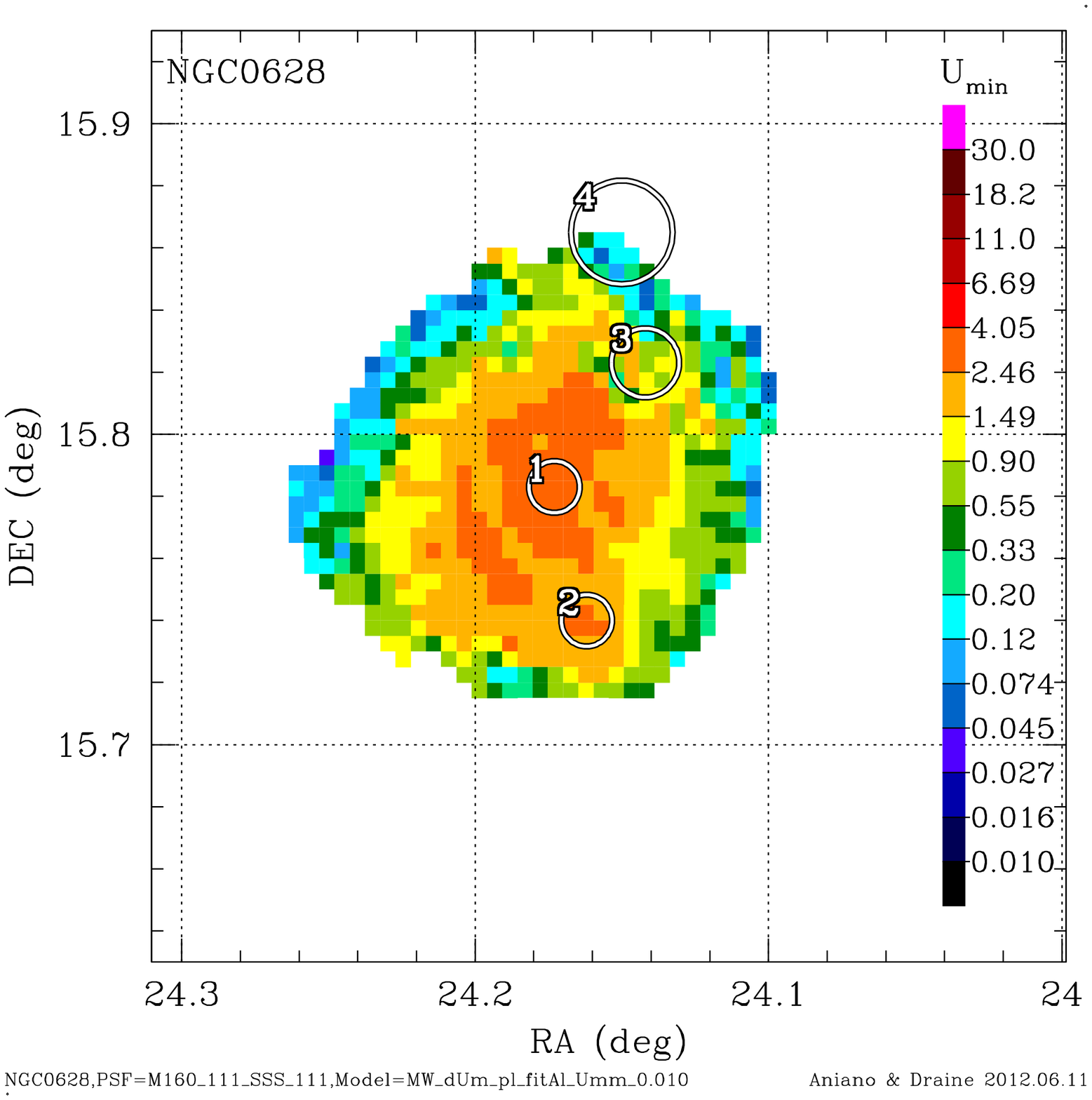}
\renewcommand \RthreeCone {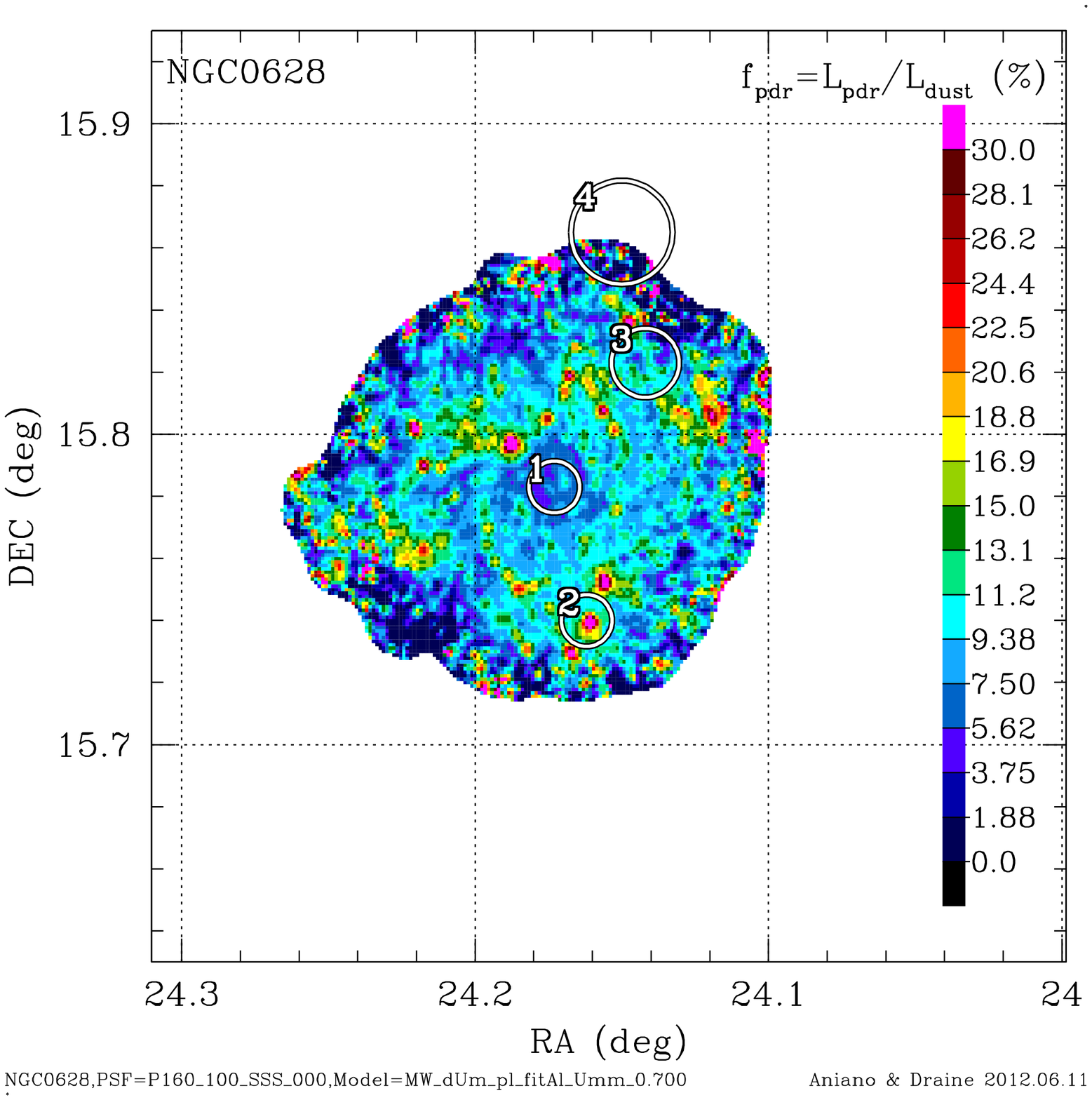}
\renewcommand \RthreeCtwo {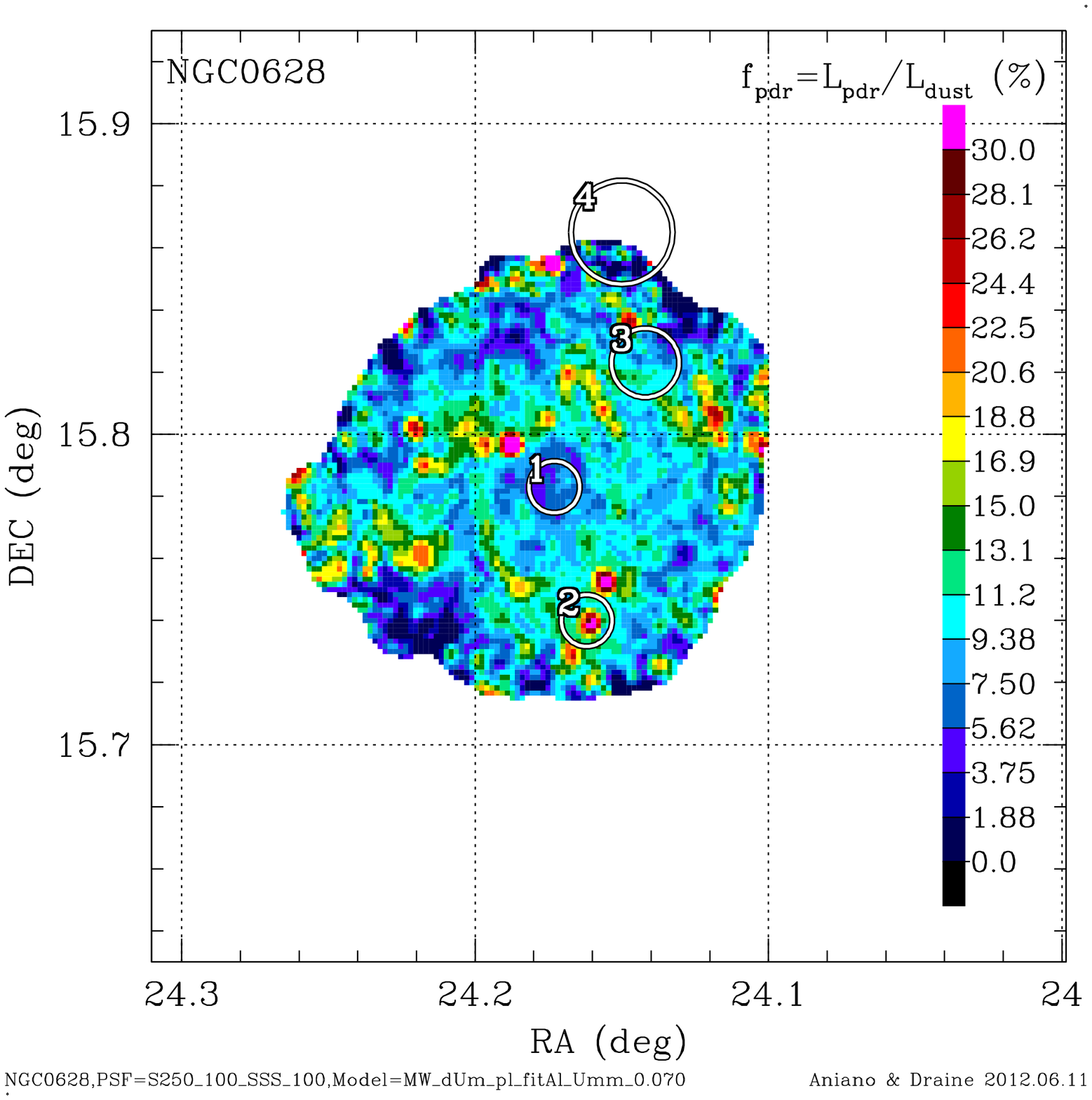}
\renewcommand \RthreeCthree {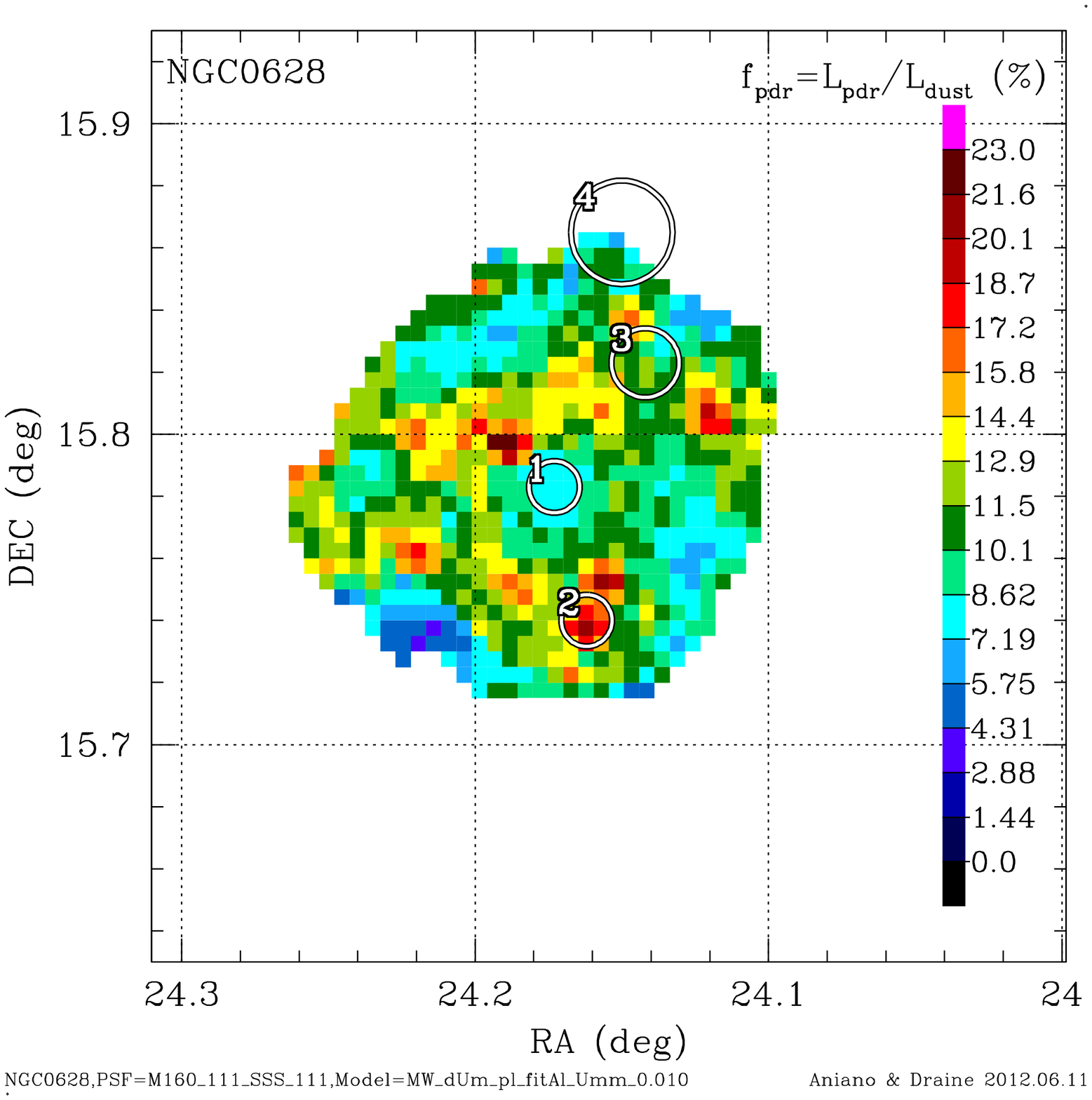}
\ifthenelse{\boolean{make_heavy}}{ }
{ \renewcommand \RoneCone    {No_image.eps}
\renewcommand \RtwoCone    {No_image.eps}
\renewcommand \RthreeCone {No_image.eps}
\renewcommand \RoneCtwo    {No_image.eps}
\renewcommand \RtwoCtwo    {No_image.eps}
\renewcommand \RthreeCtwo {No_image.eps}}
\begin{figure}[h,t]
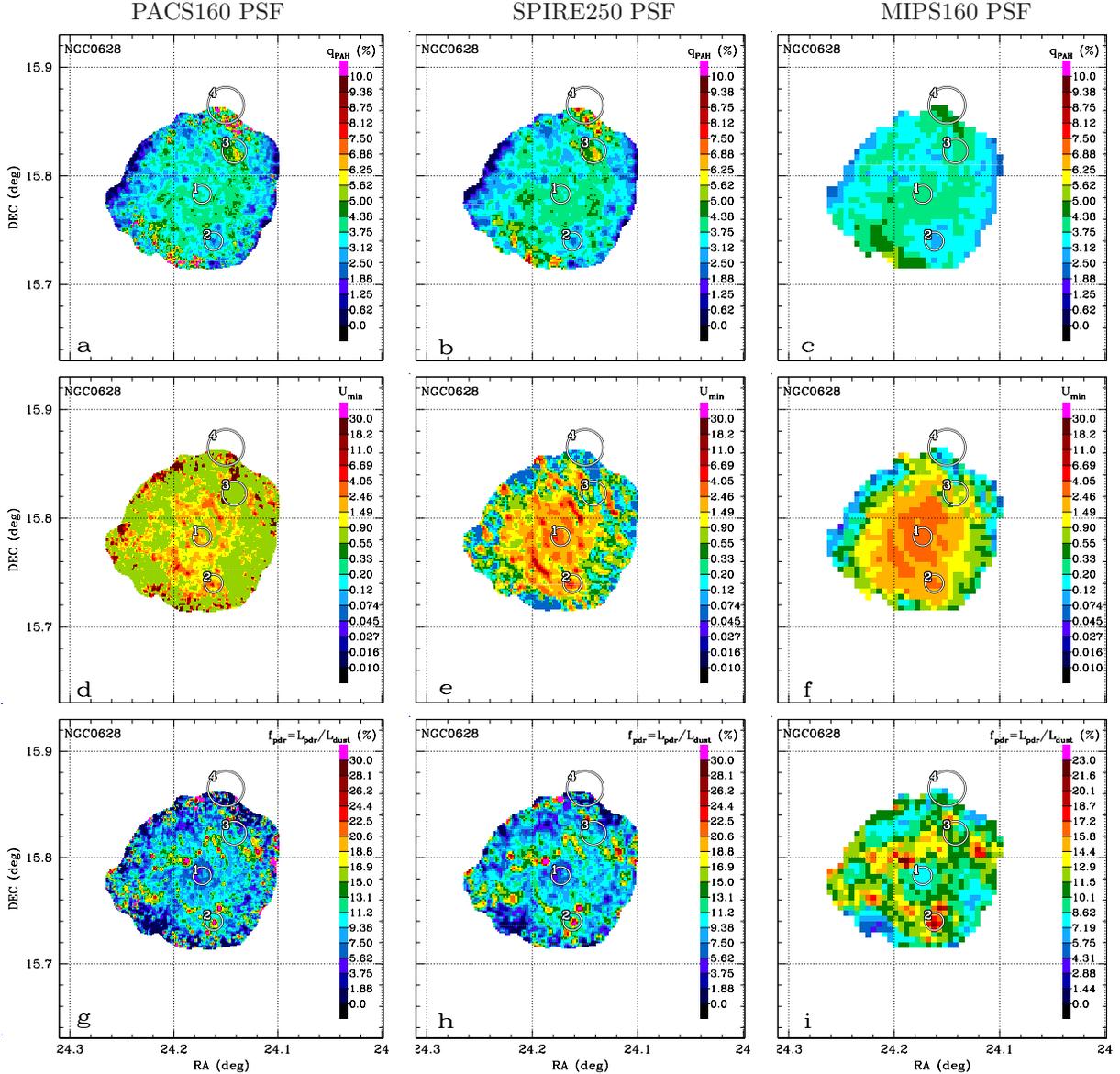
 
\centering
\begin{tabular}{c@{$\,$}c@{$\,$}c} 
\footnotesize PACS160 PSF & \footnotesize SPIRE250 PSF & \footnotesize MIPS160 PSF \\
\FirstNormal
\SecondNormal
\ThirdLast
\end{tabular}
\vspace*{-0.5cm}
\caption{\footnotesize\label{fig:ngc0628-2}
NGC~628 dust models at the resolution of PACS160 (left), SPIRE250 (center),
and MIPS160 (right).
  Top row: PAH abundance parameter $\qpah$.  
  Middle row: diffuse starlight intensity parameter $\Umin$.  
  Bottom row: PDR  fraction $f_\PDR$.}
\end{figure}


\subsubsection{Maps of $\qpah$ and Starlight Parameters}

Figure \ref{fig:ngc0628-2} shows maps of the
PAH abundance $\qpah$ (panels a-c), the starlight intensity parameter $\Umin$ (panels d-f), and the PDR fraction $f_\PDR$ (panels g-i),
over the ``galaxy mask'' region where the galaxy is well-detected.
The galaxy mask for NGC~628 has a diameter of $\sim$0.16$^\circ$, or
20~kpc $@$ 7.2~Mpc.

The PAH abundance parameter $\qpah$, shown in Figure
\ref{fig:ngc0628-2}a-c is remarkably uniform over the region where it
can be reliably estimated.  In Figure \ref{fig:ngc0628-2}c, $\qpah$
varies from a high of $\sim0.05$ a few kpc from the center down to
$\sim0.035$ near the edge of the galaxy mask.  If there is a radial
gradient in $\qpah$, it is weak, consistent with the weak gradients
found for the SINGS sample (including NGC~628) by
\citet{Munoz-Mateos+Gil_de_Paz+Boissier+etal_2009}.

At PACS160 resolution, the starlight intensity parameter $\Umin$ (see
Fig.\ \ref{fig:ngc0628-2}d) varies between $\sim$0.6 and $\sim$3 over
most of NGC\,628, following the galaxy structures, but near the edges
of the galaxy mask $\Umin$ appears to rise.  This is because the
reduced signal/noise results in PACS70/PACS160 ratios that appear to
be anomalously high, leading to high inferred values of $\Umin$ for
some pixels.  This is probably the result of low PACS160 fluxes for
those pixels, making it appear that the dust is rather warm.  We see
that when SPIRE250 data is introduced (Fig.\ \ref{fig:ngc0628-2}e), we
have many fewer high values of $\Umin$ near the edge of the galaxy
mask, and the $\Umin$ values in the brighter regions appear
well-behaved.  This continues when the MIPS160, SPIRE350, and SPIRE500 data are
brought into the fit in the MIPS160 resolution image
(Fig.\ \ref{fig:ngc0628-2}f).

The bottom row of Figure \ref{fig:ngc0628-2} (panels g-i) shows $f_\PDR$,
the fraction of the dust luminosity coming from dust heated by
starlight intensities $U>100$, which we expect to be associated with
star-forming regions.
At PACS160 resolution, most of the galaxy has $f_\PDR \approx 0.03$ and some small
bright regions (for example the center of aperture 2 in Figure
\ref{fig:ngc0628-5}) have higher values, up to $f_\PDR \approx 0.3$.
These regions with high values of $f_\PDR$ generally 
coincide with H$\alpha$ peaks, and also with the regions of the highest
dust luminosity per area (see Fig.\ \ref{fig:ngc0628-1}g).
As we shift to coarser modeling (i.e, using PSFs with larger FWHM), the
$f_\PDR$ peaks are smoothed, and the general galaxy pixels tend to
have larger $f_\PDR$ values: the dynamic range of values of $f_\PDR$
decreases.  At MIPS160 resolution, most of the galaxy has $f_\PDR > 0.05$.
The overall (dust luminosity-weighted) mean for the galaxy is
$\langle f_\PDR\rangle = 0.116$: $11.6\%$ of the total IR power is radiated by dust in regions where $U>100$.

Unfortunately, at MIPS160 resolution the arm structure of both
galaxies is not clearly resolved (see the MIPS160 images of the
galaxies in panel (f) of Figures \ref{Fig_0628_ori} and \ref{Fig_6946_ori}).
 Therefore we cannot reliably study variation of the dust parameters between arm and
interarm regions. A subsequent work (Hunt et al.\ 2012, in prep.) will study radial variations in the model
parameters.

\subsubsection{Comparison Between Observed and Modeled Flux Densities}

\renewcommand \RoneCone {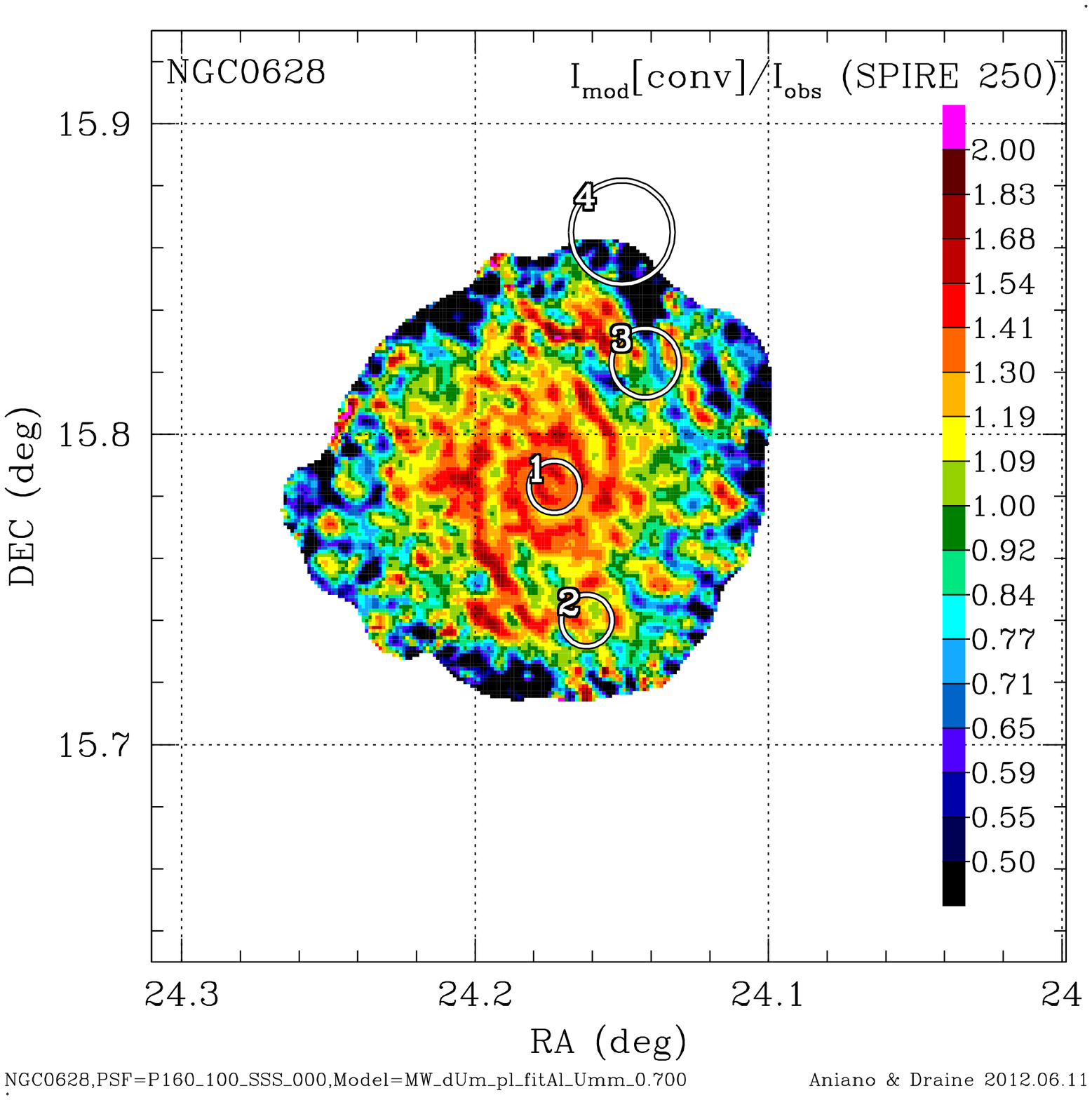}
\renewcommand \RoneCtwo {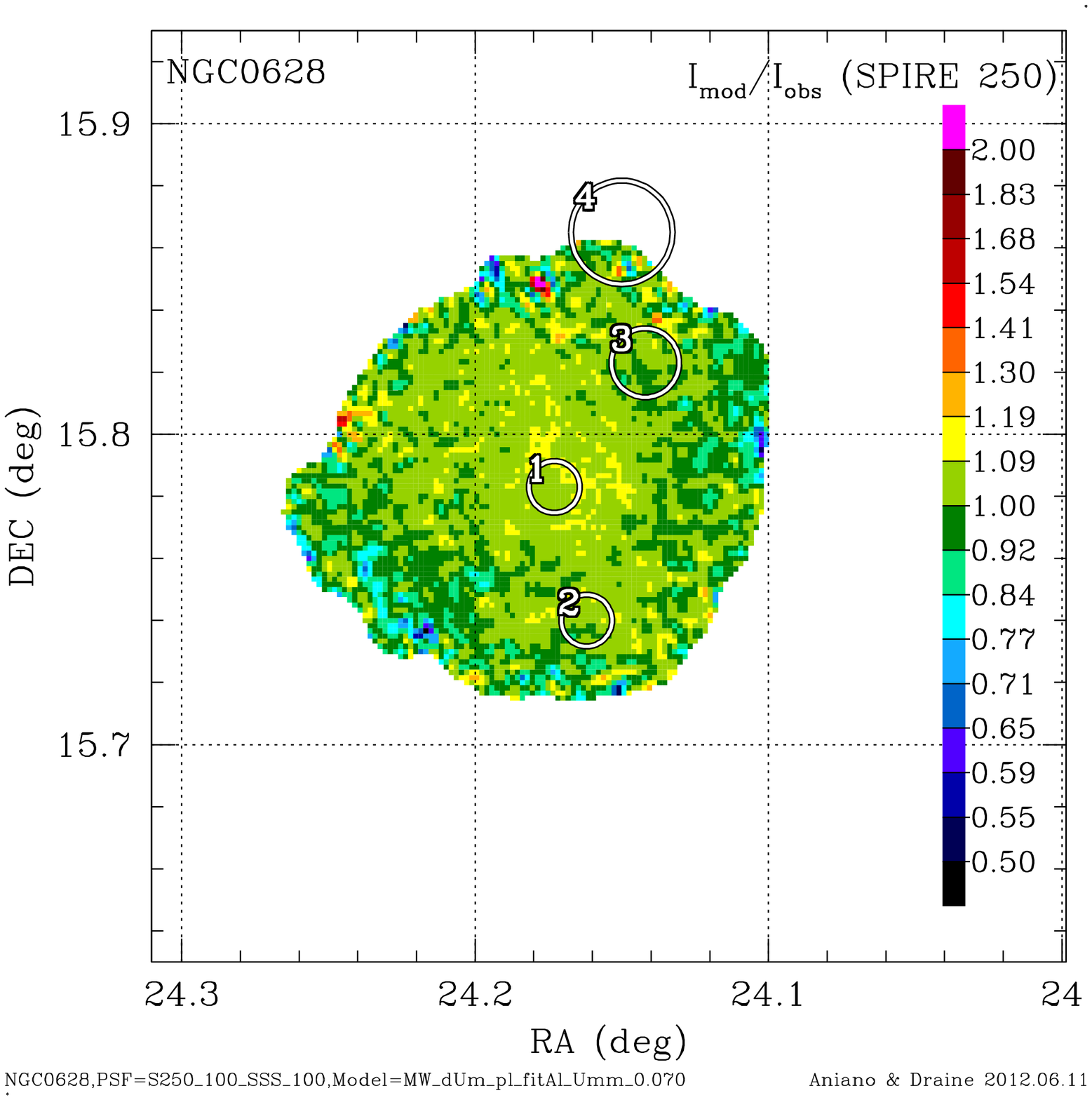}
\renewcommand \RoneCthree {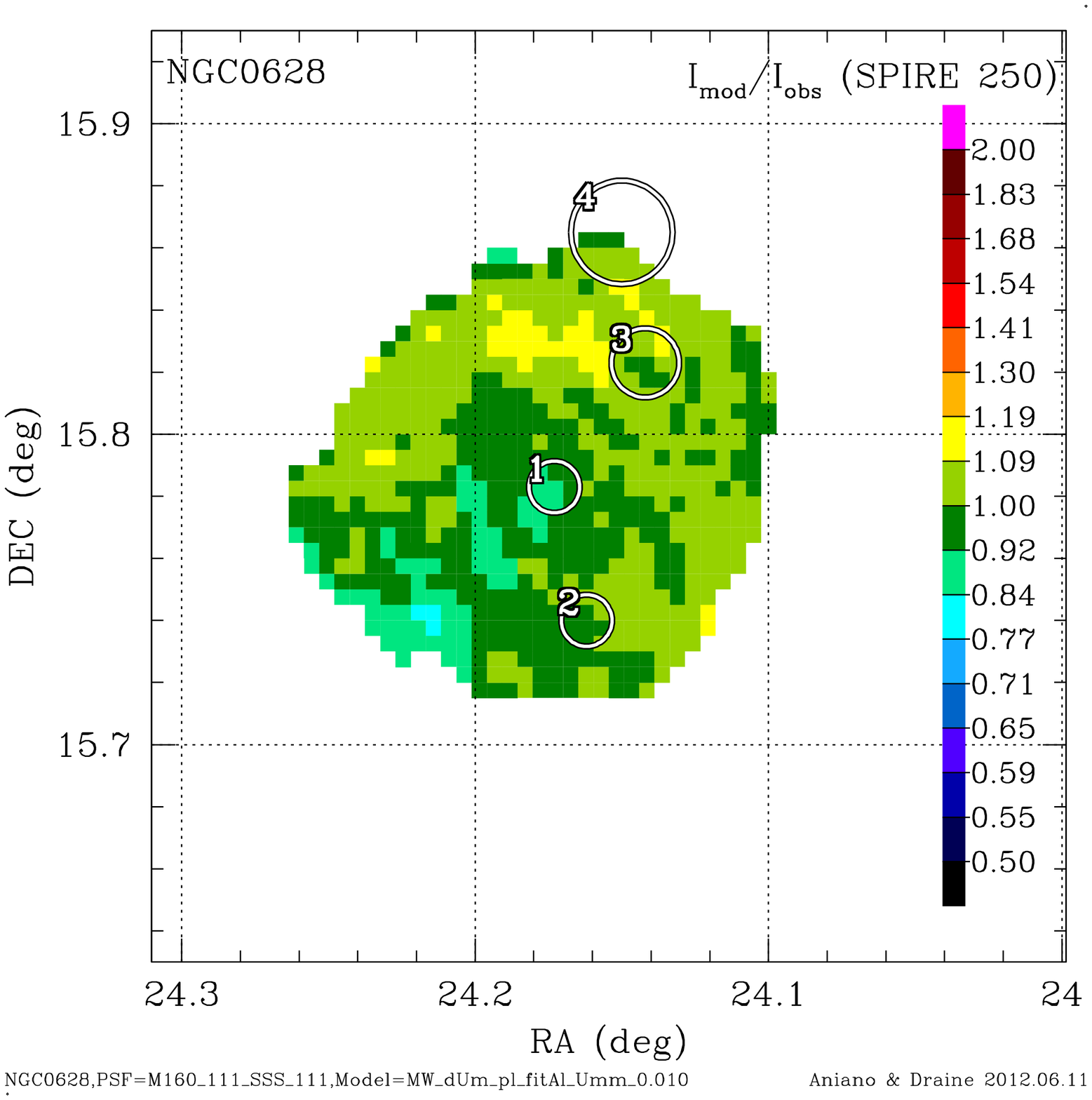}
\renewcommand \RtwoCone {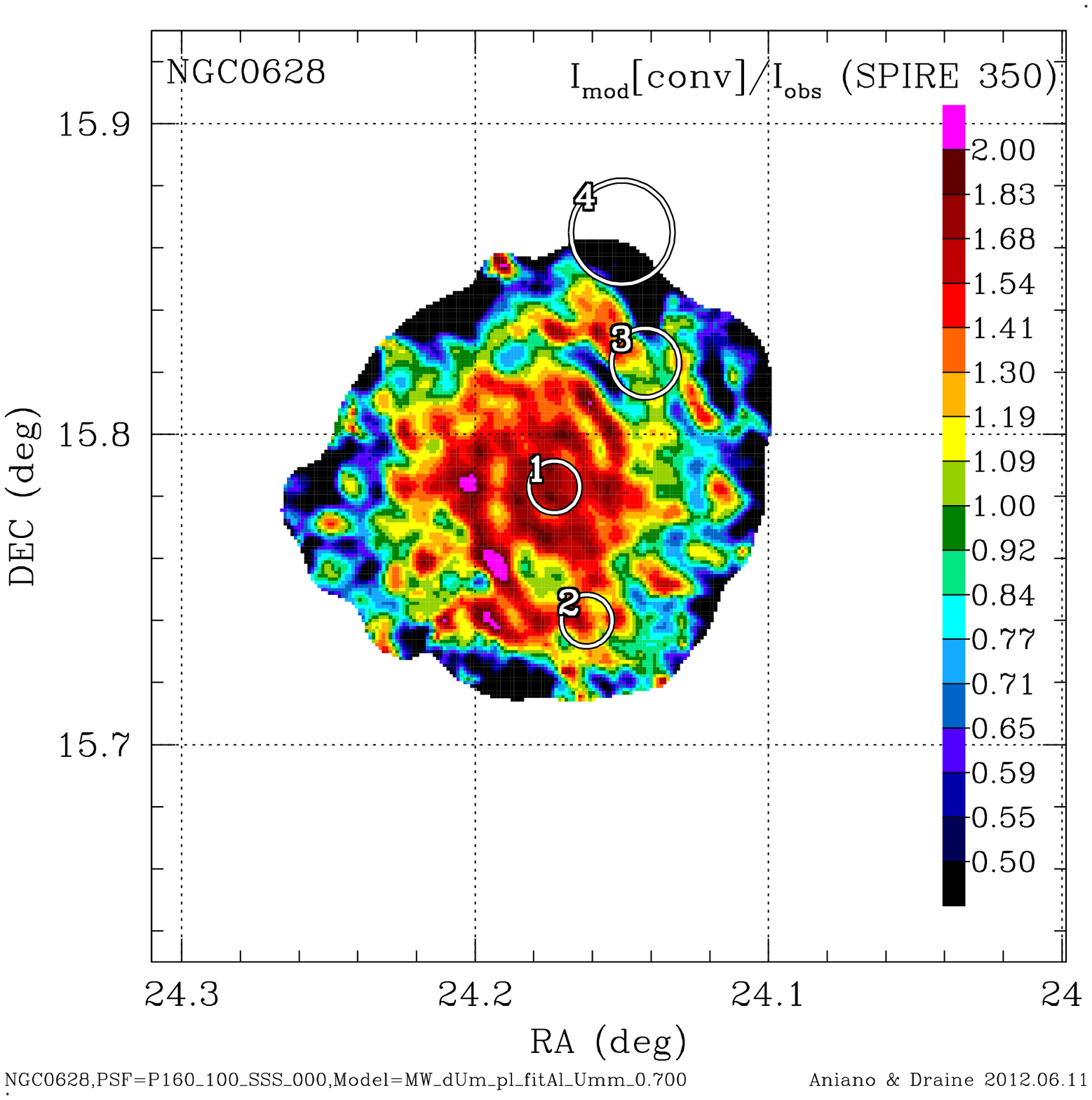}
\renewcommand \RtwoCtwo {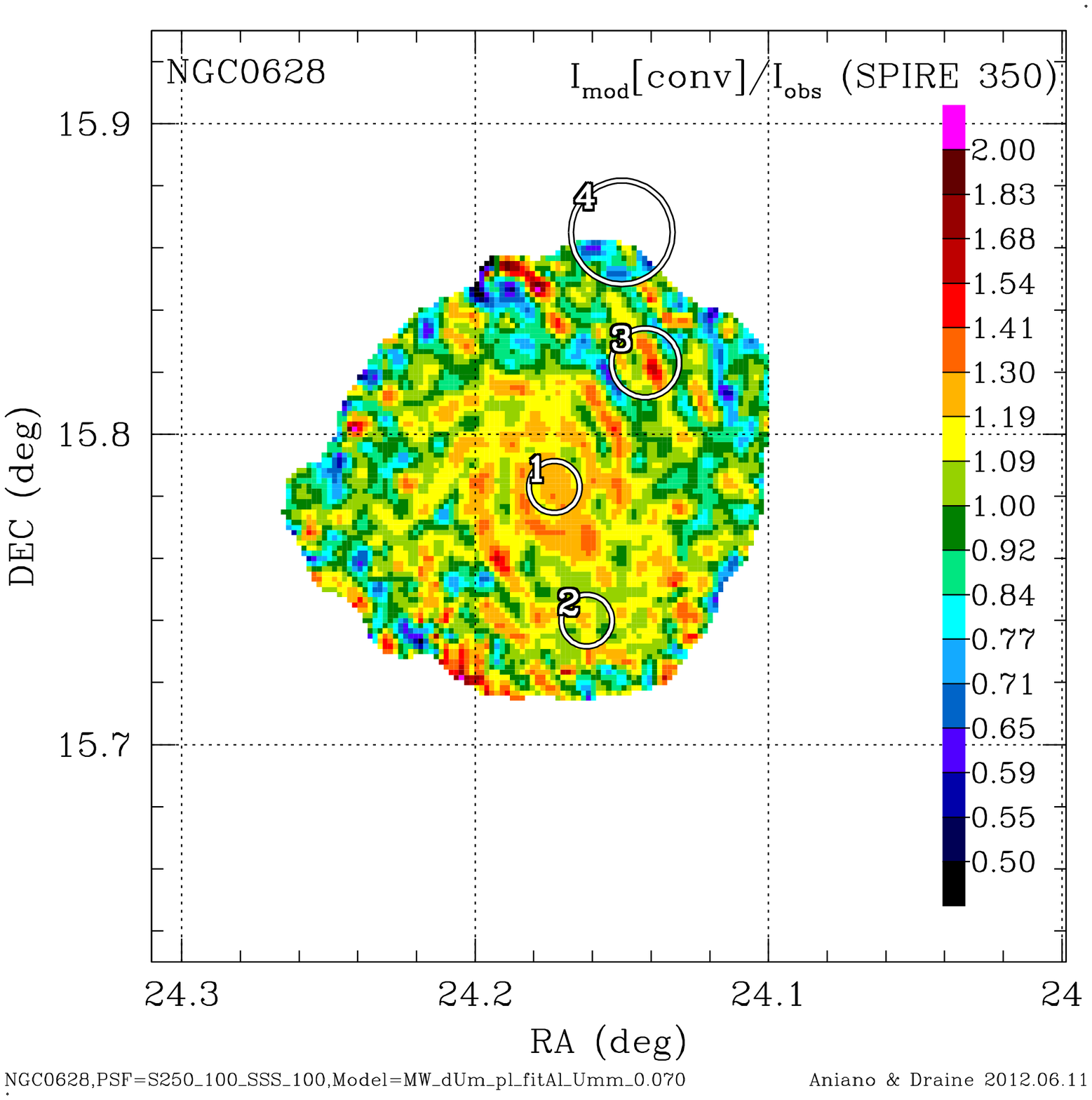}
\renewcommand \RtwoCthree {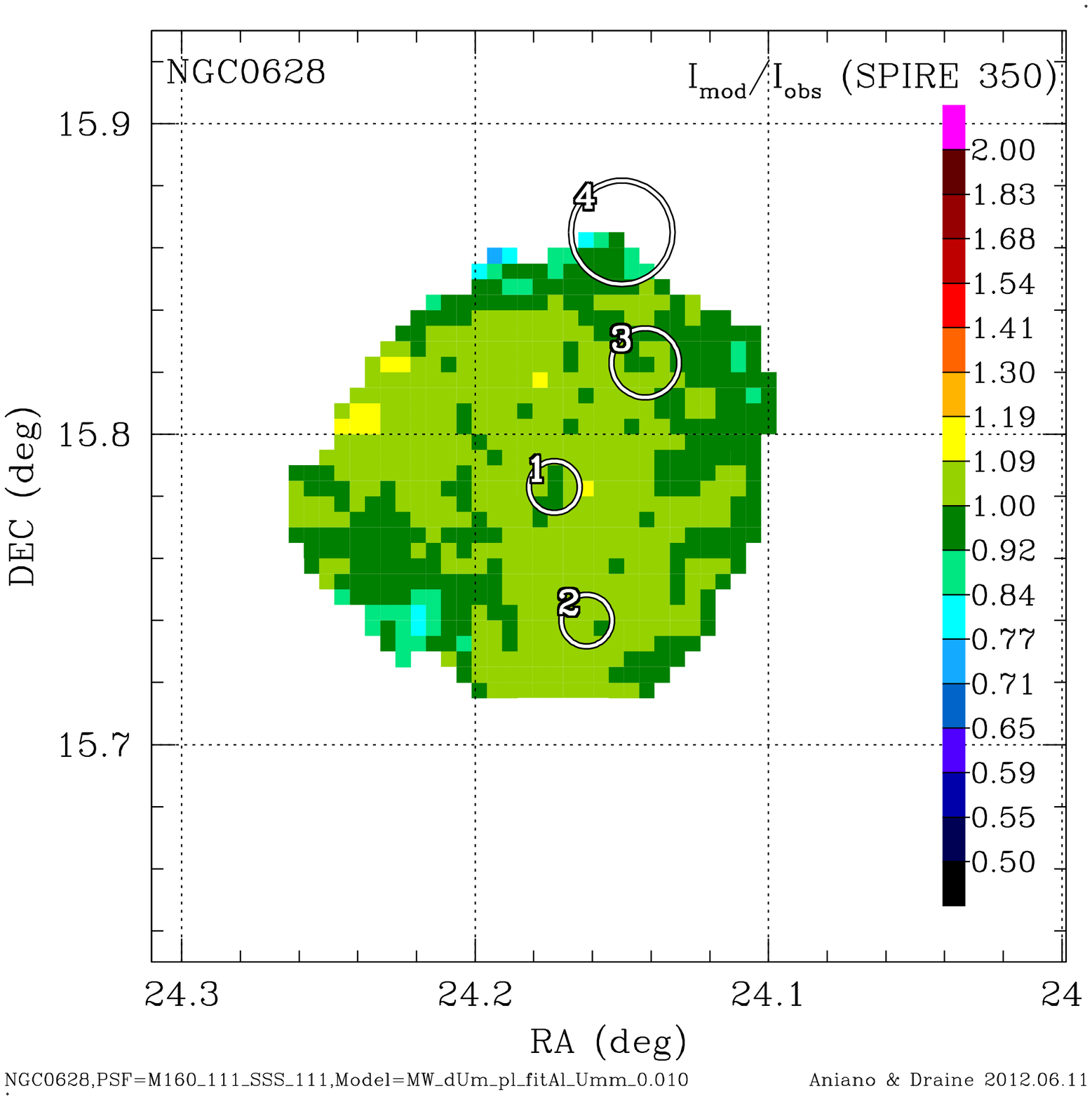}
\renewcommand \RthreeCone {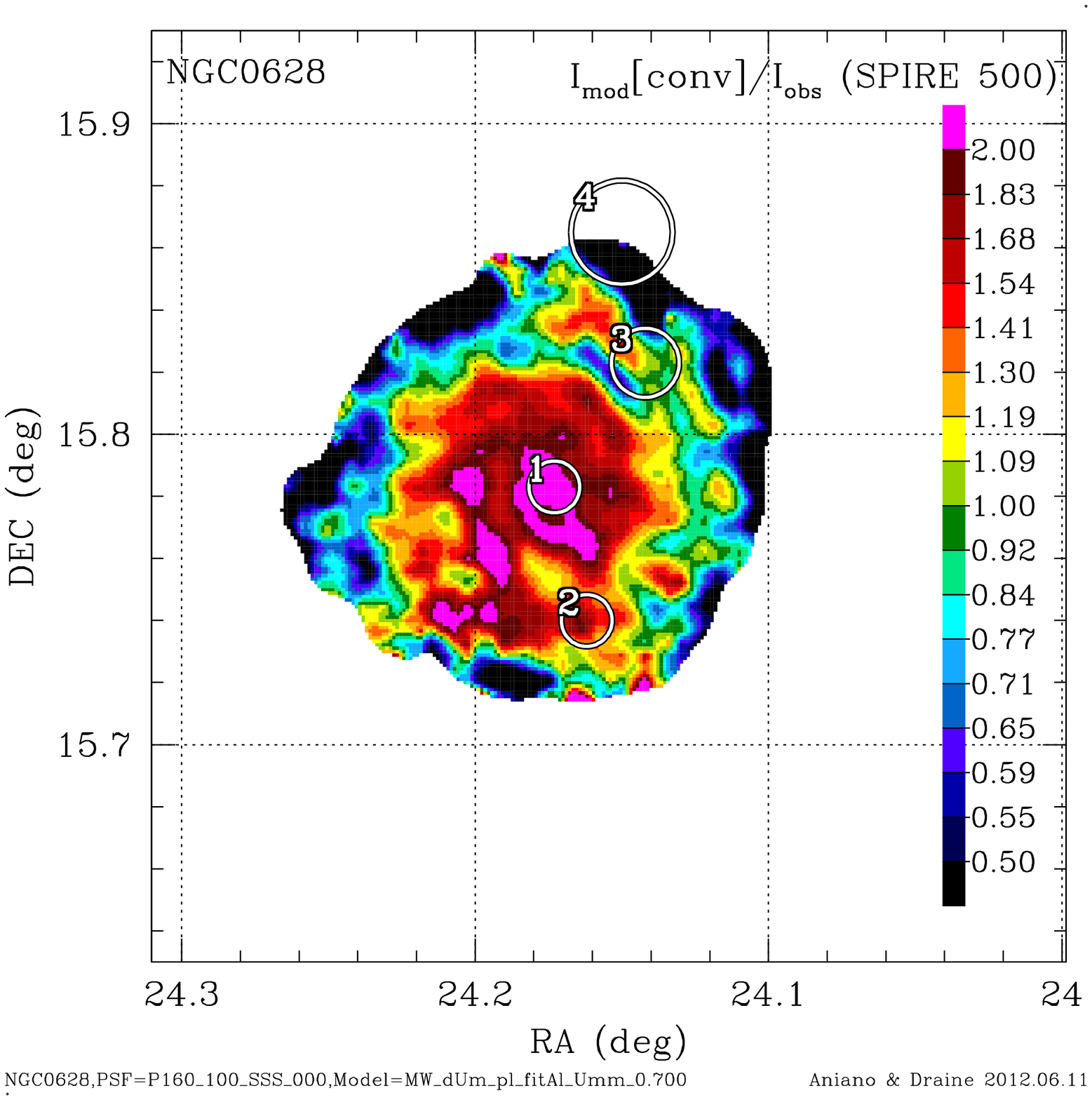}
\renewcommand \RthreeCtwo {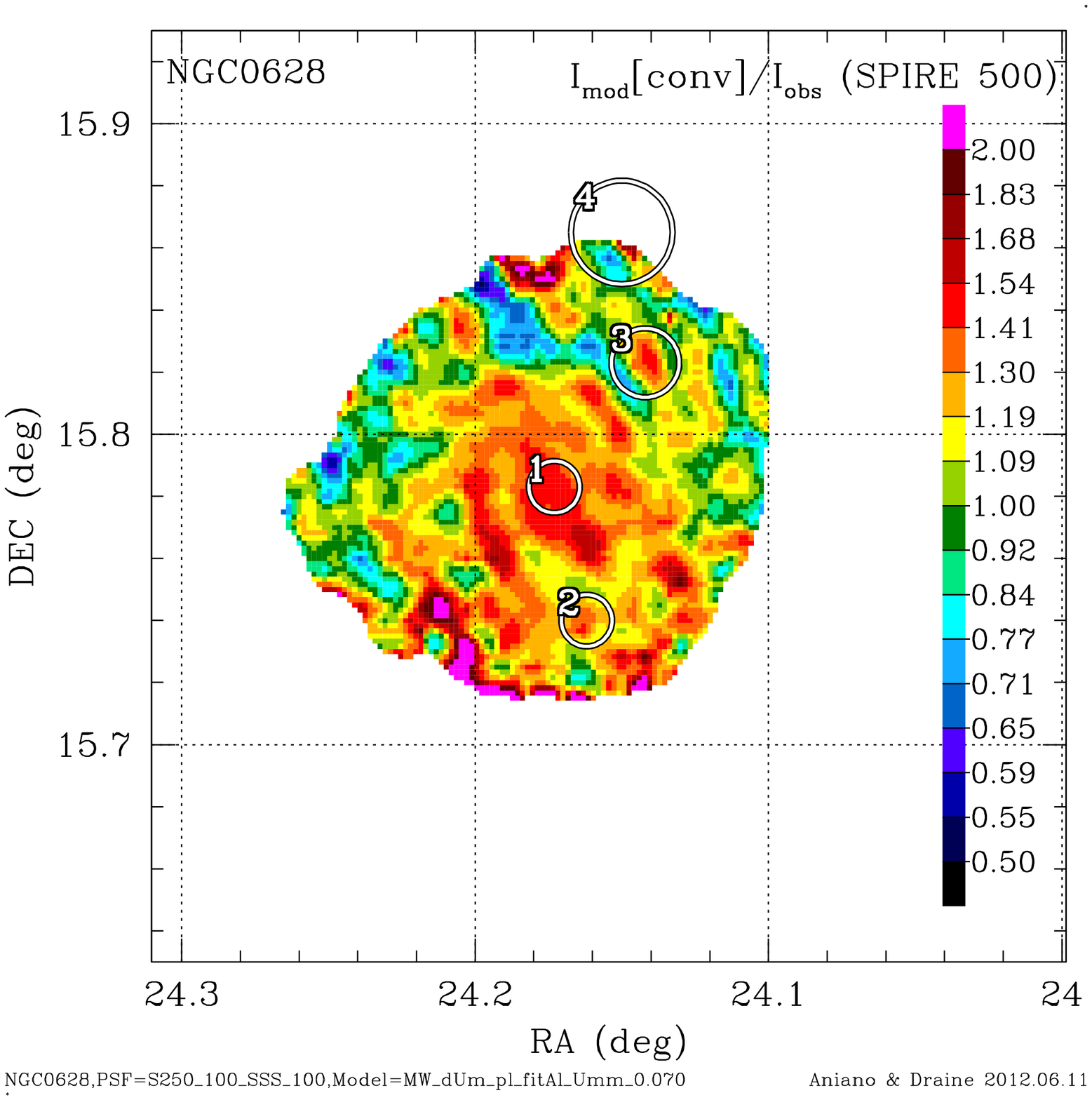}
\renewcommand \RthreeCthree {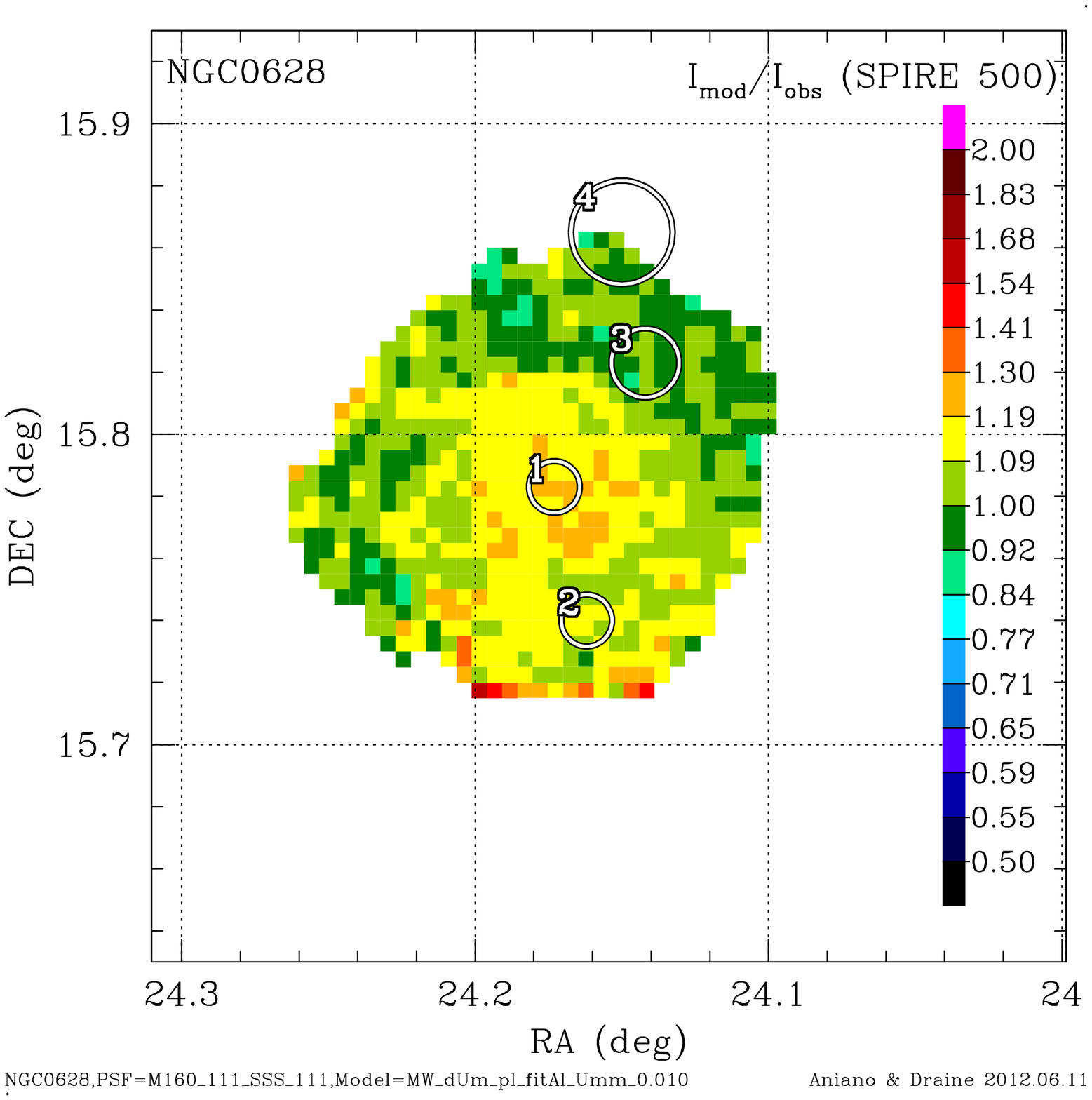}
\ifthenelse{\boolean{make_heavy}}{ }
{ \renewcommand \RoneCone    {No_image.eps}
\renewcommand \RtwoCone    {No_image.eps}
\renewcommand \RthreeCone {No_image.eps}
\renewcommand \RfourCone {No_image.eps}
\renewcommand \RoneCtwo    {No_image.eps}
\renewcommand \RtwoCtwo    {No_image.eps}
\renewcommand \RthreeCtwo {No_image.eps}
\renewcommand \RfourCtwo {No_image.eps}}
\begin{figure}[h]
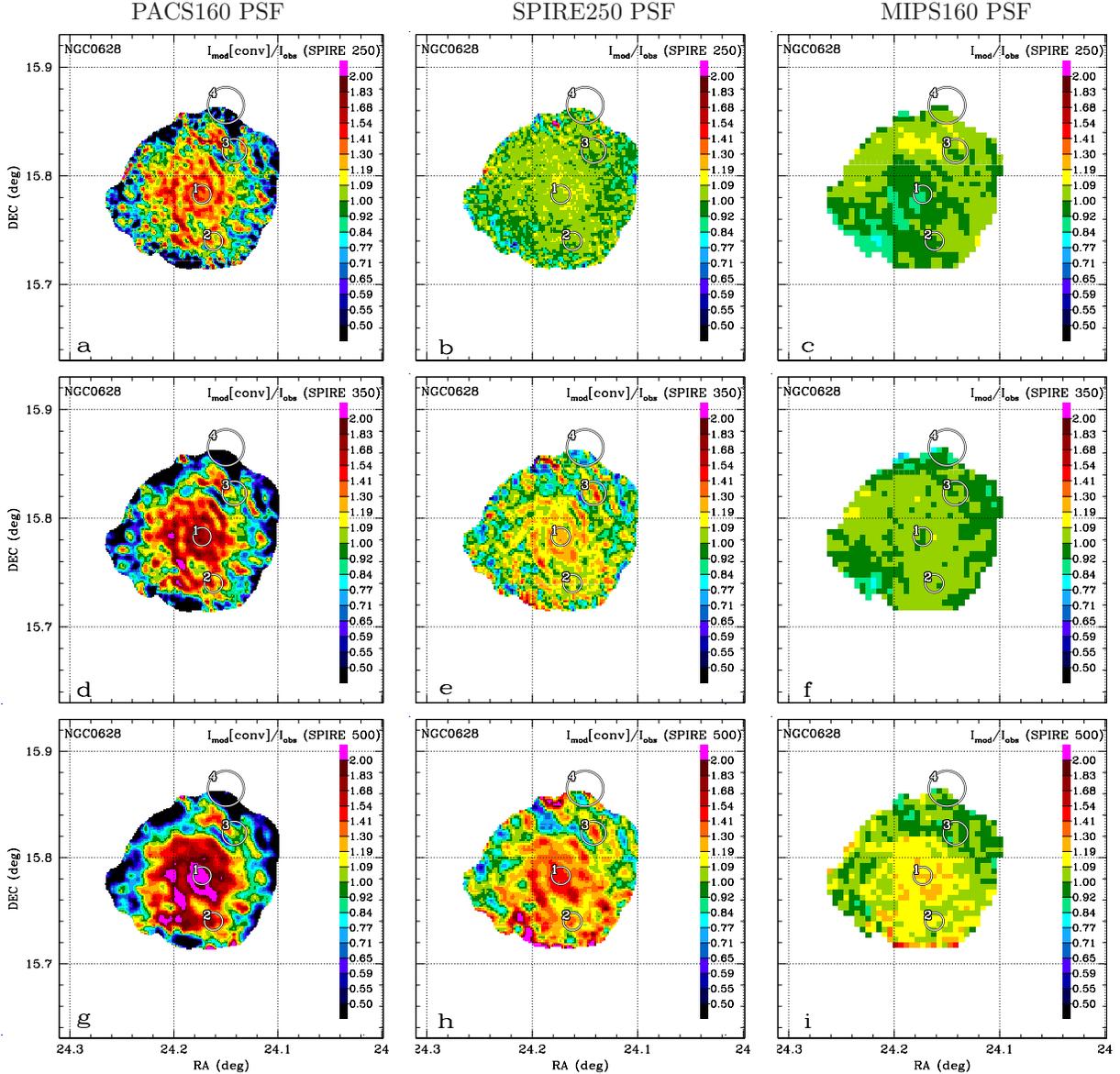
 
\centering
\begin{tabular}{c@{$\,$}c@{$\,$}c} 
\footnotesize PACS160 PSF & \footnotesize SPIRE250 PSF & \footnotesize MIPS160 PSF \\
\FirstNormal
\SecondNormal
\ThirdLast
\end{tabular}
\vspace*{-0.5cm}
\caption{\footnotesize\label{fig:ngc0628-3} 
Ratio of model/observed intensity at $\lambda$=250\um (top row), $\lambda$=350\um (middle row), and $\lambda$=500\um (bottom row) for NGC~628.
 Left column: PACS160 PSF, model constrained only by IRAC, MIPS24 and PACS.
 Center column: SPIRE250 PSF, model constrained by IRAC, MIPS24, PACS, and SPIRE250.
 Right column: MIPS160 PSF, model constrained by all 13 cameras.
 The model in the center column does a fairly good job in predicting the emission at $\lambda$=350\um, and $\lambda$=500\um, except near the edge where the S/N is low.
 The model in the right column is in excellent agreement with all three SPIRE bands. In all panels the model predicted and observed intensities have been convolved to a common PSF before taking the ratios.
}
\end{figure}

How well does the dust model reproduce the SPIRE photometry?  Figure
\ref{fig:ngc0628-3} shows the ratios of model-predicted intensity to
observed intensity at $\lambda$=250, 350 and 500\um for dust models
obtained by fitting photometry with PACS160, SPIRE250, and MIPS160
resolution.  In order to make the comparison, we degrade either the
observed image or the model-predicted image to a common resolution
(i.e., when the modeling is done at PACS160 resolution, we convolve
the model-predicted SPIRE250 image to the SPIRE250 PSF, and when the
modeling is done at MIPS160 resolution, we convolve the observed
SPIRE250 image to the MIPS160 PSF).  Except near the edge of the
galaxy mask (where the low S/N in the PACS data becomes an issue), the
modeling tends to {\it over}predict the SPIRE500 photometry -- there
is no evidence for a significant mass of very cold dust radiating only
at the longer wavelengths.

When we attempt to model at PACS160 resolution (using only IRAC,
MIPS24, and PACS data to constrain the model), the model predictions at
SPIRE250, SPIRE350, and SPIRE500 do not agree very well with
observations (see Fig.\ \ref{fig:ngc0628-3}a, d, and g).  If we
coarsen the modeling to SPIRE250 resolution and add SPIRE250 to the
model constraints, we now reproduce the SPIRE250 image (not
surprising) and the model predictions at 350\um and 500\um are within
$\pm 30\%$ and $\pm 50\%$ of the SPIRE observations respectively over
most of the galaxy mask.  If we further coarsen the modeling to the
MIPS160 PSF, and use all the data to constrain the model, we find good
agreement at all SPIRE bands (see Fig.\ \ref{fig:ngc0628-3}c, f, and
i), with only SPIRE500 slightly over-predicted by $\approx 10 \%$ in some regions.

Figure \ref{fig:ngc0628-3}c, f, and i show that when all of the
cameras are used to constrain the modeling (and with the improved S/N
of the larger MIPS160 pixels), the model emission is generally in good
agreement with the SPIRE imaging -- for all three SPIRE bands the
ratio of model/observation is close to 1 over most of the galaxy, with
significant departures only just at the edge of the galaxy mask, where
observational errors are likely to be the explanation.

\subsubsection{Global SED Fitting}

\renewcommand \RoneCone {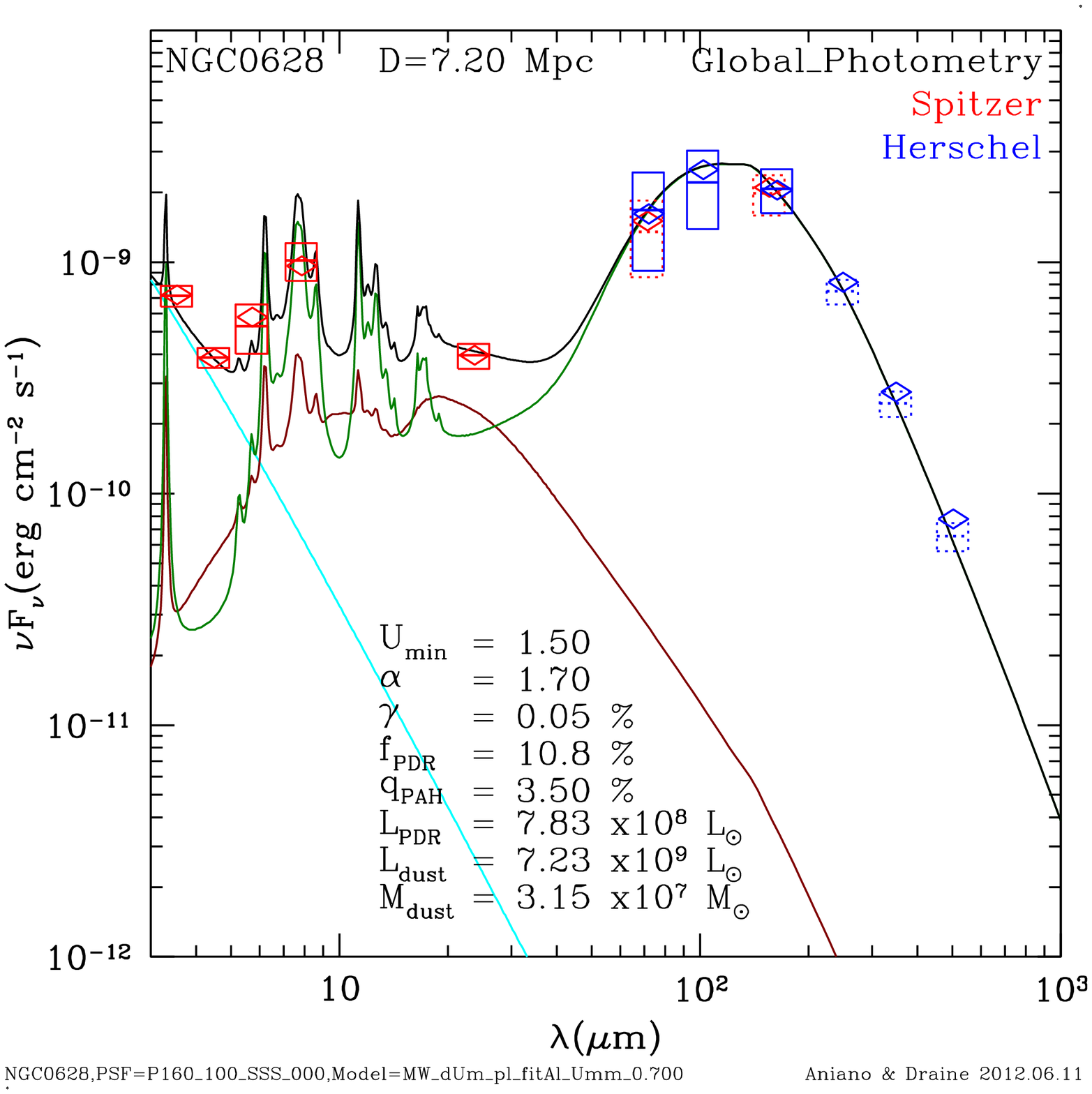}
\renewcommand \RoneCtwo {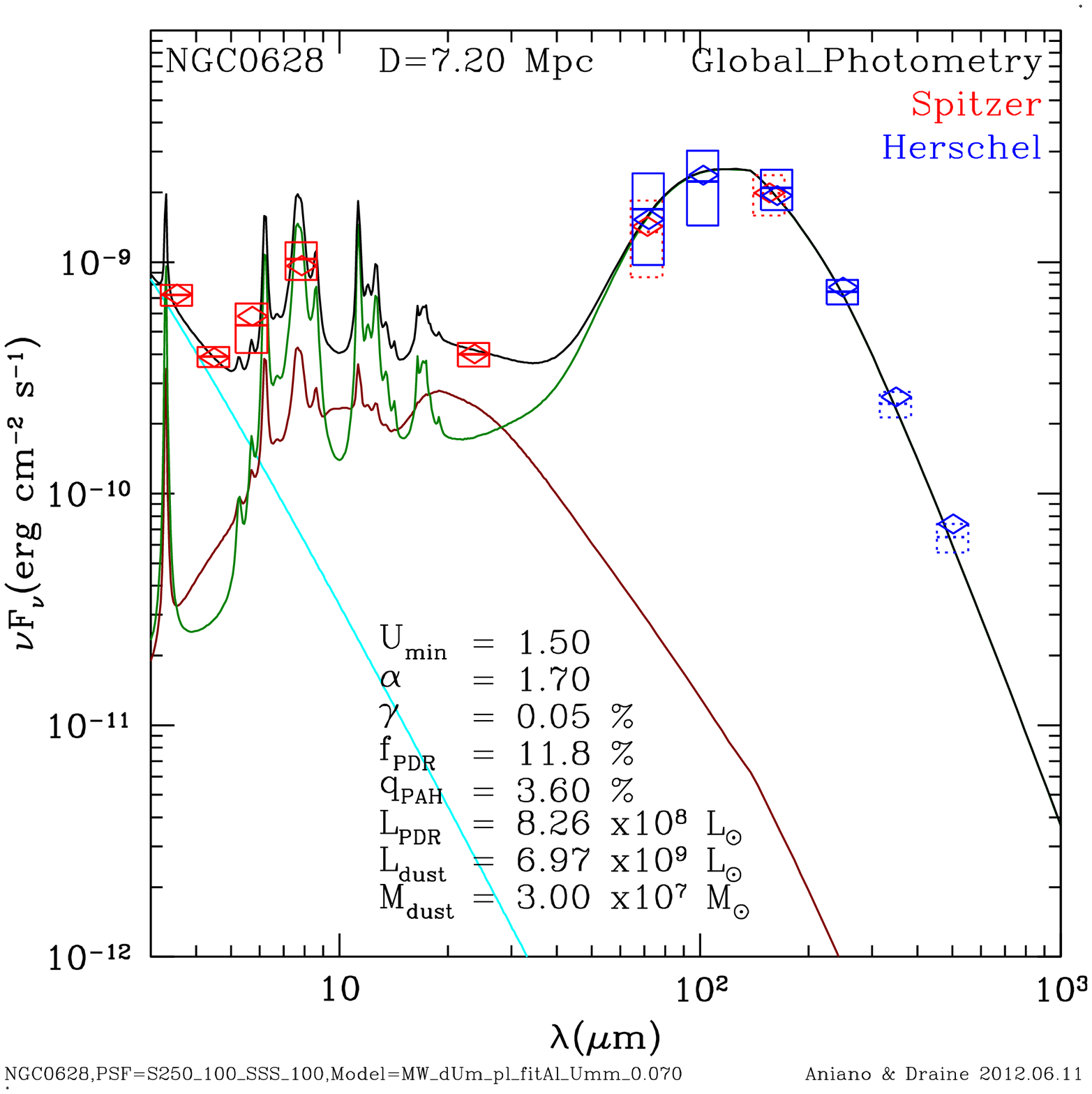}
\renewcommand \RoneCthree {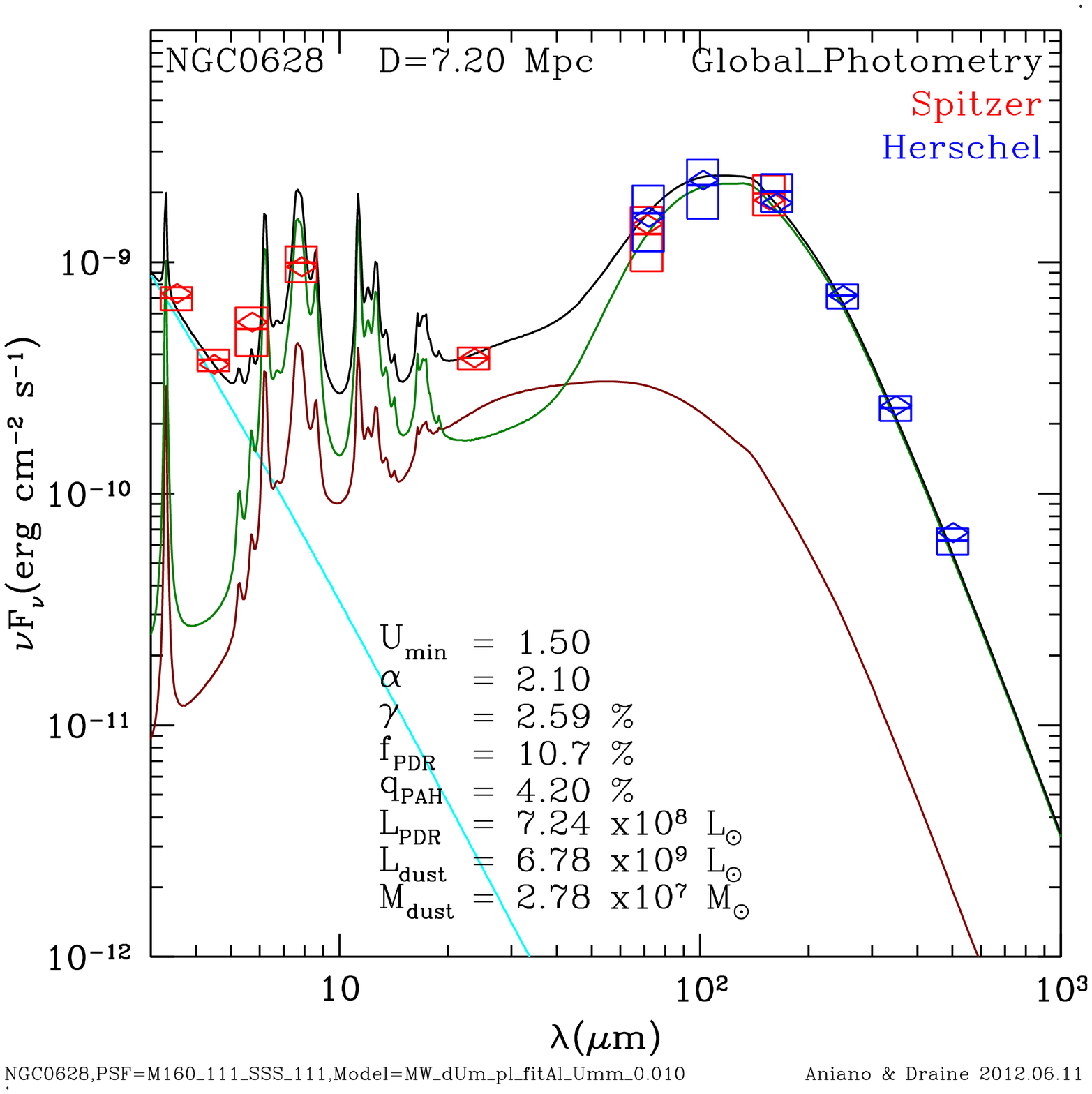}
\renewcommand \RtwoCone {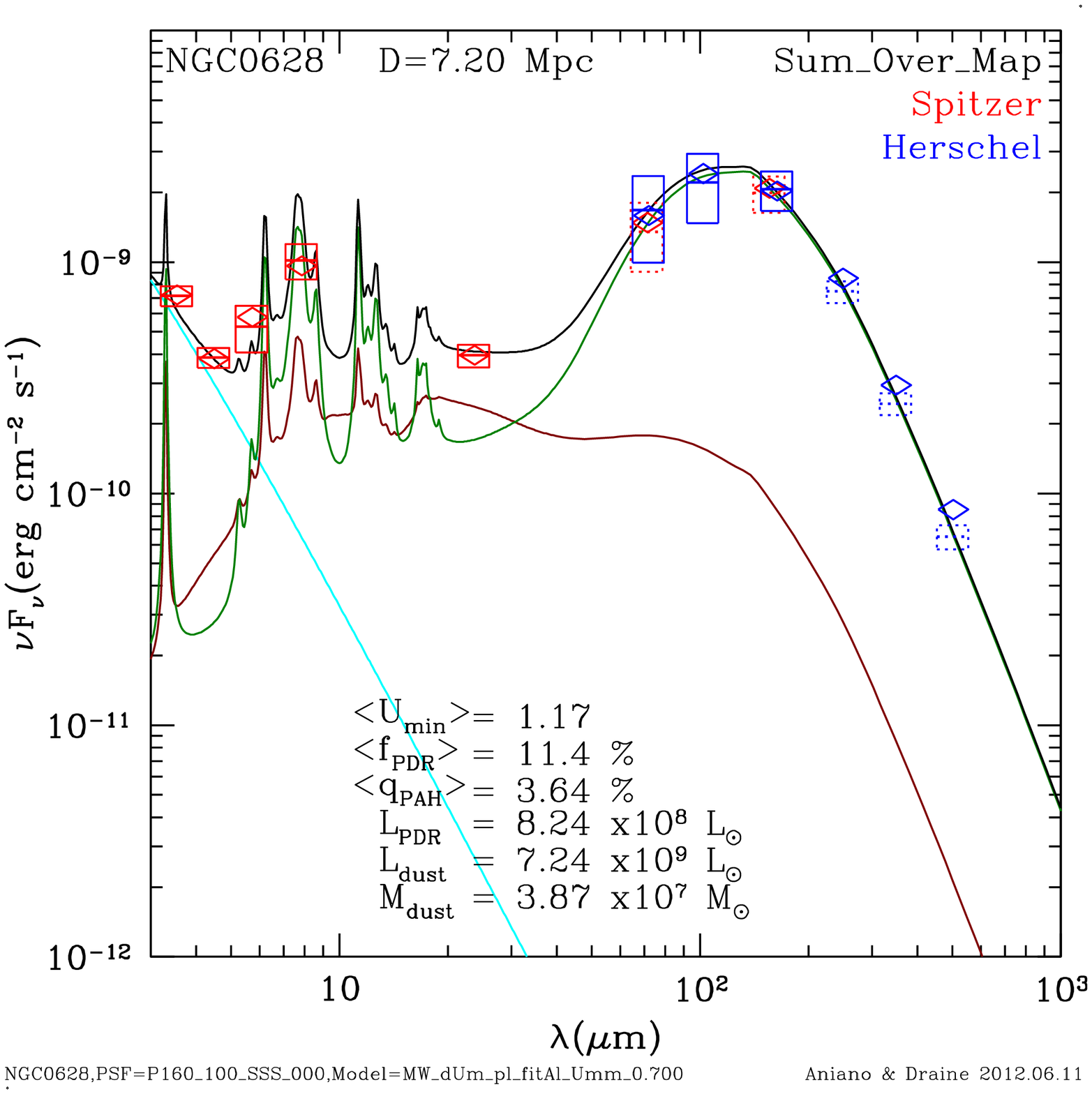}
\renewcommand \RtwoCtwo {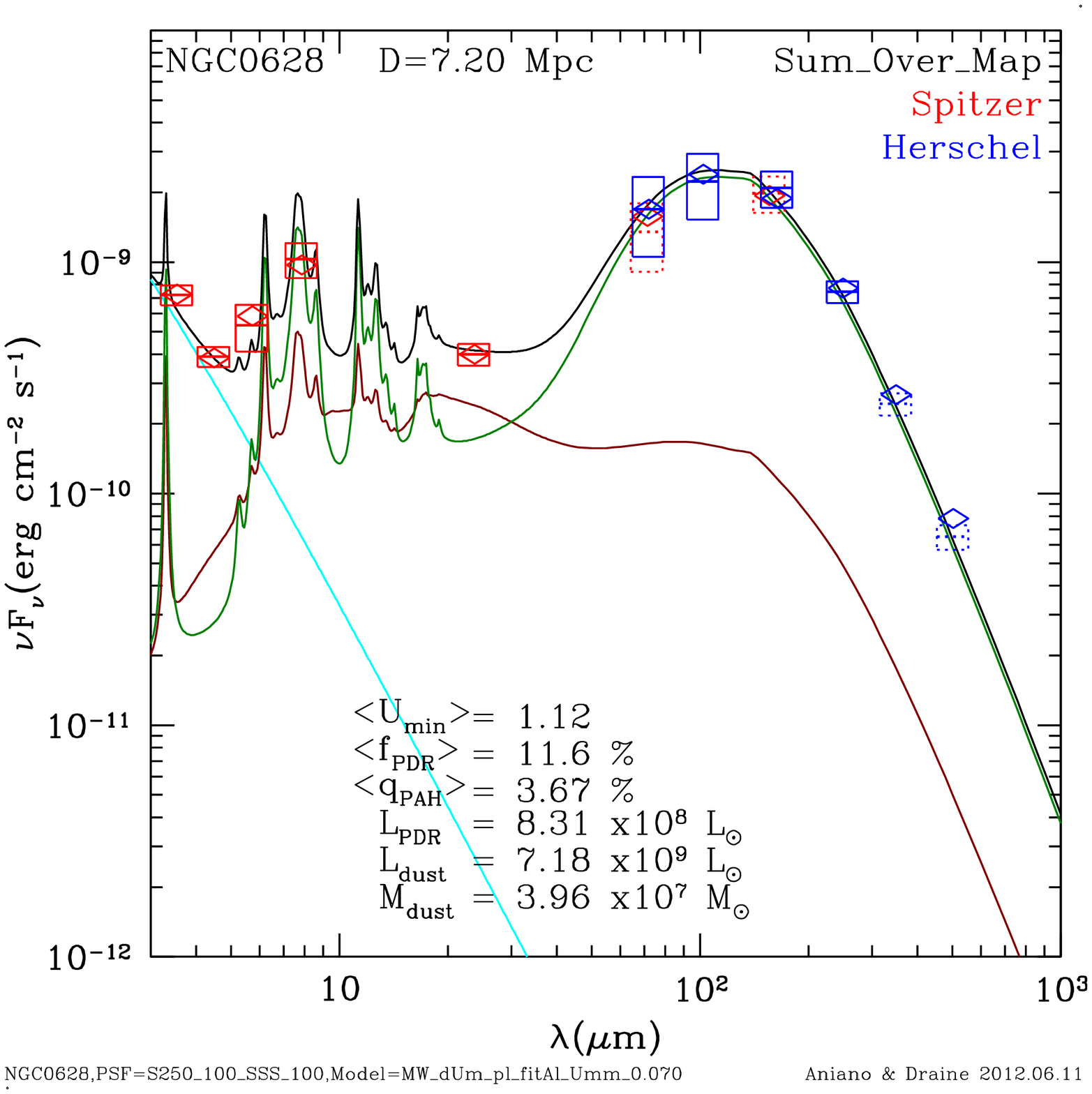}
\renewcommand \RtwoCthree {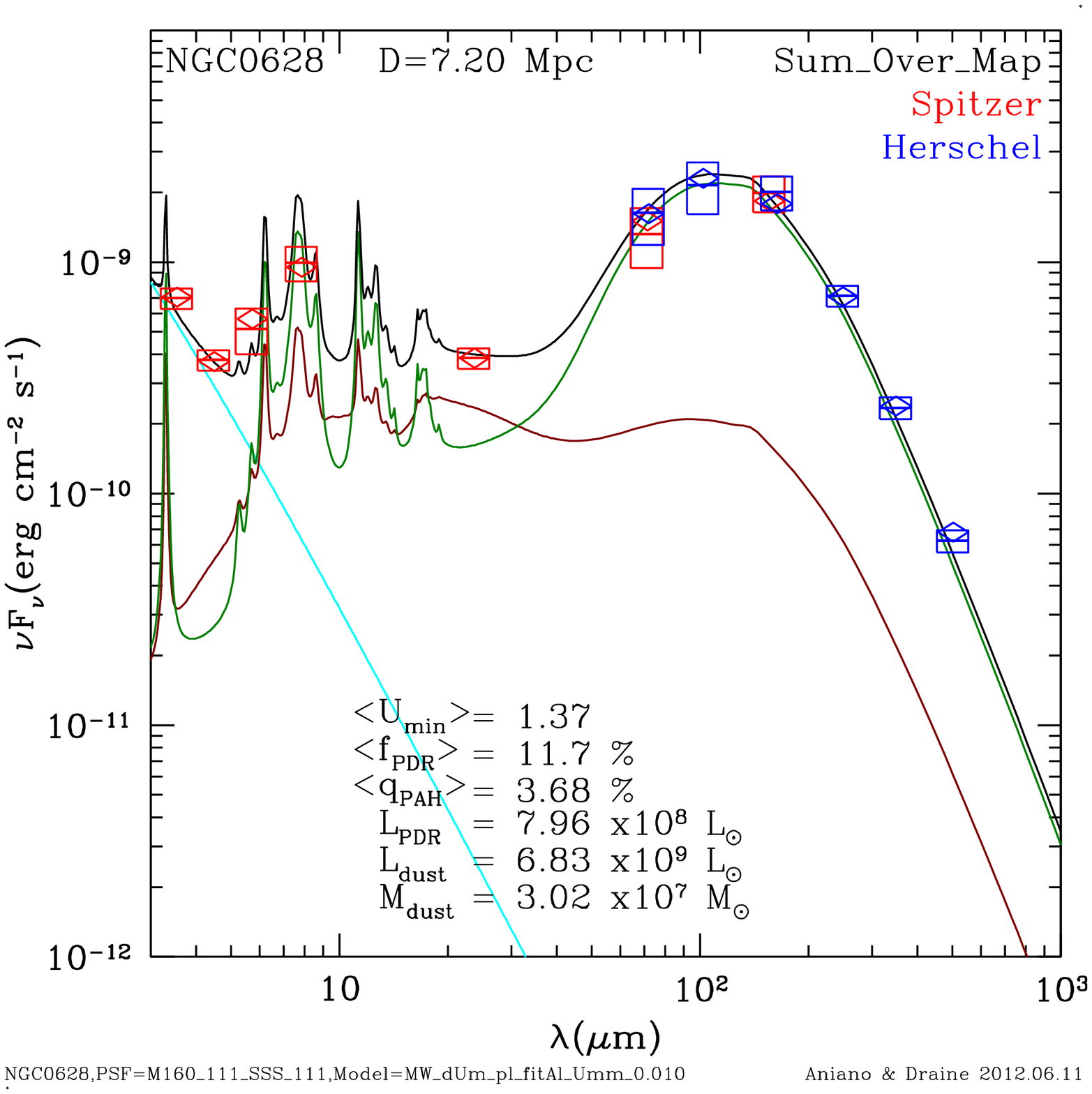}
\begin{figure}
\centering
\begin{tabular}{c@{$\,$}c@{$\,$}c} 
\footnotesize IRAC, MIPS24, PACS &\footnotesize{IRAC, MIPS24, PACS, SPIRE250}& \footnotesize IRAC, MIPS, PACS, SPIRE\\
\FirstNormal
\SecondLast
\end{tabular}
\vspace*{-0.5cm}
\caption{\footnotesize\label{fig:ngc0628-4} Model SEDs for NGC~628.
  Black line: total model spectra.  
  Cyan line: stellar contribution.  
  Dark red line: emission from dust heated by the power-law $U$ distribution.
  Dark green line: emission from dust  dust heated by $U=\Umin$.
 Solid-line rectangles: observations used in the fit (red: Spitzer (IRAC, MIPS); blue: Herschel (PACS(S), SPIRE)).  
 Dashed-line rectangles: observations not used in the fit (same color scheme as solid-line rectangles).  
  Diamonds: model convolved with camera response function (i.e., expected camera
  photometry of the model) (red: Spitzer; blue: Herschel).  
  Top row:
  Global SED compared with single-pixel models.  
  Second row:
  Global SED, for multi-pixel model.  In the left column,
  only IRAC, MIPS24, and PACS are used in the fit (MIPS70,160 and
  SPIRE are not used), yet the model nevertheless falls close to the
  observed SPIRE fluxes.
  In the center column, IRAC, MIPS24, PACS, and
  SPIRE250 are used (i.e., MIPS70,160 and SPIRE350, 500 are not used),
  and the agreement with SPIRE is improved.
  In the right column, all IRAC, MIPS, PACS, and SPIRE data are used
  to constrain the model, and the agreement with all SPIRE bands is excellent,
  although the model slightly overpredicts the emission at 350\um and 500\um.
  }
\end{figure} 

Figure \ref{fig:ngc0628-4} shows global SEDs for NGC~628.  The
observed photometry is represented by rectangular boxes (Spitzer
(IRAC, MIPS) in red; Herschel (PACS(S), SPIRE) in blue).  The model
convolved with camera response function (i.e., expected camera
photometry for the model) is represented by diamonds (Spitzer in red,
Herschel in blue).  The two rows show the global SED for NGC~628.  The
observed global photometry for NGC~628 is the same across the two top
rows, but the models differ.  Each black curve is a fitted model
spectrum.  The diamonds show the fitted model spectrum convolved with
the instrumental response function for each camera, i.e, the expected
photometry for the fitted model.

The top row shows ``single pixel'' models for the galaxy based on (a)
IRAC, MIPS24, and PACS data only, (b) IRAC, MIPS24, PACS, and
SPIRE250, and (c) IRAC, MIPS, PACS and SPIRE (all the cameras).

The second row shows the predicted model SED obtained by fitting a
dust model to each resolved pixel, and then summing the emission over
all the pixels.  These predicted SEDs differ from the previous
``single pixel'' predicted SED because the dust modeling is a
non-linear process.  The total dust mass predicted in this way is,
however, similar to the ``single pixel'' prediction (see \S 8 for
details).  When the modeling is done at lower resolutions (MIPS160)
the ``single pixel'' and map-averaged quantities are in closer
agreement.

\subsubsection{Fitting in Selected Apertures}

\renewcommand \RoneCone {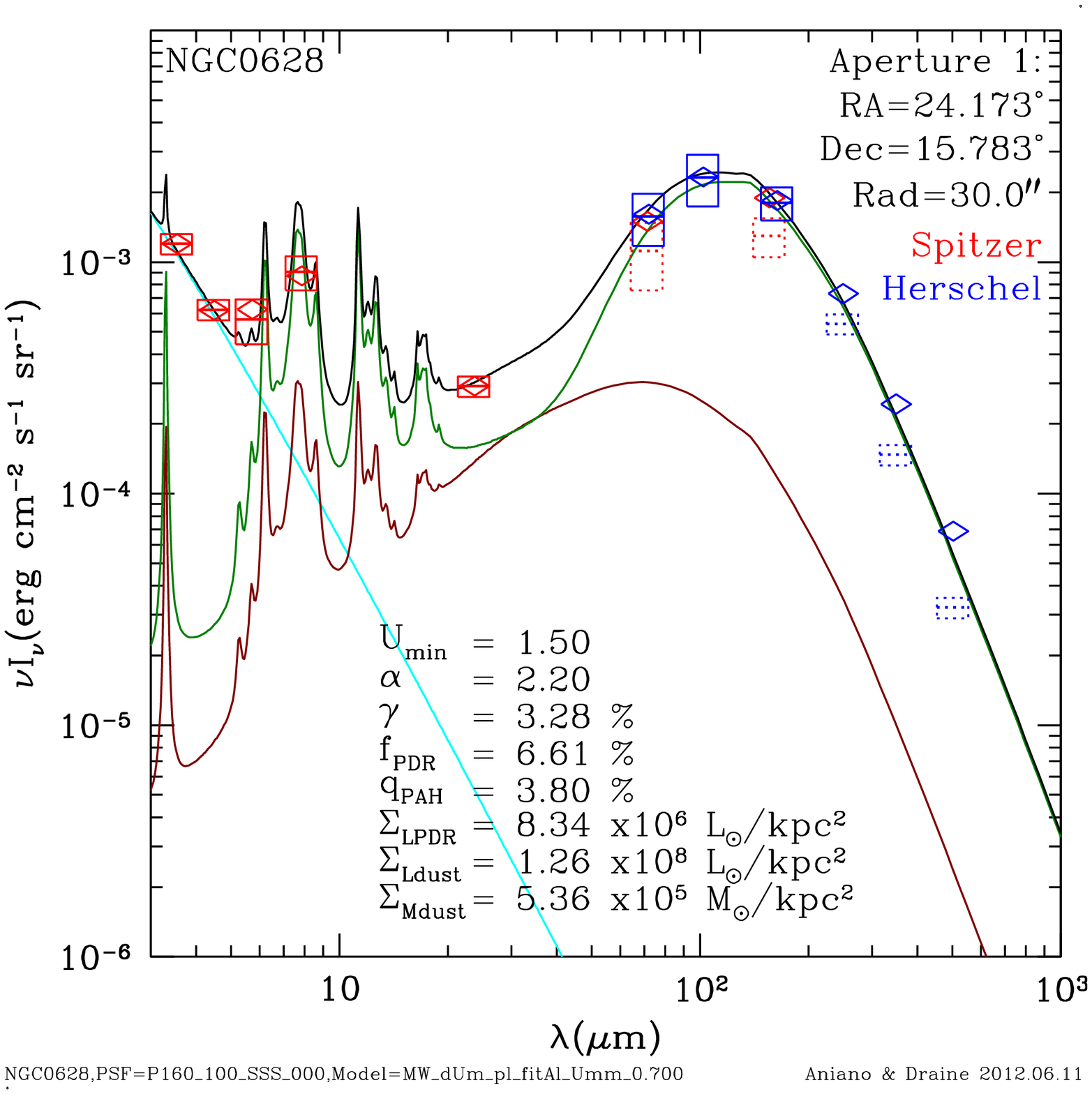}
\renewcommand \RoneCtwo {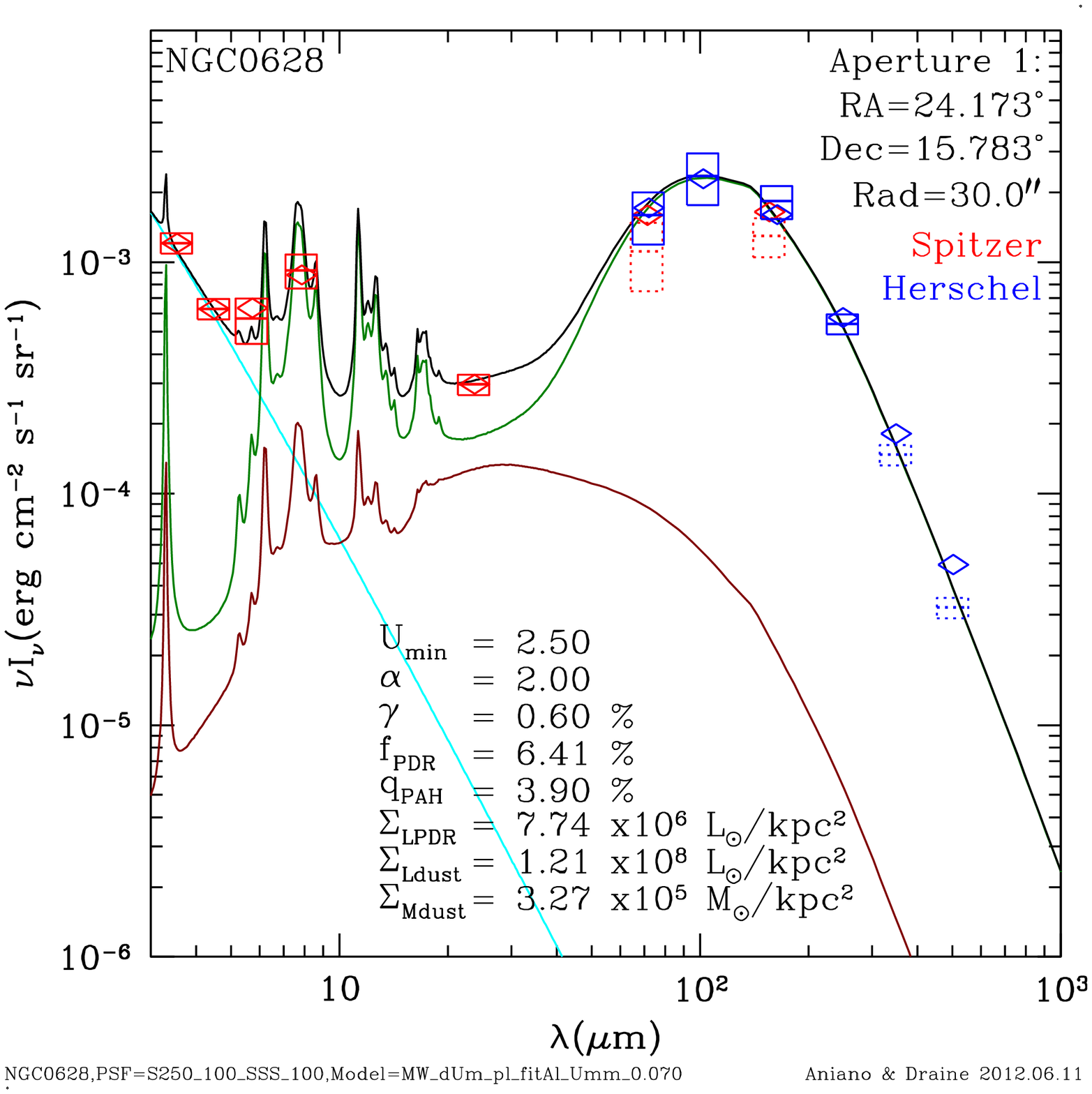}
\renewcommand \RoneCthree {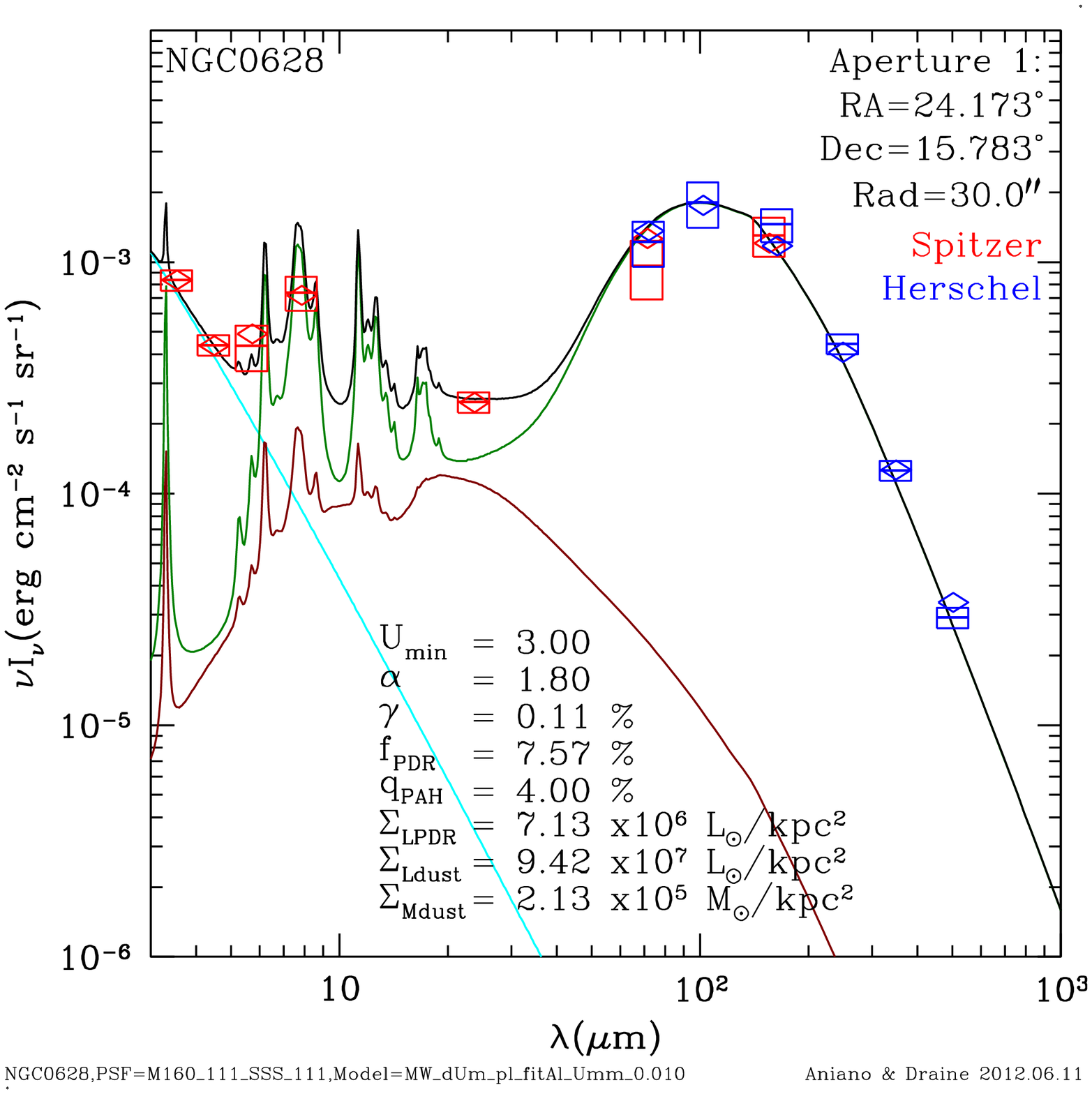}
\renewcommand \RtwoCone {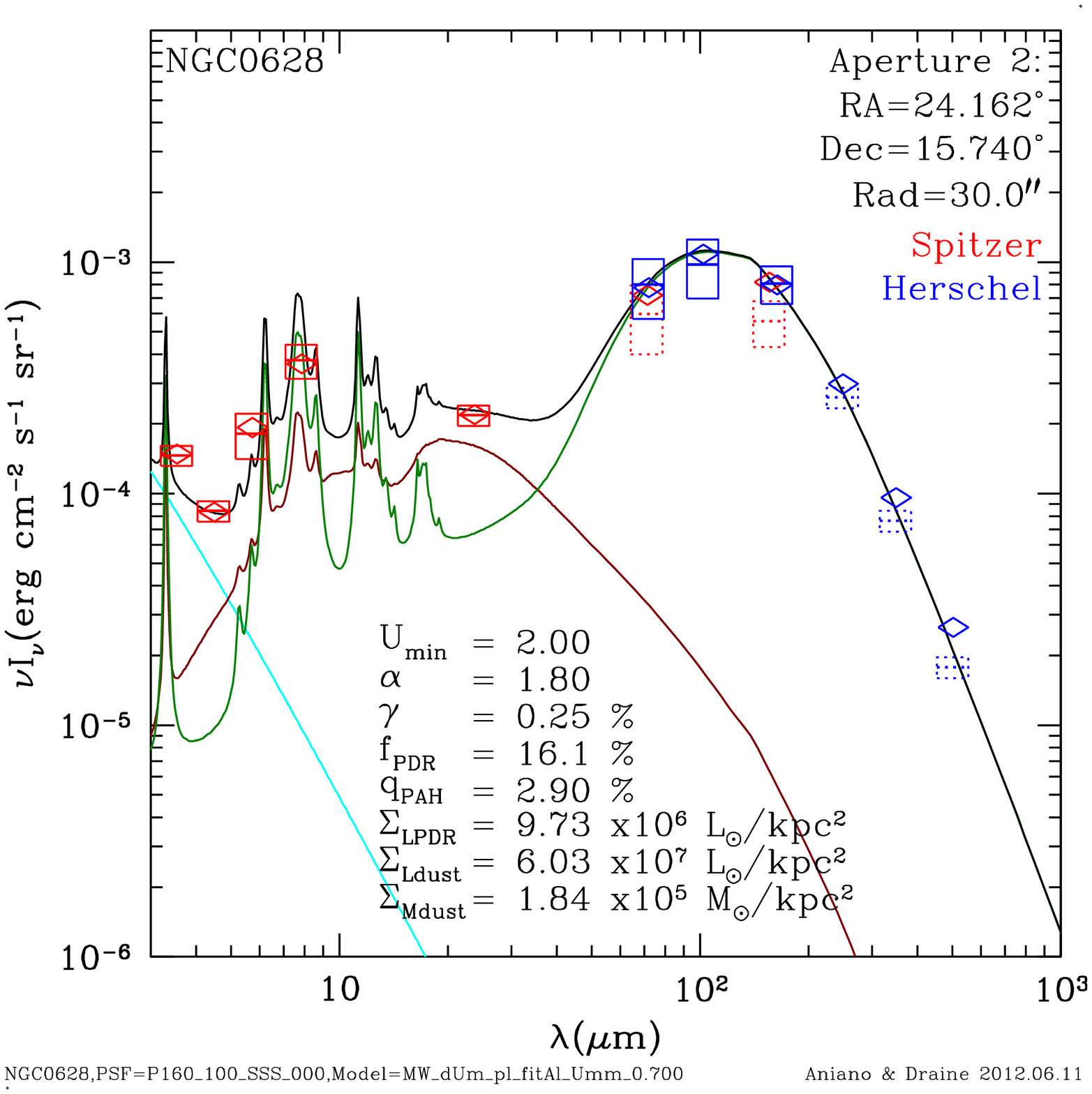}
\renewcommand \RtwoCtwo {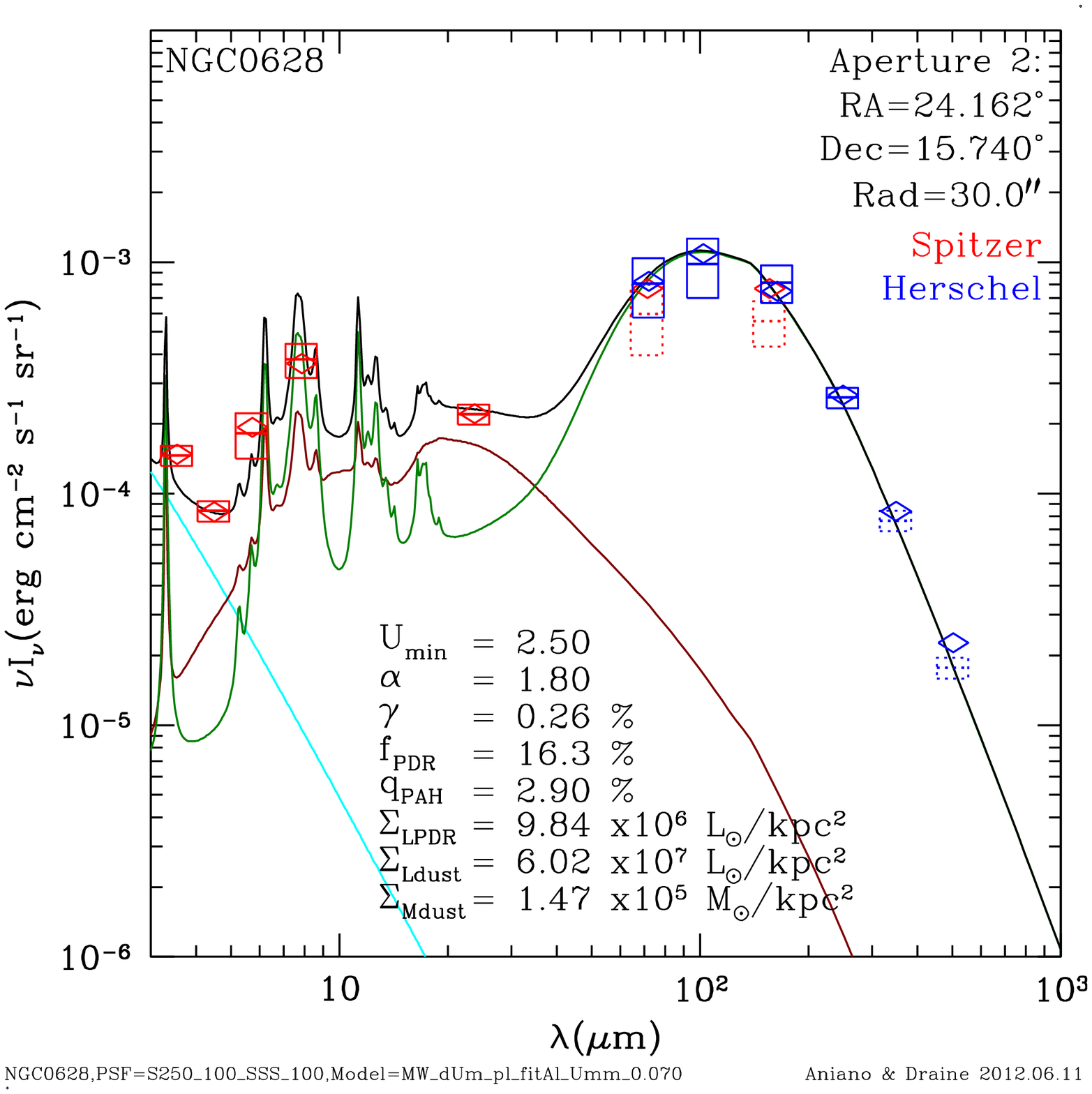}
\renewcommand \RtwoCthree {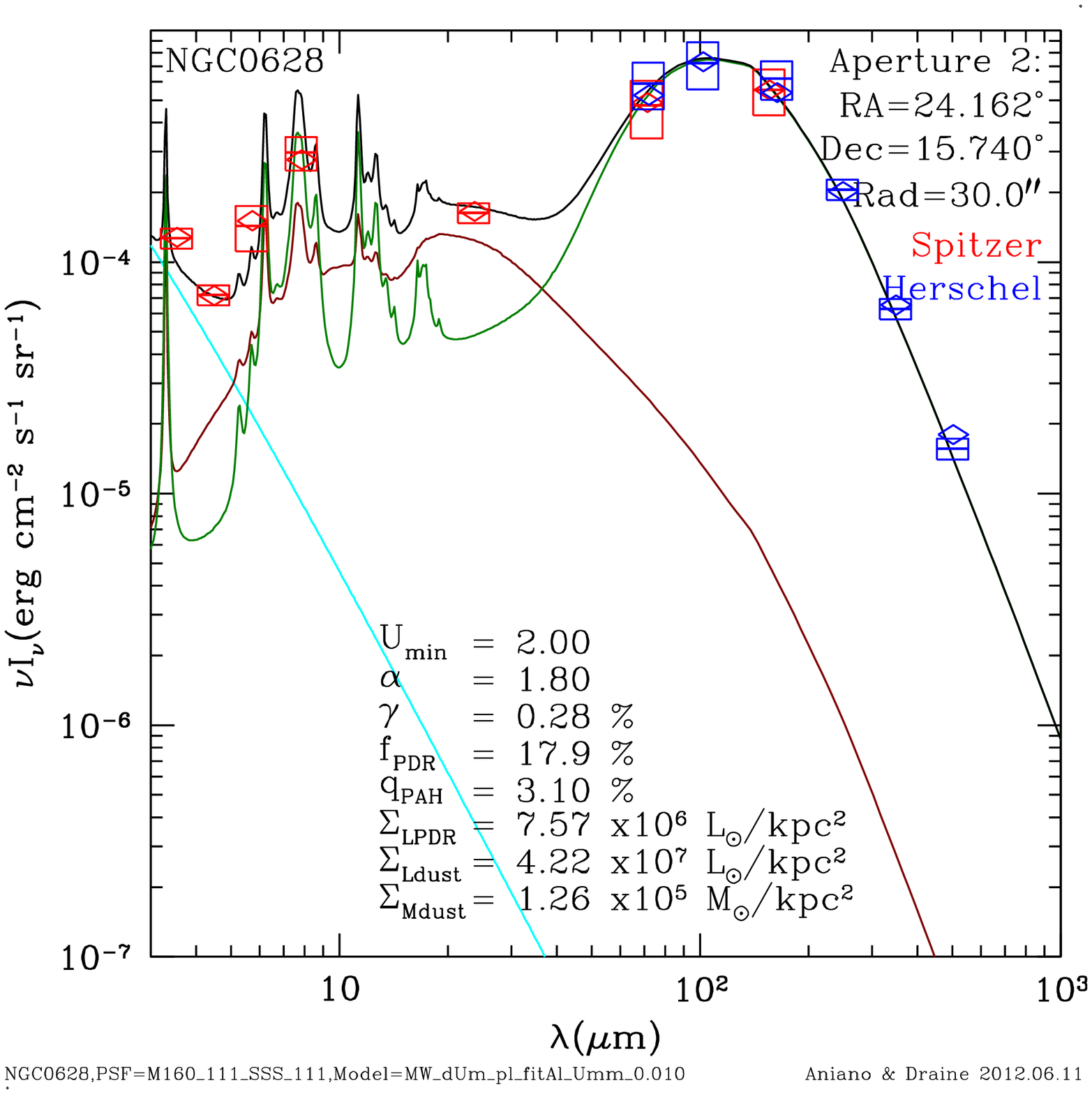}
\renewcommand \RthreeCone {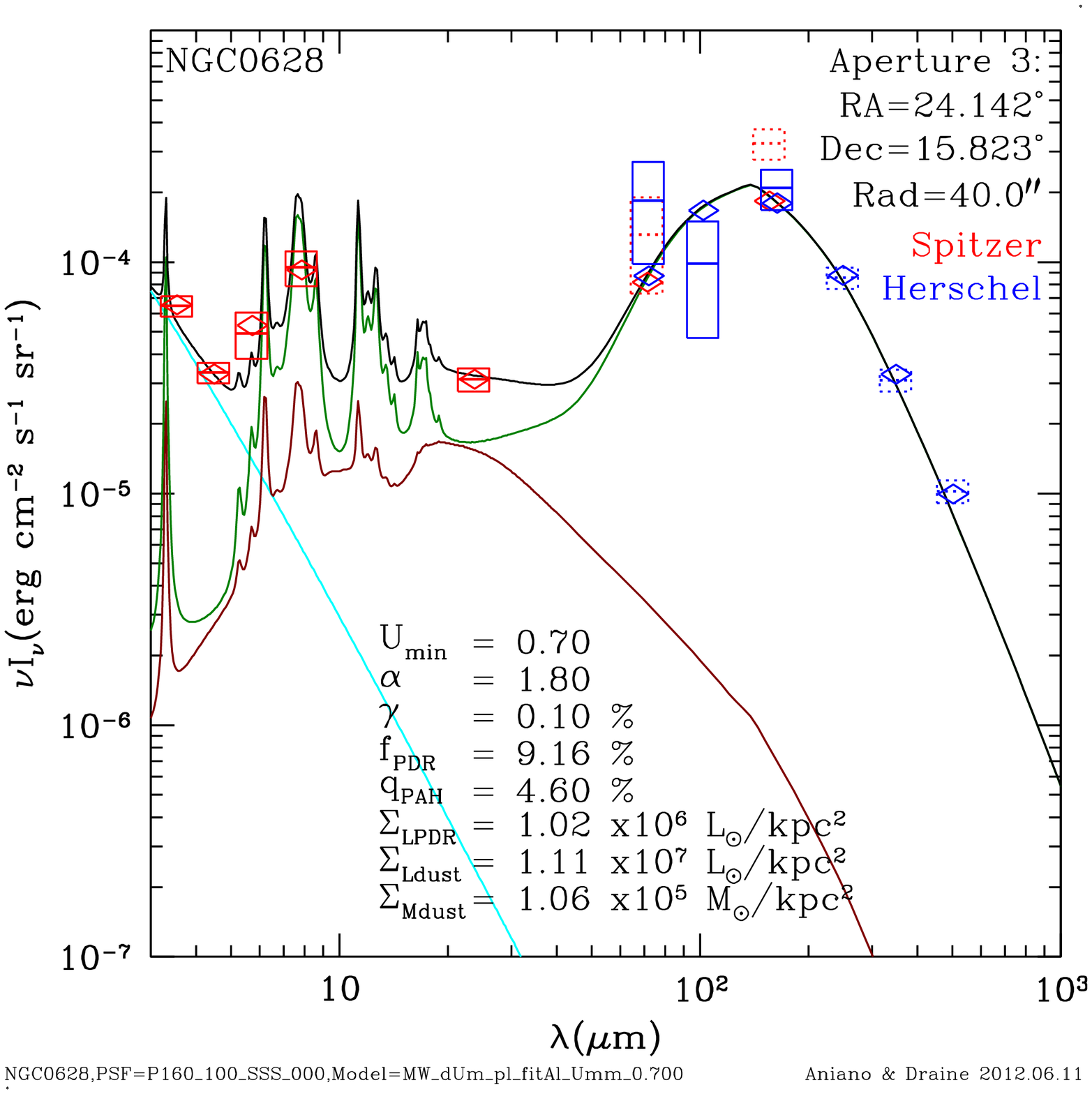}
\renewcommand \RthreeCtwo {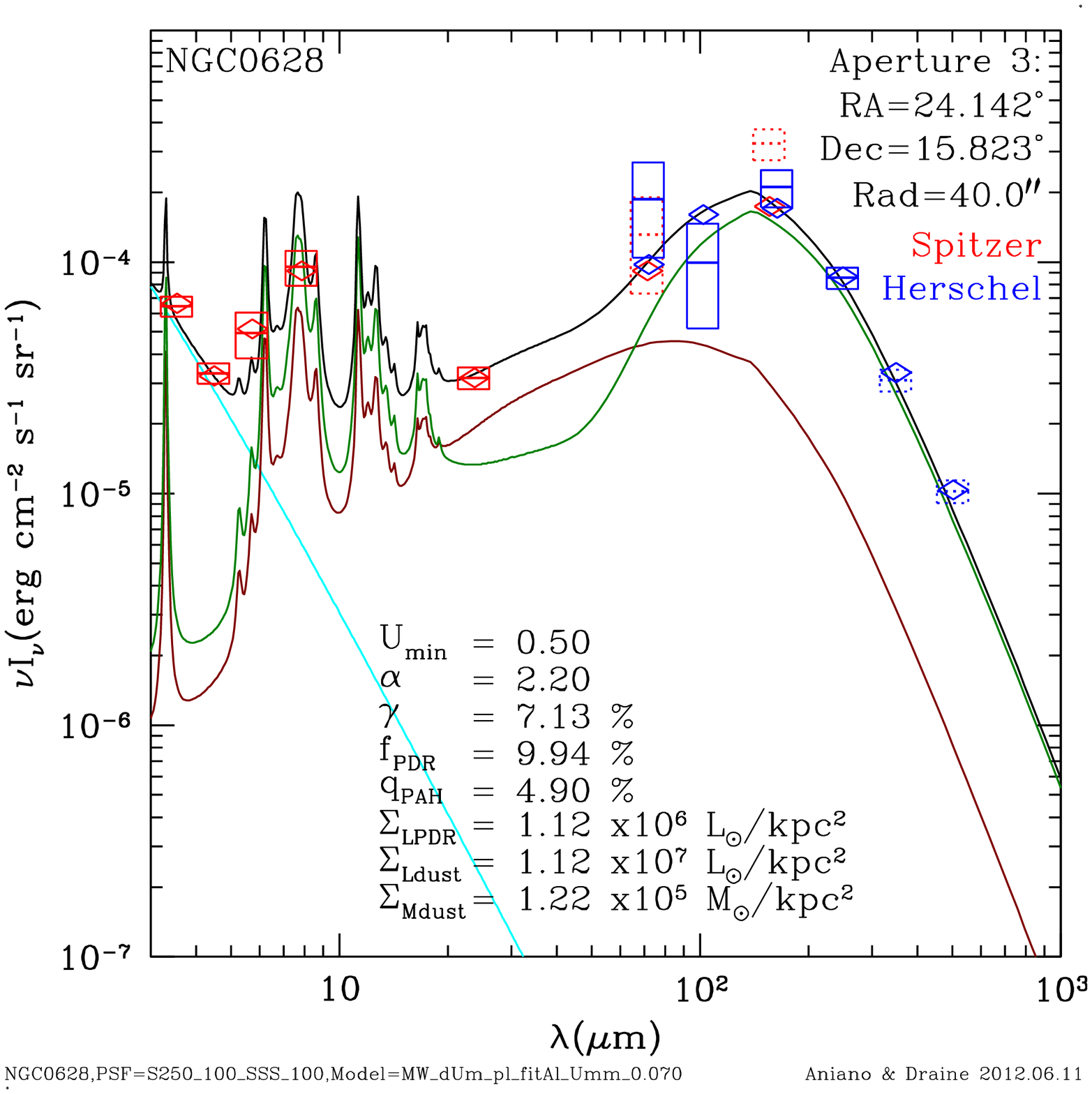}
\renewcommand \RthreeCthree {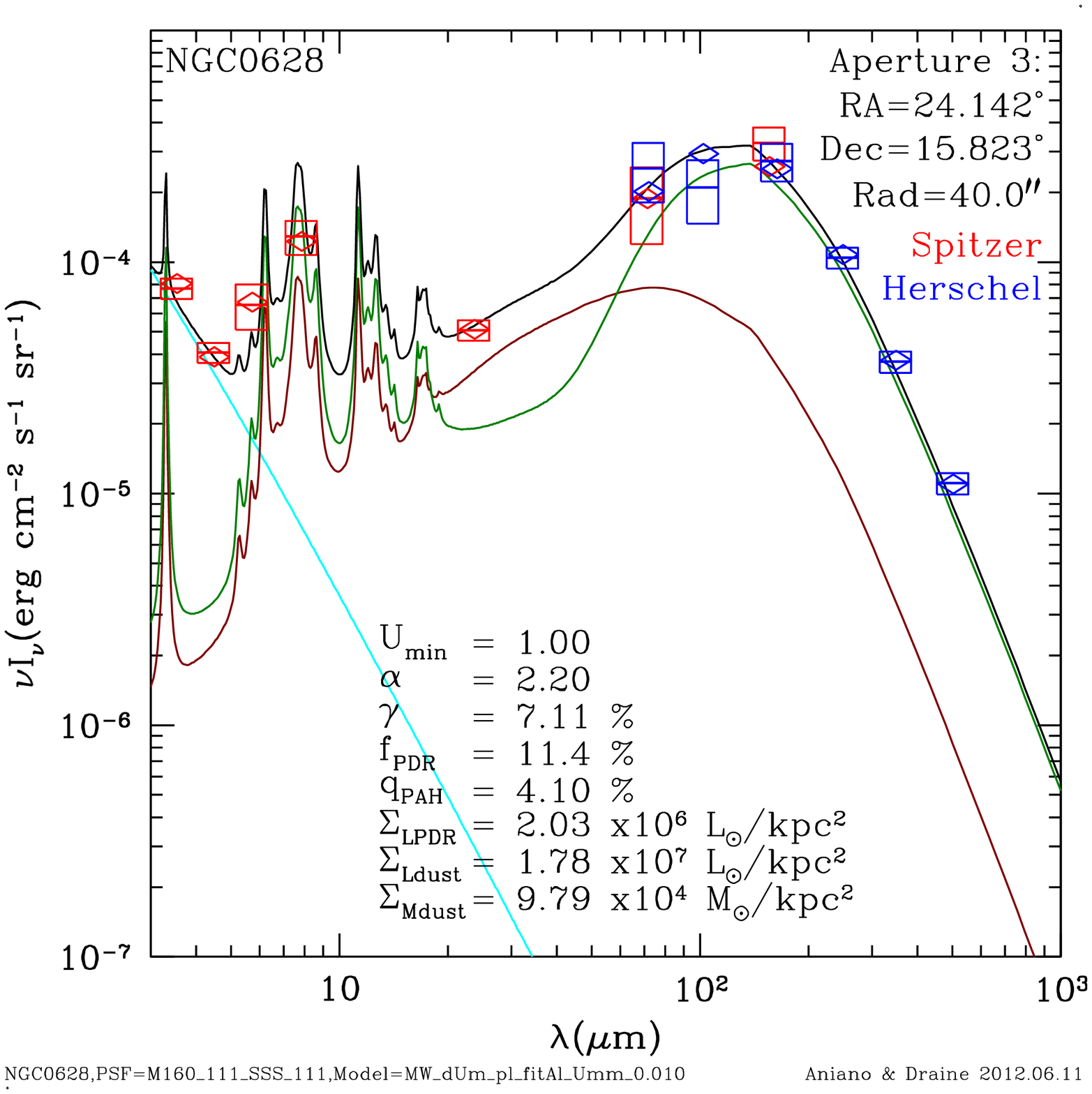}
\renewcommand \RfourCone {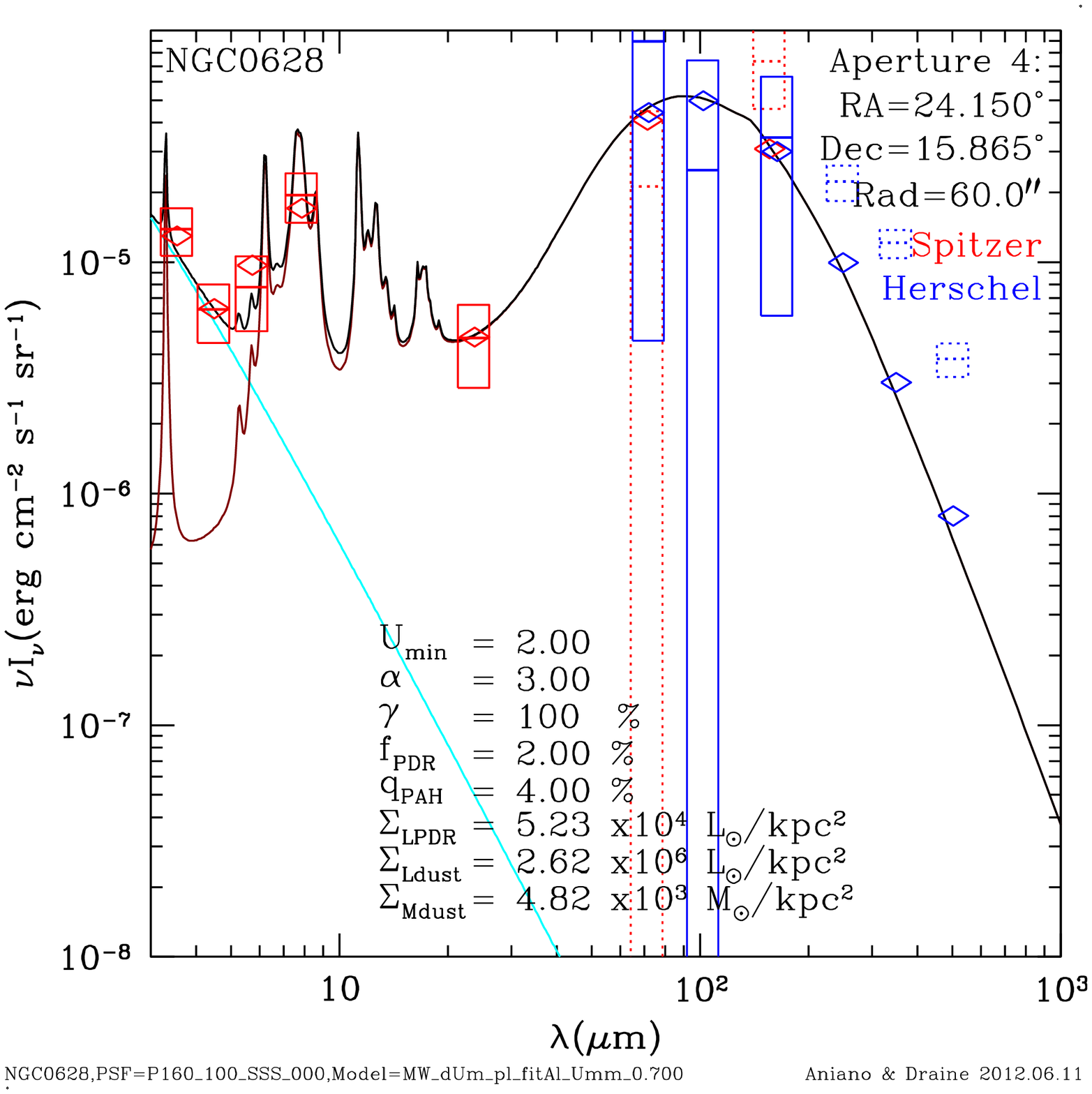}
\renewcommand \RfourCtwo {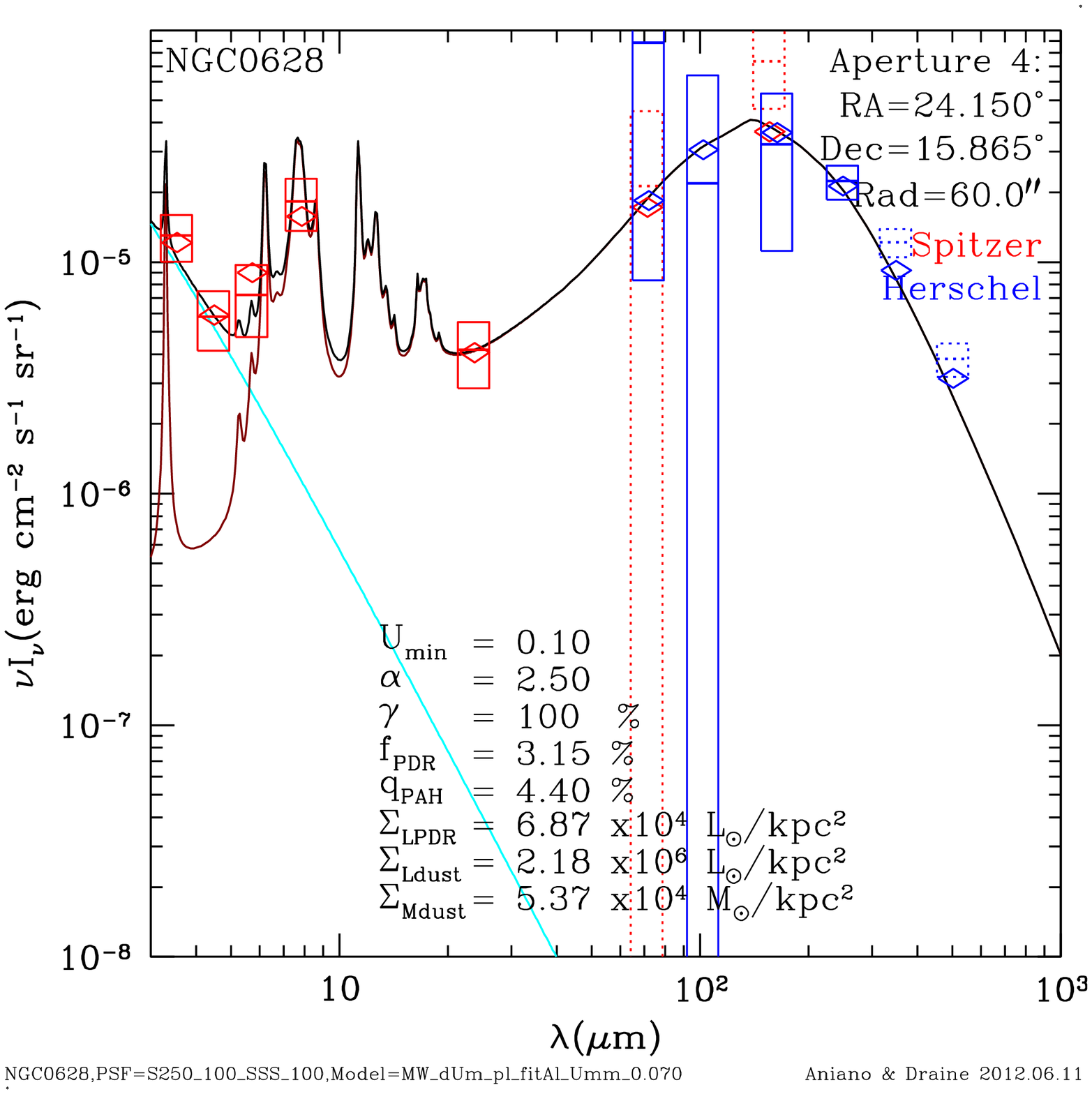}
\renewcommand \RfourCthree {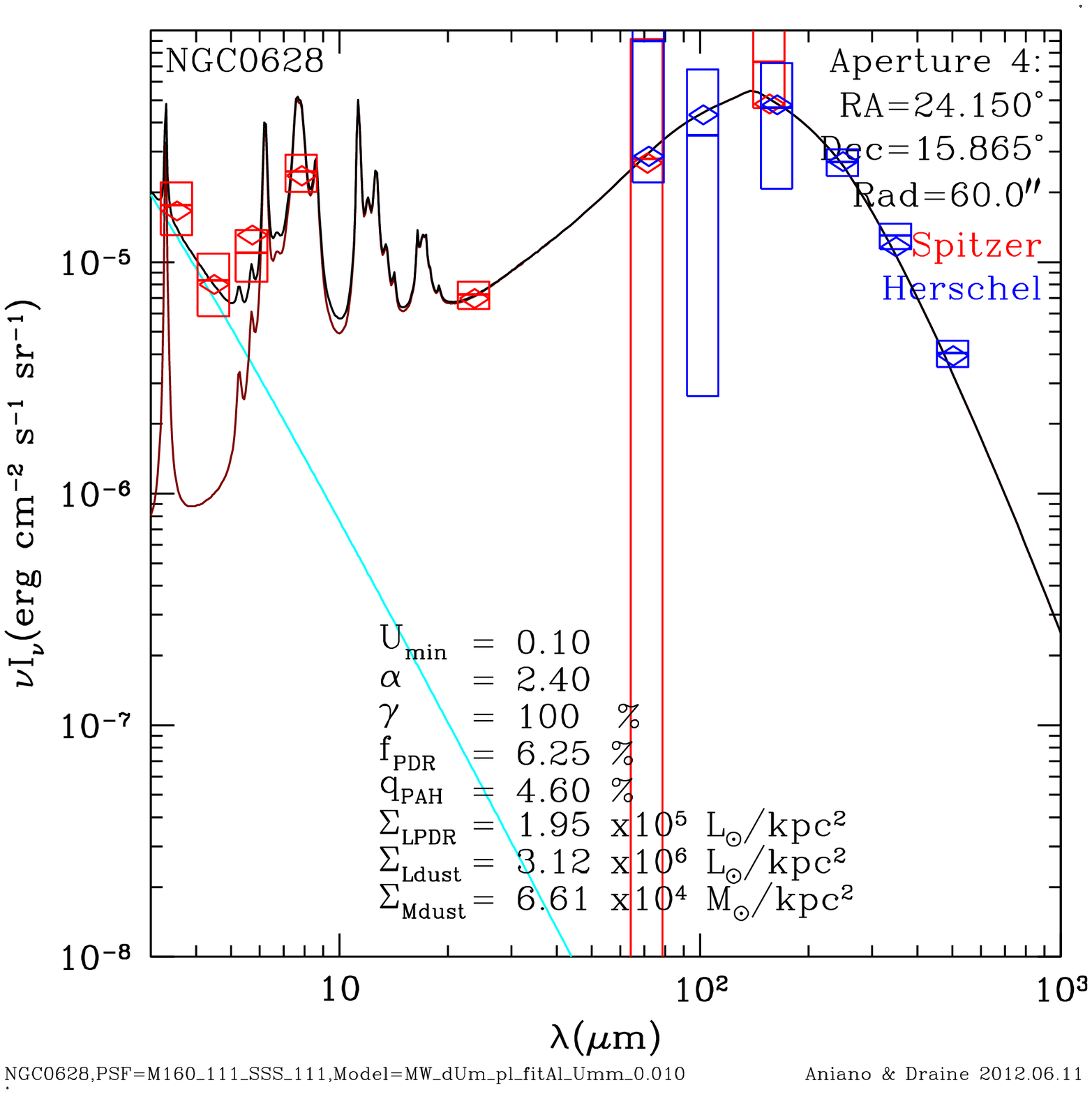}
\begin{figure}
\centering
\begin{tabular}{c@{$\,$}c@{$\,$}c} 
\footnotesize PACS160 PSF & \footnotesize SPIRE250 PSF & \footnotesize MIPS160 PSF \\
\FirstNormal
\SecondNormal
\ThirdNormal
\FourthLast
\end{tabular}
\vspace*{-0.5cm}
\caption{\footnotesize\label{fig:ngc0628-5}
Model SEDs for four selected apertures on NGC~628. 
See Fig.\ \ref{fig:ngc0628-1} for aperture location, and Fig.\ \ref{fig:ngc0628-4} for an explanation of the color coding.
Top row: $60\arcsec$ (circular) aperture centered on the galaxy nucleus.
Second row: $60\arcsec$ aperture located on a bright spot on the spiral arms.
Third row: $80\arcsec$ aperture in a mid-luminosity region.
Bottom row: $120\arcsec$ aperture in a low-luminosity region.}
\end{figure} 

Figure \ref{fig:ngc0628-5} shows SEDs for four circular apertures
located on NGC~628, sampling a wide range of surface brightnesses, 
$\Sigma_{L_\dust}\approx (0.3-9)\times10^{7}\Lsol\kpc^{-2}$ 
(see Fig.\ \ref{fig:ngc0628-1} for aperture locations). 
Aperture 4 is located partially outside the galaxy mask, where the single pixel dust modeling is not reliable, but the improved S/N of a large aperture allows reliable determination of the dust model parameters.
We fit the DL07 model to the summed flux within each aperture.
The top row shows the SED for a $60\arcsec$ (diameter) circular aperture
centered on the galaxy nucleus.  The second row shows the SED for a
$60\arcsec$ aperture located on a bright spot on the
spiral arms.  
We observe that the PACS photometry is larger
than the corresponding MIPS photometry in the high surface brightness apertures 1 and 2. 
When both data sets are
included in the modeling, the dust model gives values intermediate
between PACS and MIPS.  The third and fourth rows show
SEDs in $80\arcsec$ and $120\arcsec$ apertures further from the
center. We employ larger apertures in order to obtain reasonable S/N
in these fainter regions.  In aperture 4, the MIPS and PACS photometry
differ significantly.  
When SPIRE, MIPS70, and MIPS160 are not used
(e.g., in the PACS160 resolution modeling, Fig.\ \ref{fig:ngc0628-5}j)
the high PACS70/PACS160 ratio causes the model to infer very high
values of $\Umin$ and low values of $\Sigma_{M_\dust}$, and hence to
underpredict the SPIRE photometry.  When we include SPIRE250
(Fig.\ \ref{fig:ngc0628-5}k) the model can reproduce SPIRE250, but
continues to underpredict SPIRE350 and SPIRE500.
Finally, when SPIRE350,500 and MIPS70,160
are added as constraints (Fig.\ \ref{fig:ngc0628-5}l), 
the modeling improves dramatically in
aperture 4 and other low-brightness areas, reproducing most of the data,
and making it clear that PACS70 is an outlier.

The astute reader will note that the estimated uncertainties for,
e.g., the PACS100 global photometry differ between the columns (i.e.,
for fluxes extracted after convolving to different PSFs).  For each
PSF, we estimate the noise per pixel based on the pixel statistics in
the background region (see Appendix \ref{app:unc_estimation}).  We
then make a simple assumption concerning the pixel-to-pixel noise
correlation.  The fact that the uncertainty estimates for the aperture
fluxes depend on the PSF is an indication that our assumption about
the correlated component of the noise is imperfect.  This is only an
issue for the faint, low S/N data, such as the PACS fluxes in
Apertures 3 and 4.

In aperture 4 (Fig.\ \ref{fig:ngc0628-5} last row), the model uses
$\gamma=1$, i.e., the dust is heated almost entirely by a power law
$U$ distribution, but with very high values of $\gamma$. This corresponds to a
broad distribution of starlight intensities in the $U \gtrsim \Umin$
range, with very little power in the high intensity range (PDR). This
may be an artifact arising from the large photometric uncertainties.


\subsection{\label{sec:6946}NGC~6946}

NGC\,6946, an active star-forming galaxy, is classified as SABcd.  At
a distance $D=6.8\Mpc$, the total \ion{H}{1} mass is 
$M({\rm H\,I})=(5.5\pm0.3)\times10^9\Msol$
\citep{Walter+Brinks+deBlok+etal_2008}, 
with 42\% of the \ion{H}{1}
falling within the galaxy mask.  The H$_2$ mass within the galaxy mask
is $M(\HH)=(8.6\pm0.6)\times10^9(\XCOxx/4)\Msol$
\citep{Leroy+Walter+Bigiel+etal_2009}.  As discussed previously, we
adopt $\XCOxx=4$ as a global estimate.  The star formation rate
is estimated to be $4.5\Msol\yr^{-1}$
\citep{Calzetti+Wu+Hong+etal_2010}.  NGC\,6946 is remarkable for
hosting at least 9 supernovae over the past century
\citep{Prieto+Kistler+Thompson+etal_2008}.

\begin{figure}[h]
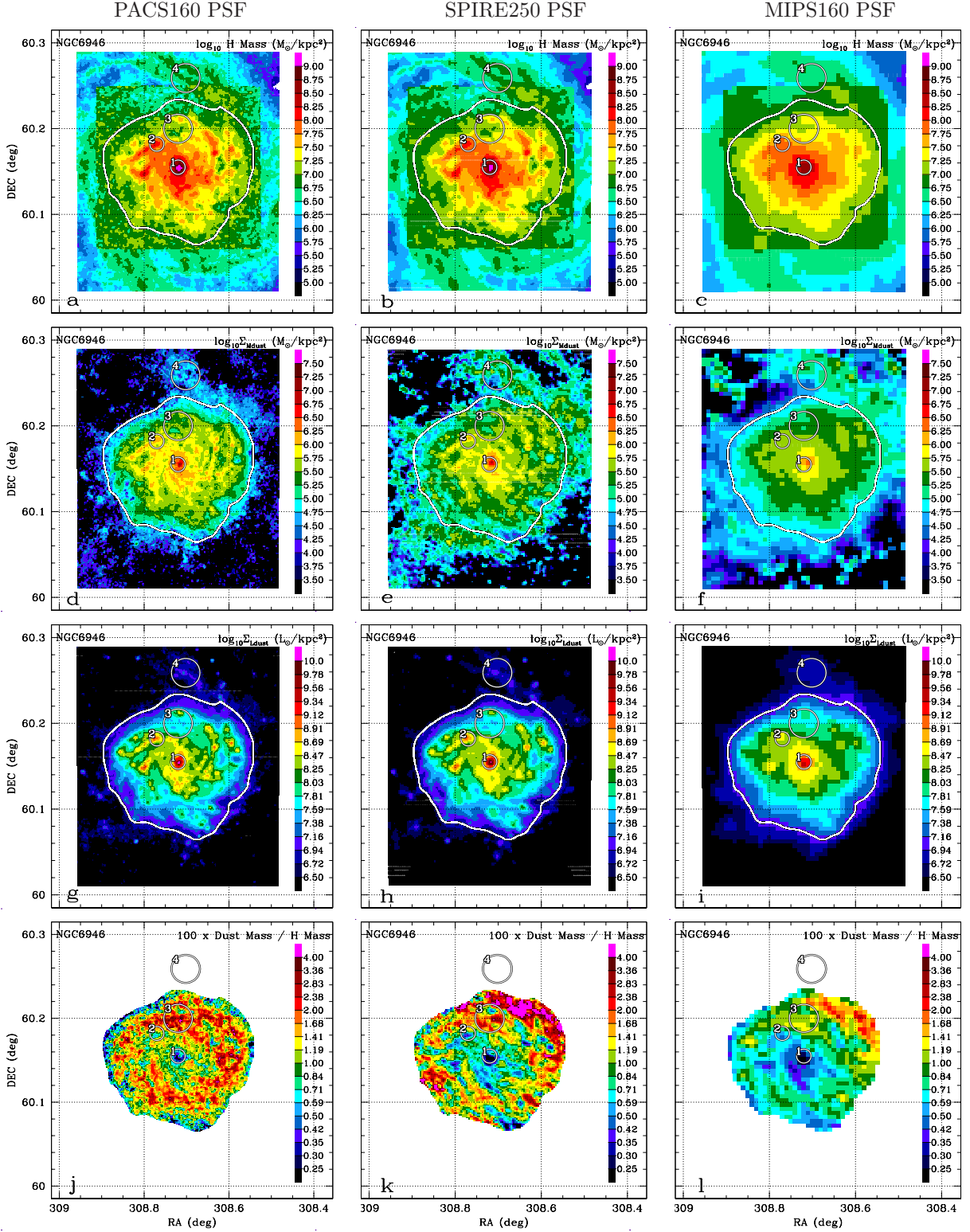
 
\centering 
\begin{tabular}{c@{$\,$}c@{$\,$}c} 
\footnotesize PACS160 PSF & \footnotesize SPIRE250 PSF & \footnotesize MIPS160 PSF \\
\FirstNormal
\SecondNormal
\ThirdNormal
\FourthLast
\end{tabular}
\vspace*{-0.5cm}
\caption{\footnotesize\label{fig:ngc6946-1}
NGC~6946 at the resolution of PACS160 (left), SPIRE250 (center), and MIPS160 (right).
Row 1: surface density of H\,I and $\HH$ for $\XCOxx=3$ (see text).
Row 2: dust surface density $\Sigma_{M_\dust}$.
Row 3: dust luminosity surface density $\Sigma_{L_\dust}$.
Row 4: dust \!/\! H ratio over the main galaxy.
The irregular white contour is the boundary of the ``galaxy mask'', within
which the signal/noise ratio is high enough to obtain reliable estimates of
dust and starlight parameters.
White circles are selected apertures (see Fig.\ \ref{fig:ngc6946-5}). }
\end{figure}

The variable carbon star V0778 Cyg,
located near RA=309.044, Dec= 60.082 (slightly off the
bottom left corner of the maps shown in figures \ref{fig:ngc6946-1} -
\ref{fig:ngc6946-3}), saturates the IRAC and
MIPS detectors. 
The associated image artifacts affect the background estimation and subtraction, making the modeling less reliable in the bottom left corner of the maps.

\renewcommand \RoneCone {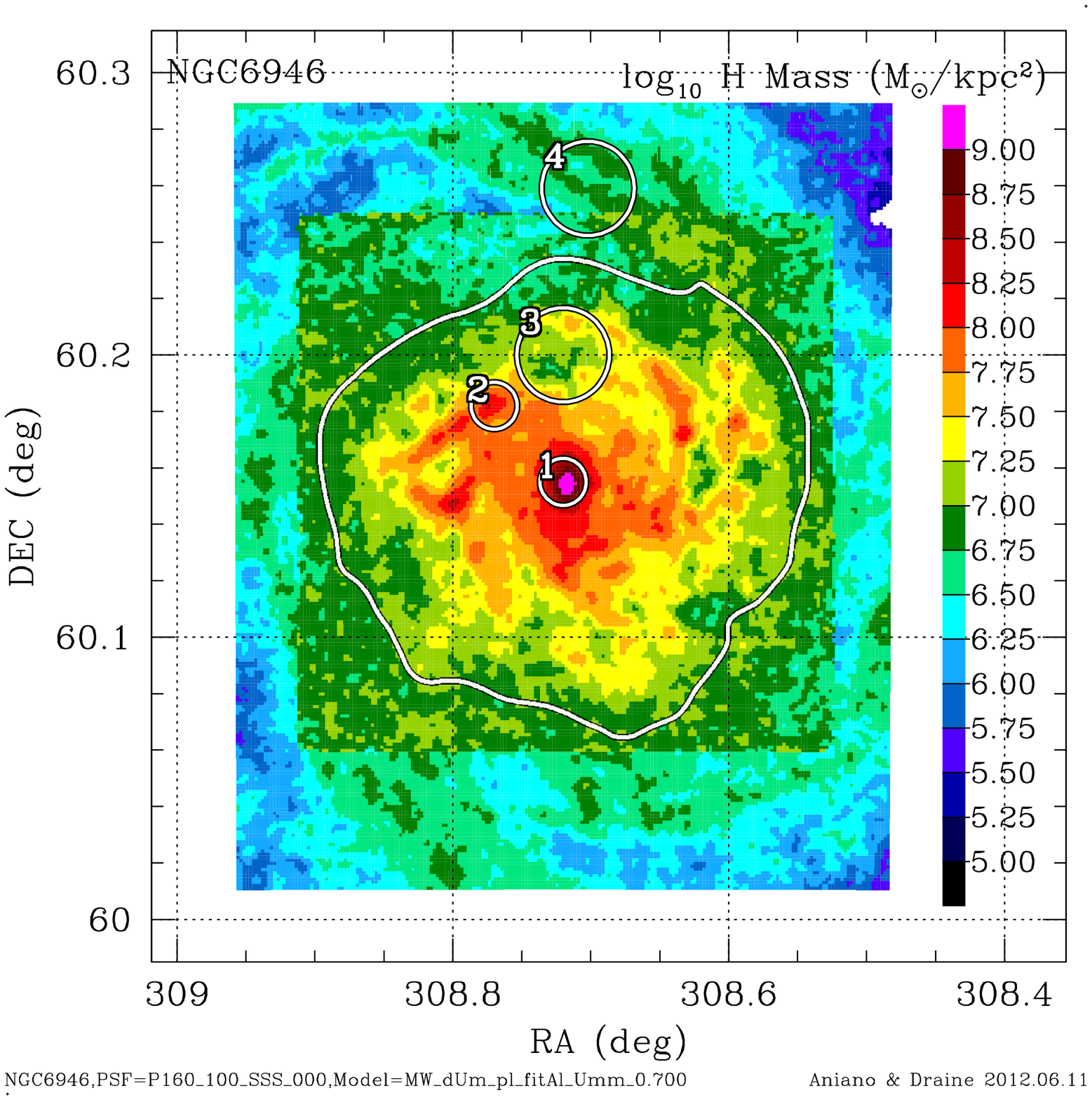}
\renewcommand \RoneCtwo {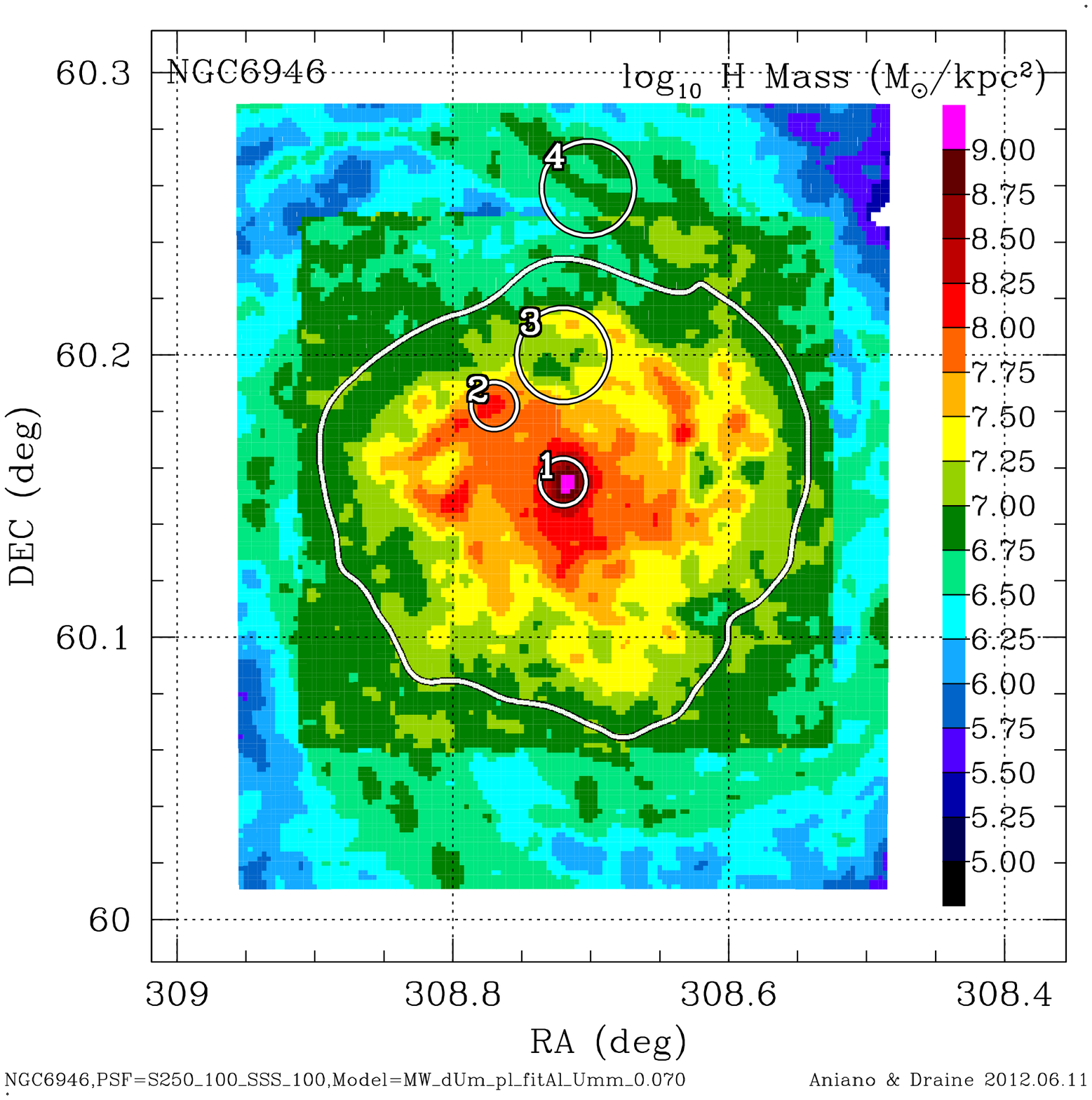}
\renewcommand \RoneCthree {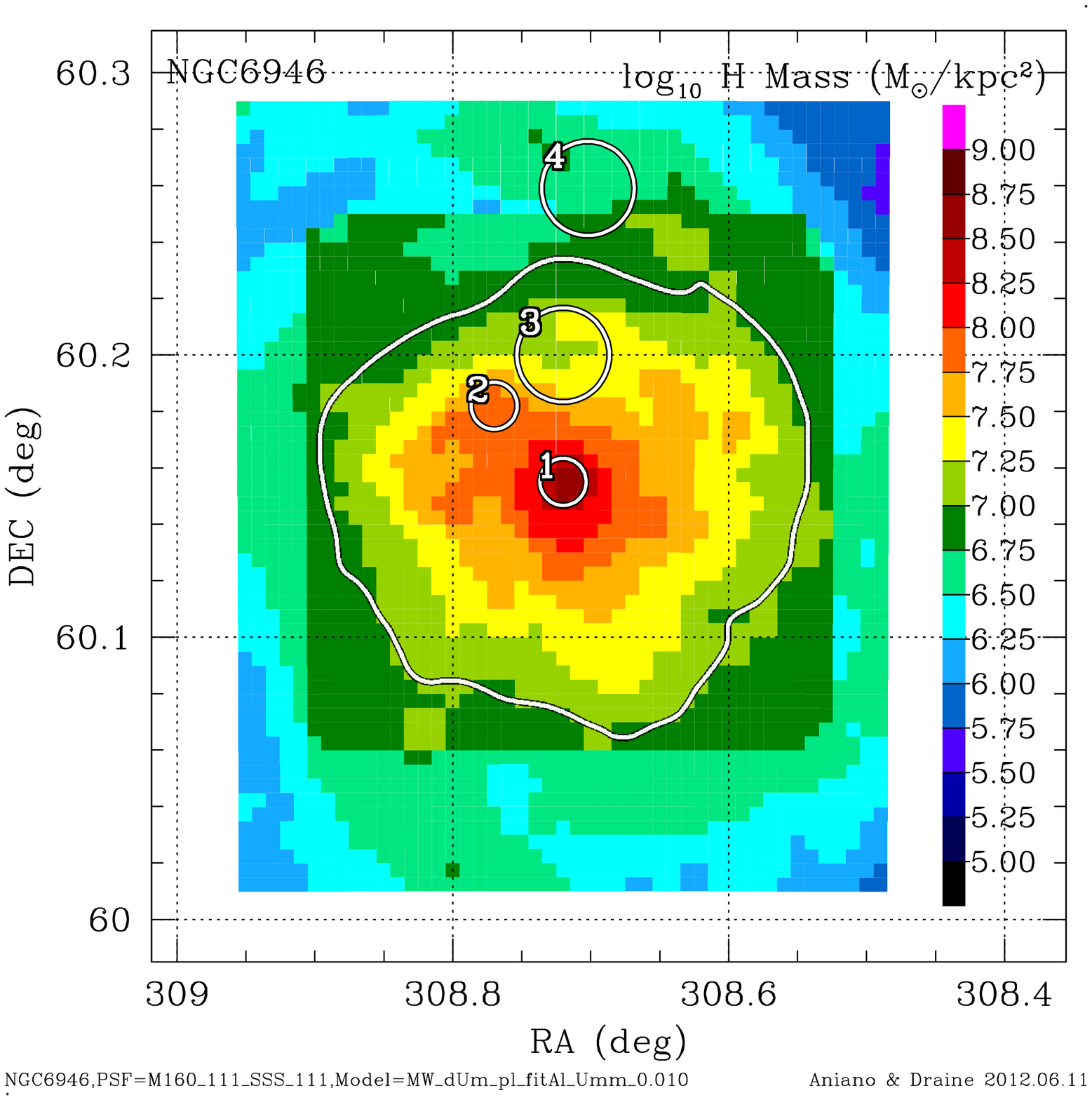}
\renewcommand \RtwoCone {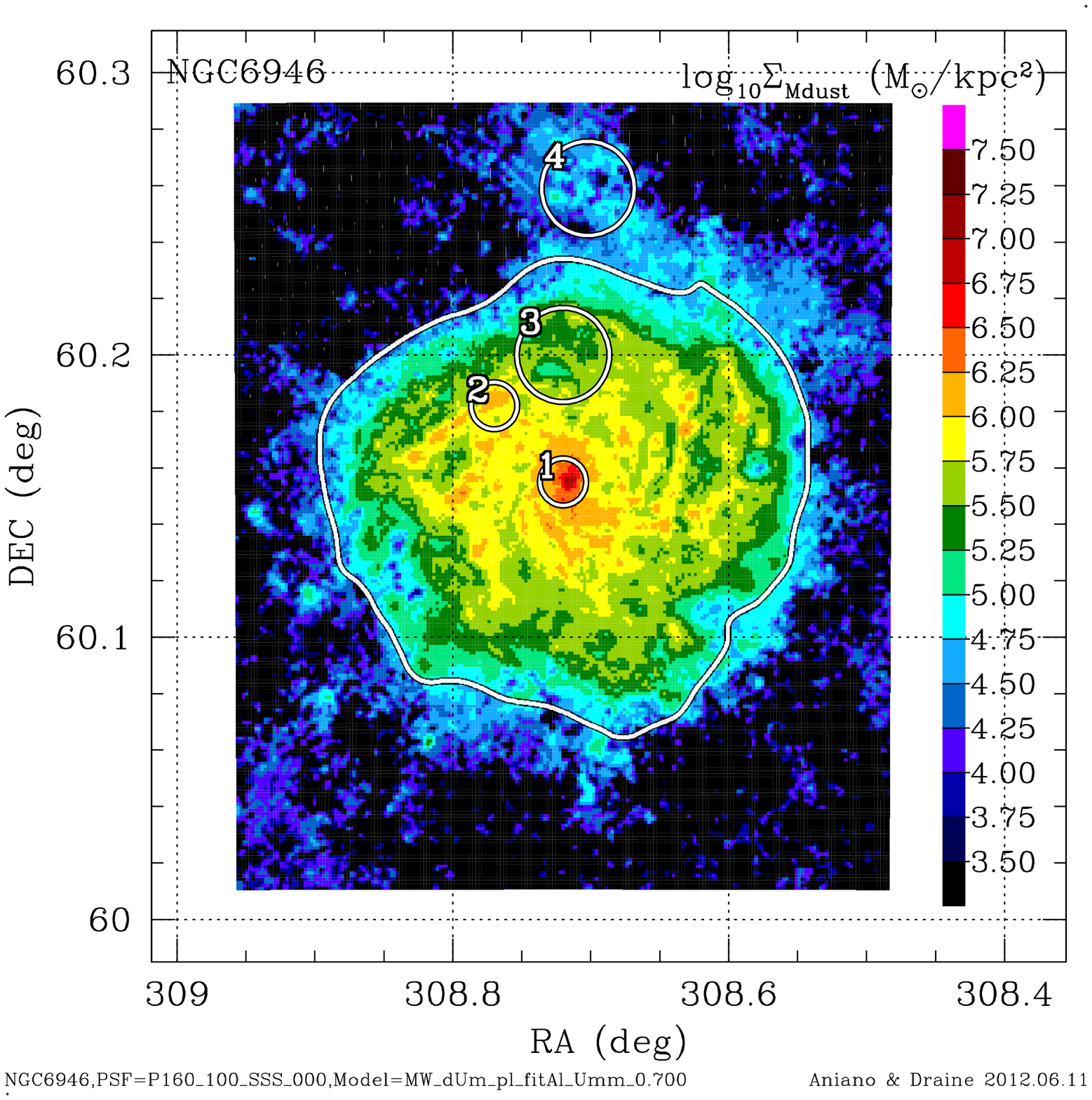}
\renewcommand \RtwoCtwo {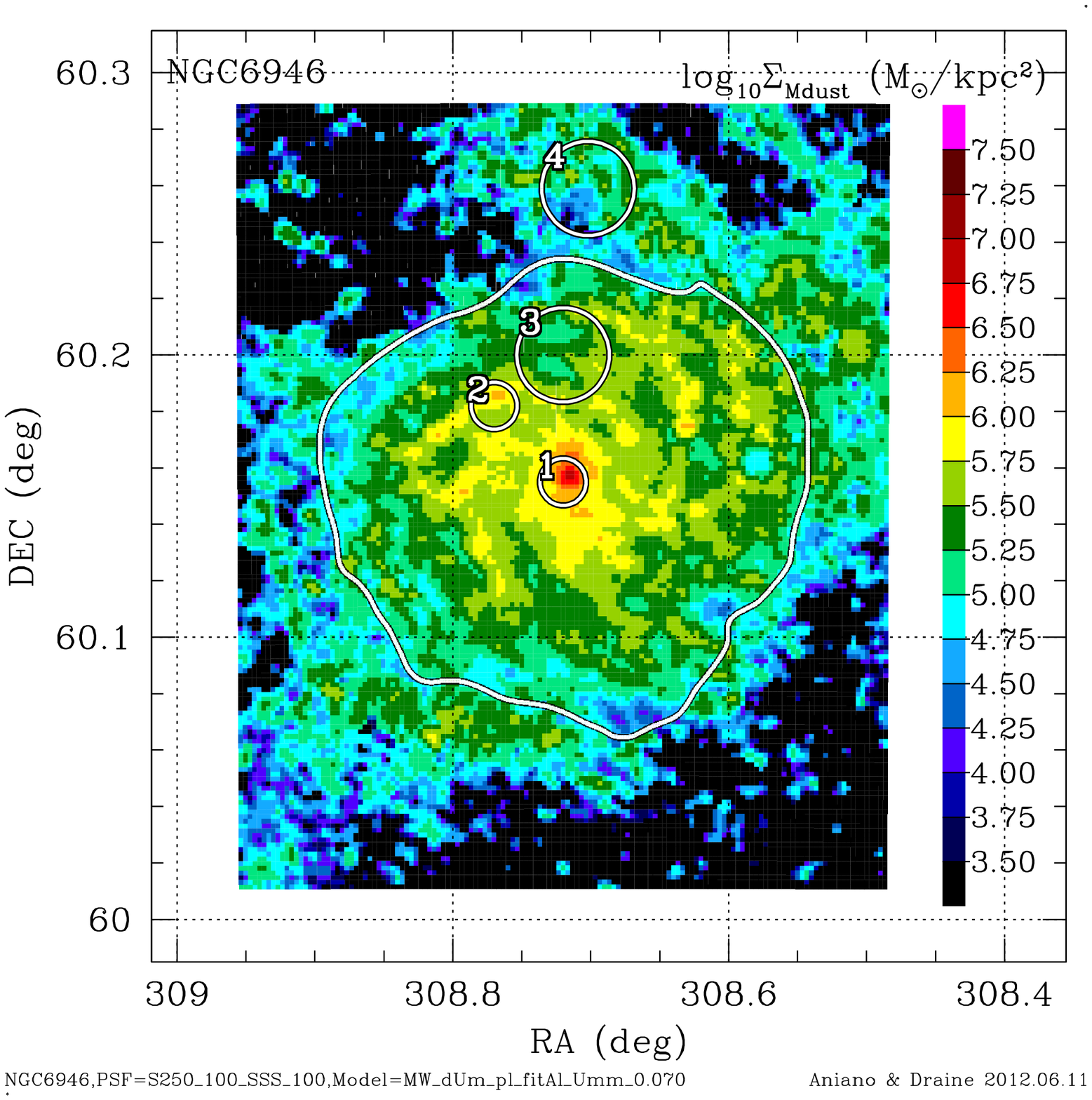}
\renewcommand \RtwoCthree {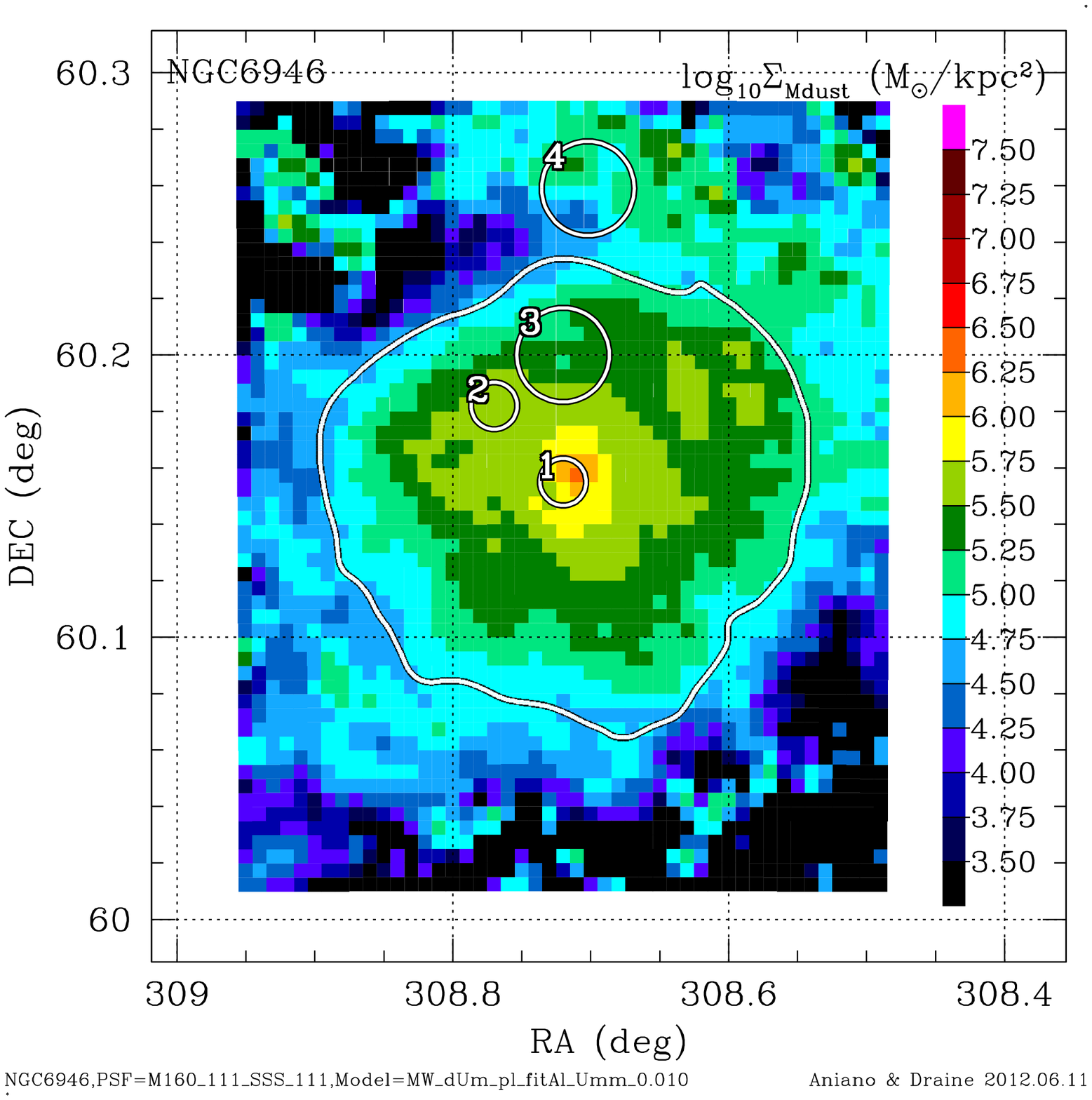} 
\renewcommand \RthreeCone {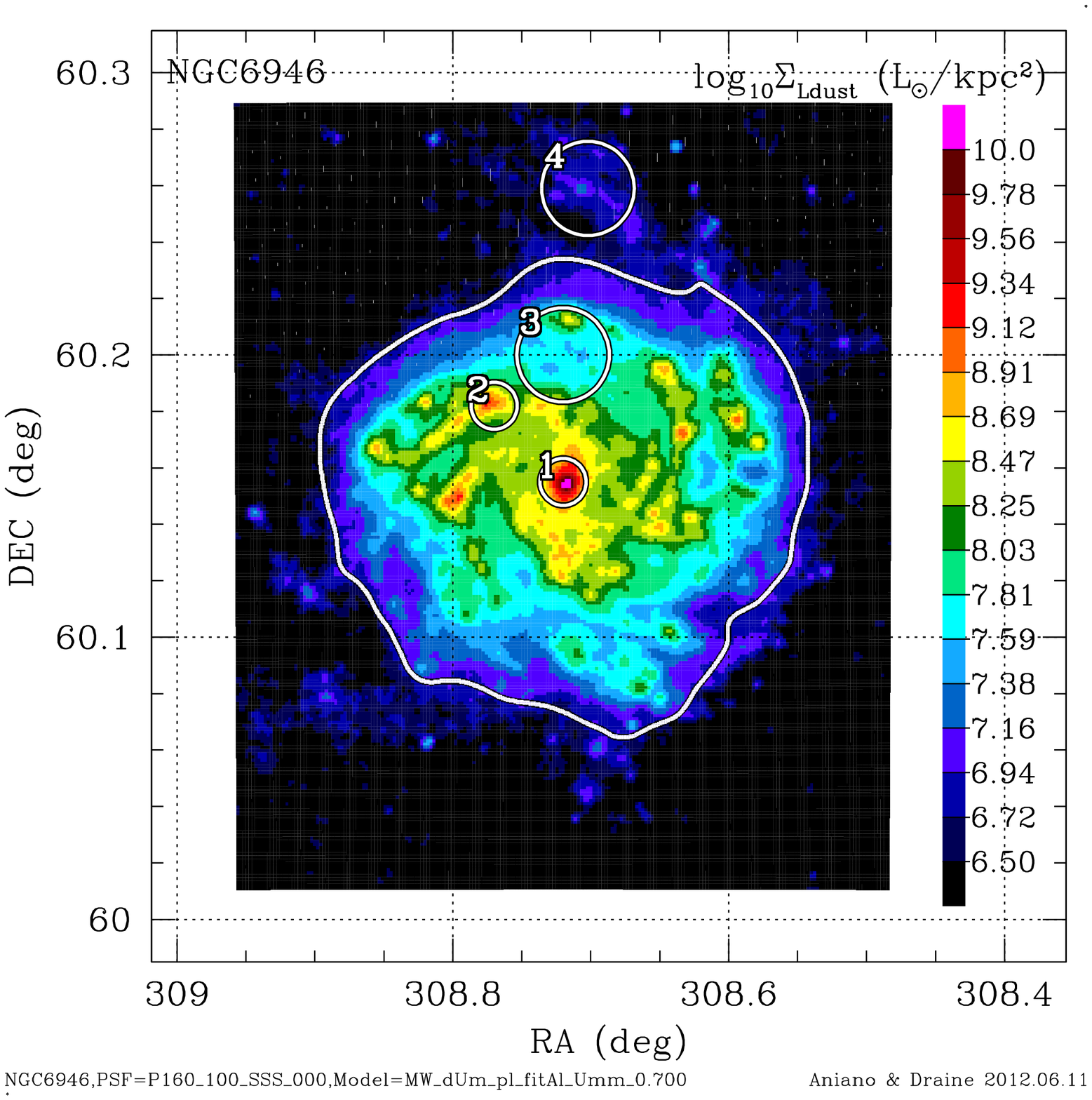}
\renewcommand \RthreeCtwo {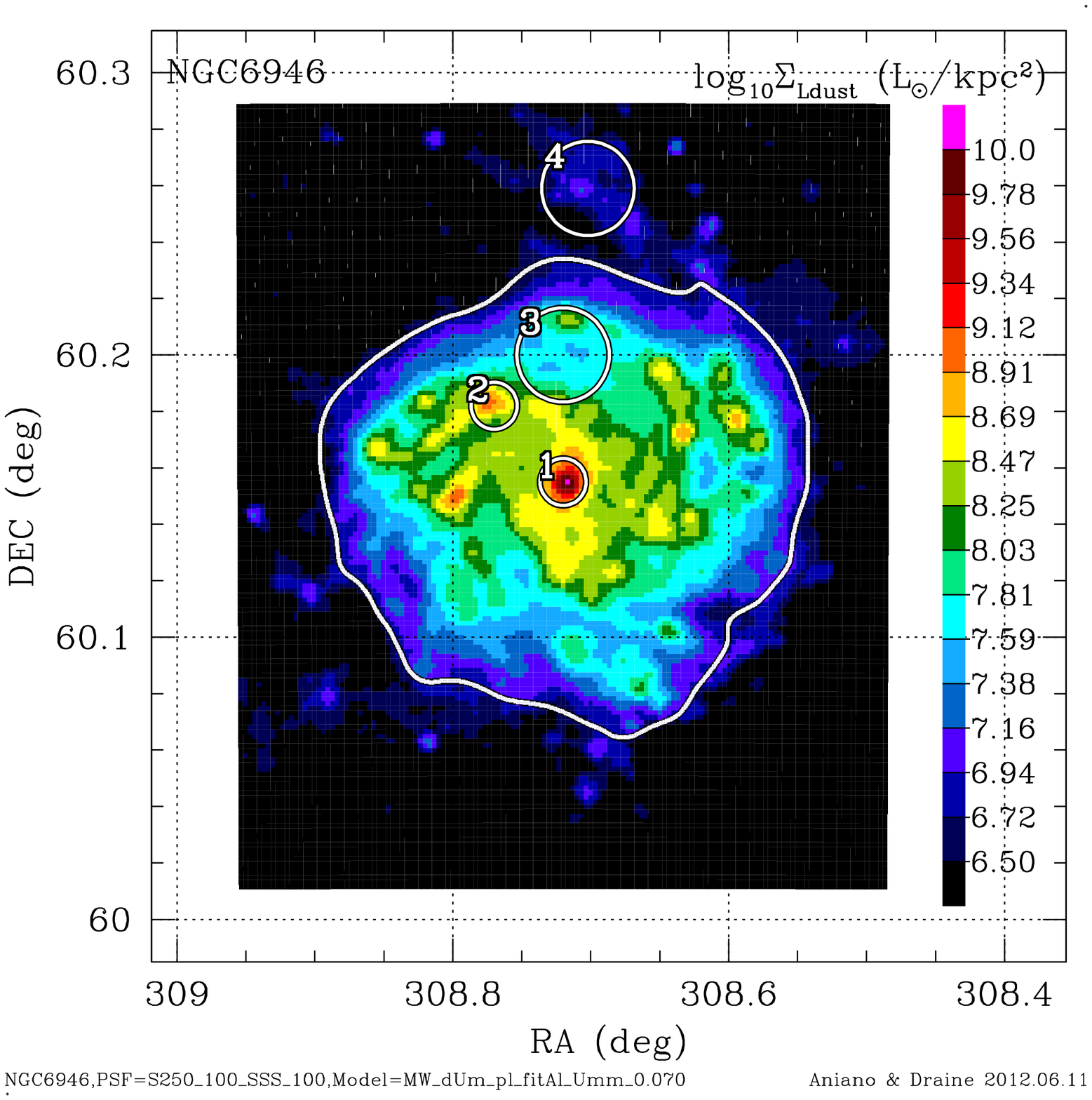}
\renewcommand \RthreeCthree {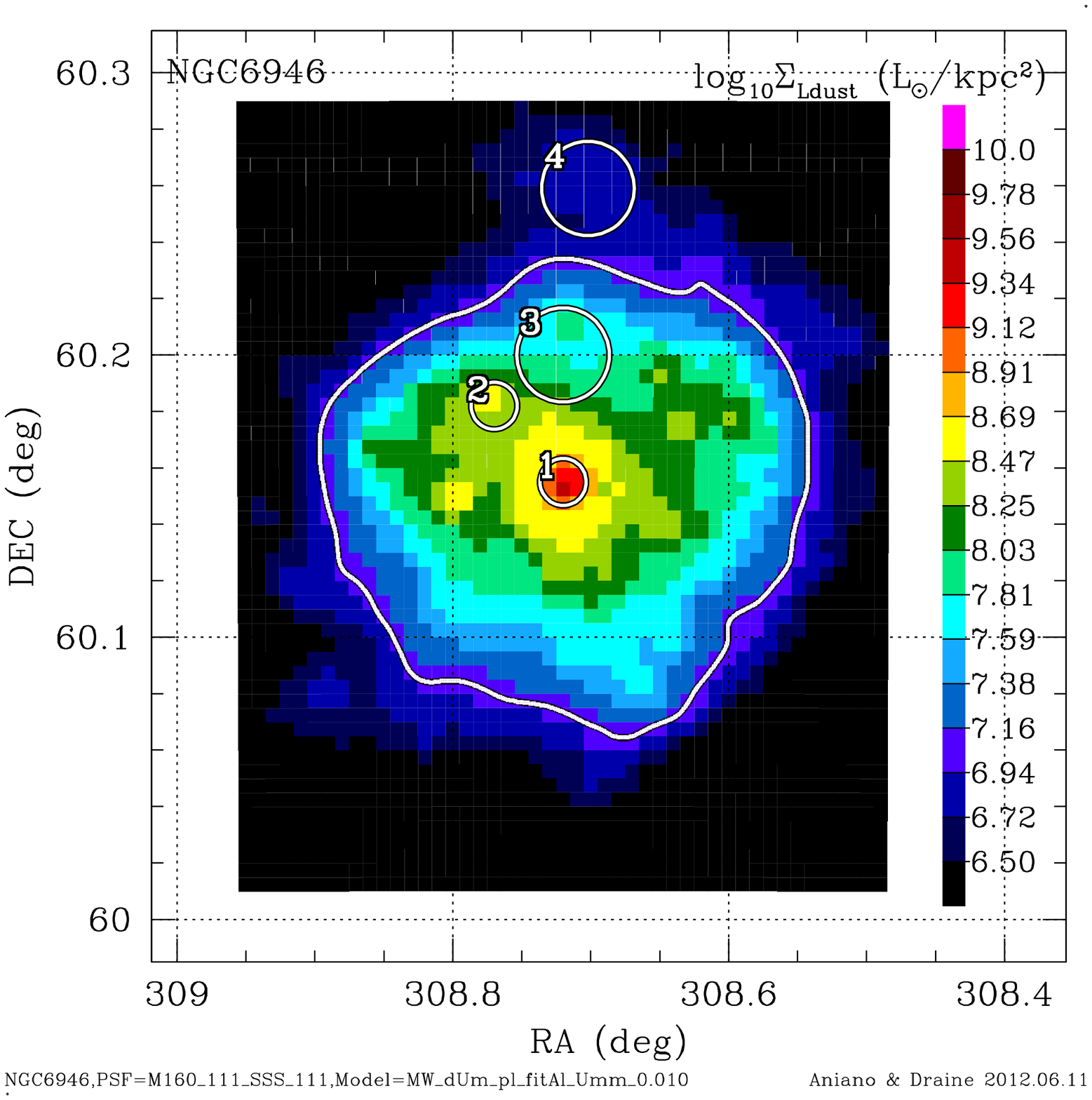}
\renewcommand \RfourCone {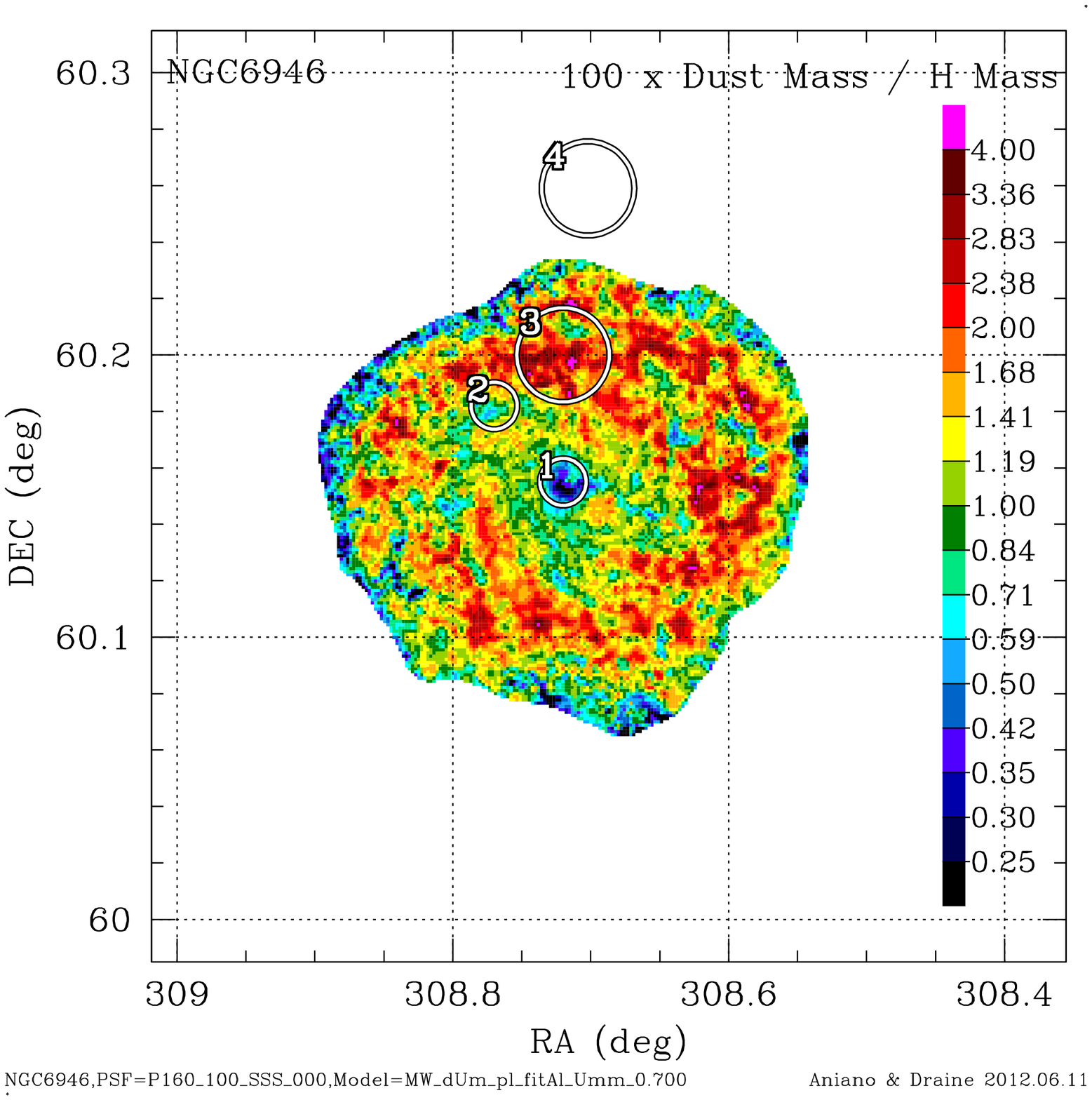}
\renewcommand \RfourCtwo {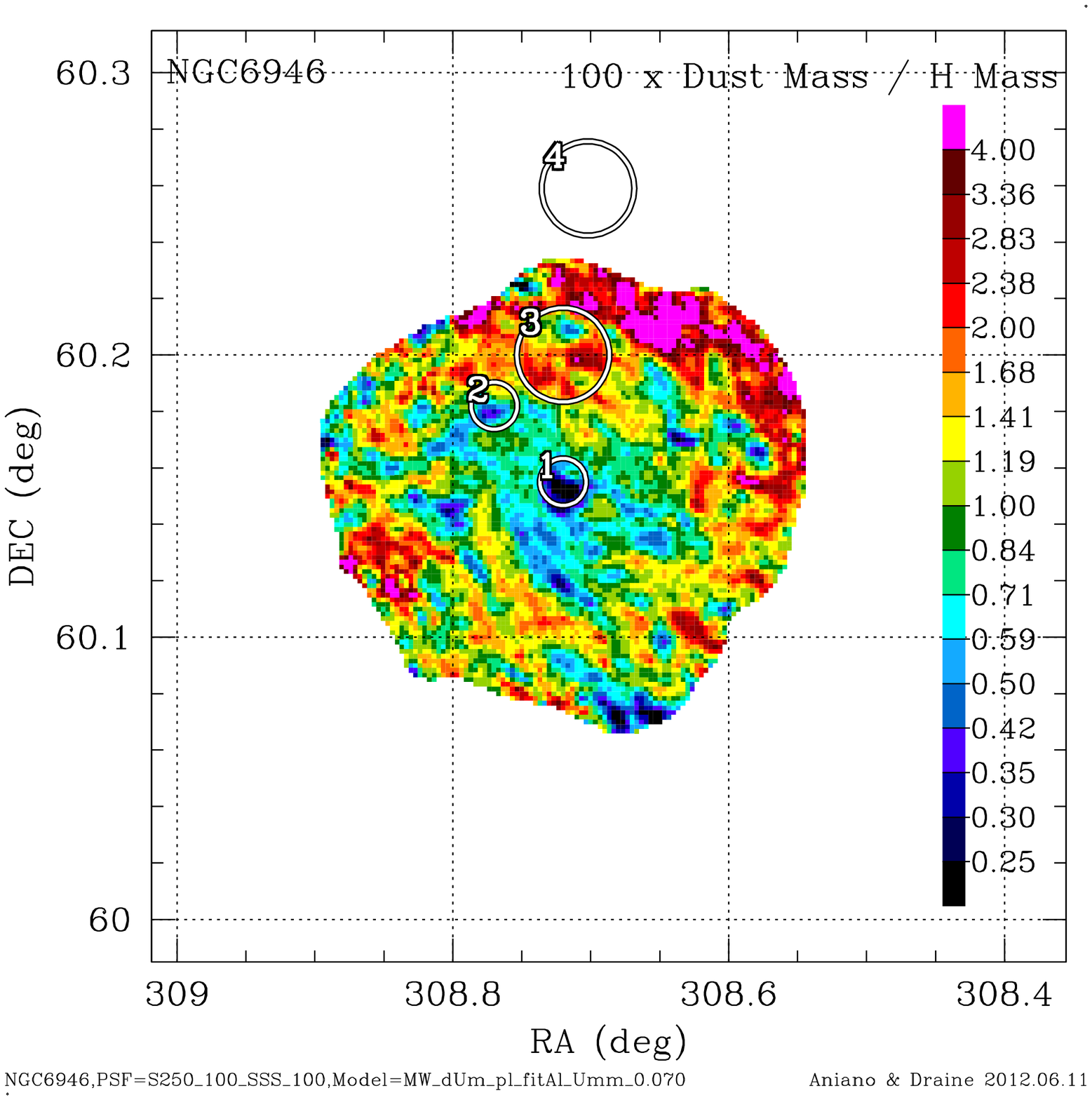}
\renewcommand \RfourCthree {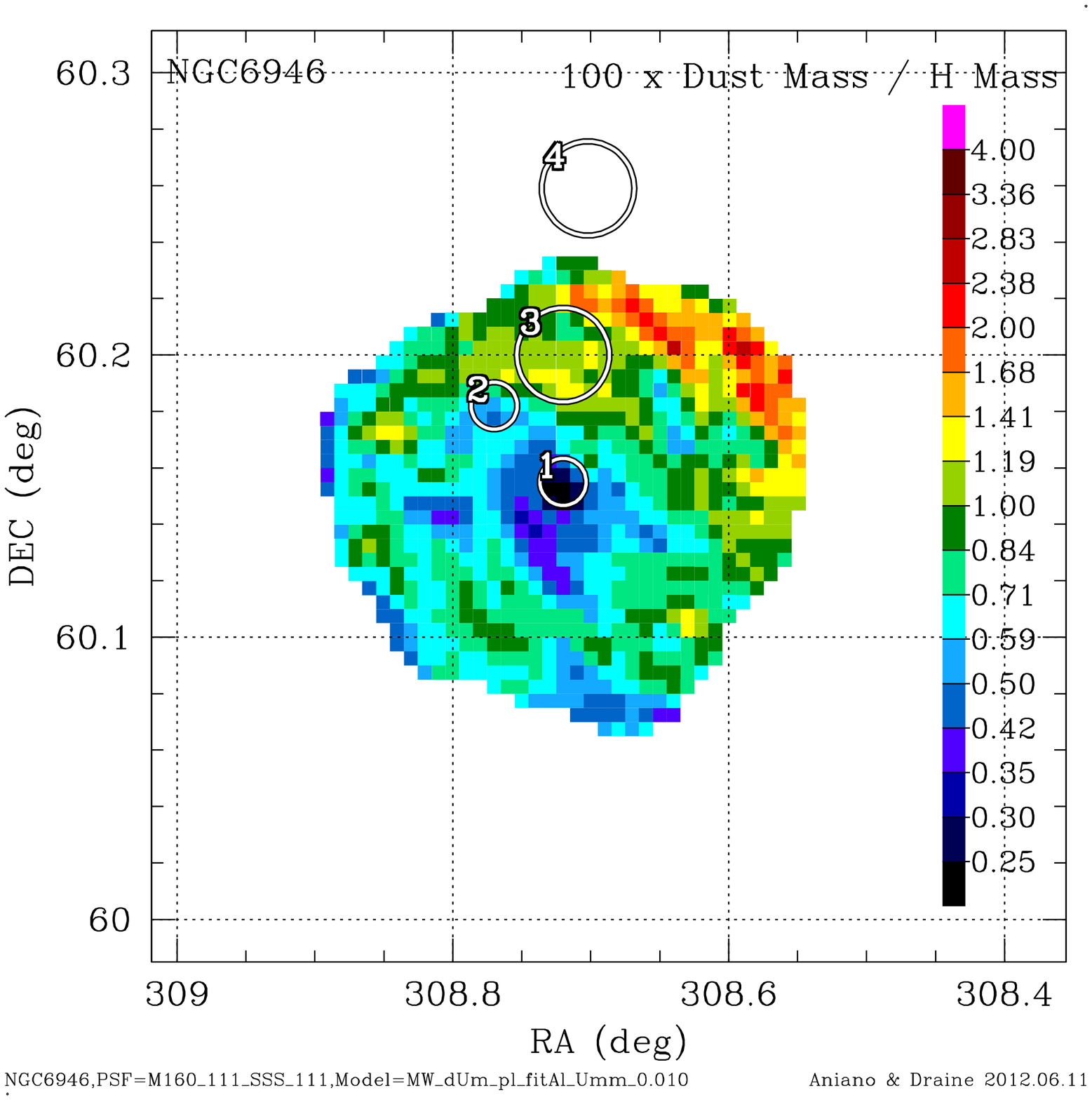}
\ifthenelse{\boolean{make_heavy}}{ }
{ \renewcommand \RoneCone    {No_image.eps}
\renewcommand \RtwoCone    {No_image.eps}
\renewcommand \RthreeCone {No_image.eps}
\renewcommand \RfourCone {No_image.eps}
\renewcommand \RoneCtwo    {No_image.eps}
\renewcommand \RtwoCtwo    {No_image.eps}
\renewcommand \RthreeCtwo {No_image.eps}
\renewcommand \RfourCtwo {No_image.eps}}


\subsubsection{\label{sec:n6946_dustmaps}
               Dust and H Mass Maps}

Figure \ref{fig:ngc6946-1} shows maps of the gas in NGC\,6946 obtained
from the THINGS 21-cm map \citep{Walter+Brinks+deBlok+etal_2008} and
the HERACLES CO\,2--1 map \citep{Leroy+Walter+Bigiel+etal_2009},
convolved to the resolution of PACS160, SPIRE250, and MIPS160.
The second row shows dust maps obtained by the present study.  
The dust map obtained using the PACS160 PSF has dust clearly visible only out to a radius of
$\sim$$250\arcsec$ ($8\kpc$ @ $6.8\Mpc$), but the $11\arcsec$
(500$\pc$) FWHM of the beam resolves the spiral structure.  The
arm/interarm contrast in dust surface density appears to be
approximately a factor of $\sim$2 in this image.  Introducing SPIRE
250$\micron$ data requires the PSF to be broadened, making the spiral
arms less apparent, but allows dust to be detected out to larger
radii, notably in the northern spiral arm.  The MIPS160 resolution
image clearly shows dust in the northern spiral arm out to a
distance of $\sim$$15\kpc$ from the nucleus.

The dust luminosity/area $\Sigma_{L_\dust}$
(Figure \ref{fig:ngc6946-1}, bottom row) peaks at
$\sim$$10^{10.1}\Lsol\kpc^{-2}$ in the nucleus 
(see the PACS160 resolution map),
and 
can be followed down to 
$\Sigma_{L_\dust}\approx10^{7.0}\Lsol\kpc^{-2}$ in the MIPS160
resolution map.
Similarly, the IR luminosity/area contributed by PDRs peaks at
$\Sigma_{L_\PDR}\approx10^9\Lsol\kpc^{-2}$ near the center, and
is reliably measured down to $\sim10^{5.6}\Lsol\kpc^{-2}$.

The dust/H mass ratio ratio shown in Figure \ref{fig:ngc6946-1}j-l,
calculated assuming $\XCOxx=4$, shows a pronounced minimum of $\sim
0.005$ in the central $\kpc$.  In this region the gas is primarily
molecular, and the estimated dust/H mass ratio is therefore sensitive
to the value assumed for $\XCO$.  NGC\,6946 has approximately solar
metallicity, and we expect the dust/H mass ratio to be $\sim$0.01.
The surprisingly low dust/H mass ratios found in the center
of NGC\,6946 indicate that the value of $\XCO$ near the center
should be about a factor of $\sim$ 2 smaller than the value $\XCOxx=4$
adopted for the bulk of the galaxy.  This conclusion is consistent
with interferometric studies of GMCs which indicate $\XCOxx \approx
1.25$ near the center of NGC\,6946
\citep{Donovan_Meyer+Koda+Momose+etal_2011}, using virial mass
estimates for individual GMCs.

\renewcommand \RoneCone {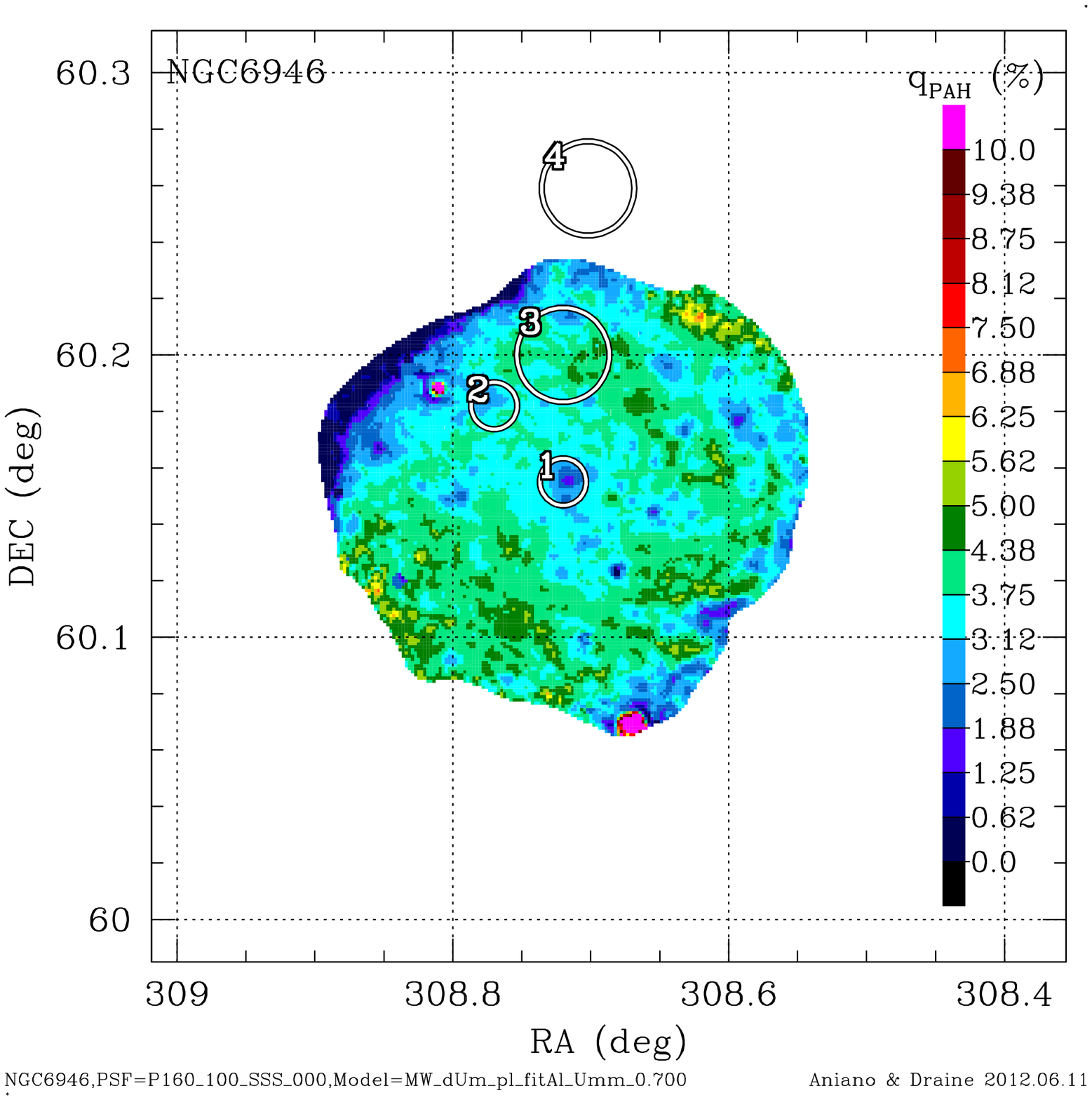}
\renewcommand \RoneCtwo {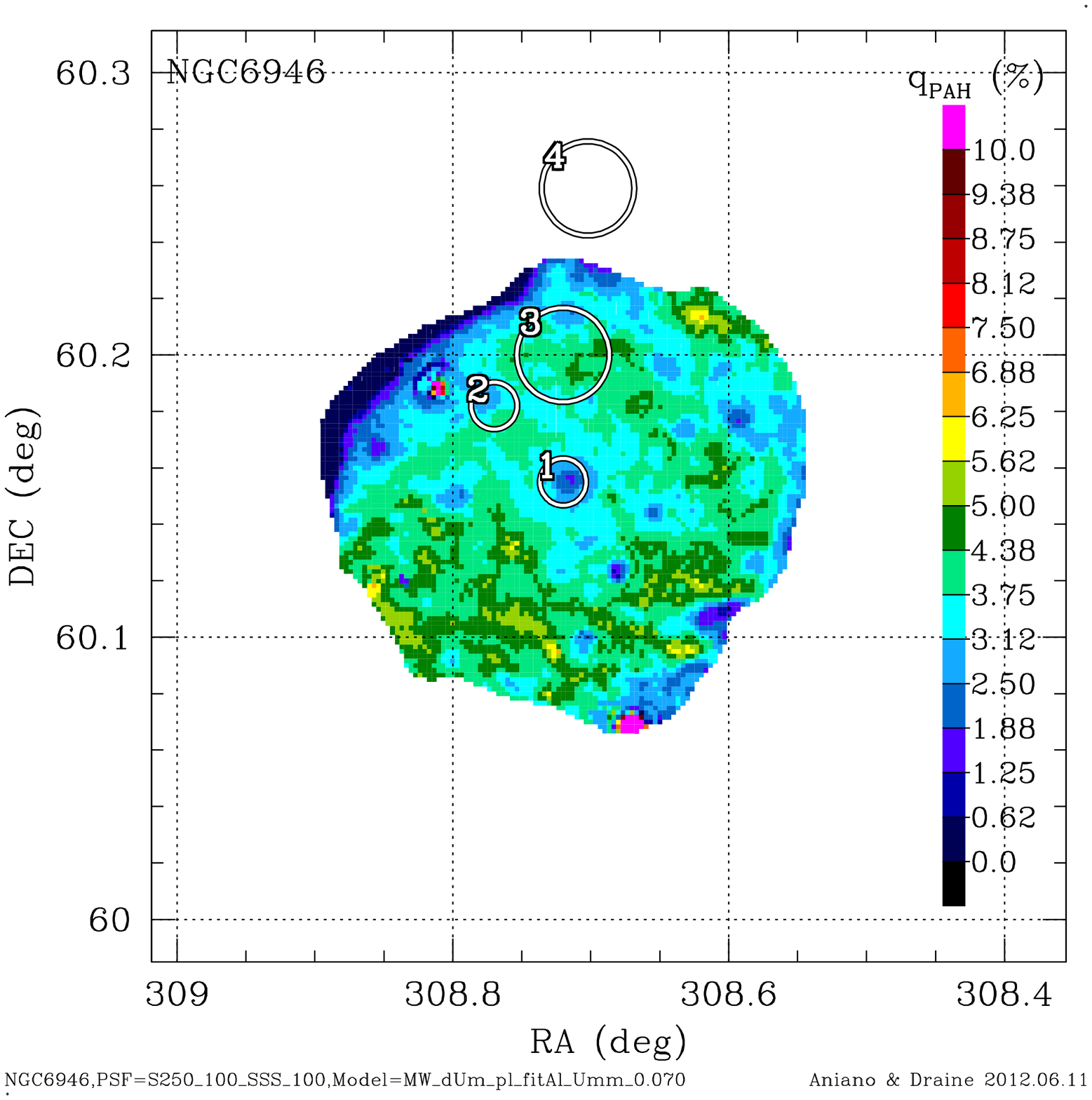}
\renewcommand \RoneCthree {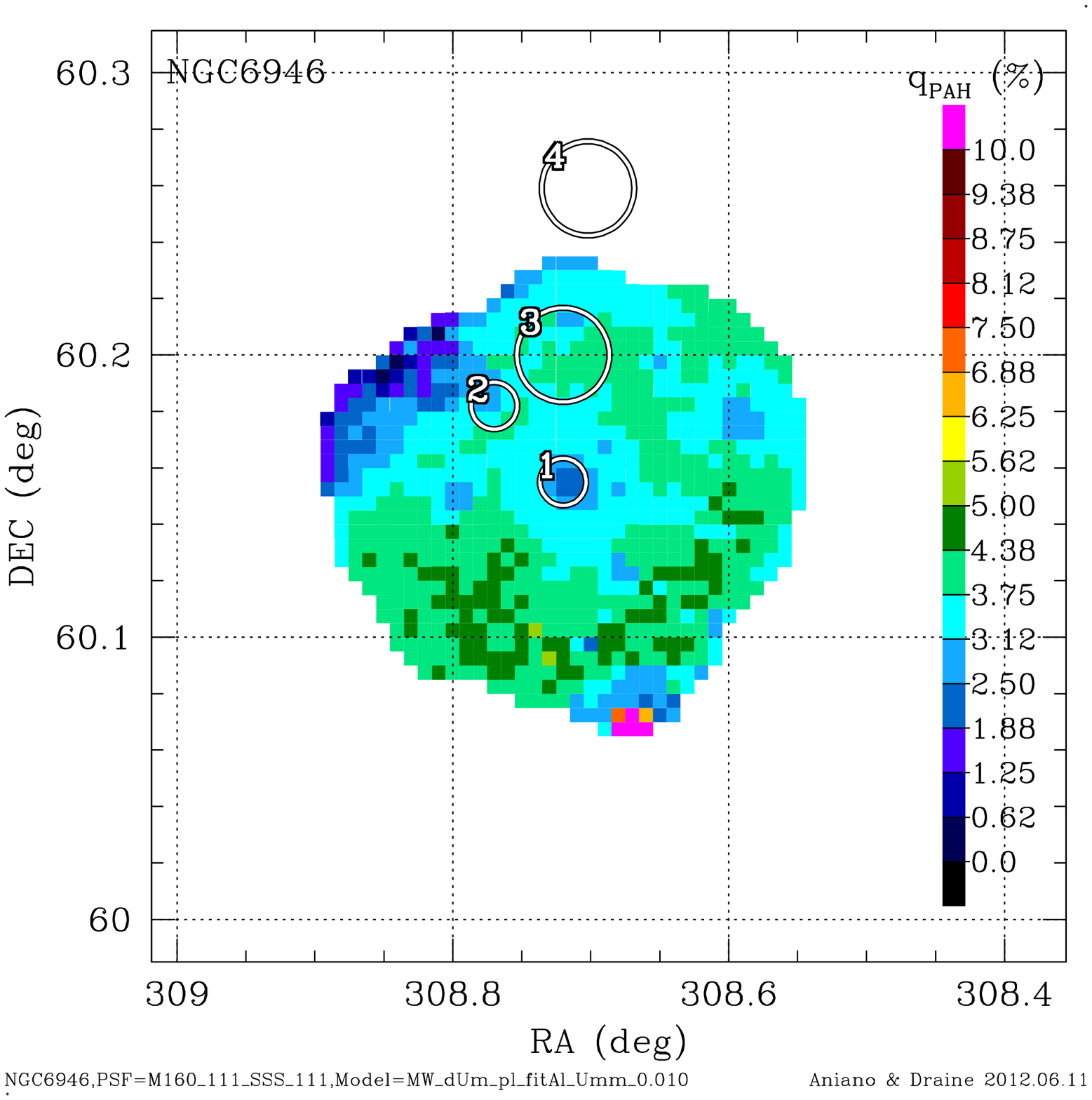}
\renewcommand \RtwoCone {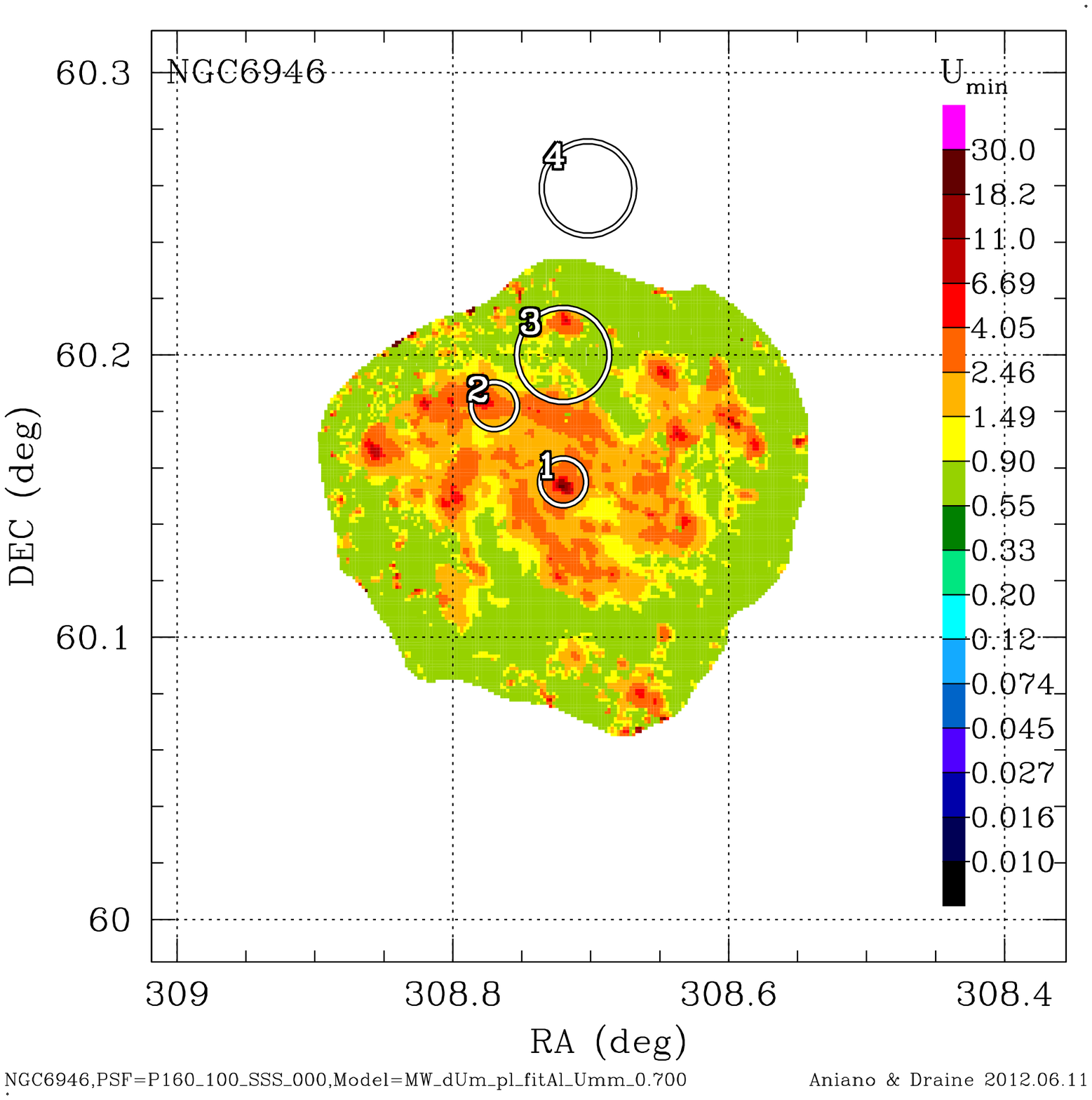}
\renewcommand \RtwoCtwo {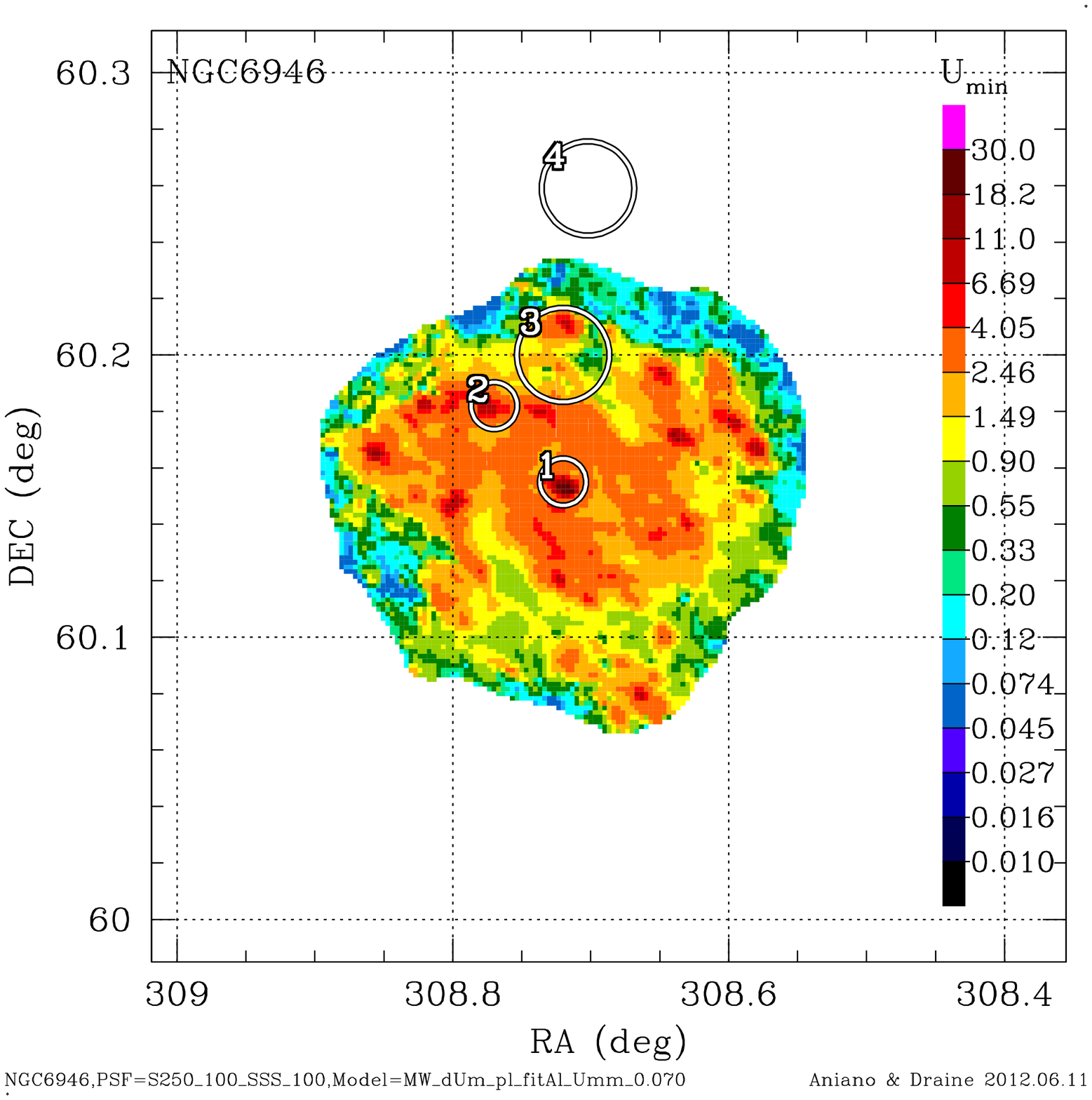}
\renewcommand \RtwoCthree {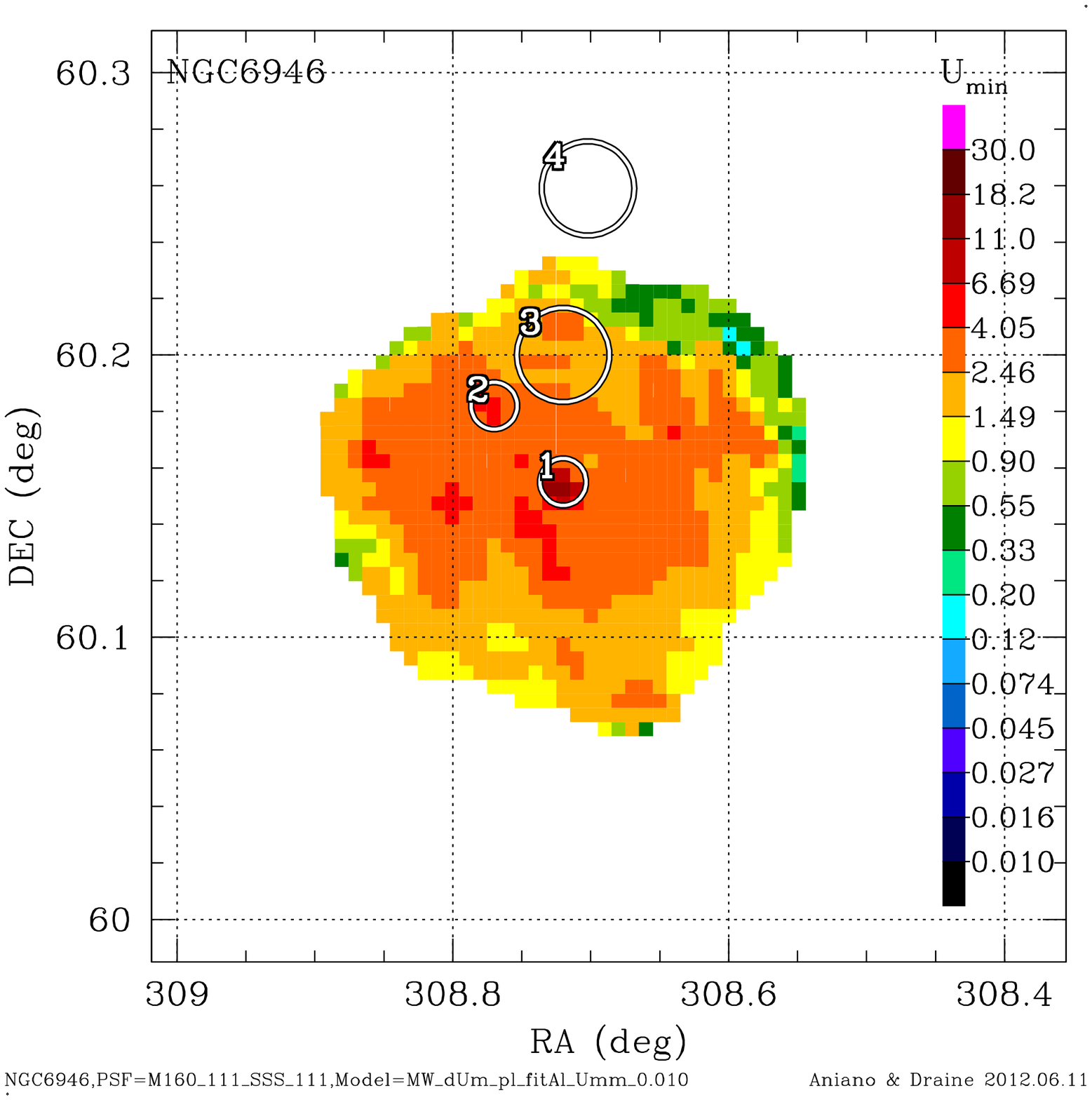}
\renewcommand \RthreeCone {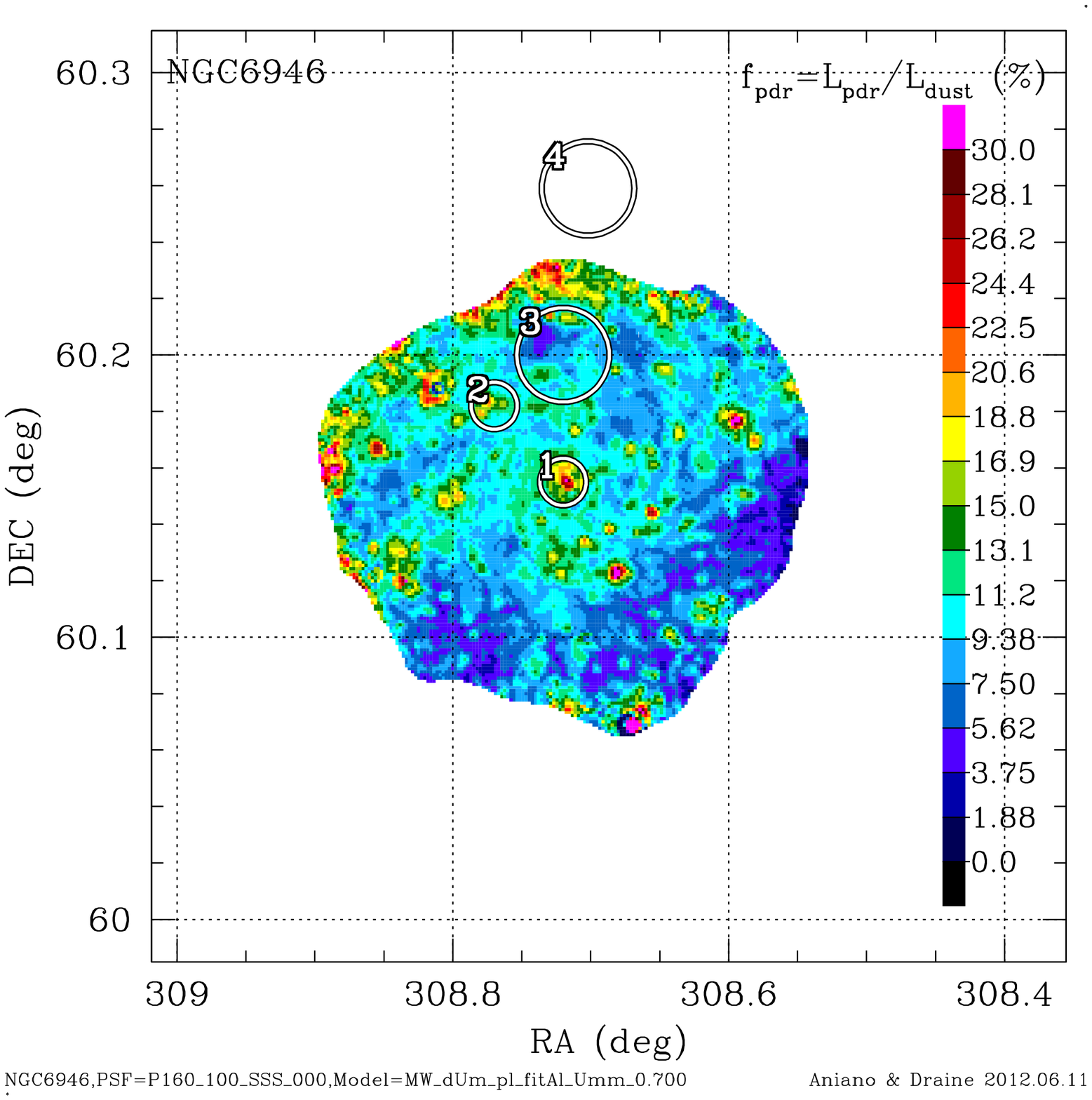}
\renewcommand \RthreeCtwo {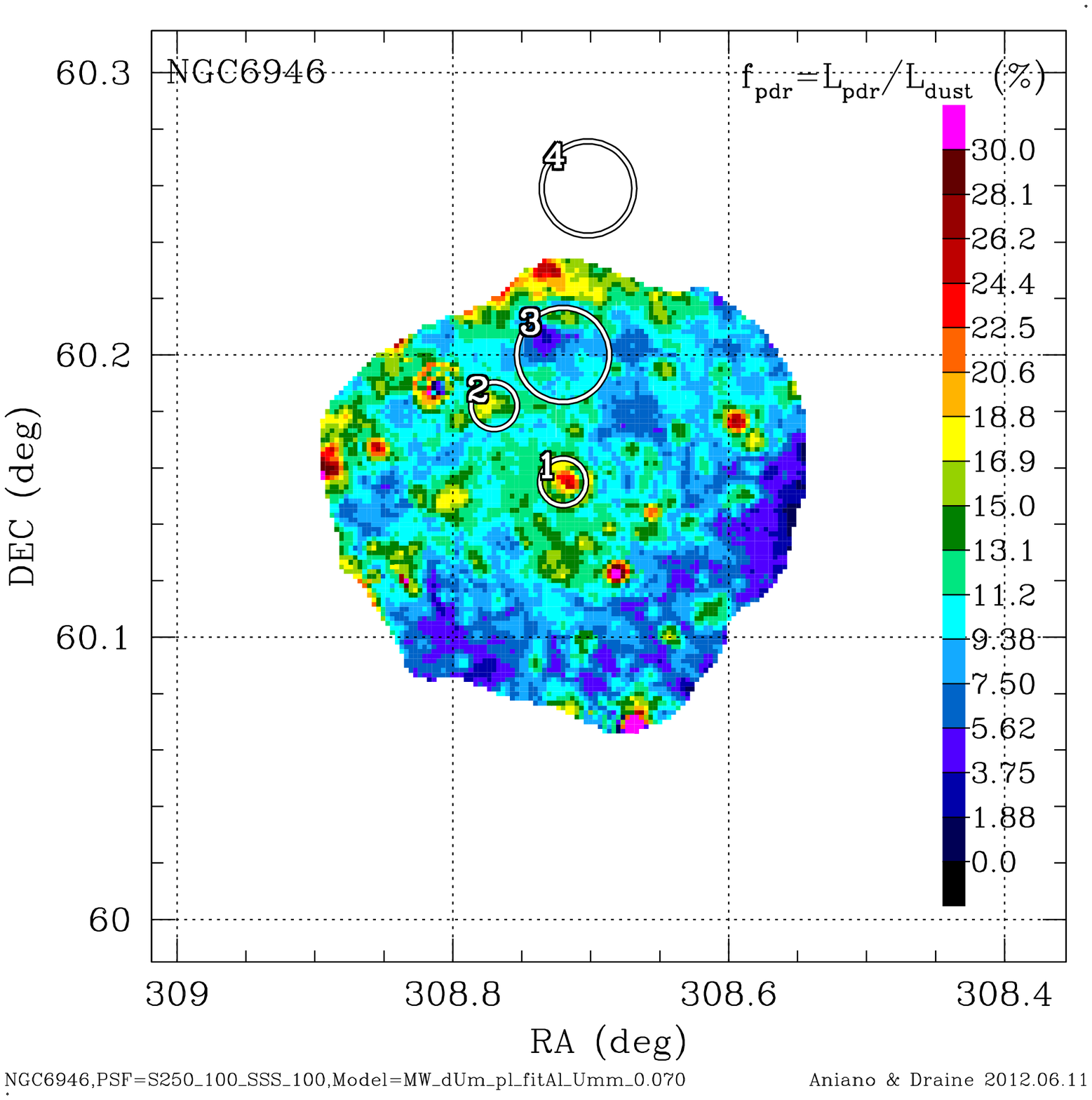}
\renewcommand \RthreeCthree {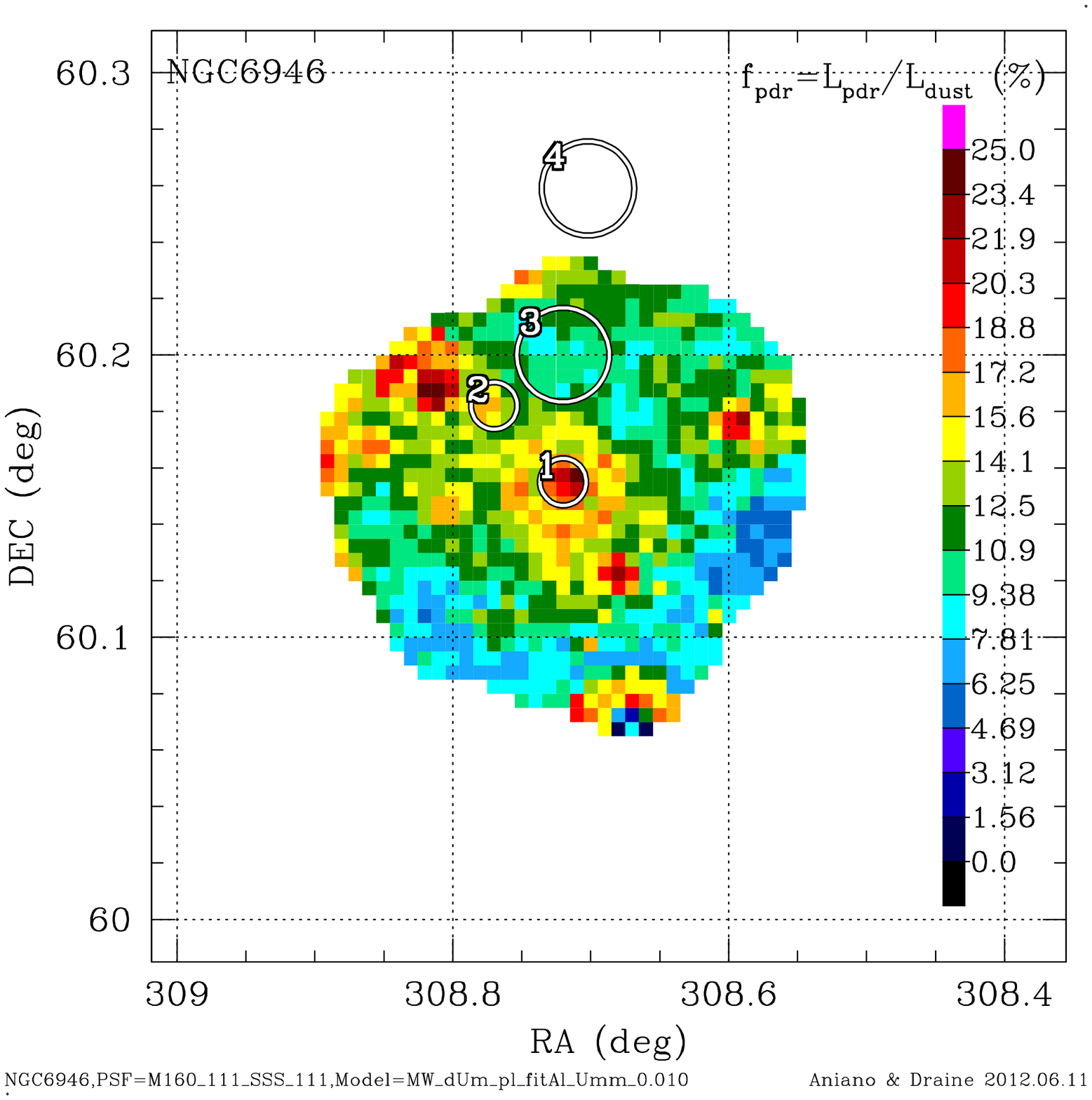}
\ifthenelse{\boolean{make_heavy}}{ }
{ \renewcommand \RoneCone    {No_image.eps}
\renewcommand \RtwoCone    {No_image.eps}
\renewcommand \RthreeCone {No_image.eps}
\renewcommand \RoneCtwo    {No_image.eps}
\renewcommand \RtwoCtwo    {No_image.eps}
\renewcommand \RthreeCtwo {No_image.eps}}
\begin{figure}[h] 
\centering 
\begin{tabular}{c@{$\,$}c@{$\,$}c} 
\footnotesize PACS160 PSF & \footnotesize SPIRE250 PSF & \footnotesize MIPS160 PSF \\
\FirstNormal
\SecondNormal
\ThirdLast
\end{tabular}
\vspace*{-0.5cm}
\caption{\footnotesize\label{fig:ngc6946-2} NGC~6946 dust models at
  the resolution of PACS160 (left), SPIRE250 (center column), and
  MIPS160 (right).  
  Top row: PAH abundance parameter $\qpah$.  
  Middle row: diffuse starlight intensity parameter $\Umin$.  
  Bottom row: PDR  fraction $f_\PDR$.}
\end{figure}


\subsubsection{Maps of $\qpah$ and Starlight Parameters}

The PAH abundance $\qpah$ in NGC\,6946 rises from a minimum of $\ltsim
1\%$ near the nucleus, reaching levels of $\sim$$(4\pm1)\%$ at
distances $\sim$1--8$\kpc$ from the center, ultimately appearing to
decline at the edge of the detectable region.  The central minimum in
$\qpah$ is real (it is independent of uncertainties in the appropriate
value of $\XCO$).  The local minimum of $\qpah$ at the center may be
the result of destruction of PAHs by processes associated with star
formation (there is no evidence of AGN activity in NGC~6946).  We note
that there are a number of other local minima of $\qpah$ evident in
Figure \ref{fig:ngc6946-2}a-c; just as for the central minimum, these
extranuclear minima tend to coincide with peaks in dust luminosity
surface brightness $\Sigma_{L_\dust}$ (see Figure
\ref{fig:ngc6946-1}g,h) and peaks in $f_\PDR$ (see Figure
\ref{fig:ngc6946-2}j,k,l).  These are both indicators of luminous
star-forming regions, which can be expected to coincide with
\ion{H}{2} regions around hot stars, which appear to destroy PAHs
\citep[e.g.,][]{Povich+Stone+Churchwell+etal_2007}.  Supernova
blastwaves are presumed to also destroy PAHs.

Because the outer falloff in $\qpah$ is occurring as the signal/noise
ratio is falling to low values, it is not certain whether the observed
decline is real or an artifact of different levels of background
subtraction at different wavelengths.  However, the decline in $\qpah$
in the outer regions persists in the MIPS160 resolution map.  These
variations in $\qpah$ will be examined in more detail in future
studies.

The diffuse starlight intensity parameter $\Umin$ 
(Fig.\ \ref{fig:ngc6946-2}, middle row) shows a general
decline with increasing distance from the center.
The PACS160 resolution map of $\Umin$ is somewhat noisy,
particularly in the outer regions, but the MIPS160 resolution map of
$\Umin$ shows a systematic decline from values as high as $\Umin\approx 10$
in the central $\sim500\pc$ down to $\Umin\approx 0.15$ in the outer regions,
$\sim 10\kpc$ from the center.  Even lower values of $\Umin$ are estimated
for some pixels, but this is seen only at the lowest signal/noise levels.

The bottom row of Figure \ref{fig:ngc6946-2} shows $f_\PDR$.  The
behavior of $f_\PDR$ in NGC~6946 is similar to NGC~628.  Bright
complexes coincide with local maxima of $f_\PDR$.  Outside very bright
complexes $f_\PDR$ is quite smooth across the galaxy.

\subsubsection{Comparison Between Observed and Modeled Flux Densities}

\renewcommand \RoneCone {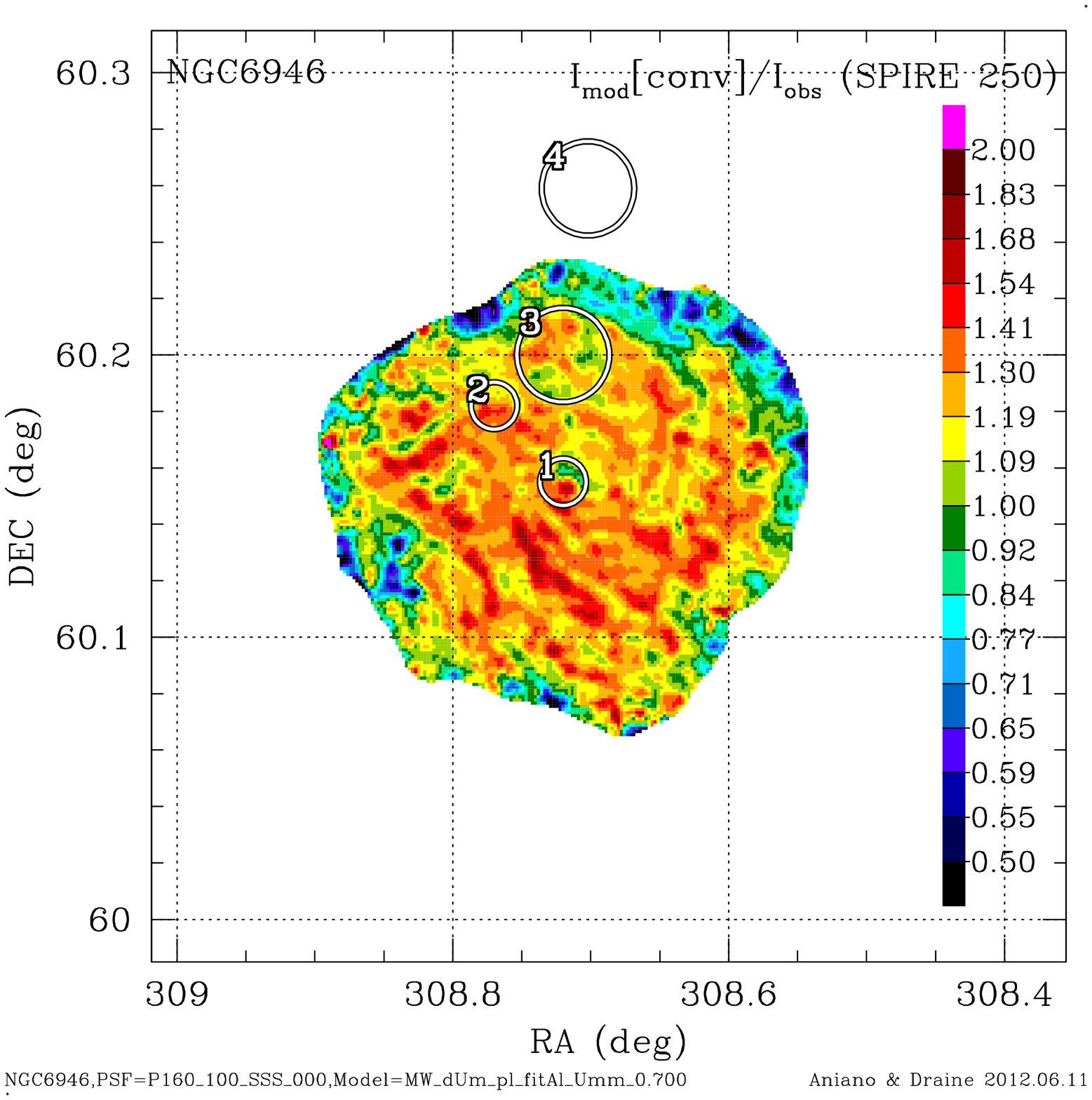}
\renewcommand \RoneCtwo {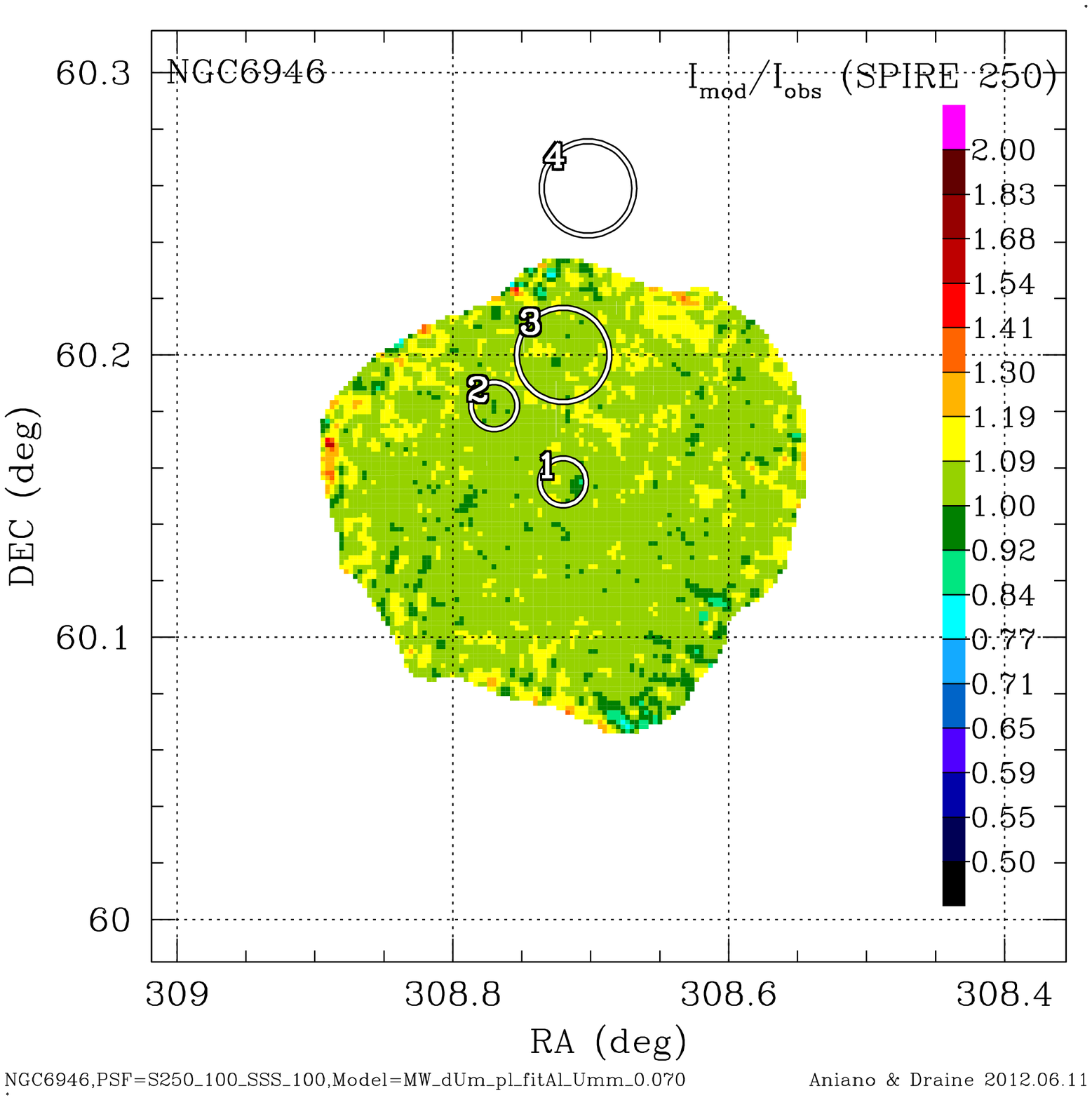}
\renewcommand \RoneCthree {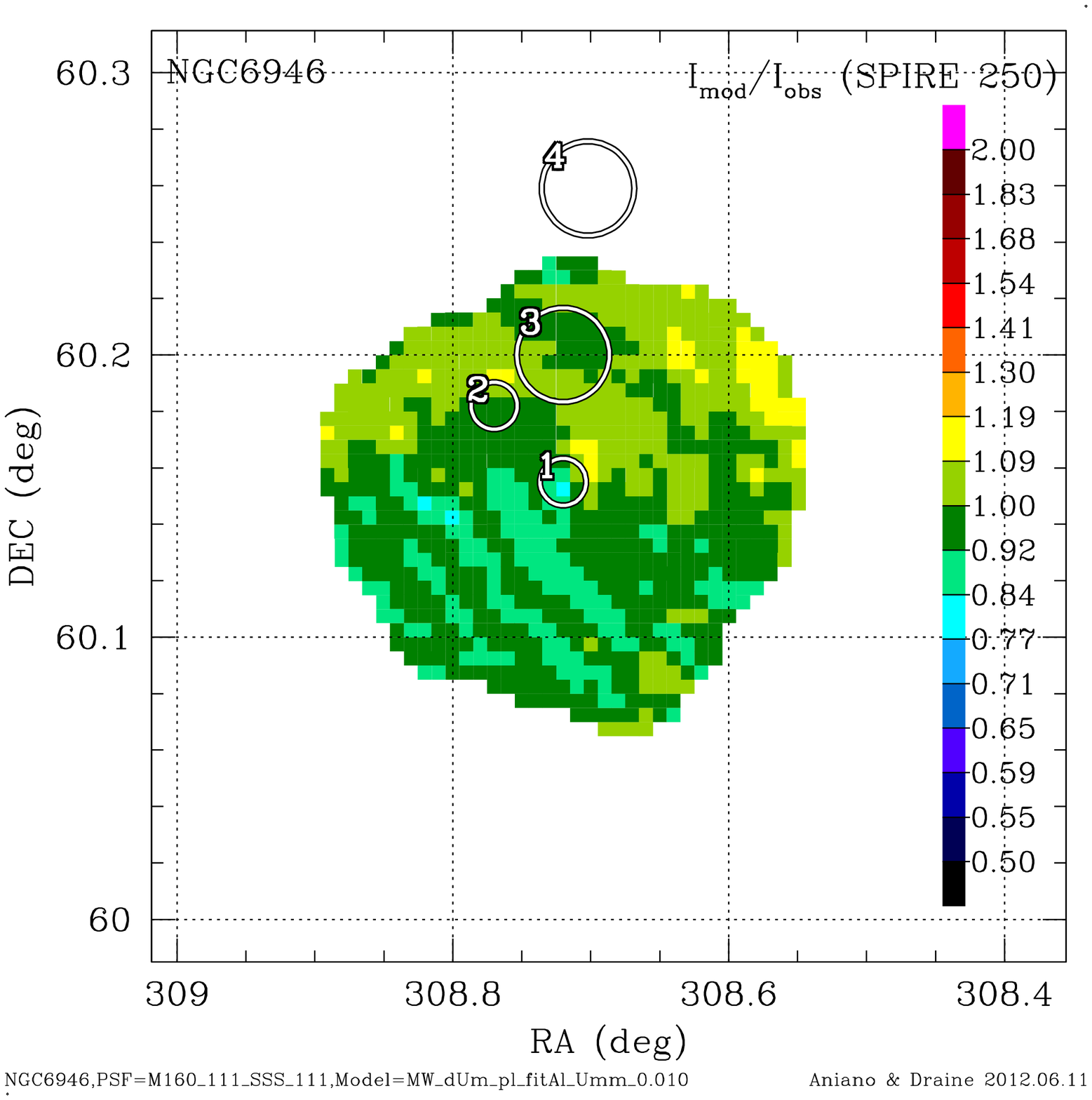}
\renewcommand \RtwoCone {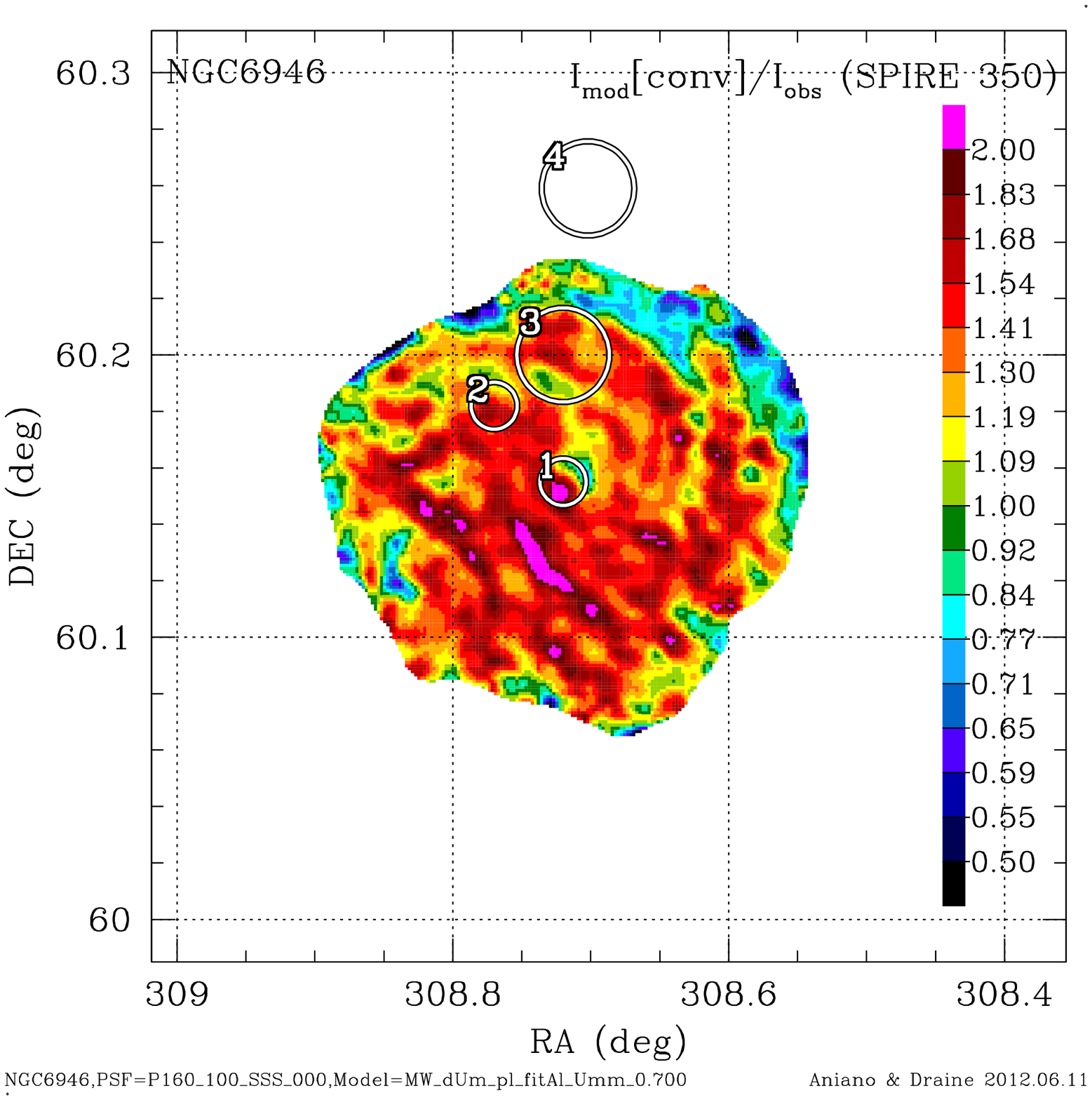}
\renewcommand \RtwoCtwo {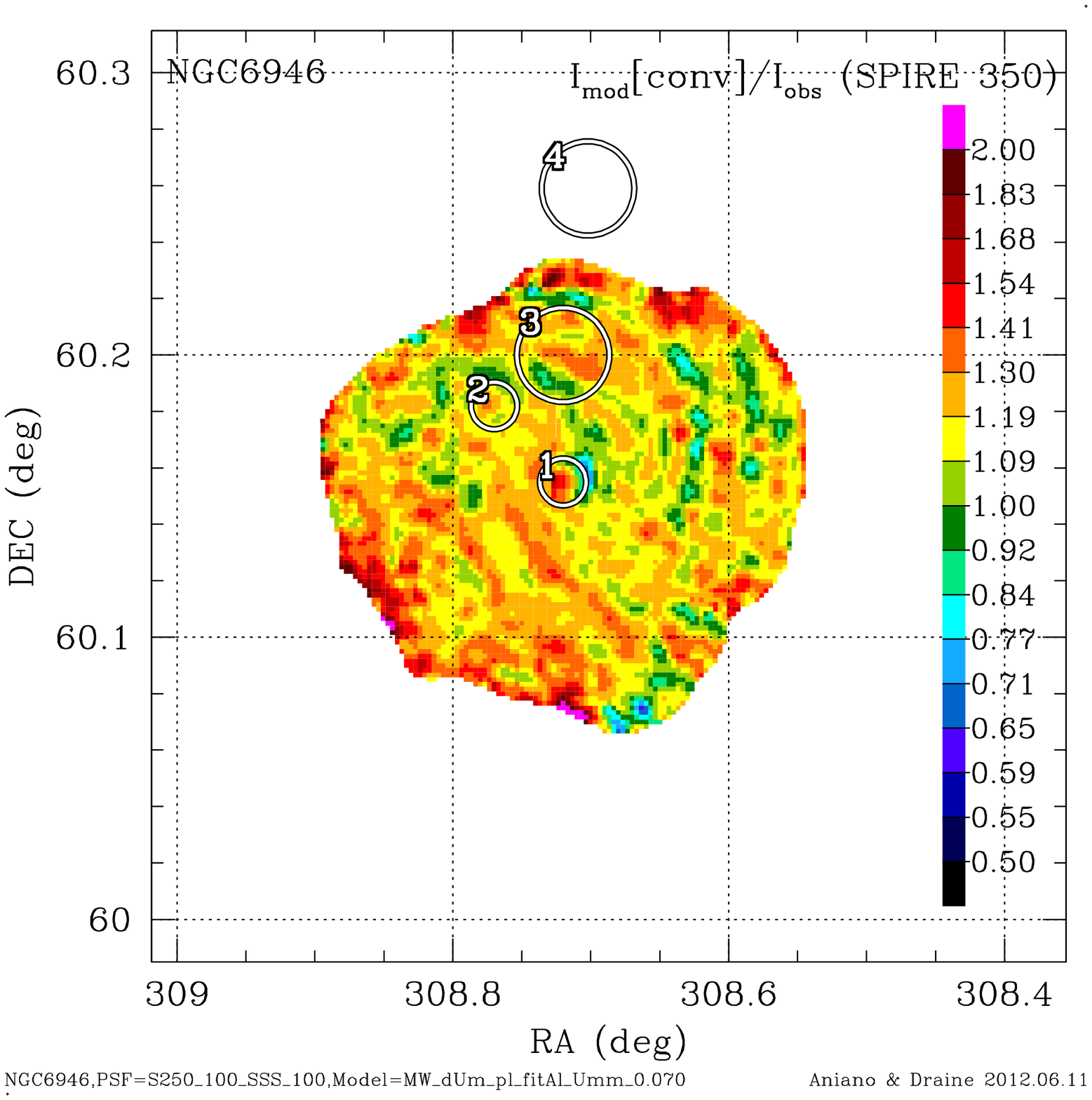}
\renewcommand \RtwoCthree {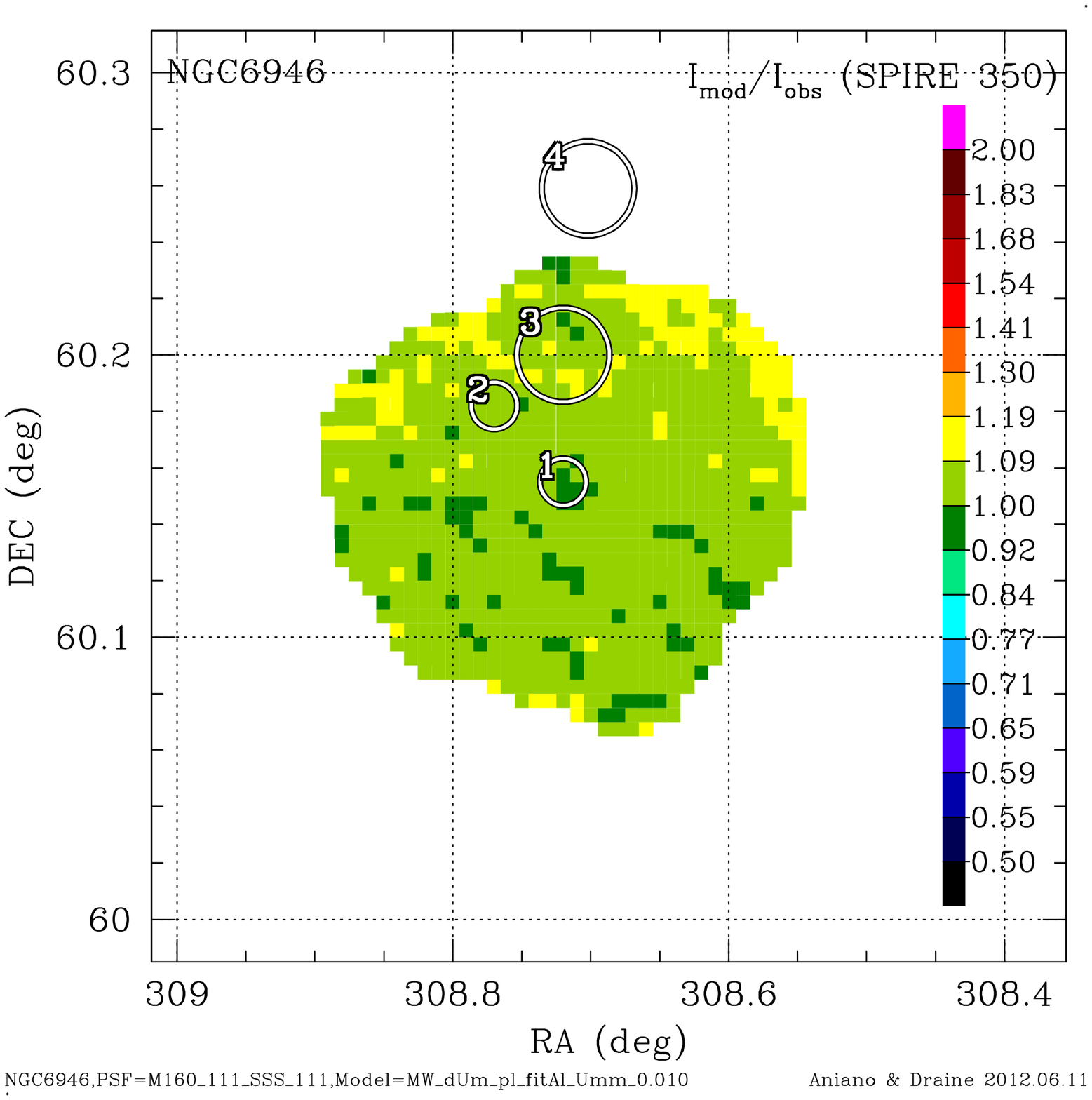}
\renewcommand \RthreeCone {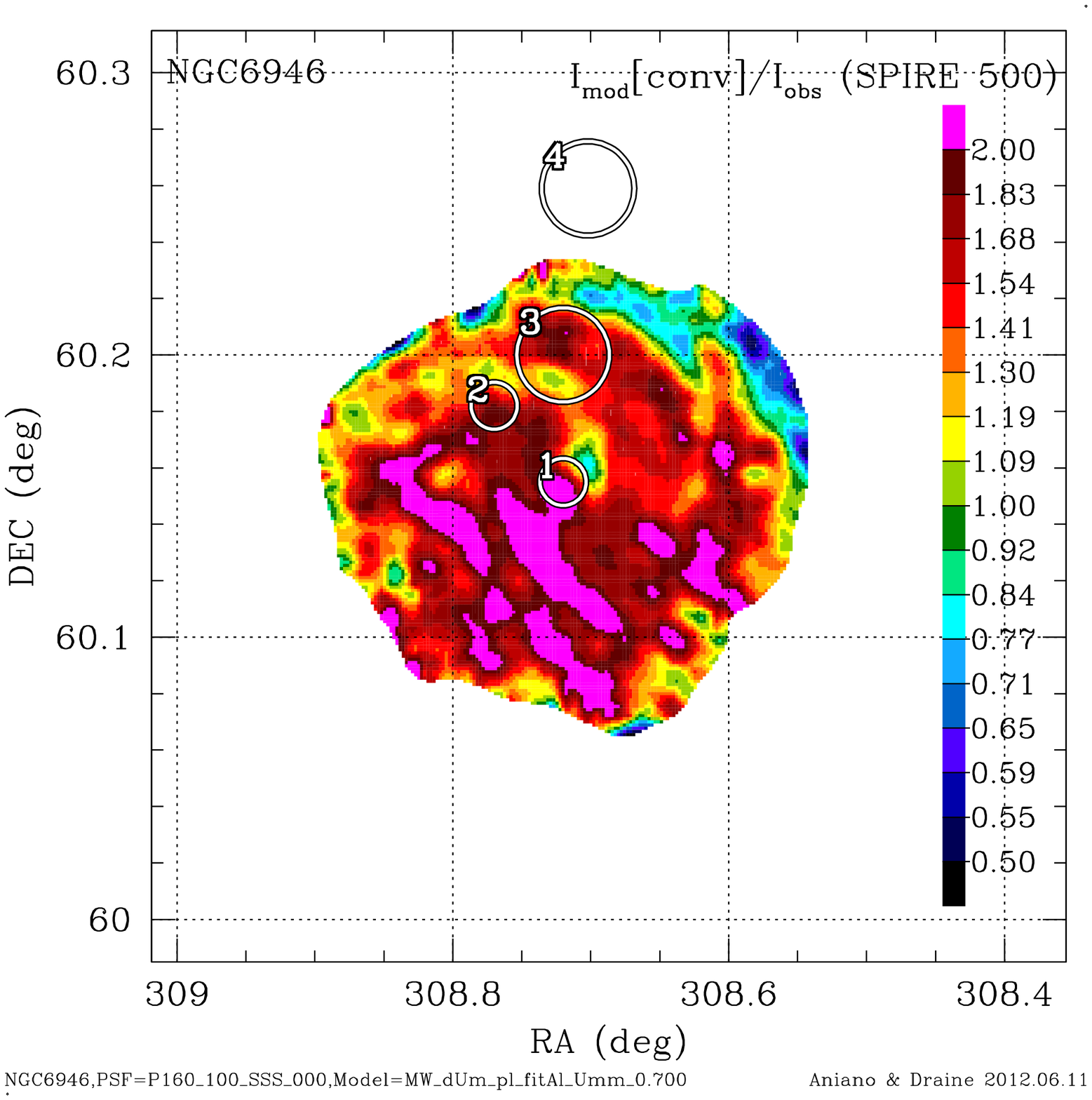}
\renewcommand \RthreeCtwo {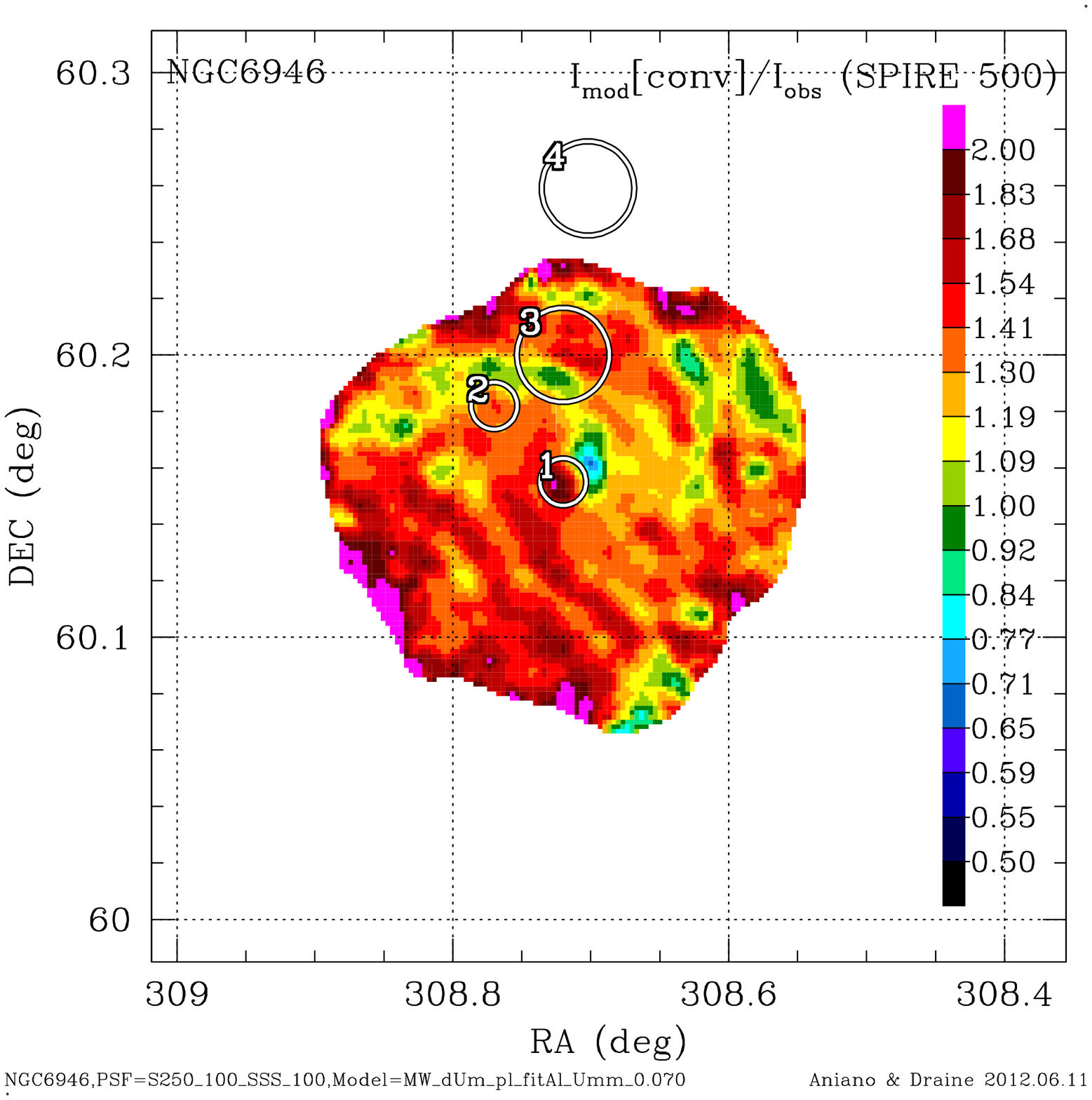}
\renewcommand \RthreeCthree {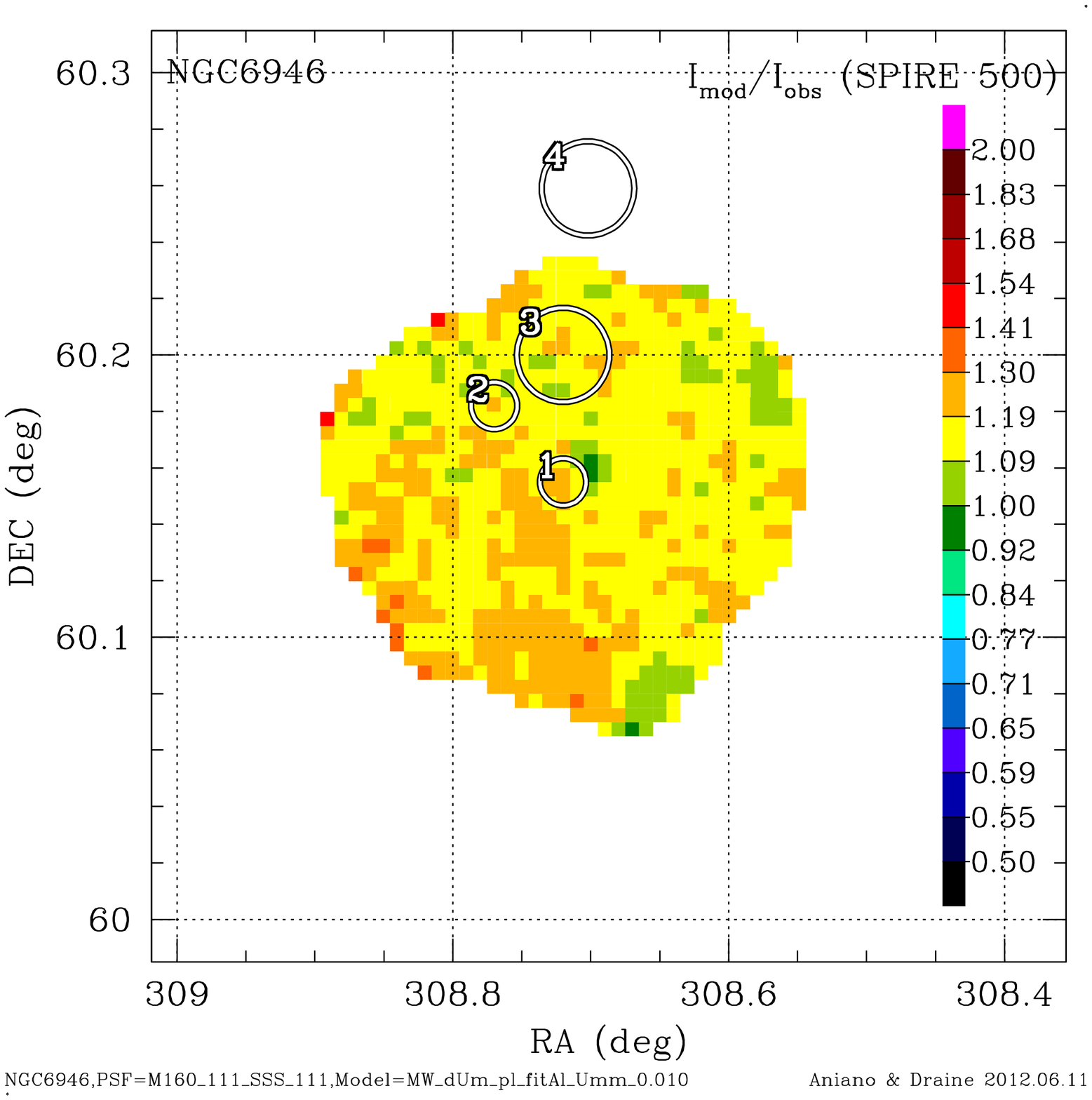}
\ifthenelse{\boolean{make_heavy}}{ }
{ \renewcommand \RoneCone    {No_image.eps}
\renewcommand \RtwoCone    {No_image.eps}
\renewcommand \RthreeCone {No_image.eps}
\renewcommand \RfourCone {No_image.eps}
\renewcommand \RoneCtwo    {No_image.eps}
\renewcommand \RtwoCtwo    {No_image.eps}
\renewcommand \RthreeCtwo {No_image.eps}
\renewcommand \RfourCtwo {No_image.eps}}
\begin{figure}[h] 
\centering
\begin{tabular}{c@{$\,$}c@{$\,$}c} 
\footnotesize PACS160 PSF & \footnotesize SPIRE250 PSF & \footnotesize MIPS160 PSF \\
\FirstNormal
\SecondNormal
\ThirdLast
\end{tabular}
\vspace*{-0.5cm}
\caption{\footnotesize\label{fig:ngc6946-3} Similar to
  Fig.\ \ref{fig:ngc0628-3}, but for NGC~6946.  Ratio of model
  intensity \!/\! observed intensity for NGC~6946 at $\lambda$=250\um
  (top row), $\lambda$=350\um (middle row), and $\lambda$=500\um
  (bottom row).  Left column: PACS160 PSF, model constrained only by
  IRAC, MIPS24 and PACS.  Center column: SPIRE250 PSF, model
  constrained by IRAC, MIPS24, PACS, and SPIRE250.  Right column:
  MIPS160 PSF, model constrained by all 13 cameras.  The model in the
  center column does a fairly good job in predicting the emission at
  $\lambda$=350\um, and $\lambda$=500\um, except near the edge where
  the S/N is low.  The model in the right column is in excellent
  agreement with all three SPIRE bands.
  }
\end{figure}

The model-predicted/observed flux ratios for NGC~6946 (shown in Figure
\ref{fig:ngc6946-3}) behave similarly to those of NGC~628 (see Figure
\ref{fig:ngc0628-3}).  When long wavelength data is not used in the
dust model fit, the modeling tends to overpredict the intensity at the
longer wavelengths.  However, when the SPIRE photometry is included
in the model constraints, the overprediction is greatly reduced. In
the case of MIPS160 PSF (all bands used), 
the SPIRE250 photometry is reproduced by the
model within 10\%, and SPIRE500 tends to be overpredicted by less
than 15\%.

\subsubsection{Global SED Fitting}

\renewcommand \RoneCone {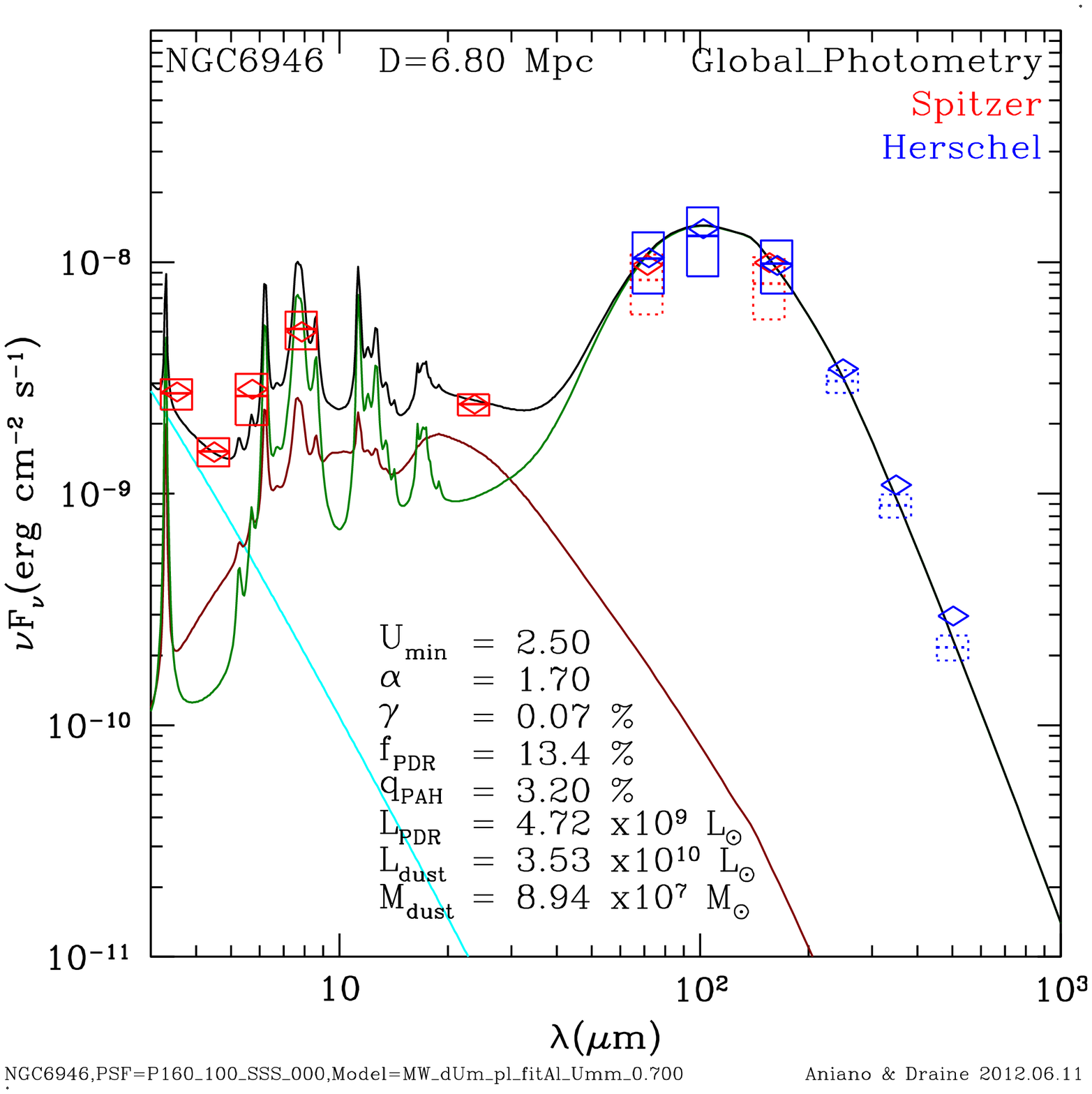}
\renewcommand \RoneCtwo {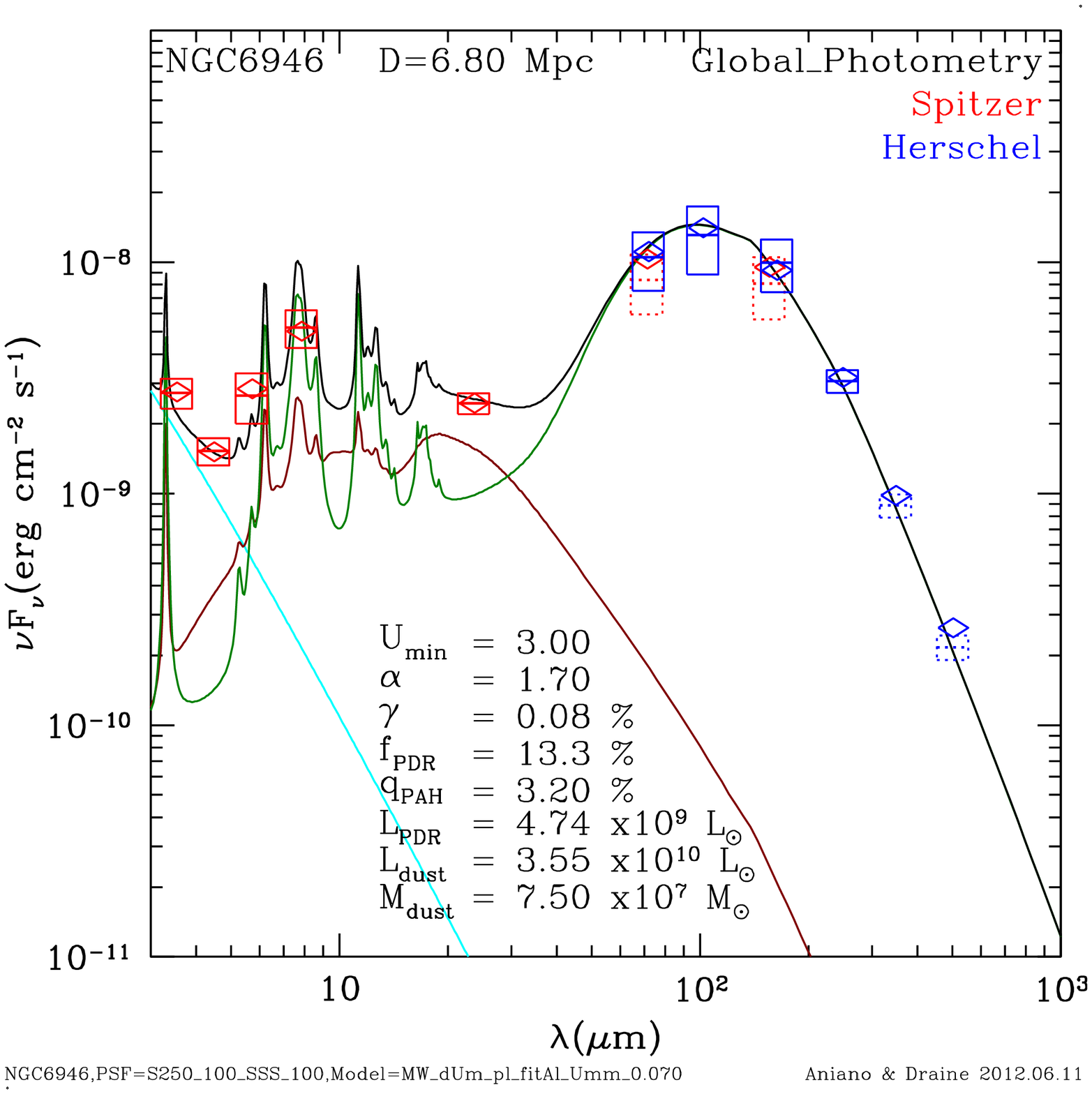}
\renewcommand \RoneCthree {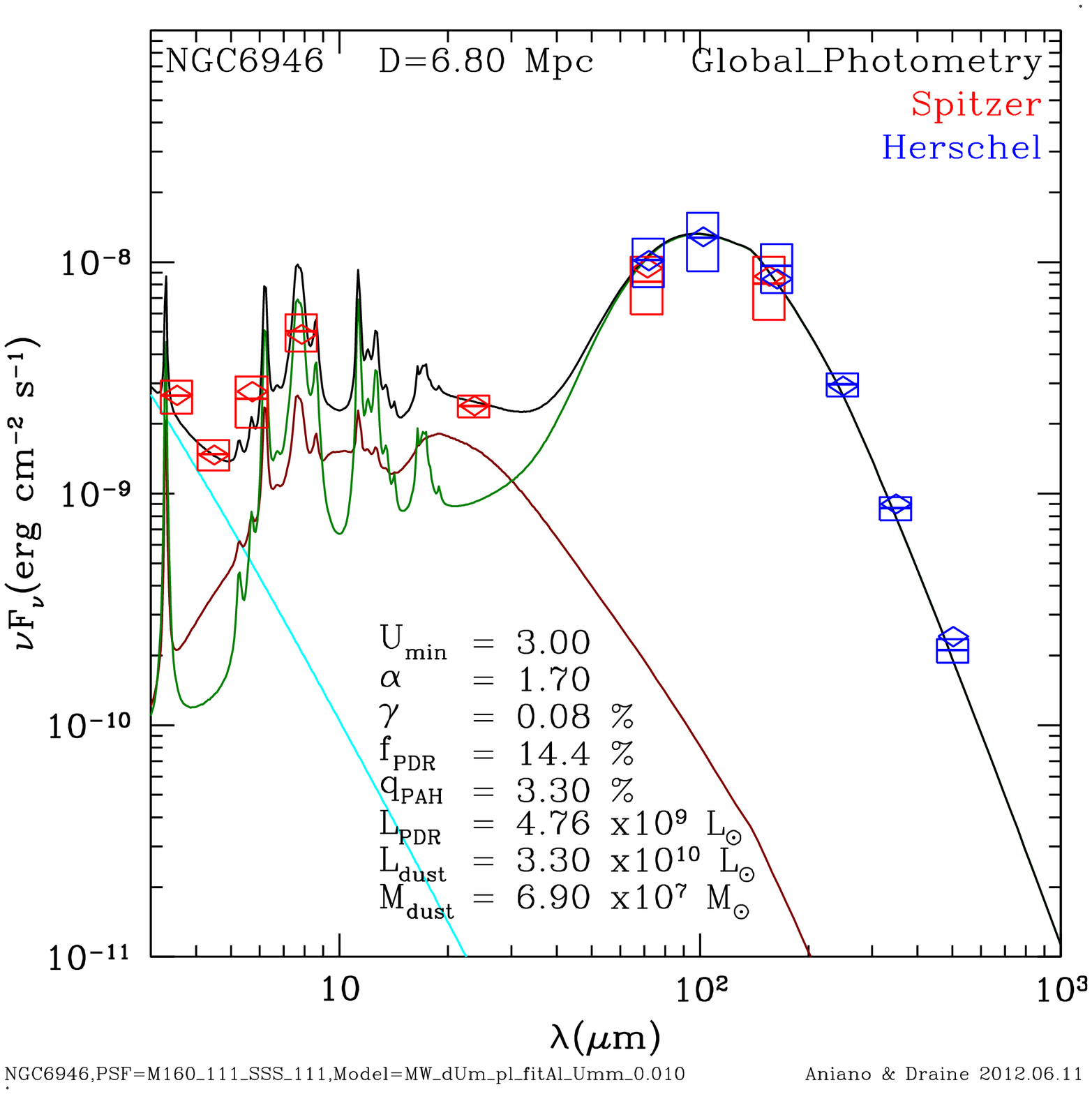}
\renewcommand \RtwoCone {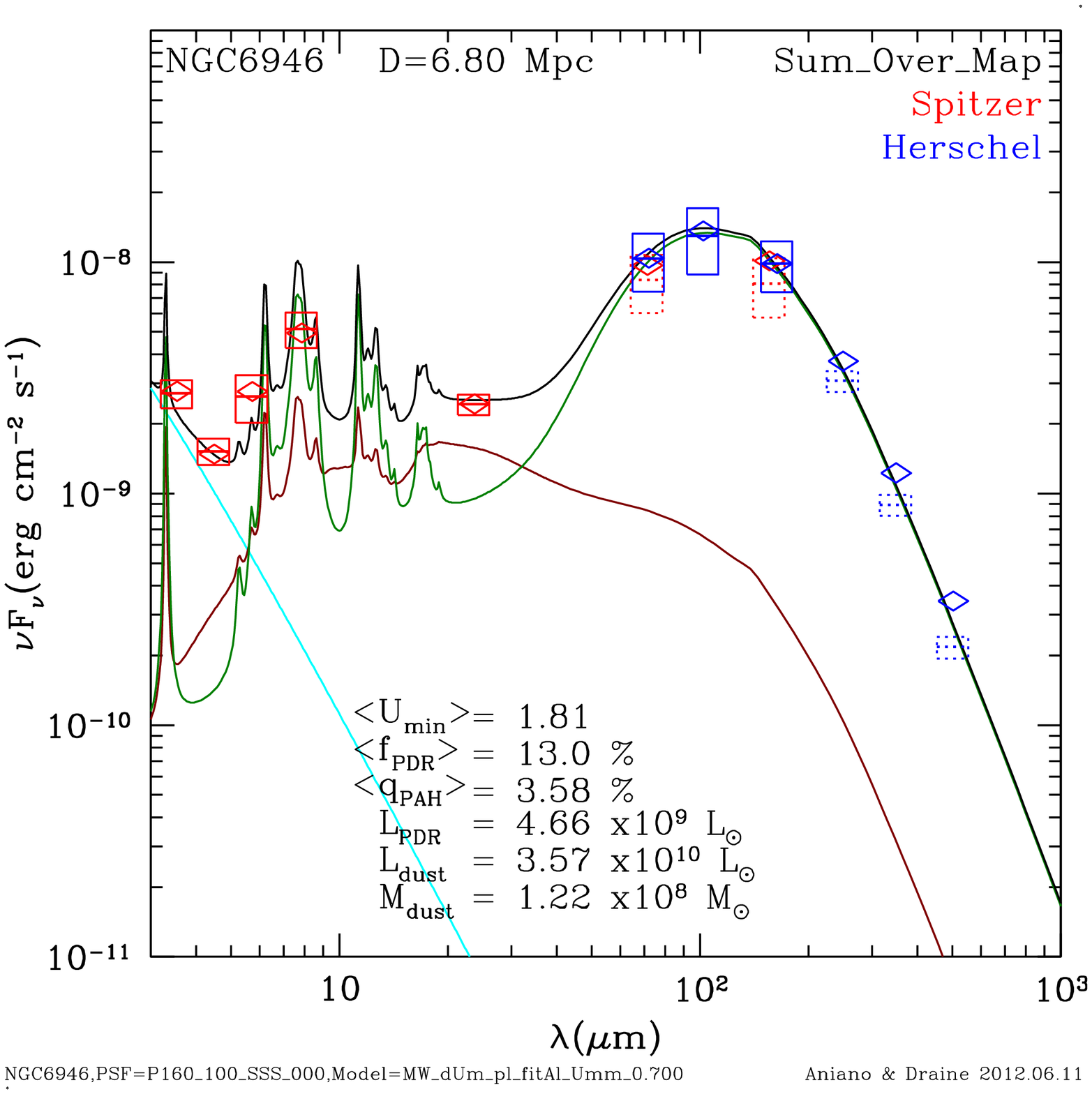}
\renewcommand \RtwoCtwo {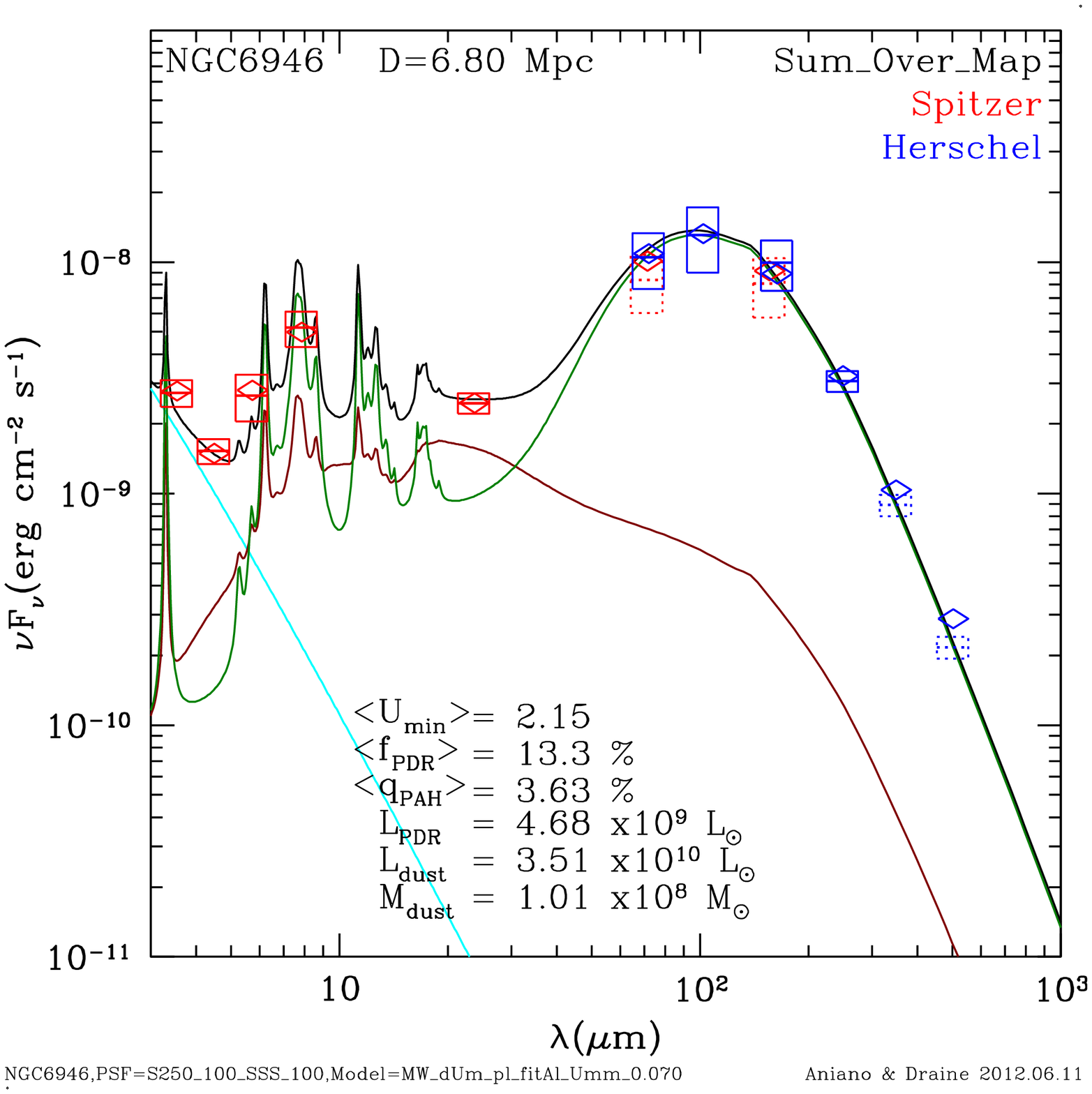}
\renewcommand \RtwoCthree {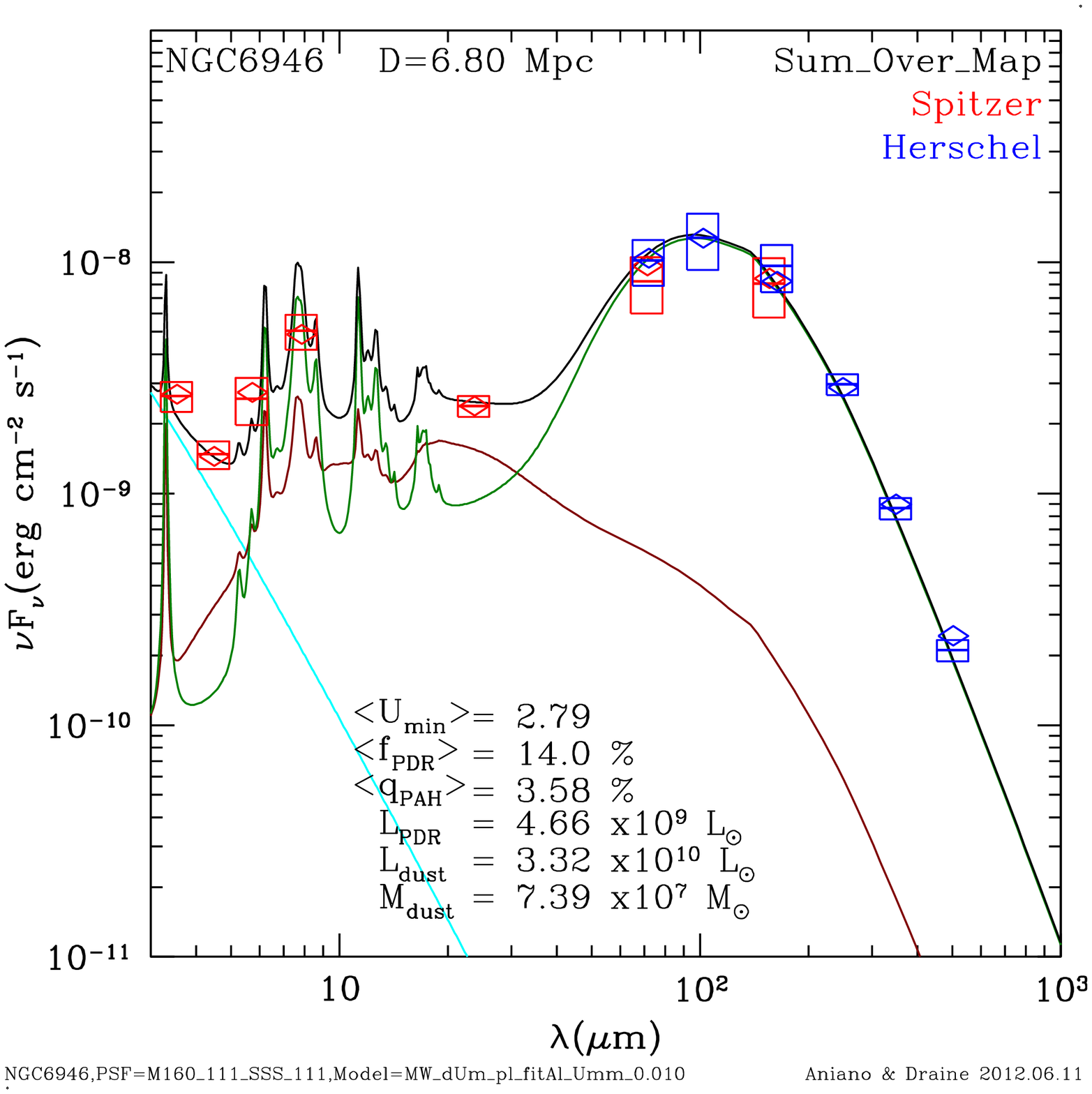}
\begin{figure}
\centering
\begin{tabular}{c@{$\,$}c@{$\,$}c} 
\footnotesize IRAC, MIPS24, PACS &\footnotesize{IRAC, MIPS24, PACS, SPIRE250}& \footnotesize IRAC, MIPS, PACS, SPIRE\\
\FirstNormal
\SecondLast
\end{tabular}
\vspace*{-0.5cm}
\caption{\footnotesize\label{fig:ngc6946-4} Model SEDs for NGC~6946.
  The color coding and columns are similar to
  Fig.\ \ref{fig:ngc0628-4}.  Top row: Global SED compared with
  single-pixel models.  Second row: Global SED, compared to
  multi-pixel model.  Even when SPIRE data are employed, the model
  tends to overpredict (slightly) the global photometry at
  $500\micron$.
  }
\end{figure}

Figure \ref{fig:ngc6946-4} shows the global SED for the galaxy
NGC~6946 (similar to Figure \ref{fig:ngc0628-4} for NGC~628).  The top
row shows the SED for a ``single pixel'' model and the second row
shows the predicted model SED obtained by summing over the model SED
for individual pixels.  Again, we observe small differences between
the global modeling (top row) and resolved studies (bottom row) due to
the non-linear behavior of the modeling process.  It is seen that when the
modeling is done using the MIPS160 PSF, the dust mass estimated from
the multi-pixel model is in good agreement with that obtained for a
single-pixel model.

\subsubsection{Fitting in Selected Apertures}

\renewcommand \RoneCone {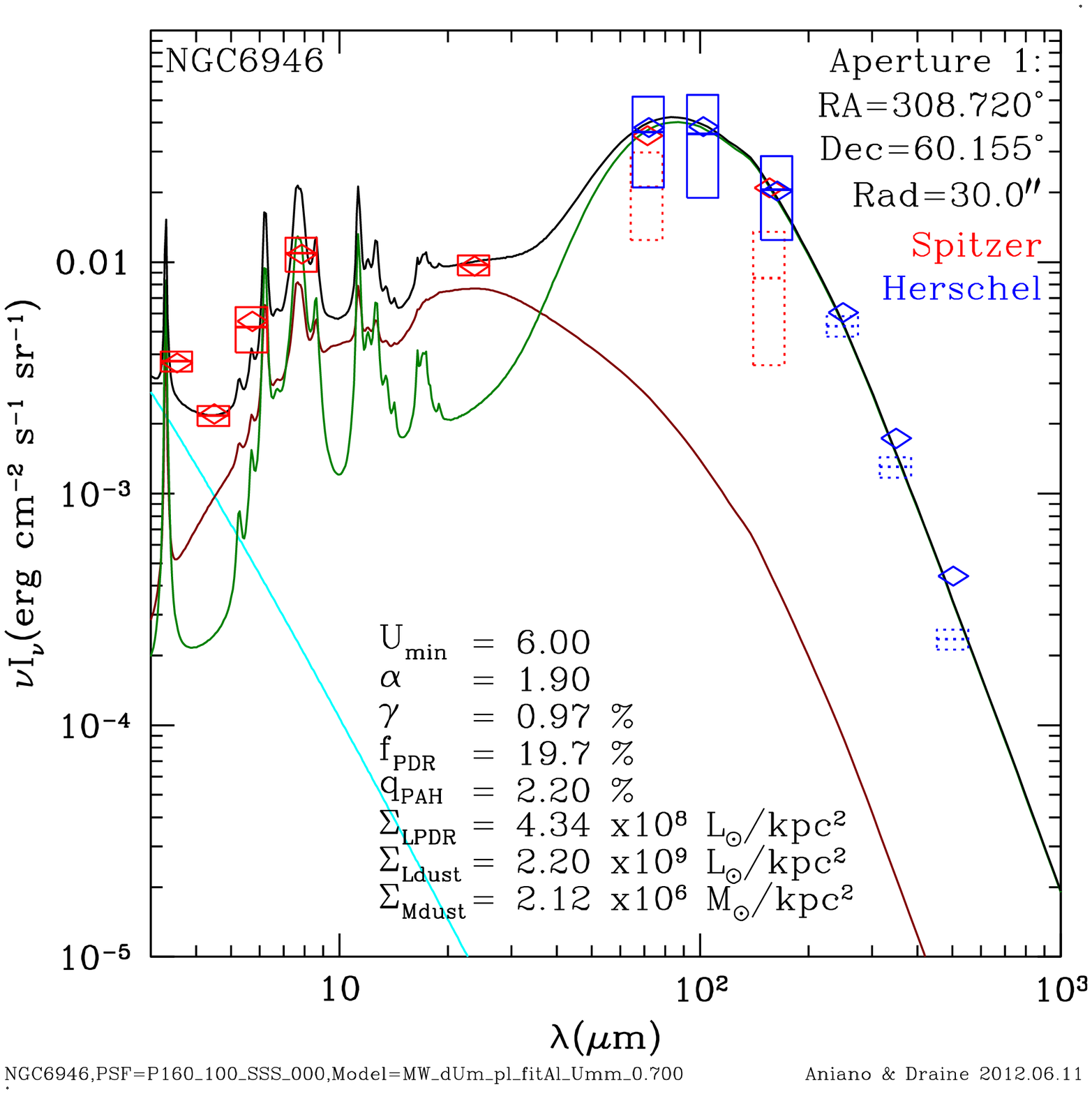}
\renewcommand \RoneCtwo {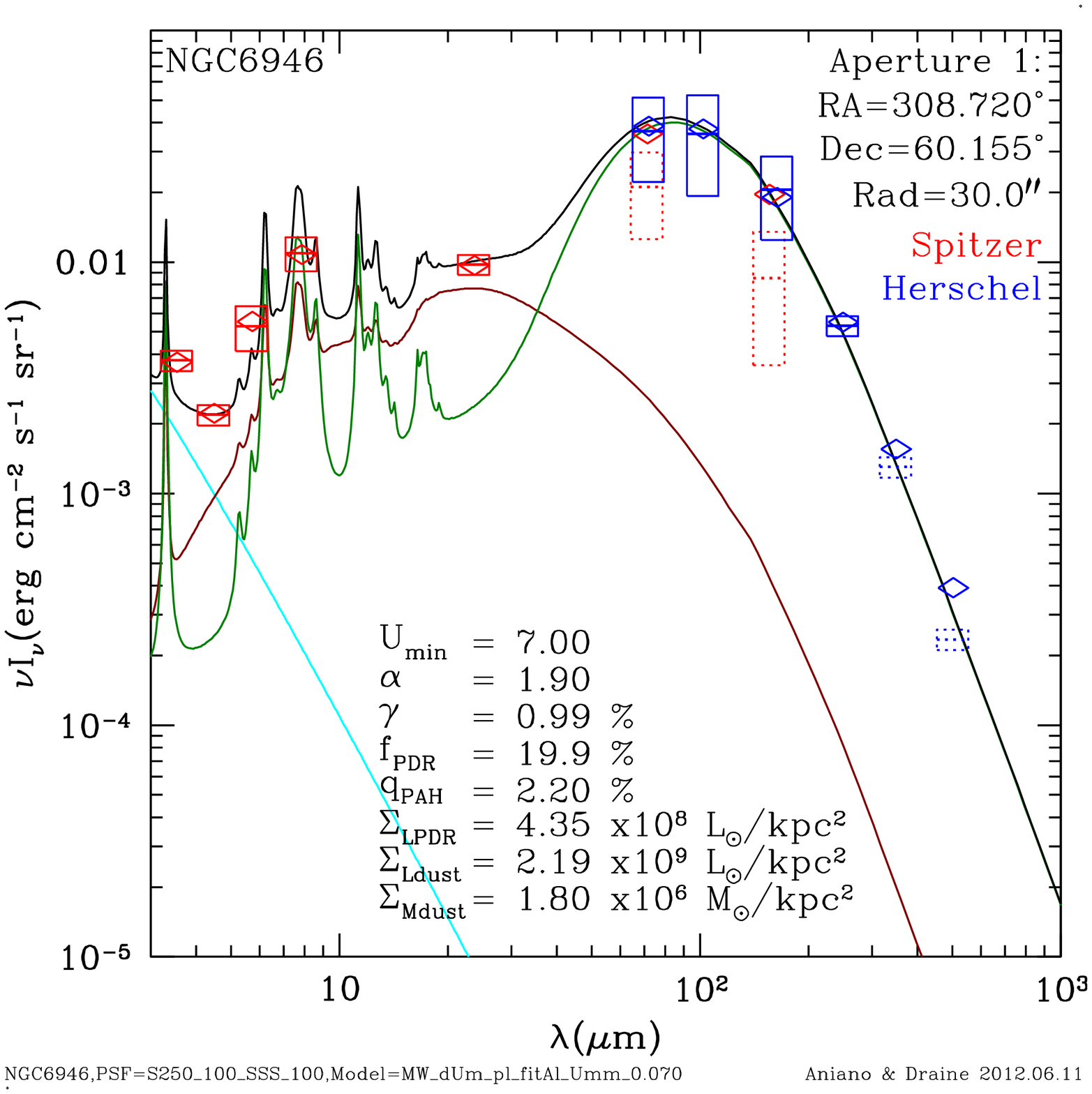}
\renewcommand \RoneCthree {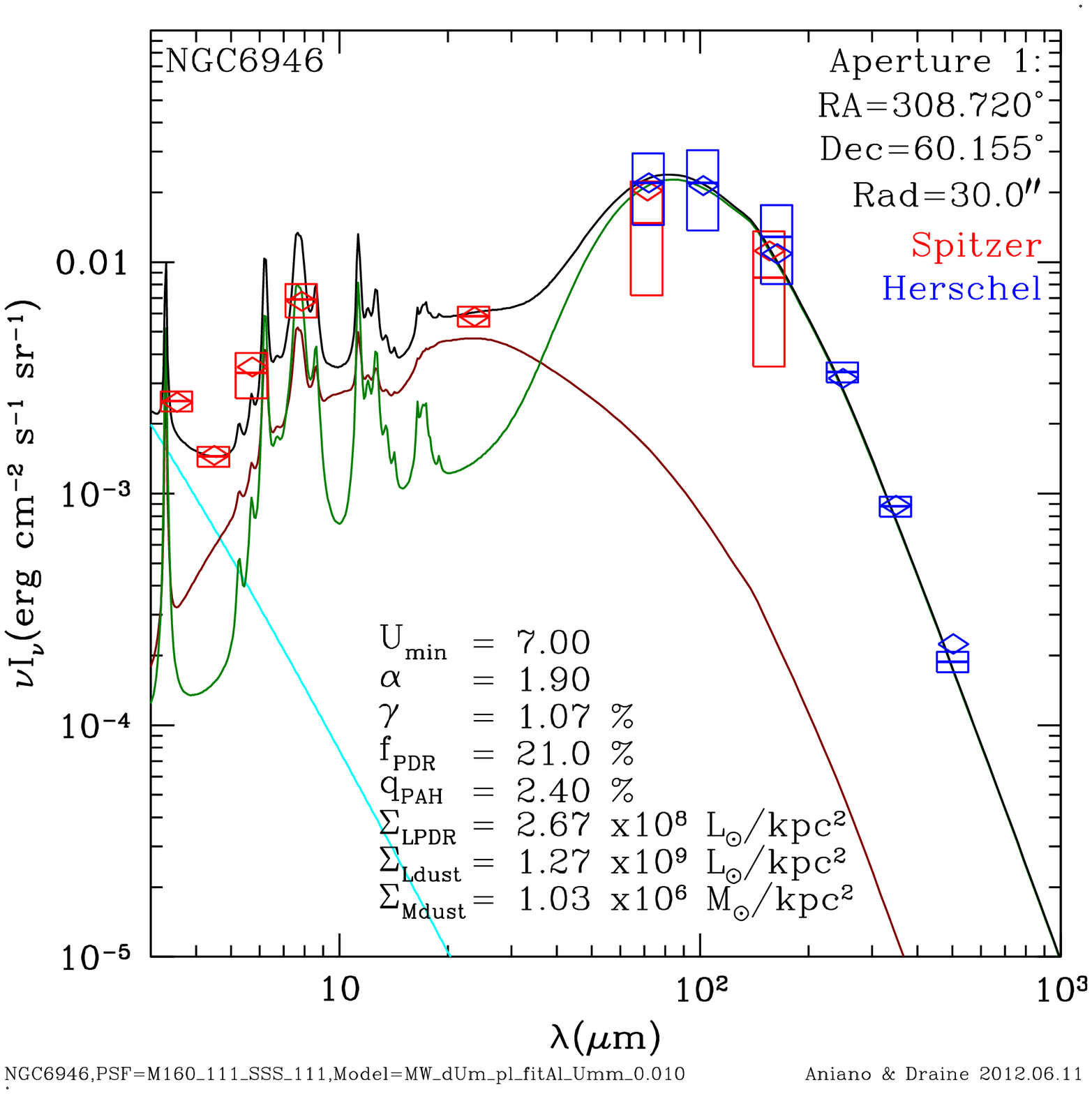}
\renewcommand \RtwoCone {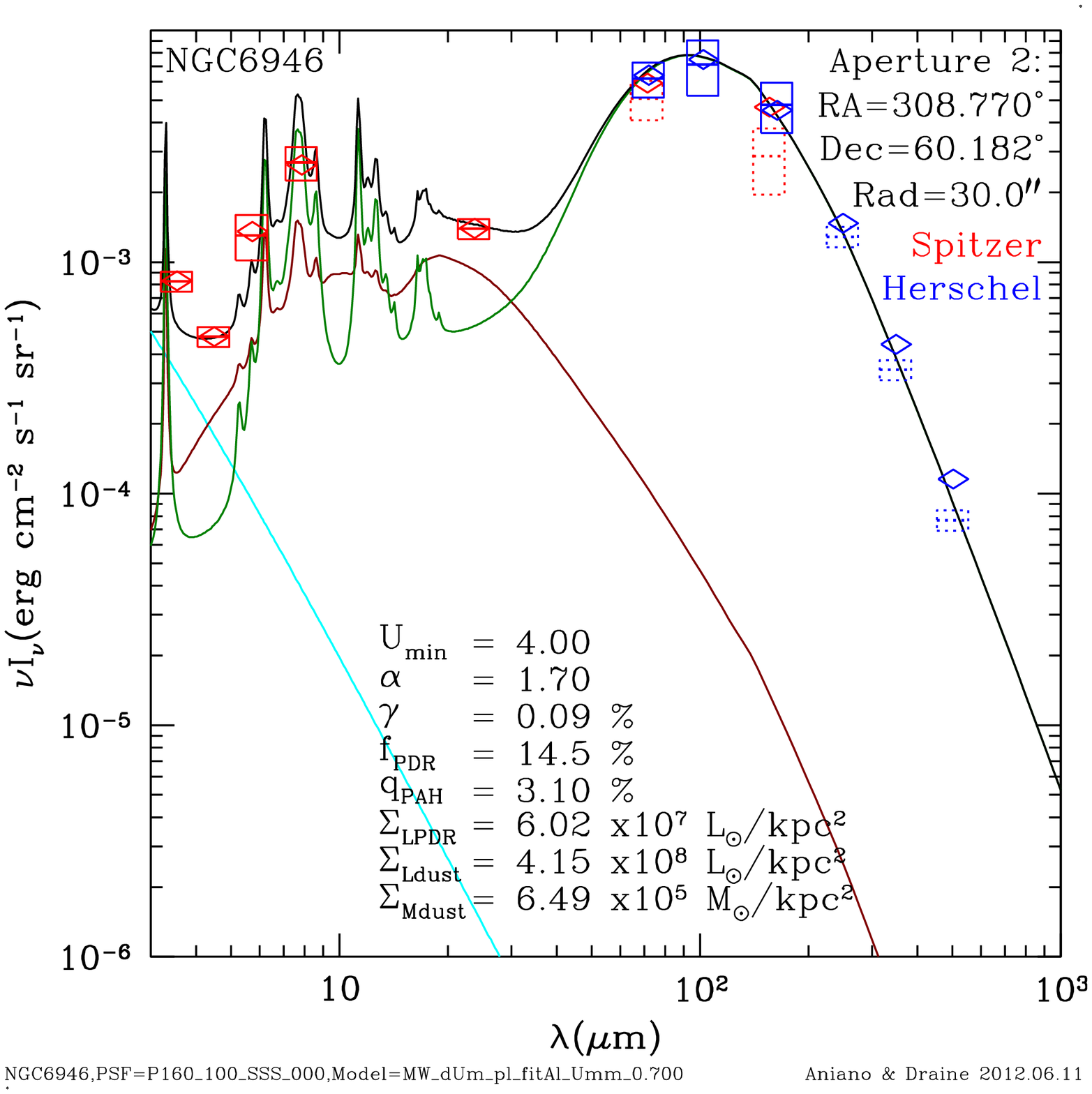}
\renewcommand \RtwoCtwo {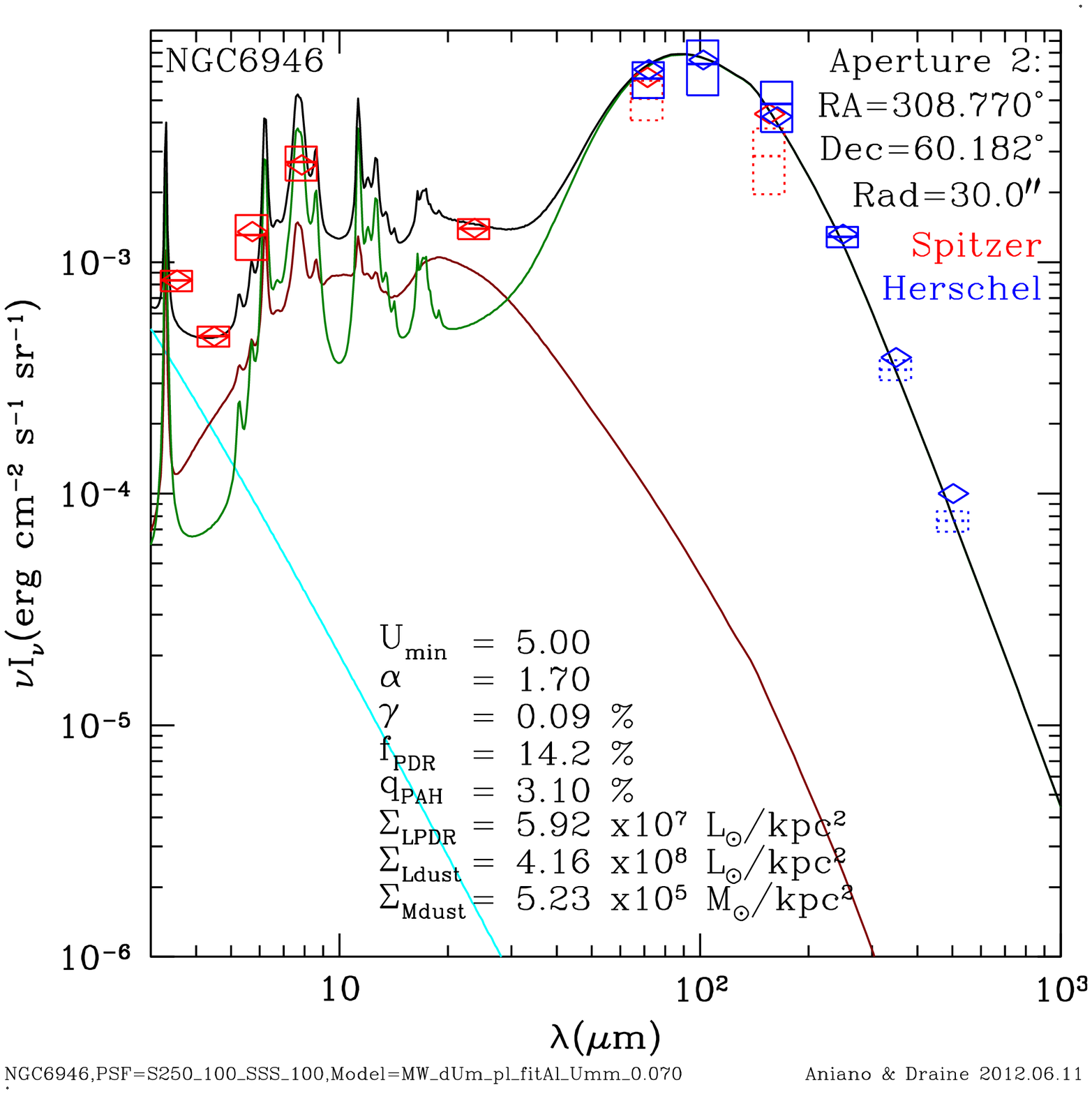}
\renewcommand \RtwoCthree {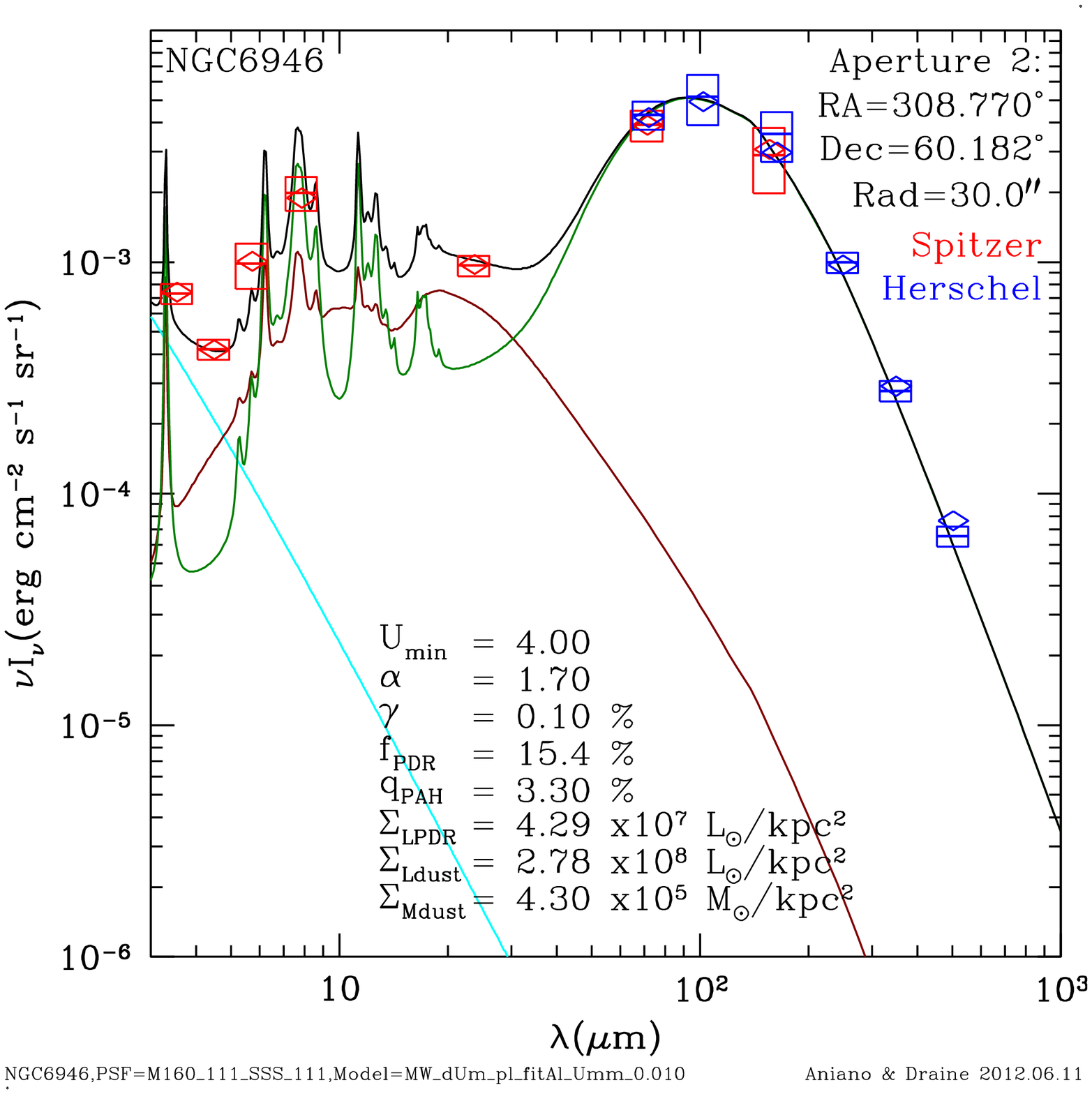}
\renewcommand \RthreeCone {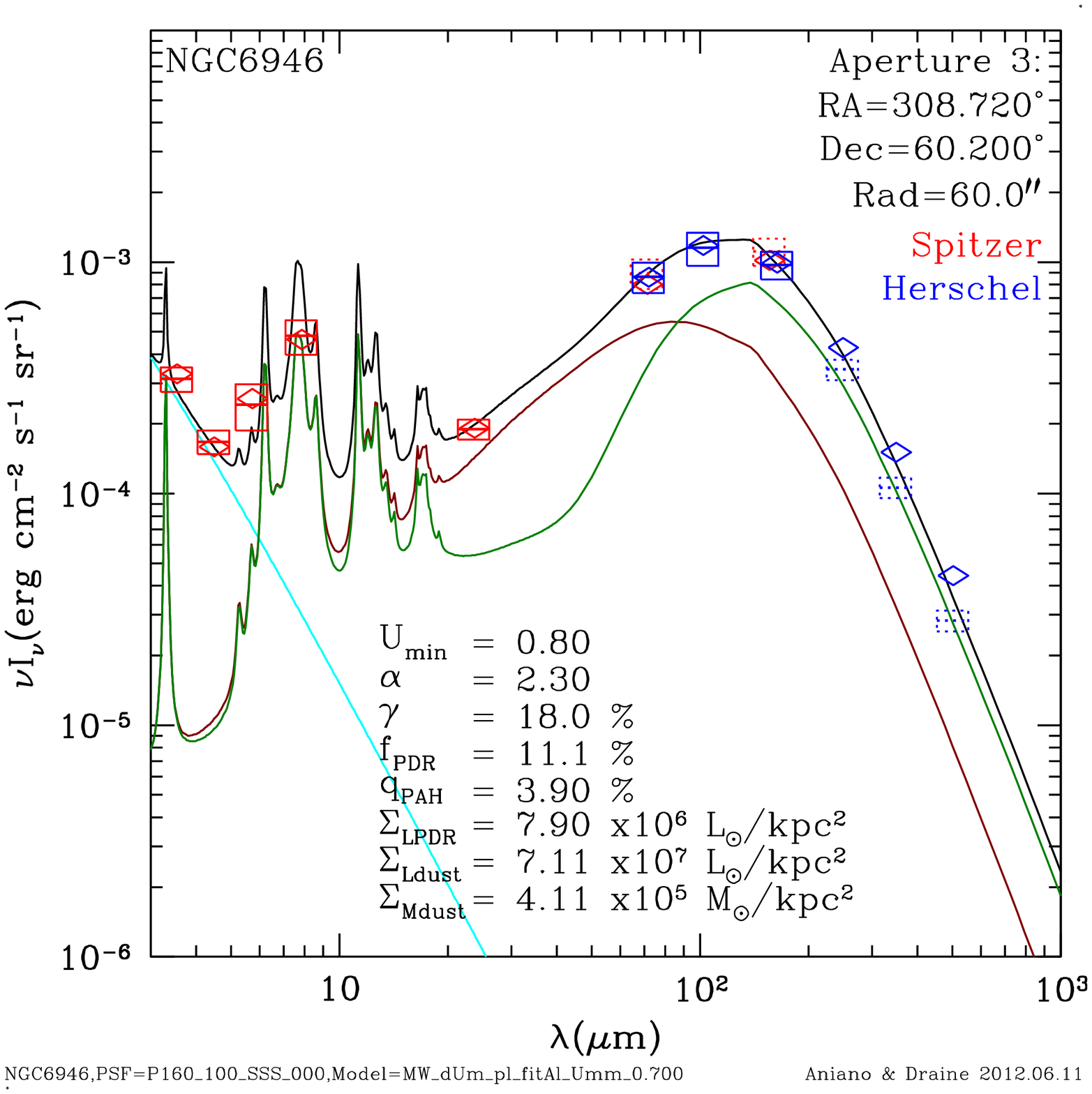}
\renewcommand \RthreeCtwo {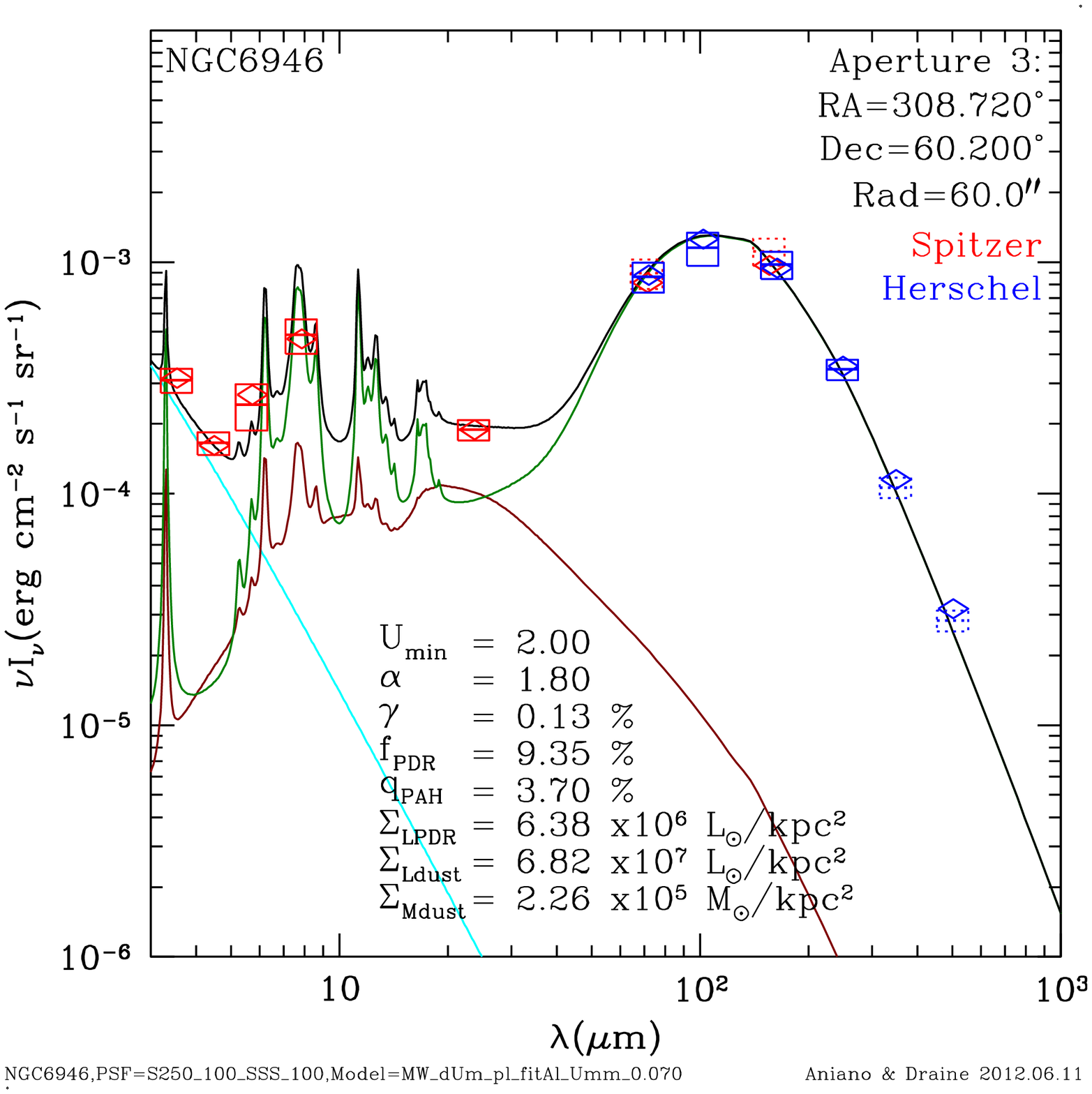}
\renewcommand \RthreeCthree {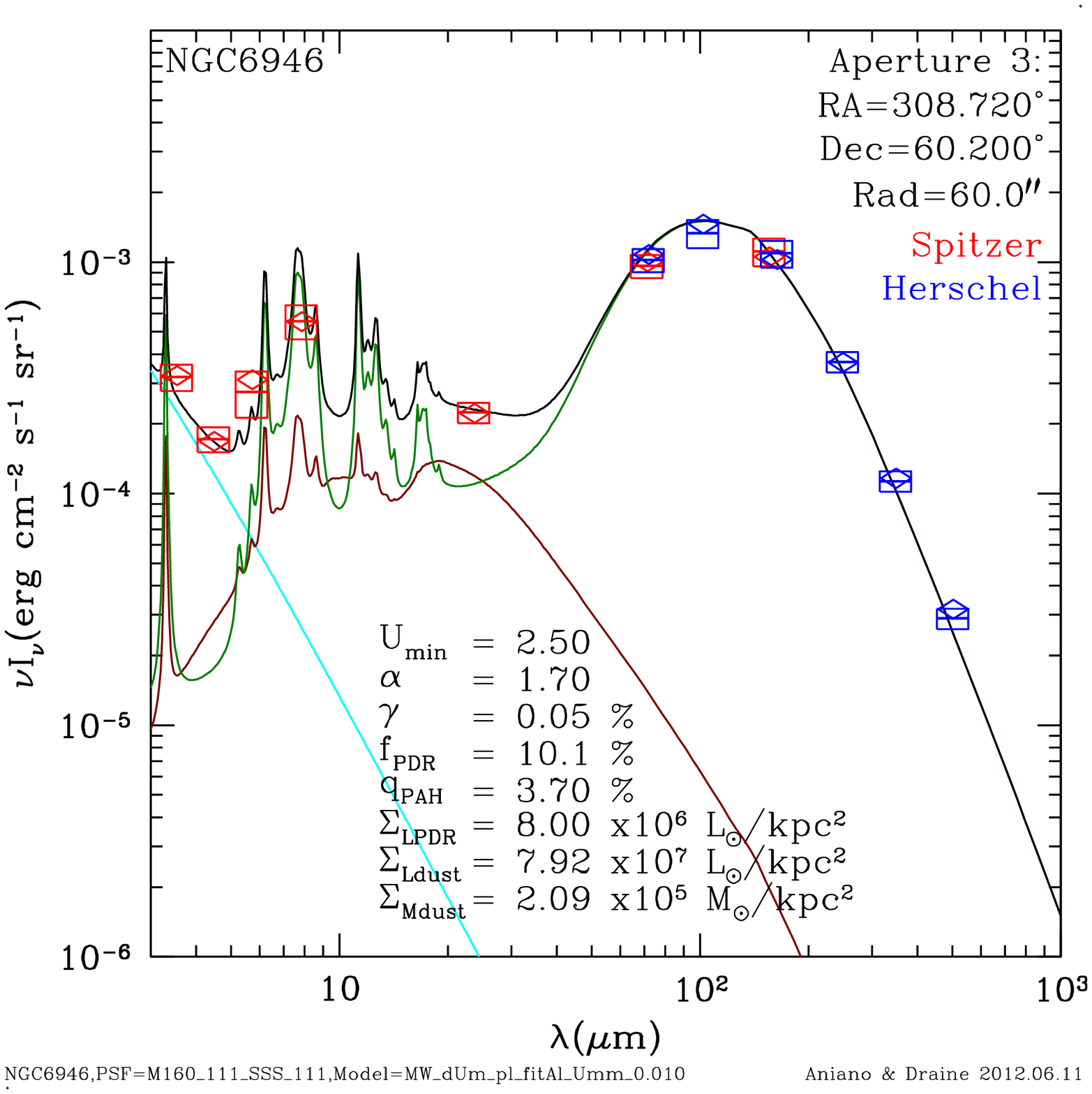}
\renewcommand \RfourCone {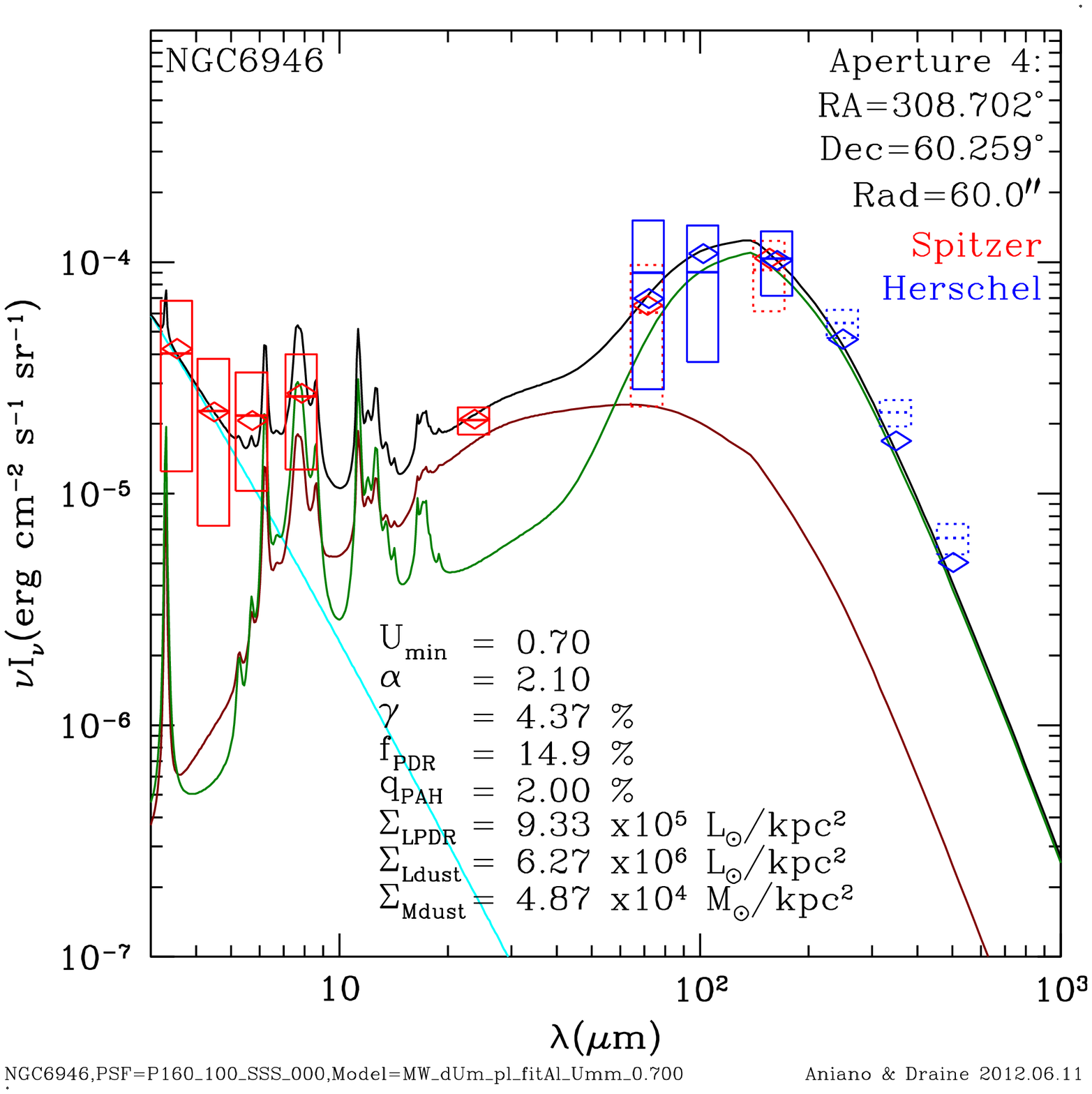}
\renewcommand \RfourCtwo {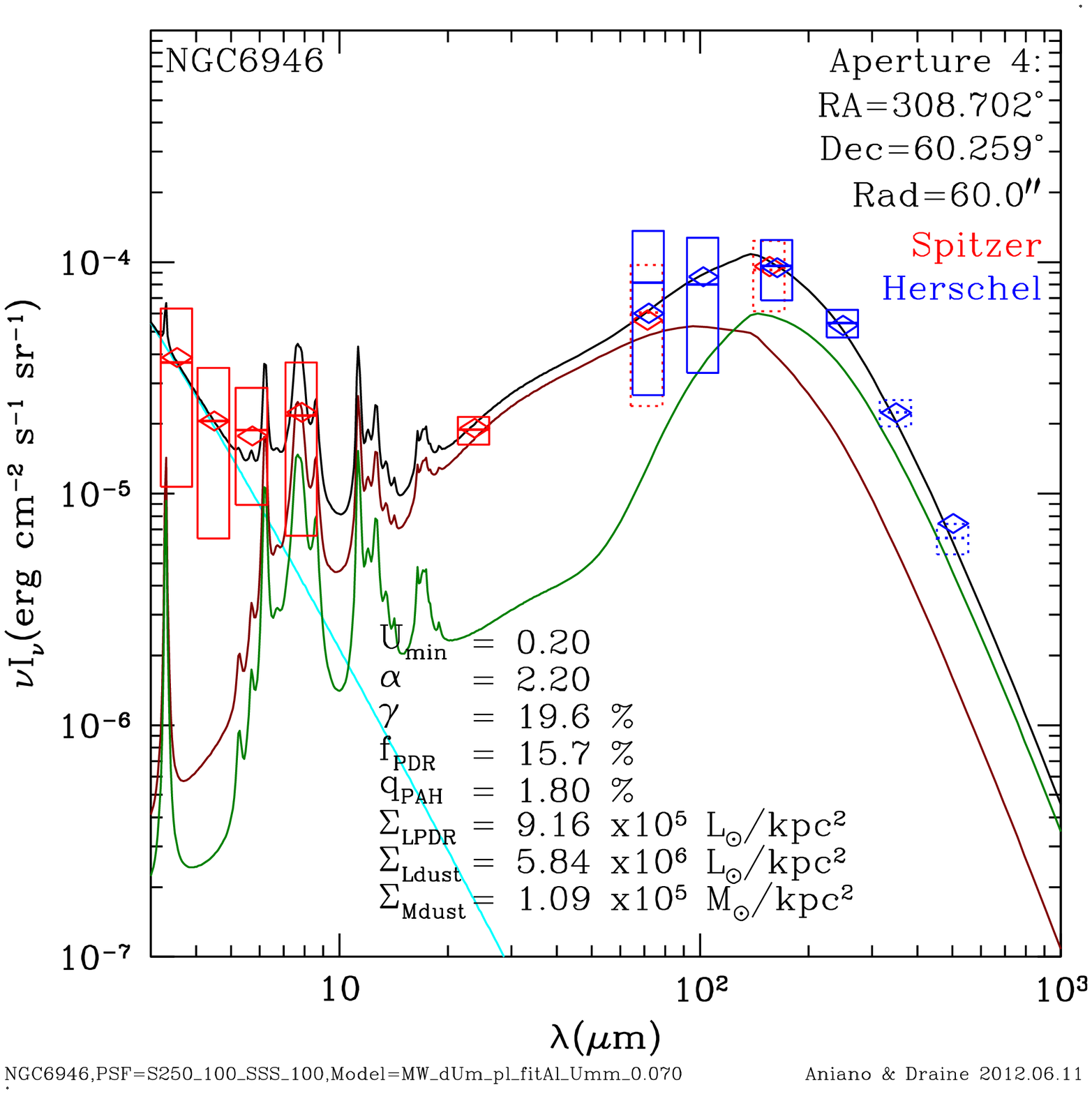}
\renewcommand \RfourCthree {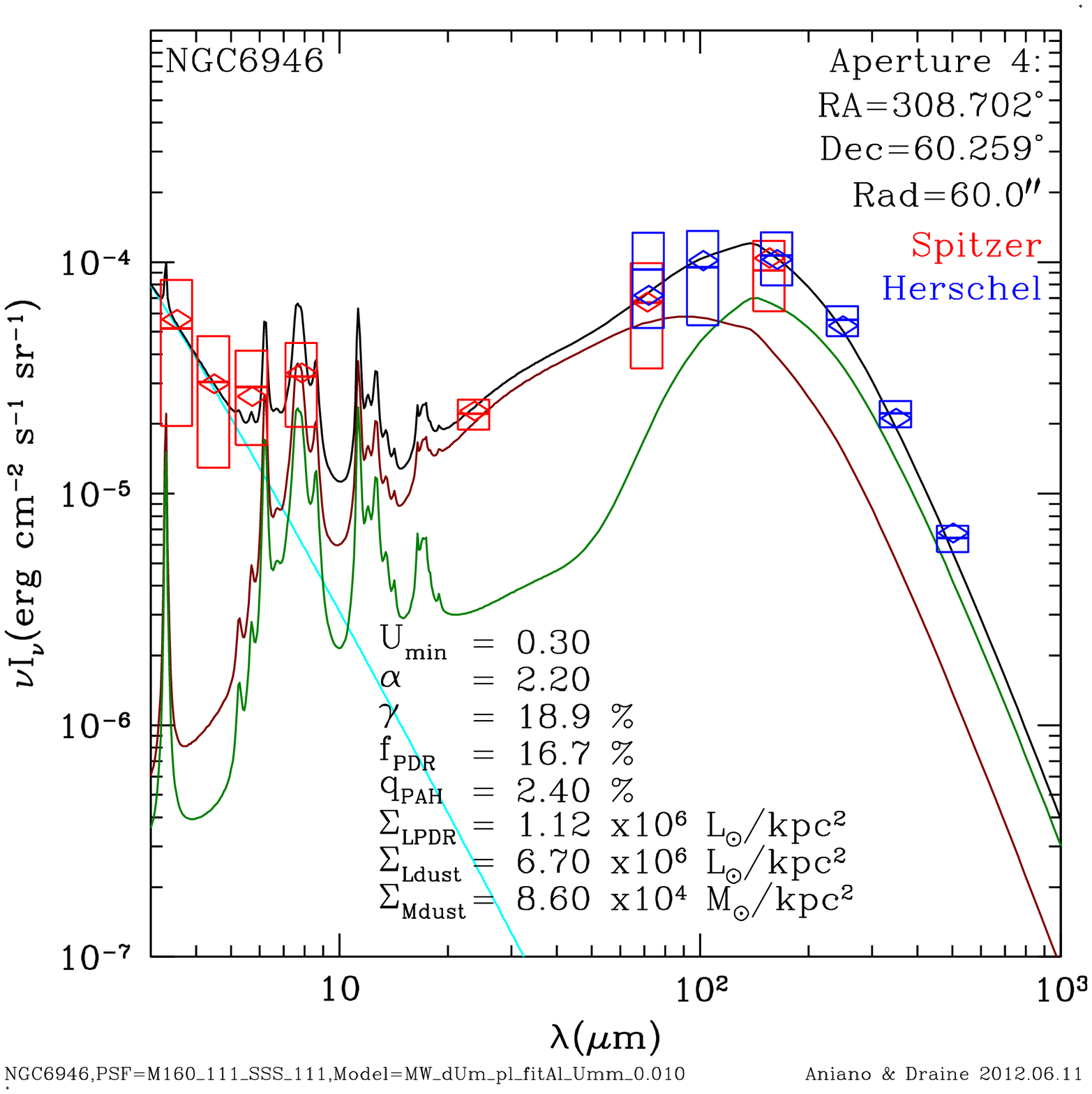}
\begin{figure}
\centering
\begin{tabular}{c@{$\,$}c@{$\,$}c} 
\footnotesize PACS160 PSF & \footnotesize SPIRE250 PSF & \footnotesize MIPS160 PSF \\
\FirstNormal
\SecondNormal
\ThirdNormal
\FourthLast
\end{tabular}
\vspace*{-0.5cm}
\caption{\footnotesize\label{fig:ngc6946-5}
Model SEDs for four selected apertures on NGC~6946.
See Fig.\ \ref{fig:ngc6946-1} for aperture location, and Fig.\ \ref{fig:ngc0628-4} for an explanation of the color coding.
Top row: $60\arcsec$ (circular) aperture centered on the galaxy nucleus.
Second row: $60\arcsec$ aperture located on a bright spot in a spiral arms.
Third row: $120\arcsec$ aperture in a mid-luminosity region.
Bottom row: $120\arcsec$ aperture in a low-luminosity region.}
\end{figure}

Figure \ref{fig:ngc6946-5} shows SEDs for four circular apertures
located on NGC~6946, sampling a wide range of surface brightnesses,
$\Sigma_{L_\dust}\approx (0.07-13)\times 10^{8}\Lsol\kpc^{-2}$,
similar to Figure \ref{fig:ngc0628-5} for NGC~628.  The top row shows
the SED for a $60\arcsec$ diameter circular aperture centered on the
galaxy nucleus.  The second row shows the SED for a $60\arcsec$
aperture located on a bright spot on the spiral arms.  The third and
fourth rows show SEDs in $120\arcsec$ apertures further from the
center. 
Aperture 4 is located completely outside the galaxy mask, but the S/N of a large aperture allows a reliable determination of the dust model parameters.

The nuclear emission is strongly peaked.  The Herschel photometry for
the central aperture noticeably decreases when the image is convolved
to MIPS160 resolution, and the estimated dust luminosity surface
brightness in the $60\arcsec$ extraction aperture drops from
$2.2\times10^9\Lsol\kpc^{-2}$ (Figs.\ \ref{fig:ngc6946-5}a,b) to
$1.3\times10^9\Lsol\kpc^{-2}$ (Fig.\ \ref{fig:ngc6946-5}c).

We observe that the PACS photometry in the high surface-brightness
nucleus tends to generally be larger than the corresponding MIPS
photometry -- compare the PACS and MIPS fluxes in Figure
\ref{fig:ngc6946-5}c.  In apertures 2, 3, and 4, the PACS and MIPS
photometry (at common resolution -- see Figures \ref{fig:ngc6946-5}f,
i, and l), the PACS and MIPS photometry is in fairly good agreement.
The disagreement in the PACS vs.\ MIPS photometry is larger for
NGC~6946 than for NGC~628 (see Table \ref{tab:ratio} below), presumably
related to the much higher peak surface brightnesses present in
NGC~6946.
At MIPS160 resolution,
$\Sigma_{L_\dust}=9\times10^7\Lsol\kpc^{-2}$ in NGC~628 aperture 1 is
only 7\% of $\Sigma_{L_\dust}$ in NGC~6946 aperture 1.


\clearpage
\section{\label{sec:importance of SPIRE}Importance of SPIRE Photometry}

\begin{figure}[h] 
\begin{center}
\includegraphics[width=8cm,height=7cm]{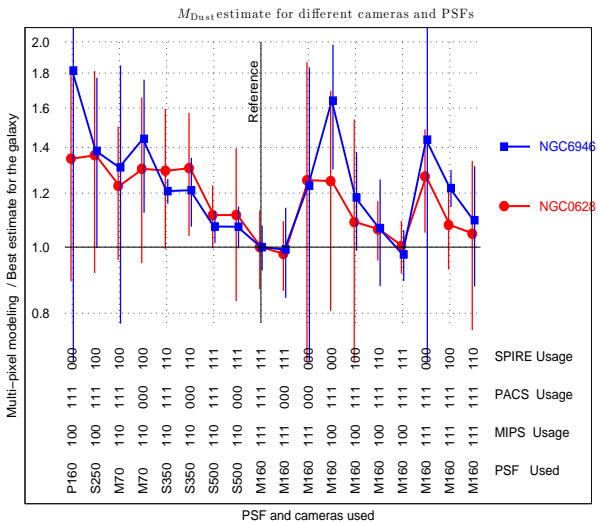}
\vspace*{-0.5cm}
\caption{\footnotesize\label{fig:dustmass} Total dust mass $M_\dust$
  (summed over all pixels) estimated for NGC\,628 and NGC\,6946 using
  different camera/PSF combinations (see text).  Masses are given
  relative to the ``best estimate'' $M_\dust$ obtained summed over all pixels using all
  wavelengths and MIPS160 resolution. }
 \end{center}
\end{figure}

We investigate the effect of only using certain sets of cameras in our
resolved (multi-pixel) modeling, since this situation will arise in
galaxies that were only observed with some of the instruments.  Dust
estimates for NGC~628 and NGC~6946 based on different camera
combinations are shown in Figure \ref{fig:dustmass}, where we compare
the dust mass estimate with our ``gold standard'', the dust mass
estimate using all cameras, at MIPS160 resolution\footnote{Even though
  MIPS160 PSF is our lowest resolution modeling, PACS - MIPS
  discrepancies made it important to be able to retain the MIPS160
  camera in the modeling, and, as shown in \S 8, using lower angular
  resolution does not introduce a major bias in the modeling.}.

\begin{figure}[h] 
\begin{center}
\includegraphics[width=9cm]{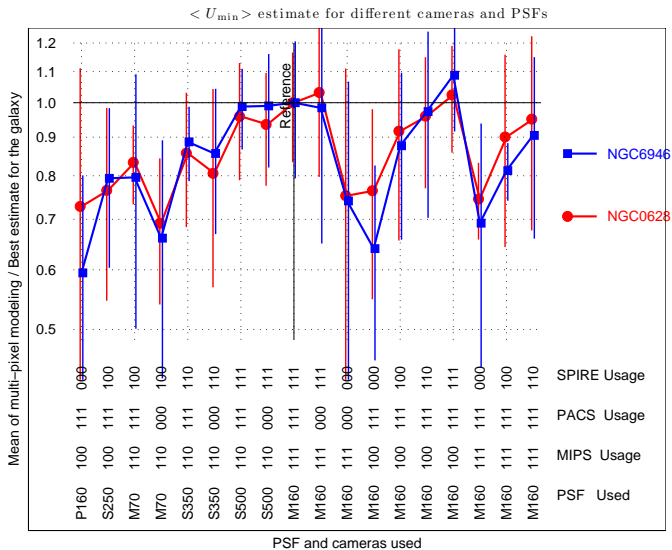}
\vspace*{-0.5cm}
\caption{\footnotesize\label{fig:umin} $\langle \Umin\rangle $
  (weighted mean over the pixels) estimated using different
  cameras and PSFs relative to the best estimate (obtained with all cameras
  and MIPS160 PSF).}
\end{center}
\end{figure}

In Figures \ref{fig:dustmass}-\ref{fig:dustmassv.s.map} the horizontal axis
represents different combinations of PSF and cameras. The nomenclature
convention is as follows: The first characters describe the PSF used;
P160, S250, S350, S500, M70, and M160 refer to PACS160,
SPIRE250, SPIRE350, SPIRE500, MIPS70, and MIPS160 respectively.

The first group of 3 digits refers to the MIPS24,
MIPS70, and MIPS160 cameras, with 1 indicating usage.  The
next set of three digits similarly indicates usage of the three PACS
bands, and the final three digits correspond to usage of the SPIRE
cameras.  For example, S350 110 111 110 uses MIPS24, MIPS70, PACS70,
PACS100, PACS160, SPIRE250, and SPIRE350 cameras at SPIRE350
PSF. We always use the Scanamorphos data reduction for PACS.

The ten right-most columns correspond to the MIPS160 PSF, and the
remaining eight left-most columns have different PSFs (sorted by
ascending FWHM left to right).  The modeling done at MIPS160
resolution uses different camera combinations, allowing one to
examine, at fixed angular resolution, the effects of different
wavelength coverage and PACS/MIPS discrepancies.

Figure \ref{fig:dustmass} shows that the dust mass estimates for these
two galaxies appear to be quite good when we use the three SPIRE bands: the total dust mass estimated without
using MIPS160 (at MIPS160 or SPIRE500 PSF) is off by only $\sim$$10\%$.
It therefore appears that resolved
maps of dust in nearby galaxies can be reliably obtained at
the resolution of SPIRE500.
The dust mass estimates appear to be good provided at least two SPIRE bands
(SPIRE250 and SPIRE350) are employed: 
the total dust mass estimated without
using SPIRE500 and MIPS160 is off by 30\% for
NGC\,628, and 22\% for NGC\,6946.  
The S250 100 111 100 dust map yields a global dust
mass that exceeds the best estimate by $\sim$$38\%$.
For some purposes this $\sim$$38\%$ loss of accuracy will be acceptable, as it makes possible a dust map with FWHM=$18.2\arcsec$.

In order to map the dust at higher angular resolution (PACS160 PSF), 
we can only use a subset of all the cameras available (IRAC, MIPS24, PACS).
Dust maps made at the resolution of PACS160 are less reliable, for
three reasons: (1) with smaller pixels, the signal/noise of the image
is poor in the faint regions, (2) there seem to be significant systematic
uncertainties associated with the PACS photometry, (3) the dust
models are unconstrained at $\lambda>160\micron$, and (4) our PACS maps are less sensitive than the MIPS counterparts.
The non-linearity of the dust model fitting procedure can result in
large errors in dust mass in the low S/N regions.
Dust masses estimated using maps at PACS160 resolution (using IRAC, MIPS24, and
PACS data only) can overestimate the dust masses by factors $\approx
1.4\,-\,1.8$.  If greater accuracy is required, PACS160 resolution should
only be used in the highest surface brightness regions, where
signal/noise issues should be less
critical.

It has sometimes been argued that dust masses estimated using only
data at wavelengths $\lambda\leq 160\micron$ (e.g., those based on
MIPS photometry only) are suspect because the observations are
insensitive to ``cold'' dust at temperatures $T_\dust\ltsim 12\K$.
The SINGS sample included 17 galaxies with SCUBA 850\um, and
\citet{Draine+Dale+Bendo+etal_2007} found that reliable dust mass
estimates could be obtained with the DL07 dust model {\it without}
using data longward of 160\um: these galaxies did {\it not} appear to
contain significant masses of ``cold dust''.  The SPIRE photometry for
NGC\,628 and NGC\,6946 confirms this: these galaxies do not appear to
contain substantial masses of cold dust.  In fact, we find that models
fitted to the PACS photometry alone tend to predict {\it more}
emission at long wavelengths than is observed.  This is evident in
Figures \ref{fig:ngc0628-3}a and \ref{fig:ngc0628-3}d, and Figures
\ref{fig:ngc6946-3}a and \ref{fig:ngc6946-3}d, which show that DL07
dust models based only on $\lambda\leq160\micron$ data (IRAC, MIPS24,
and PACS) tend to {\it overpredict} the actual fluxes at 250 and
500\um in the central 6 arcmin (12kpc).  This is also seen in Figures
\ref{fig:ngc0628-4}d and \ref{fig:ngc6946-4}d, which compare the
observed global SED with the sum of the SEDs from a multipixel model
for each galaxy.  The model (the diamonds) overpredicts the SPIRE data
by factors of $\sim$1.1, 1.2, and 1.2 at 250, 350, and 500\um.  This
is rather good agreement, and indicates that there is no substantial
amount of very cold dust present in either galaxy.  In fact, when the
SPIRE data are used to constrain the fit, the best-fit models actually
{\it raise} the starlight intensities slightly in order to match both
the PACS, MIPS, and SPIRE photometry, reducing the dust mass estimate.
Note the very good agreement between observed and model SEDs in
Figures \ref{fig:ngc0628-4}f and \ref{fig:ngc6946-4}f, where all data
are used to constrain the multipixel fit.

The main difference in the estimated dust masses in the different
PSF/cameras combination is due to the different values of $\Umin$
found when we do not use all the cameras available.  Figure
\ref{fig:umin} shows $\langle \Umin\rangle $ (defined in
eq. \ref{eq:mean_2}), the dust-mass weighted value of the starlight
intensity parameter $\Umin$ relative to the best estimate for $\langle
\Umin\rangle $ -- the estimate obtained using all the data at MIPS160
resolution.

\section{\label{sec:maps vs global}Maps v.s. Global Photometry}

The KINGFISH sample consists of galaxies that are near enough to be
well-resolved, enabling us to make maps of dust and starlight
properties.  However, because the dust modeling used here is a
nonlinear procedure, the total dust mass estimate, for example, will depend on
what resolution is employed.  In the preceding sections, we usually made dust
maps using the best angular resolution possible with each chosen camera set
used, with the limiting resolution therefore determined by the longest
wavelength data used in the fit (or the inclusion of MIPS160).

Herschel is frequently used to observe distant galaxies that are
unresolved at the longest wavelengths.  For the KINGFISH galaxies, we
will therefore also estimate dust masses using the global photometry (the flux from within the galaxy mask),
to see how dust mass estimates based on global photometry compare with
the best estimates for a well-resolved galaxy. 

\begin{figure}[h]
\begin{center}
\includegraphics[width= 10cm]{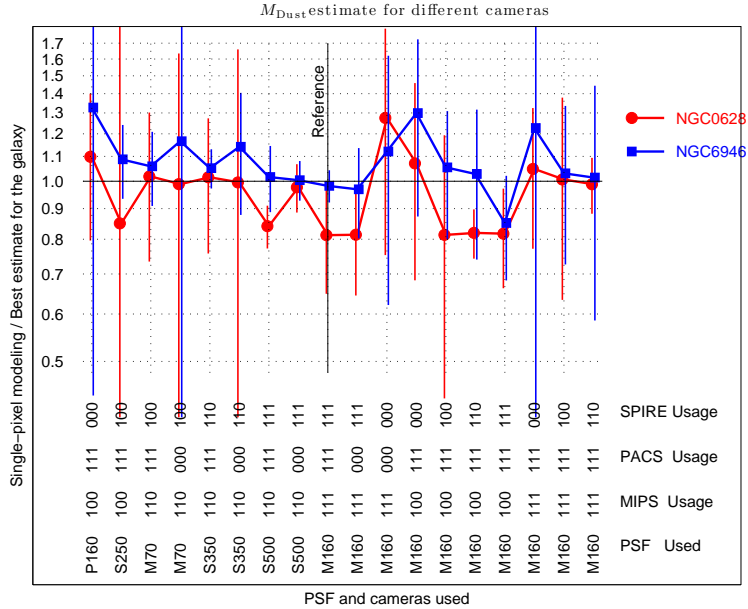}
\vspace*{-0.5cm}
\caption{\footnotesize\label{fig:dustmassv.s.best} Dust mass
  estimates based on global photometry for NGC~628 and NGC~6946,
  using different camera combinations, relative to the sum of the dust
  mass (obtained from pixel-by-pixel modeling of the resolved galaxy)
  in the MIPS160 PSF using all the cameras (i.e., relative to the best total dust
  estimate for the galaxies). 
  In principle estimates based on global photometry should not depend on the chosen PSF, only on the camera combination. 
  In practice, flux scattered out of the galaxy mask (into the PSF's extended wings) changes the global photometry estimates, leading to small differences in the parameter estimates   (e.g., see differences between P160 100 111 000 and M160 100 111 000 modeling).}
\end{center}
\end{figure}

Figure \ref{fig:dustmassv.s.best} shows the dust mass estimated using
the DL07 dust models to fit the global SED, divided by the ``gold
standard'', i.e., the sum of the dust mass obtained from
pixel-by-pixel modeling of the resolved galaxy in the MIPS160 PSF
using all the cameras.  Recall that the DL07 model assumes that most
of the dust is exposed to a single starlight intensity $\Umin$.  A
multi-pixel model with $\Uminj$ varying from pixel to pixel cannot be
exactly reproduced by a model that assumes a single value of $\Umin$
for the entire galaxy, hence fitting the summed photometry with a
single-pixel model will in general give a dust mass that will differ
from that obtained by summing over a multipixel model.

Figure \ref{fig:dustmassv.s.best} shows that if all (IRAC, MIPS, PACS,
SPIRE) data are used, the dust mass obtained by fitting the global SED
is very close to that obtained from a multipixel fit of the resolved
galaxy: the global dust mass is within 18\% of the best estimate for
NGC\,628, and within 3\% of the best estimate for NGC\,6946.  If data
from some cameras are not used, the single-pixel estimate for the dust
mass will of course change, but we see that for NGC\,628 and NGC\,6946
the resulting dust masses vary suprisingly little provided we have
SPIRE data or MIPS160.  If we use only PACS for $\lambda \ge$ 70\um,
the dust mass estimate for NGC~628 is high by $\sim$$10\%$.
The M160$\_$111$\_$000$\_$000
column in Figure \ref{fig:dustmassv.s.best} compares the global dust
mass estimated from a single-pixel fit using only Spitzer (IRAC and MIPS) data
versus our ``gold standard'' best estimate (multipixel model using all cameras).
The dust masses estimated using only
the global photometry from Spitzer are within $\sim$$30\%$ of our
present best estimates.  
Thus, at least for
normal-metallicity star-forming galaxies such as NGC\,628 and
NGC\,6946, dust mass estimates based on IRAC and MIPS photometry,
e.g., the dust masses estimated for the SINGS sample
\citep{Draine+Dale+Bendo+etal_2007} appear to be reliable.

\begin{figure}[h]
\begin{center}
\includegraphics[width=10cm]{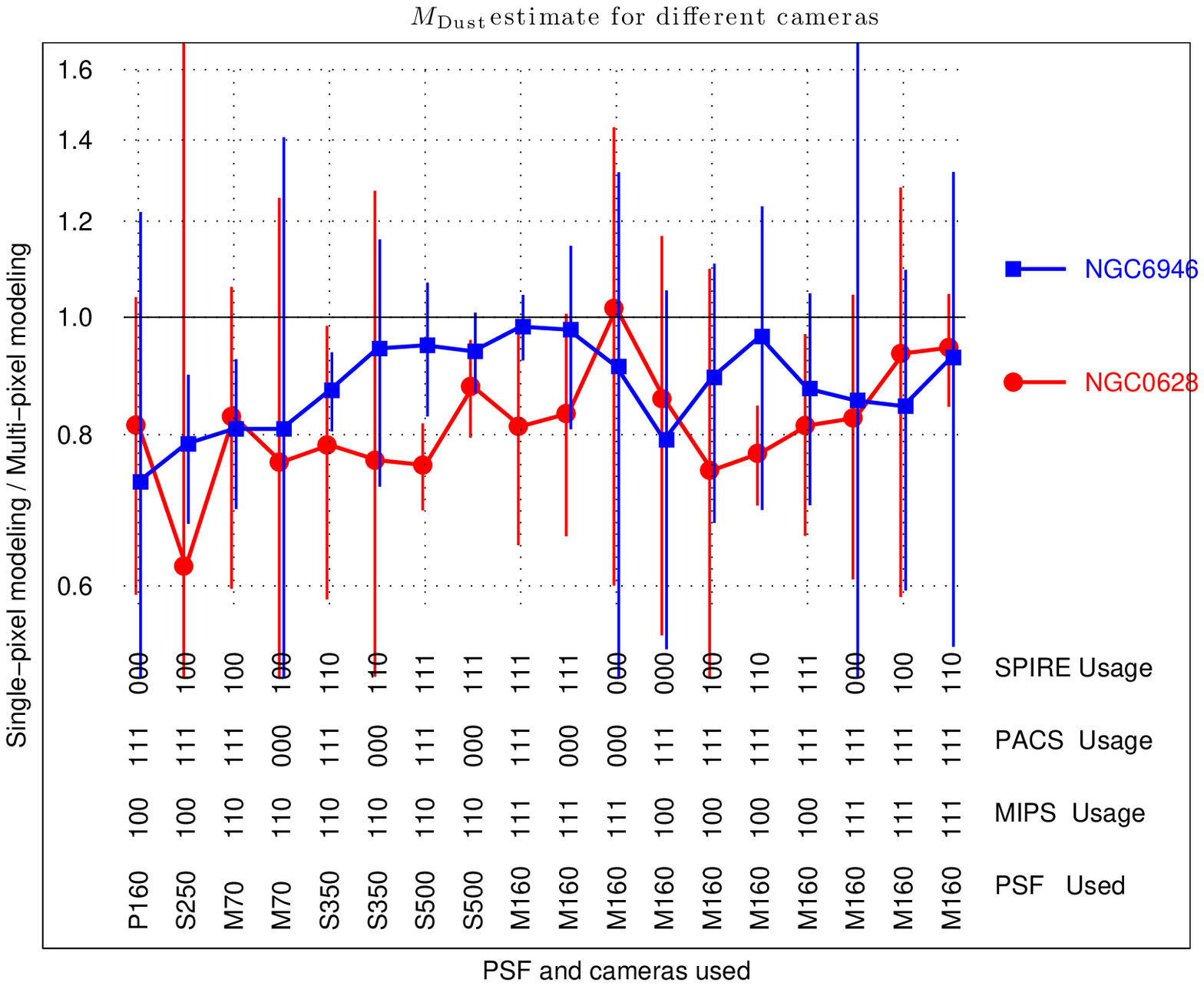}
\footnotesize
\vspace*{-0.5cm}
\caption{\footnotesize\label{fig:dustmassv.s.map}
  Dust mass estimates based on global photometry for NGC\,628 and NGC\,6946,
  using different camera combinations, 
relative to the sum of the dust mass obtained from pixel-by-pixel modeling
  of the resolved galaxy using the same camera set.}
\end{center}
\end{figure}

Figure \ref{fig:dustmassv.s.map} compares the
global dust mass estimated using global fluxes to the dust masses estimated
summing the dust masses over each map, for different sets of cameras. 
Essentially it is a measure of the non-linear behavior of the dust modeling.
For each camera and PSF combination, performing resolved modeling allows the dust to be colder in some regions, and this introduces more dust mass. 
Modeling based on global photometry under-predicts the dust masses by 0-20\% in most cases. 

In principle, using the MIPS160 PSF results as our ``gold standard''
best estimate for the dust parameters may also suffer from the
above-mentioned non-linear behavior: each MIPS160 pixel covers a
rather large area of the galaxy, in which the dust properties (e.g.,
$\qpah$) or radiation field (e.g., $\Umin$) may have variations.
Unfortunately, the discrepancy of PACS160 and MIPS160 fluxes makes it
important to retain MIPS160 photometry, and, therefore, to use the
broader MIPS160 PSF.

Table \ref{tab:summary} summarizes the main parameters obtained by
single-pixel and multi-pixel modeling of the galaxies at MIPS160
resolution, using all the cameras.

\begin{table}[h]
\begin{center}
\caption{\label{tab:summary}Multi-pixel and single-pixel best-fit model parameters at MIPS160 resolution.}
\begin{tabular}{|c|cc|cc|}
\hline
                    & \multicolumn{2}{|c|}{ NGC~628} & \multicolumn{2}{|c|}{ NGC~6946}\\
Parameter           &  Resolved$^a$ & Global$^b$     &  Resolved$^a$ & Global$^b$\\
\hline
$M_\dust$  ($\Msol$)              & 2.87$\times10^7$ & 2.33$\times10^7$ & 6.74$\times10^7$    & 6.62$\times10^7$    \\
$L_\dust $  ($\Lsol$)             & 6.83$\times10^9$ & 7.08$\times10^9$ & 3.31$\times10^{10}$ & 3.20$\times10^{10}$ \\
$\Umin$                           & 1.46             & 2.00             & 3.05                & 3.00                \\
$\overline{U}$                    & 1.75             & 2.23             & 3.61                & 3.55                \\
$100 \times \fpdr$                & 11.6             & 10.2             & 14.2                & 15.3                \\
$100 \times \qpah$                & 3.67             & 3.70             & 3.55                & 3.30                \\
$100\times M_\dust/M_{\rm H}$$^c$ & 0.82             & 0.66             & 0.63                & 0.61                \\
\hline
\multicolumn{5}{l}{$^a$ Total or mean values are defined in equations (\ref{eq:mean_mdust} - \ref{eq:mean_fpdr}).}\\
\multicolumn{5}{l}{$^b$ Values are from a model fit to the global photometry. }\\
\multicolumn{5}{l}{$^c$ For $\XCOxx=4$.}\\
\end{tabular}
\end{center}
\end{table}


\section{\label{sec:chi2}Alternative parameterizations of the starlight intensity distribution.}

Recently \citet{Galliano+Hony+Bernard+etal_2011}, in a resolved study
of the dust in the Large Magellanic Cloud, advocated a
starlight heating intensity distribution function different from what
is employed in the current work. They recommended using the starlight
distribution function of \citet{Dale+Helou_2002}, with a power law
distribution between two adjustable parameters $\Umin$ and $\Umax$,
with adjustable power law exponent $\alpha$.  We note that
\citet{Galliano+Hony+Bernard+etal_2011} uses a slightly different
nomenclature: $\Umax \equiv \Umin + \Delta U$.  The distribution used
in the present work (eq. \ref{eq:dMd/dU} and \ref{eq:dMd/dU1})
incorporates an additional ``delta function'' component at $\Umin$,
but fixes the value of $\Umax=10^7$.

\citet{Galliano+Hony+Bernard+etal_2011} claimed that satisfactory fits
can be obtained without the ``delta function'' component.  We note
that while we fix $\Umax$ to a given value in our model,
\citet{Galliano+Hony+Bernard+etal_2011} use $\Umax$ as a free
parameter. Both models have thus the same number of degrees of
freedom.  \citet{Galliano+Hony+Bernard+etal_2011} based their claim on
maps of the LMC using IRAC, MIPS and SPIRE. Here we test this claim
using our maps of NGC~628 and NGC~6946.

Figure \ref{fig:chicomp} shows maps of $\chi^2$ (goodness of fit)
obtained for three different starlight distribution schemes, all done
at MIPS160 resolution, using all 13 cameras.  The top row images
correspond to NGC 0628 and the bottom row images correspond to
NGC~6946.  In the left column the starlight distribution is a power
law between $\Umin$ and $\Umax$ with power law exponent $\alpha$ (all
adjusted) without the delta function contribution at $\Umin$, as
proposed by \citet{Galliano+Hony+Bernard+etal_2011}. It has 6
adjustable parameters $(\Omega_\star,\Mdust,\qpah, \Umin, \Umax,
\alpha)$, and therefore $13-6=7$ degrees of freedom per pixel.  The
fit for this starlight distribution is poor, giving $\chi^2$ in the
range $10-20$ in the bright regions of the galaxies.  In the middle
column the starlight distribution is a power law between $\Umin$ and
fixed $\Umax=10^7$, with power law exponent $\alpha$ (only $\Umin$
and $\alpha$ are adjusted), plus a delta function contribution at
$\Umin$ . This fit produces excellent results, and is the one
recommended in the present work. We note that this fit has the same
number of degrees of freedom as the ones in the left column panels;
the adjustable parameter $\Umax$ is replaced by the normalization
of the ``delta function'' via the parameter $\gamma$.  In the right
column the starlight distribution is a power law between $\Umin$ and
$\Umax$ with power law exponent $\alpha$ (all adjusted), plus a delta
function contribution at $\Umin$. Although the right panels have one
extra adjustable parameter ($\Umax$) with respect to the recommended fit,
the fit quality is similar to the middle panels. If $\Umax$ were a
relevant parameter, one would expect $\chi^2$ to decrease
significantly (of order 1). In fact, in the right column fit, for most
of the bright pixels the best-fit $\Umax$ value is $\Umax=10^7$ and
thus the value of $\chi^2$ is for those pixels exactly the same as in
the center column fit.

\renewcommand \RoneCone {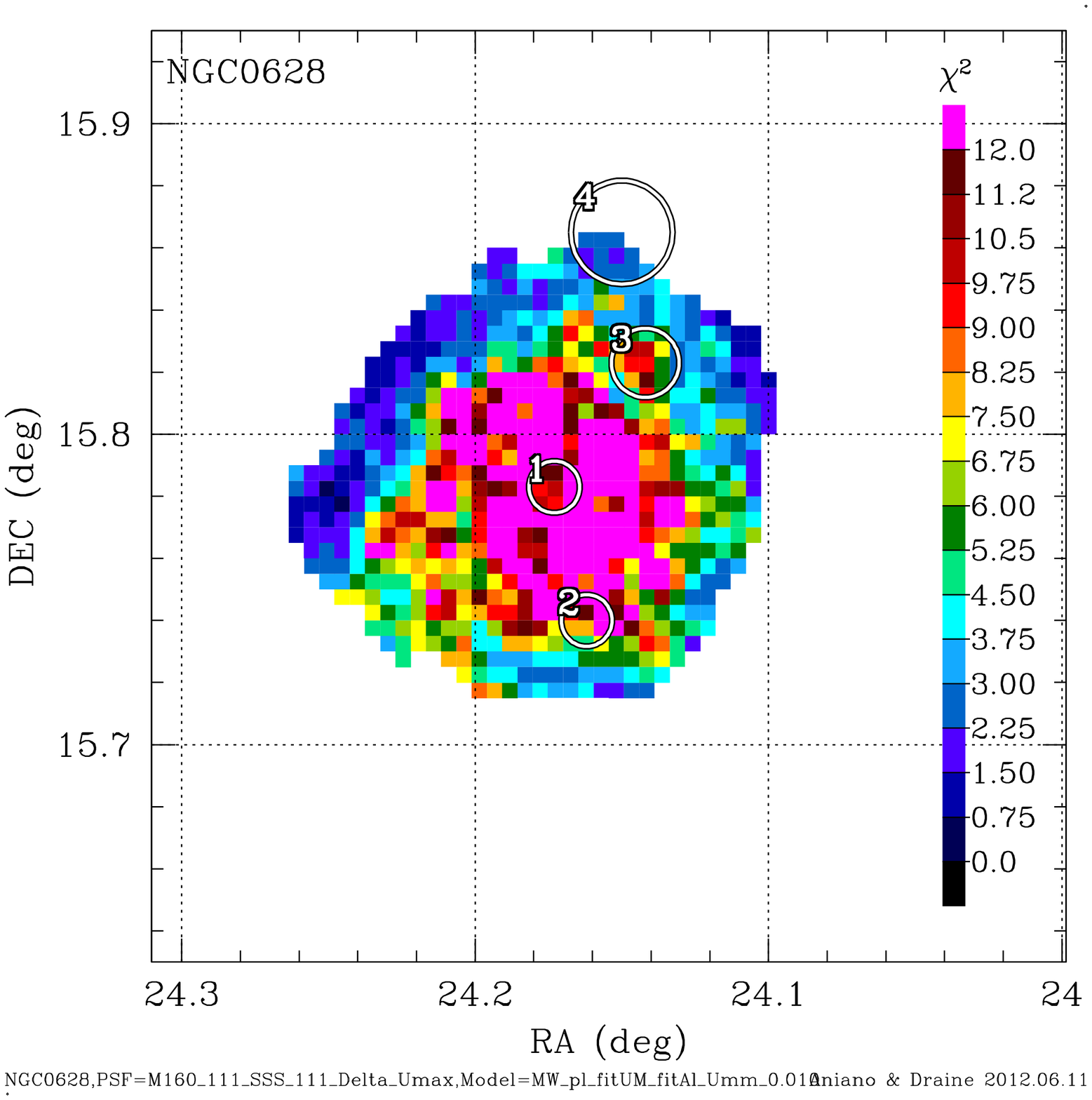}
\renewcommand \RoneCtwo {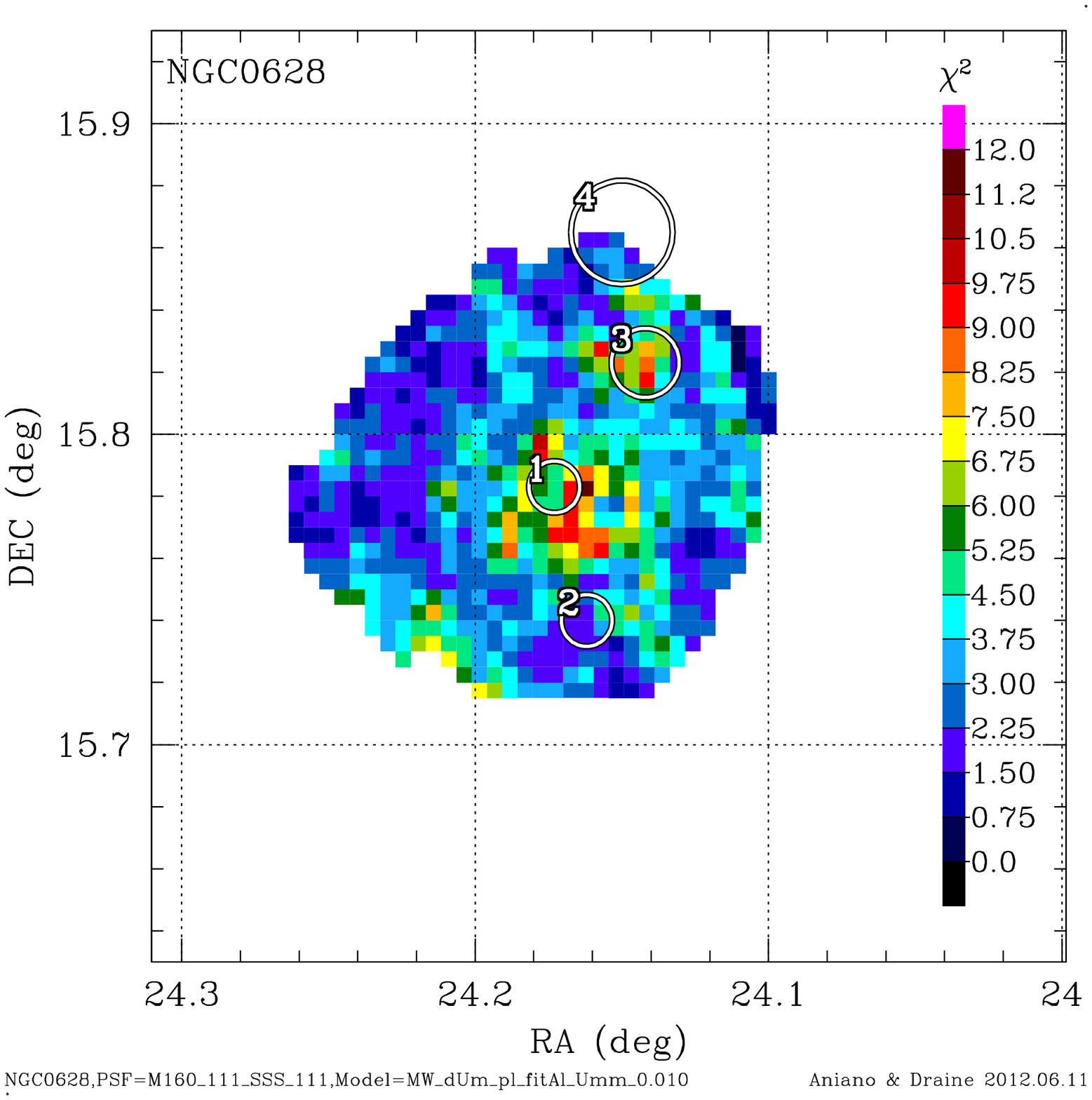}
\renewcommand \RoneCthree {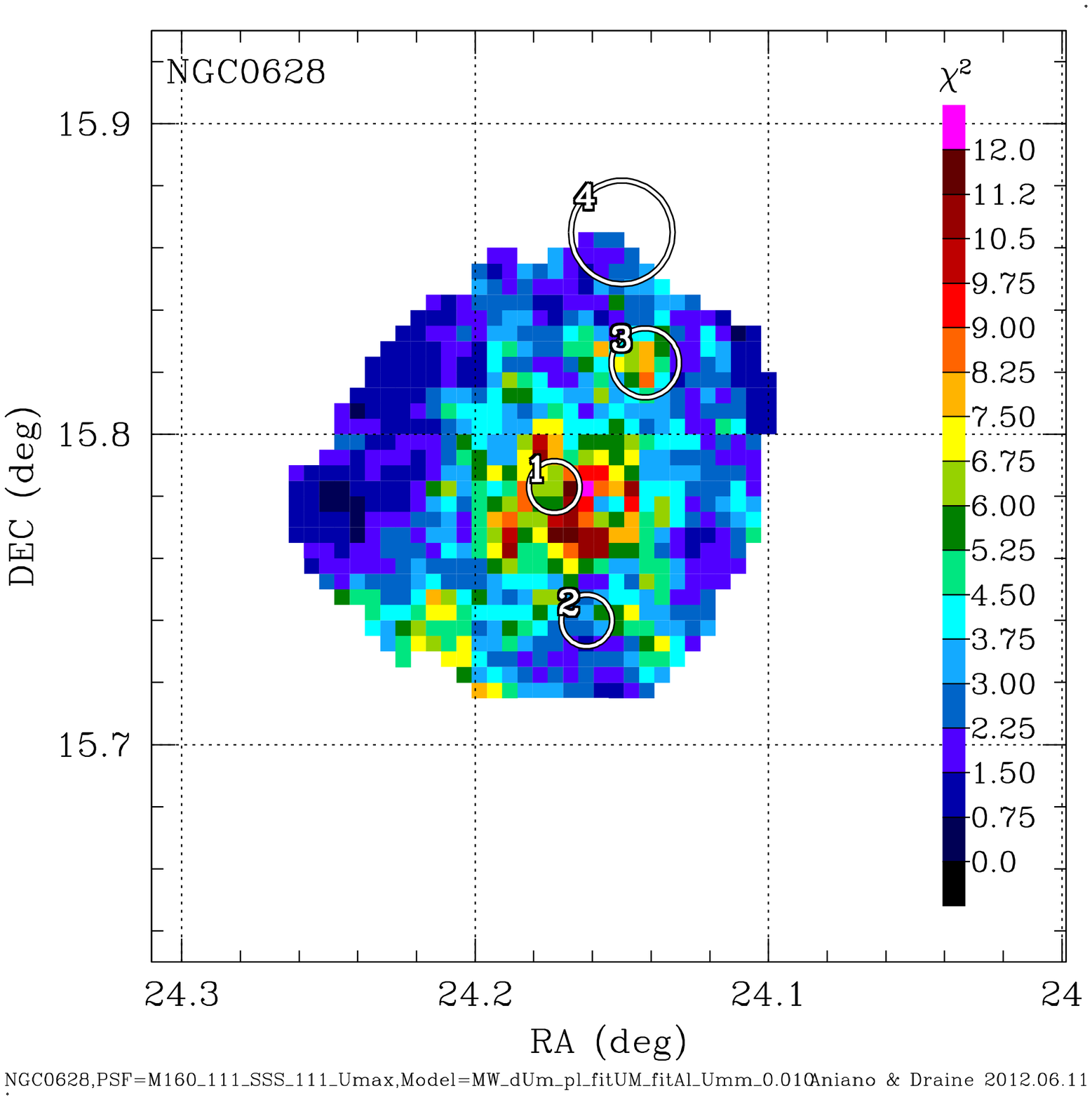}
\renewcommand \RtwoCone {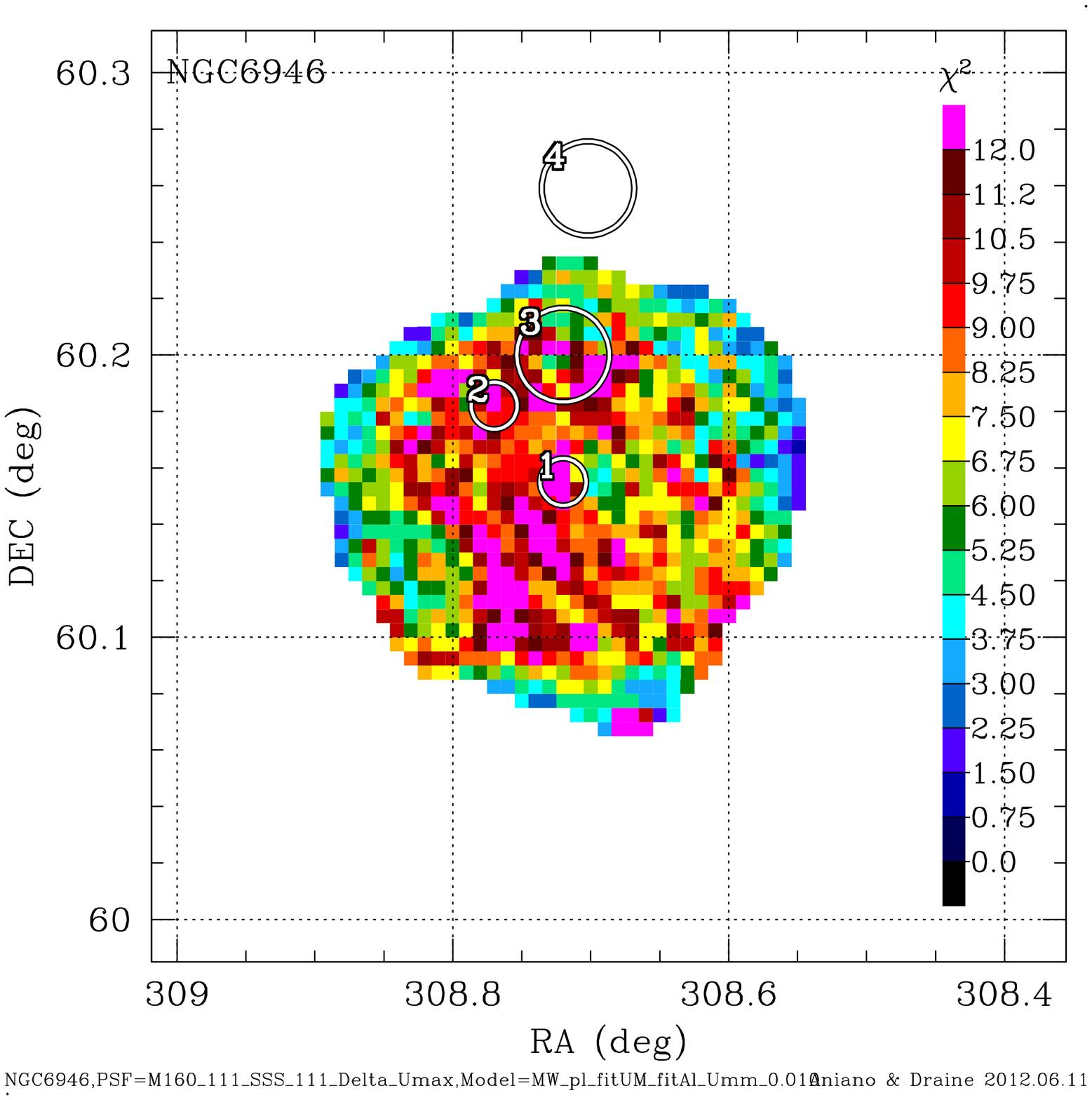}
\renewcommand \RtwoCtwo {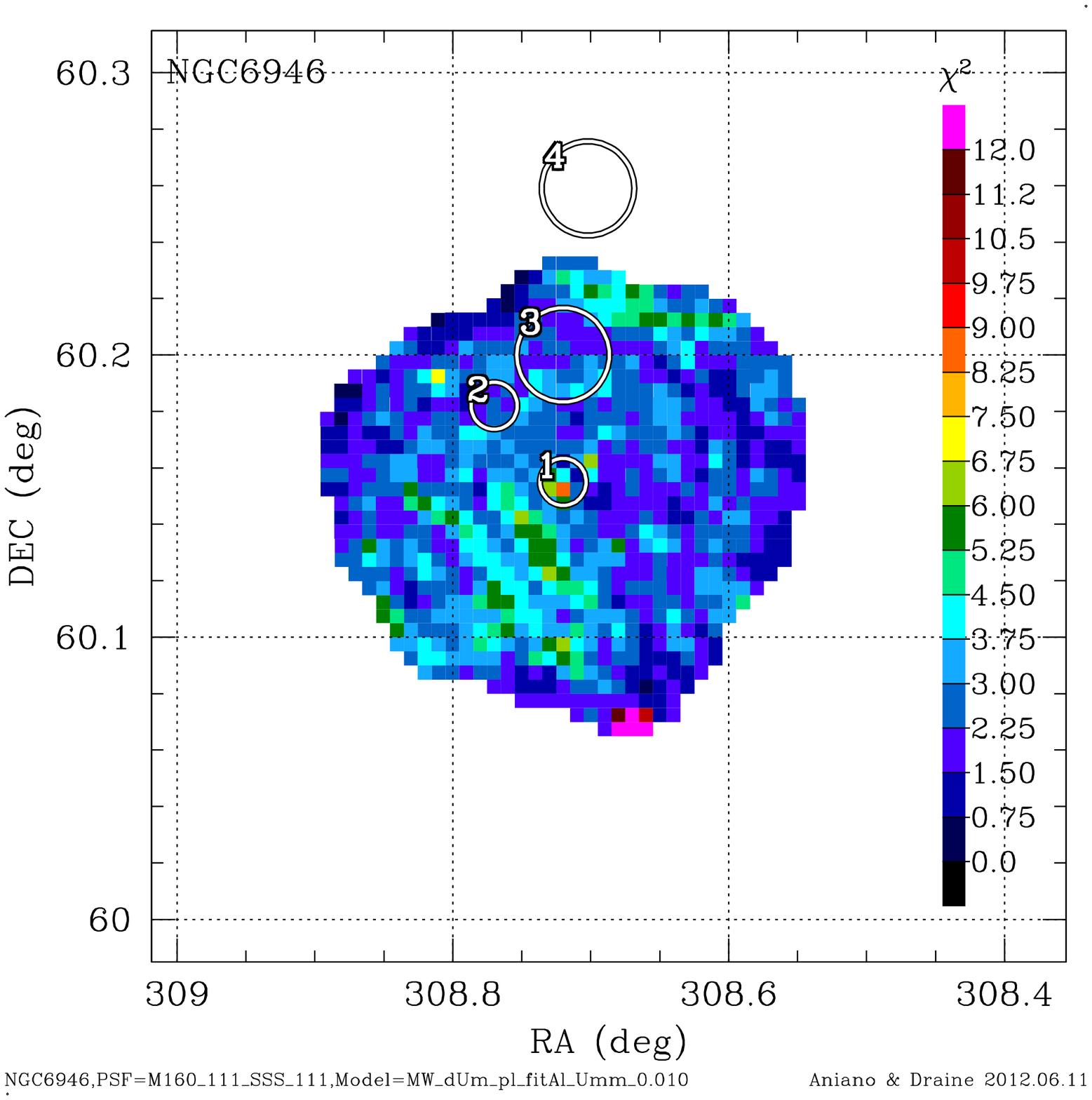}
\renewcommand \RtwoCthree {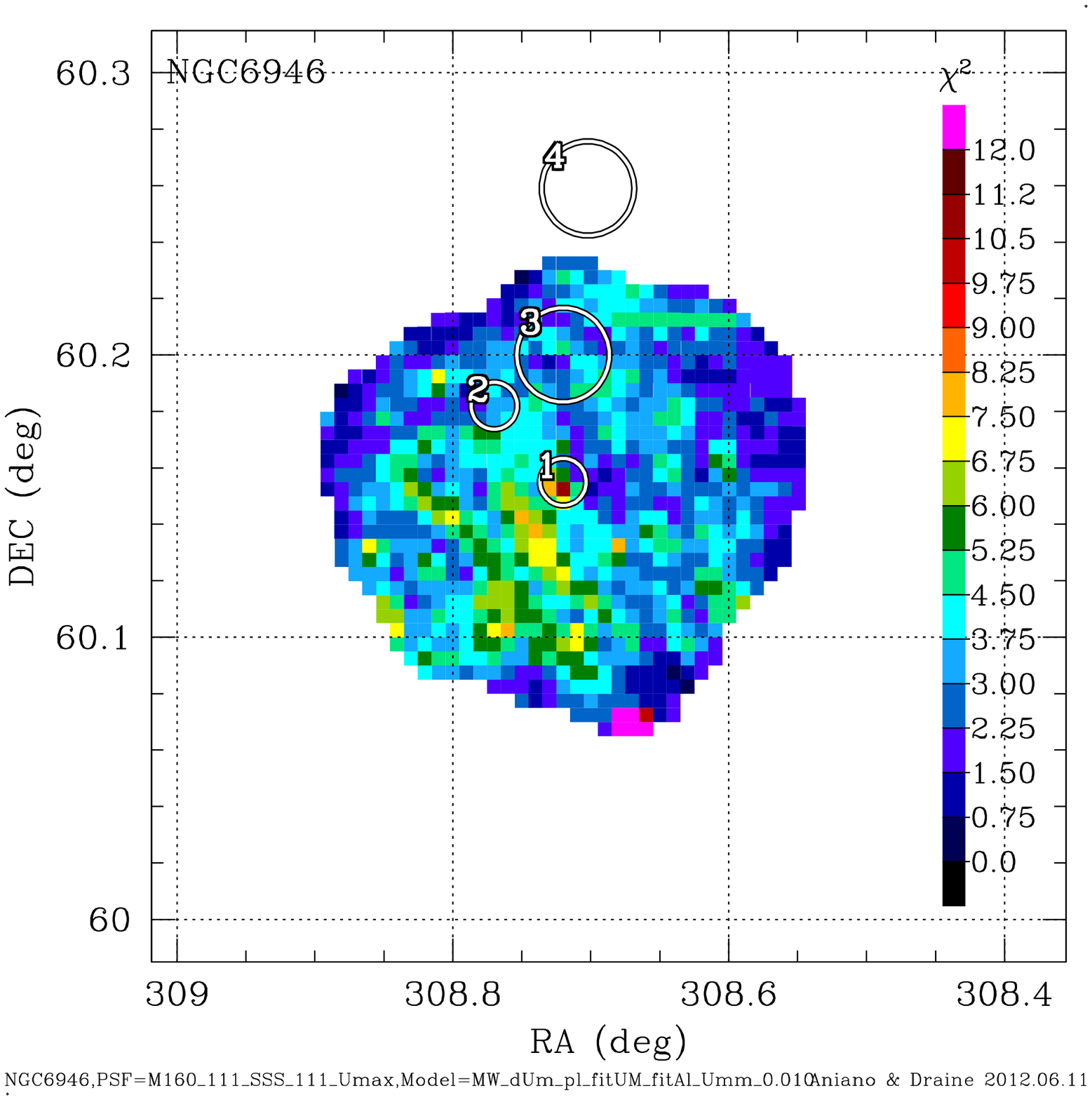}
\begin{figure} 
\centering 
\begin{tabular}{c@{$\,$}c@{$\,$}c} 
\FirstLast
\SecondLast
\end{tabular} 
\vspace*{-0.5cm}
\caption{\footnotesize Comparison of $\chi^2$ (goodness of fit) of
  three different starlight distribution schemes.  In the left column
  the starlight distribution is of the form proposed by
  \citet{Dale+Helou_2002} and recently favored by
  \citet{Galliano+Hony+Bernard+etal_2011}: a power law between $\Umin$
  and $\Umax$ (both adjusted) without the delta function contribution
  at $\Umin$. The fit for this starlight distribution is poor.  In the
  center column the starlight distribution is of the form used in the
  present modeling: a power law between $\Umin$ and $\Umax=10^7$, plus
  a delta function contribution at $\Umin$.  This fit produces
  excellent results.  In the right column the starlight distribution
  is a power law between $\Umin$ and $\Umax$ (both adjusted),
  including a delta function contribution at $\Umin$. Although the
  right panels have one extra degree of freedom ($\Umax$), the fit
  quality is similar to the middle panels.  The top row is for NGC~628
  and the bottom row is for NGC~6946.  The power law exponent $\alpha$
  is treated as an adjustable parameter in all the fits.  \label{fig:chicomp}}
\end{figure} 

We found that the $\chi^2$ values of the distribution used in the
current work are significantly smaller than the ones found using the
\citet{Galliano+Hony+Bernard+etal_2011} distribution.  Therefore, we
recommend adding the ``delta function'' component at $\Umin$. Since
the fit does not improve significantly by allowing $\Umax$ to be
fitted, we recommend fixing it to $\Umax=10^7$, leading to a simpler,
more robust fit.  It should be noted that the
\citet{Galliano+Hony+Bernard+etal_2011} study is based on the dust in
the Large Magellanic Cloud, and in our present work we use NGC~628 and
NGC~6946, two galaxies with metallicities closer to the Milky Way
metallicity. Additionally, due to the proximity of the LMC, each pixel
covers a physical area much smaller than our modeling pixels.


\section{\label{sec:discussion}Discussion}

\subsection{Dust Mass Estimation}

Dust mass estimates are, of course, model-dependent.  Above we have
estimated the dust mass using a specific dust model, the DL07
silicate-graphite-PAH model, with an assumed parametric form for the
starlight intensities heating the dust.  
The physical dust model and the ansatz for the distribution of starlight intensities
together appear to 
successfully reproduce the observed SEDs, and to
give reliable estimates of the dust masses.  

As discussed in \S \ref{sec:dustmodel}, the far-infrared opacity of
the amorphous silicate originally put forward by
DL84 was subsequently adjusted slightly
\citep{Li+Draine_2001b} to improve agreement with the far-infrared and
submm emission observed for the local high-latitude dust.  Thus the
DL07 model uses dust properties that can reproduce the observed
FIR-submm emission from the Milky Way (MW) cirrus with a single
starlight heating intensity -- no ``cold dust'' is needed for the MW
cirrus.  In this dust model, the opacities are assumed to be
independent of the dust temperatures.

Here we see that the same dust model, with suitable adjustment of the
starlight assumed to be heating the dust, is able to reproduce the
emission from NGC\,628 and NGC\,6946 out to 500\um, without
introducing any ``cold dust'' component.  The modeling performed at
MIPS160 resolution, using all the IRAC, MIPS, PACS, and SPIRE
cameras, in addition to being able to reproduce the observed SED,
gives dust masses that are in line with the expected dust/H
mass ratios for these galaxies: the dust/H mass ratio images in
Figures \ref{fig:ngc0628-2} and \ref{fig:ngc6946-2}, at MIPS160
resolution, are smooth, and the global dust/H mass ratios
are $0.0082\pm0.0017$ and $0.0063\pm0.0009$ for NGC~628 and NGC~6946,
respectively.

Observed depletions in the local Milky Way indicate a dust/H mass
ratio $M_{\rm dust}/M_{\Ha} = 0.0091\pm0.0006$ \citep[][Table
  23.1]{Draine_2011a}.  Thus, a galaxy with a similar fraction of
interstellar heavy elements in dust would be expected to have \beq
\label{eq:Md/MH expected}
\frac{M_{\rm dust}}{M_{\Ha}} \approx 0.0091 \times 10^{[{\rm O}]}~~~,
\eeq
where $[{\rm O}]\equiv \log_{10}[({\rm O/H)}/({\rm O/H})_\odot]$.
Thus galaxies with heavy element abundances (and depletions)
similar to the local Milky Way
should have $M_{\rm dust}/M_\Ha\approx 0.009$.

NGC~628 and NGC~6946 appear to be mature star-forming galaxies that
would be expected to have interstellar metallicities similar to the
local Milky Way.  If so, then the values of $M_{\rm dust}/M_{\rm H}$
found by fitting a dust model to the observations appear to be in
excellent agreement with expectations.  We note that the H\,II region
oxygen abundances found by
\citet{Moustakas+Kennicutt+Tremonti+etal_2010} together with 
$(A_{\rm O})_\odot=12+\log_{10}({\rm O/H})_\odot=8.73$
\citep{Asplund+Grevesse+Sauval+Scott_2009} yield [O]$\,=-0.46$ and
$-0.36$ for NGC~628 and NGC~6946, and Eq.\ (\ref{eq:Md/MH expected})
would predict $M_\dust/M_\Ha=0.0032$ and 0.0040 for NGC~628 and
NGC~6946, respectively.  However, we suspect that the PT05 oxygen
abundance estimates may be biased low:
\citet{Moustakas+Kennicutt+Tremonti+etal_2010} list PT05 HII-region
oxygen abundances for 38 galaxies; the highest oxygen abundance is
$A_{\rm O}=8.59\pm0.11$ for NGC~4826.  It seems unlikely that none of
the galaxies in their sample have oxygen abundances that are solar or
supersolar.

\subsection{Dust Opacity in Molecular Gas and the Value of $\XCO$}

Both NGC~628 and NGC~6946 are rich in molecular gas, and the estimated
gas mass depends on the adopted value of $\XCO$.  In the present study
we have adopted $\XCOxx=4$, which, as discussed in
Section \ref{sec:gas}, is significantly larger than the
value $\XCOxx\approx2$ found for resolved CO clouds in the Milky Way.
The larger value of $\XCO$ adopted here may reflect the presence of
so-called ``dark gas'', diffuse $\HH$ with very low CO abundances,
which does not radiate effectively in either \ion{H}{1} 21-cm or
CO$\,J = 2 \rightarrow1$
\citep{Wolfire+Hollenbach+McKee_2010,Leroy+Bolatto+Gordon+etal_2011}.

However, if the dust opacity in molecular regions is actually larger
than the opacity in \ion{H}{1} gas, then the present approach -- where
we favor a value of $\XCO$ that minimizes small-scale structure in
maps of dust optical depth/H surface density -- will overestimate
$\XCO$.  \citet{Planck_molecular_clouds_2011}, using
measurements of submm emission by Planck, and $\NH$ inferred from
NIR reddening of stars
\citep{Pineda+Goldsmith+Chapman+etal_2010}, conclude
that the dust opacity per H nucleus in the Taurus molecular cloud is
larger than in the local diffuse \ion{H}{1} by a factor $R\approx
2.0\pm0.4$.  Such an enhancement in the far-infrared and submm opacity
might be a consequence of coagulation, which is expected to
increase the far-infrared and submm opacity 
\citep{Ossenkopf+Henning_1994,Stognienko+Henning+Ossenkopf_1995}. 
However, grain coagulation
could also flatten the NIR extinction curve, so that the value
of $\NH$ inferred from the $(J-H)$ and $(H-K)$ stellar colors might be an
underestimate.  It should also be noted that studies of the
Corona Austrina molecular cloud, using MIPS160/LABOCA870$\micron$ ratios
to determine the dust temperature, found a normal ratio of 870$\micron$
optical depth
to visual extinction \citep{Juvela+Pelkonen+Porceddu_2009}.

The dust model used here has been calibrated on dust in \ion{H}{1}
regions.  If the actual dust opacity per H nucleon in molecular clouds
is larger than in \ion{H}{1} clouds by a factor $R$, then the actual
value of $\XCO$ in NGC~628 and NGC~6946 would be $\XCOxx=4/R$.  A
value of $R\approx2$ would then bring us into agreement with the
$\XCOxx\approx2$ inferred from other estimators (virial
mass estimates, $\gamma$-ray emission) of molecular mass.

\subsection{Dust Mass Estimates from Single-Temperature Fits}

\citet{Skibba+Engelbracht+Dale+etal_2011} used a single dust
temperature $T_\dust$ with an assumed
$\kappa_\nu\propto\lambda^{-1.5}$ opacity to fit the 70--500$\micron$
photometry from MIPS and SPIRE.  The opacity at $500\micron$ was taken
to be that of the DL07 dust model.  They estimated $T_\dust=24.0\K$,
$M_\dust=10^{7.03\pm0.08}\Msol$ for NGC\,628, and $T_\dust=26.0\K$,
$M_\dust=10^{7.47\pm0.08}\Msol$ for NGC\,6946.  The dust masses
estimated by \citet{Skibba+Engelbracht+Dale+etal_2011} for these two
galaxies are smaller than the dust masses found here (see Table
\ref{tab:galaxy properties}) by factors of 3.3 and 3.6, respectively.
This is the result of using a single dust temperature to try to
reproduce emission from 70--500\um.  With the more realistic
assumption of a distribution of dust temperatures, a small amount of
warmer dust can provide much of the 70\um emission, thus requiring an
increased mass of ``normal'' temperature dust to account for the
emission at $\lambda\gtsim 160\micron$.  A range of dust temperatures
is of course expected from both spatial variations in the starlight
intensity heating the dust, and the fact that a grain model with more
than one dust composition, and a broad range of grain sizes, will have
a distribution of temperatures even in a single radiation field.
Single-temperature dust fits, if constrained by emission at 70\um,
will not provide a reliable estimate of the dust mass.
In a separate work (Aniano \& Draine 2012, in prep.) 
we discuss the bias introduced
when the DL07 SED is approximated by a (single or dual) temperature
blackbody multiplied by a power law opacity.

\subsection{Radial Gradients in Dust and Starlight}

In both NGC~628 and NGC~6946, we find that the dust/H mass
ratios outside the nucleus vary slowly, with little indication of a
decrease in the dust/H  ratio as one moves outward (see
Figures \ref{fig:ngc0628-1}l and \ref{fig:ngc6946-1}l).  In NGC~6946, however,
the dust/H mass ratio appears to have a pronounced minimum at
the center.  We interpret this as due to overestimation of the gas
mass in this region: we employ a single value of $\XCOxx=4$ for this
galaxy, but \citet{Donovan_Meyer+Koda+Momose+etal_2011}, using virial
mass estimates, find $\XCOxx=1.2$ for the giant molecular clouds
(GMCs) in the central 5 kpc.  Because the molecular gas dominates near
the center, our use of a higher value of $\XCO$ implies an
overestimate of the gas mass, resulting in an underestimate of the dust/H
ratio.  We therefore suspect that the central minimum in the
dust/H ratio in Figure \ref{fig:ngc6946-2} is entirely an artifact
of using a value of $\XCO$ that is too large for the central
region.  If $\XCOxx=1.2$ is
appropriate in the center of NGC~6946, the gas surface density will be
lowered by a factor $\sim 1.2/4$, and the dust/H mass ratio
increased by a factor $\sim3$, from the central value $\sim0.004$ in
Figure 7 to the expected value $\sim 0.012$.  The present study of the
dust surface density therefore supports the finding by
\citet{Donovan_Meyer+Koda+Momose+etal_2011} of a low value of $\XCO$
in the center of NGC~6946.

In both NGC~628 and NGC~6946, the PAH abundance parameter $\qpah$ appears to
be quite uniform, with little evidence for
a radial decline (see Figures \ref{fig:ngc0628-2} and \ref{fig:ngc6946-2}).
Previous studies have shown that low-metallicity galaxies have low values of
$\qpah$ \citep[e.g.,][]{Engelbracht+Gordon+Rieke+etal_2005,
Draine+Dale+Bendo+etal_2007}; a galaxy with a negative radial gradient
in metallicity might then show a radial decline in $\qpah$.
\citet{Munoz-Mateos+Gil_de_Paz+Boissier+etal_2009} found radial declines in
$\qpah$ for a number of SINGS galaxies.
For both NGC~628 and NGC~6946, $\qpah$ appeared to be quite uniform out to a radius of
$\sim$$10\kpc$, in qualitative agreement with our findings here.
Future papers will examine radial trends in the KINGFISH sample.

The starlight intensity parameter $\Umin$ is presumed to characterize the
average starlight intensity in the diffuse interstellar medium.
The maps of $\Umin$ (Figs.\ \ref{fig:ngc0628-2}ghi and \ref{fig:ngc6946-2}ghi)
have a peak $\Umin\approx 4$ and $6$ 
in the central regions of
NGC~628 and NGC~6946, respectively, declining to
$\Umin\approx 0.3$ near the edge of the galaxy mask ($\sim$10\,kpc from the
center).  It is gratifying that $\Umin\approx 1$ at
$\sim$8\,kpc from the center, just as for our location in the Galaxy. 

The parameter $\fpdr$ is the fraction of the dust heating that is
produced by starlight intensities $U>100$.  Averaged over the full
galaxy, $\langle f_\PDR\rangle =0.12$ and 0.14 
for NGC~628 and NGC~6946,
respectively, but
both galaxies have hot spots where $f_\PDR$ is much higher, with peak
values of $\sim$$0.30$ (see Figs.\ \ref{fig:ngc0628-2}jkl and
\ref{fig:ngc6946-2}jkl).  With the $\sim$635\,pc resolution of the SPIRE250
camera at the 7.2 Mpc distance of NGC~628, these hot spots presumably
correspond to regions of very active star formation.

\subsection{Interpretation of the Starlight Heating Parameters $\alpha$ and $\gamma$}

\renewcommand \RoneCone {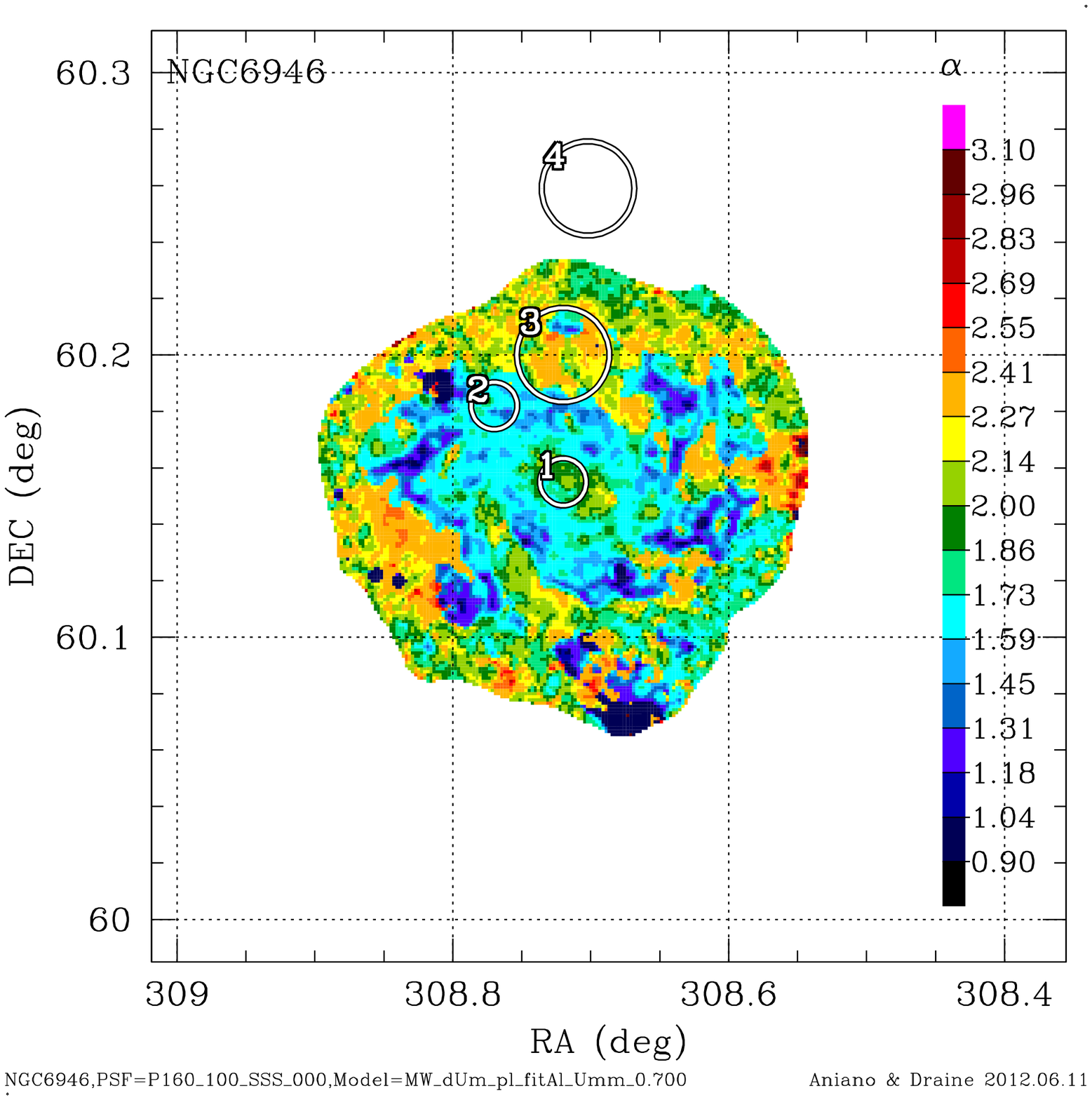}
\renewcommand \RoneCtwo {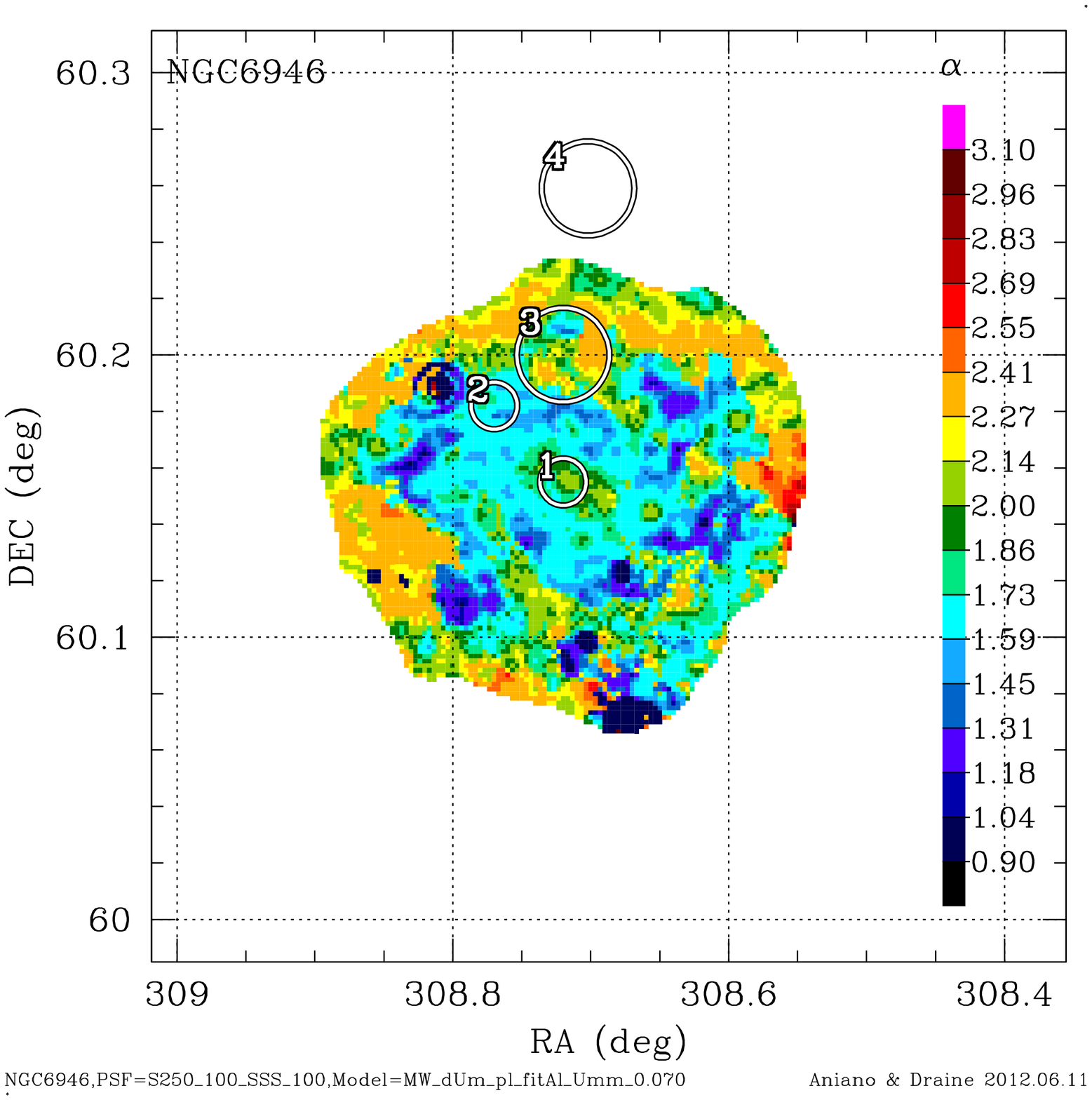}
\renewcommand \RoneCthree {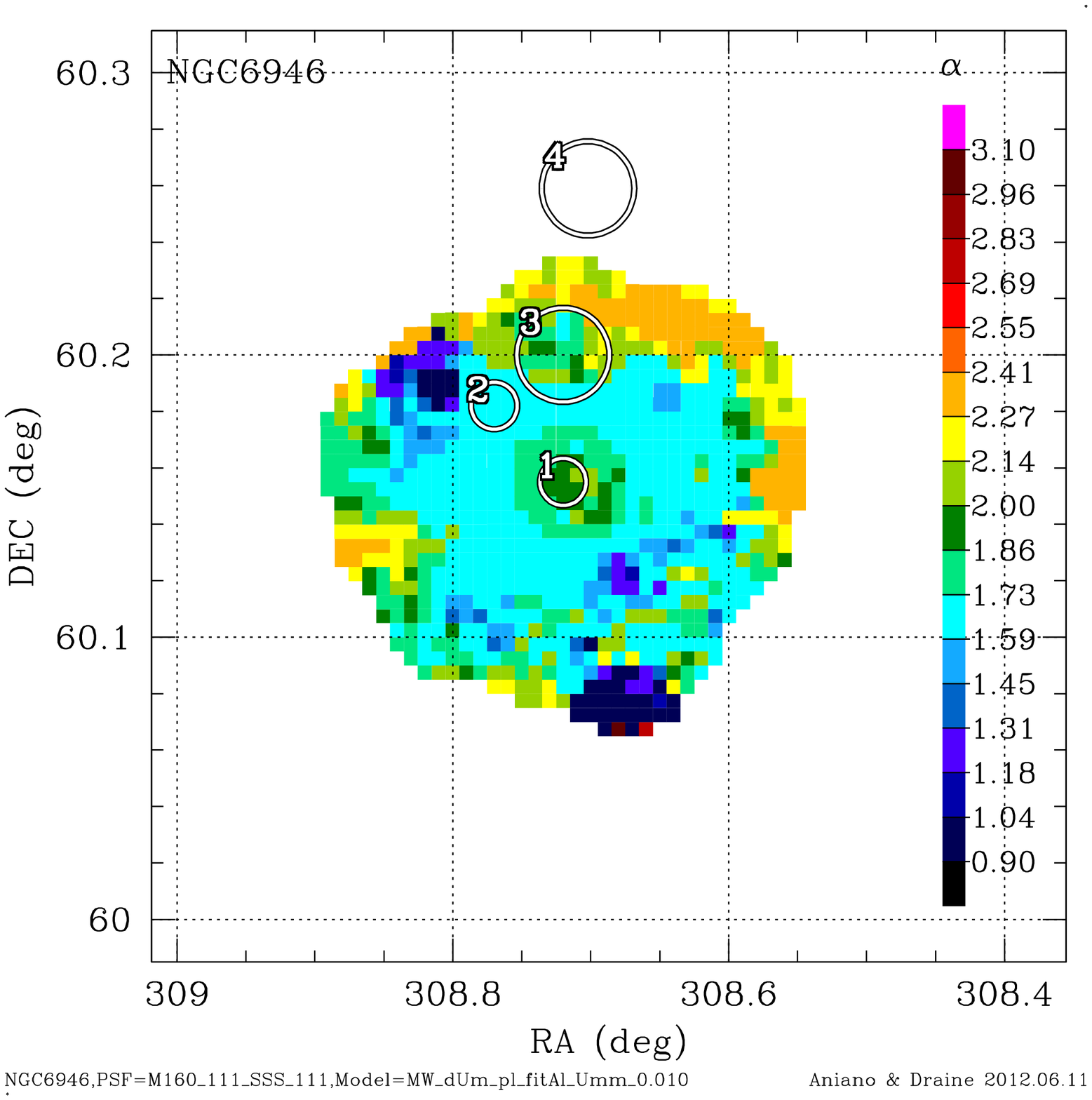}
\renewcommand \RtwoCone {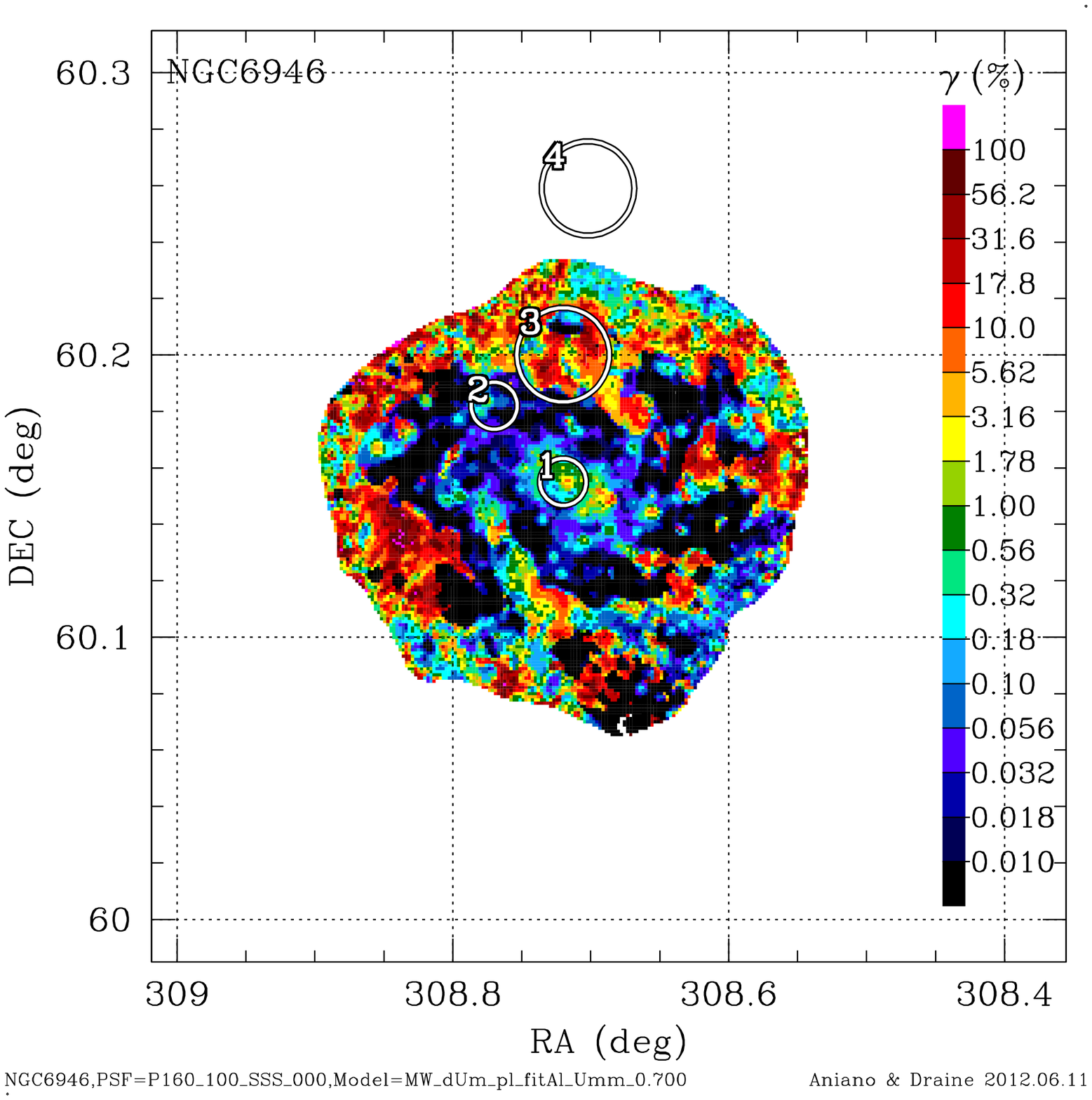}
\renewcommand \RtwoCtwo {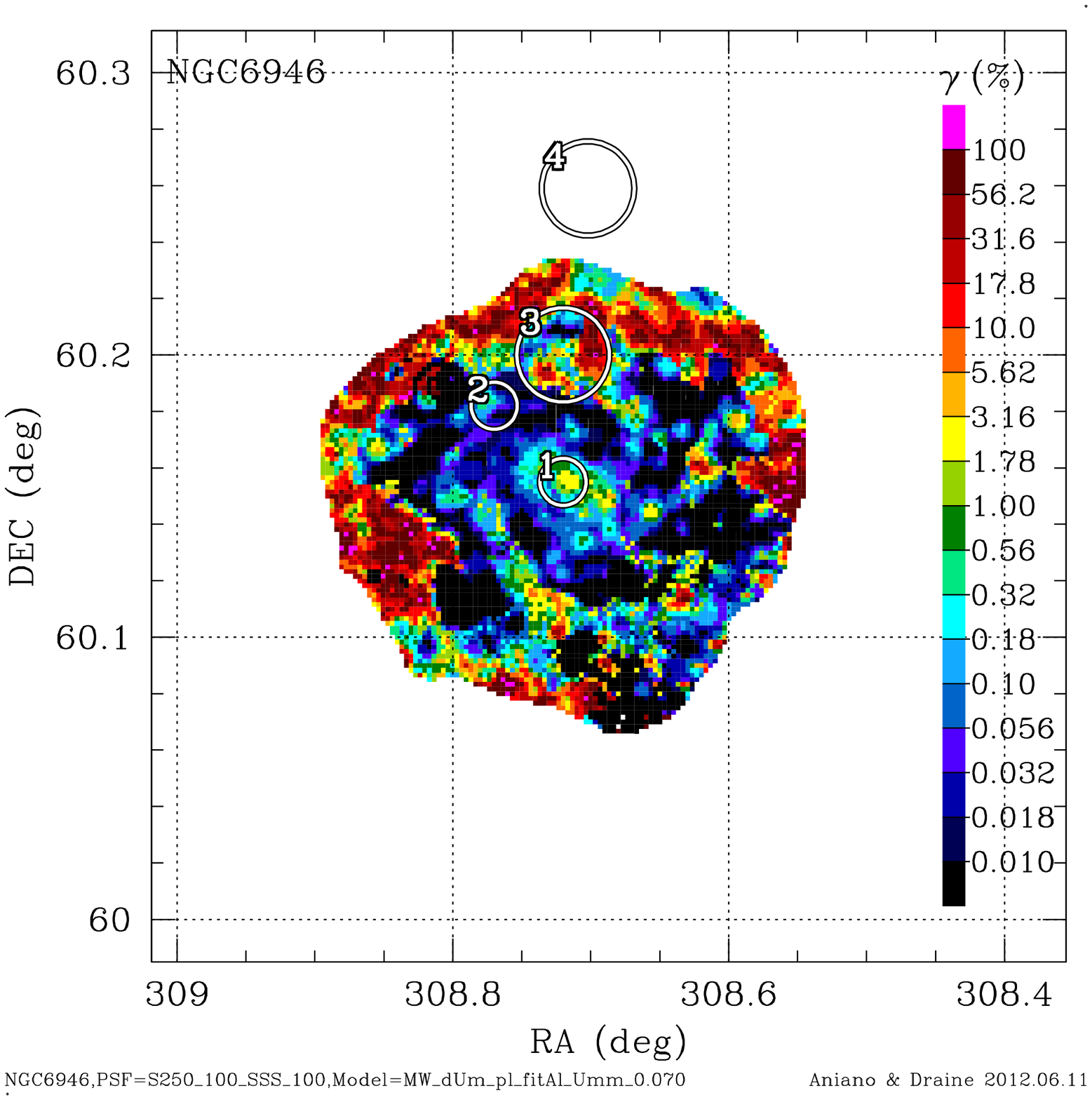}
\renewcommand \RtwoCthree {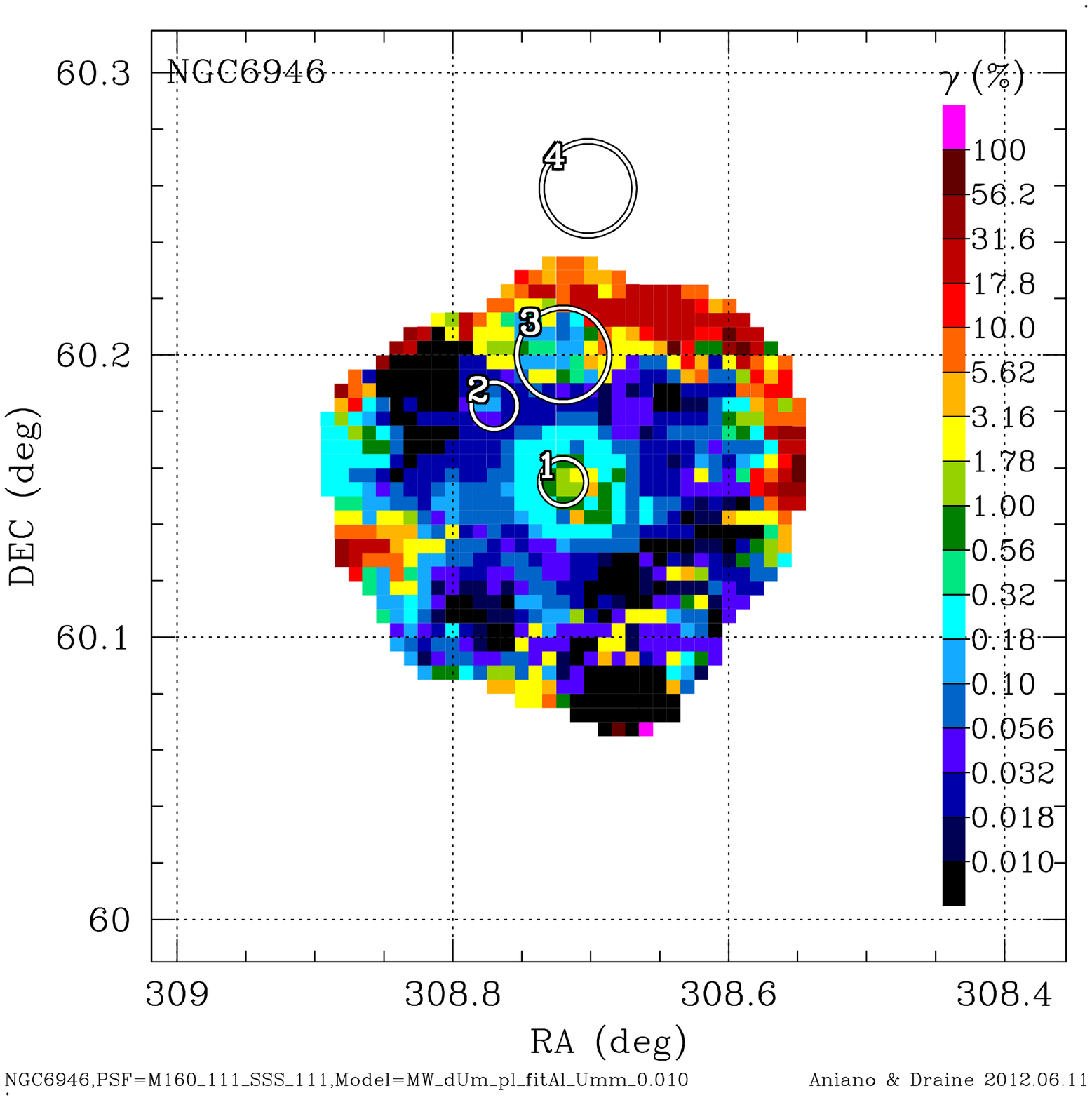}
\renewcommand \RthreeCone {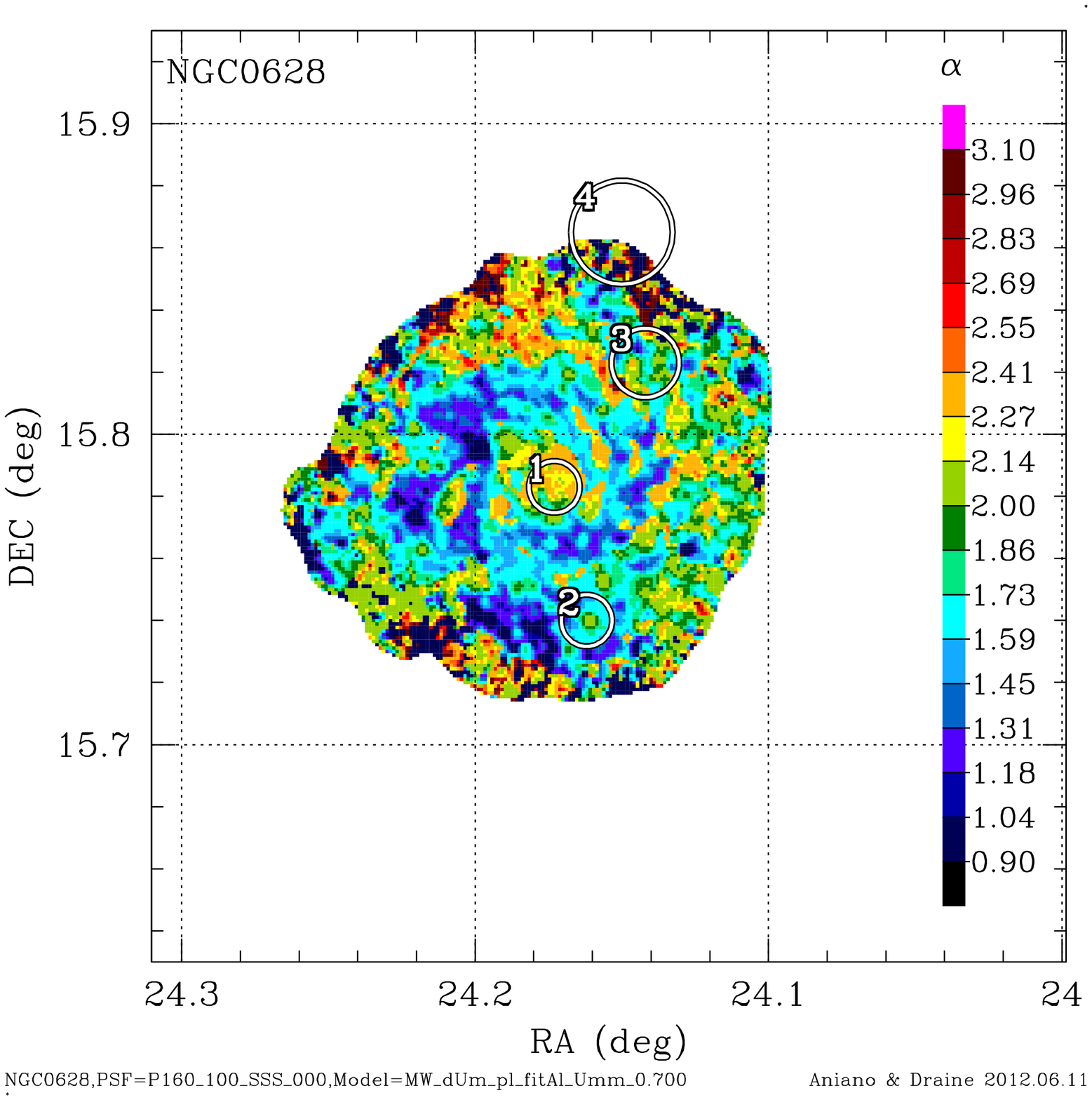}
\renewcommand \RthreeCtwo {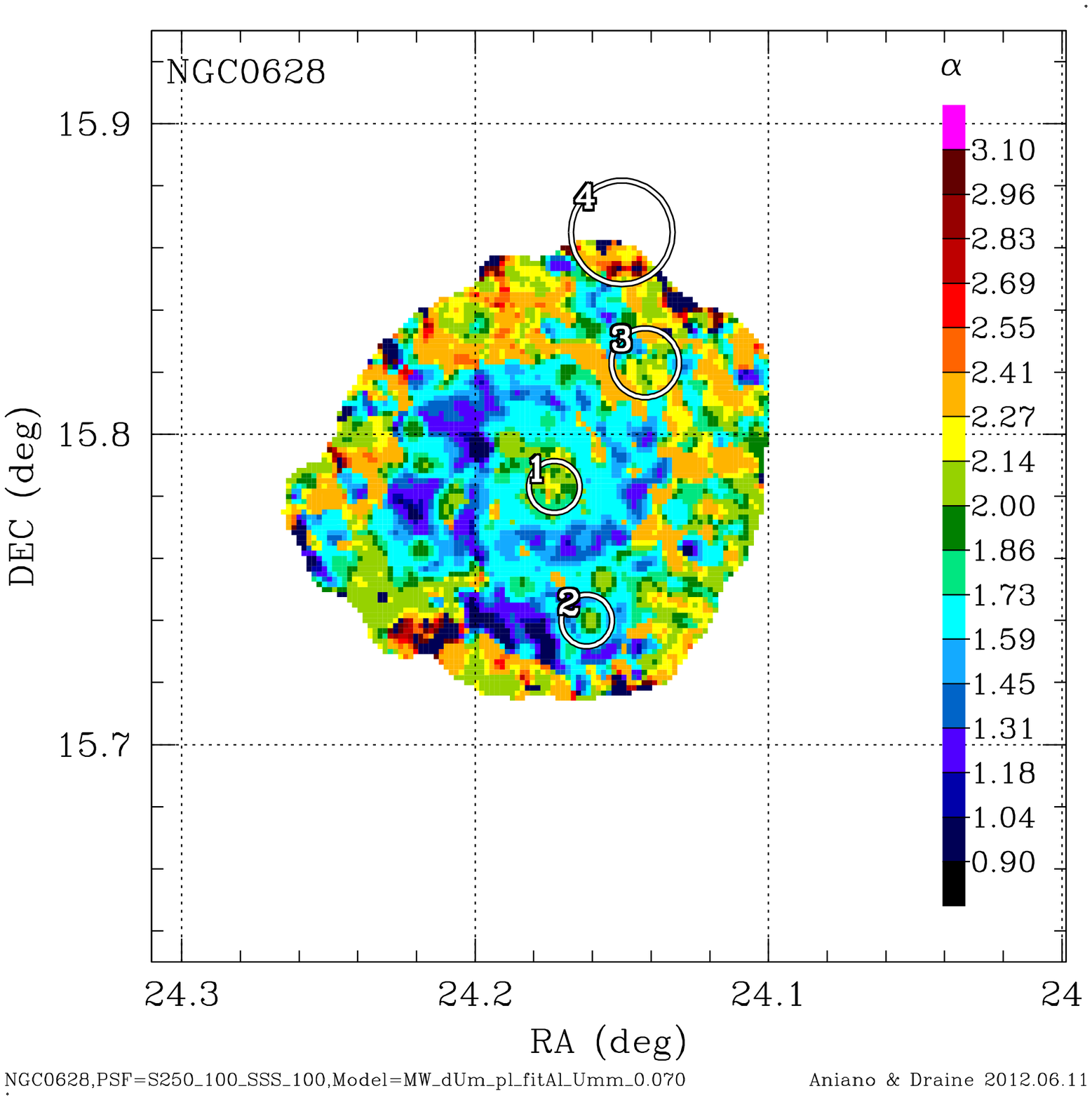}
\renewcommand \RthreeCthree {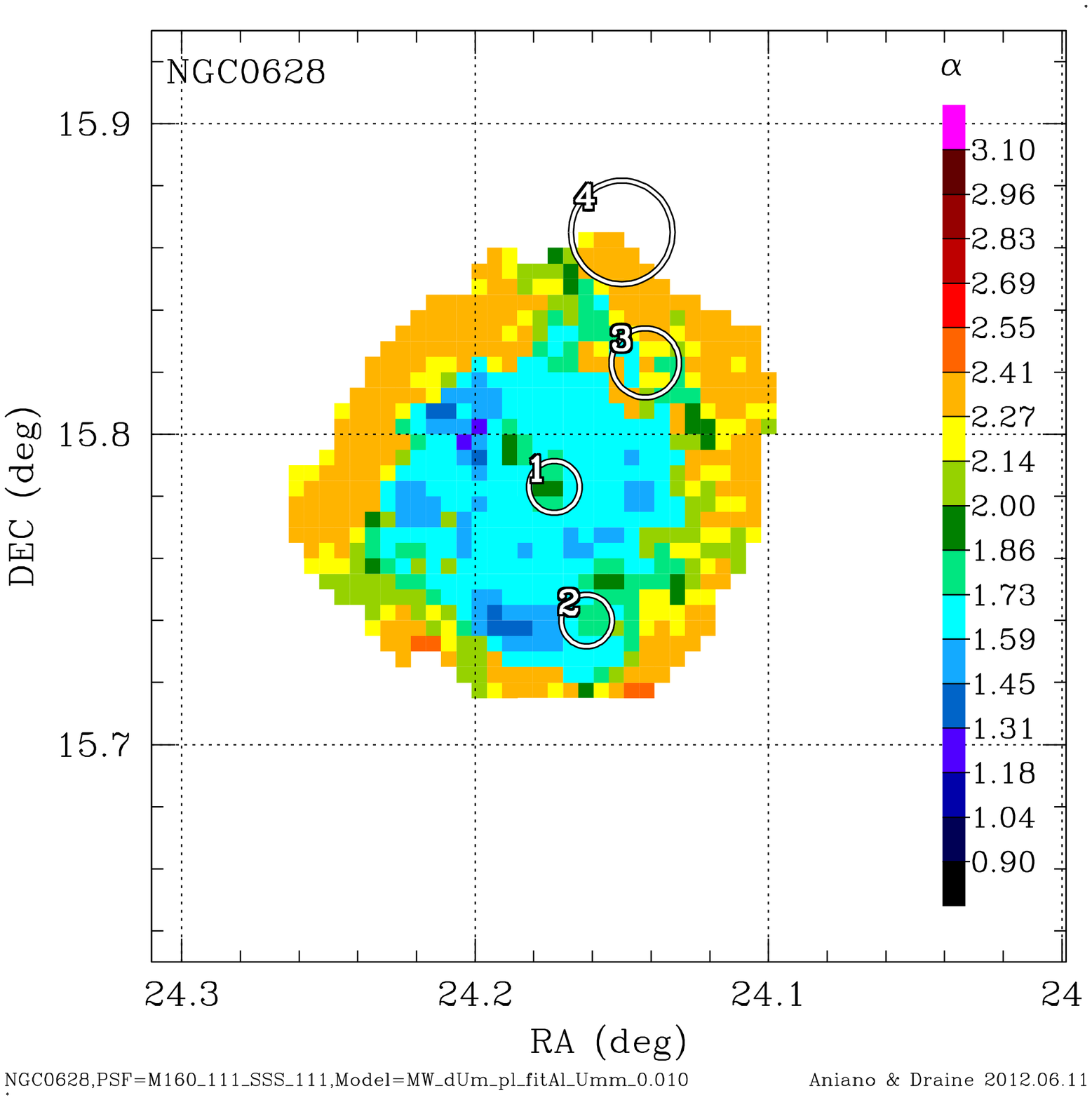}
\renewcommand \RfourCone {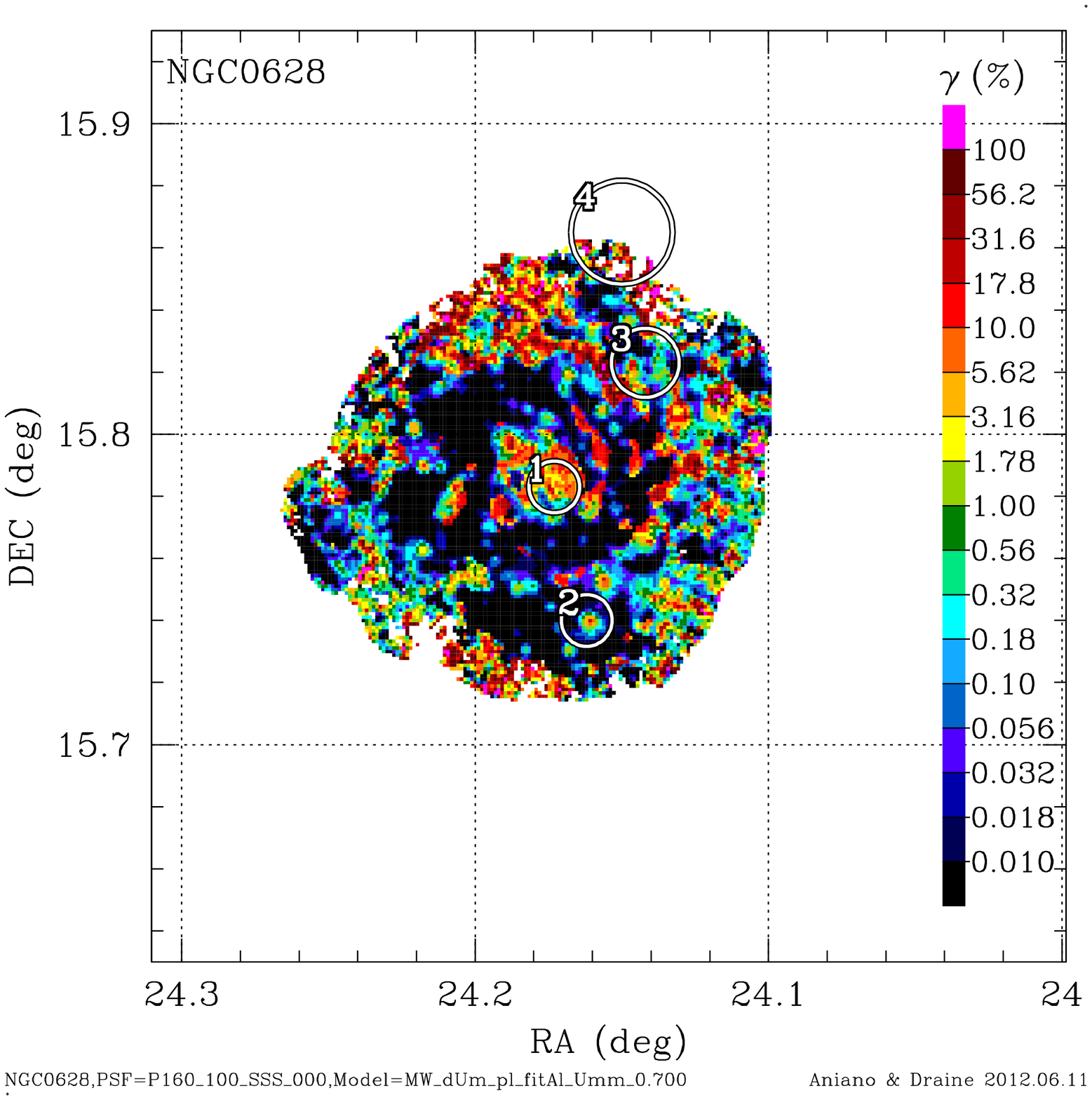}
\renewcommand \RfourCtwo {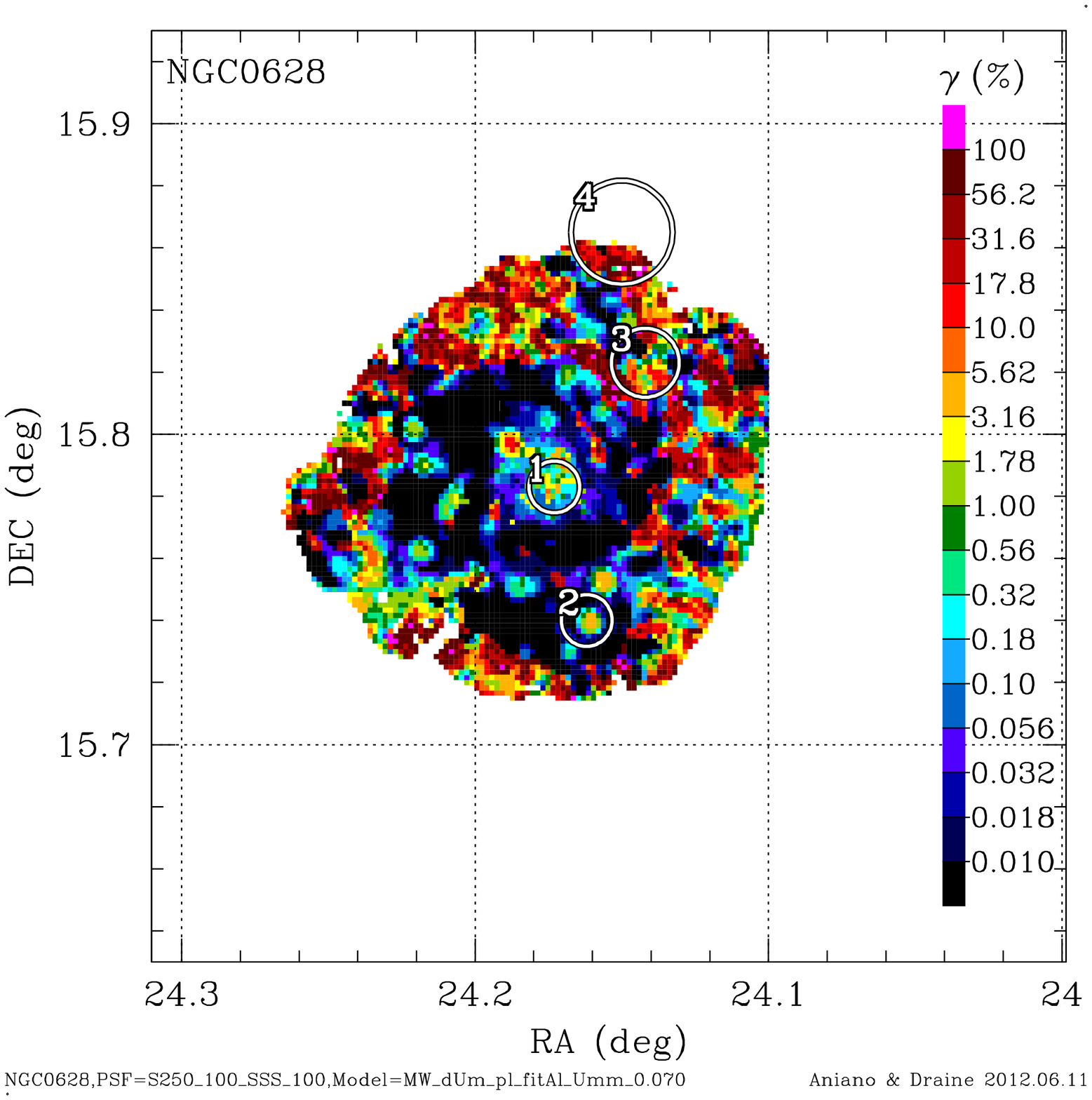}
\renewcommand \RfourCthree {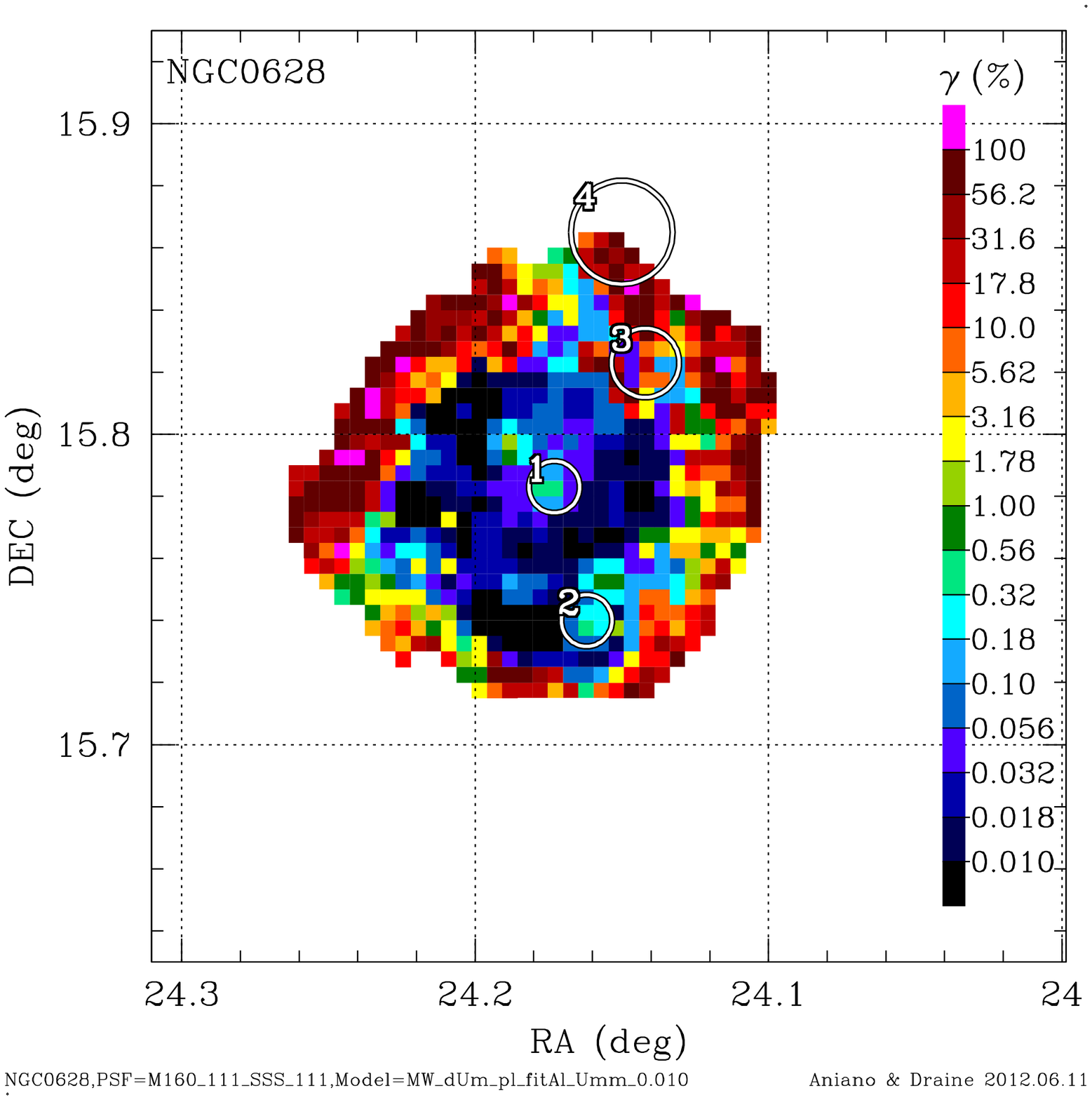}
\ifthenelse{\boolean{make_heavy}}{ }
{ \renewcommand \RoneCone    {No_image.eps}
\renewcommand \RtwoCone    {No_image.eps}
\renewcommand \RthreeCone {No_image.eps}
\renewcommand \RfourCone {No_image.eps}
\renewcommand \RoneCtwo    {No_image.eps}
\renewcommand \RtwoCtwo    {No_image.eps}
\renewcommand \RthreeCtwo {No_image.eps}
\renewcommand \RfourCtwo {No_image.eps}}
\begin{figure}
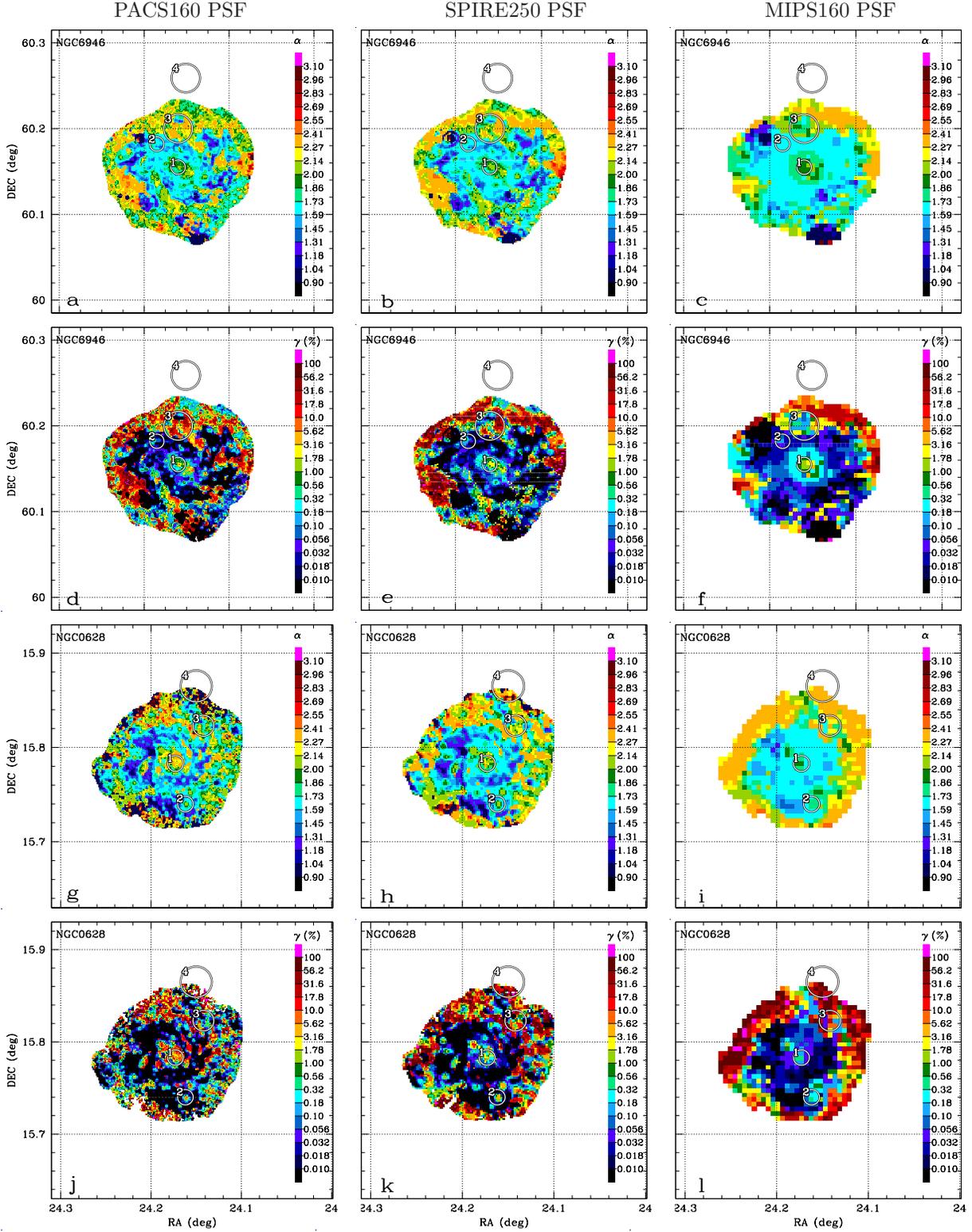
 
\centering 
\begin{tabular}{c@{$\,$}c@{$\,$}c} 
\footnotesize PACS160 PSF & \footnotesize SPIRE250 PSF & \footnotesize MIPS160 PSF \\
\FirstNormal
\SecondNormal
\ThirdNormal
\FourthLast
\end{tabular}
\vspace*{-0.5cm}
\caption{\footnotesize\label{fig:alpha-gamma}
Characterization of the starlight power-law component, at the resolution of PACS160 (left), SPIRE250 (center), and MIPS160
  (right).  
Row 1 and 3: power-law exponent $\alpha$ map for NGC~628 and NGC~6946 respectively.  
Row 2 and 4: dust mass fraction in the power-law component $\gamma$ map for NGC~628 and NGC~6946 respectively.}
\end{figure} 

As discussed in \S \ref{DustHeating}, a fraction $\gamma$ of the dust mass is taken to be
heated by a power-law distribution of starlight intensities between
$\Umin$ and $\Umax$ with $dM/dU\propto U^{-\alpha}$, 
and the remaining fraction  $(1-\gamma)$ of the dust mass is heated by a radiation field with intensity $\Umin$.
Dust heated by this simple parameterization of the starlight intensities is quite successful in reproducing the observed
  SED (see the SEDs in Figures \ref{fig:ngc0628-4}c,f;
  \ref{fig:ngc0628-5}c,f,i,l; \ref{fig:ngc6946-4}c,f; and \ref{fig:ngc6946-5}c,f,i,l).
  
 \citet{Dale+Helou+Contursi+etal_2001} discussed two ideal cases that have analytical predictions for the value of $\alpha$: 
 a single star in a homogeneous diffuse medium, and a dark cloud.
The first case, with $U \propto r^{-2}$, has $dM_{\dust}/{dU} \propto U^{-2.5}$, i.e., $\alpha = 2.5$. 
In this case, $\Umax$ would be very large, $\gtsim 10^{10}$, the heating rate at which dust grains would vaporize.
In the second case, a slab where the heating intensity is primarily attenuated by dust absorption,
 one would get $dM_{\dust}/{dU} \propto U^{-1}$, i.e., $\alpha = 1$, and $\Umax$ would be set by the starlight intensity at the edge of the slab. 
This case might correspond to a photodissociation region, the edge of a dark cloud.
Different clouds in the pixel would have different values of $\Umax$, so their superposition could be approximated as a power-law with a slightly larger value of $\alpha$.
We expect dust in a real galaxy to have $1 \lesssim \alpha \lesssim 2.5$.

Consider for the moment the fraction $\gamma$ of dust with $dM/dU\propto U^{-\alpha}$. 
The power radiated is $dL\,\propto\, U\, dM\, \propto\, U^{2-\alpha}\, d\log U$.
A value of $\alpha = 2$ would have uniform dust luminosity per unit interval in $\log U$. 
A value $\alpha<2$ would concentrate the dust luminosity $L$ in the high radiation field regions $(U\approx \Umax)$, 
and a value $\alpha >2$ would have $L$ dominated by dust with $U\approx \Umin$.

Figure \ref{fig:alpha-gamma} shows the starlight power-law component parameters $\alpha$ and $\gamma$.
The power-law index $\alpha$ has a preferred
value $\alpha \approx 1.6$ in the bright regions. 
In these regions, the power-law distribution is representing the heating of dust in high-$U$ regions, e.g., photodissociation regions near OB stars.
There are also regions  with $\alpha \approx 2.4$. 
In these regions, most of the dust luminosity is concentrated in regions with $U \approx \Umin$, 
and the power law component is essentially making the $U=\Umin$ component broader, i.e., is allowing for variations in the starlight intensities in the diffuse ISM.

We do not attach great physical significance to the parameters $\gamma$ and $\alpha$. 
The more physically meaningful quantities are the mass-weighted mean starlight intensity $\Ubar$, and $f_\PDR$, the fraction of the dust luminosity that originates in regions with $U>10^2$.
In Figures \ref{fig:ngc0628-2} and \ref{fig:ngc6946-2}, one sees that $f_\PDR > 0.05$ over most of the galaxy mask for both galaxies.


\section{\label{sec:summary}Summary}

Using Spitzer and Herschel data, we perform a resolved study of the 
dust physical parameters, and of the starlight heating the dust,
in the nearby galaxies NGC~628 and NGC~6946.
We employ the DL07 dust model, having amorphous silicate grains and carbonaceous
grains, including PAHs (see \S \ref{sec:dustmodel}). 
The model has a distribution of grain sizes, and we allow for a
distribution of starlight intensities heating the dust within each pixel.
 The model assumes a
frequency-dependent opacity that reproduces the observed emission
spectrum of high-latitude dust.

In order to perform resolved studies of the galaxies, it is important
to convolve all the images into a common PSF. This is achieved using
the convolution kernels generated by
\citet{Aniano+Draine+Gordon+Sandstrom_2011}. We perform our dust
modeling using all appropriate combinations of PSFs and cameras. Table
\ref{tab:resolutions} lists some of the resolutions and cameras used
in this work.

Our principal findings are as follows:
\begin{enumerate}

\item The dust model is quite successful in reproducing the observed
  SED over the full wavelength range 6--500$\micron$ where dust
  emission dominates (see the SEDs in Figs.\ \ref{fig:ngc0628-4}c,f;
  \ref{fig:ngc0628-5}c,f,i,l; \ref{fig:ngc6946-4}c,f; and
  \ref{fig:ngc6946-5}c,f,i,l; and the model/observed maps in
  Figs.\ \ref{fig:ngc0628-3}c,f and \ref{fig:ngc6946-3}c,f).  The dust
  model reproduces the emission out to 500$\micron$ without
  introduction of a ``cold dust'' component, and with no allowance for
  possible temperature-dependent dust opacities.  The DL07 dust
  opacities therefore appear to be consistent with both the observed
  emission from local ``cirrus'' at high galactic latitudes, and the
  observed SED from 500\,pc-sized regions of the ISM in NGC~628 and
  6946.  There is no indication that $T$-dependent opacities are
  needed to reproduce these data.

\item Maps of the dust/H mass ratio show it to be
  relatively uniform over both galaxies, outside of the nucleus (see
  Figs.\ \ref{fig:ngc0628-1}l and \ref{fig:ngc6946-1}l).  The derived
  dust/H mass ratio for these galaxies is consistent with
  that expected if the interstellar abundances in NGC~628 and NGC~6946
  are close to solar.  As near-solar abundances seem likely, this is
  strong support for the quantitative accuracy of the dust masses
  obtained by modeling the IR and FIR emission.

\item Figure \ref{fig:dustmass} shows how dust mass estimates depend
  on the camera and PSF used.  The ``gold standard'' is taken to be
  dust modeling done at MIPS160 resolution, using all cameras, with
  Scanamorphos \citep{Roussel_2012} processing of the PACS data.
  Relative to this standard, mass estimates done with smaller pixels
  (giving up the lowest resolution cameras) tend to be biased slightly
  high.  At SPIRE250 resolution, the bias is $\sim$$38\%$.
   Working at SPIRE350 resolution
  provides a good compromise between resolution and accuracy, with
  $\sim$$30\%$ errors for the total dust mass
  estimated for NGC~628 and NGC~6946.

\item Even though the dust modeling process is non-linear, resolved
  modeling is compatible with modeling using the
  global photometry of the galaxy. Differences in the inferred
  parameters are in most cases under 20\%.

\item In NGC~6946 the dust/H mass ratio calculated with a single value of
  $\XCOxx=4$ appears to have a minimum at the center (see
  Fig.\ \ref{fig:ngc6946-2}a,b,c).  This minimum is interpreted as an
  artifact due to overestimation of the molecular mass.  The dust
  surface densities found here therefore support the finding by
  \citet{Donovan_Meyer+Koda+Momose+etal_2011} of $\XCOxx\approx 1.2$
  in the center of NGC~6946.
  
\item The present fitting procedures employ six adjustable parameters
  for each pixel: $\Omega_\star$, $\Mdust$, $\qpah$, $\Umin$,
  $\alpha$, and $\gamma$ (we fix $\Umax=10^7$).  For NGC~628 and
  NGC~6946 we find that this approach provides substantially better
  fits than the approach advocated by
  \citet{Galliano+Hony+Bernard+etal_2011} who treat $\Omega_\star$,
  $\Mdust$, $\qpah$, $\Umin$, $\alpha$, and $\Umax$, as adjustable
  parameters and fix $\gamma=1$: compare the $\chi^2$ maps in
  Fig.\ \ref{fig:chicomp}a with \ref{fig:chicomp}b, and
  \ref{fig:chicomp}d with \ref{fig:chicomp}e.
  
\item The starlight heating the dust within a single pixel is
  characterized by three adjustable parameters: $\Umin$, $\gamma$, and
  $\alpha$ (see eq. \ref{eq:dMd/dU} and \ref{eq:dMd/dU1}).  $\Umin$ is
  interpreted as the intensity of the diffuse starlight, responsible
  for the bulk of the dust heating.  Maps of $\Umin$ (see Figs.\ 2 and
  7) show significant structure: there is a tendency of $\Umin$ to be
  higher in spiral arms, as well as a significant radial decline in
  $\Umin$.  The overall mass-weighted mean value of $\Umin=1.5$ for
  NGC~628 and $\Umin=3.1$ for NGC~6946, but $\Umin$ declines to values
  of 0.3 or lower in the outer regions of both galaxies.

\item The parameter $\fpdr$ is the fraction of the dust heating power
  that is contributed by starlight intensities $U > 100$. The overall
  value is $\langle \fpdr\rangle =0.12$ for NGC~628 and $\langle
  \fpdr\rangle =0.14$ for NGC~6946, but in both galaxies $\fpdr$ peaks
  at local maxima of the dust surface brightness, which are associated
  with star-forming regions. At SPIRE250 resolution, $\fpdr$ reaches
  values as high as 0.25 in both galaxies.

\item Current PACS photometry disagrees with MIPS photometry (see Fig.\ \ref{fig:color_0628} and \ref{fig:color_6946}). 
  PACS70 photometry is up to 80\% higher than MIPS70, and PACS160  is up to 50\% higher than MIPS70 photometry in the bright areas of the galaxies.

\item The PACS-MIPS photometry disagreement induces a bias when trying
  to model dust at high spatial resolutions (PACS160 PSF), where
  MIPS70 and MIPS160 cannot be used.  We do not recommend modeling
  dust at PACS160 resolution in low surface brightness areas.
  Modeling done at SPIRE250 resolution is more reliable than modeling
  at PACS160, but the inferred dust masses still disagree with our ``gold
  standard'' (i.e., modeling done at MIPS160 resolution using all the
  IRAC, MIPS, PACS and SPIRE cameras) by up to $\approx 30\%$.

\end{enumerate}

\acknowledgements

 Subsequent work (Aniano et al.\ 2012, in preparation) will extend
  this study to all 61 galaxies in the full KINGFISH sample
  \citep{Kennicutt+Calzetti+Aniano+etal_2011}.
  
  We are grateful to R.H. Lupton for availability of the SM graphics
program, and to the anonymous referee for helpful suggestions. 

This research was supported in part by JPL grants 1329088 and 1373687,
and by NSF grant AST1008570. 


\bibliography{btdrefs.bib}

\begin{thebibliography}{}

\bibitem[\protect\citeauthoryear{{Aniano} et~al.}{{Aniano}
  et~al.}{2011}]{Aniano+Draine+Gordon+Sandstrom_2011}
{Aniano}, G., {Draine}, B.~T., {Gordon}, K.~D.,  \& {Sandstrom}, K.~M. 2011,
  \pasp, 123, 1218

\bibitem[\protect\citeauthoryear{{Arab} et~al.}{{Arab}
  et~al.}{2012}]{Arab+Abergel+Habart+etal_2012}
{Arab}, H., {Abergel}, A., {Habart}, E., et~al. 2012, ArXiv e-prints, 1202.1624

\bibitem[\protect\citeauthoryear{{Asplund} et~al.}{{Asplund}
  et~al.}{2009}]{Asplund+Grevesse+Sauval+Scott_2009}
{Asplund}, M., {Grevesse}, N., {Sauval}, A.~J.,  \& {Scott}, P. 2009, \araa,
  47, 481

\bibitem[\protect\citeauthoryear{{Bendo} et~al.}{{Bendo}
  et~al.}{2006}]{Bendo+Dale+Draine+etal_2006}
{Bendo}, G.~J., {Dale}, D.~A., {Draine}, B.~T., et~al. 2006, \apj, 652, 283

\bibitem[\protect\citeauthoryear{{Blitz} et~al.}{{Blitz}
  et~al.}{2007}]{Blitz+Fukui+Kawamura+etal_2007}
{Blitz}, L., {Fukui}, Y., {Kawamura}, A., et~al. 2007, in Protostars and
  Planets V, ed. B.~{Reipurth}, D.~{Jewitt}, \& K.~{Keil} (Tucson: University
  of Arizona Press), 81

\bibitem[\protect\citeauthoryear{{Calzetti} et~al.}{{Calzetti}
  et~al.}{2010}]{Calzetti+Wu+Hong+etal_2010}
{Calzetti}, D., {Wu}, S., {Hong}, S., et~al. 2010, \apj, 714, 1256

\bibitem[\protect\citeauthoryear{{Dale} et~al.}{{Dale}
  et~al.}{2005}]{Dale+Bendo+Engelbracht+etal_2005}
{Dale}, D.~A., {Bendo}, G.~J., {Engelbracht}, C.~W., et~al. 2005, \apj, 633,
  857

\bibitem[\protect\citeauthoryear{{Dale} et~al.}{{Dale}
  et~al.}{2009}]{Dale+Cohen+Johnson+etal_2009}
{Dale}, D.~A., {Cohen}, S.~A., {Johnson}, L.~C., et~al. 2009, \apj, 703, 517

\bibitem[\protect\citeauthoryear{{Dale} \& {Helou}}{{Dale} \&
  {Helou}}{2002}]{Dale+Helou_2002}
{Dale}, D.~A.,  \& {Helou}, G. 2002, \apj, 576, 159

\bibitem[\protect\citeauthoryear{{Dale} et~al.}{{Dale}
  et~al.}{2001}]{Dale+Helou+Contursi+etal_2001}
{Dale}, D.~A., {Helou}, G., {Contursi}, A., {Silbermann}, N.~A.,  \&
  {Kolhatkar}, S. 2001, \apj, 549, 215

\bibitem[\protect\citeauthoryear{{Dame}, {Hartmann}, \& {Thaddeus}}{{Dame}
  et~al.}{2001}]{Dame+Hartmann+Thaddeus_2001}
{Dame}, T.~M., {Hartmann}, D.,  \& {Thaddeus}, P. 2001, \apj, 547, 792

\bibitem[\protect\citeauthoryear{{Donovan Meyer} et~al.}{{Donovan Meyer}
  et~al.}{2012}]{Donovan_Meyer+Koda+Momose+etal_2011}
{Donovan Meyer}, J., {Koda}, J., {Momose}, R., et~al. 2012, \apj, 744, 42

\bibitem[\protect\citeauthoryear{{Draine}}{{Draine}}{2003}]{Draine_2003a}
{Draine}, B.~T. 2003, \araa, 41, 241

\bibitem[\protect\citeauthoryear{{Draine}}{{Draine}}{2011a}]{Draine_2011b}
{Draine}, B.~T. 2011a, in EAS Publications Series, Vol.~46, PAHs and the
  Universe, ed. C.~{Joblin} \& A.~G.~G.~M. {Tielens}, 29

\bibitem[\protect\citeauthoryear{{Draine}}{{Draine}}{2011b}]{Draine_2011a}
{Draine}, B.~T. 2011b, {Physics of the Interstellar and Intergalactic Medium}
  (Princeton, NJ: Princeton Univ.\ Press)

\bibitem[\protect\citeauthoryear{{Draine} et~al.}{{Draine}
  et~al.}{2007}]{Draine+Dale+Bendo+etal_2007}
{Draine}, B.~T., {Dale}, D.~A., {Bendo}, G., et~al. 2007, \apj, 663, 866

\bibitem[\protect\citeauthoryear{{Draine} \& {Lee}}{{Draine} \&
  {Lee}}{1984}]{Draine+Lee_1984}
{Draine}, B.~T.,  \& {Lee}, H.~M. 1984, \apj, 285, 89

\bibitem[\protect\citeauthoryear{{Draine} \& {Li}}{{Draine} \&
  {Li}}{2001}]{Draine+Li_2001}
{Draine}, B.~T.,  \& {Li}, A. 2001, \apj, 551, 807

\bibitem[\protect\citeauthoryear{{Draine} \& {Li}}{{Draine} \&
  {Li}}{2007}]{Draine+Li_2007}
{Draine}, B.~T.,  \& {Li}, A. 2007, \apj, 657, 810

\bibitem[\protect\citeauthoryear{{Engelbracht} et~al.}{{Engelbracht}
  et~al.}{2007}]{Engelbracht+Blaylock+Su+etal_2007}
{Engelbracht}, C.~W., {Blaylock}, M., {Su}, K.~Y.~L., et~al. 2007, \pasp, 119,
  994

\bibitem[\protect\citeauthoryear{{Engelbracht} et~al.}{{Engelbracht}
  et~al.}{2005}]{Engelbracht+Gordon+Rieke+etal_2005}
{Engelbracht}, C.~W., {Gordon}, K.~D., {Rieke}, G.~H., et~al. 2005, \apjl, 628,
  L29

\bibitem[\protect\citeauthoryear{{Fazio} et~al.}{{Fazio}
  et~al.}{2004}]{Fazio+Hora+Allen+etal_2004}
{Fazio}, G.~G., {Hora}, J.~L., {Allen}, L.~E., et~al. 2004, \apjs, 154, 10

\bibitem[\protect\citeauthoryear{{Finkbeiner}, {Davis}, \&
  {Schlegel}}{{Finkbeiner} et~al.}{1999}]{Finkbeiner+Davis+Schlegel_1999}
{Finkbeiner}, D.~P., {Davis}, M.,  \& {Schlegel}, D.~J. 1999, \apj, 524, 867

\bibitem[\protect\citeauthoryear{{Galliano} et~al.}{{Galliano}
  et~al.}{2011}]{Galliano+Hony+Bernard+etal_2011}
{Galliano}, F., {Hony}, S., {Bernard}, J.~P., et~al. 2011, \aap, 536, A88

\bibitem[\protect\citeauthoryear{{Gordon} et~al.}{{Gordon}
  et~al.}{2007}]{Gordon+Engelbracht+Fadda+etal_2007}
{Gordon}, K.~D., {Engelbracht}, C.~W., {Fadda}, D., et~al. 2007, \pasp, 119,
  1019

\bibitem[\protect\citeauthoryear{{Gordon} et~al.}{{Gordon}
  et~al.}{2011}]{Gordon+Meixner+Meade+etal_2011}
{Gordon}, K.~D., {Meixner}, M., {Meade}, M.~R., et~al. 2011, \aj, 142, 102

\bibitem[\protect\citeauthoryear{{Griffin} et~al.}{{Griffin}
  et~al.}{2010}]{Griffin+Abergel+Abreu+etal_2010}
{Griffin}, M.~J., {Abergel}, A., {Abreu}, A., et~al. 2010, \aap, 518, L3

\bibitem[\protect\citeauthoryear{{Ishihara} et~al.}{{Ishihara}
  et~al.}{2010}]{Ishihara+Onaka+Kataza+etal_2010}
{Ishihara}, D., {Onaka}, T., {Kataza}, H., et~al. 2010, \aap, 514, A1

\bibitem[\protect\citeauthoryear{{Juvela}, {Pelkonen}, \& {Porceddu}}{{Juvela}
  et~al.}{2009}]{Juvela+Pelkonen+Porceddu_2009}
{Juvela}, M., {Pelkonen}, V.-M.,  \& {Porceddu}, S. 2009, \aap, 505, 663

\bibitem[\protect\citeauthoryear{{Kennicutt} et~al.}{{Kennicutt}
  et~al.}{2003}]{Kennicutt+Armus+Bendo+etal_2003}
{Kennicutt}, R.~C., {Armus}, L., {Bendo}, G., et~al. 2003, \pasp, 115, 928

\bibitem[\protect\citeauthoryear{{Kennicutt} et~al.}{{Kennicutt}
  et~al.}{2011}]{Kennicutt+Calzetti+Aniano+etal_2011}
{Kennicutt}, R.~C., {Calzetti}, D., {Aniano}, G., et~al. 2011, \pasp, 123, 1347

\bibitem[\protect\citeauthoryear{{Kennicutt} et~al.}{{Kennicutt}
  et~al.}{2008}]{Kennicutt+Lee+Funes+etal_2008}
{Kennicutt}, R.~C., Jr., {Lee}, J.~C., {Funes}, S.~J., Jos{\'e}~G., {Sakai},
  S.,  \& {Akiyama}, S. 2008, \apjs, 178, 247

\bibitem[\protect\citeauthoryear{{Klein}}{{Klein}}{1962}]{Klein_1962}
{Klein}, C.~A. 1962, \japplphys, 33, 3338

\bibitem[\protect\citeauthoryear{{Leroy} et~al.}{{Leroy}
  et~al.}{2011}]{Leroy+Bolatto+Gordon+etal_2011}
{Leroy}, A.~K., {Bolatto}, A., {Gordon}, K., et~al. 2011, \apj, 737, 12

\bibitem[\protect\citeauthoryear{{Leroy} et~al.}{{Leroy}
  et~al.}{2009}]{Leroy+Walter+Bigiel+etal_2009}
{Leroy}, A.~K., {Walter}, F., {Bigiel}, F., et~al. 2009, \aj, 137, 4670

\bibitem[\protect\citeauthoryear{{Li} \& {Draine}}{{Li} \&
  {Draine}}{2001}]{Li+Draine_2001b}
{Li}, A.,  \& {Draine}, B.~T. 2001, \apj, 554, 778

\bibitem[\protect\citeauthoryear{{Low} et~al.}{{Low}
  et~al.}{1984}]{Low+Young+Beintema+etal_1984}
{Low}, F.~J., {Young}, E., {Beintema}, D.~A., et~al. 1984, \apjl, 278, L19

\bibitem[\protect\citeauthoryear{{Mathis}, {Mezger}, \& {Panagia}}{{Mathis}
  et~al.}{1983}]{Mathis+Mezger+Panagia_1983}
{Mathis}, J.~S., {Mezger}, P.~G.,  \& {Panagia}, N. 1983, \aap, 128, 212

\bibitem[\protect\citeauthoryear{{Meny} et~al.}{{Meny}
  et~al.}{2007}]{Meny+Gromov+Boudet+etal_2007}
{Meny}, C., {Gromov}, V., {Boudet}, N., et~al. 2007, \aap, 468, 171

\bibitem[\protect\citeauthoryear{{Moustakas} et~al.}{{Moustakas}
  et~al.}{2010}]{Moustakas+Kennicutt+Tremonti+etal_2010}
{Moustakas}, J., {Kennicutt}, R.~C., {Tremonti}, C.~A., et~al. 2010, \apjs,
  190, 233

\bibitem[\protect\citeauthoryear{{Mu{\~n}oz-Mateos} et~al.}{{Mu{\~n}oz-Mateos}
  et~al.}{2009}]{Munoz-Mateos+Gil_de_Paz+Boissier+etal_2009}
{Mu{\~n}oz-Mateos}, J.~C., {Gil de Paz}, A., {Boissier}, S., et~al. 2009, \apj,
  701, 1965

\bibitem[\protect\citeauthoryear{{M\"uller} et~al.}{{M\"uller}
  et~al.}{2011}]{Muller+Nielbock+Balog+etal_2011}
{M\"uller}, T., {Nielbock}, M., {Balog}, Z., {Klaas}, U.,  \& {Vilenius}, E.
  2011, PACS ICC Document, PICC-ME-TN-037 v1.0

\bibitem[\protect\citeauthoryear{{Okumura}, {Kamae}, \& {for the Fermi LAT
  Collaboration}}{{Okumura} et~al.}{2009}]{Okumura+Kamae+FLAT_2009}
{Okumura}, A., {Kamae}, T.,  \& {for the Fermi LAT Collaboration}. 2009,
  \arxiv, 0912.3860

\bibitem[\protect\citeauthoryear{{Ossenkopf} \& {Henning}}{{Ossenkopf} \&
  {Henning}}{1994}]{Ossenkopf+Henning_1994}
{Ossenkopf}, V.,  \& {Henning}, T. 1994, \aap, 291, 943

\bibitem[\protect\citeauthoryear{{Ott}}{{Ott}}{2010}]{Ott_2010}
{Ott}, S. 2010, in Astronomical Society of the Pacific Conference Series, Vol.
  434, Astronomical Data Analysis Software and Systems XIX, ed. {Y.~Mizumoto,
  K.-I.~Morita, \& M.~Ohishi}, 139

\bibitem[\protect\citeauthoryear{{Paradis} et~al.}{{Paradis}
  et~al.}{2011}]{Paradis+Bernard+Meny+Gromov_2011}
{Paradis}, D., {Bernard}, J.~P., {M{\'e}ny}, C.,  \& {Gromov}, V. 2011, \aap,
  534, A118

\bibitem[\protect\citeauthoryear{{Paradis} et~al.}{{Paradis}
  et~al.}{2010}]{Paradis+Veneziani+Noriega-Crespo+etal_2010}
{Paradis}, D., {Veneziani}, M., {Noriega-Crespo}, A., et~al. 2010, \aap, 520,
  L8

\bibitem[\protect\citeauthoryear{{Pilbratt} et~al.}{{Pilbratt}
  et~al.}{2010}]{Pilbratt+Riedinger+Passvogel+etal_2010}
{Pilbratt}, G.~L., {Riedinger}, J.~R., {Passvogel}, T., et~al. 2010, \aap, 518,
  L1

\bibitem[\protect\citeauthoryear{{Pineda} et~al.}{{Pineda}
  et~al.}{2010}]{Pineda+Goldsmith+Chapman+etal_2010}
{Pineda}, J.~L., {Goldsmith}, P.~F., {Chapman}, N., et~al. 2010, \apj, 721, 686

\bibitem[\protect\citeauthoryear{{Planck Collaboration} et~al.}{{Planck
  Collaboration} et~al.}{2011a}]{Planck_molecular_clouds_2011}
{Planck Collaboration}, {Abergel}, A., {Ade}, P.~A.~R., et~al. 2011a, \aap,
  536, A25

\bibitem[\protect\citeauthoryear{{Planck Collaboration} et~al.}{{Planck
  Collaboration} et~al.}{2011b}]{Planck_dust_2011}
{Planck Collaboration}, {Ade}, P.~A.~R., {Aghanim}, N., et~al. 2011b, \aap,
  536, A19

\bibitem[\protect\citeauthoryear{{Poglitsch} et~al.}{{Poglitsch}
  et~al.}{2010}]{Poglitsch+Waelkens+Geis+etal_2010}
{Poglitsch}, A., {Waelkens}, C., {Geis}, N., et~al. 2010, \aap, 518, L2

\bibitem[\protect\citeauthoryear{{Povich} et~al.}{{Povich}
  et~al.}{2007}]{Povich+Stone+Churchwell+etal_2007}
{Povich}, M.~S., {Stone}, J.~M., {Churchwell}, E., et~al. 2007, \apj, 660, 346

\bibitem[\protect\citeauthoryear{{Prieto} et~al.}{{Prieto}
  et~al.}{2008}]{Prieto+Kistler+Thompson+etal_2008}
{Prieto}, J.~L., {Kistler}, M.~D., {Thompson}, T.~A., et~al. 2008, \apjl, 681,
  L9

\bibitem[\protect\citeauthoryear{{Primak}}{{Primak}}{1956}]{Primak_1956}
{Primak}, W. 1956, \pr, 103, 544

\bibitem[\protect\citeauthoryear{{Reach} et~al.}{{Reach}
  et~al.}{1995}]{Reach+Dwek+Fixsen+etal_1995}
{Reach}, W.~T., {Dwek}, E., {Fixsen}, D.~J., et~al. 1995, \apj, 451, 188

\bibitem[\protect\citeauthoryear{{Rieke} et~al.}{{Rieke}
  et~al.}{2004}]{Rieke+Young+Engelbracht+etal_2004}
{Rieke}, G.~H., {Young}, E.~T., {Engelbracht}, C.~W., et~al. 2004, \apjs, 154,
  25

\bibitem[\protect\citeauthoryear{{Roussel}}{{Roussel}}{2012}]{Roussel_2012}
{Roussel}, H. 2012, arXiv:1205.2576

\bibitem[\protect\citeauthoryear{{Sandstrom} et~al.}{{Sandstrom}
  et~al.}{2010}]{Sandstrom+Bolatto+Draine+etal_2010}
{Sandstrom}, K.~M., {Bolatto}, A.~D., {Draine}, B.~T., {Bot}, C.,  \&
  {Stanimirovic}, S. 2010, \apj, 715, 701

\bibitem[\protect\citeauthoryear{{Skibba} et~al.}{{Skibba}
  et~al.}{2011}]{Skibba+Engelbracht+Dale+etal_2011}
{Skibba}, R.~A., {Engelbracht}, C.~W., {Dale}, D., et~al. 2011, \apj, 738, 89

\bibitem[\protect\citeauthoryear{{Smith} et~al.}{{Smith}
  et~al.}{2007}]{Smith+Draine+Dale+etal_2007}
{Smith}, J.~D.~T., {Draine}, B.~T., {Dale}, D.~A., et~al. 2007, \apj, 656, 770

\bibitem[\protect\citeauthoryear{{Stansberry} et~al.}{{Stansberry}
  et~al.}{2007}]{Stansberry+Gordon+Bhattacharya+etal_2007}
{Stansberry}, J.~A., {Gordon}, K.~D., {Bhattacharya}, B., et~al. 2007, \pasp,
  119, 1038

\bibitem[\protect\citeauthoryear{{Stognienko}, {Henning}, \&
  {Ossenkopf}}{{Stognienko} et~al.}{1995}]{Stognienko+Henning+Ossenkopf_1995}
{Stognienko}, R., {Henning}, T.,  \& {Ossenkopf}, V. 1995, \aap, 296, 797

\bibitem[\protect\citeauthoryear{{Totani} et~al.}{{Totani}
  et~al.}{2011}]{Totani+Takeuchi+Nagashima+etal_2011}
{Totani}, T., {Takeuchi}, T.~T., {Nagashima}, M., {Kobayashi}, M.~A.~R.,  \&
  {Makiya}, R. 2011, \pasj, 63, 1181

\bibitem[\protect\citeauthoryear{{Walter} et~al.}{{Walter}
  et~al.}{2008}]{Walter+Brinks+deBlok+etal_2008}
{Walter}, F., {Brinks}, E., {de Blok}, W.~J.~G., et~al. 2008, \aj, 136, 2563

\bibitem[\protect\citeauthoryear{{Weingartner} \& {Draine}}{{Weingartner} \&
  {Draine}}{2001}]{Weingartner+Draine_2001a}
{Weingartner}, J.~C.,  \& {Draine}, B.~T. 2001, \apj, 548, 296

\bibitem[\protect\citeauthoryear{{Werner} et~al.}{{Werner}
  et~al.}{2004}]{Werner+Roellig+Low+etal_2004}
{Werner}, M.~W., {Roellig}, T.~L., {Low}, F.~J., et~al. 2004, \apjs, 154, 1

\bibitem[\protect\citeauthoryear{{Wolfire}, {Hollenbach}, \& {McKee}}{{Wolfire}
  et~al.}{2010}]{Wolfire+Hollenbach+McKee_2010}
{Wolfire}, M.~G., {Hollenbach}, D.,  \& {McKee}, C.~F. 2010, \apj, 716, 1191

\bibitem[\protect\citeauthoryear{{Wright} et~al.}{{Wright}
  et~al.}{1991}]{Wright+Mather+Bennett+etal_1991}
{Wright}, E.~L., {Mather}, J.~C., {Bennett}, C.~L., et~al. 1991, \apj, 381, 200

\bibitem[\protect\citeauthoryear{{Yamamura} et~al.}{{Yamamura}
  et~al.}{2010}]{Yamamura+Makiuti+Ikeda+etal_2010}
{Yamamura}, I., {Makiuti}, S., {Ikeda}, N., et~al. 2010, VizieR Online Data
  Catalog, 2298

\end{thebibliography}

\clearpage
\appendix


\section{\label{app:masks}Separation of the Target Galaxy and ``Non-Background'' Regions from Background Regions}

In order to backgound-subtract an image, one needs to identify the
image areas without any emission, i.e., recognize all the
``non-background'' areas to exclude them in any background estimation
procedure.  Different cameras have different sensitivities to extended
emission and other sources, so if one attempts to background-subtract
an image using only the image itself, one may under- or
over-overestimate the background level. As an example, the PACS
cameras are not very sensitive to the extended emission in the North
part of NGC~6946, (see Fig.\ \ref{Fig_6946_ori}), so, unless these
areas are recognized by other cameras, one could erroneously consider
them as background areas leading to over-estimation of background
level.  We proceed using all the information available: i.e., using
all the cameras, to recognize the ``non-background''
areas.\footnote{``Non-background'' areas will include the target
  galaxy, and any other recognizable emission structures in the field}

We developed an algorithm (described in Appendix B) that, given an
image $I$ and a set of ``non-background'' areas, identifies
``background'' areas and estimates the variance in the ``background''
area (i.e., a measure of the image noise). To obtain the most accurate
background estimation, we generate background masks for each camera
and combine them to generate a mask of the non-background areas,
proceeding in several steps as follows.

\begin{figure}[h]
\begin{center}
\begin{tabular}{cc}
\includegraphics[width=8.2cm,height=7cm,,clip=true,trim=2.0cm 0.0cm 0.0cm 0.0cm]{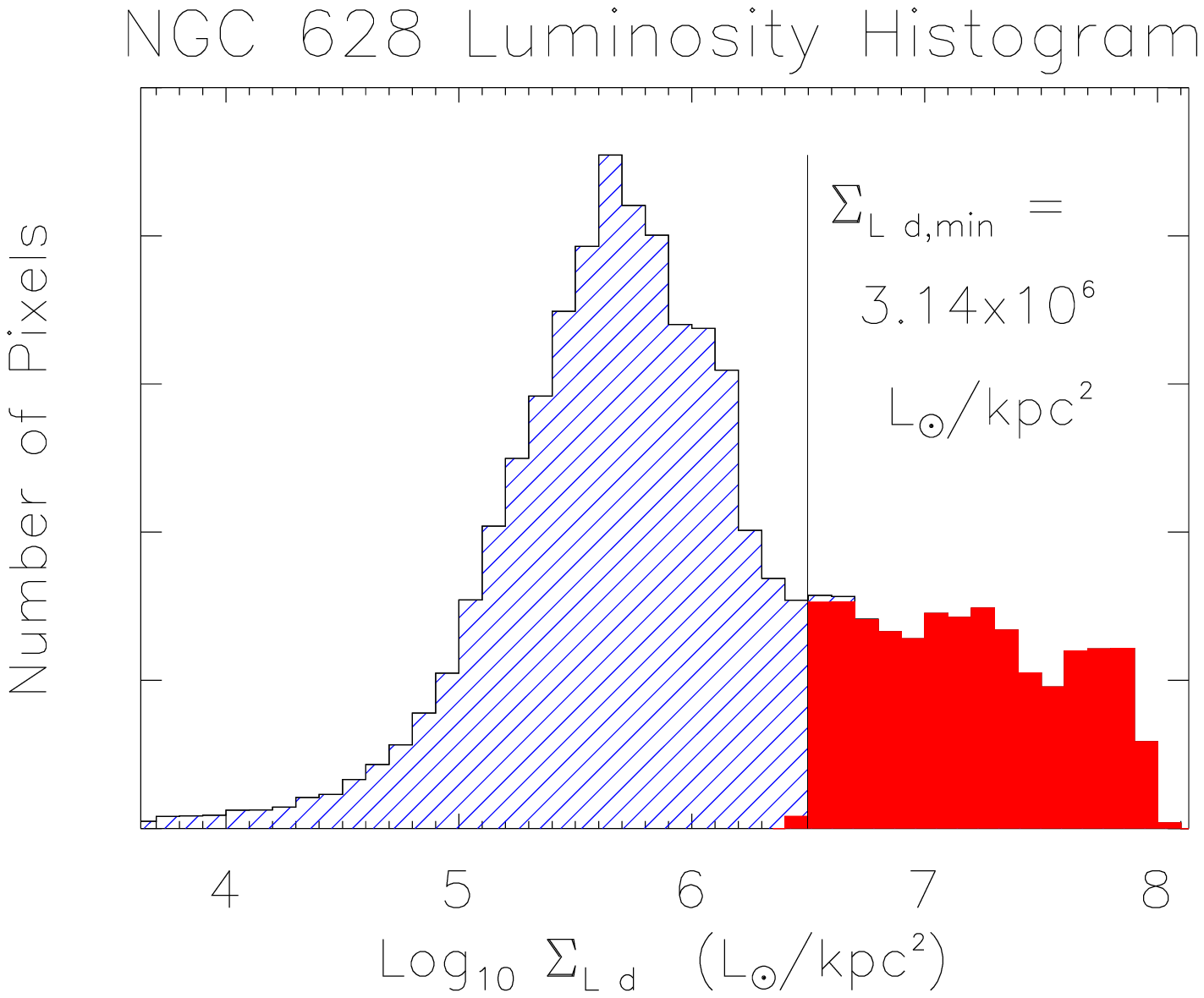}
\includegraphics[width=7.5cm,height=7cm,clip=true,trim=3.3cm 0.0cm 0.0cm 0.0cm]{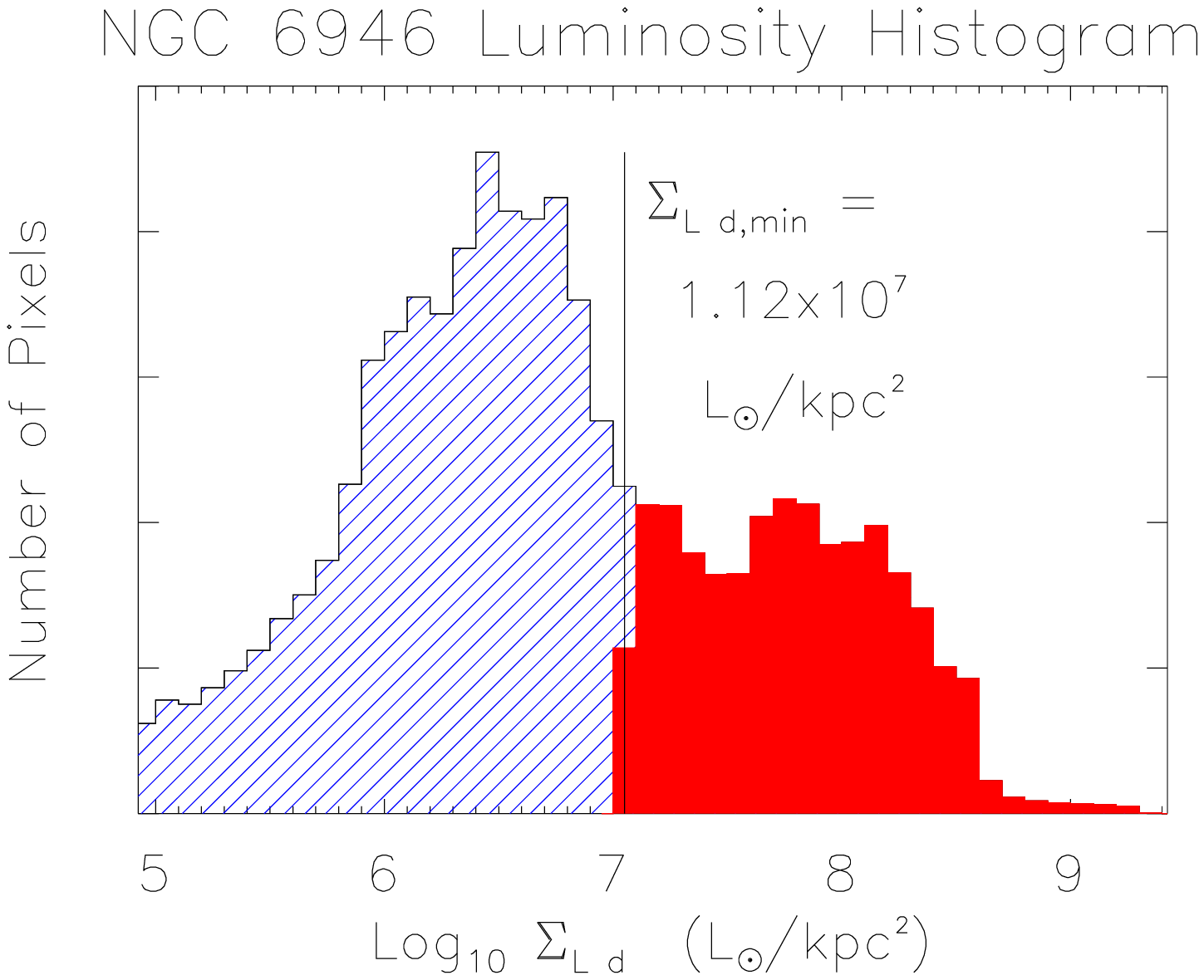}
\end{tabular}
\vspace*{-0.5cm}
\caption{\footnotesize\label{fig:histo} Dust luminosity surface
  density $\Sigma_{L_{\dust}}$ histogram of a region containing NGC~628
  (left) and NGC~6946 (right) shown in Figs.\ 4. and 10. The modeling
  is done at MIPS160 resolution (18\as pixels) using all the cameras
  available.  The shaded (red) area corresponds to the pixels
  identified as ``galaxy pixels'' in the galaxy mask.  The line-shaded
  (blue) area corresponds to ``non-galaxy pixels'' (i.e. background
  pixels and other sources in the field not connected to the
  galaxies).}
\end{center}
\end{figure}

\begin{enumerate}

\item
As a preliminary identification of the ``non-background'' areas we select
the sky regions with $S/N>2$ in the most sensitive cameras. In our
study, for this purpose we use IRAC5.8, IRAC8.0, MIPS24, MIPS70, and
MIPS160.  We estimate the noise in the images using the iterative
algorithm described in Appendix B, using an empy set as starting
``non-background'' areas.  The ``non-background'' regions obtained
contain the target galaxy and bright background sources.  We consider
the Starting Non-Background Mask (SNBM) as those regions in which at
least 3 of the 5 cameras have $S/N>2$.  The SNBM include the target
galaxy and some foreground bright sources.

\item
For each band $b$, we obtain
a {\em Preliminary Individual Background Mask} (PIBM$_b$)
with the iterative algorithm described in Appendix B. 
For this construction, we consider the pixels in the SNBM as ``non-background'' pixels (i.e., they were not considered as background pixel candidates).

\item
We construct a {\em Preliminary Background Mask} (PBM) as follows: We
compute the average $A$ of the PIBM$_b$s, where only the cameras
observing the pixel are taken into account.  At each pixel, the value
of $A$ will be the fraction of cameras that do not detect any emission
with $S/N > 1.8\times0.4202 = 0.76$ .  The outer regions of the field
(that may be observed by a small number of cameras), can still be
considered as background if the observing cameras do not show
emission.  We smooth the image $A$ by convolving it with a
(normalized) gaussian kernel $\propto \exp[-(r/r_0)^2]$, with $r_0$
equal to 0.5 times the width of the pixel. The convolution lowers the
value of $A$ near sources, reducing the background areas.  We consider
the PBM as the pixels where the smoothed $A>0.35$, i.e., fewer than
35\% of the cameras detect emission.  With 13 cameras, PBM corresponds
to the areas with $S/N > 0.76$ detection in no more than 4 cameras.

\item
Using the complement of PBM as ``non-background areas'' and the
algorithm described in Appendix B, we background-subtract each
individual image. 
The complement of PBM masks all recognized sources in the background estimate for each individual image.

\item
Using the background-subtracted images, we proceed to perform a
preliminary dust modeling at MIPS160 PSF (FWHM = $38.8\mas$) using all
the cameras available.  We obtain a preliminary estimate of the dust
luminosity surface density $\Sigma_{L_{\dust}}$ in each pixel.  We
will construct a Preliminary Galaxy Mask (PGM) based on the dust
luminosity surface density obtained.  In order to obtain smoother
galaxy borders, we smooth the $\Sigma_{L_{\dust}}$ image by convolving
it with a (normalized) circular kernel of $20\arcsec$ radius.  We
consider the PGM as the regions with (smoothed) $\Sigma_{L_{\dust}}
\ge \Sigma_{L_{\dust},min}$.  We use the threshold values $
\Sigma_{L_{\dust},min} = 3.14 \times10^{6} \Lsol\kpc^{-2}$ for NGC~628
and $\Sigma_{L_{\dust},min} = 1.12 \times10^{7} \Lsol\kpc^{-2}$ for
NGC~6946.  The thresholds $\Sigma_{L_{\dust},min}$ were manually
chosen so that the FGM covers the area where can reliably estimate
$\qpah, \Umin, \gamma,$ and $\alpha$, i.e., so that the inferred dust
parameters have $S/N\gtsim1$.  Regions not connected to the main
galaxy (i.e., other sources) are now omitted from the mask.  Figure
\ref{fig:histo} shows the histogram of the dust luminosity
$\Sigma_{L_{\rm d}}$ for a region containing NGC~628 (left) and NGC~6946
(right).  The solid shaded (red) area corresponds to the pixels
identified as ``galaxy pixels'' (i.e., in the galaxy mask), and the
line-shaded (blue) pixels correspond to ``non-galaxy pixels''
(i.e., background pixels and other sources in the field not connected
to the galaxies).  We note that the galaxy cut-off is performed in the
smoothed image, and thus, the boundaries in the non-smoothed
luminosity histograms shown in Figure \ref{fig:histo} are not sharp.

\end{enumerate}

At this point, we have a well-determined separation of the image into
(preliminary) ``galaxy pixels'' and the non-galaxy pixels. We now
continue with steps 6-9, where we essentially repeat steps 2-5,
starting with the PGM as our SNBM.  The new masks generated in steps
6-9 are very similar to the corresponding ones previously generated on
steps 2-5, with only small differences in the low surface brightness
areas. The combined procedure is robust and precise.

\begin{enumerate}
\setcounter{enumi}{5}
\item 
For each band $b$, we obtain
an {\em Individual Background Mask} (IBM$_b$)
with the iterative algorithm described in Appendix B. 
For this construction, we consider the pixels in the PGM as ``non-background'' pixels (i.e., they were not considered as background pixel candidates).

\item
We construct a {\em Background Mask}, (BM) using the same procedure as in step 3, but using the IBM$_b$s instead of the PIBM$_b$s.

\item
Using the the complement of the BM ``non-background
areas'' and the algorithm described in Appendix B, we
background-subtract each individual image.
The complement of BM masks all recognized sources in the background estimate for each individual image.

\item
Using the background-subtracted images, we proceed to perform the
final dust modeling at MIPS160 PSF using all the cameras available.
We construct a Galaxy Mask (GM) in the same way as we constructed
the PGM (using the same values of $ \Sigma_{L_{\dust},min}$).  The
final GM and PGM will differ only in a few boundary pixels.

\end{enumerate}

In 
Aniano et al. (2012, in preparation)
the same masking method is
applied to the full sample of 61 KINGFISH galaxies. Background
companion galaxies are manually identified and removed from the masks
when possible.


\section{\label{app:background_estimation}Algorithm for Background Recognition (Image Segmentation) and Subtraction.}

Given an image $I$ expressed in its native pixel grid $(x,y)$, and a
(potentially empty) set $N\!B_{Std}$ of ``non-background'' sky regions
expressed in a standard $4\mas \times 4\mas$ grid (a grid common to
all the images), we iteratively construct a sequence of ``background''
pixels masks $B_{n}, n=0,1,2,...$, and a sequence of tilted planes
$plane_{n}, n=0,1,2,...$ that approximate the image values over the
sets $B_{n}, n=0,1,2,...$. The sets $B_{n}, n=0,1,2,...$ will exclude
the ``non-background'' $N\!B_{Std}$ regions, i.e., it is assumed that
the pixel in $N\!B_{Std}$ are not background-candidate pixels.

Throughout the algorithm, the given ``non-background'' regions will be
masked and not considered in the fitting.  We start by bi-linearly
interpolating or averaging (depending on the grid relative pixel
sizes) the set $N\!B_{Std}$ into the image $I$ grid $(x,y)$ to obtain
$N\!B_{Prev}$.

We first smooth the original image
by convolving it with a (normalized) gaussian kernel $\propto
\exp[-(r/r_0)^2]$, with $r_0$ equal to 0.56 times the width of
the native pixel for each camera. 
The factor 0.56 is chosen so that each pixel contributes half of its flux to the corresponding pixel in the convolved image.
This step provides a less noisy image, in which we can more robustly identify
the galaxy and the external sources. 

We proceed as follows.  We start by setting $B_0$ to be all pixels in
the grid $(x,y)$, except for the pixels in $N\!B_{prev}$ (if any).  We
fit a plane $plane_0$ to the image values over the $B_0$ pixels.  This
is achieved by finding the $plane_0$ that minimizes $\epsilon_0$
defined as: 
\beq
\label{eq:epsilon}
\epsilon_n \equiv \sum_{(x,y) \in B_n} {[I(x,y)-plane_n(x,y)]^2},
\eeq
where $n=0$.
We compute the dispersion $\theta_0$ defined as:
\beq 
\label{eq:sigma}
\theta_n \equiv
\sqrt{\frac{\epsilon_n}{\#\{(x,y)\in B_n\}-3}}, 
\eeq
where $n=0$, so the sum is over the $B_0$ (original ``background'' candidate) pixels.

We take the first ($n=1$) set of ``background pixels''\footnote{
  ``Background pixels'' consist of pixels that are not part of the
  target galaxy, and any other recognizable emission structures (e.g.,
  background galaxies) in the field, removing also any pixels that are
  ``outliers'', either above or below the general background.} $B_1$
to be all pixels that do not belong to $N\!B_{prev}$, or lie more than
$\kappa \, \theta_0$ above $plane_0$ ($\kappa$ is a constant, and we
use $\kappa\equiv1.8$, see below for details): 
\beq (x,y)\in B_1 \iff
\left\{
\begin{array}{l}
      (I(x,y)- plane_0 )< \kappa \,  \theta_o \\
      (x,y) \notin N\!B_{prev}
\end{array}
\right.
\eeq
In the first step, these pixels typically
correspond to the brightest point sources and central pixels of the galaxy.

We now proceed iteratively by:
\begin{enumerate}

\item Fitting a tilted plane $plane_n$ to all the candidate ``background
pixels'' $B_n$, by minimizing $\epsilon_n$ defined by eq.\ (\ref{eq:epsilon}).

\item Computing $\theta_n$, defined by eq.\ (\ref{eq:sigma}), where the sum extends over the $B_n$ pixels.

\item Identifying a new set of ``background pixels'' $B_{n+1}$ to be those pixels which lie inside the interval $ (plane_n-\kappa \, \theta_n,plane_n+\kappa \, \theta_n)$, and do not belong to $N\!B_{prev}$:
\beq
(x,y)\in B_n \iff
\left\{
\begin{array}{l}
      |I(x,y)- plane_n |< \kappa \,  \theta_o \\
      (x,y) \notin N\!B_{prev}
\end{array}
\right.
\eeq
\end{enumerate}

This iteration will generate a sequence of tilted planes $plane_n$, and a (generally decreasing) set of ``background pixels'' $B_n$.
The iteration will converge, i.e., it reaches a step $M$ where $B_{M+1} = B_M$), typically in $M=20-250$ iterations.

Once the iteration converges, we have set of ``background pixels''
$B_{M}$, and a fitting plane $plane_{M}$ (all the computed sets $B_n$
and planes $plane_n$ will depend on the choice of $\kappa$, see below
for a discussion).  We bi-linearly interpolate or average $B_{M}$ to
the standard $4\arcsec\times4\arcsec$ grid.  This will generate a mask
of the ``background regions'' in the standard grid, so it can be
compared with the results of the algorithm from images of different
cameras.  We can generate standard mask of ``non-background regions''
by taking the complement of the ``background regions'' mask.  The left
column of Figures \ref{fig:resi_0628} and \ref{fig:resi_6946} show the
resulting residuals for NGC~628 and NGC~6946 respectively over
$B_{M}$, for $\kappa = 1.8$.  Rows 1--4 correspond to IRAC4.5, MIPS24,
PACS160, and SPIRE350, respectively.  It can be seen that the
algorithm masks out all the point sources, extended sources and
several background regions. Residuals slightly positive are found in
the edges of the extended emission areas, although most of the
extended emission areas are recognized and excluded.

After the iteration converges to a set of ``background pixels''
$B_{M}$, if we desire to perform a background subtraction to the
images, we run the iterative steps again.  In the second iteration, we
use the original (non-smoothed) image, and the complement of the set
$B_{M}$ found as the starting set of ``non-background pixels''
$N\!B_{prev}$, i.e., the pixels which were recognized as having
emission in the first iteration are masked out.  The second run of the
iteration will converge to a new set of ``background pixels'' $B_{N}$,
and a new fitting plane $plane_{N}$, typically in $N=20-250$
iterations.  Clearly, by construction, we always have $B_{N} \subseteq
B_{M}$.  Finally, the last plane $plane_{N}$ is subtracted from $I$.

The reason to perform the second iteration, with the original image
and a larger set of starting ``non-backgrund pixels'' is so we can
remove noisy pixels from the background pixels, recognize the edges of
the emission areas more precisely, and perform a more precise
background level estimation.  Moreover, if we try to perform a single
iteration (i.e., by going directly to the second iteration), then the
larger image noise will prevent us from recognizing the dimmer sources
and extended emission areas.  The center column of Figures
\ref{fig:resi_0628} and \ref{fig:resi_6946} show the resulting
residuals for NGC~628 and NGC~6946 respectively, over the sets
$B_{N}$.

\begin{figure}[h]
\centering\includegraphics[width=10cm]{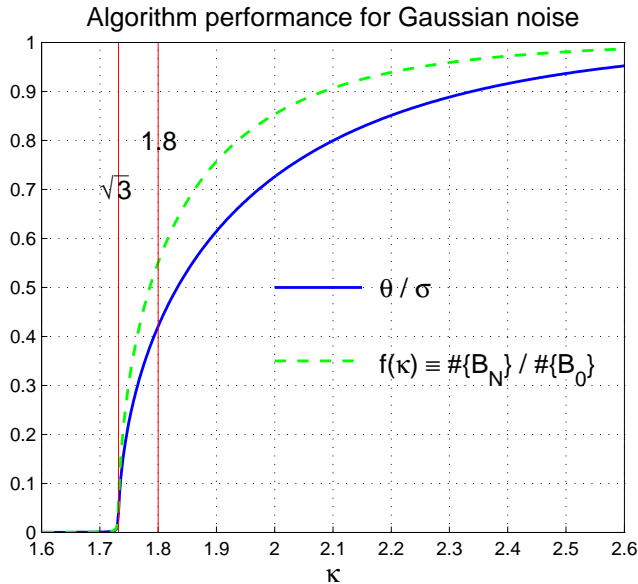}
\footnotesize
\vspace*{-0.5cm}
\caption{\footnotesize\label{fig:kappa} Theoretical performance of the
  iterative algorithm for gaussian random noise. The horizontal axis
  correspond to the cutoff threshold $\kappa$.  The dotted green line
  shows $\theta/\sigma$, the ratio between the dispersion of the final
  set $B_N$ and the dispersion of the original distribution. The solid
  blue line shows the fraction of the original points that survive in
  $B_N$. The vertical red lines correspond to the critical value
  $\kappa = \sqrt{3}$ and the used value $\kappa = 1.8$.}
\end{figure}

\begin{figure}
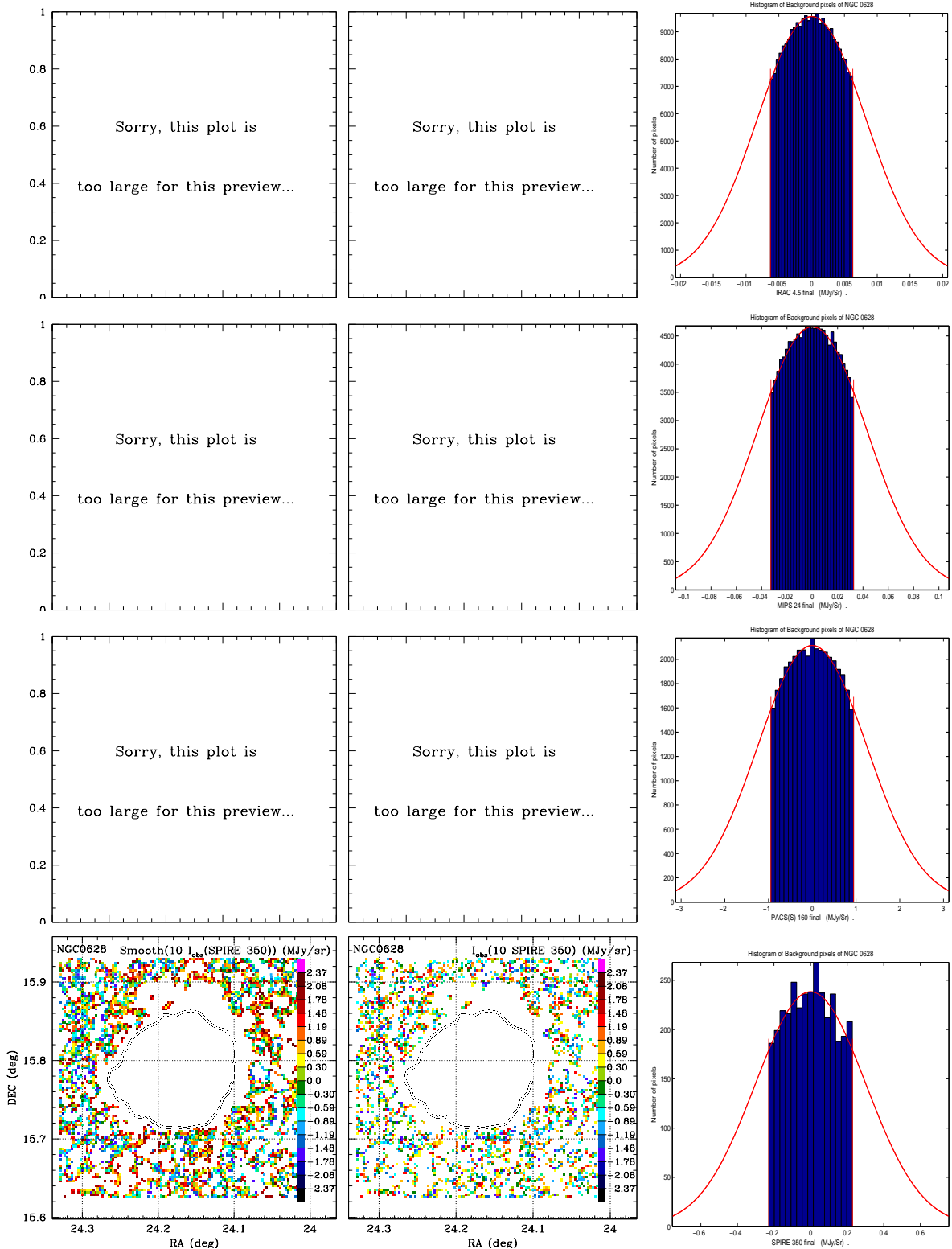

\centering
\begin{tabular}{c@{$\,$}c@{$\,$}c} 
\FirstNoTrim
\SecondNoTrim
\ThirdNoTrim
\FourthNoTrim
\end{tabular}
\vspace*{-0.5cm}
\caption{\footnotesize\label{fig:resi_0628}
Analysis of the residuals in NGC~0628. 
Row 1: IRAC4.5. 
Row 2: MIPS24.
Row 3: PACS160.
Row 4: SPIRE350. 
Left column: residuals on the set $B_{M}$ (after the first iteration of the algorithm on the smooth image). 
Center column: residuals on the set $B_{N}$ (after the second iteration on the original image).
Right column: histograms of the pixels on $B_{N}$. The (red) solid line is a gaussian with $\sigma = \theta / 0.4202$ (see text for details).
}
\end{figure}

\begin{figure}
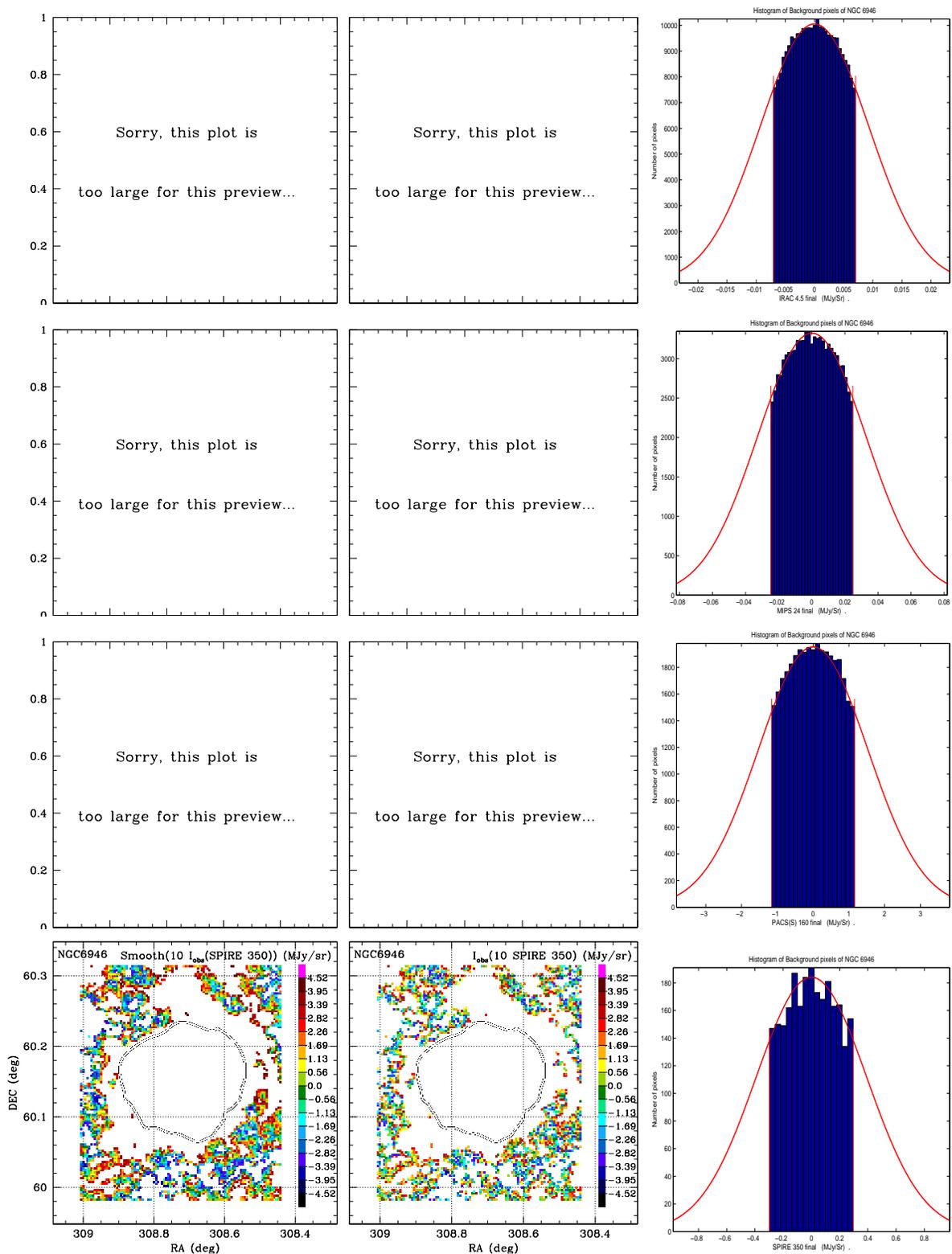

\centering
\begin{tabular}{c@{$\,$}c@{$\,$}c} 
\FirstNoTrim
\SecondNoTrim
\ThirdNoTrim
\FourthNoTrim
\end{tabular}
\vspace*{-0.5cm}
\caption{\footnotesize\label{fig:resi_6946}
Same as Fig.\ \ref{fig:resi_0628}, but for NGC~6946.
The bright carbon star V0778 Cyg,
located near RA=309.044, Dec= 60.082 (slightly off the
bottom left edge of the maps shown) produces strong artifacts in the IRAC and MIPS cameras, reducing the useful background area of the images.}
\end{figure}

It can be shown that if there exist an infinite set of background
points $B_0$ in which the intensities are independent noise drawn from
a distribution with probability density $\rho$, then the algorithm
will asymptotically converge: $B_n \xrightarrow [n\rightarrow
  \infty]{} B_{N} \subseteq B_0$, that will depend on $\kappa$. We
define $f(\kappa)$, the fraction of the original background-candidate
pixels that are asymptotically considered background: \beq f(\kappa)
\equiv \#\{B_{N}\} / \#\{B_0\} \eeq For noise distribution function
densities $\rho$ which are analytical around their expectation values,
it can be shown that if $\kappa < \sqrt{3}$ then $B_n$ will converge
to an empty set $\emptyset$; and $B_n$ will converge to a non-empty
set provided that $\kappa > \sqrt{3}$, i.e., $f(\kappa <
\sqrt{3})=0,\,\,\,\,f(\kappa > \sqrt{3})>0$.  In particular, for a top
hat distribution ($\rho(x)=1/2 \iff |x|\leqslant 1$), $B_{N} = B_0$ if
$\kappa > \sqrt{3}$, i.e., $f(\kappa > \sqrt{3})=1$.  If the noise has
a gaussian distribution with standard deviation $\sigma$, we can
numerically compute $f(\kappa)$, and the actual $\sigma / \theta$ (the
ratio between the distribution standard deviation $\sigma$ and the
computed final dispersion $\theta$).  Figure \ref{fig:kappa} shows the
values of $f(\kappa)$ (dotted green line) and $\theta / \sigma$ (solid
blue line) for $1.6 < \kappa < 2.6$ for a gaussian distribution. The
vertical red lines correspond to the threshold value $\kappa =
\sqrt{3}$, and the used value $\kappa = 1.8$

Our choice of $\kappa$ is dictated by (1) the need to have $\kappa>
\sqrt{3}$ in order for the procedure to converge to a non-empty set,
(2) the desire to minimize inclusion of outliers that may be due to
rare events (e.g., cosmic rays, image artifacts), and (3) the desire
to be as sensitive as possible to dim sources. We choose $\kappa =
1.8$ as a good compromise. If the background noise were gaussian, for
our choice $\kappa = 1.8$, we have $f(\kappa = 1.8) = 0.5506$, $ \theta / \sigma = 0.4202$, (i.e., the final iteration should fit only
$\approx55\%$ of the background points, and their measured dispersion
should be 0.4203 $\sigma$).

How well does the iterative procedure behave with real images?
Typically the iterative procedure converges in 20--250 steps.  The
right column of Figures \ref{fig:resi_0628} and \ref{fig:resi_6946}
show the resulting histograms of the background-subtracted intensities
in $B_{N}$ for NGC~628 and NGC~6946 respectively.  The solid red curve
is a gaussian distribution function with $\sigma = \theta / 0.4202$,
i.e., with the standard deviation expected from the measured
dispersion. The fit of the gaussian curve to the histograms 
in Figures \ref{fig:resi_0628} and \ref{fig:resi_6946} is very
good, supporting the validity of the presented algorithm.

This background recognition algorithm has several advantages over
other approaches.  Firstly, the algorith is very robust, automatic,
and quick, and produces superior results when dealing with all the
$\approx 1400$ images of the KINGFISH sample.  Secondly, by keeping
only a subset $(\approx55\%)$ of the background pixels, the last
computed dispersion is small enough that even dim sources are easily
recognized and masked.  This allows us to exclude virtually all
detectable foreground or background sources from the subset of points
used to determine the background of the images.  Thirdly this
algorithm also recognizes areas with dim extended emission.  Fourthly,
image artifacts over the background areas are easily identified,
allowing us to minimize bias induced by them in the background
estimation.  This allows us to avoid over- or under-subtraction of the
background in each image, which is critical in order to model the
galaxy low surface brightness areas.

\begin{table}[h] 
\caption{Background information for the different Cameras} 
\label{tab_bck}
\centering 
\begin{tabular}{|l|ccc|c|ccc|c|} 
\hline \hline 
                &  \multicolumn{4}{|c|}{NGC~0628}&  \multicolumn{4}{|c|}{NGC~6946}\\
Camera & \multicolumn{3}{|c|}{Background area} & Image & \multicolumn{3}{|c|}{Background area} & Image  \\
               & \# of    & percent$^b$ & $\sigma$$^c$ &   max.    &  \# of    & percent$^b$ &$\sigma$$^c$ &   max.     \\
               &  pixels$^a$ & (\%) & MJy/sr &   MJy/sr   & pixels$^a$ & (\%) & MJy/sr &   MJy/sr    \\
\hline \hline 
IRAC3.6         &  234851  &  29.7   & 0.0052  &  160    &  291159  &   30.4   & 0.0093  &  183    \\ 
IRAC4.5         &  253580  &  31.1   & 0.0083  &  152    &  267213  &   28.2   & 0.0093  &  232    \\
IRAC5.8         &  260864  &  32.6   & 0.0285  &  59.6   &  301640  &   31.5   & 0.0274  &  617  \\
IRAC8.0         &  264790  &  32.6   & 0.0350  &  50.9   &  286791  &   30.3   & 0.0356  &  1441   \\
\hline
MIPS 24         &  124131  &   36.8   & 0.0432  &   88.0  &  87694   &   34.1   & 0.0328  &   2192   \\
MIPS 70         &   14264   &   41.1   & 0.439  &   70.9  &   8632     &   35.1   & 0.334  &  1645    \\
MIPS 160       &   3059     &   31.7   & 0.396  &   79.9  &   2996     &   40.8   & 1.02  &   715      \\
\hline
PACS(H)70   &  315588  &  43.3    &  4.55  &  605   &  231316  &   42.7    & 4.50  &  24106    \\
PACS(H)100 & 311051   &  42.7    & 4.40   &  635   & 226840   &  42.5     & 4.28   &  20227   \\
PACS(H)160 & 36836     &  45.3    & 2.35   &  226   &  264298   &  49.0    & 3.12   &  7090      \\
\hline
PACS(S)70    & 155675   &  40.1   &  2.50   &  590    & 126604   &  40.7   &  2.52   &   23115 \\
PACS(S)100  & 104132   &  39.5   &  2.30   &  506    & 82813     &  39.3   &  2.22   &  19275  \\
PACS(S)160  &  51115    &  39.3   &  1.25    &  237   &  34252    &  44.3   &  1.53    &  7439   \\
\hline
SPIRE250      &  7433     &  32.8   &   0.499    &  76.9 &  7393     &  40.9   &   0.873    &  1584   \\
SPIRE350      &   3306    &  39.8   &   0.303    &  21.7 &   2470    &  38.1   &   0.392    &  358   \\
SPIRE500      &   1230    &   29.5  &   0.122    &  6.92 &   1151    &   33.7  &   0.186    &  84.1   \\
\hline \hline 
\multicolumn{9}{l}{$^a$ Number of native pixels that are considered ``background''.}\\
\multicolumn{9}{l}{$^b$ Percent of the non-galaxy pixels that are considered ``background'' pixels.}\\
\multicolumn{9}{l}{$^c$ Standard deviation of the Gaussian fit of the ``background'' pixels.}\\
\end{tabular}
\end{table}

\renewcommand \RoneCone {NGC0628_Bck_Removed_IRAC_4oo5_Residuals_smooth.eps}
\renewcommand \RoneCtwo {NGC0628_Bck_Removed_IRAC_4oo5_Residuals.eps}
\renewcommand \RoneCthree {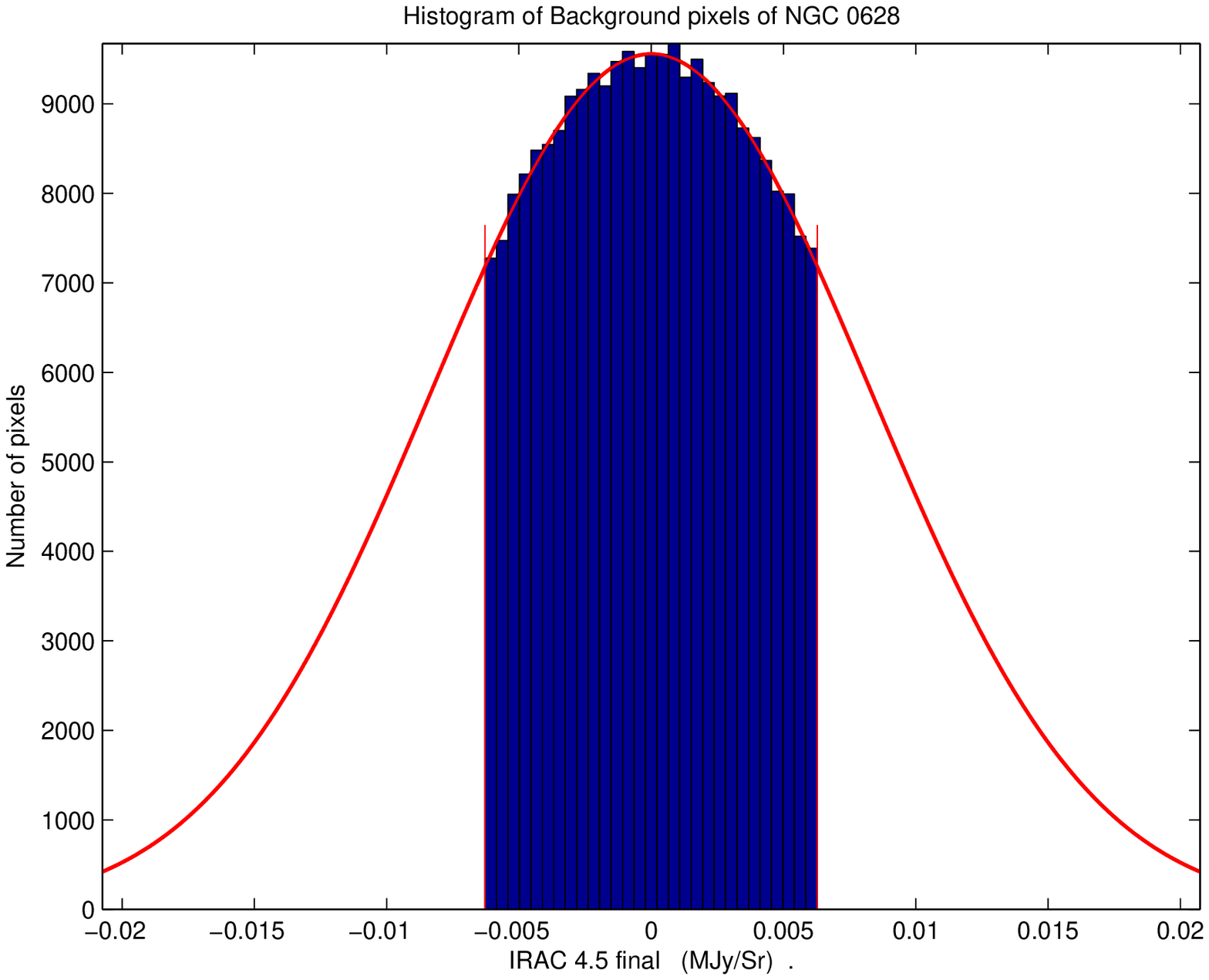}
\renewcommand \RtwoCone {NGC0628_Bck_Removed_MIPS_24_Residuals_smooth.eps}
\renewcommand \RtwoCtwo {NGC0628_Bck_Removed_MIPS_24_Residuals.eps}
\renewcommand \RtwoCthree {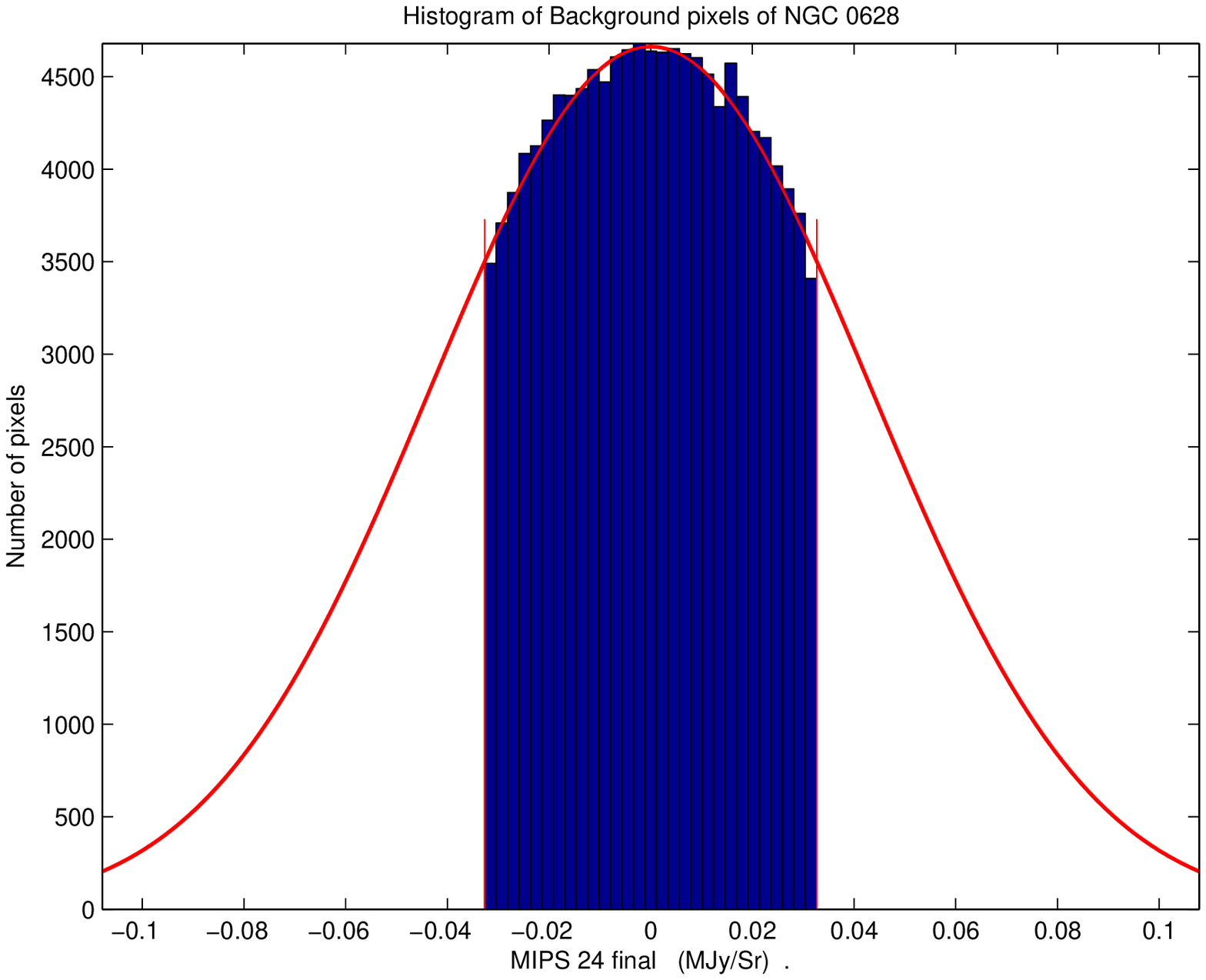}
\renewcommand \RthreeCone {NGC0628_Bck_Removed_PACSS_160_Residuals_smooth.eps}
\renewcommand \RthreeCtwo {NGC0628_Bck_Removed_PACSS_160_Residuals.eps}
\renewcommand \RthreeCthree {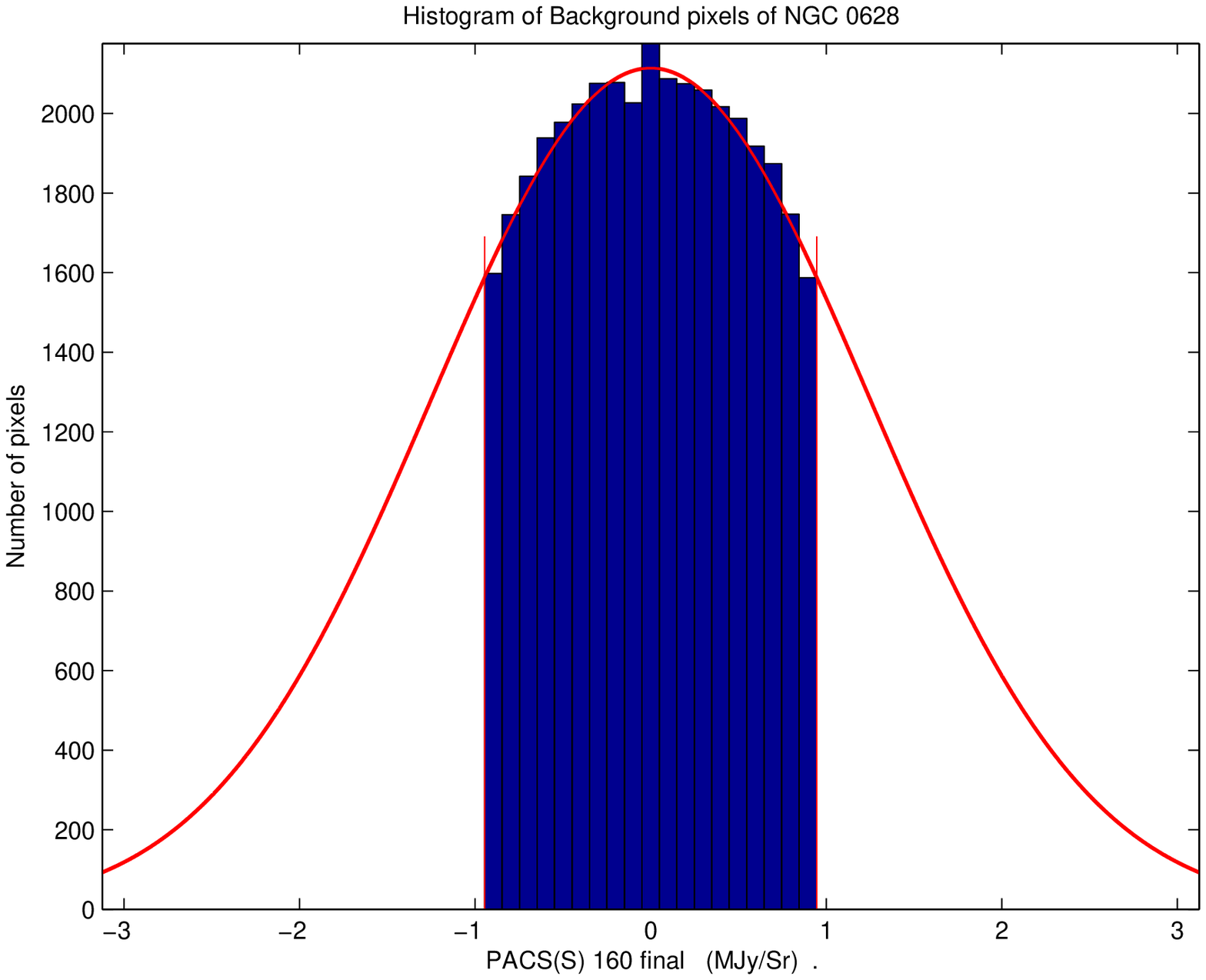}
\renewcommand \RfourCone {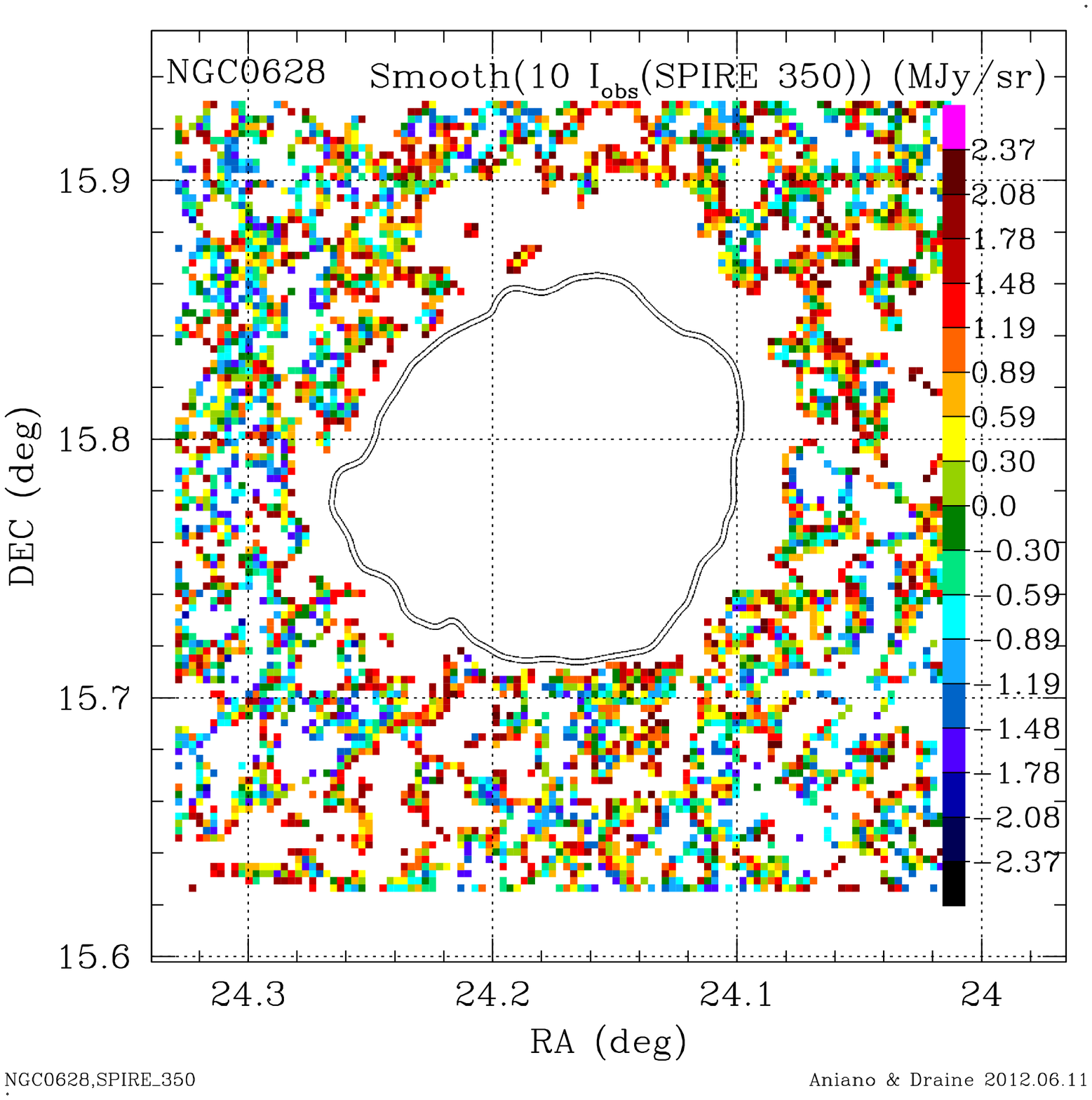}
\renewcommand \RfourCtwo {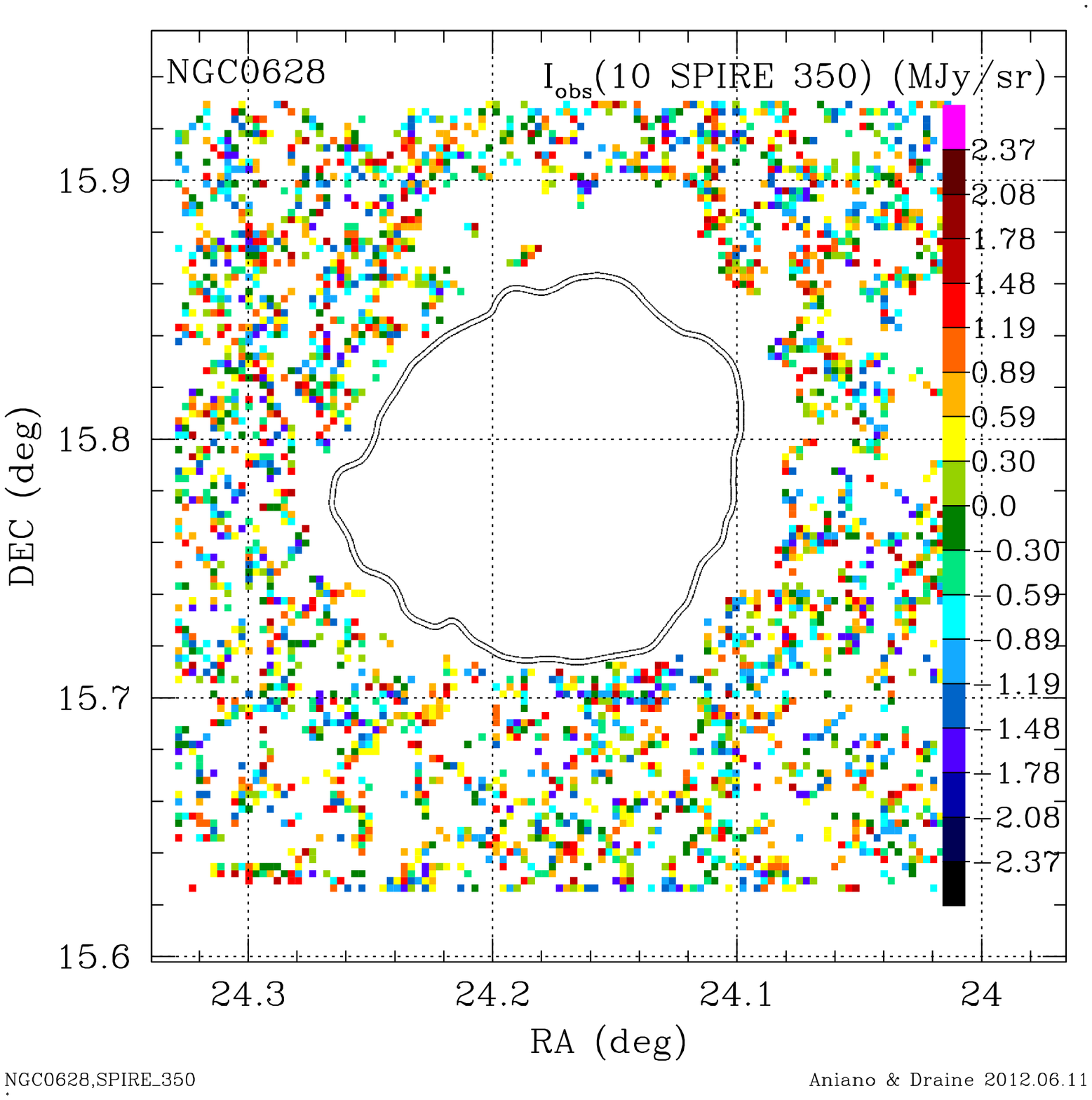}
\renewcommand \RfourCthree {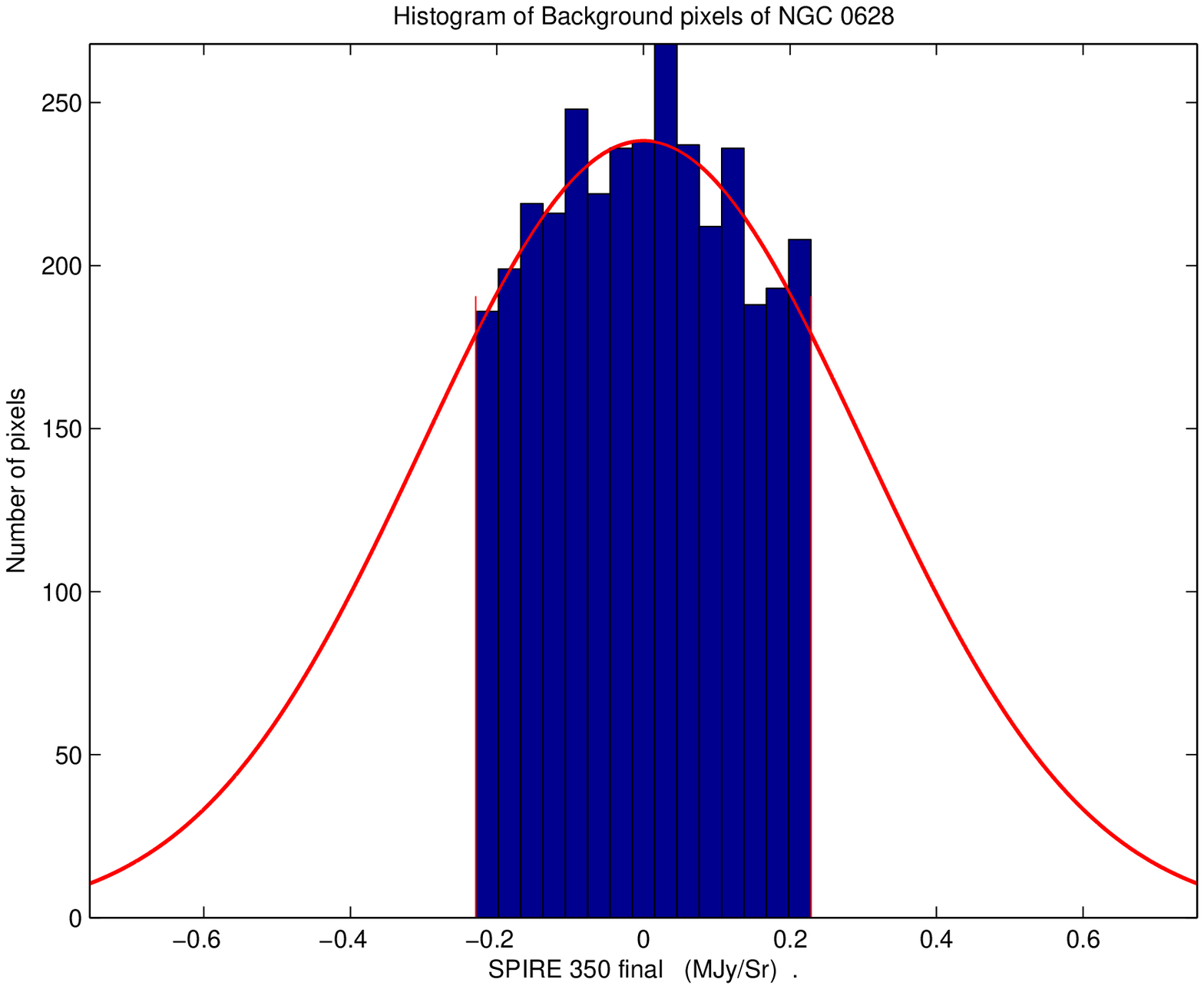}
\ifthenelse{\boolean{make_very_heavy}}{ }
{ \renewcommand \RoneCone    {No_image.eps}
\renewcommand \RoneCtwo    {No_image.eps}
\renewcommand \RtwoCone    {No_image.eps}
\renewcommand \RtwoCtwo    {No_image.eps}
\renewcommand \RthreeCone    {No_image.eps}
\renewcommand \RthreeCtwo    {No_image.eps}}

\renewcommand \RoneCone {NGC6946_Bck_Removed_IRAC_4oo5_Residuals_smooth.eps}
\renewcommand \RoneCtwo {NGC6946_Bck_Removed_IRAC_4oo5_Residuals.eps}
\renewcommand \RoneCthree {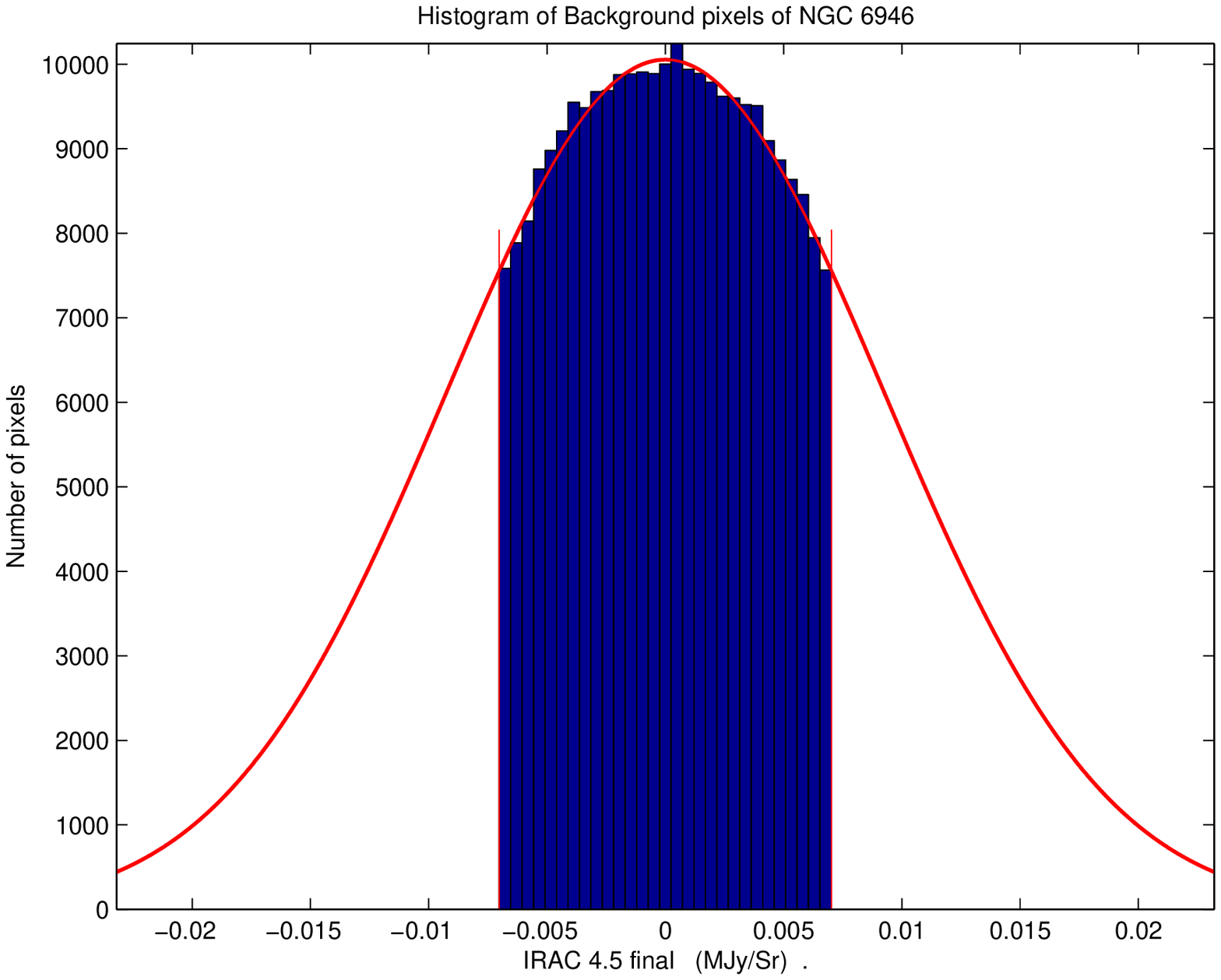}
\renewcommand \RtwoCone {NGC6946_Bck_Removed_MIPS_24_Residuals_smooth.eps}
\renewcommand \RtwoCtwo {NGC6946_Bck_Removed_MIPS_24_Residuals.eps}
\renewcommand \RtwoCthree {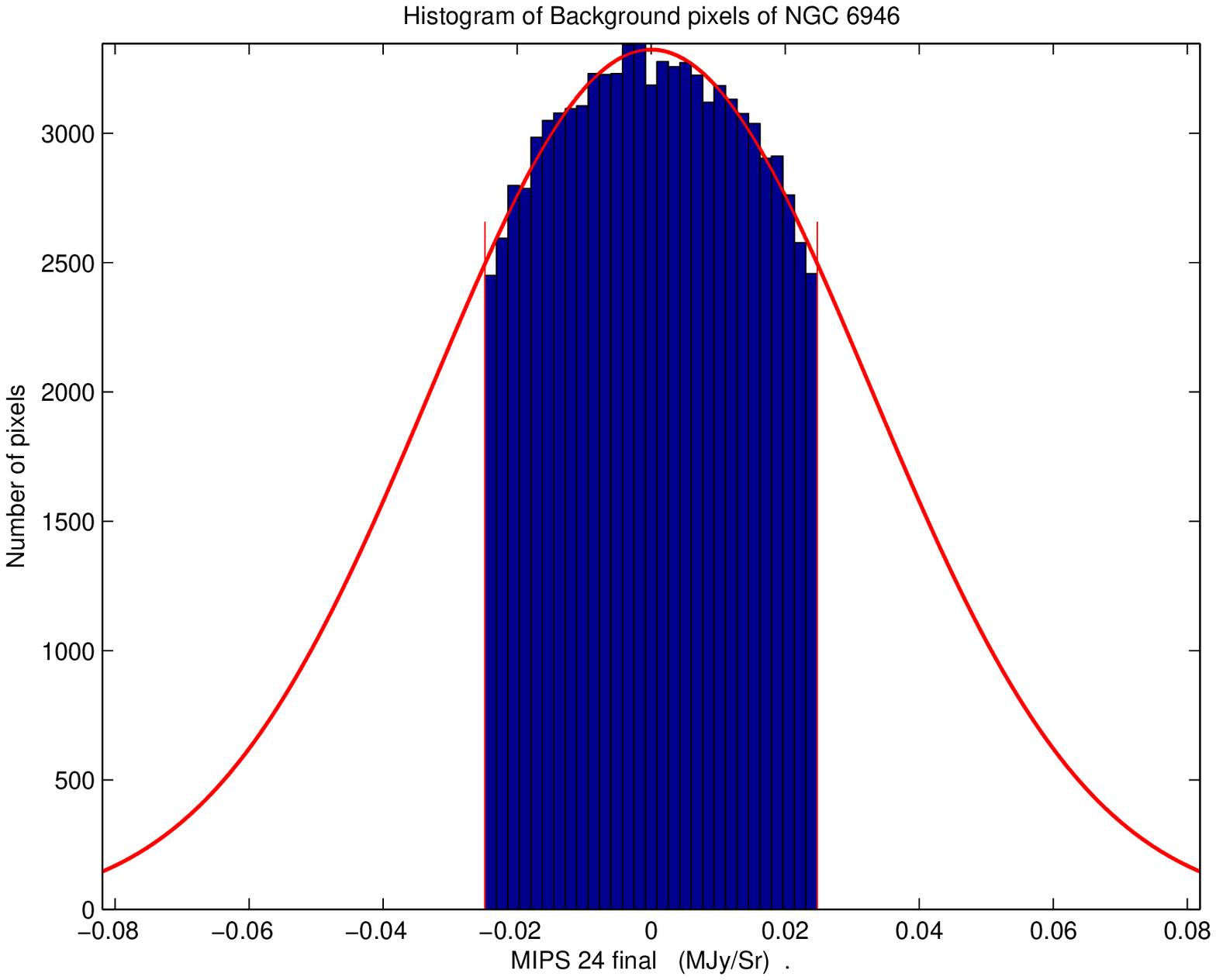}
\renewcommand \RthreeCone {NGC6946_Bck_Removed_PACSS_160_Residuals_smooth.eps}
\renewcommand \RthreeCtwo {NGC6946_Bck_Removed_PACSS_160_Residuals.eps}
\renewcommand \RthreeCthree {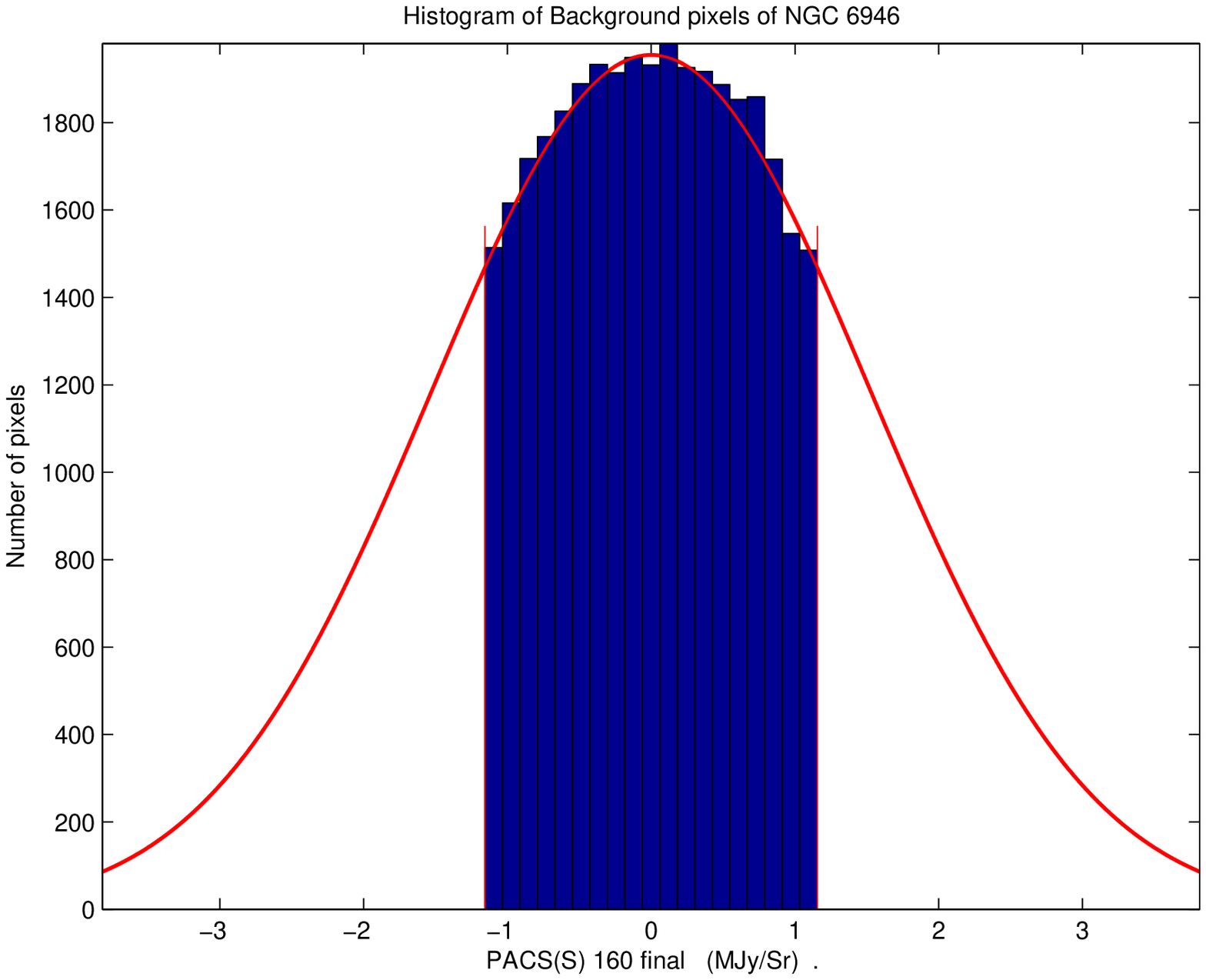}
\renewcommand \RfourCone {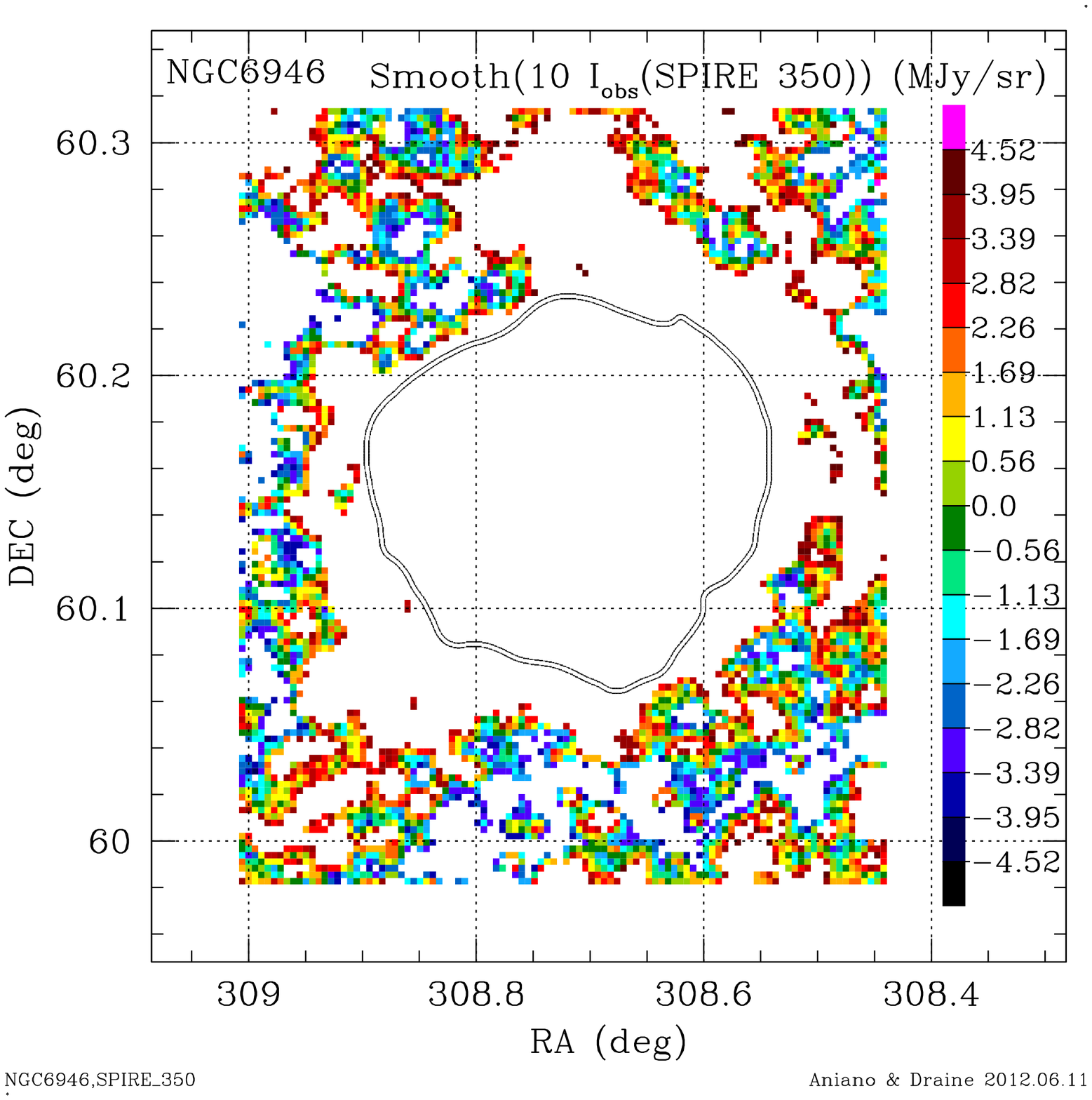}
\renewcommand \RfourCtwo {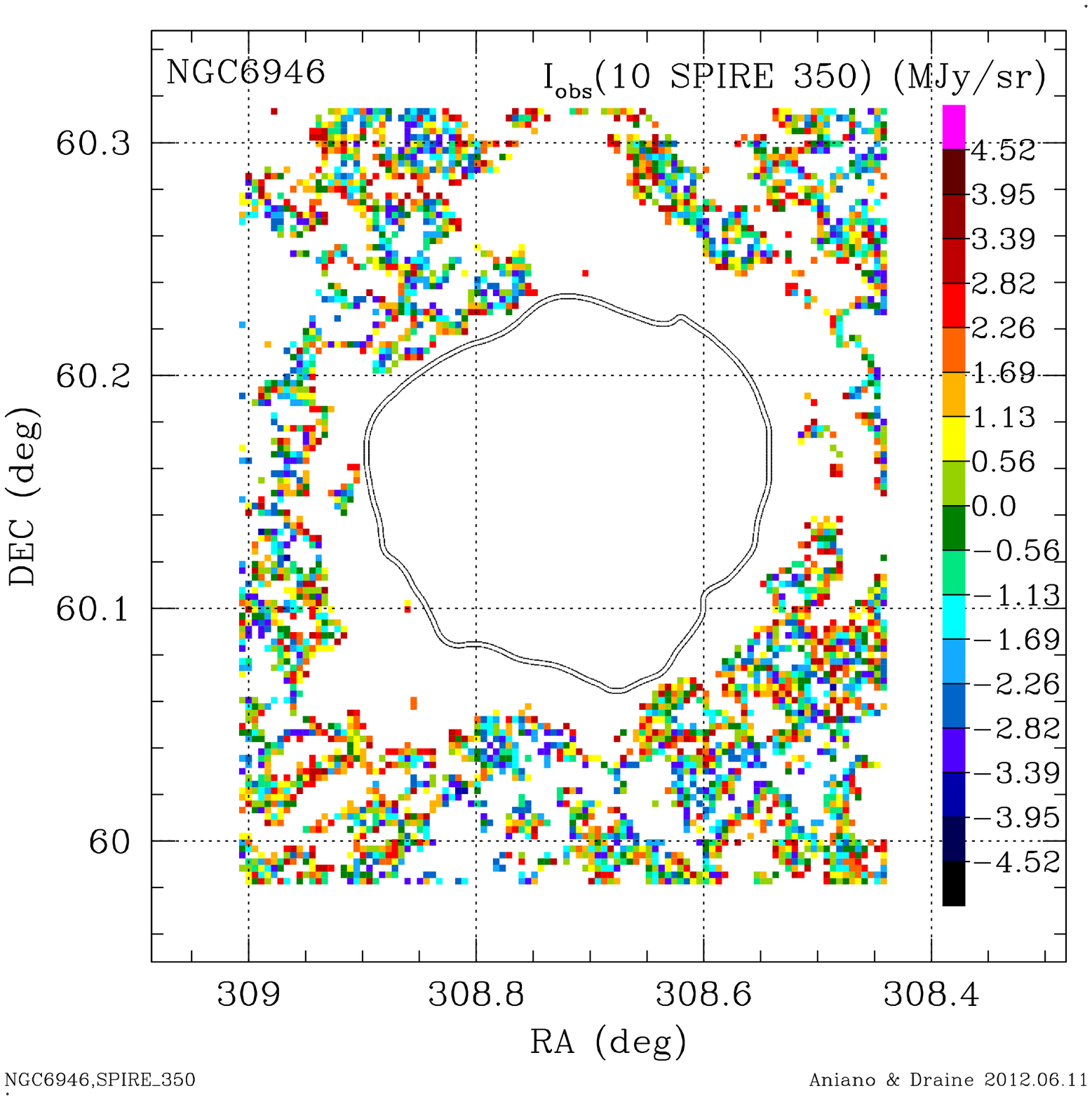}
\renewcommand \RfourCthree {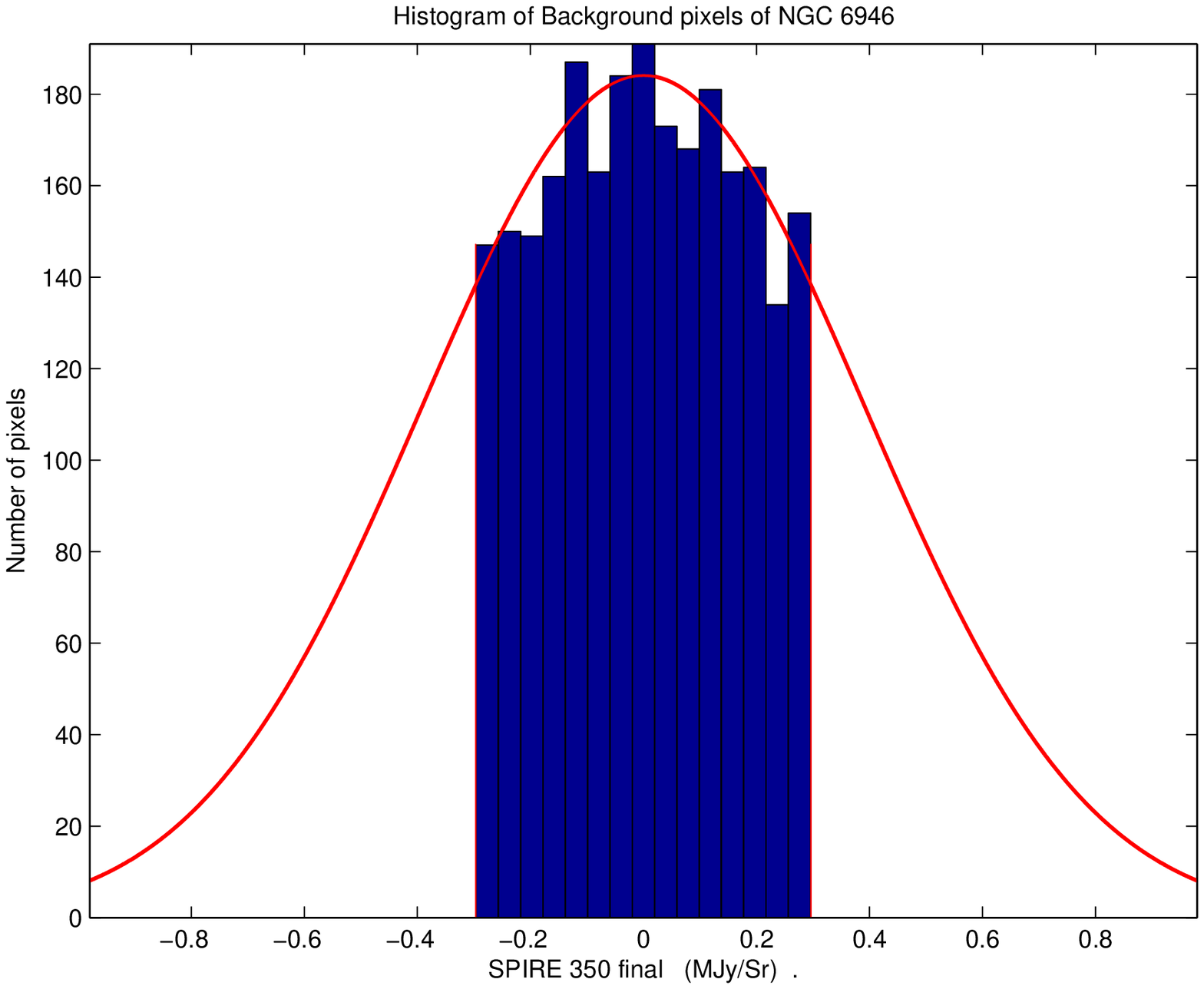}
\ifthenelse{\boolean{make_very_heavy}}{ }
{ \renewcommand \RoneCone    {No_image.eps}
\renewcommand \RoneCtwo    {No_image.eps}
\renewcommand \RtwoCone    {No_image.eps}
\renewcommand \RtwoCtwo    {No_image.eps}
\renewcommand \RthreeCone    {No_image.eps}
\renewcommand \RthreeCtwo    {No_image.eps}}


\section{\label{app:missing_data}Correction for Missing Data (Bad Pixels) in the Original Images}

Some of the original images include bad pixels due to detector
saturation, incomplete scanning, or other problems.  This effect is
not important when present in the pixels far away from the target
galaxy, but may have undesired consequences if not treated
appropriately when present near or within the target galaxy.  A
similar situation arises when observations are made over a small part
of a galaxy, for example with the IRS instrument onboard the Spitzer
Space Telescope. We develop a method for correcting this problem.

In order to estimate (and correct) the influence of bad pixels of a
camera $A$, we start by choosing a second camera $B$. We require that
the sky coverage of camera $B$ be larger than camera $A$, or at least
that does not present bad pixels in the same areas as $A$. From all
the possible candidates for being the correcting camera $B$, we choose
the camera with wavelength most similar to $A$. For example, we
typically correct the MIPS160 camera with the SPIRE250
camera\footnote{ We avoid correcting the MIPS160 camera with the
  PACS160 camera due to the inability of PACS160 camera to correctly
  measure the flux in extended emission regions. Moreover, the SPIRE
  maps extend to a larger area than the PACS maps.}.

We divide the original image grid of the camera $A$, $OG_A$, into two
set of pixels: $OG_{A,good}$, the pixels of $OG_A$ with good $A$ data,
and $OG_{A,bad}$, the pixels of $OG_A$ in which the camera $A$ has bad
data.

We interpolate the background-subtracted image $B$ from its original
grid $OG_B$ into the $OG_A$ grid, obtaining $B_{interp\,A}$, which
looks morphologically similar to $A$. We construct 2 images $B_{good}$
and $B_{bad}$, both defined on $OG_A$, as follow:

\beq 
B_{good}(x,y)  \in OG_A = 
\left \{ \begin{array}{lcc} 
B_{interp\,A} &\iff&(x,y) \in OG_{A,good} \\ 
0                                      &\iff&(x,y) \in OG_{A,bad} 
\end{array} \right.
\eeq

\beq 
B_{bad}(x,y)  \in OG_A = 
\left \{ \begin{array}{lcc} 
0                                     &\iff&(x,y) \in OG_{A,good} \\ 
B_{interp\,A} &\iff&(x,y) \in OG_{A,bad} 
\end{array} \right.
\eeq

We process the images $B_{good}$ and $B_{bad}$ exactly as we do with
the image $A$; we rotate them, convolve with the kernel that
corresponds to the image $A$, and interpolate them to the final grid
($FG$).  A given output pixel in $FG$ will have flux $flux_{good}B$,
and $flux_{bad}B$ in the final stage of processing the $B_{good}$ and
$B_{bad}$ images respectively.  We define $F_{A,B}(x,y)\in FG$, the
fraction of the total flux in the final grid coming from areas with
good $A$ data, as:

\beq
F_{A,B}={flux_{good}B\over flux_{good}B +flux_{bad}B}
\eeq

$F_{A,B}$ provides an estimate of how much flux we are missing in
each output pixel due to the missing data in the original image $A$.
In the pixels for which $F_{A,B} > 0.65$, (i.e, those where most of
the flux is actually coming from regions where we know $A$) we can
further compensate the final stage of the $A$ image processing by
multiplying the value obtained by $1/F$.  In the points $F_{A,B} < 0.65$
a significant amount of the flux (more than 35\%) is coming from
regions where $A$ is not known, so we mask and ignore them.

This correction algorithm would be exact if the color $A/B$ were
constant, and the PSFs of both cameras coincided.  We test the algorithm
by purposely removing some part of an image, and correcting it with
itself or another image.  In the former case, the results are exact
within numerical noise.  In the latter case, because we are correcting
an image $A$ with a different one $B$, the method is not exact, but in
most cases still provides a quite accurate result.


\section{\label{app:unc_estimation}Uncertainty Estimates in Common-Resolution Images}

The $\chi^2$ maps also shed light on the estimated flux uncertainties.
The value of 
\beq 
\chi^2 \equiv \sum_{k}
\frac{[F(\lambda_k)-F(\model,\lambda_k)]^2} {\sigma_{\lambda_k}^2}
\eeq 
depends not only on the ability of the model to fit the data, but
also on the adopted uncertainties $\sigma_{\lambda_k}$.  For a given
data set and fitted model, underestimation of the uncertainties
$\sigma_{\lambda_k}^2$ would lead to higher vales of $\chi^2$.  Recall
that the $\sigma_{\lambda_k}$ were estimated by studying the
pixel-to-pixel variations in the background regions outside the galaxy
mask. If part of the background variations actually comes from
real variations in emission from foreground dust, or dusty background
galaxies, then our dust model procedure may actually be able to
partially fit the background variations, leading to a low value of
$\chi^2$.  On the other hand, image artifacts and departures of the
real data from the models will lead to higher values of $\chi^2$. It
is clear, for example, that the differences in the PACS and MIPS
photometry must lead to an increase in the $\chi^2$ values, since the
model cannot fit simultaneously two different flux values.

If we have $N$ measures from an ideal model with $M$ adjustable
parameters, each measure having independent gaussian noise with
variance $\sigma_{\lambda_k}^2$, then the $\chi^2$ of the fit will
have a chi-square distribution with $\rm{dof}$ degrees of freedom,
where $\rm{dof} = N-M$. The $\chi^2/\rm{dof}$ map will have mean 1,
and dispersion $\sqrt{2/\rm{dof}}$.

\begin{figure}[h]
\centering\includegraphics[width=10cm]{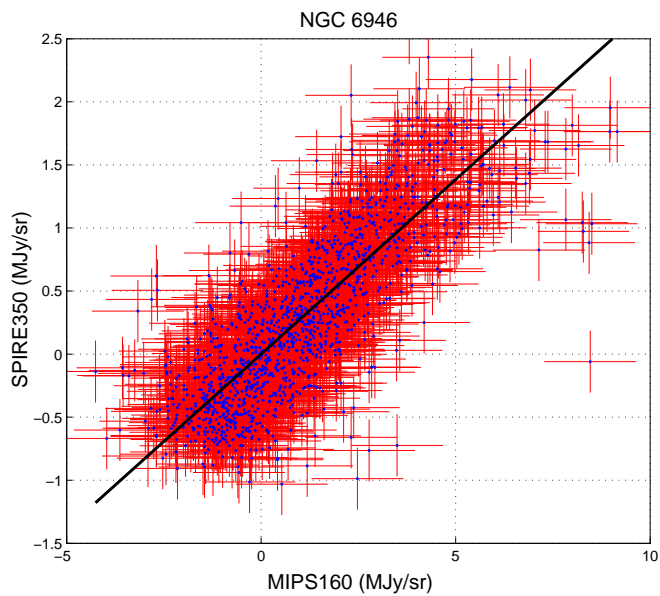}
\footnotesize
\vspace*{-0.5cm}
\caption{\footnotesize\label{fig:bck_sct} Background-subtracted
  intensities in the MIPS160 and SPIRE350 cameras, for background
  pixels around NGC~6946.  There is a strong correlation between the
  MIPS160 \um and SPIRE350 \um fluxes, indicating that the dispersion
  has a component coming from unresolved dim astrophysical sources.}
\end{figure}

In our background noise estimation procedure, after we recognize and
mask all the background sources and subtract a ``tilted plane'', in
each image there remains a zero-mean background due to unresolved dim
sources. We use the dispersion of the background pixels (with respect
to the best fit plane) to estimate the stochastic background
uncertainty.  This provides a valid uncertainty estimate since the
same background structures will be present over the galaxy, so our
galaxy fluxes will be uncertain (at least) at this level.  However,
this background dispersion is not only due to random noise in the
detectors, it also includes a contribution from foreground and
background sources. Figure \ref{fig:bck_sct} shows the correlation of
the dispersion of the background pixels from the MIPS160 (horizontal
axis) and SPIRE350 (vertical axis) cameras.  Although the residual
distributions have zero mean, there is a strong correlation between
the MIPS160 and SPIRE350 flux in the background pixels, demostrating a
component coming from real astrophysical sources. Other cameras show
similar correlations.  The modeling can fit these correlated
departures from zero by adding positive or negative amounts of real
dust, reproducing the departures and therefore decreasing the $\chi^2$
of the fit. We therefore expect to have $\chi^2/\rm{dof} < 1$ in areas
where the flux uncertainties have an important contribution from
unresolved dim sources (i.e., in the background pixels or low surface
brightness areas).

Figure \ref{fig:chi_0628} shows the contribution of different cameras
to $\chi^2$ for NCG 628. The fit is done at MIPS160 resolution, using
all 13 cameras available.  We employ the recommended dust model with 6
adjusted parameters $(\Omega_\star,\Mdust,\qpah,\gamma, \Umin,
\alpha)$ leading to a fit with $\rm{dof}=7$.  The upper left panel (a)
shows the $\chi^2/\rm{dof}$ map. If the data were artifact-free,
perfectly modeled, with independent gaussian noise, the map would have
mean 1, and dispersion $\sqrt{2/7}\approx0.53$. The actual
$\chi^2/\rm{dof}$ values are slightly above 1 in the bright regions
and below 1 in the outer parts of the galaxy. In the outer parts of
the galaxies, the correlation of the noise between the cameras drives
the $\chi^2/\rm{dof}$ below the expected value 1. In the bright spots
of the galaxy, the differences in PACS and MIPS photometry decrease
the quality of the fit. We estimate the PACS and MIPS photometry
uncertainties in a way that their contribution to the $\chi^2$ maps is
controlled and consistent, but the model will try to fit some
intermediate (probably non-physical) value which may impair its
ability of accurately fitting  the remaining cameras.

In order to remedy this situation, uncertainties for the PACS and MIPS
photometry are estimated by comparing the PACS and MIPS images (see
below). The uncertainties are slightly overestimated, thus giving more
weight to the remaining artifact-free cameras. Figure
\ref{fig:chi_0628}b shows the PACS(S)70/MIPS70 ratio map. Stripes
aligned with the MIPS sky scanning directions near the galaxy center
suggest the presence of MIPS image artifacts. Discrepancies in the low
surface brightness areas (outer parts of the galaxy) suggest
differences in the ability of the telescopes to measure the low
surface brightness areas correctly. The top right panel (c) shows the
assigned MIPS70 signal/noise map. The procedure described in Appendix
D.1 assigns large uncertainties to the areas where the PACS and MIPS
images have discrepancies, leading to reduced S/N. This is a critical
step toward getting $\chi^2$ maps that really measure the goodness of
the model fit, not only the data discrepancies.

The lower row of Figure \ref{fig:chi_0628} shows the contribution
$\Delta\chi^2$ of selected cameras to the total $\chi^2$ for NGC~628.
The lower left panel (d) shows the contribution of MIPS24 camera. We
note that this contribution is extremely low, under 0.05 for most of
the galaxy.  This is because the dust model can adjust the 24\um flux
relatively independently of the other observed bands by adjusting
$\gamma$ and $\alpha$. The 24\um band is unique in lying a factor of 3
in wavelength away from the nearest other bands (8\um and 70\um).
Figure \ref{fig:chi_0628}e is the contribution of MIPS70\um to the
total $\chi^2$.  Even though the MIPS70 camera has severe artifacts,
its $\Delta\chi^2$ near the center is controlled by the large values
of $\sigma$ assigned.  Figure \ref{fig:chi_0628}f shows $\Delta\chi^2$
for SPIRE350.  The camera behaves exactly as expected, contributing
$\Delta\chi^2\approx 0.6$ near the galaxy nucleus and smaller
$\Delta\chi^2$ in the low surface brightness areas. The behavior for
NGC~6946 is similar.

\renewcommand \RoneCone {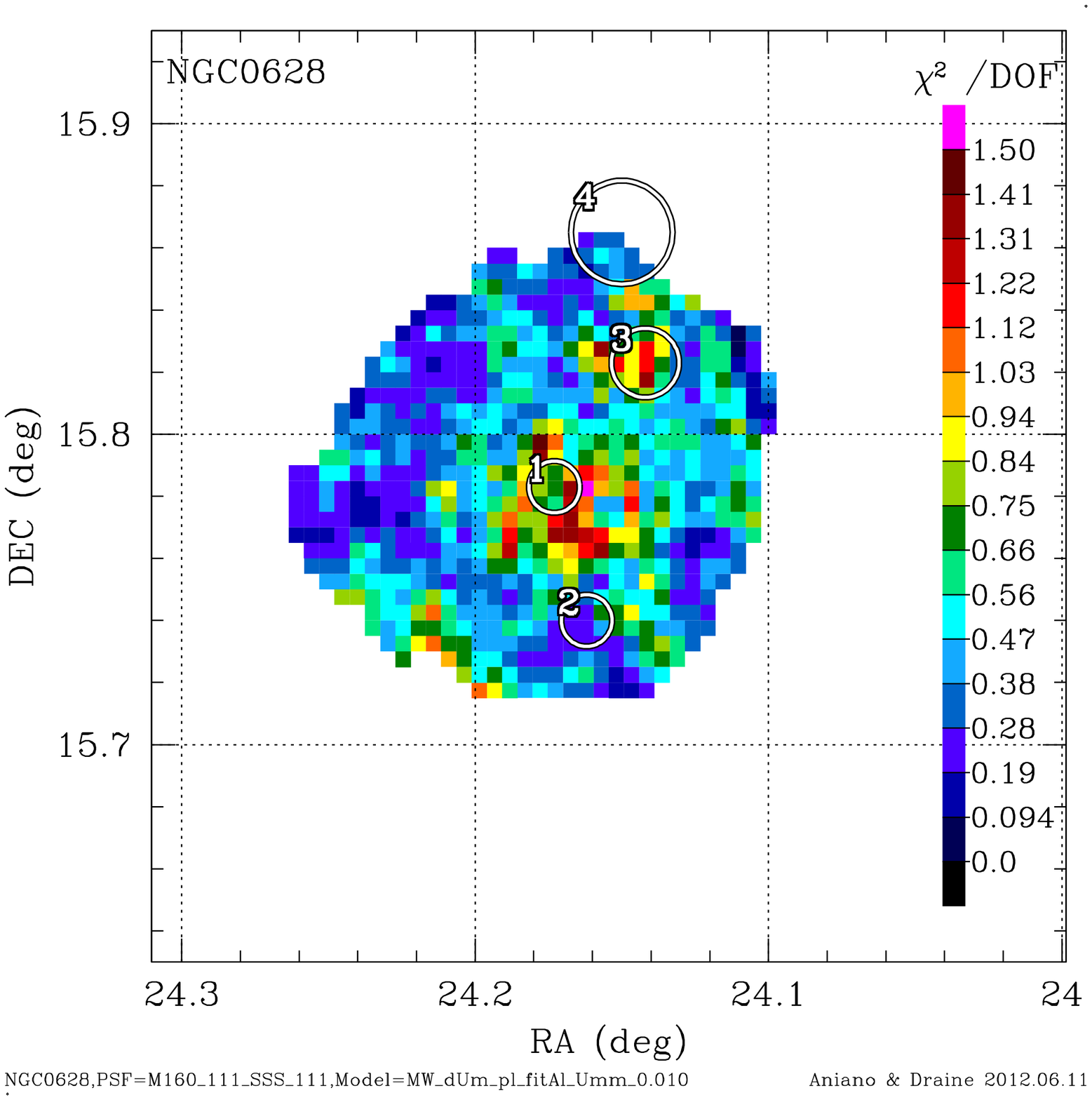}
\renewcommand \RoneCtwo {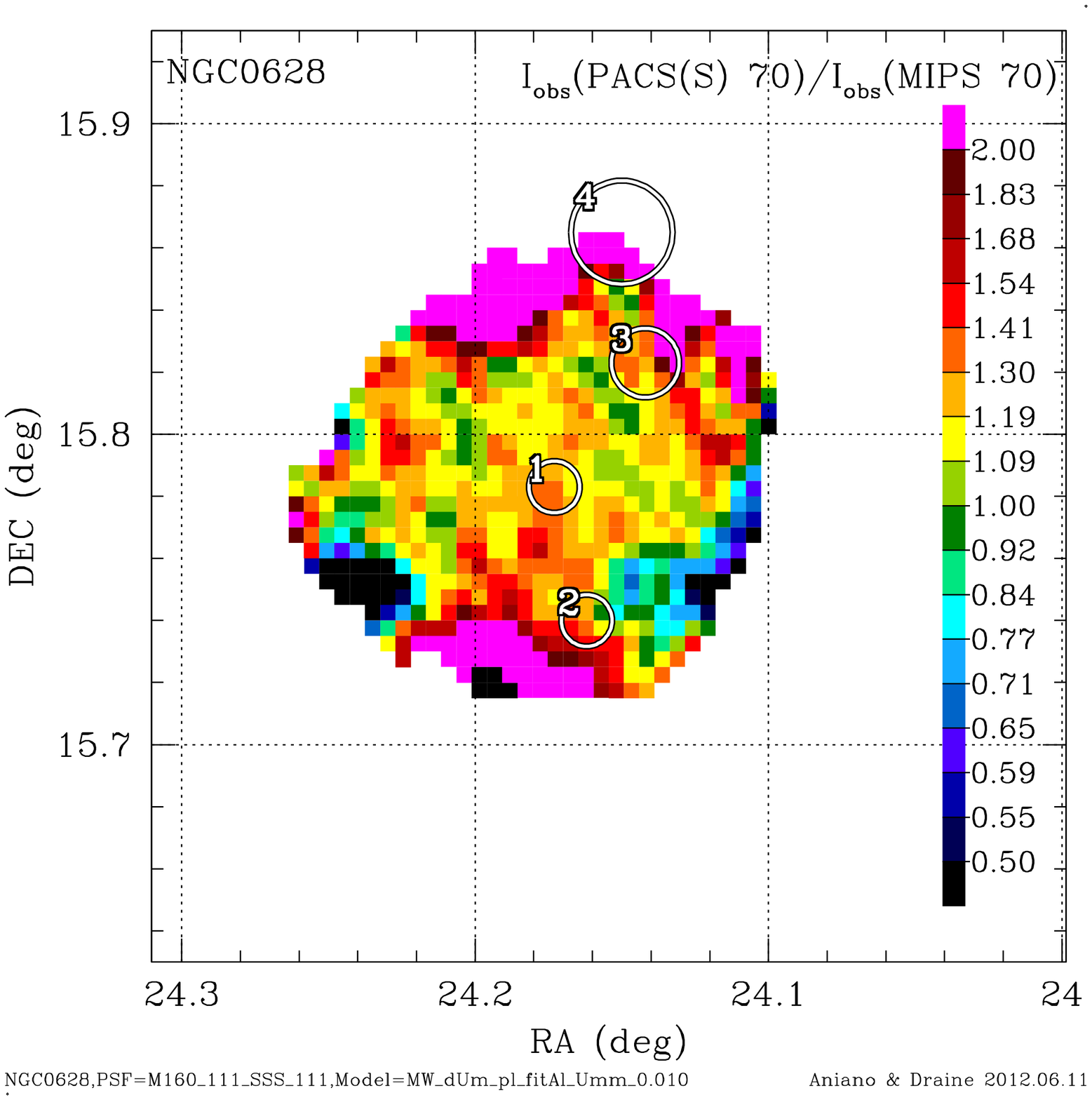}
\renewcommand \RoneCthree {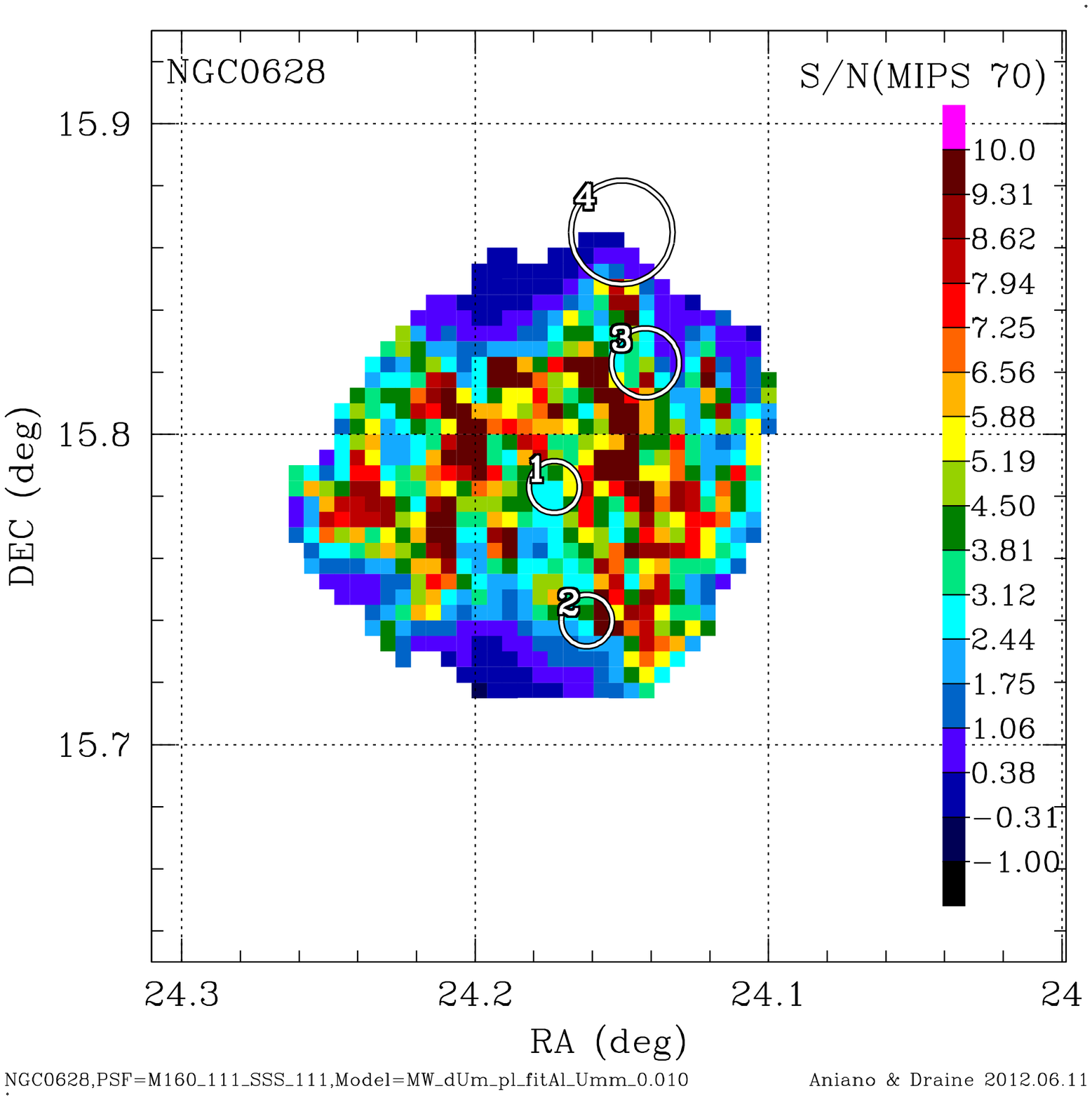} 
\renewcommand \RtwoCone {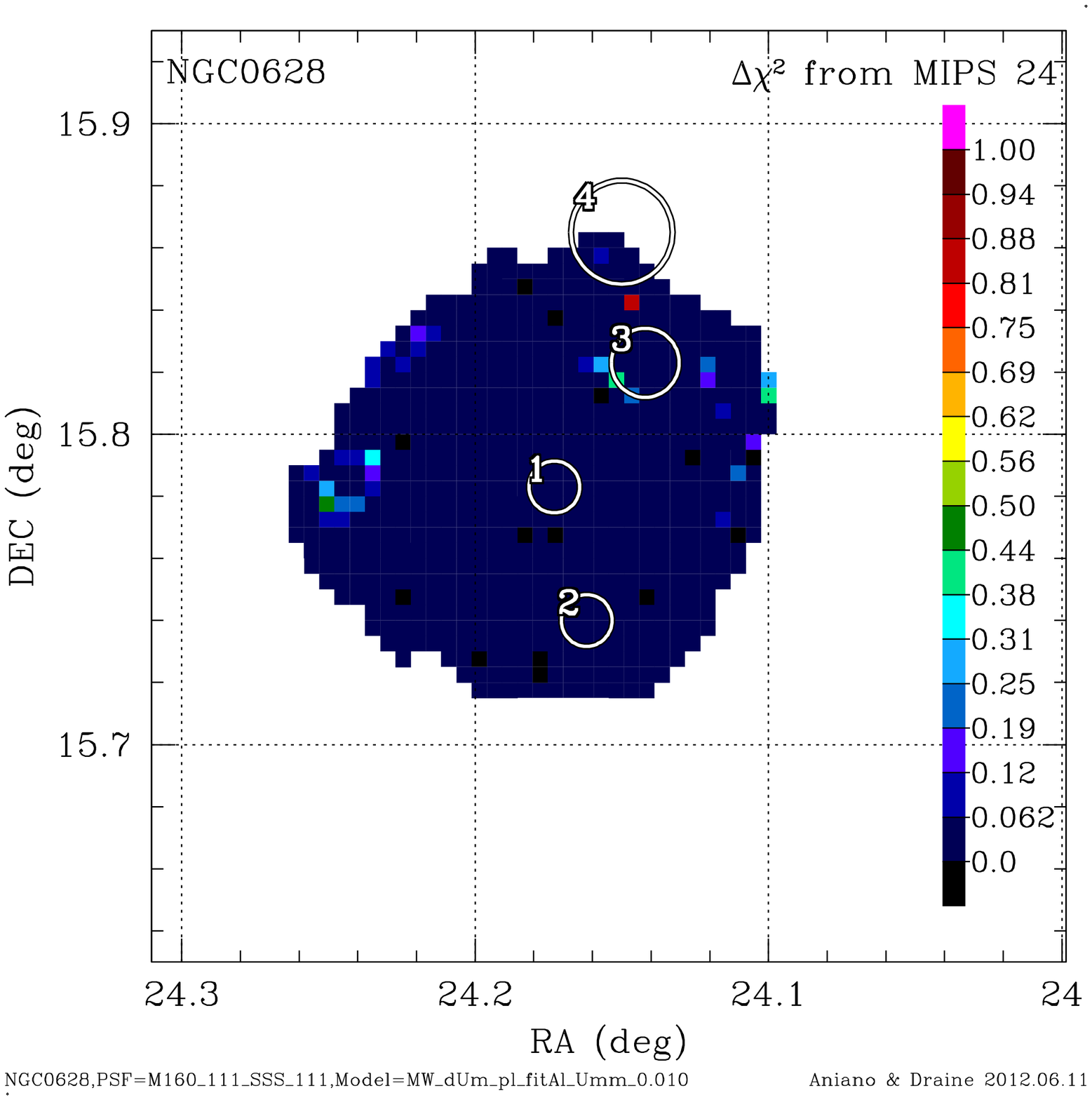}
\renewcommand \RtwoCtwo {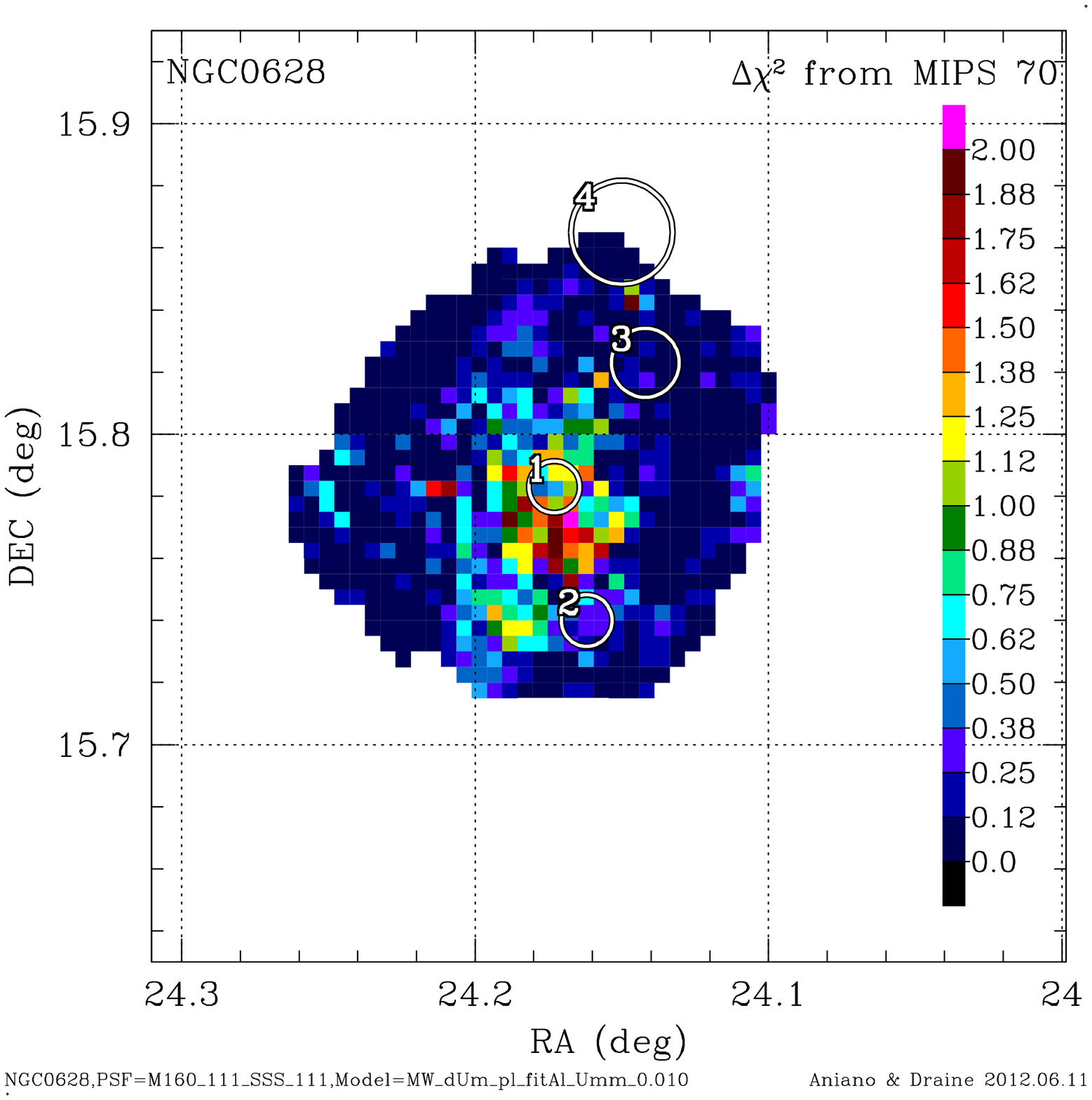}
\renewcommand \RtwoCthree {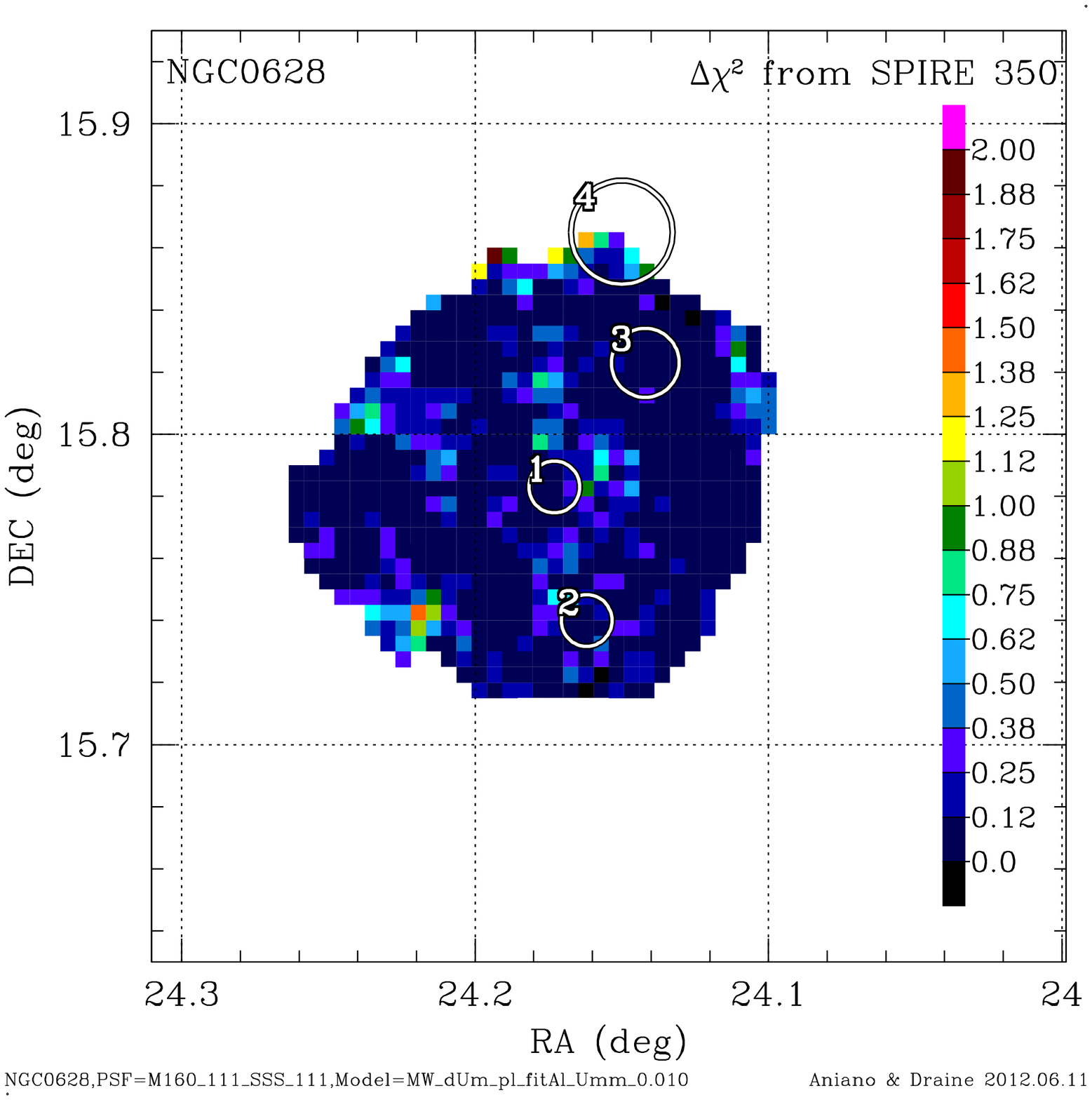} 
\begin{figure} 
\centering 
\begin{tabular}{c@{$\,$}c@{$\,$}c} 
\FirstNormal
\SecondLast
\end{tabular} 
\vspace*{-0.5cm}
\caption{\footnotesize Analysis of the $\chi^2$ contributions for
  NCG~628. The fit is done at MIPS160 resolution, using all 13 cameras
  with 6 adjusted parameters $(\Omega_\star,\Mdust,\qpah,\gamma,
  \Umin, \alpha)$.  Top left panel: $\chi^2/\rm{dof}$ map.  Top center
  panel: PACS70/MIPS70 ratio map.  Top right panel: estimated
  signal/noise map for MIPS70.  The lower row shows the contribution
  $\Delta\chi^2$ of selected cameras to the total $\chi^2$: left panel
  shows MIPS24 camera, center panel MIPS70, and right panel shows
  SPIRE350.  The $\Delta\chi^2$ behaves as expected. 
  \label{fig:chi_0628}}
\end{figure} 

\subsection{Uncertainty Estimate Procedure}

In order to estimate uncertainties for the different cameras, we
proceed as follows.

For each camera of wavelength $\lambda_k$, after the image processing
(rotation to RA-Dec, background subtraction, convolution to a common
PSF, correction for boundary effects and missing data, and resampling
to the final grid) the flux in each final pixel is a (known) linear
combination of the flux of (in principle) all the original pixels of
the camera.  If the statistical properties of the uncertainties in the
original pixel fluxes were known, it would be possible to infer the
uncertainties (and their statistical properties) in each final pixel.
The original maps oversample the beam, and have artifacts that extend
over several pixels, so realistic statistical properties of the
uncertainties are difficult to determine.  We therefore estimate the
uncertainties directly in the final (post-processed) image.

We always assume that the pixel noise is not correlated between different cameras, so in what follows we work in each image independently, and omit the subindex identifying the camera in the following discussion.
For each pixel, we will consider two independent sources of uncertainties $\delta_{sto}$ and $\delta_{sys}$. 
The term $\delta_{sto}$ will be the uncertainty due to stochastic pixel noise and the existence of (unresolved) background dim sources. 
We estimate it by measuring the flux fluctuations in the background pixels.  
The term $\delta_{sys}$ will include systematic and calibration uncertainties.

In order to estimate the term $\delta_{sto}$, the stochastic pixel noise, we proceed as follows.
We re-grid the {\em background mask} (BM) to the final-map grid to
define a set of $N_{\rm bg}$ ``background pixels''.  For each camera
$k$, we compute the background dispersion $\delta_{sto,\lambda_k}$ as: 
\beq
\delta_{sto,\lambda_k} = \sqrt{\frac{1}{(N_{\rm bg}-3)}\sum_{{\rm (x,y)\in
      \rm{(BM)}}}{\left[I_{\lambda_k}^{obs}(x,y)\right]^2}}, 
\label{eq_sto}
\eeq
where $I_{\lambda_k}^{obs}$ is the background-subtracted flux of the camera of wavelength
$\lambda_k$ in the pixel $(x,y)$ in the last stage of the image
processing (i.e., after the image has been rotated to RA-Dec, 
background plane subtracted, convolved to a common PSF, corrected for
boundaries effects and missing data, and resampled to the final grid).
 The background dispersion $\delta_{sto,\lambda_k}$ includes the
propagation of original pixel noise and artifacts into the background
pixels, possible non-ideal background subtraction and the
contributions of unidentified faint background
sources. $\delta_{sto,\lambda_k}$ does not contain information on the
correlation of the noise among the different pixels or calibration
uncertainties.

For each camera, we consider the pixel-dependent uncertainty
$\delta_{sys,j}$ as 
\beq 
\delta_{sys,j} =\max \{ \epsilon \times
I_j^{obs},K_j\} ~~~,
\label{eq_sys}
\eeq 
where $\epsilon$ is the calibration uncertainty for the camera;
$K$ is 0 for IRAC, MIPS24 and SPIRE cameras and an estimate of
systematic uncertainties for MIPS70,160 and PACS cameras (where such
estimation is possible).  The main difficulty in estimating $\epsilon$
is that the cameras are typically calibrated using point sources.  We
are performing resolved studies of an extended sources, so we have
additional uncertainties due to the extended source corrections.

For the IRAC cameras, the standard calibration uncertainty is
$\epsilon = 0.05$ for point sources
\citep{Dale+Bendo+Engelbracht+etal_2005}, but we will adopt a larger
value to account for the uncertainties in the extended emission
correction.  The IRAC images are calibrated for point sources.  We
multiplied the intensities by the asymptotic (infinite radii) value of
the aperture correction factors $C= 0.91, 0.94, 0.66, 0.74$ for the
3.6\um, 4.5\um, 5.8\um and 8.0\um bands respectively.  This correction
would lead to a correctly calibrated image if the source had constant
surface brightness.  Since galaxies contain nonuniform emission, the
aperture correction factor should, in principle, be different in each
region.  We multiply the entire image by the factors $C$, and
arbitrarily assign 1/3 of the correction as uncertainty.  We use $
\epsilon = 0.05 + 1/3 \times \left( 1/C - 1\right) = 0.083, 0.071,
0.221, 0.167$ for the 3.6\um, 4.5\um, 5.8\um and 8.0\um bands,
respectively.  These corrections will overestimate the uncertainties
in the diffuse regions and underestimate the uncertainties near bright
point sources.

For the MIPS, PACS and the SPIRE cameras we adopt
$\epsilon=0.10$. This value is larger than the estimated $4\%$
calibration uncertainty for point sources for MIPS24 \citep{Engelbracht+Blaylock+Su+etal_2007}, $\approx 4\%$ for  PACS \citep{Muller+Nielbock+Balog+etal_2011},
and $7\%$ for SPIRE\footnote{ See the SPIRE ObserversÕ Manual 
HERSCHEL-DOC-0798, version 2.4}, and
comparable to estimates for MIPS70 \citep{Gordon+Engelbracht+Fadda+etal_2007} and MIPS160 \citep{Stansberry+Gordon+Bhattacharya+etal_2007}.  
For the MIPS70,160 and PACS cameras, however, $\delta_{sys}$ will be dominated
by $K$ in most of the pixels.

For MIPS70 and MIPS160, we can estimate the systematic uncertainties
$K$ by comparing the PACS and MIPS resolved photometry. In most of the
pixels, the observed differences are larger than expected due to the
differences in the camera spectral responses or calibration
uncertainties. For PACS70 and PACS160 we can estimate the systematic
uncertainties $K$ by comparing the PACS and MIPS resolved photometry
For the PACS100 camera we can estimate the systematic uncertainties
$K$ by extrapolating the uncertainties estimated for the PACS70, and
PACS160 cameras. We proceed as follows.

\subsubsection {Systematic Uncertainties for MIPS70 and MIPS160 Cameras.\label{PACS-MIPS-1}}

Since for the 70\um and 160\um bands the PACS FWHM are smaller than
the corresponding MIPS FWHM (see Table \ref{tab:resolutions}), in all
the common-PSF maps where we employ MIPS data we have PACS data as
well.  Therefore, we can compare PACS and MIPS images after the image
processing. The comparison is straightforward since both images are in
the same final-map grid and PSF.

We consider the difference: 
\beq 
D_{{\rm PM},j} \equiv | I_{{\rm
    PACS(S)},j}^{obs} - I_{{\rm MIPS},j}^{obs}| ~~~, 
\eeq 
where
$I_{{\rm MIPS},j}^{obs}$ and $I_{{\rm PACS(S)},j}^{obs}$ are the
observed MIPS and PACS (Scanamorphos) flux, in the same band (70\um or
160\um).  We consider the Scanamorphos data reduction more reliable
than the HIPE, so we only employ the Scanamorphos images in the PACS -
MIPS comparison.  The term $D_{{\rm PM},j}$ captures both image
artifacts and differences in camera performance and calibration.
Unfortunately, artifacts in the PACS cameras will induce uncertainties
in the MIPS fluxes and vice-versa, but there is no robust, automatic
way of isolating the image artifacts.  We take 
\beq 
K_{{\rm MIPS},j} =
D_{{\rm PM},j}\,.  
\eeq

\subsubsection{Systematic Uncertainties for PACS70 and PACS160 Cameras.\label{PACS-MIPS-2}}

Based on our dust modeling, we consider the Scanamorphos data
reduction to be more reliable than the HIPE-only data reduction. When
the MIPS and PACS cameras for a band (70\um or 160\um) are used in a
final-map PSF (e.g., MIPS70 and PACS70 can be used if the final-map
PSF is SPIRE350, SPIRE500 or MIPS160) we proceed in a similar way as
for the MIPS bands, taking 
\beq 
K_{{\rm PACS},j} = D_{{\rm PM},j} \,.
\eeq

When the MIPS camera for a band (70\um or 160\um) is not used in a
given final-map PSF (e.g., MIPS70 cannot be used if the modelling is
done at PACS160 PSF), the comparison between PACS and MIPS is more
complex.  We start by convolving the PACS image into the corresponding
MIPS PSF, and re-binning it into the MIPS native pixel grid, denoted
as $I_{{\rm PACS@M}}^{obs}$.  We can compare $I_{{\rm PACS@M}}^{obs}$
and $I_{{\rm MIPS}}^{obs}$ since both images are expressed at the same
grid and have the same PSF.  We construct $\epsilon_{{\rm PM}}$, the
fractional PACS-MIPS difference (with respect to the PACS image), at
the MIPS PSF, as 
\beq 
\epsilon_{{\rm PM},J} = {| I_{{\rm
      PACS(S)@M},J}^{obs} - I_{{\rm MIPS},J}^{obs}| \over | I_{{\rm
      PACS(S)@M},J}^{obs} |}\,.  
\eeq 
We re-bin $\epsilon_{{\rm PM}}$
into the final-map grid, and we take 
\beq 
E_{{\rm PM},j}' =
\epsilon_{{\rm PM},J} \times | I_{{\rm PACS(S)},j}^{obs}| ~~~, 
\eeq 
where
$J$ is the pixel in the ${\rm PACS@M}$ image grid that contains pixel
$j$.  This procedure produces an uncertainty image $E_{{\rm PM},j}$ at
the final-map grid and PSF, that if degraded into the corresponding
MIPS PSF, would be $\sim | I_{{\rm PACS(S)@M}}^{obs} - I_{{\rm
    MIPS}}^{obs}|$.  In this situation, we finally take 
\beq 
K_{{\rm PACS},j} =E_{{\rm PM},j}\,.  
\eeq

\subsubsection{Systematic Uncertainties for PACS100 Camera.}

\begin{figure} 
\centering 
\begin{tabular}{c@{$\,$}c@{$\,$}c} 
\includegraphics[width=5.7cm,height=5.3cm,clip=true,trim=0.3cm 0.7cm 0.0cm 0.3cm]
{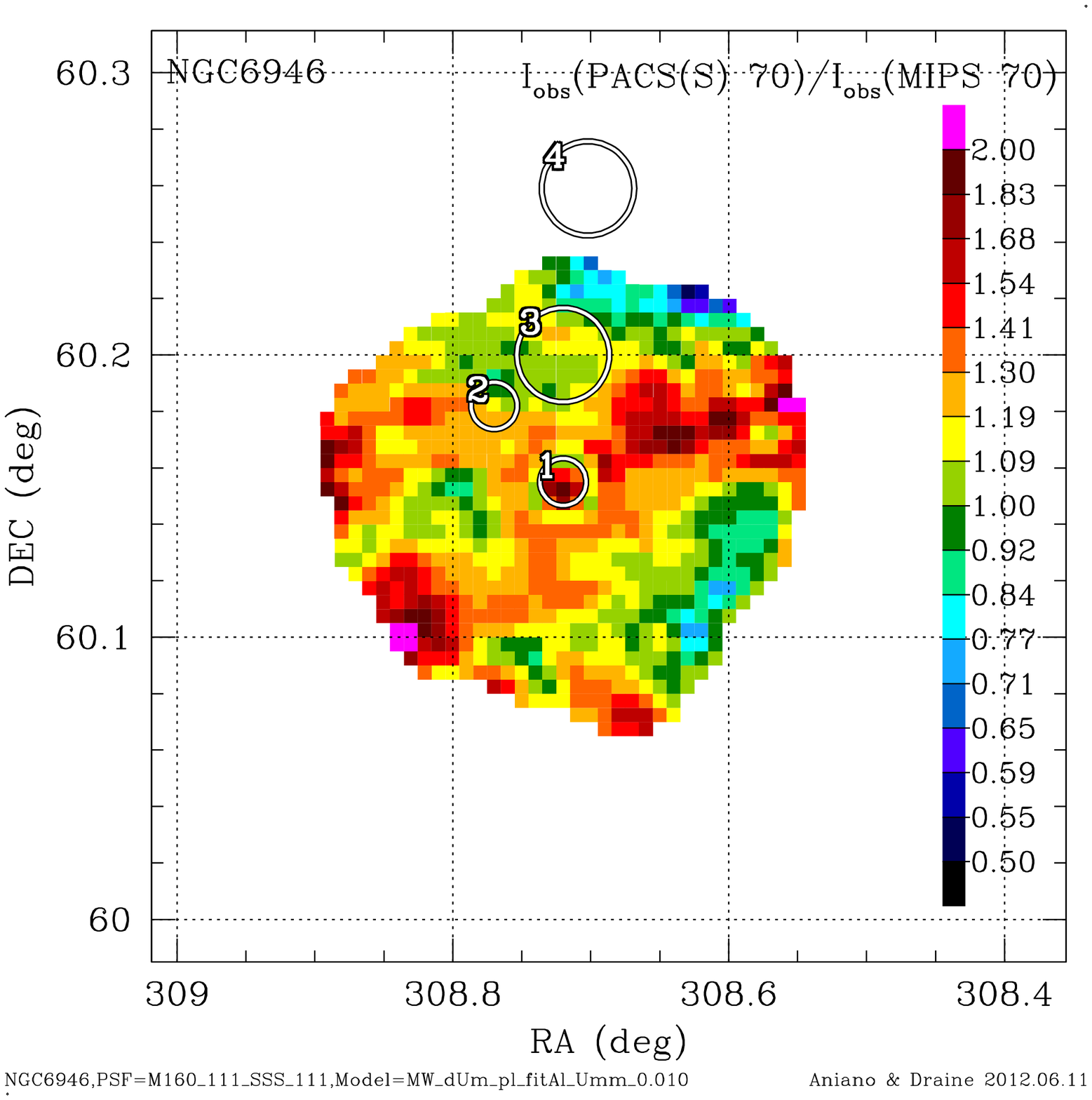}
&
\includegraphics[width=4.9cm,height=5.3cm,clip=true,trim=2.8cm 0.7cm 0.0cm 0.3cm]
{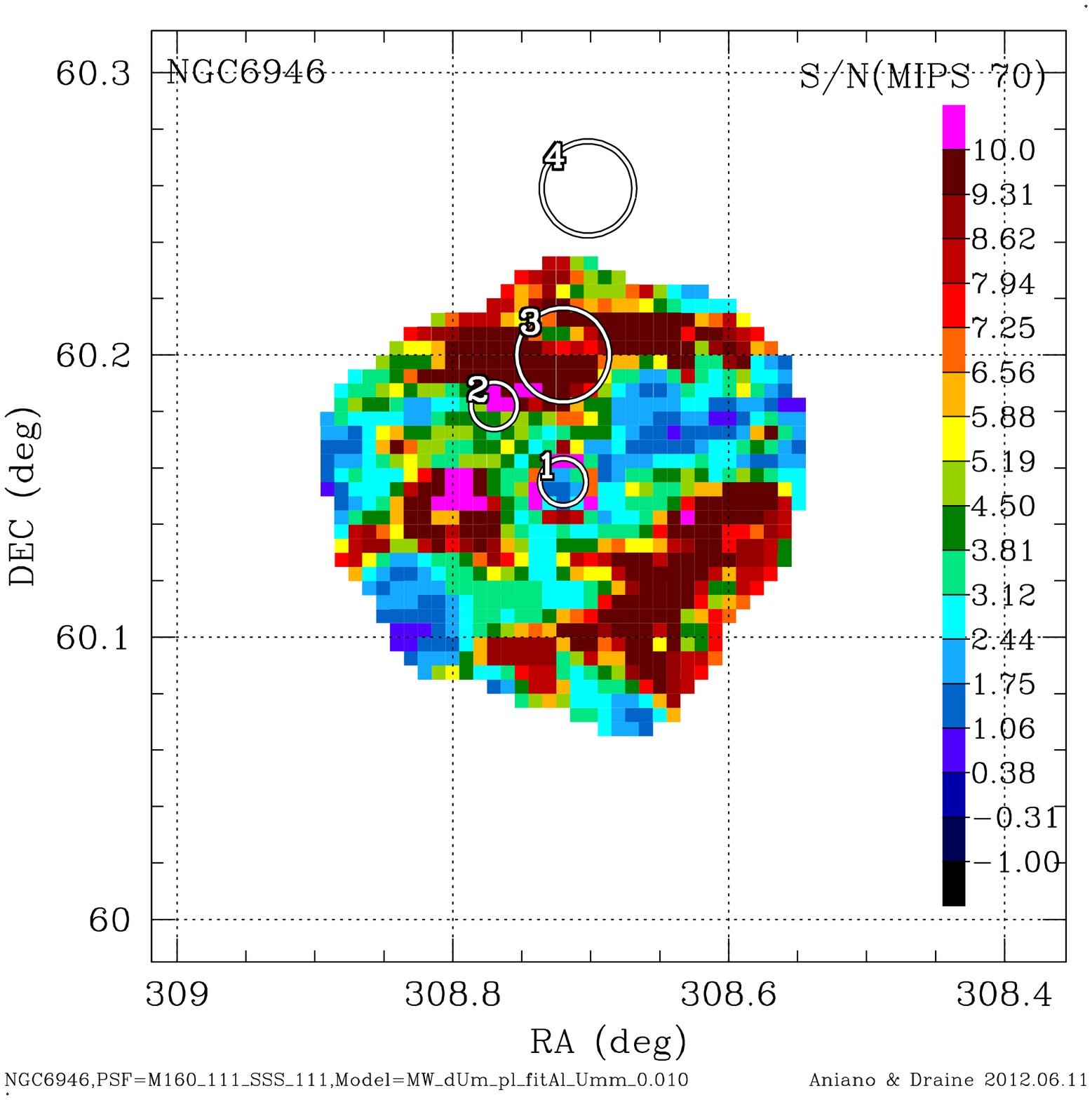}
&
\includegraphics[width=4.9cm,height=5.3cm,clip=true,trim=2.8cm 0.7cm 0.0cm 0.3cm]
{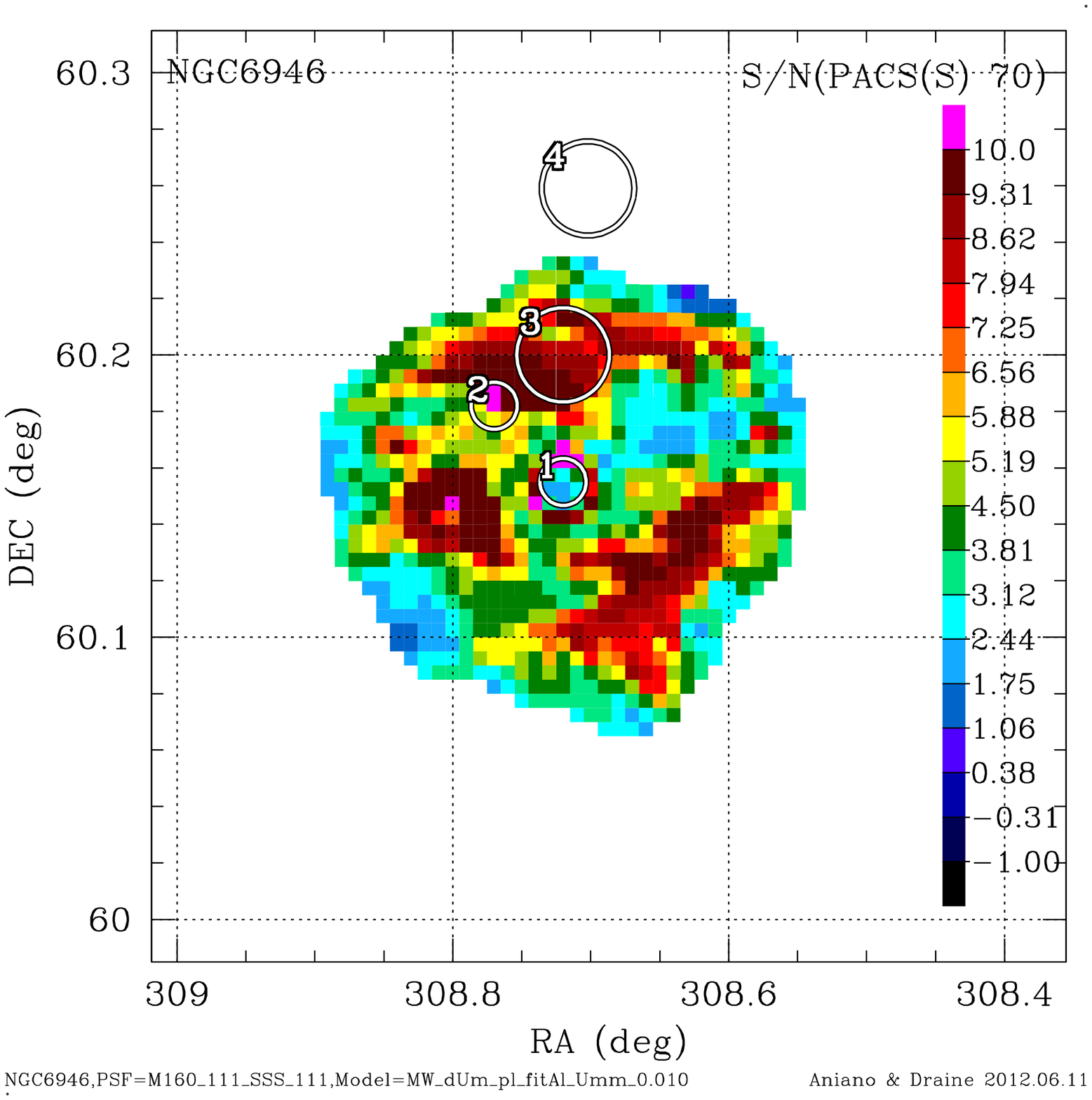}
\end{tabular}
\vspace*{-0.5cm}
\caption{\footnotesize 70\um images of NGC~6946 convolved into MIPS160
  PSF. Left: map of PACS70/MIPS70. Center: S/N map of MIPS70. Right:
  $S/N$ map of PACS70. The image artifact that extends from the
  galaxy nucleus toward the WNW induces a low $S/N$ area in the
  cameras. This artifact, aligned with the MIPS sky scanning
  direction, is probably due to saturation, latency or non-linearities
  in the MIPS detector.  In other areas with PACS70/MIPS70
  $\not\approx 1$ it is not clear which camera is responsible for the
  bad behaviour.}
\label{correction}
\end{figure}

In the case of PACS100, we do not have any other camera available to
perform a direct comparison.  We therefore estimate the uncertainties
by extrapolating the uncertainties previously calculated for PACS70
and PACS160 as follows. We therefore will assume:
\beq
K_{{\rm PACS100},j} = \max\{ K_{70,j}',K_{160,j}' \} ~~~,
\eeq
where
\beqa
K_{70,j}'   &=&{K_{{\rm PACS70},j} \over  | I_{{\rm PACS(S)70},j}^{obs}| } \times| I_{{\rm PACS(S)100},j}^{obs}|
\\
K_{160,j}' &=&{K_{{\rm PACS160},j} \over  | I_{{\rm PACS(S)160},j}^{obs}| } \times| I_{{\rm PACS(S)100},j}^{obs}|~~~.
\eeqa
Figure \ref{correction} shows the resulting $S/N$ estimates for MIPS70 and PACS70 observations of NGC~6946.

\section{\label{app:unc_parameters}Uncertainty Estimates for Model Parameters}

In order to estimate the model parameter uncertainties, we generate a
set $r=1,2,...,N_{r}$ realizations $I_{\lambda_k,j,r}$ including
random noise, for each pixel $j$, modeling the noise
$I_{\lambda_k,j,r}-I_{\lambda_k,j}^{obs}$ in pixel $j$ as a sum of 3
components:
\beq
I_{\lambda_k,j,r}=I_{\lambda_k,j}^{obs} + \delta I_{\lambda_k,j,r,ind}+\delta I_{\lambda_k,r,cor}+\delta I_{\lambda_k,j,r,sys}.
\eeq
$\delta I_{\lambda_k,j,r,ind}$ represents the non-correlated noise
component of the pixel $j$, while $\delta I_{\lambda_k,r,cor}$ and
$\delta I_{\lambda_k,j,r,sys}$ represent the correlated noise
component and systematic/calibration uncertainties, assumed to be
(completely) correlated among the different pixels of the camera.
$\delta I_{\lambda_k,j,r,ind}$ is drawn from an independent gaussian
distribution for each pixel, while $\delta I_{\lambda_k,r,cor}$ and
$\delta I_{\lambda_k,j,r, sys}$ are drawn from a global gaussian
distribution, properly normalized for each pixel.

For each camera, we generate a set of $N\times N_{r}$ ``pixel''
gaussian variables, with zero mean and variance 1,
$\eta_{j,r},j=1,2,...N,\,r=1,2,...,N_{r}$, where $N$ is the total
number of pixels in the image, and a set of $2N_{r}$ ``global''
gaussian variables, with zero mean and variance 1,
$\psi_1,\psi_2...,\psi_{N_{r}},\phi_1,\phi_2,...,\phi_{N_{r}}$.

We set:
\beqa 
\delta I_{\lambda_k,j,r,ind}&= &\sqrt{1/2} \times \delta_{sto,\lambda_k}  \times \eta_{j,r}\\
\delta I_{\lambda_k,r,cor}   &=& \sqrt{1/2} \times \delta_{sto,\lambda_k}  \times \psi_r\\
\delta I_{\lambda_k,j,r,sys}        &=& \delta_{sys,\lambda_k}  \times \phi_r
~~~, 
\eeqa
where $\delta_{sto,\lambda_k}$ and $\delta_{sys,\lambda_k}$ were defined by eq. \ref{eq_sto} and eq. \ref{eq_sys}.
The division of the dispersion $\delta_{sto,\lambda_k}$  into equal correlated and uncorrelated
components $\delta I_{\lambda_k,j,r,ind}$ and $\delta
I_{\lambda_k,r,cor}$ is arbitrary, but seems to capture the
main features of the noise.
A more sophisticated treatment would use the statistic of the background
region to characterize the correlation of the noise in pixels as a function
of their separation, but this is beyond the scope of the present work.

For each pixel, the 1-$\sigma$ uncertainty in the measured flux
density $\sigma_{\lambda_k,j}$ used to calculate the $\chi_j^2$ of the
model (see equation \ref{eq:chi2}) is given by:
\beq
\sigma_{\lambda_k,j} =  
\sqrt{ (\delta_{sto,\lambda_k})^2+ (\delta_{sys,\lambda_k})^2} ~~~.
\eeq
Uncertainties for the inferred (pixel) model parameters are computed
using equation \ref{eq.var1}.
In order to compute the uncertainties in the global quantities, we
use equation \ref{eq.var1} on the global
quantities, i.e., for each random realization we compute the global
inferred quantity and finally we compute the dispersion on the
inferred global quantities. This approach preserves the correlation in
the pixel noise consistently into the global uncertainties.


\section{\label{sec:importance of MIPS}Instrument Comparison: PACS-MIPS Photometry disagreement}

In \S 7 we show that maps made at PACS160 resolution (using IRAC, MIPS24, and
PACS data only) are less reliable. 
Furthermore, maps using IRAC and PACS data at MIPS160 resolution, provide less accurate\footnote{
That is, the resulting dust/H mass ratios seem less ``reasonable''} 
results than maps computed with IRAC and MIPS data using the same PSF.
We therefore proceed to analyze the differences in the PACS and MIPS photometry, to shed light onto the roots of such discrepancies.

\renewcommand \RoneCone {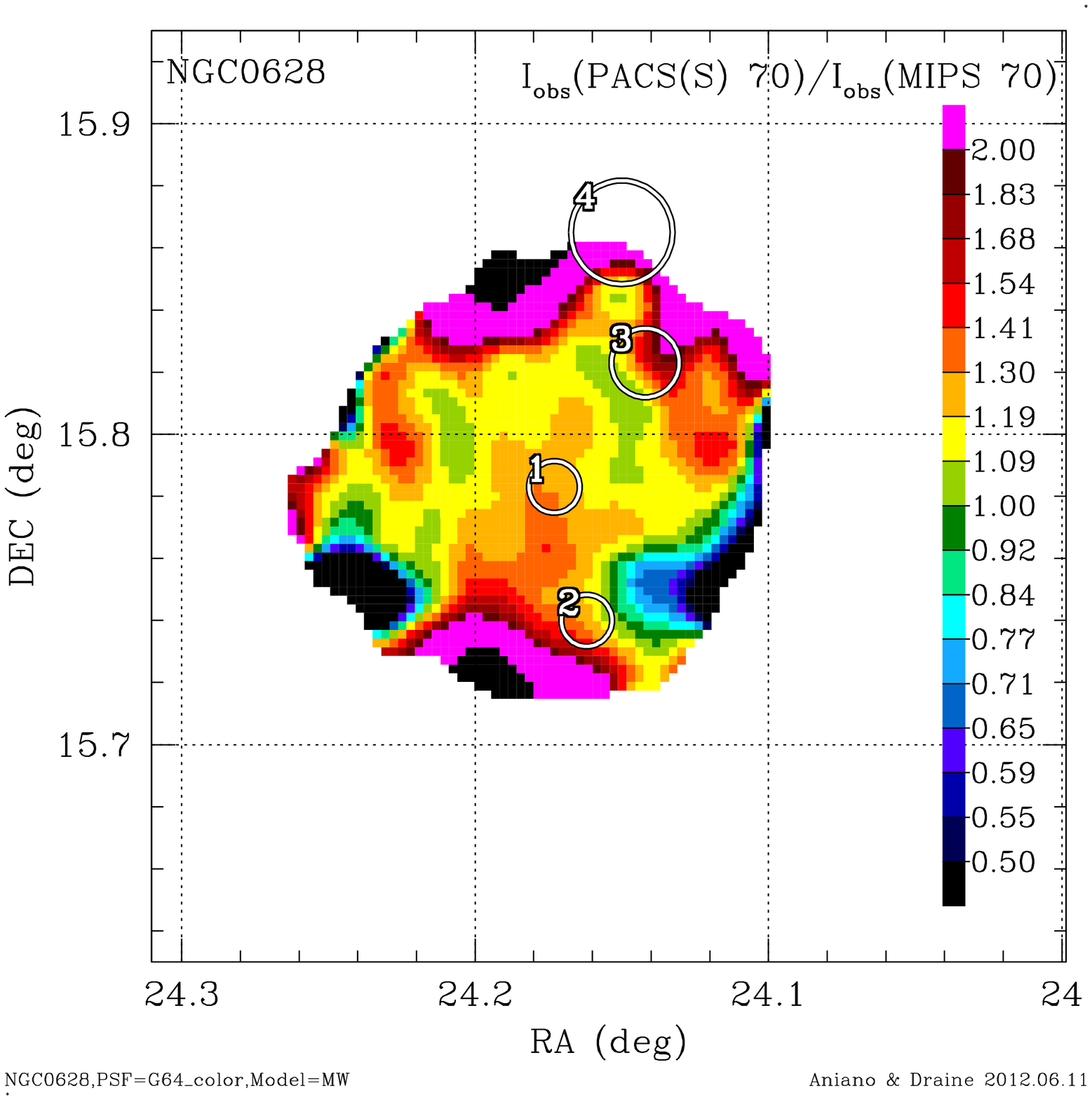}
\renewcommand \RoneCtwo {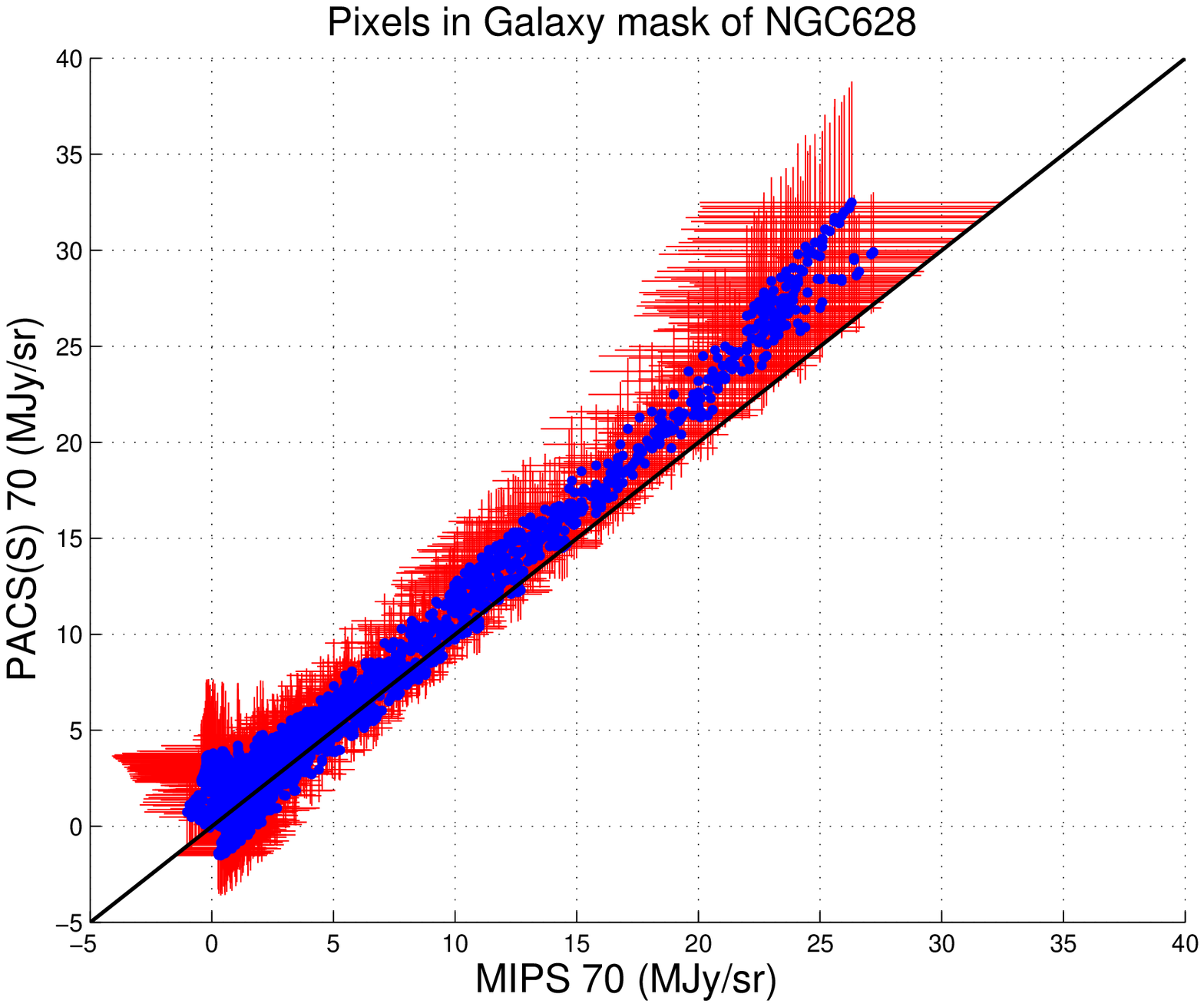}
\renewcommand \RoneCthree {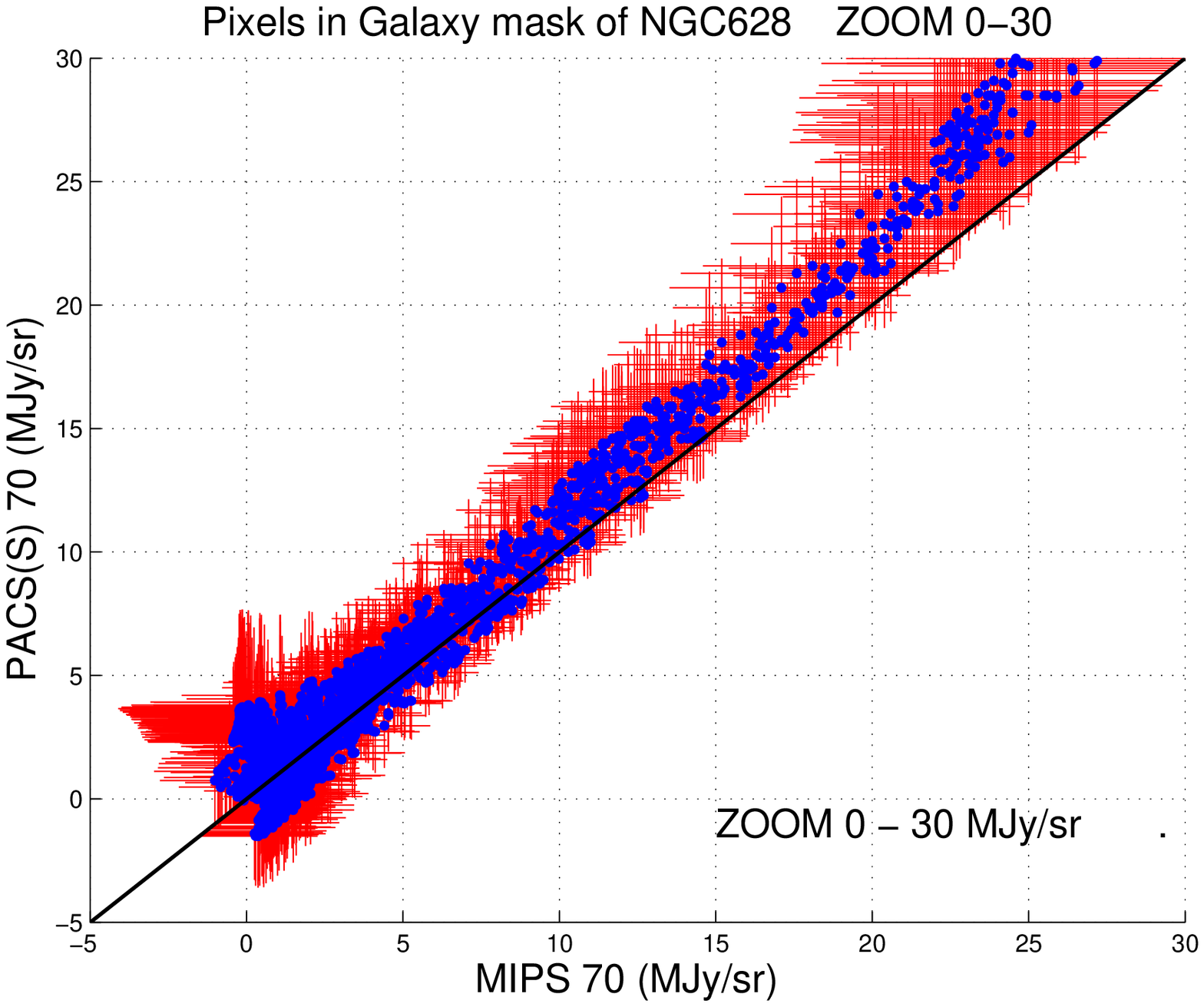}
\renewcommand \RtwoCone {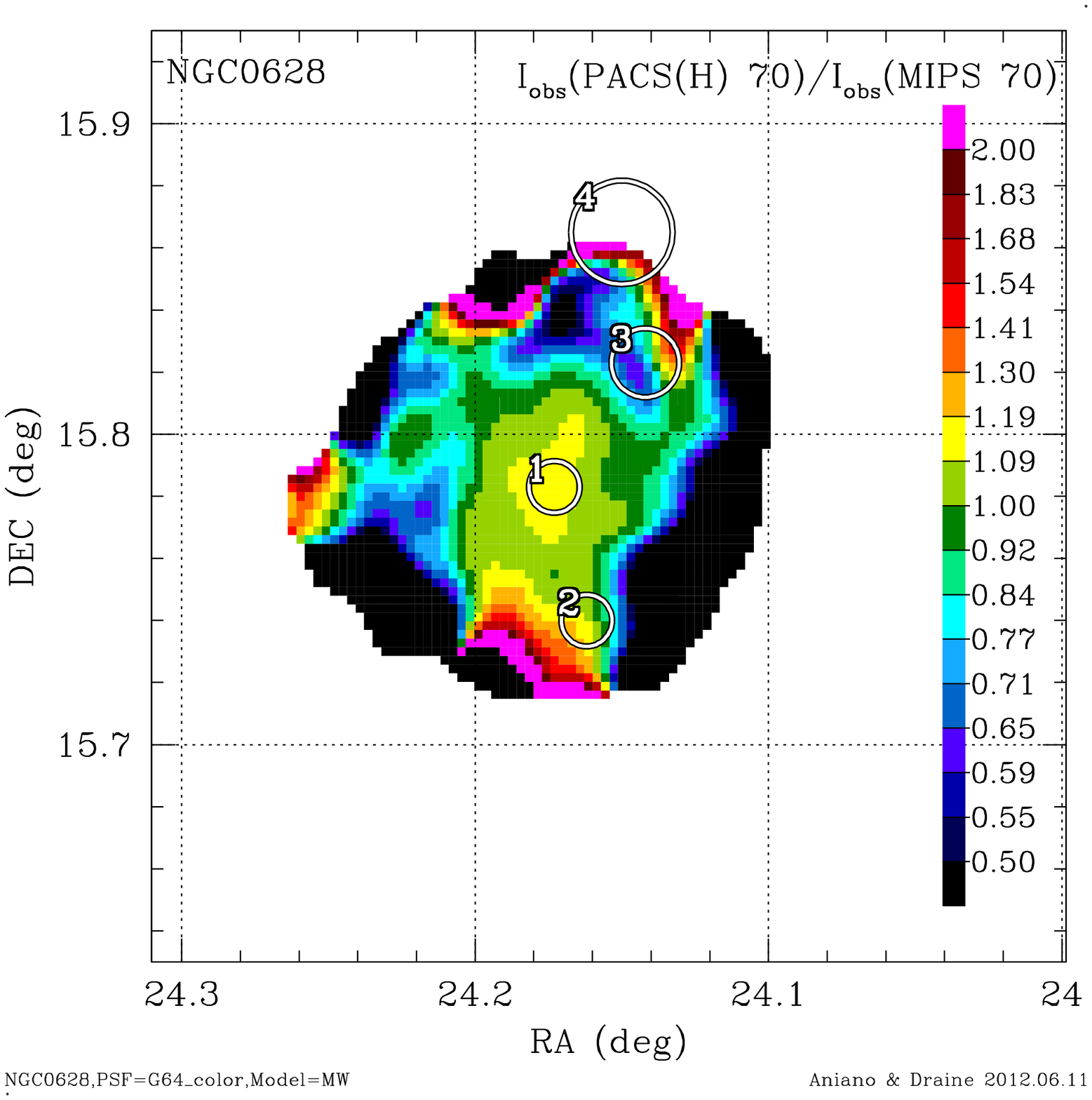}
\renewcommand \RtwoCtwo {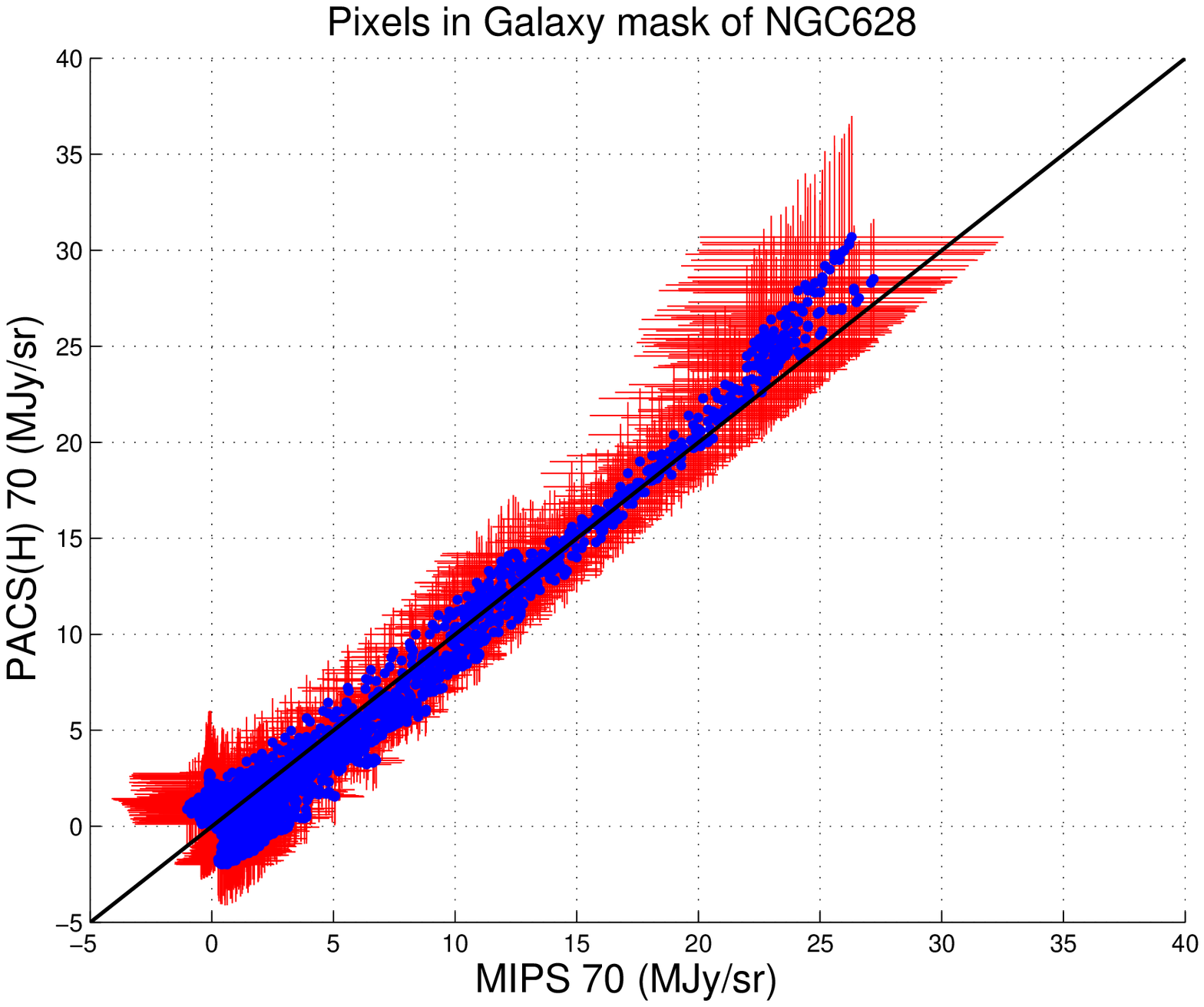}
\renewcommand \RtwoCthree {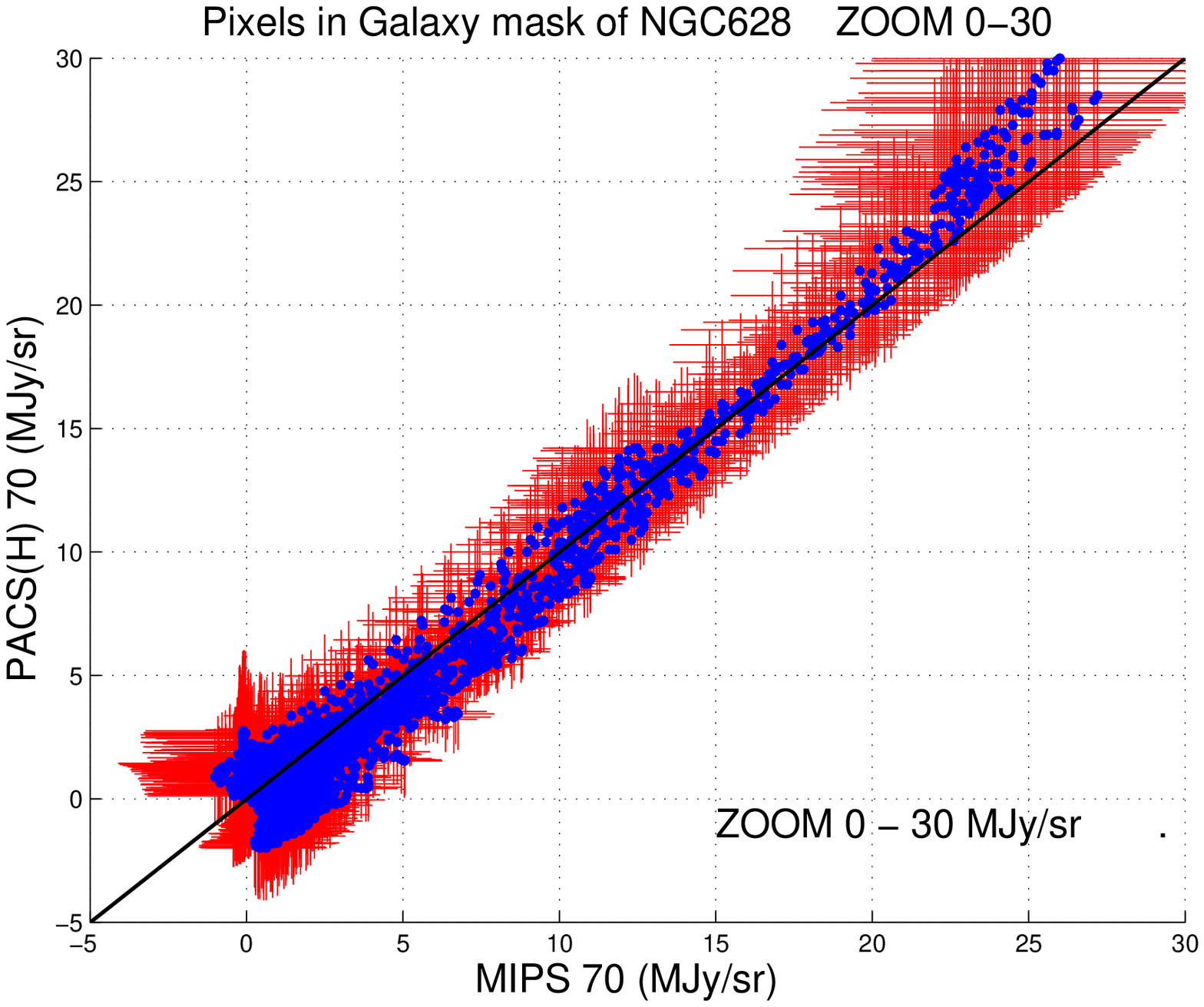}
\renewcommand \RthreeCone {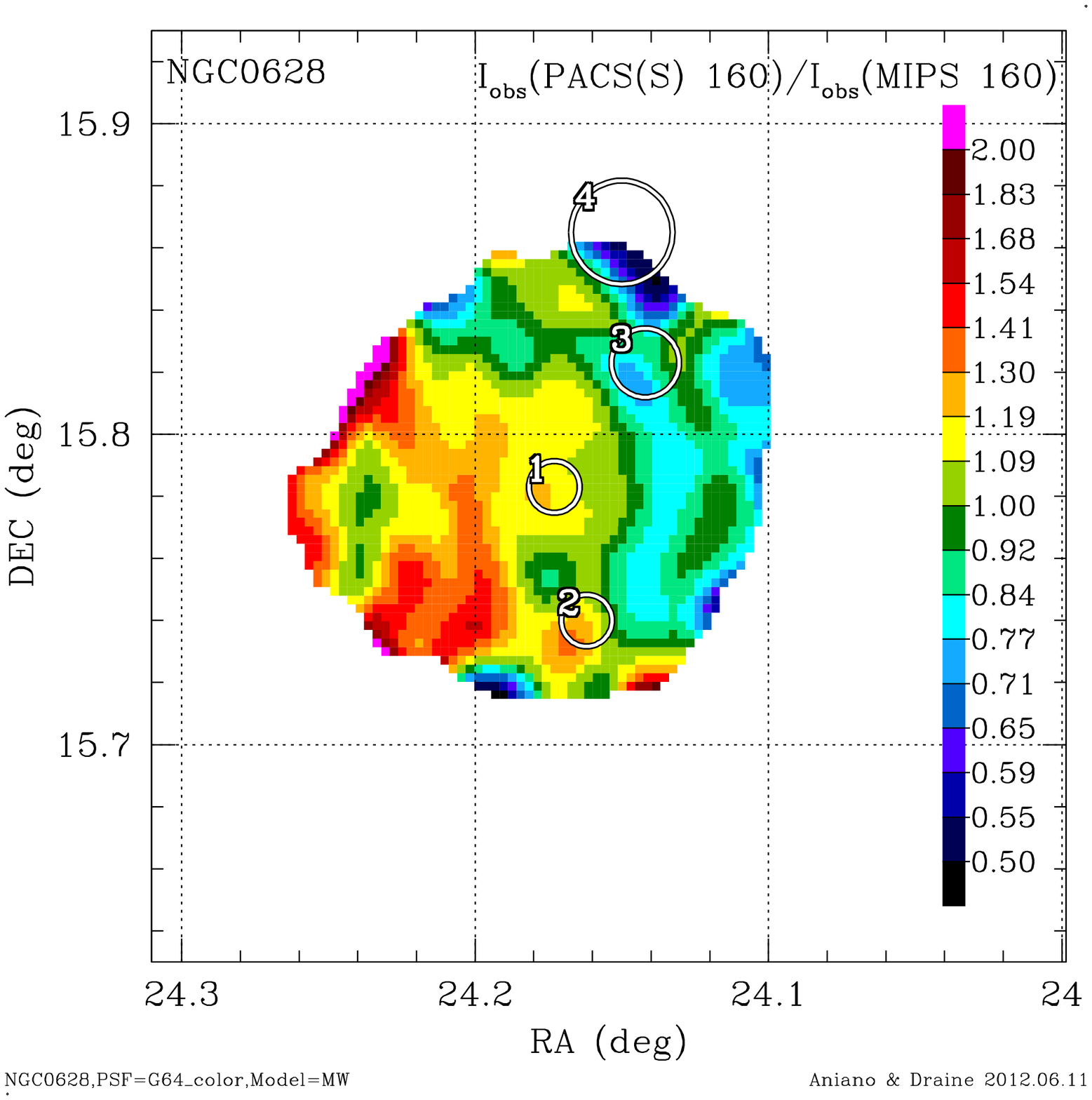}
\renewcommand \RthreeCtwo {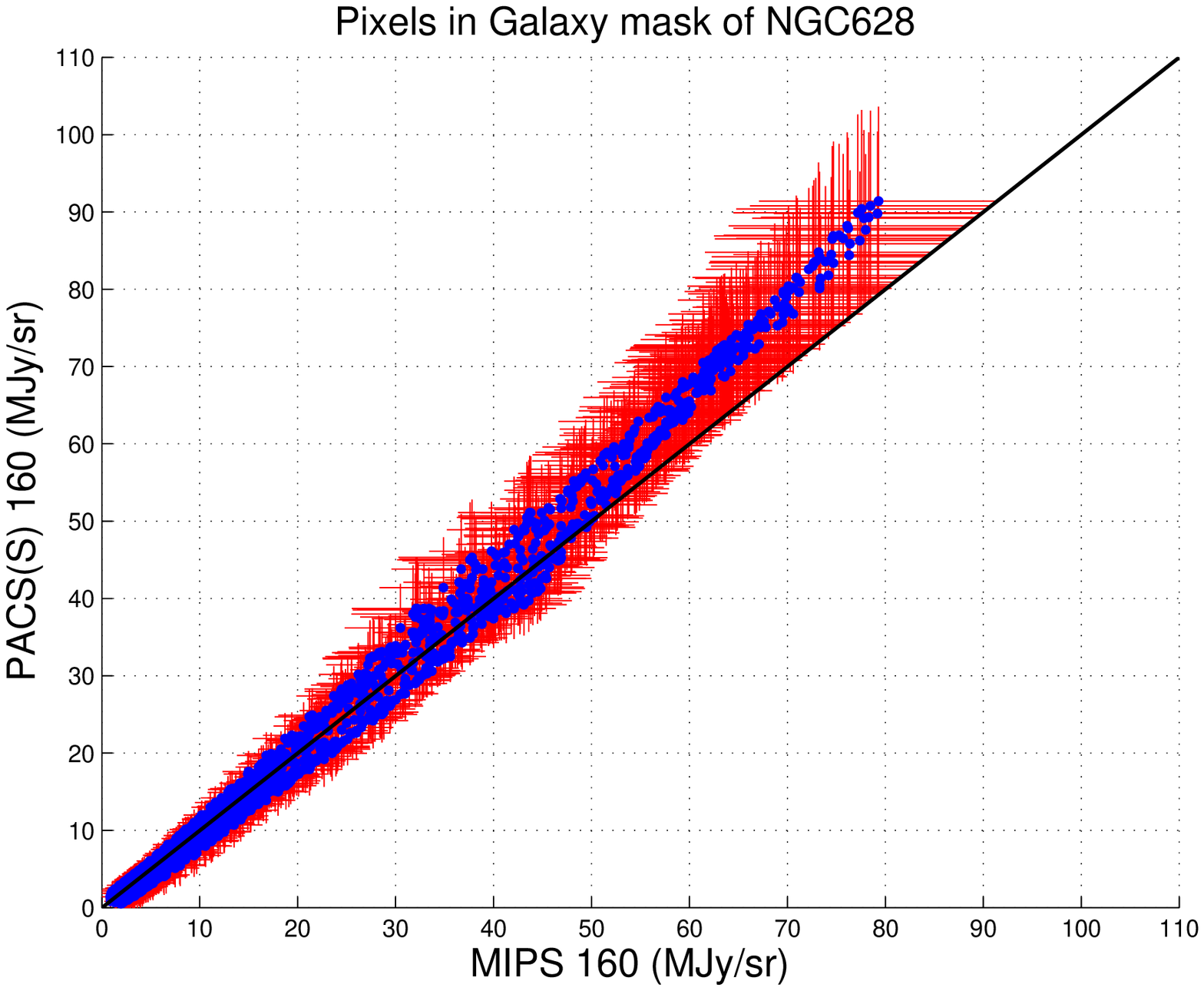}
\renewcommand \RthreeCthree {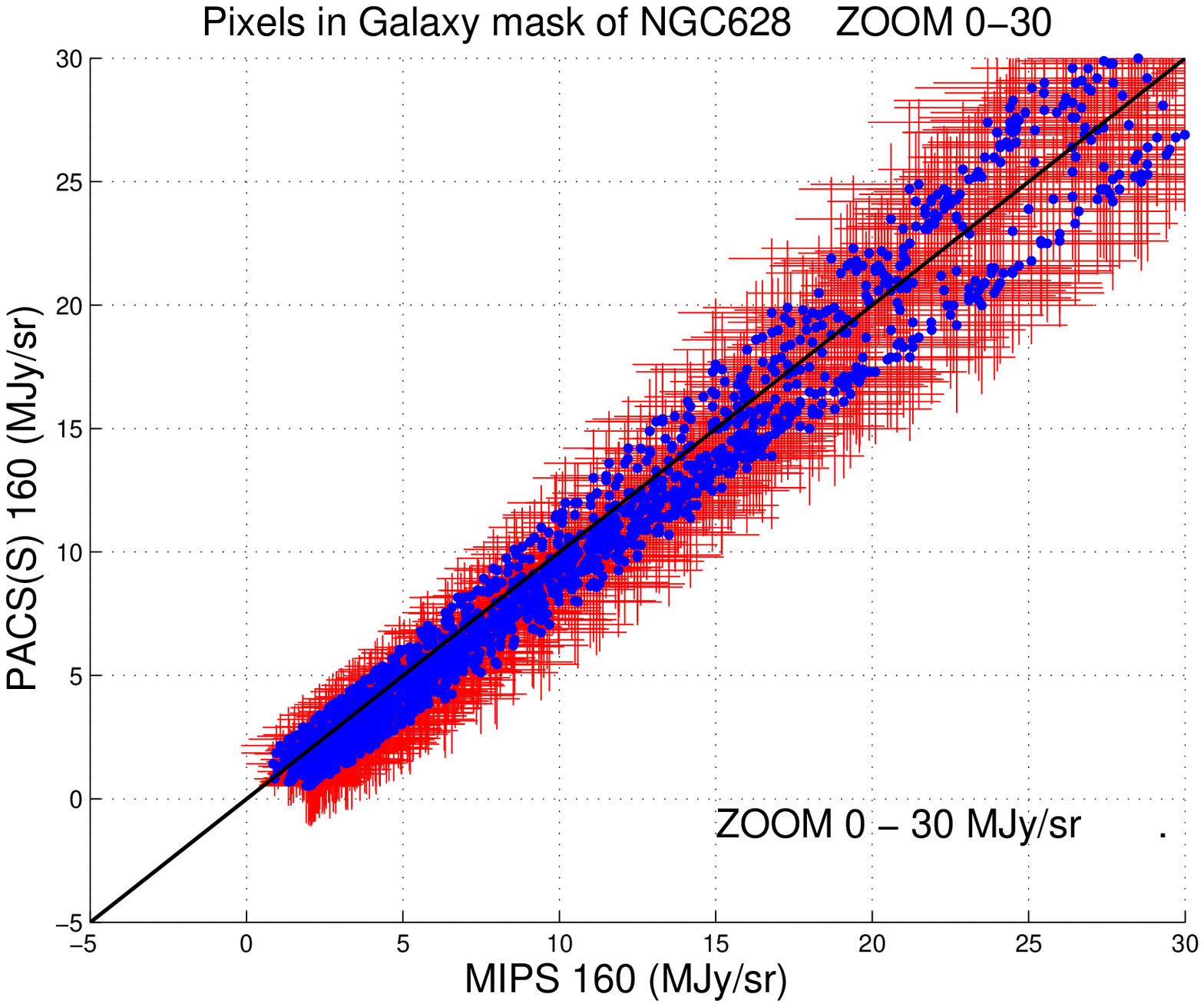}
\renewcommand \RfourCone {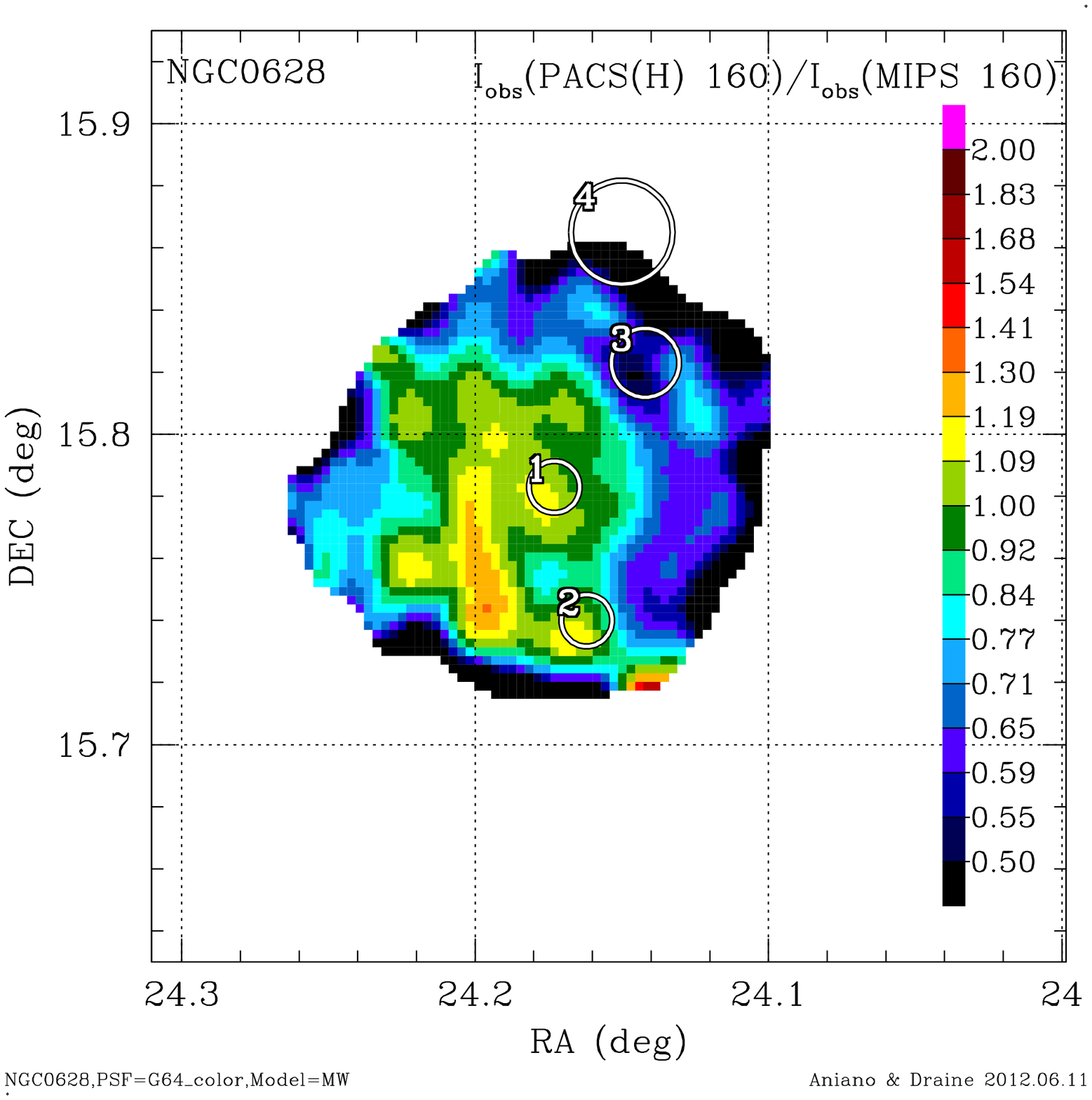}
\renewcommand \RfourCtwo {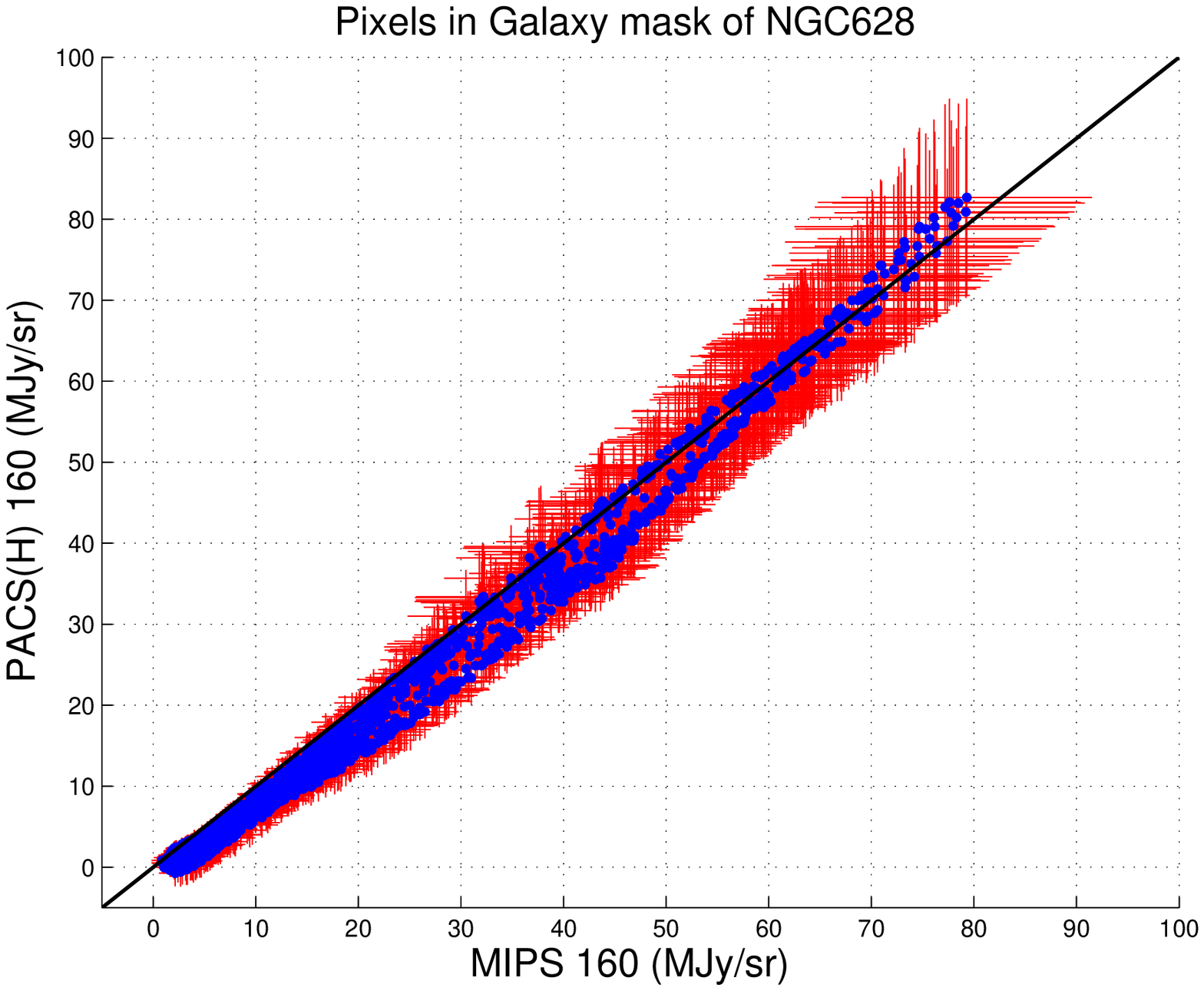}
\renewcommand \RfourCthree {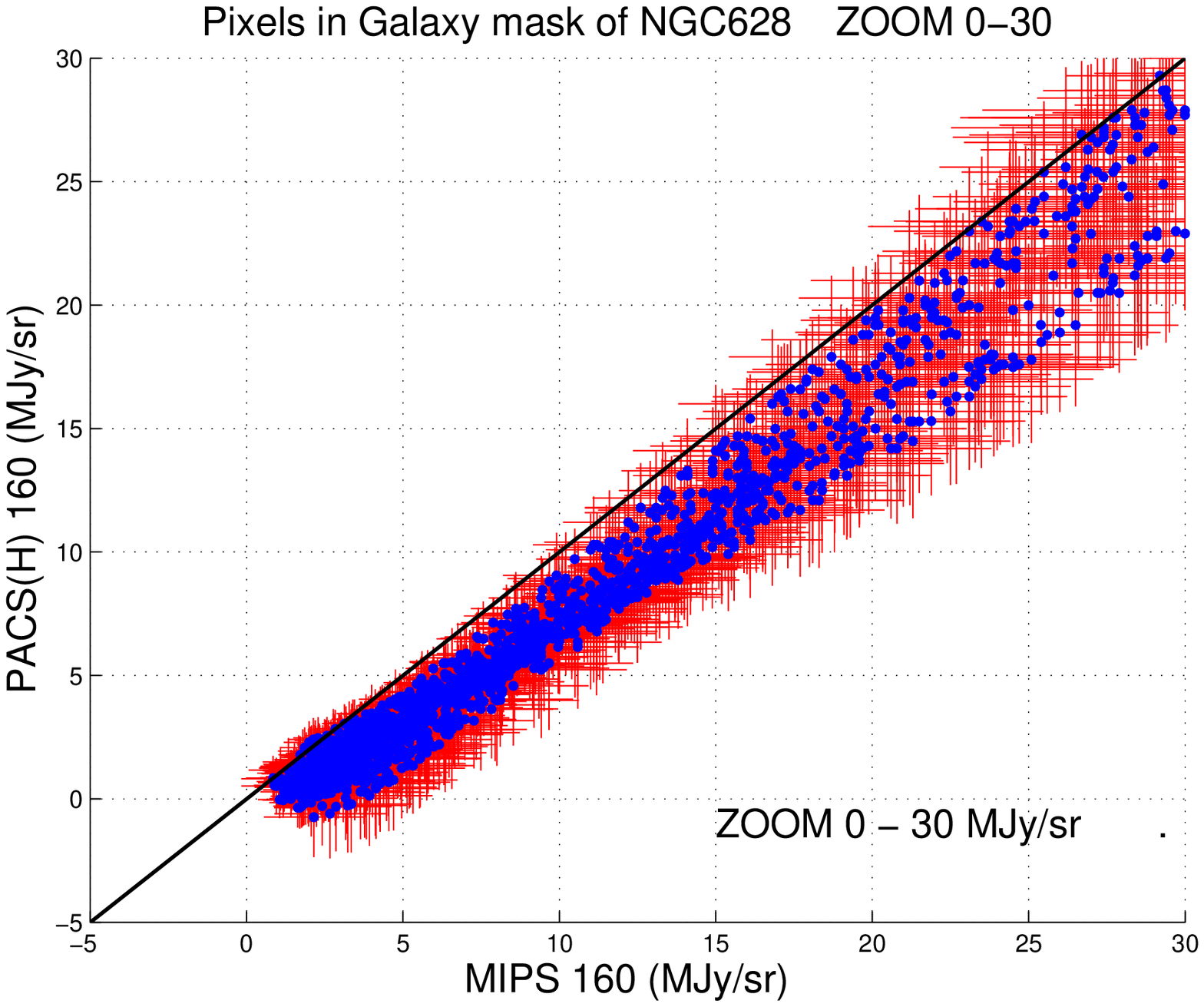}
\begin{figure} 
\centering 
\begin{tabular}{c@{$\,$}c@{$\,$}c} 
\FirstColor
\SecondColor
\ThirdColor
\FourthColor
\end{tabular} 
\vspace*{-0.5cm}
\caption{\footnotesize
  Comparison of the different 70\um and 160\um data available for NGC~628, after background subtraction, and
  convolved into a Gaussian PSF with 64$\arcsec$ FWHM.
  The left column shows the PACS - MIPS ratio maps over the galaxy mask.
  The center column shows scatter plot of the camera intensity of the galaxy pixels.
  The right column is a zoom of the -5 -- 30 MJy/sr region of the center column plots.
  Top row: PACS70 Scanamorphos and MIPS70.
  Second row: PACS70 HIPE and MIPS70.
  Third row: PACS160 Scanamorphos and MIPS160.
  Bottom row: PACS160 HIPE and MIPS160.
    The ratio maps shows strong discrepancies in the three data sets (i.e. the band ratios differ from $\approx1$ in most of the galaxy).
   \label{fig:color_0628}}
\end{figure} 

\renewcommand \RoneCone {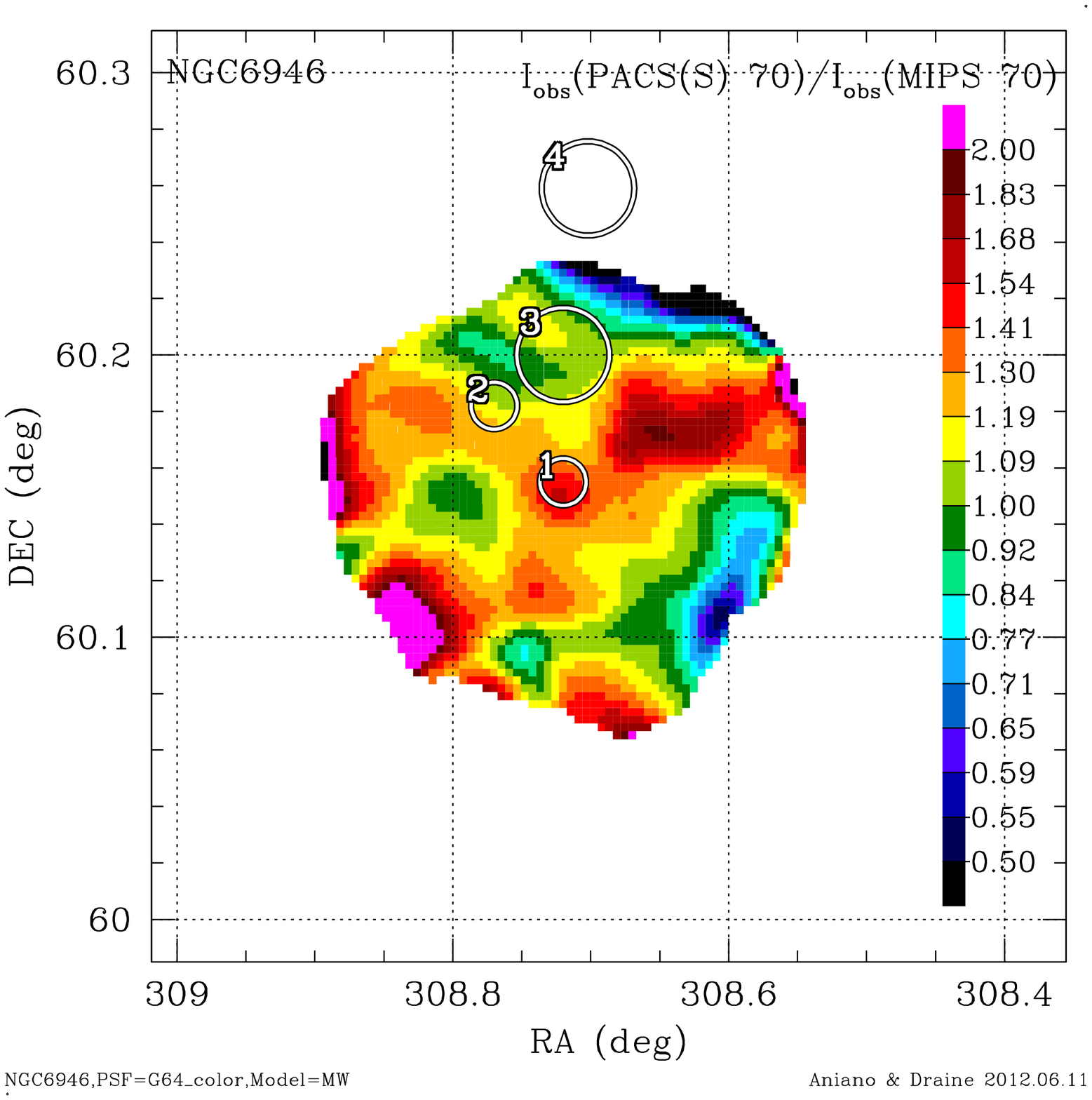}
\renewcommand \RoneCtwo {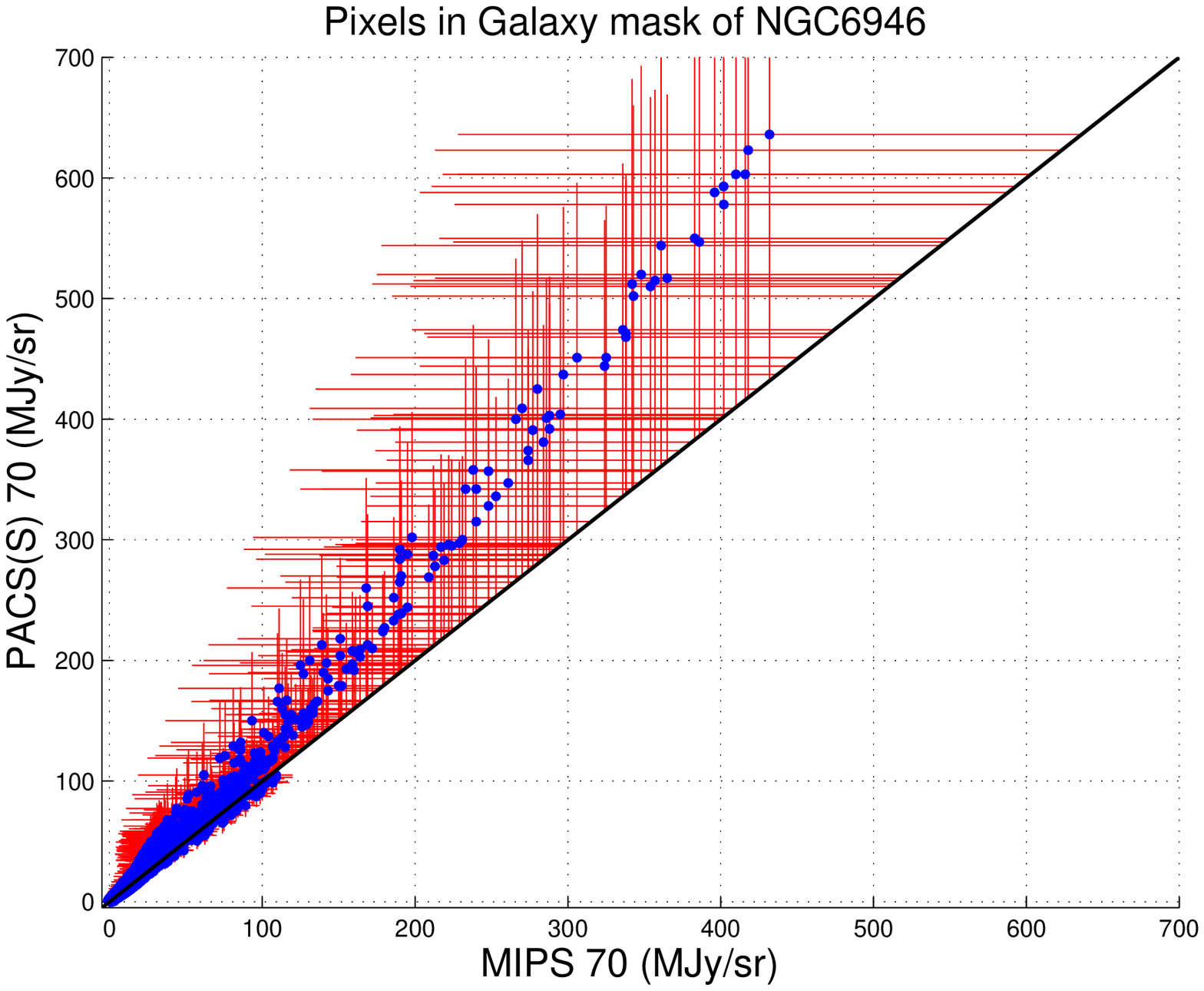}
\renewcommand \RoneCthree {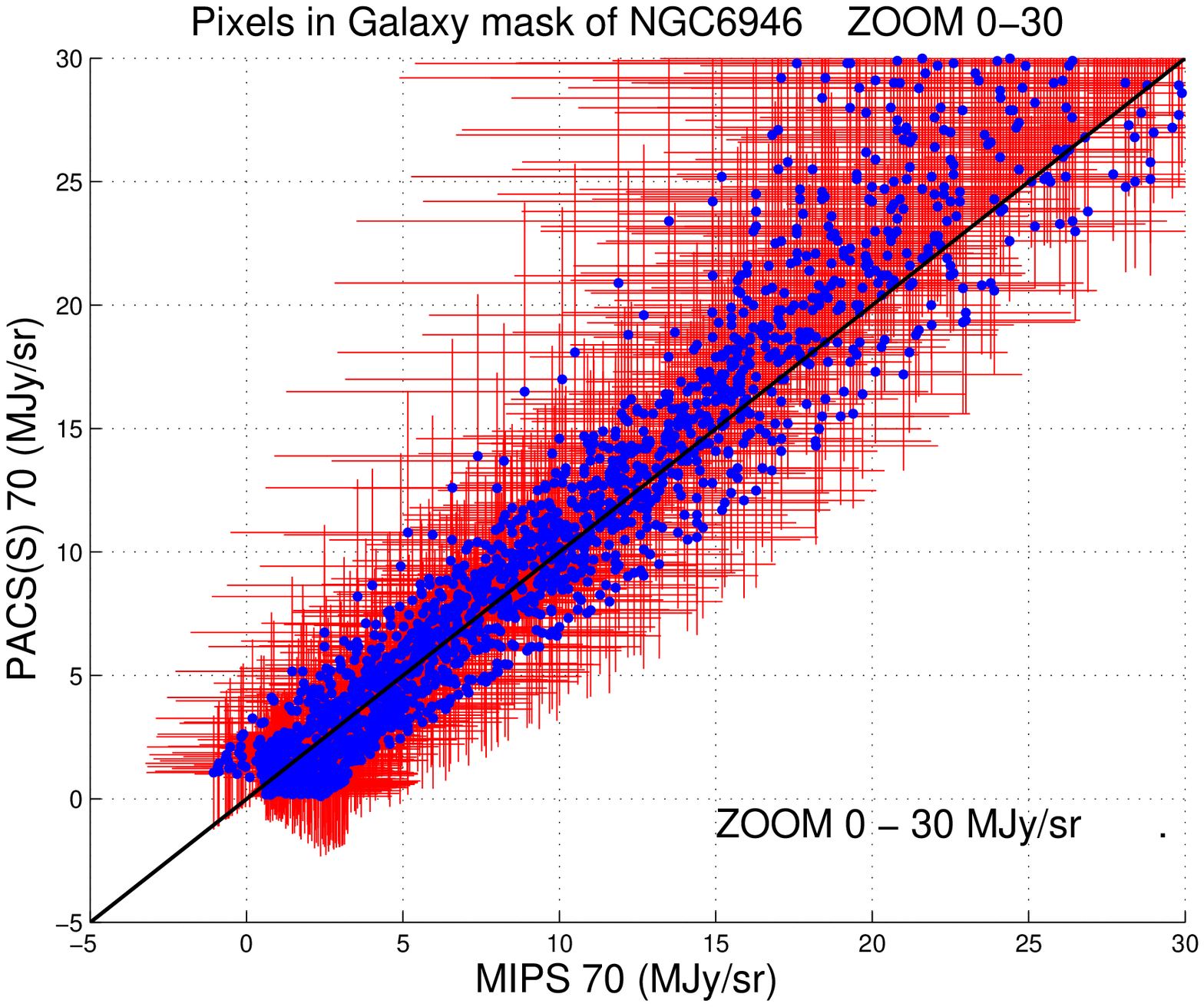}
\renewcommand \RtwoCone {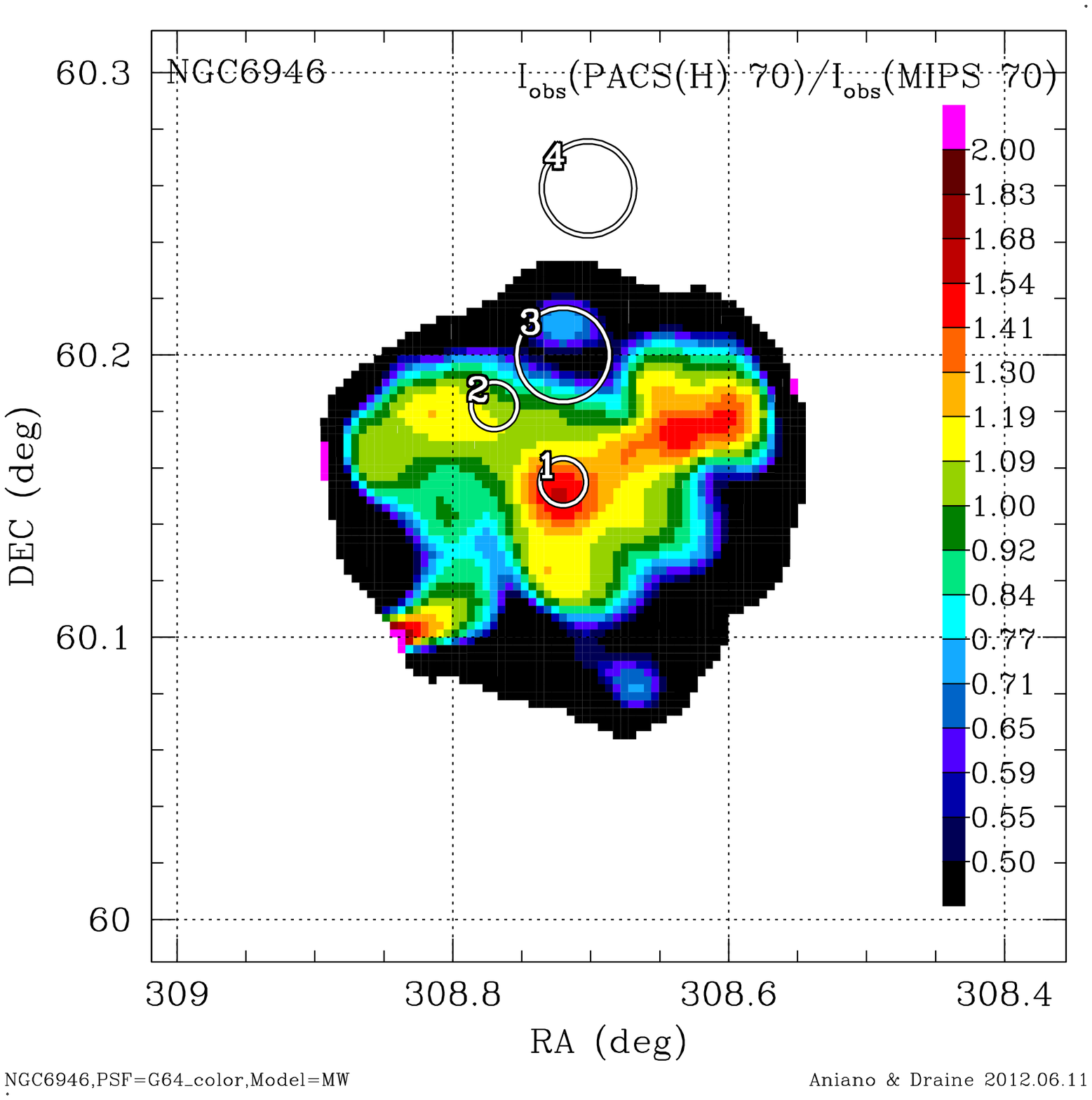}
\renewcommand \RtwoCtwo {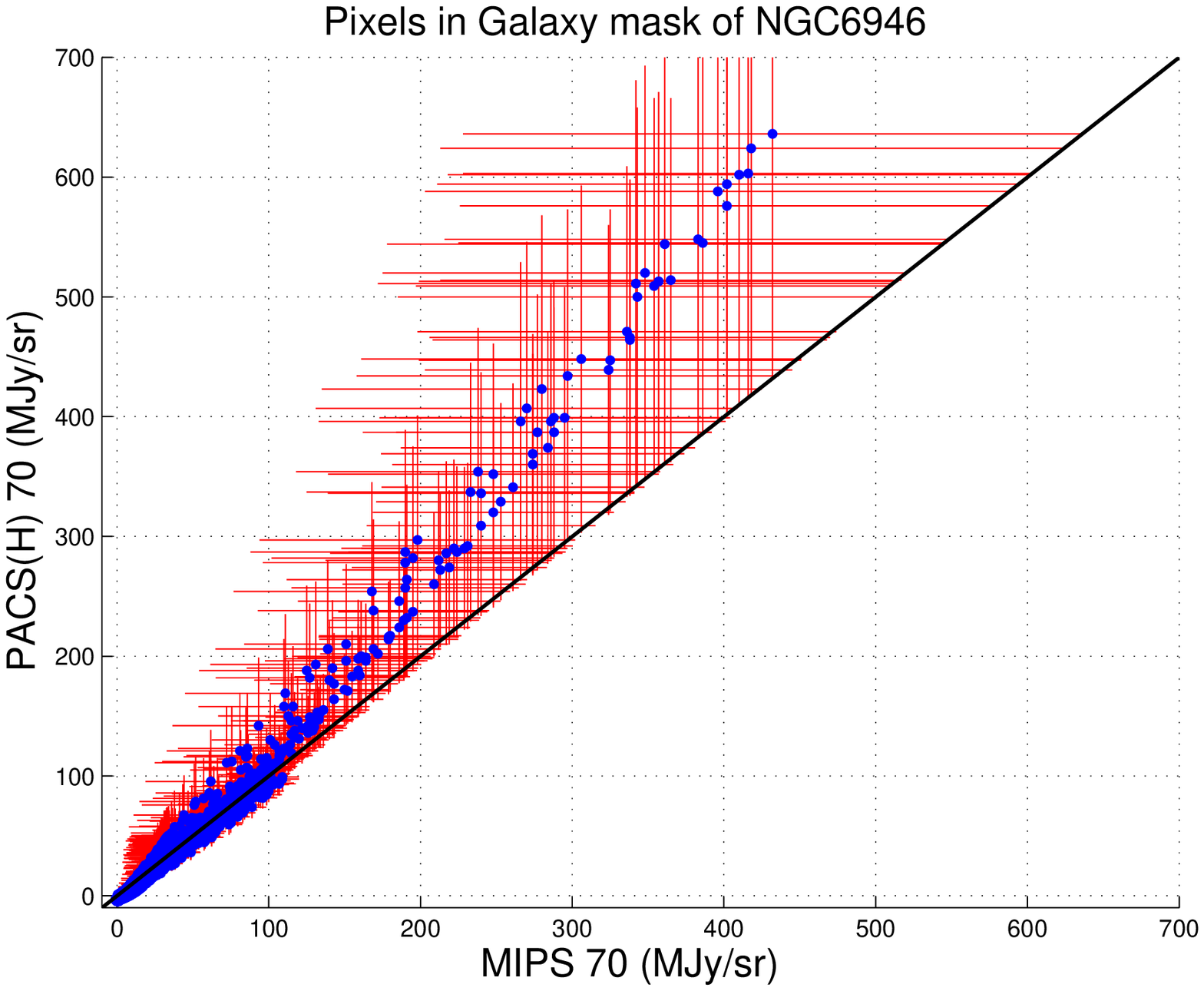}
\renewcommand \RtwoCthree {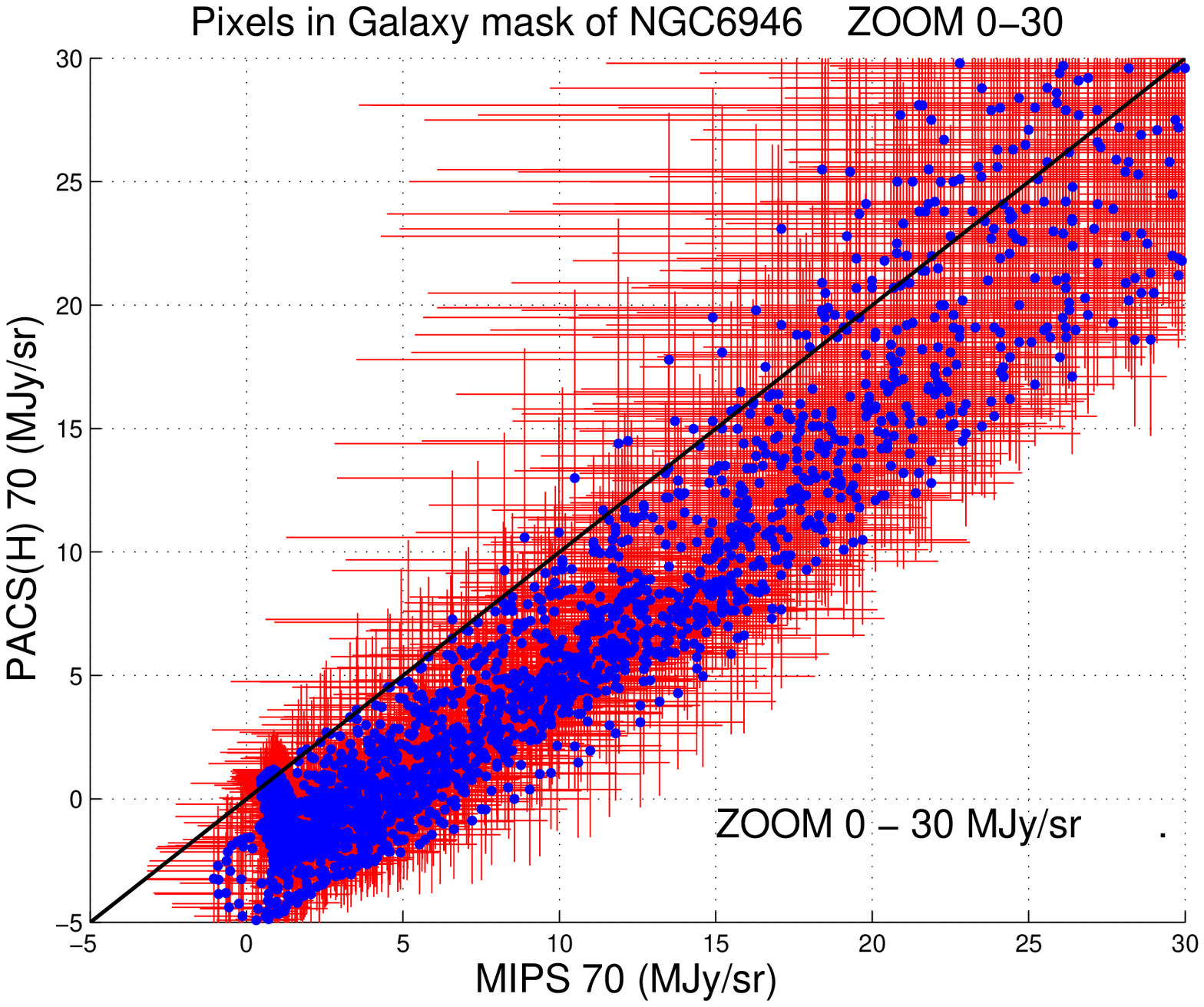}
\renewcommand \RthreeCone {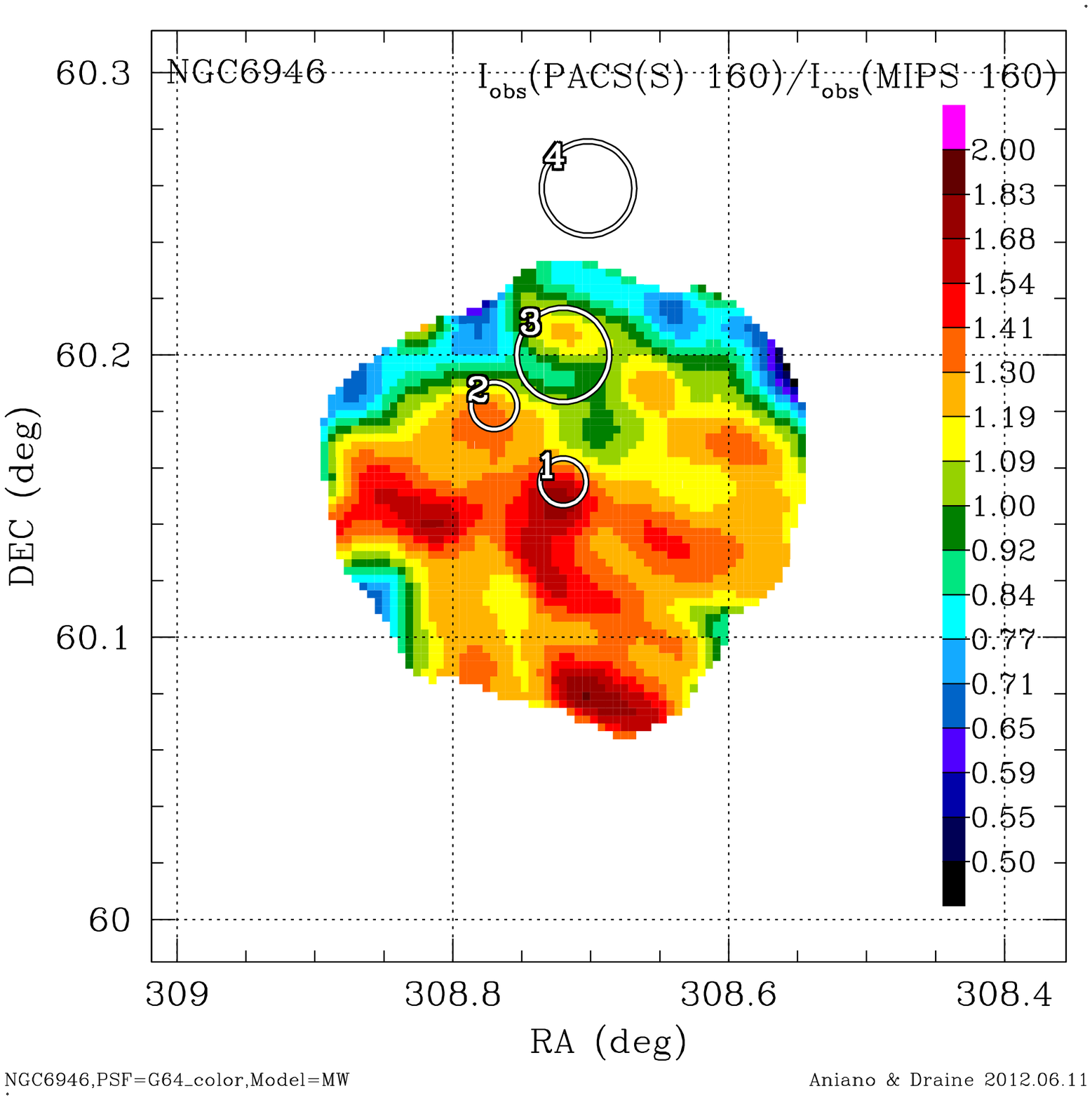}
\renewcommand \RthreeCtwo {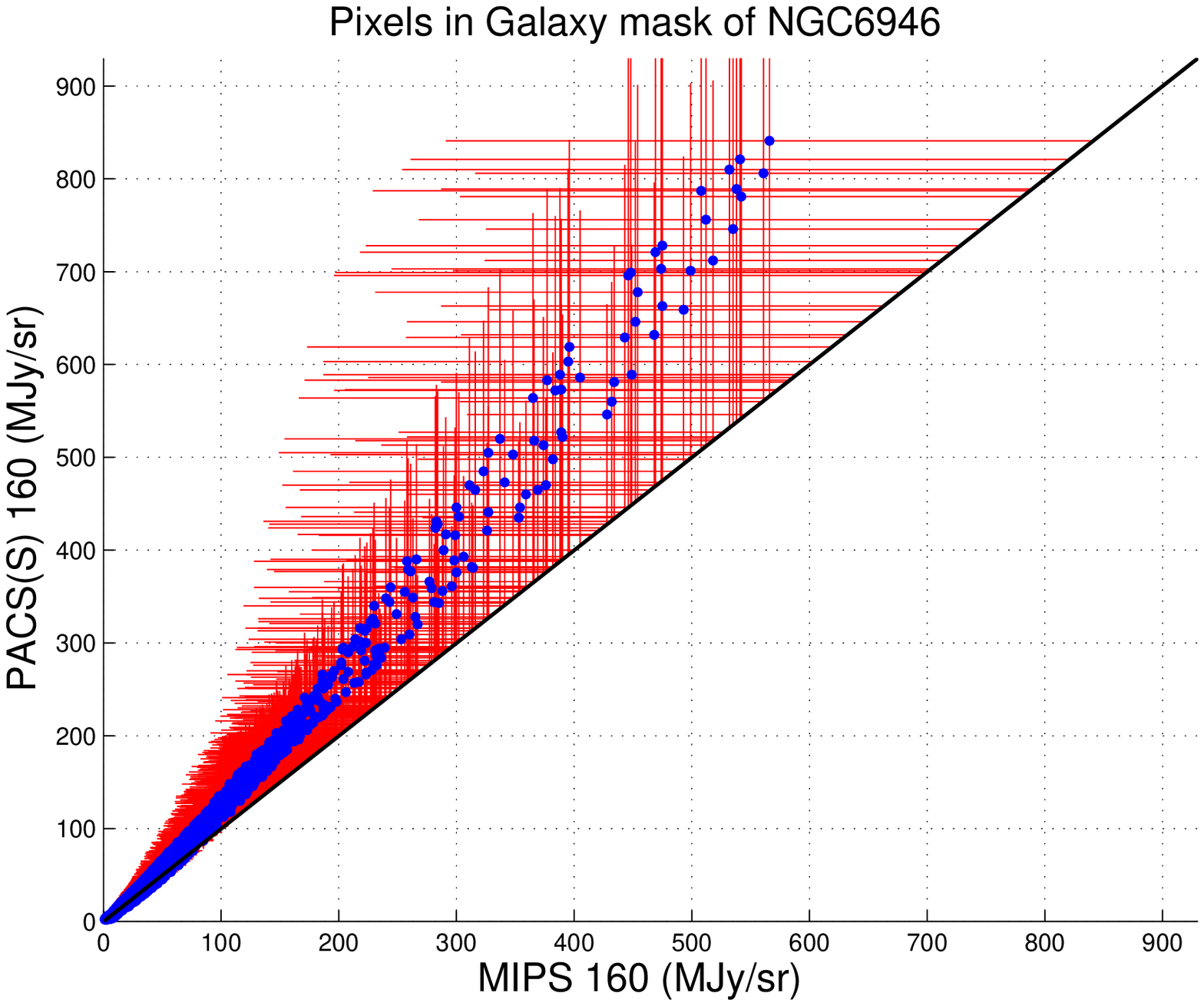}
\renewcommand \RthreeCthree {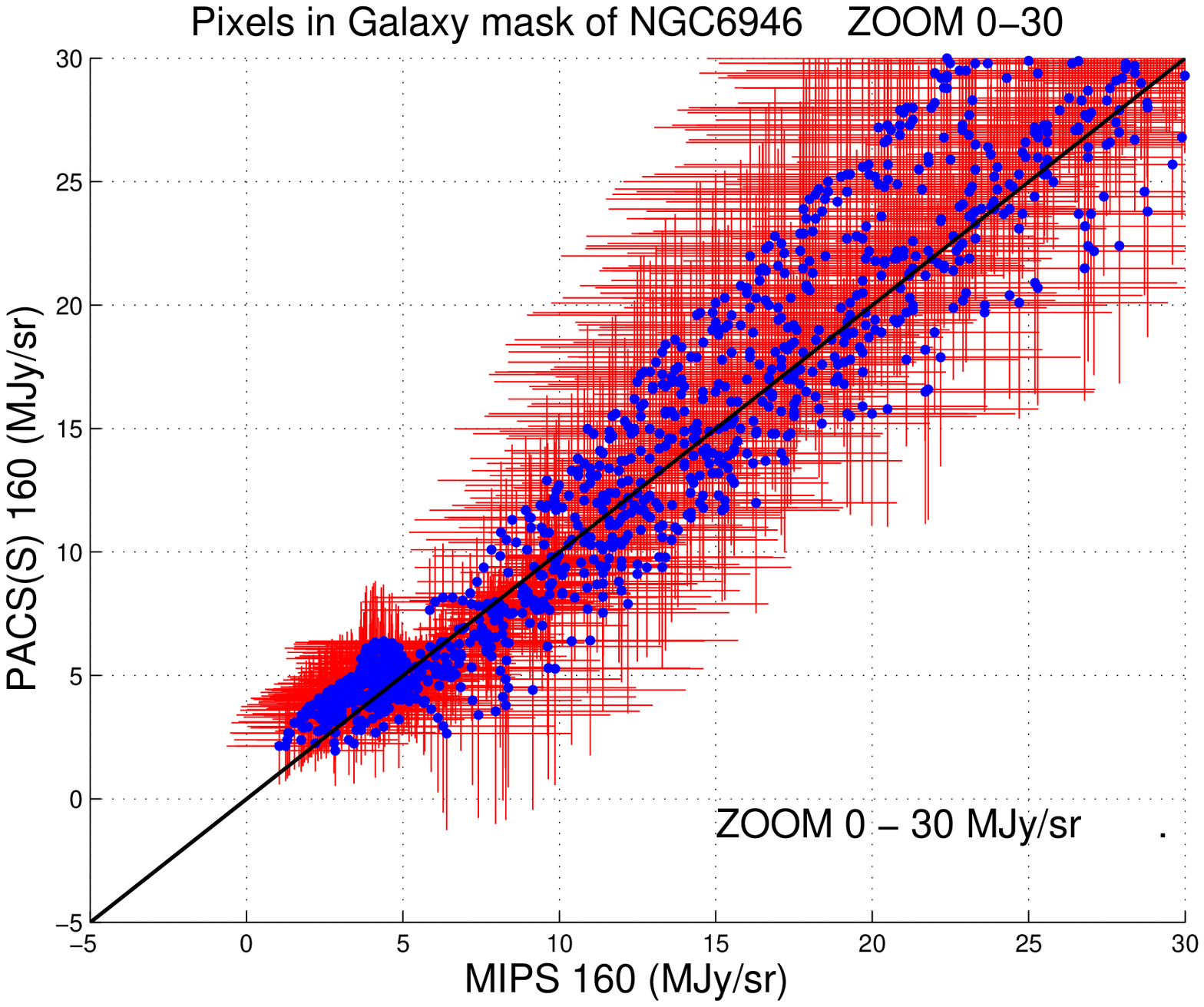}
\renewcommand \RfourCone {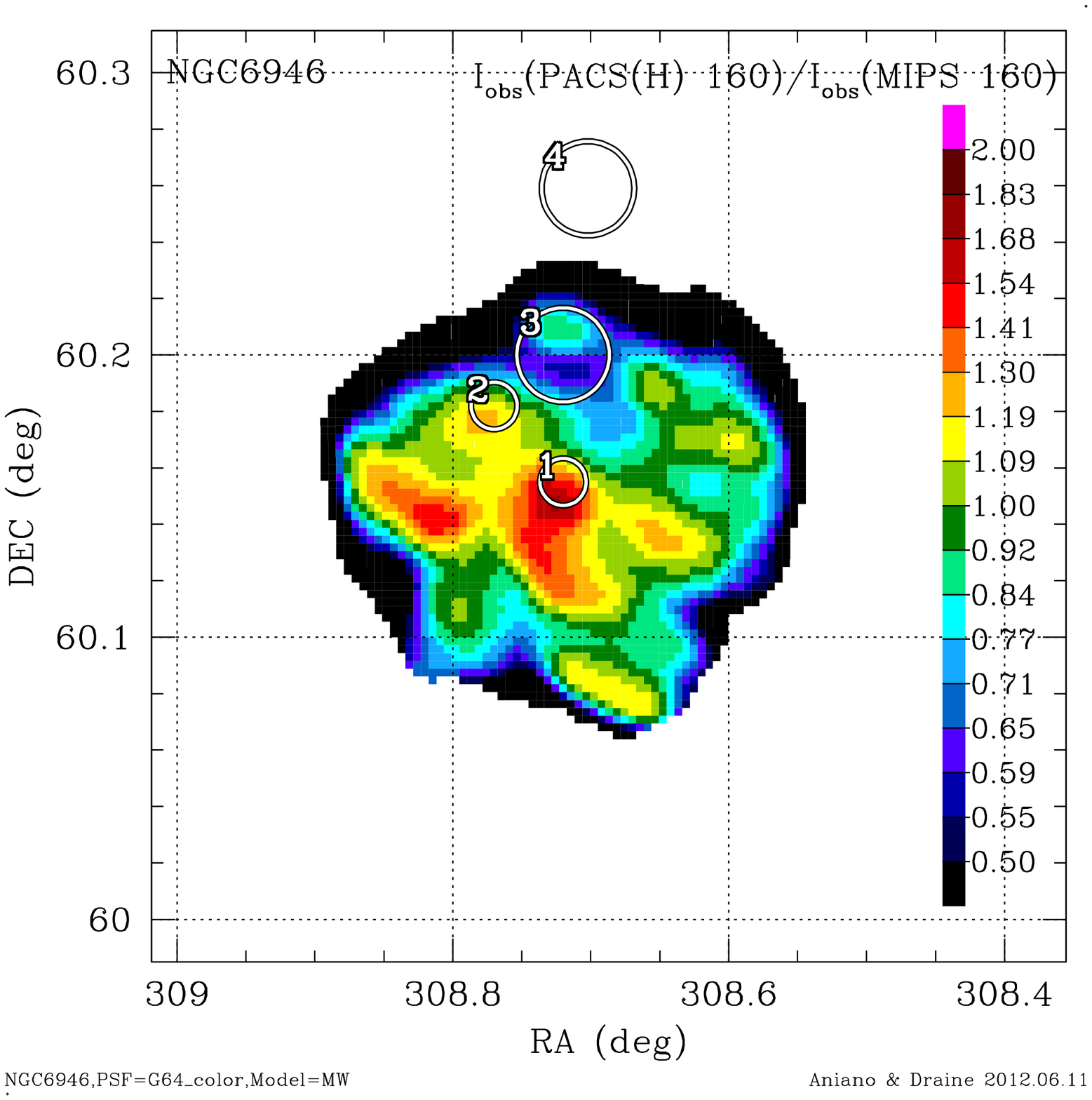}
\renewcommand \RfourCtwo {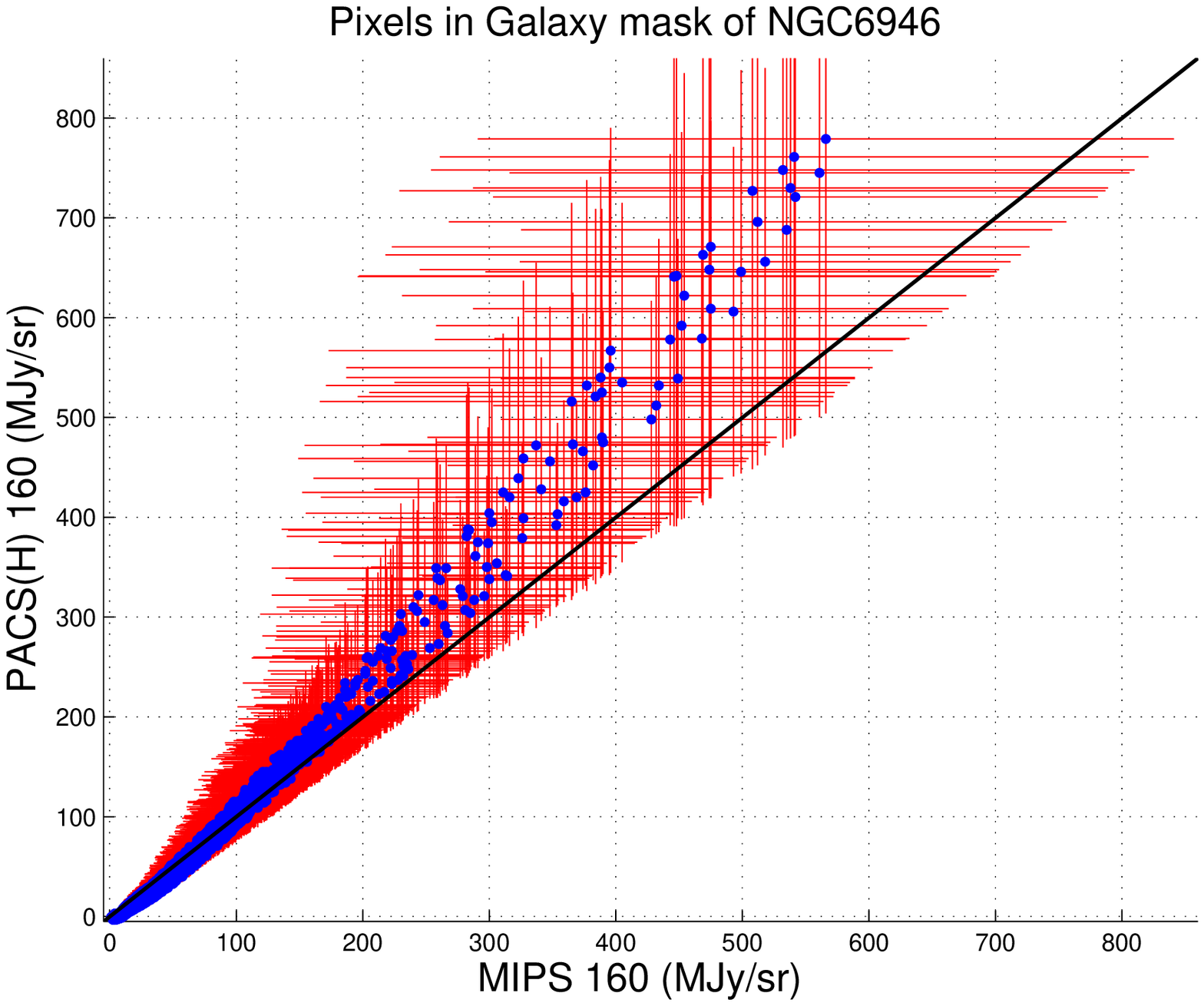}
\renewcommand \RfourCthree {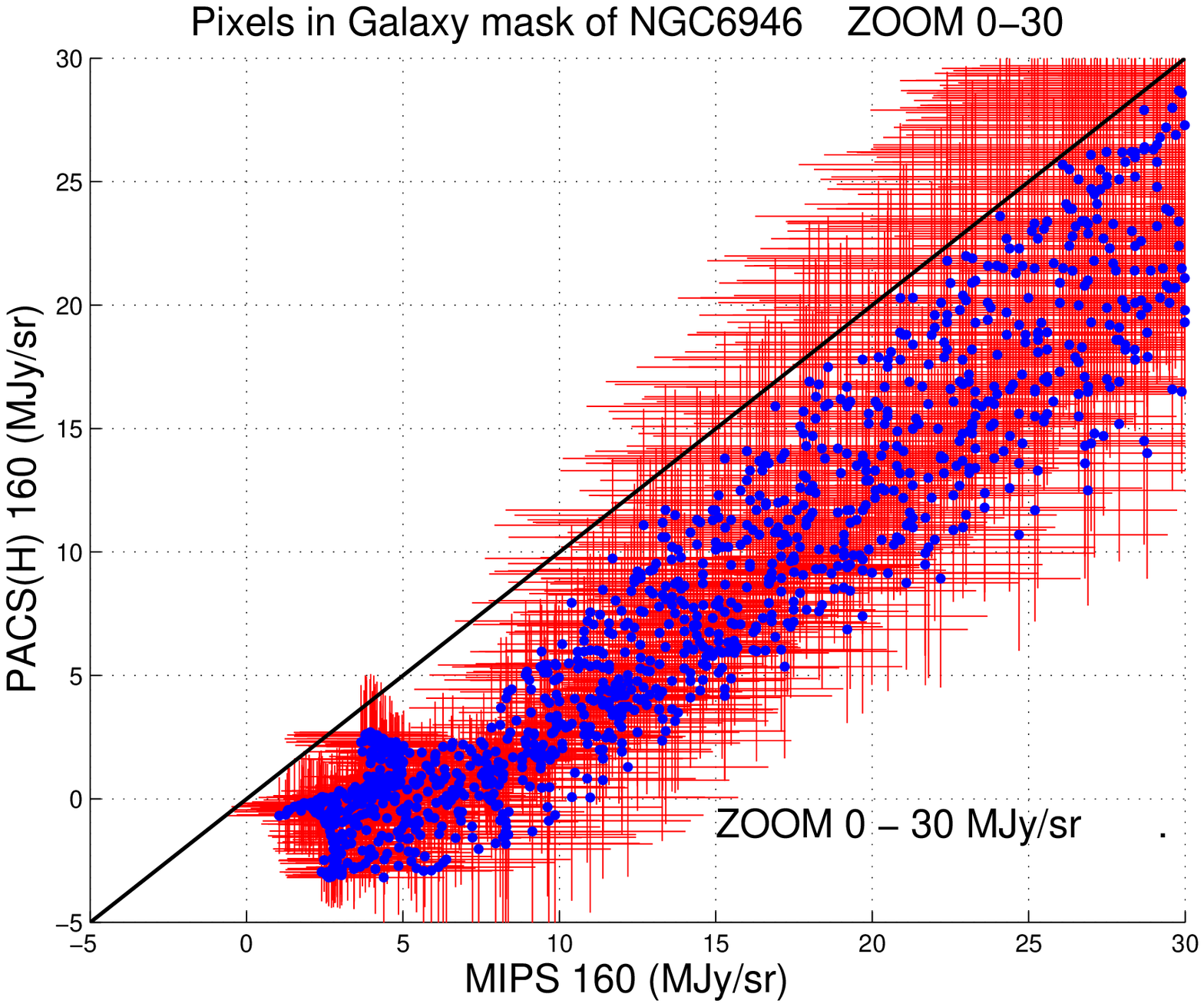}
\begin{figure} 
\centering 
\begin{tabular}{c@{$\,$}c@{$\,$}c} 
\FirstColor
\SecondColor
\ThirdColor
\FourthColor
\end{tabular} 
\vspace*{-0.5cm}
\caption{\footnotesize
\label{fig:color_6946}
  Same as Fig.\ \ref{fig:color_0628}, but for NGC~6946.
}
\end{figure} 

Figures \ref{fig:color_0628} and \ref{fig:color_6946} compare the two different PACS data reduction schemes (Scanamorphos and HIPE) with  the MIPS camera for
NGC~628 and NGC~6946 respectively.  All images were convolved into a
Gaussian PSF with 64$\arcsec$ FWHM, a PSF broad enough that any camera
PSF mismatch should not produce any artifact. 
  The left column show the PACS/MIPS ratio maps over the galaxy mask.
  The center columns show an intensity-intensity scatter plot for the galaxy pixels (the pixels in the left column).
  The right column has a zoom of the -5 -- 30 MJy/sr region of the center column plots.
  We compare PACS70 Scanamorphos and MIPS70, 
  PACS70 HIPE and MIPS70,
  PACS160 Scanamorphos and MIPS160, and PACS160 HIPE and MIPS160.
 In the center and right column, the horizontal and vertical red lines correspond to the $1\sigma$ error estimates.
By the inclusion of the PACS(S)-MIPS difference term in the uncertainty estimates (see Appendices \ref{PACS-MIPS-1} and \ref{PACS-MIPS-2}), the $1\sigma$ error estimates will intersect the PACS=MIPS line.

The Scanamorphos and HIPE pipelines give similar results in the high surface brightness
areas, having slightly larger intensities than MIPS (see figures \ref{fig:color_0628} and \ref{fig:color_6946}).
  The main difference between the two data reduction schemes is in the low surface
brightness areas: Scanamorphos intensities are larger than 
HIPE intensities.

Even though the PACS and MIPS cameras have different spectral
response, smooth spectra appropriate to star-forming galaxies should
result in nearly the same measured intensity for both MIPS70 and
PACS70, and likewise for MIPS160 and PACS160.  
For our best-fit SEDs for NGC~628 and NGC~6946, 
we expect PACS70/MIPS70 $\sim$1.085, and
PACS160/MIPS160 $\sim$ 1.022 (see Table \ref{tab:ratio}).  
However, the measured PACS global flux densities do not agree so well with MIPS
values, as seen in Table \ref{tab:ratio}.
The situation is more critical in resolved studies.
Even though the global PACS and MIPS fluxes agree within $\approx 20\%$, the ratio maps shows strong discrepancies in the three data sets: the band ratios differ from $\approx1 $ in most of the galaxy.

\begin{table}[h]
\begin{center}
\caption{\label{tab:ratio}PACS70/MIPS70 flux ratio predicted by the model and 
observed.}
\begin{tabular}{|c|cc|cc|}
\hline
                     & \multicolumn{2}{|c|}{ NGC~628} & \multicolumn{2}{|c|}{ NGC~6946}\\
Band ratio           &  Model & Observed & Model & Observed \\
\hline
PACS(H)70/MIPS70      &  1.083 & 0.919   & 1.088 & 1.040\\
PACS(S)70/MIPS70      &  1.083 & 1.243   & 1.088 & 1.246\\
\hline
PACS(H)160/MIPS160    &  1.024 & 0.886    & 1.020 & 1.016\\
PACS(S)160/MIPS160    &  1.024 & 1.058    & 1.020 & 1.252\\
\hline
\end{tabular}
\end{center}
\end{table}

\renewcommand \RoneCone {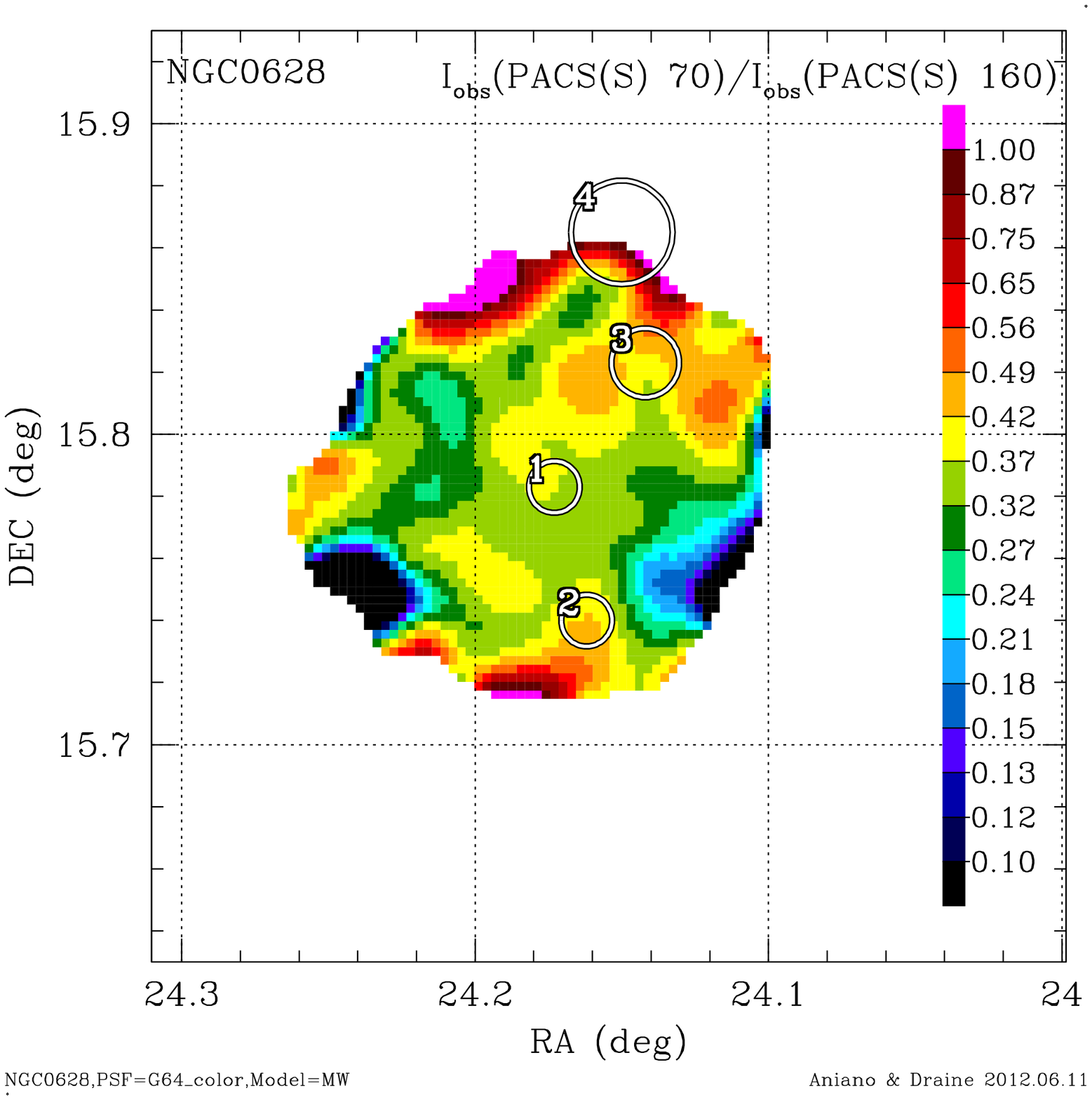}
\renewcommand \RoneCtwo {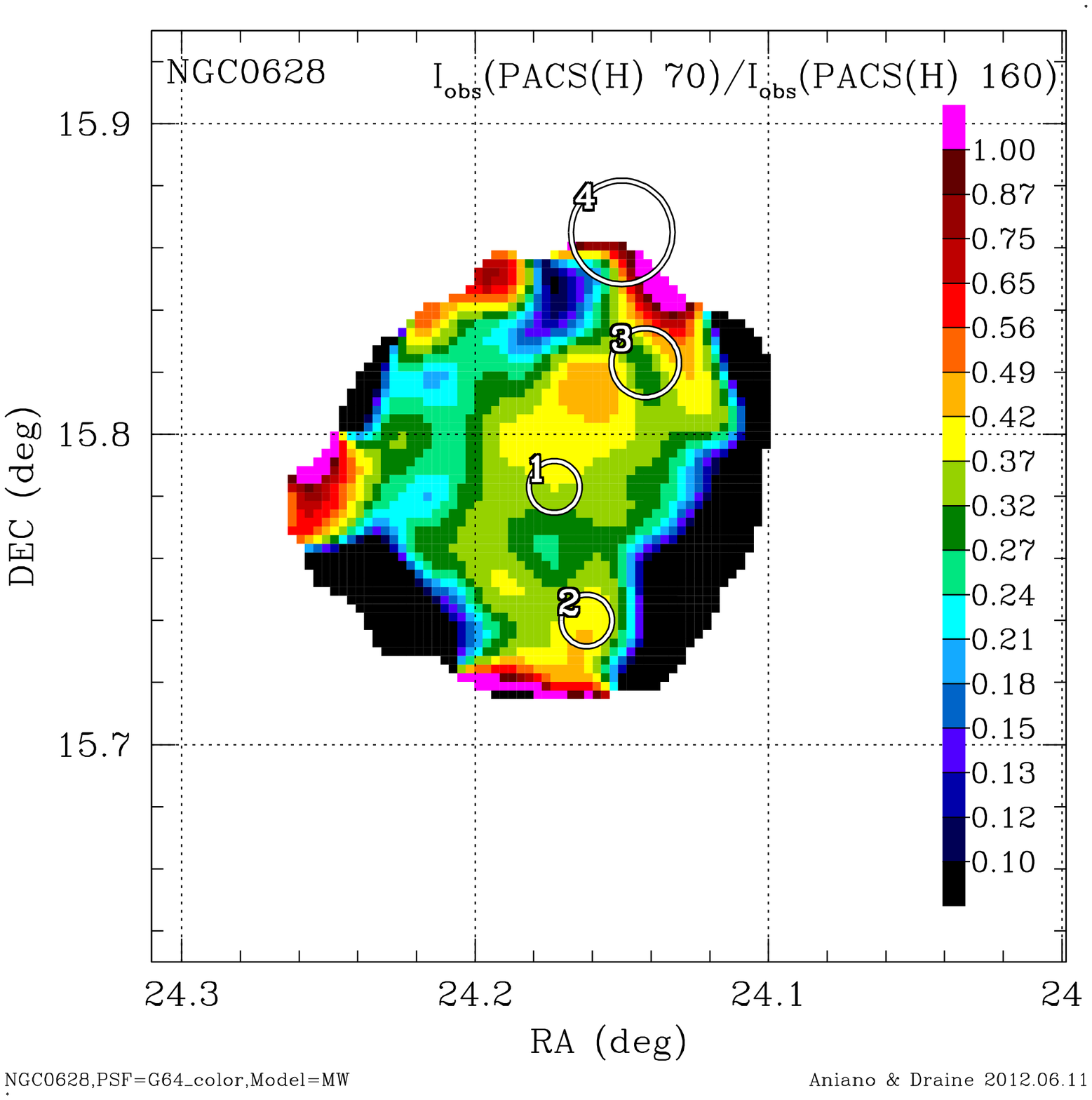}
\renewcommand \RoneCthree {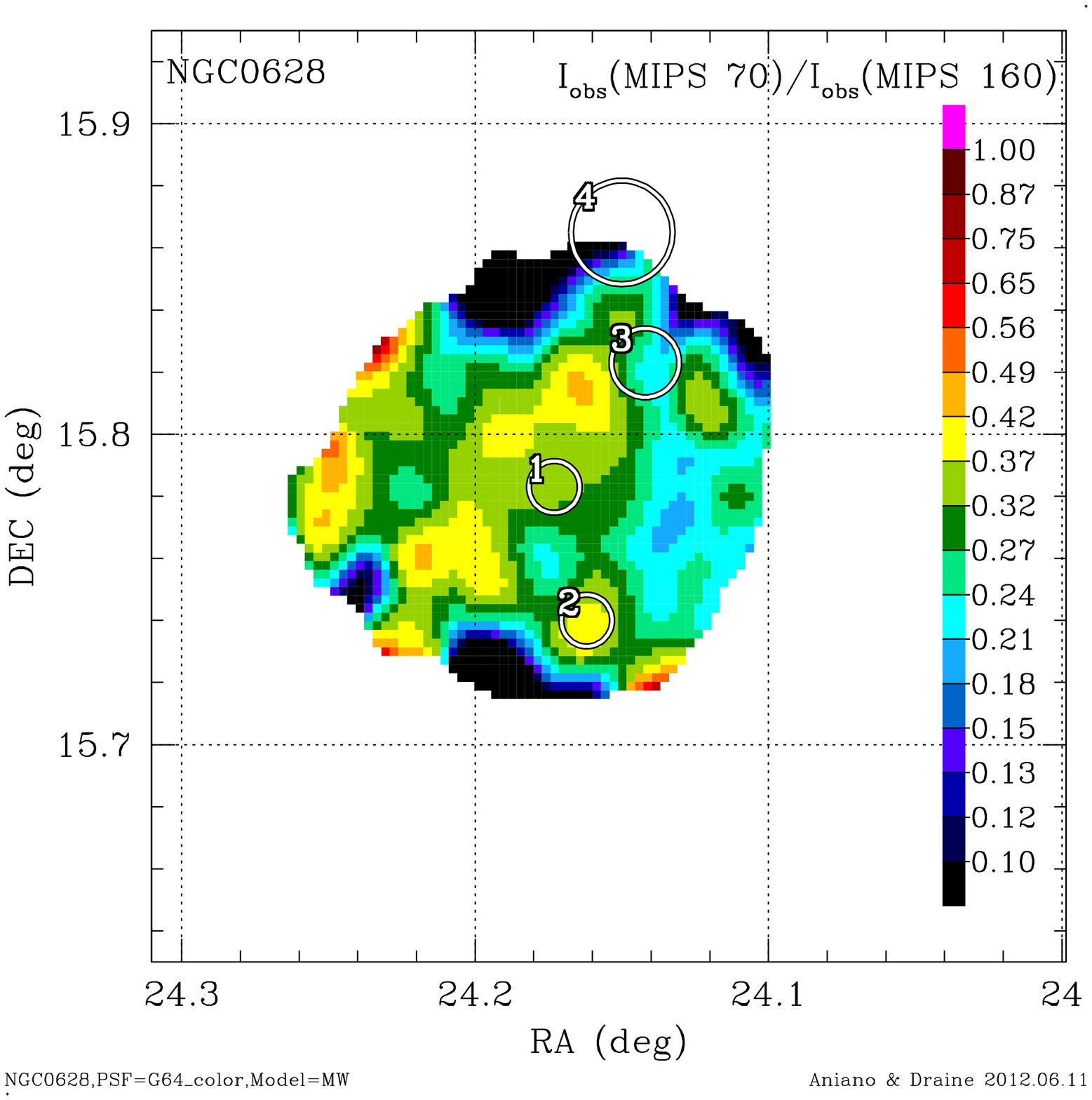}
\renewcommand \RtwoCone {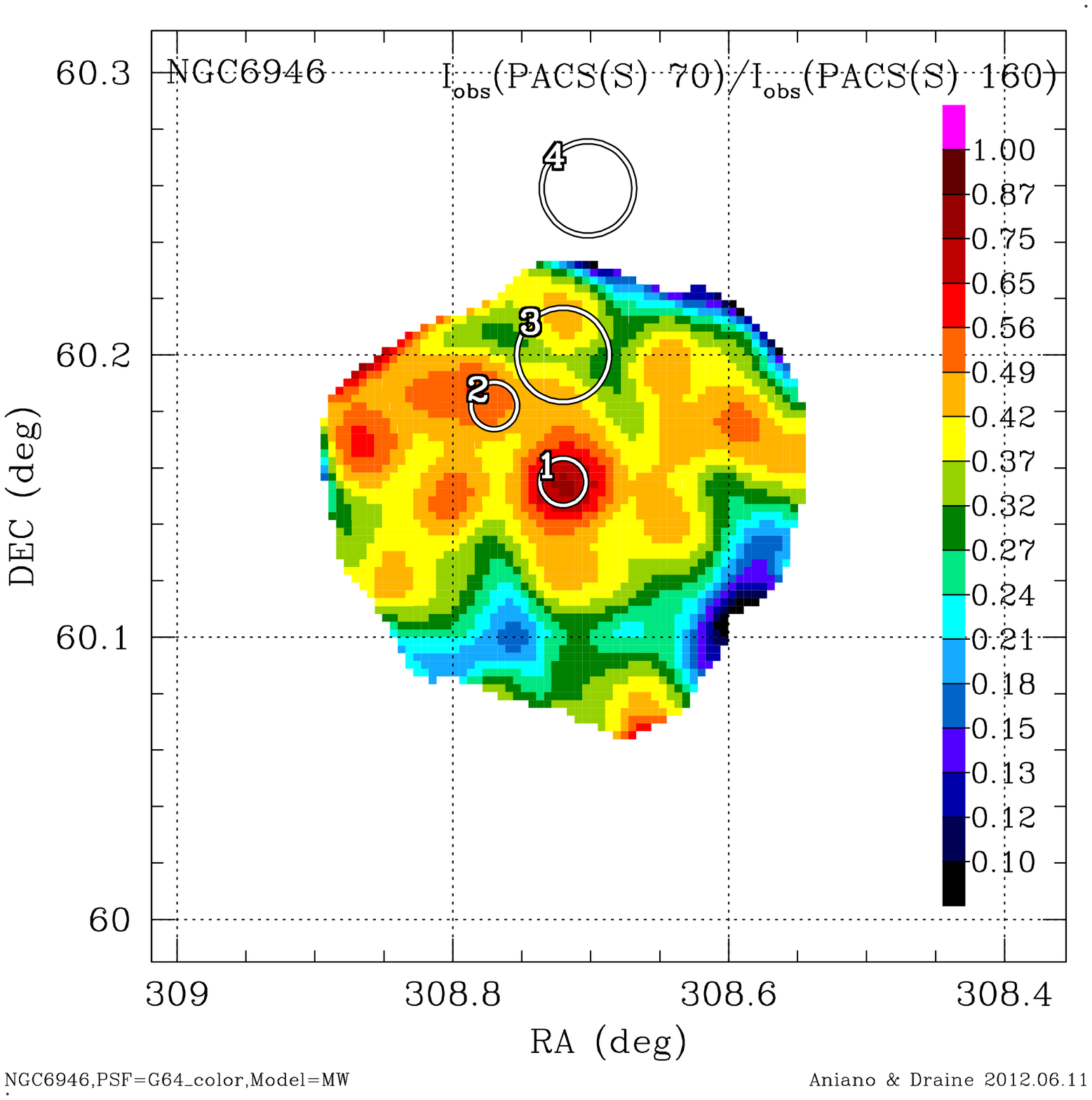}
\renewcommand \RtwoCtwo {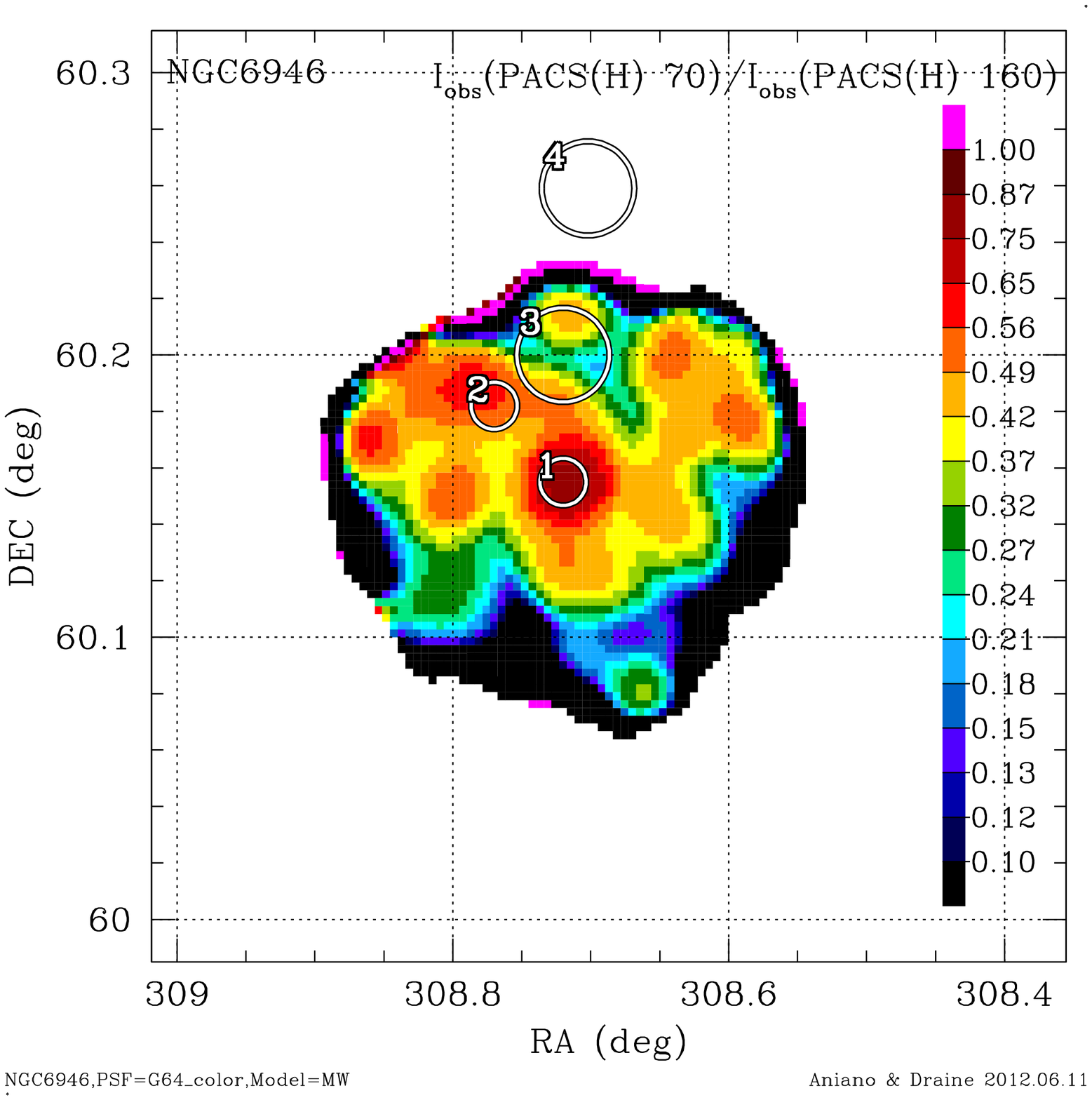}
\renewcommand \RtwoCthree {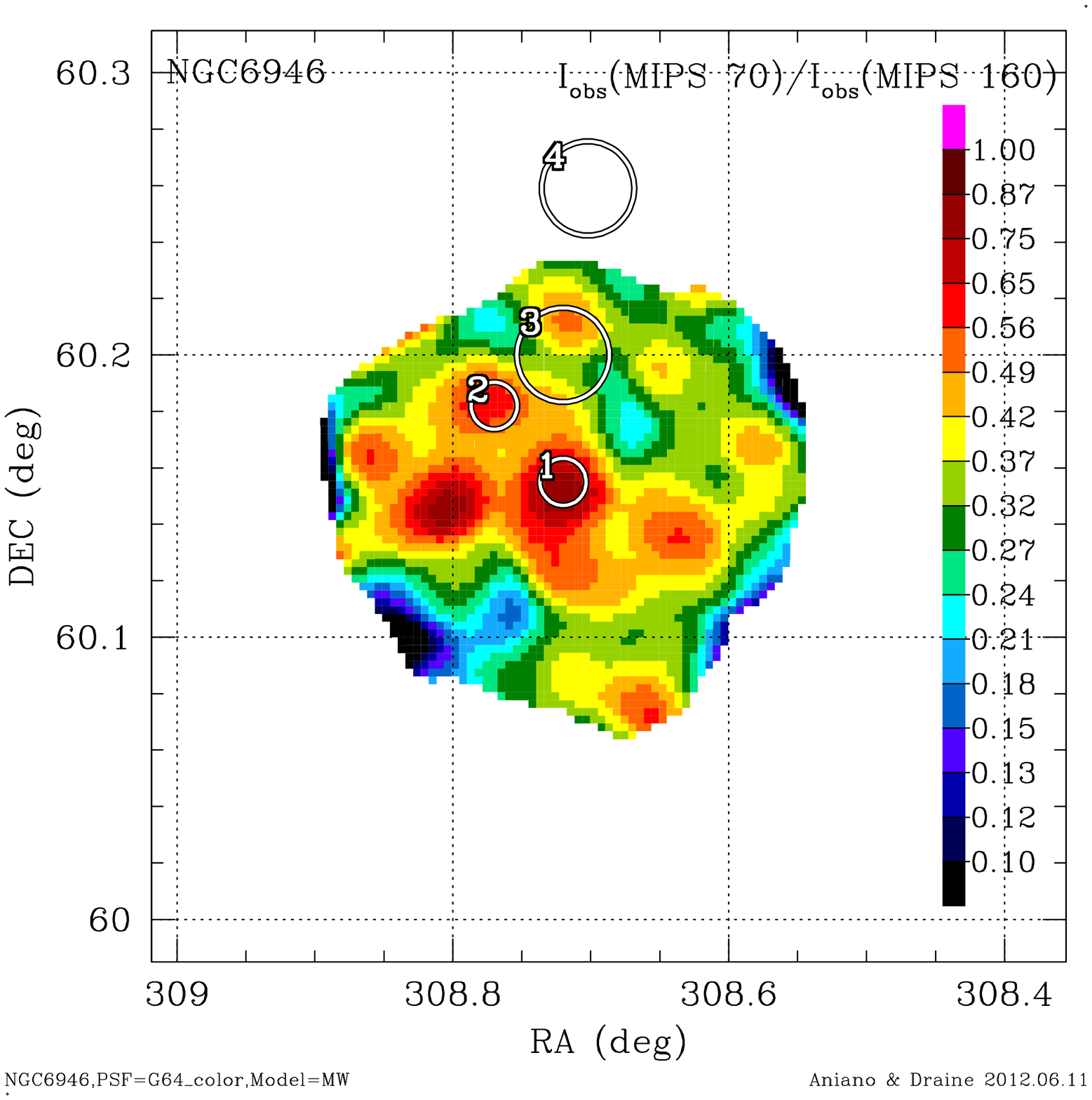}
\begin{figure} 
\centering 
\begin{tabular}{c@{$\,$}c@{$\,$}c} 
\FirstLast
\SecondLast
\end{tabular} 
\vspace*{-0.5cm}
\caption{\footnotesize
Comparison of the 70\um/160\um color inferred from the different
  data sets. All images were previously convolved into a Gaussian PSF with 64$\arcsec$ FWHM.
  The top row corresponds to NGC~628 and the bottom row to NGC~6946.
  Left column: PACS Scanamorphos color.
  Center column: PACS HIPE.
  Right column: MIPS.
   The 70\um/160\um colors inferred with the different cameras are quite different, specially in the outer parts of the galaxies.
\label{fig:color_color}}
\end{figure} 

Figure \ref{fig:color_color} shows the 70\um/160\um color inferred from the different
  data sets. Again, all images were convolved into a Gaussian PSF with 64$\arcsec$ FWHM.
  The top row corresponds to NGC~628 and the bottom row to NGC~6946.
  The left column is the PACS Scanamorphos color, PACS HIPE is shown in the center column, and, MIPS is shown in the the right column.
   The 70\um/160\um colors inferred with the different cameras are also quite different, especially in the outer parts of the galaxies.

The colors inferred from PACS(S)70/PACS(S)160 and MIPS70/MIPS160 are
similar for most of the galaxy, and thus the inferred dust parameters ($\Umin$) obtained at PACS160
resolution are similar to the ones obtained at MIPS160 resolution. 
The large PACS(S)160/MIPS160 flux ratio makes the
modeling at PACS160 resolution add more dust over all the galaxy, compared to our modeling
at MIPS160 resolution, when MIPS constraints are present.

The colors inferred from PACS(H)70/PACS(H)160, and MIPS70/MIPS160 are
very different in the outer parts of the galaxy, potentially giving different
best-fit parameters in the dust modeling.  The HIPE pipeline uses a
high-pass filter in the time-line observing series before constructing
the two dimensional image map. In studies of HIPE performance
using simulated data with realistic noise statistics, it was shown
that HIPE intensities are always equal to or smaller than the ``true'' 
intensities in
the extended low surface brightness regions 
(Sauvage 2011, private communication).
When only one SPIRE band
(SPIRE250) is included, then the extremely low flux of PACS(H)160 in
the low surface brightness regions of the galaxy makes the dust appear to 
be very cold, and thus increases the amount of dust needed to reproduce the SPIRE250 flux in those regions.
We therefore do not recommend to use the HIPE data reduction pipeline for the PACS cameras.

We do not recommend working at PACS160 resolution, unless there is reason
to think that high signal/noise has been achieved.  We note, however, 
that even in regions of moderately high surface brightness, the PACS and
MIPS photometry often disagree at the 30\% level (see Figs.\ \ref{fig:color_0628} and \ref{fig:color_6946}),
making dust modeling risky if PACS provides 
the only data longward of 24$\micron$.
If SPIRE250 data are available, dust modeling can be done at SPIRE250
resolution with results that are reasonably reliable.
We note in Fig.\ \ref{fig:dustmass} that total dust masses obtained using
IRAC, MIPS24, PACS, and SPIRE250 as constraints are in error by only $\approx 35\%$,
 provided the PACS-Scanamorphos data are used.
 
We further note that NGC~6946 is a particularly challenging galaxy to model: it
has over 4 decades of dynamic range in the PACS160 band, including
an extremely high surface brightness nucleus.

\end{document}